\documentclass[a4paper, 9pt, twocolumn, openany]{extbook}

\usepackage[british]{babel}
\usepackage[T1]{fontenc}
\usepackage[ansinew]{inputenc}

\usepackage[outline, light, all]{draftcopy}
\newif\ifdraft

\newif\iffullversion

\ifdefined \pdfsuppresswarningpagegroup
\fi

\usepackage{silence}
\usepackage{color}

\usepackage{hyperref}
\usepackage{bookmark}

\usepackage{lmodern} 

\usepackage{graphicx} 
\usepackage{pdfpages}

\usepackage{amsmath}
\usepackage{amsthm}
\usepackage{amsfonts}
\usepackage{amssymb}
\usepackage{epsfig}
\usepackage{upgreek}
\usepackage{relsize}

\usepackage{cite}
\usepackage{xspace}

\usepackage{fancyhdr}
\usepackage{titlesec}

\usepackage{enumitem}
\usepackage{subcaption}
\usepackage{lineno}

\usepackage{placeins}

\usepackage[toc]{multitoc}

\usepackage{booktabs}		
\usepackage{multirow}   
\usepackage{siunitx}		

\DeclareSIUnit[number-unit-product = {}]\c{c}
\DeclareSIUnit[number-unit-product = {}]\bit{bit}
\DeclareSIUnit[number-unit-product = {}]\byte{byte}
\DeclareSIUnit[number-unit-product = {}]\muon{\ensuremath{\mu}}

\usepackage{caption}		
\usepackage{subcaption}

\usepackage{balance}

\newcommand{\mte}{$\mu \rightarrow eee$\xspace}
\newcommand{\mteSigned}{$\mu^+ \rightarrow e^+e^-e^+$\xspace}
\newcommand{\mtenunu}{$\mu \rightarrow eee\nu\nu$\xspace}
\newcommand{\mtenunuSigned}{$\mu^+ \rightarrow e^+e^-e^+\nu\nu$\xspace}

\newcommand{\mupix}{{\sc MuPix}\xspace}
\newcommand{\mutrig}{{\sc MuTRiG}\xspace}

\titleformat{\chapter}[display]{\sc}{Chapter \thechapter}{10.0pt}{\Huge}

\titleformat{\part}[display]{\sc \begin{center}}{\Large Part \thepart}{35.0pt}{\includegraphics[width=0.3\textwidth]{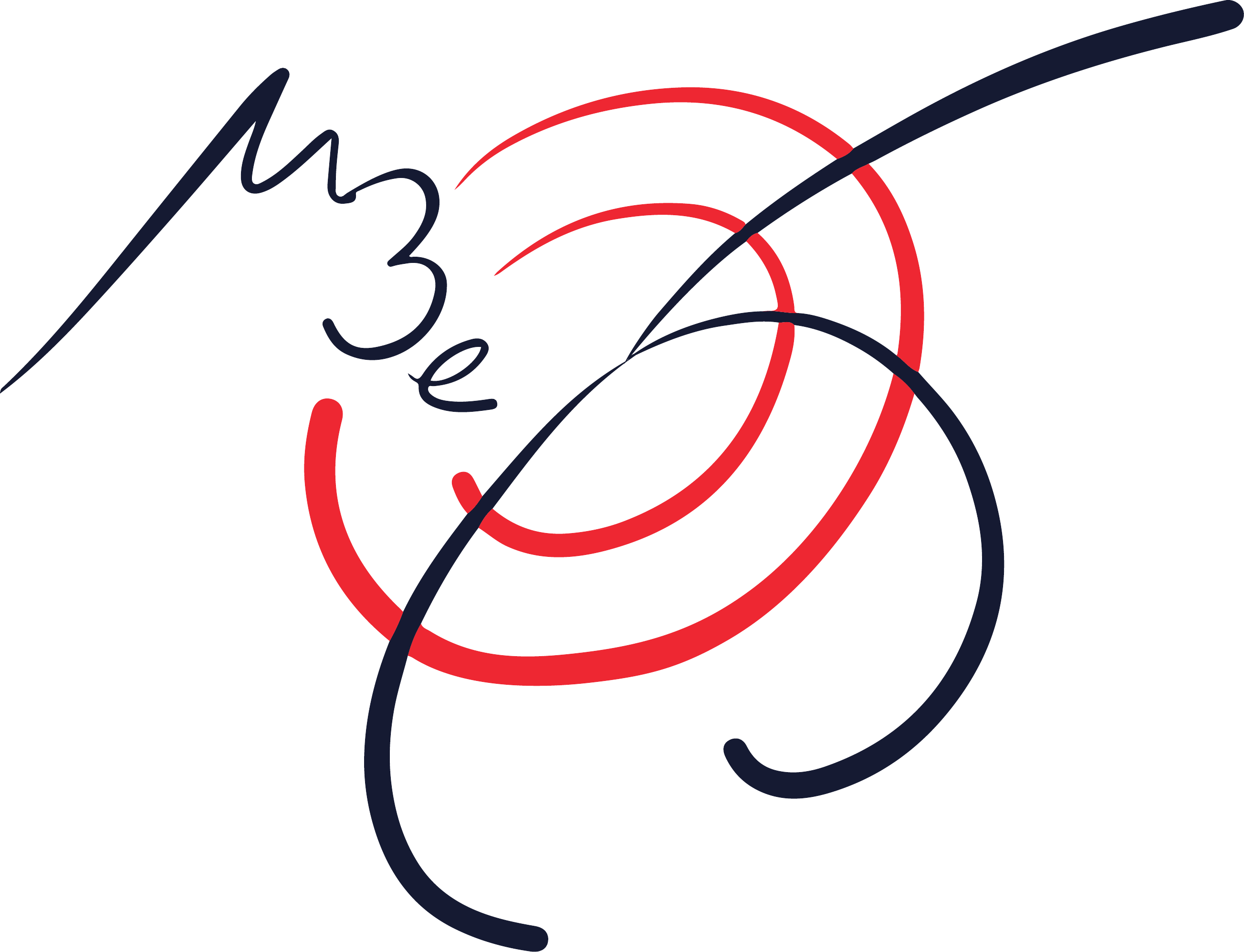}\\ \vspace{1cm} \Huge}[\end{center}]

\titleformat{\section}{}{\large \thesection~}{0cm}{\Large}[]

\titleformat{\subsection}{\sc}{\thesubsection~}{0cm}{}[]

\titleformat{\subsubsection}{\it}{\thesubsubsection~}{0cm}{}[]

\newcommand{\chapterresponsible}[1]{\ifdraft \begin{center}{\sc Responsibility for chapter: }\\ {\bf #1}\end{center} \vspace{1cm} \fi}

\definecolor{seaborndeep1}{rgb}{0.298, 0.447, 0.690}
\definecolor{seaborndeep2}{rgb}{0.333, 0.659, 0.408}
\definecolor{seaborndeep3}{rgb}{0.769, 0.306, 0.322}
\definecolor{seaborndeep4}{rgb}{0.506, 0.447, 0.698}
\definecolor{seaborndeep5}{rgb}{0.800, 0.725, 0.455}
\definecolor{seaborndeep6}{rgb}{0.392, 0.710, 0.804}

\makeatletter

\newcommand\Autoref[1]{\@first@ref#1,@}
\def\@throw@dot#1.#2@{#1}
\def\@set@refname#1{
    \edef\@tmp{\getrefbykeydefault{#1}{anchor}{}}%
    \xdef\@tmp{\expandafter\@throw@dot\@tmp.@}%
    \ltx@IfUndefined{\@tmp autorefnameplural}%
         {\def\@refname{\@nameuse{\@tmp autorefname}s}}%
         {\def\@refname{\@nameuse{\@tmp autorefnameplural}}}%
}
\def\@first@ref#1,#2{%
  \ifx#2@\autoref{#1}\let\@nextref\@gobble
  \else%
    \@set@refname{#1}
    \@refname~\ref{#1}
    \let\@nextref\@next@ref
  \fi%
  \@nextref#2%
}
\def\@next@ref#1,#2{%
   \ifx#2@ and~\ref{#1}\let\@nextref\@gobble
   \else, \ref{#1}
   \fi%
   \@nextref#2%
}

\makeatother

\begin{document}
\let\WriteBookmarks\relax
\def\floatpagepagefraction{1}
\def\textpagefraction{.001}

\onecolumn
\includepdf{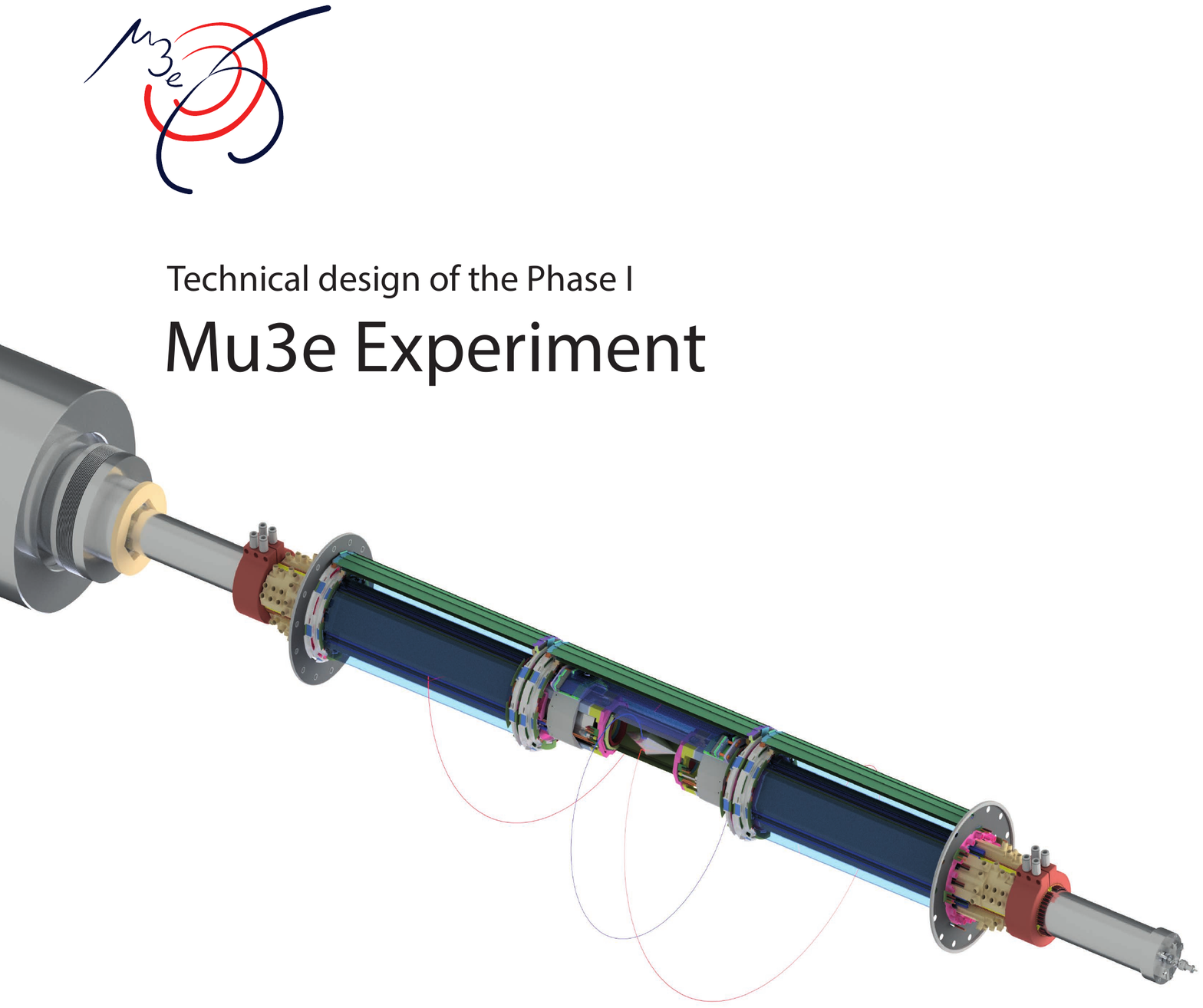}
\clearpage
\begin{center}
{\Huge Technical design of the phase I Mu3e experiment}
\end{center}

\vspace{0.7cm}

\noindent
K.~Arndt$^{\dagger a}$,
H.~Augustin$^b$,
P.~Baesso$^c$,
N.~Berger$^d$,
F.~Berg$^e$,
C.~Betancourt$^f$,
D.~Bortoletto$^a$,
A.~Bravar$^g$,
K.~Briggl$^{g,}$\footnote{Now at Kirchhoff-Institut, Universit\"at Heidelberg},
D.~vom~Bruch$^{d,}$\footnote{Now at Universit\'e de Marseille, France}
% ...

\vspace{0.7cm}

{\footnotesize
\begin{flushleft}
\noindent
$^a $Department of Physics, University of Oxford, Denys Wilkinson Building, Keble Road, Oxford~OX1~3RH, United Kingdom

\noindent
$^b $Physikalisches Institut, Universit\"at Heidelberg, Im Neuenheimer Feld~226, 69120~Heidelberg, Germany

\noindent
$^c $University of Bristol, H.H. Wills Physics Laboratory, Tyndall Avenue, Bristol~BS8~1TL, United Kingdom

\noindent
$^d $Institut f\"ur Kernphysik und Exzellenzcluster PRISMA$^+$, Johannes Gutenberg-Universit\"at Mainz, Johann-Joachim-Becher-Weg~45, 55128~Mainz, Germany

\noindent
$^e $Institute for Particle Physics and Astrophysics, Eidgen\"ossische Technische Hochschule Z\"urich, Otto-Stern-Weg 5, 8093~Z\"urich, Switzerland

\noindent
$^f $Physik-Institut, Universit\"at Z\"urich, Winterthurerstrasse 190, 8057~Z\"urich, Switzerland

\noindent
$^g $D\'epartement de physique nucl\'eaire et corpusculaire, Universit\'e de Gen\`eve, 24, quai Ernest-Ansermet, 1211~Gen\`eve 4, Switzerland

\noindent
$^h $Department of Physics, University of Liverpool, The Oliver Lodge Laboratory, Liverpool~L69~7ZE, United Kingdom

\noindent
$^i $Kirchhoff-Institut f\"ur Physik, Universit\"at Heidelberg, Im Neuenheimer Feld 227, 69120~Heidelberg, Germany

\noindent
$^j $Institut f\"ur Prozessdatenverarbeitung und Elektronik, Karlsruhe Institut f\"ur Technologie, Hermann-von-Helmholtz-Platz 1, 76344~Eggenstein-Leopoldshafen, Germany

\noindent
$^k $Laboratory for Particle Physics, Paul Scherrer Institut, Forschungsstrasse 111, 5232~Villigen, Switzerland

\noindent
$^l $Department of Physics and Astronomy, University College London, Gower Street, London~WC1E~6BT, United Kingdom

\end{flushleft}
}
\vspace{0.5cm}

\par\noindent\rule{\textwidth}{0.5pt}

\medskip

\noindent
\textbf{Abstract}

The Mu3e experiment aims to find or exclude the lepton flavour violating decay
\mte at branching fractions above \num{e-16}. A first phase of the experiment
using an existing beamline at the Paul Scherrer Institute (PSI) is designed
to reach a single event sensitivity of \num{2e-15}. We present an overview
of all aspects of the technical design and expected performance of the phase~I Mu3e detector.
The high rate of up to \num{e8} muon decays per second and the low momenta of the decay electrons
and positrons pose a unique set of challenges, which we tackle using an
ultra thin tracking detector based on high-voltage monolithic active pixel sensors
combined with scintillating fibres and tiles for precise timing measurements.



\medskip

\par\noindent\rule{\textwidth}{0.5pt}

\vspace{1cm}
\noindent
Spokespersons:
S.~Ritt (\url{stefan.ritt@psi.ch}),
A.~Sch\"oning (\url{schoning@physi.uni-heidelberg.de})

\noindent
Technical coordinator and editor:
F.~Meier~Aeschbacher (\url{frank.meier@psi.ch}),

\twocolumn
\newpage


\onecolumn
\vspace*{\fill}

\par\noindent\rule{\textwidth}{2pt}

\begin{center}
\textbf{Kirk Arndt}

\bigskip

\emph{Silicon Detector Development Engineer  (1959-2019)}
\end{center}

\bigskip

Kirk Arndt was a silicon detector development engineer. He came to Europe in 2014 after an illustrious career in the U.S. where he became known as ``the best pair of hands in the silicon business''. He built some remarkable silicon digital cameras for particle physics at the University of California Santa Barbara, Purdue University, and Oxford including one that glimpsed the Higgs particle for the first time at the LHC at CERN in 2012. These were the Silicon Micro-Vertex Detector for the CLEO II.V experiment at the Cornell Electron Storage Ring with UCSB, the CLEO III Silicon Vertex Detector and the CMS Phase 0 and Phase 1 Silicon Forward Pixel Detectors for the CMS experiment at the LHC at CERN with Purdue, and he was working on the ATLAS Upgrade Silicon Forward Pixel Detector at Oxford, the Mu3e tracker and detectors for photon science at the time of his death. For over a decade he also played an important role in the design of the 3.2~Gigapixel camera for the Large Synoptic Survey Telescope both at Purdue and Oxford.

Kirk was a very highly valued and widely appreciated colleague. He was always ready to help colleagues in the CLEO, CMS, LSST, ATLAS and Mu3e collaborations and many others in the international community who sought him out for advice. His positive can do attitude, exacting professional standards, dedication, willingness to nurture younger colleagues and his kindness are an example to us all. Shortly before his unexpected death at the age of 59, he hosted a Mu3e collaboration meeting in Oxford. It was a privilege to work with Kirk. He will live on in the hearts and minds of all that knew him.

\par\noindent\rule{\textwidth}{2pt}

\vspace*{\fill}


\twocolumn
\pagestyle{fancy}

\setcounter{tocdepth}{1}
\tableofcontents{}

\newpage

\chapter[Muon decay to three electrons and the Experimental Challenge]{The Decay {\mte} and the Experimental Challenge}
\label{sec:DecayMu3e}

\chapterresponsible{Nik, Gavin}
	
\section{Goals of the Experiment}
\label{sec:GoalsOfTheExperiment}
	
	The goal of the Mu3e experiment is to observe the process \mteSigned
	\footnote{We will refer to this decay as \mte in this document; the
	charge of the muon is always positive.}
        if its branching fraction is larger than $\num{e-16}$, or
        otherwise to exclude a branching fraction of $>\num{e-16}$ at
        the $\SI{90}{\percent}$ confidence level.  In order to achieve
        these goals, more than $\num{1e17}$
muons have to be stopped in the detector (assuming a total reconstruction
 efficiency of $\SI{20}{\%}$) and any
background mimicking the signal process suppressed to below the
$\num{e-16}$ level. The additional requirement of achieving these
goals within a reasonable measurement time of one year of data taking
dictates a muon stopping rate of $\SI{2e9}{\Hz}$, along with a high
geometrical acceptance and efficiency for the experiment.

The current best source of low-energy muons in the world, the
$\pi$E5 beam line at PSI, provides muon rates up to $\SI{1e8}{\Hz}$.
Higher intensities are possible and currently under study in the
\emph{high intensity muon beam} (HiMB) project.  However, the new
beamlines will not be available before 2025.  In order to establish
the novel technologies for Mu3e, set first competitive limits and
prepare for the very high intensity running, we plan to run a phase~I
experiment at $\pi$E5. The aim of this phase~I experiment is a single
event sensitivity of $\num{2e-15}$ on the branching fraction, which
would require $>\num{2.5e15}$ stopped
muons\footnote{$N_{\textrm{required}} = 1/(s \cdot \epsilon)$ for a
  sensitivity $s$ and a total efficiency $\epsilon \approx
  \SI{20}{\%}$ (phase~I)} or $\SI{2.5e7}{\s}$ (290~days) of run time
at $\SI{1e8}{\Hz}$ stopping rate. The present document describes the technical design of this
phase~I detector.

 For more on the physics motivation and theory predictions, please
 consult the Mu3e letter of intent \cite{LOI} and research proposal
 \cite{RP}. This chapter describes the kinematics of the signal and the
 main background sources, and how these motivate the design of the
 experiment. Running with $\SI{1e8}{\Hz}$ of muon decays also poses
 challenges for the detectors, the data acquisition and the
 readout, which will be discusses in later chapters.

\section{Signal Kinematics}
To discriminate the signal from the background, energy and momentum conservation
are exploited. 
The decaying muons are at rest and the three decay particles are emitted at the same time.
The vectorial sum of all decay particle momenta $\vec{p}_i$ 
should therefore vanish:
\begin{linenomath*}
\begin{equation}
\left|\vec{p}_{tot}\right| \; = \; \left|\sum \vec{p}_i\right| \; = \; 0  \quad
\end{equation}
\end{linenomath*}
and the invariant mass, which is equal to the sum of the energies in the case of vanishing momentum, be equal to the muon mass:
\begin{linenomath*}
\begin{equation}
m_{inv} \; = \; \left|\sum p_i\right| \;  = \; \sum E_i \; = \; m_\mu . \quad
\end{equation}
\end{linenomath*}

The energies of the decay particles range from the electron mass
up to half the muon mass, which is about $\SI{53}{\mega\electronvolt}$.
All decay particles must lie in a plane. Therefore, the  decay is described by two
independent variables in addition to three global rotation angles describing 
the orientation in space.

\section{Modelling of the Signal}
\label{sec:ModellingOfTheSignal}

The decay dynamics for the \mte signal are dependent on the unknown lepton
flavour violating (LFV) mechanism.
We typically assume a phase-space distribution for the signal electrons in our
simulations, if not stated otherwise.
In order to study effects of different decay dynamics, we utilise the general
parametrised Lagrangian proposed by Kuno and Okada \cite{Kuno:1999jp}:
\begin{linenomath*}
\begin{equation} \label{eq:lagrangian}
\begin{split}
L_{\mu \rightarrow eee} \;  = & \; \; 
- \frac{4G_F}{\sqrt{2}}   \left[ 
  m_\mu A_R \; \overline{\mu_R} \sigma^{\mu \nu} e_L F_{\mu \nu} \right.\\
& + m_\mu A_L \; \overline{\mu_L} \sigma^{\mu \nu} e_R F_{\mu \nu}  \\
 & +  \;g_1 \; (\overline{\mu_R} e_L) \; (\overline{e_R} e_L) \\
 & + g_2 \; (\overline{\mu_L} e_R)\; (\overline{e_L}e_R)  \\
 & +  \;g_3 \;  (\overline{\mu_R} \gamma^\mu e_R) \; (\overline{e_R} \gamma_\mu e_R) \\
 & + g_4 \; (\overline{\mu_L} \gamma^\mu e_L) \; (\overline{e_L} \gamma_\mu e_L)  \\
 & + \; g_5 \;  (\overline{\mu_R} \gamma^\mu e_R) \; (\overline{e_L} \gamma_\mu e_L) \\
 & + g_6 \;\left. (\overline{\mu_L} \gamma^\mu e_L) \; (\overline{e_R} \gamma_\mu e_R) \; + \; H.c.
\; \right] 
\end{split}
\end{equation}
\end{linenomath*}
The form factors $A_{R,L}$ describe tensor type (dipole) couplings, mostly acquiring 
contributions from the photon penguin diagram, whereas the scalar-type ($g_{1,2}$) and
vector-type ($g_{3}-g_{6}$) form factors can be regarded as four fermion contact
interactions, to which the tree diagram contributes at leading order.
We generate different signal models by varying the relative strengths of the
$A_{R,L}$ and $g_{1}-g_{6}$ parameters.

\section{Signal Acceptance}

\begin{figure}
	\centering
		\includegraphics[width=\columnwidth]{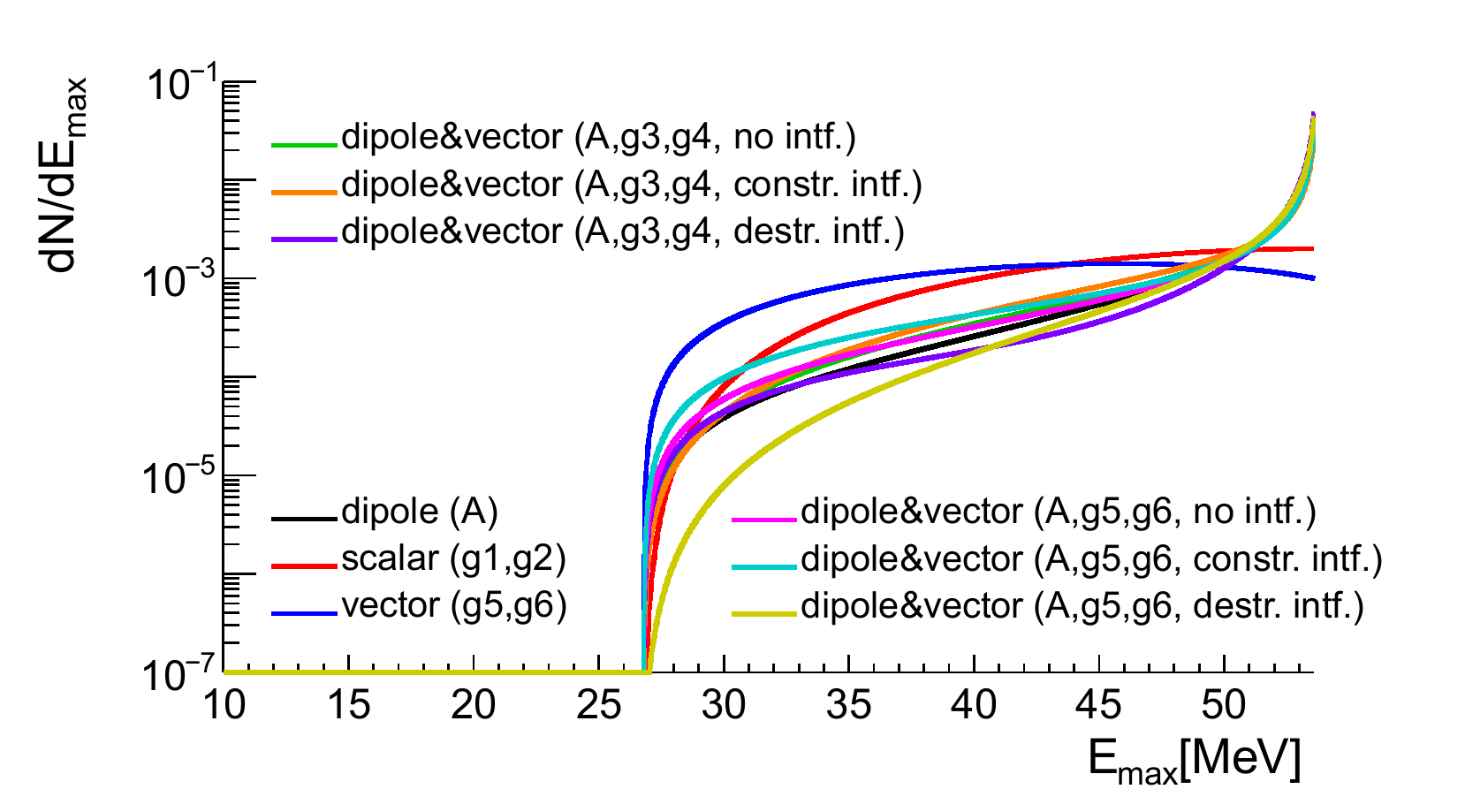}
	\caption{Energy distribution of the highest energy decay particle in the
          decay \mte for different
effective LFV models. The black line corresponds to pure dipole and the red and
blue line to pure four-fermion contact interaction models (no penguin
contribution); the other lines correspond
to a mixture of dipole and vector interactions. Based on \cite{Kuno:1999jp}.} 
	\label{fig:brmee_x1_10.norm}
\end{figure}

\begin{figure}[tb!]
	\centering
		\includegraphics[width=\columnwidth]{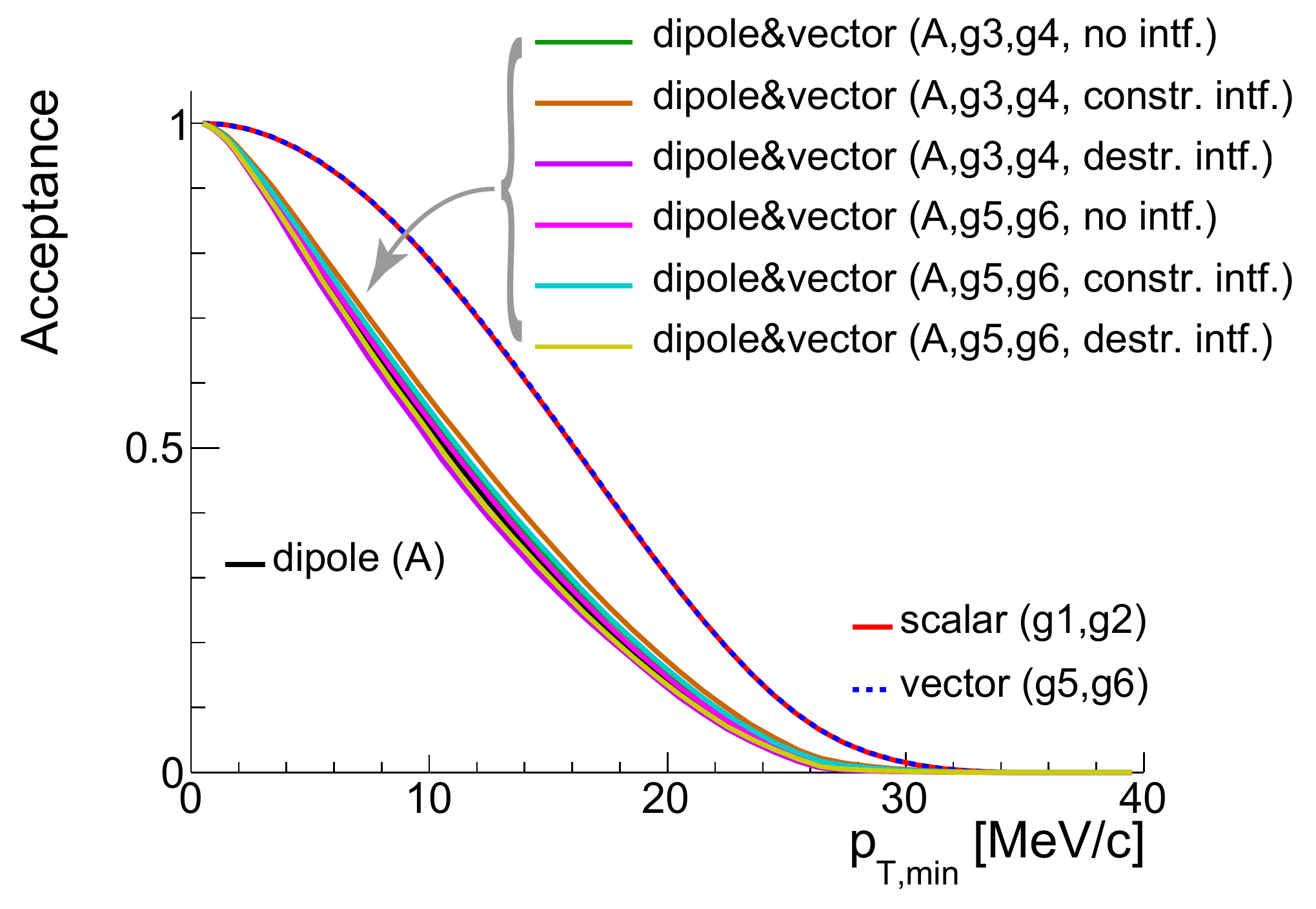}
	        \caption{The acceptance, defined as the fraction of \mte decays in which
                  all decay products have transverse momenta above
                  $p_{T,min}$, against $p_{T,min}$ for different effective LFV
                  models. The black line corresponds to pure dipole
                  and the red and blue dotted lines (on top of each other) to pure four-fermion
                  contact interaction models (no penguin
                  contribution); the other lines correspond to a
                  mixture of dipole and vector interactions. Based on
                  \cite{Kuno:1999jp}.}
	\label{fig:brmee_ptmin_10.summed}
\end{figure}

For a three-body decay with a priori unknown kinematics such as \mte,
the acceptance has to be as high as possible in order to test new
physics in all regions of phase space. To illustrate the phase space
coverage needed, the energy spectrum of the highest energy decay
particle ($E_{max}$) for various LFV coupling amplitudes is shown in
\autoref{fig:brmee_x1_10.norm}, and the fraction of events where
all decay particles have transverse momenta above $p_{T,min}$ is shown in
\autoref{fig:brmee_ptmin_10.summed}. From these figures, it can be
seen that a high acceptance for the signal is only possible if the
detector can reconstruct tracks with momenta ranging from half the
muon mass down to a few $\si{\mega\electronvolt}$. This must be
achieved with large solid angle coverage, limited by the beam entry
and exit points preventing instrumentation.

\section{Backgrounds}

\begin{figure}
	\centering
		\includegraphics[width=0.48\textwidth]{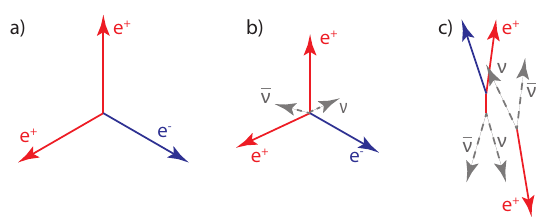}
	\caption{Topologies for a) the signal decay, b) the internal conversion decay
	and c) combinatorial background from two Michel decays with a Bhabha scattering. }
	\label{fig:topologies}
\end{figure}
	
\begin{figure}
	\centering
		\includegraphics[width=0.48\textwidth]{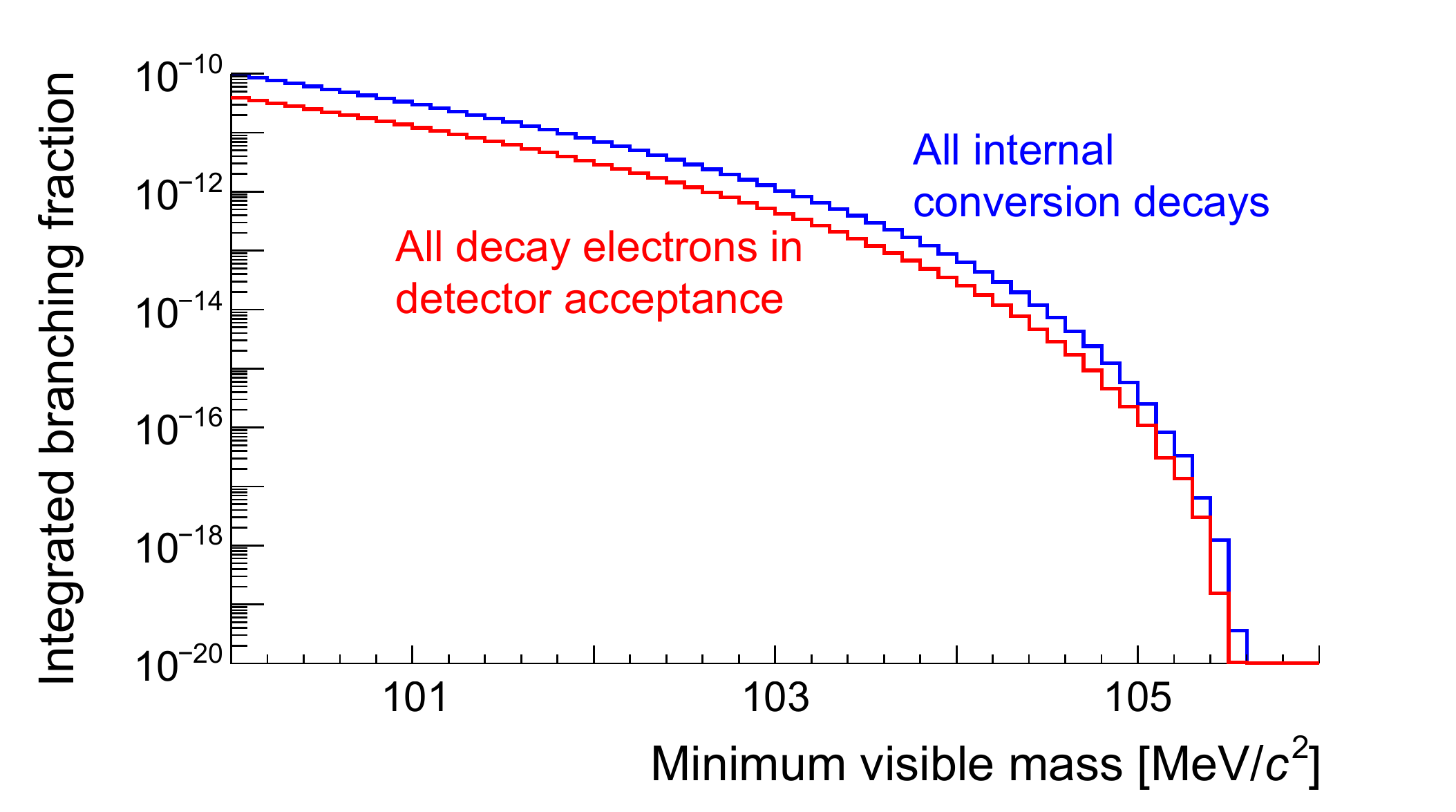}
	\caption{Integrated branching fraction of the decay \mtenunu
          for which the invariant mass of the three decay electrons
          lies above the $x$ axis value. This is shown for all
          internal conversion decays (blue line) and those with all
          three decay particles in the detector acceptance, defined as
          $E >\SI{10}{MeV}$ and $|\cos\theta| < 0.8$ (red line).  The
          matrix element was taken from \cite{Djilkibaev:2008jy}.}
	\label{fig:icfraction}
\end{figure}
	
The Standard Model branching fraction for the \mte process is
\num{2.9E-55} (normal neutrino mass ordering) or \num{4.6E-55}
(inverted ordering) \cite{Hernandez-Tome:2018fbq,Blackstone:2019njl}; the experiment
therefore has no irreducible physics backgrounds, and the final sensitivity
depends purely on the ability to reduce backgrounds in two
categories: overlays of different processes producing three tracks
resembling a \mte decay (\emph{combinatorial background}) and
radiative decays with internal conversion (\emph{internal conversion
  background}) with a small energy fraction carried away by the
neutrinos; see \autoref{fig:topologies} for the signal and background topologies.  
Combinatorial backgrounds have to be suppressed via
vertexing, timing and momentum measurement; momentum measurement is
the only handle on internal conversion. In the following sections, these
main background sources are discussed.

\subsection{Internal Conversions}

\begin{figure}
	\centering
		\includegraphics[width=0.48\textwidth]{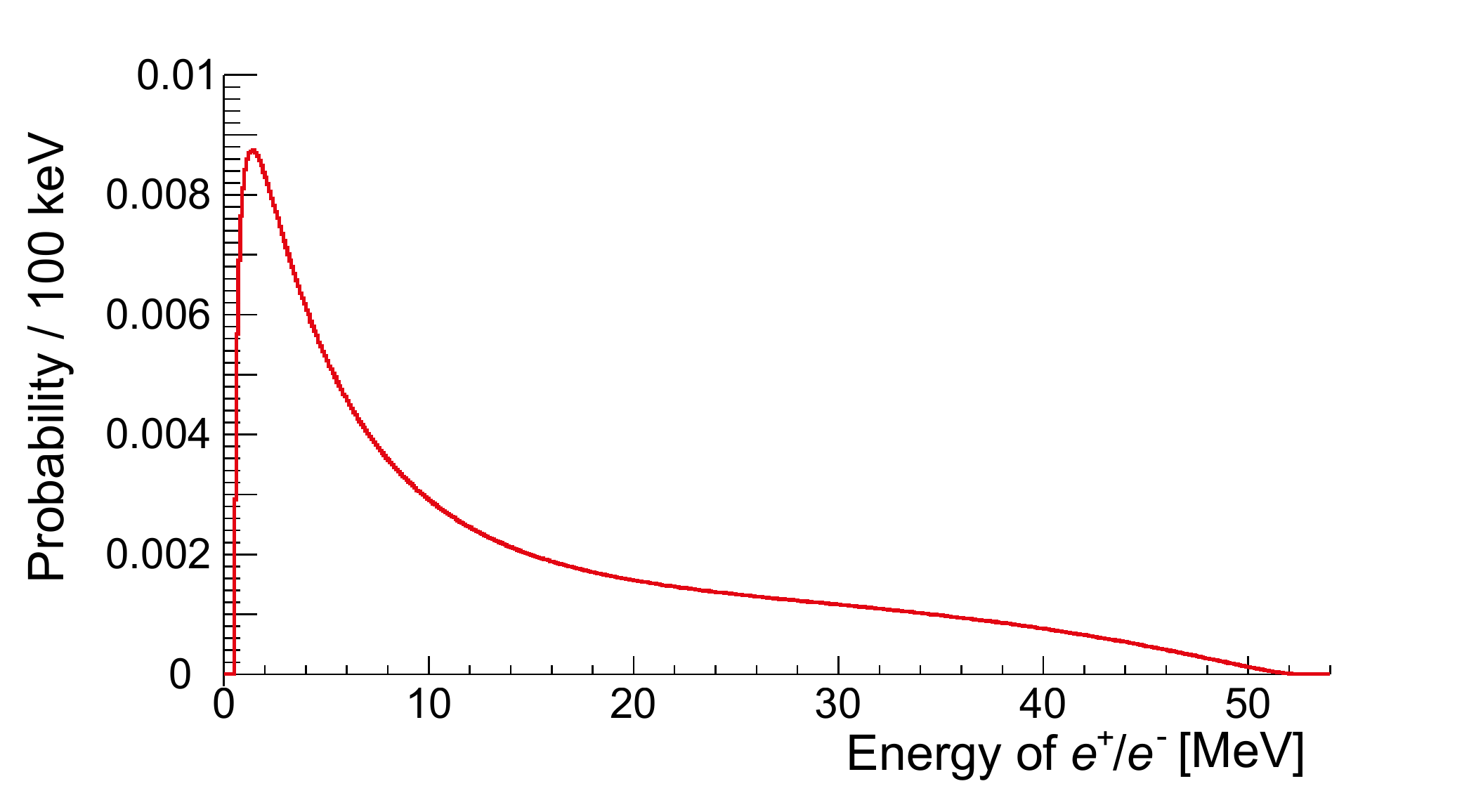}
	\caption{Energy spectrum of all electrons and positrons from internal conversion decays.}
	\label{fig:ic_e_spectrum}
\end{figure}

\begin{figure}
	\centering
		\includegraphics[width=0.48\textwidth]{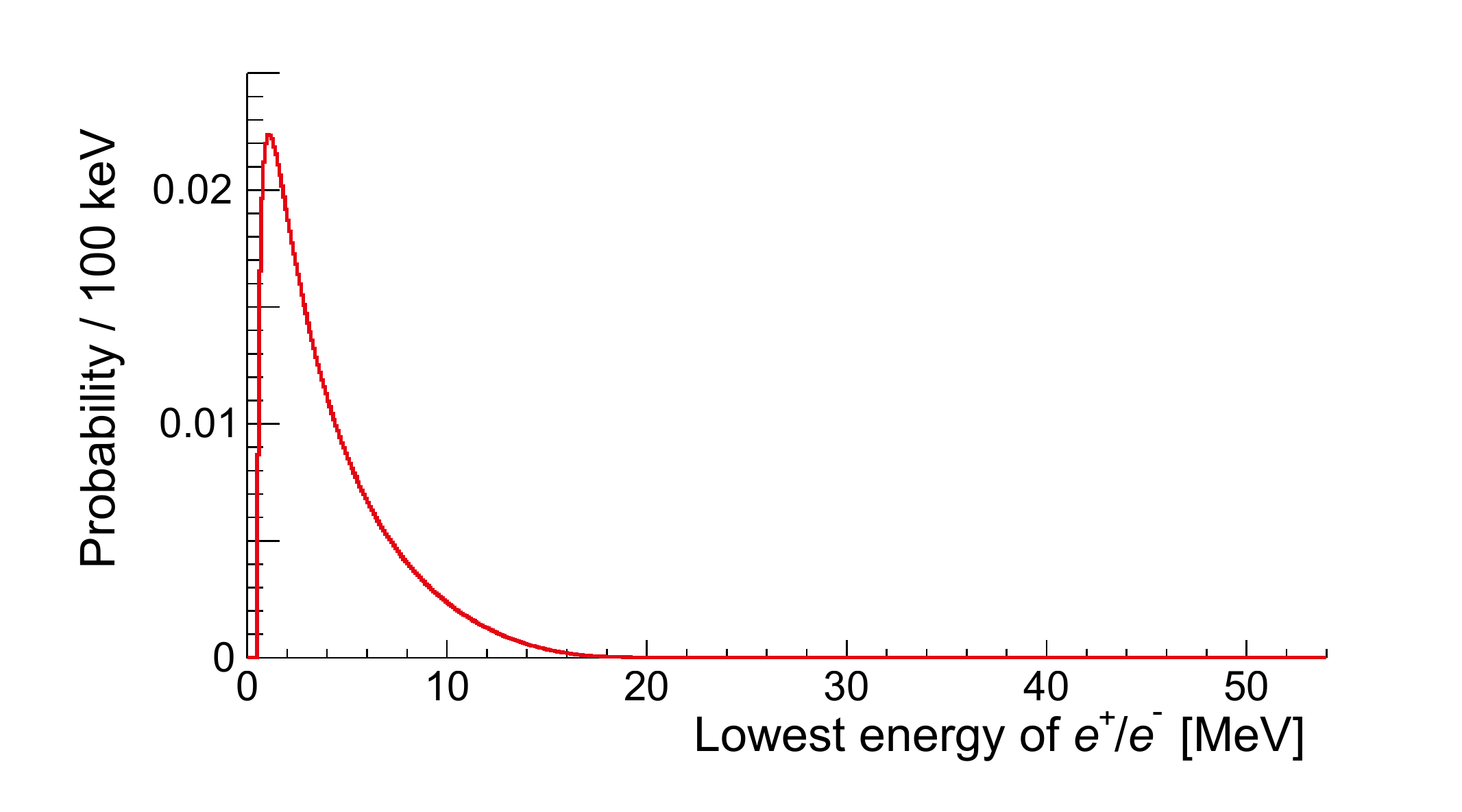}
	\caption{Energy spectrum of the electron or positron with lowest energy from internal conversion decays.}
	\label{fig:ic_emin_spectrum}
\end{figure}

The decay \mtenunu occurs with a branching fraction of $\num{3.4e-5}$
\cite{Bertl19851}. It can be distinguished from the \mte process by
making use of energy and momentum conservation to infer the presence
of the undetected neutrinos: in order to separate the \mte events from
\mtenunu events, the total momentum in the event is required to be
zero and the visible mass (defined as the invariant mass of the three
electrons) equal to the muon rest energy. The branching fraction for \mtenunu \cite{Djilkibaev:2008jy}
decays above a given visible mass value  is shown in \autoref{fig:icfraction}.
\Autoref{fig:ic_e_spectrum,fig:ic_emin_spectrum} show
the energy spectrum of all and the lowest energy electron from
\mtenunu decays
calculated with the matrix element from
\cite{Djilkibaev:2008jy}.
Recently, next-to-leading order (NLO) calculations of the internal
conversion decays have become available \cite{Pruna:2016spf,
  Fael:2016yle}; they predict branching fractions close to the
end-point that are about \SI{10}{\percent} smaller than the leading order (LO)
prediction.

\begin{figure}
	\centering
		\includegraphics[width=0.48\textwidth]{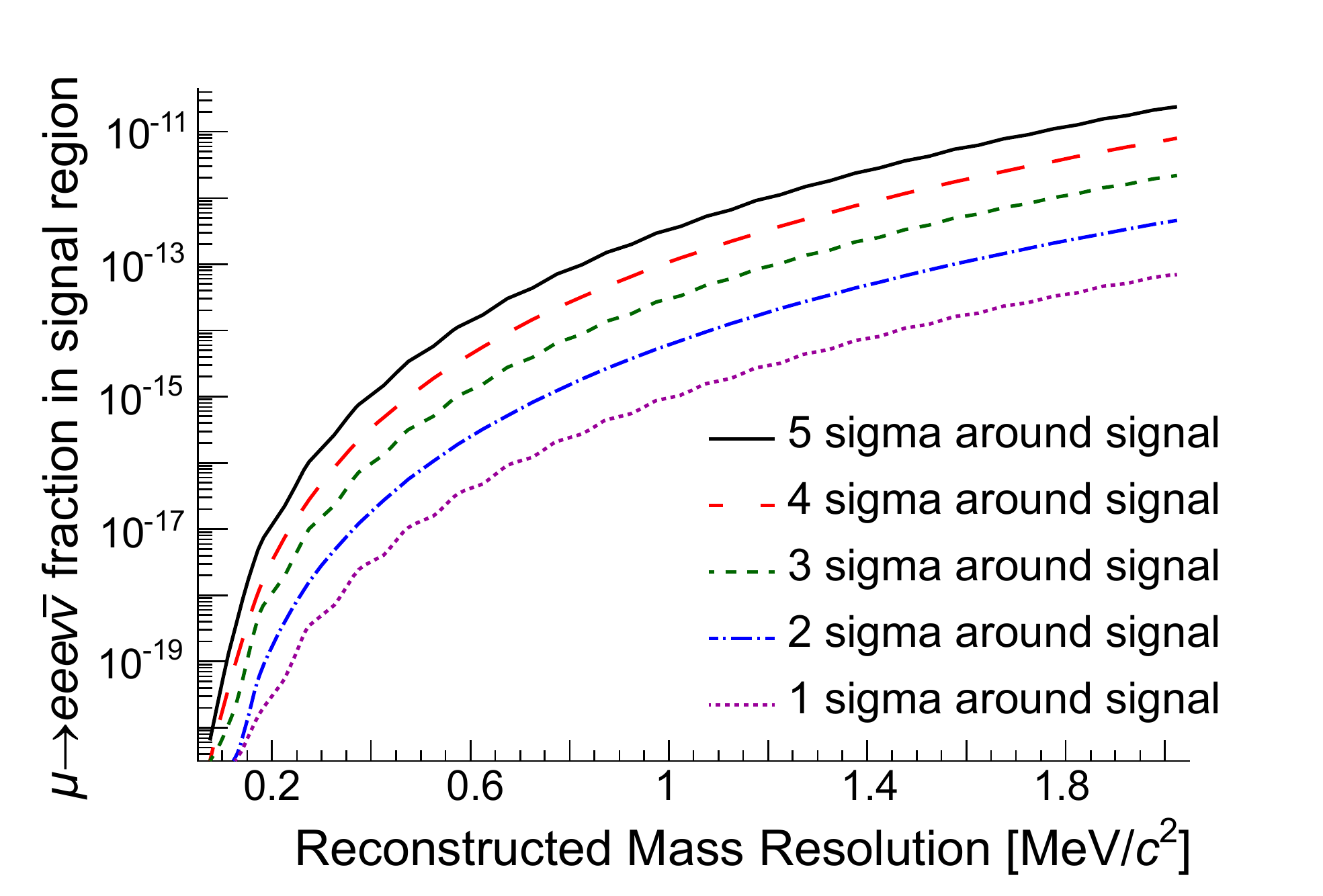}
	\caption{Contamination of the signal region (one sided cut) with internal conversion 
	events as a function of momentum sum resolution.}
	\label{fig:ic_contamination}
\end{figure}
		
Internal conversion is the most serious background for the \mte
search, and the momentum resolution directly determines to what level
it can be suppressed and thus the experiment run background free. In
order to reach a sensitivity of $\num{2e-15}$ with a $2 \sigma$ cut on
the reconstructed muon mass, the average three-particle mass resolution has to be
better than $\SI{1.0}{MeV}$, see \autoref{fig:ic_contamination}, and
is not allowed to have sizeable tails towards the high momentum side.

\subsection{Combinatorial Backgrounds}
\subsubsection{Michel Decays}

Using a beam of positive muons, one of the main processes contributing
to combinatorial background is that of the Michel decay $\mu^+
\rightarrow e^+ \nu_\mu \bar{\nu}_e$. This process does not produce a
negatively charged particle (electron), so it can only contribute as a
background in combination with an incorrectly reconstructed track, or
with other processes that ``naturally'' provide negatively charged
particles.

The rate of fake electron tracks being reconstructed from those
positrons performing complete turns in the magnetic field is reduced by a
reliable determination of the direction of motion of particles:
achieved by accurate curvature measurements and accurate time of flight
measurements. The main sources of genuine negatively charged particles
are Bhabha scattering and radiative decays.

\subsubsection{Radiative Muon Decays}

The process $\mu^+ \rightarrow e^+ \gamma \nu \bar{\nu}$ (branching
fraction $\num{1.4e-2}$ for photon energies above
$\SI{10}{\mega\electronvolt}$ \cite{Crittenden1961}) can deliver a
negatively charged electron if the photon converts either in the
target region or in the detector.  Conversions in the target region
generate an event topology similar to the radiative decay with
internal conversion \mtenunu discussed above. Contributions from
conversions outside the target region are greatly suppressed both by a
vertex constraint and by minimising the material in both the target
and detector. However, this process can still contribute to the
combinatorial background in combination with an ordinary muon decay.

As for the internal conversion background, a NLO calculation for
radiative decay has recently been published \cite{Fael:2015gua} and is
implemented in the Mu3e simulation.

\subsubsection{Bhabha Scattering}

Any positron, either from a muon decay in the target or in the beam, can
undergo Bhabha scattering with electrons in the target material,
leading to an electron-positron pair from a common vertex. In
combination with a positron from a Michel decay, this can mimic a
signal decay.  In addition, Bhabha scattering is the main source of
electrons for combinatorial background involving two Michel
decays. Similarly to the external photon conversion background, the
amount of Bhabha scattering is reduced by minimising both the amount,
and the average atomic number of the material in the target.

\subsubsection{Vertex and Timing Resolution Requirements}

Separating vertices from different muon decays is a key tool in
suppressing combinatorial background. The vertex position resolution is
essentially determined by the amount of multiple scattering (and thus
material) in the innermost detector layer and the stopping target as
well as the average distance between the vertex and the first detector
layer.

At high muon rates, good time resolution is essential for reducing
combinatorial background, while also facilitating event reconstruction.
The combinatorial background has a component scaling linearly with the
rate ($e^+e^-$ pair plus a Michel positron) and a component quadratic
in the rate (electron plus two Michel positrons).  The suppression of
these components by timing measurements is also linear and quadratic
in the timing resolution.  Simulation studies have shown that the
linear part dominates at rates at least up to \num{2e9} muon stops
per second.  The requirement of reducing the combinatorial background by at
least two orders of magnitude puts very tight demands on the
resolution of the timing detectors.  The timing resolution should be
below \SI{500}{ps} per track to allow for reliable charge
identification by time-of-flight and ideally \SI{100}{ps} or better to
identify non-synchronous muon decays.

\subsection{Other Backgrounds}
\subsubsection{Pion Decay}
\label{sec:PionDecay}

Certain pion decays, especially $\pi \rightarrow eee \nu$ (branching
fraction $\num{3.2e-9}$ \cite{Egli1989}) and $\pi \rightarrow \mu
\gamma \nu$ (branching fraction $\num{2.0e-4}$ \cite{Bressi1998}) with
subsequent photon conversion are indistinguishable from signal events
if the momenta of the final state particles fit the muon mass
hypothesis. The low pion contamination in the muon beam delivered to
the experiment ($< \num{2e-7}$ of the muon flux), in addition to the 
small branching fractions, lead to
negligible rates for these backgrounds.

\subsubsection{Mis-reconstruction}
\label{sec:MisrecBG}

\begin{sloppypar}
Mis-reconstruction of tracks (e.g.~from hits created by different
particles or noise hits) combined with real tracks from muon decays
can fake \mte decays.  Great care is taken in the track reconstruction
algorithms to keep a minimal rate of fake tracks, balanced against
reconstruction efficiency.
\end{sloppypar}

\subsection[Background Summary]{Summary}

The sensitivity aims of the Mu3e experiment place strict requirements
on the experimental design. Electrons and positrons must be
reconstructed down to a few $\si{\mega\electronvolt}$ with large solid
angle coverage, running at a rate of $\SI{1e8}{\Hz}$ of muon
decays. The material in the target and detector must be minimised, while
achieving excellent momentum, vertex and timing resolution to suppress
backgrounds to the necessary level. The following chapters will
provide a description of the experimental apparatus and procedures that will 
allow this is to be achieved.


\chapter{Experimental Concept}
\label{sec:Concept}

\chapterresponsible{Andr\'e and Nik}

\nobalance

\begin{figure}[b!]
	\centering	\includegraphics[width=0.35\textwidth]{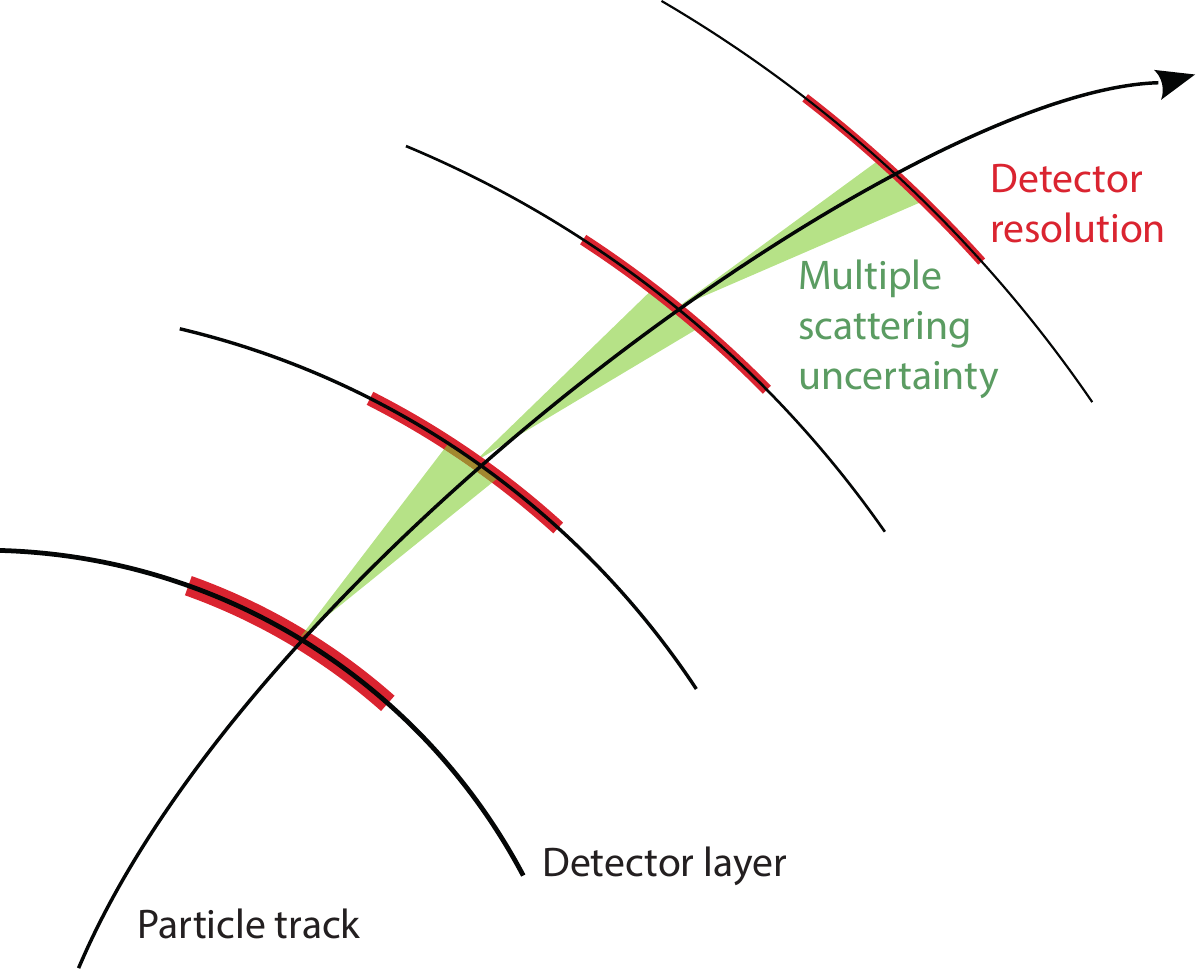}
	\caption{Tracking in the spatial resolution dominated regime}
	\label{fig:tracking_spatial_regime}
\end{figure}

Phase~I of the Mu3e experiment aims for the background free
measurement or exclusion of the branching fraction for the decay \mte
at the level of $\num{2e-15}$.  As discussed in more detail in 
\autoref{sec:DecayMu3e}, these goals require running at high muon rates,
excellent momentum resolution to suppress background from internal
conversion decay (\mtenunu), and a good vertex and timing resolution
to suppress combinatorial background. The present chapter introduces
the conceptual design of the Mu3e experiment, driven by these requirements.

The momenta of electrons and positrons from muon decays are measured using a silicon
pixel tracker in a solenoidal magnetic field.  At the
energies of interest, multiple Coulomb scattering in detector material
is the dominating factor affecting the momentum resolution.
Minimising the material in the detector is thus of the utmost
importance.

\begin{figure}[b!]
	\centering
		\includegraphics[width=0.35\textwidth]{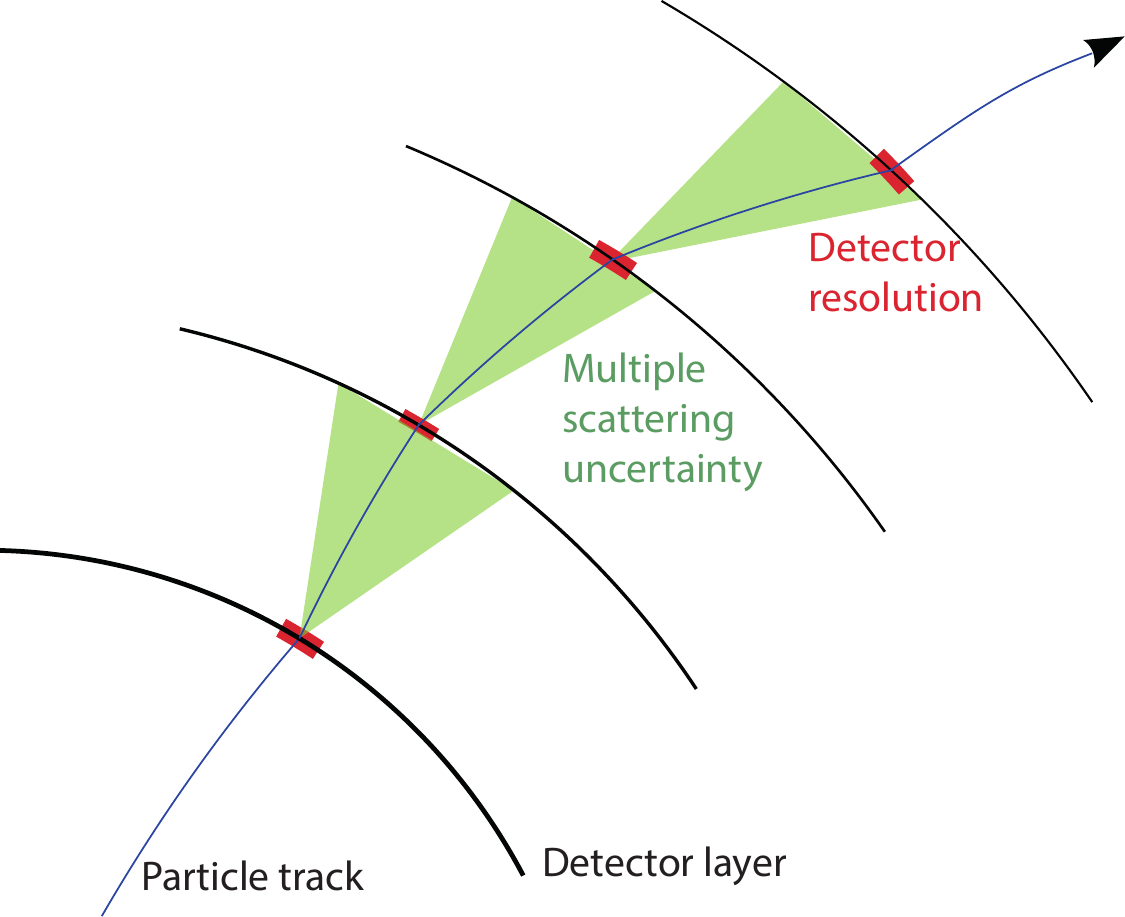}
	\caption{Tracking in the scattering dominated regime}
	\label{fig:tracking_scattering_regime}
\end{figure}

The detector consists of an ultra-thin silicon pixel tracker, made
possible by the High-Voltage Monolithic Active Pixel (HV-MAPS)
technology (see \autoref{sec:Pixel}).  Just four radial layers of
HV-MAPS sensors around a fixed target in a solenoidal magnetic field
allow for precise momentum and vertex determination. Two timing
detector systems guarantee good combinatorial background suppression
and high rate capabilities.
 
\begin{figure}[t!]
	\centering
		\includegraphics[width=0.3\textwidth]{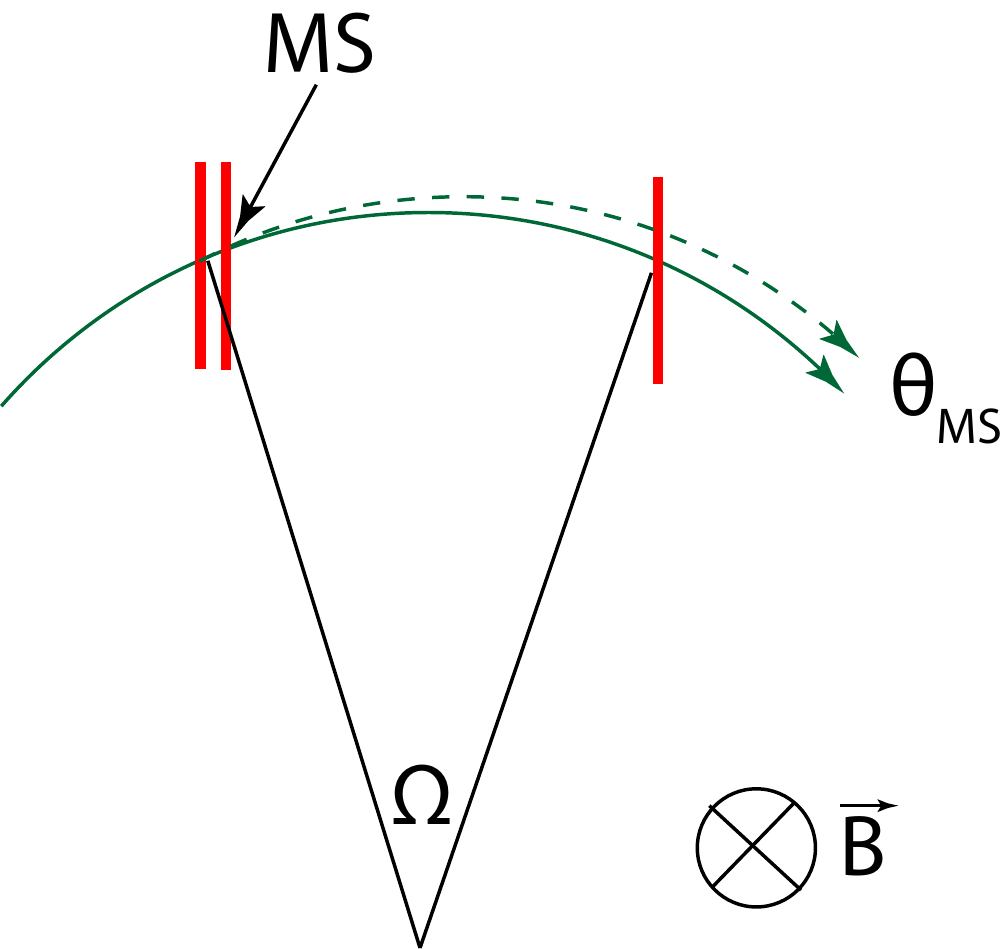}
	\caption{Multiple scattering as seen in the plane transverse to the magnetic field direction. The red lines indicate measurement planes.}
	\label{fig:MS}
\end{figure}

\begin{figure}[t!]
	\centering
		\includegraphics[width=0.3\textwidth]{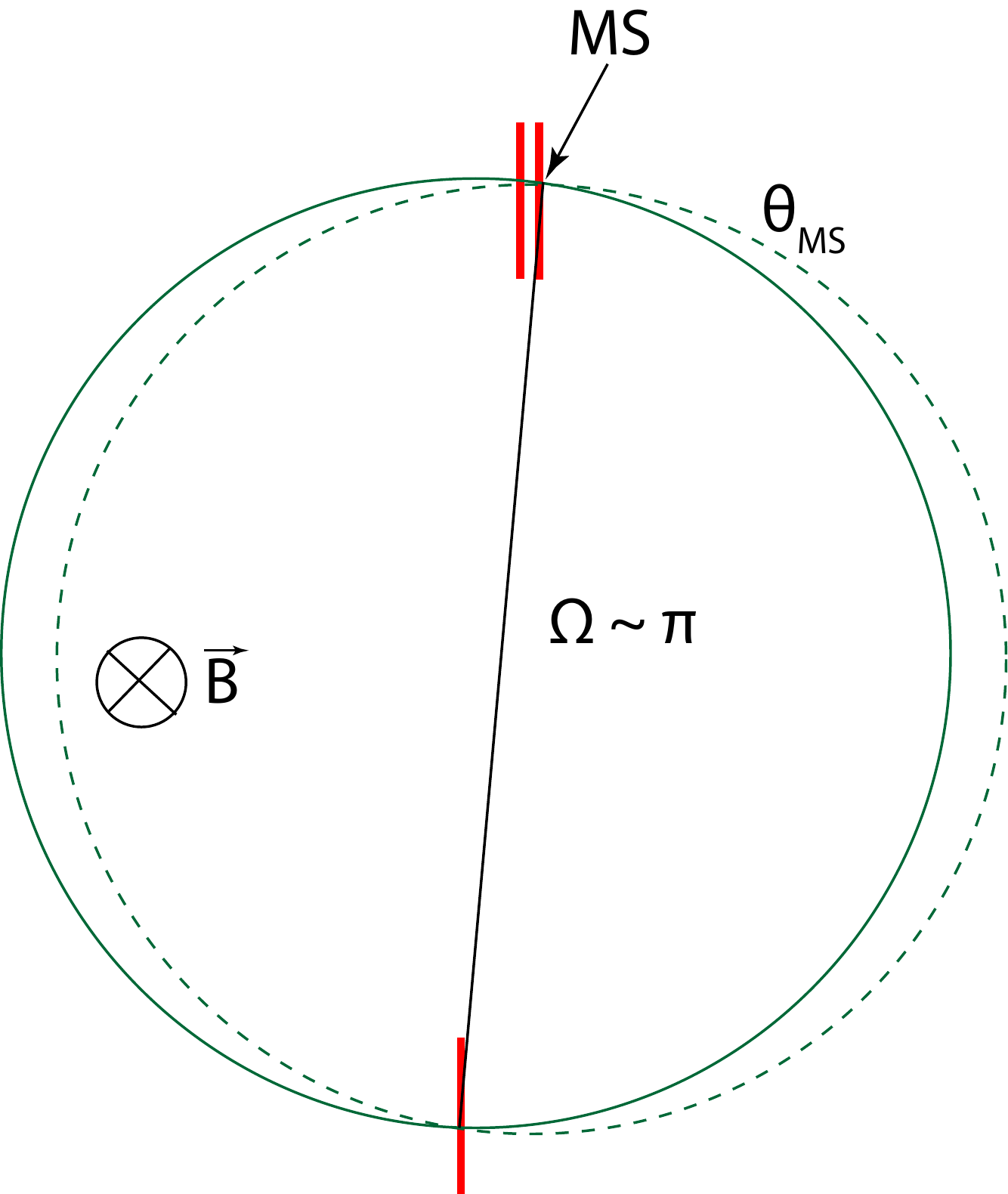}
	\caption{Multiple scattering for a semi-circular trajectory. The red lines indicate measurement planes.}
	\label{fig:MScircle}
\end{figure}

\begin{figure}[tb!]
	\centering
		\includegraphics[width=0.8\columnwidth]{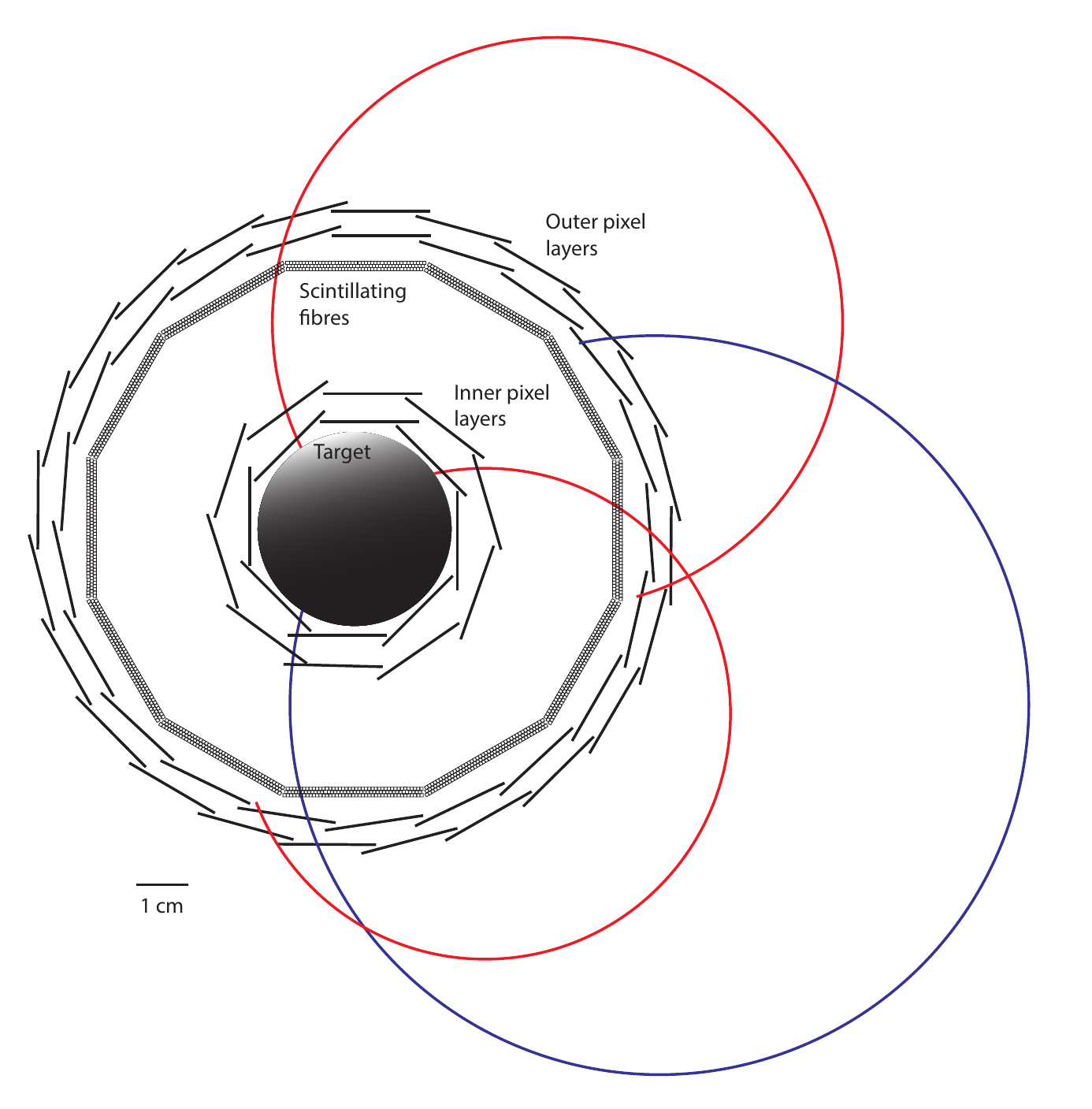}
	\caption{Schematic view of the experiment cut transverse to the beam axis. 
	Note that the fibres are not drawn to scale.}
	\label{fig:schematic_transverse}
\end{figure}

\begin{figure*}[tb!]
	\centering
		\includegraphics[width=0.9\textwidth]{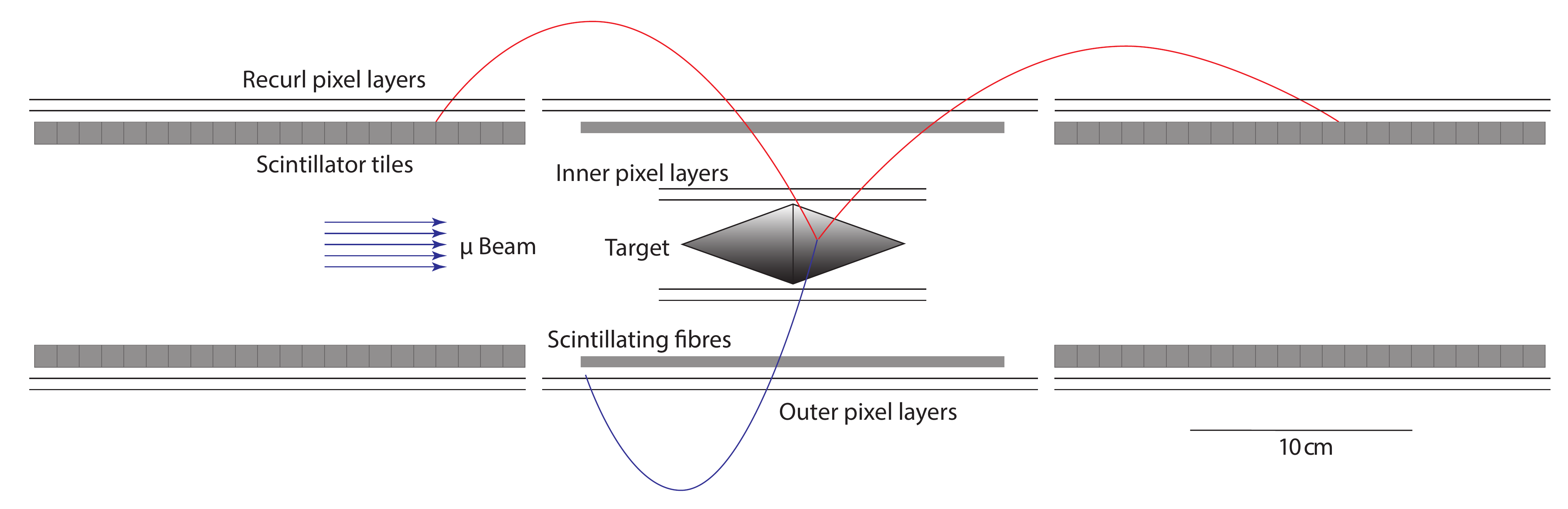}
	\caption{Schematic view of the experiment cut along the beam axis in
          the phase~I configuration.}
	\label{fig:schematic_longitudinal}
\end{figure*}

\begin{figure*}[tb!]
	\centering
		\includegraphics[width=0.9\textwidth]{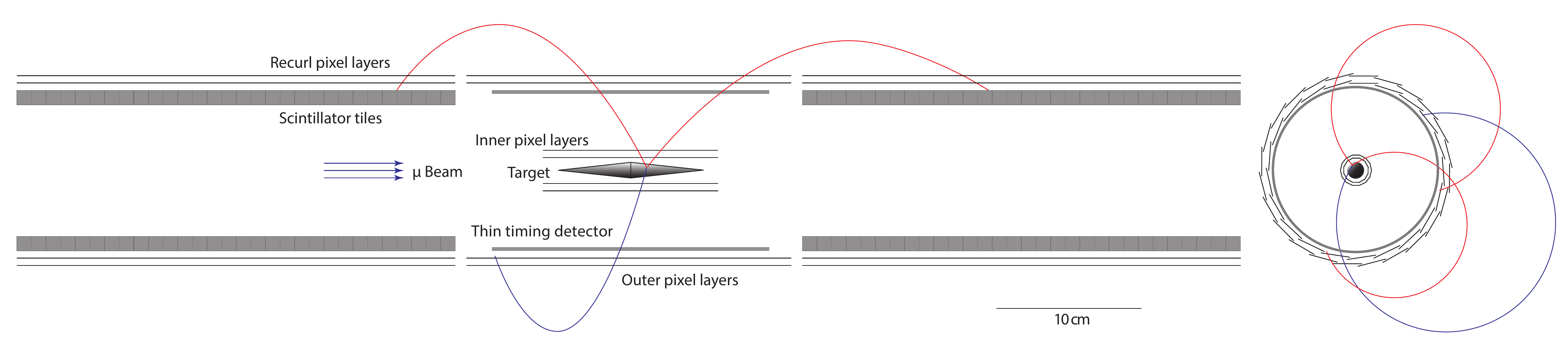}
	\caption{Possible final detector with longer recurl stations, smaller target
	and more segmented inner layers for high intensity physics runs (phase II).}
	\label{fig:Schematic10_Full}
\end{figure*}

\section[Momentum Measurement]{Momentum Measurement with Recurlers}
\label{sec:MomentumMeasurementWithRecurlers}

With a fine-grained
pixel detector, we are in a regime where multiple scattering effects dominate
over sensor resolution effects, see
\Autoref{fig:tracking_spatial_regime,fig:tracking_scattering_regime}.
Adding additional measurement
points does not necessarily improve the precision.

The precision of a momentum measurement depends on the amount of track
curvature $\Omega$ in the magnetic field $B$ and the multiple
scattering angle $\Theta_{MS}$, see \autoref{fig:MS}; to first
order:
\begin{linenomath*}
\begin{equation}
	\frac{\sigma_p}{p} \propto \frac{\Theta_{MS}}{\Omega}.
\end{equation}
\end{linenomath*}
So in order to have a high momentum precision, a large lever arm is
needed.  This can be achieved by moving tracking stations to large
radii, which would limit the acceptance for low momentum particles.
Instead, we utilise the fact that, in the case of muon decays at rest, all
track momenta are below $\SI{53}{\MeV}$ and all tracks will curl back
towards the magnet axis if the magnet bore is sufficiently large.
After half a turn, effects of multiple scattering on the momentum
measurement cancel to first order, see \autoref{fig:MScircle}.  To
exploit this feature, the experimental design is optimised
specifically for the measurement of recurling tracks, leading to a
narrow long tube layout.

Determining the momentum from a particle's trajectory outside the
tracker allows us to place thicker timing detectors on the inside both
upstream and downstream of the target without significantly affecting the
resolution, see \autoref{fig:schematic_longitudinal}.

\begin{figure}[b!]
	\centering
		\includegraphics[width=\columnwidth]{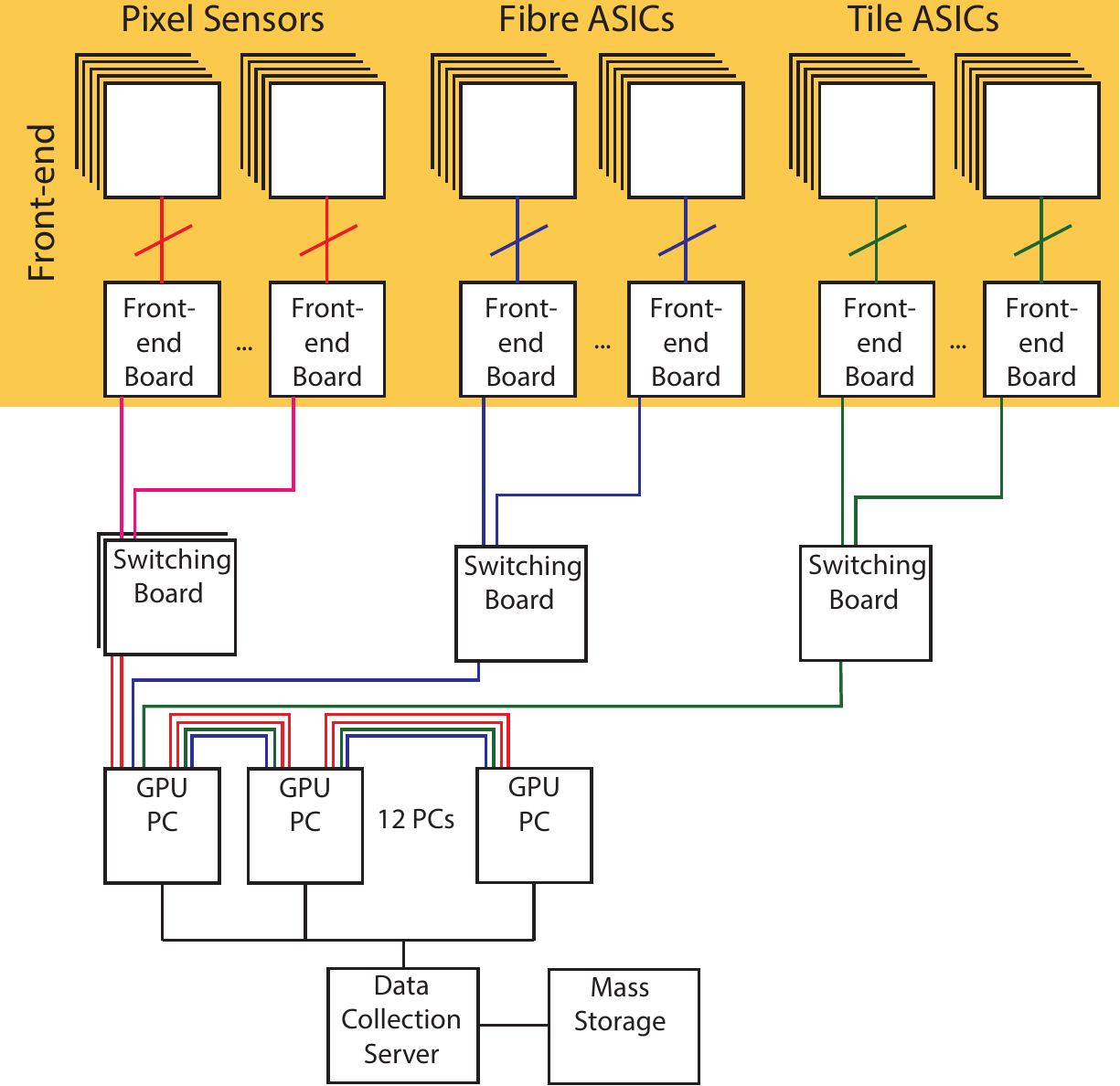}
	\caption{Schematic overview of the Mu3e readout system.}
	\label{fig:RO_Scheme_Simple}
\end{figure}

\section{Coordinate System}
\label{sec:CoordinateSystem}

The Mu3e coordinate system is centred in the muon stopping target with
the $z$ axis pointing in beam direction, the $y$ axis pointing upward
and the $x$ axis chosen to obtain a right handed coordinate system.
The polar angle measured from the $z$ axis is denoted with $\vartheta$,
and measured from the $x$-$y$ plane denoted with $\lambda$.
Azimuthal angles are denoted with $\varphi$.

\section{Baseline Design}
\label{sec:BaselineDesign}

The proposed Mu3e detector is based on two double-layers of HV-MAPS
around a hollow double cone target, see
\Autoref{fig:schematic_longitudinal,fig:schematic_transverse}.
The outer two pixel sensor layers are
extended upstream and downstream to provide precise momentum
measurements in an extended region to increase the acceptance for
recurling electrons and positrons.  The silicon detector layers (described in detail
in \autoref{sec:Pixel}) are supplemented by two timing systems, a
scintillating fibre tracker in the central part (see
\autoref{sec:Fibre}) and scintillating tiles
(\autoref{sec:Tiles}) inside the recurl layers.  Precise timing of
all tracks is necessary for event building and to suppress the
combinatorial background.

\section{Detector Readout}
\label{sec:DetectorReadout}

The Mu3e experiment will run a continuous, triggerless readout, and
employs application-specific integrated circuits (ASICs) for the pixel 
and timing detectors which stream
out zero-suppressed digital hit data.  These hits are collected by
field-programmable gate arrays (FPGAs) located on \emph{front-end boards} 
and then optically forwarded
to \emph{switching boards}, which in turn distribute them to a
computer farm.  This network makes it possible for every node in the
farm to have the complete detector information for a given time slice.
Decays are reconstructed using graphics processing units, and
interesting events are selected for storage.  A system overview is
shown in \autoref{fig:RO_Scheme_Simple} and a detailed description
can be found in \autoref{sec:DAQ}.

\section{Building up the Experiment}
\label{sec:BuildingUp}

One of the advantages of the design concept presented is its
modularity. Even with a partial detector, physics runs can be taken.
In an early commissioning phase at smaller muon stopping rates, the
detector will run with all of the timing detectors but only the
central barrel of silicon detectors. The silicon detectors of the
recurl stations are essentially copies of the central outer silicon
detector; after a successful commissioning of the latter, they can be
produced and added to the experiment as they become available.  The
configuration with two recurl stations
(\Autoref{fig:schematic_longitudinal,fig:schematic_transverse}) defines a medium-size setup, well
suited for phase~I running at the highest possible rate at the $\pi$E5
muon beam line at PSI of $\approx\SI{1e8}{\Hz}$.  The sensitivity
reach in this phase of the experiment of $\mathcal O(\num{e-15})$ will
be limited by the available muon rate.

\section{The Phase II Experiment}
	\label{sec:ThePhaseIIExperiment}

A new high intensity muon beam line, delivering $>\SI{2e9}{\Hz}$ muons
and currently under study at PSI, is crucial for Mu3e phase~II.  To
fully exploit the new beam facility the detector acceptance of phase I
will be further enhanced by longer detector stations, see
\autoref{fig:Schematic10_Full}.  These longer stations will allow
the precise measurement of the momentum of all particles in the
acceptance of the inner tracking detector.  At the same time the
longer tile detector stations with their excellent time resolution and
small occupancy will help to fight the increased combinatorial
backgrounds at very high decay rates.  The larger initial muon rate
allows for a more restrictive collimation of the beam and thus a
smaller (and potentially longer) target region leading to a much
improved vertex resolution.  The HV-MAPS technology can reach a
time resolution of $\mathcal{O}(\SI{1}{ns})$ if an adequate time-walk
correction is implemented -- this would allow to further reduce
combinatorial background without adding material and could eventually
replace the scintillating fibre detector.  Advanced wafer
post-processing technologies and chip-to-chip bonding could obviate
the need for parts of the flexprints, further reducing the multiple
scattering.  The combined performance of the enhanced detector setup
together with the high stopping rate will allow to search for the \mte
decay with a sensitivity of B(\mte)$\leq\num{e-16}$.  Whilst we always
keep this ultimate goal in mind, the rest of this document is
concerned with the phase~I detector for existing beamlines.


\chapter{Muon Beam}
\label{sec:MuonBeam}

\chapterresponsible{Peter-Raymond}

\section{Beam Requirements}
An experiment such as Mu3e, with a phase~I sensitivity goal of
\num{2e-15} while challenged by combinatorial background, not only
requires running at the intensity frontier, but also substantially
benefits from a continuous beam structure rather than a pulsed one,
allowing a lower instantaneous muon rate.  Both of these conditions
are satisfied by the high intensity proton accelerator complex (HIPA)
at PSI running at \SI{1.4}{MW} of beam power.

\begin{figure}[t]
   \centering
   \includegraphics[width=\columnwidth]{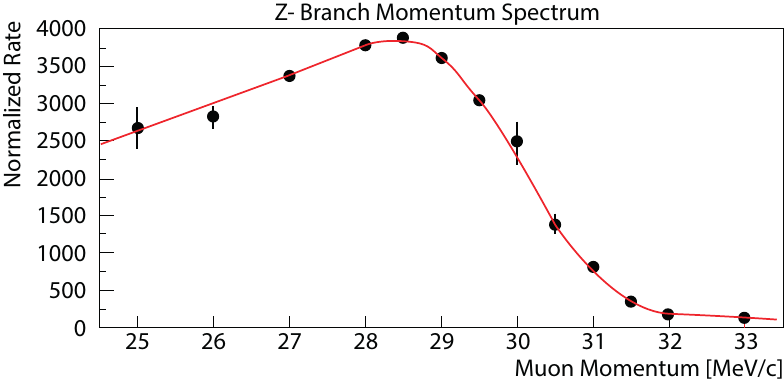}
   \caption{Measured muon momentum spectrum in $\pi$E5, with full momentum acceptance. 
   Each point is obtained by optimising the whole beam line for the 
	corresponding central momentum and measuring the full beam-spot intensity. 
   The red line is a fit to the data, based on a theoretical $p^{3.5}$ 
	behaviour, folded with a Gaussian resolution function corresponding to the 
	momentum-byte plus a constant cloud-muon background.}
   \label{fig:muon_p_spectrum}
\end{figure}

Mu3e requires a muon beam with the highest possible rate of ``surface
muons'', produced from stopped pion decay at the surface of the
primary production target~\cite{Pifer:1976ia}.  The surface muon yield
and hence beam intensity peaks at around \SI{28}{MeV/c}, close to the
kinematic edge of the two-body momentum spectrum of pion decay at rest
as can be seen from the measured momentum spectrum in
\autoref{fig:muon_p_spectrum}.

The intensity goal and low energy not only necessitates a beam line
capable of guiding these muons to a small, thin stopping target with
minimal losses but at the same time minimising beam-related
backgrounds.  The former requires a small beam emittance and a
moderate momentum-byte (full width at half maximum of the momentum acceptance $\Delta p/p$), with an achromatic final focus to balance
between beam intensity and stopping density in the target. The
minimising of beam-related backgrounds, in the form of Michel $e^{+}$
from $\mu^{+}$-decay, $e^{+}$ produced from $\pi^{0}$-decays in the
production target, or from decay-in-flight particles produced along the
beam line, puts strong restrictions on the amount of material, such as
windows and momentum moderators, that can be placed along the beam path,
requiring an extension of the vacuum system to just in front of the
target.

\begin{figure*}[t]
   \centering
   \includegraphics[width=14 cm]{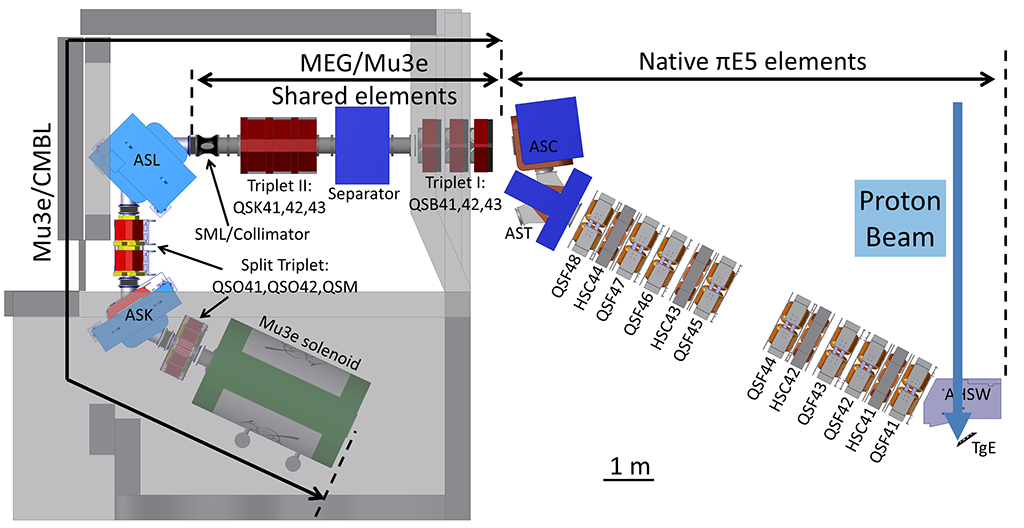}
   \caption{CAD model of the entire $\pi$E5 channel \& CMBL used as a 
	basis for the \textsc{g4bl} models.}
   \label{fig:cmbl_cad_top}
\end{figure*}

\section[Compact Muon Beam Line]{The Compact Muon Beam Line~(CMBL)}
\begin{sloppypar}
For Mu3e phase~I, muon intensities close to 10$^{8}$~muons/s will be
required, which leaves only one choice of facility in the world, PSI's
$\pi$E5 channel.  This channel will be shared with the upgrade version
of the MEG experiment -- MEG II \cite{Baldini:2013ke}, whose large
detector and infrastructure are permanently located in the rear-part
of the $\pi$E5 area.
\end{sloppypar}

The new CMBL for Mu3e, as presented in the following, not only allows
the \SI{3.2}{m} long Mu3e solenoid to be placed in the front part of
the $\pi$E5 area -- see \autoref{fig:cmbl_cad_top} -- but also allows
both experiments MEG II and Mu3e to share the front beam transport
elements required by both.  This solution allows the efficient
switching between experiments by only replacing the superconducting
beam transport solenoid of MEG~II by a dipole magnet (ASL) for Mu3e.

\begin{figure}[]
   \centering
   \includegraphics[width=\columnwidth]{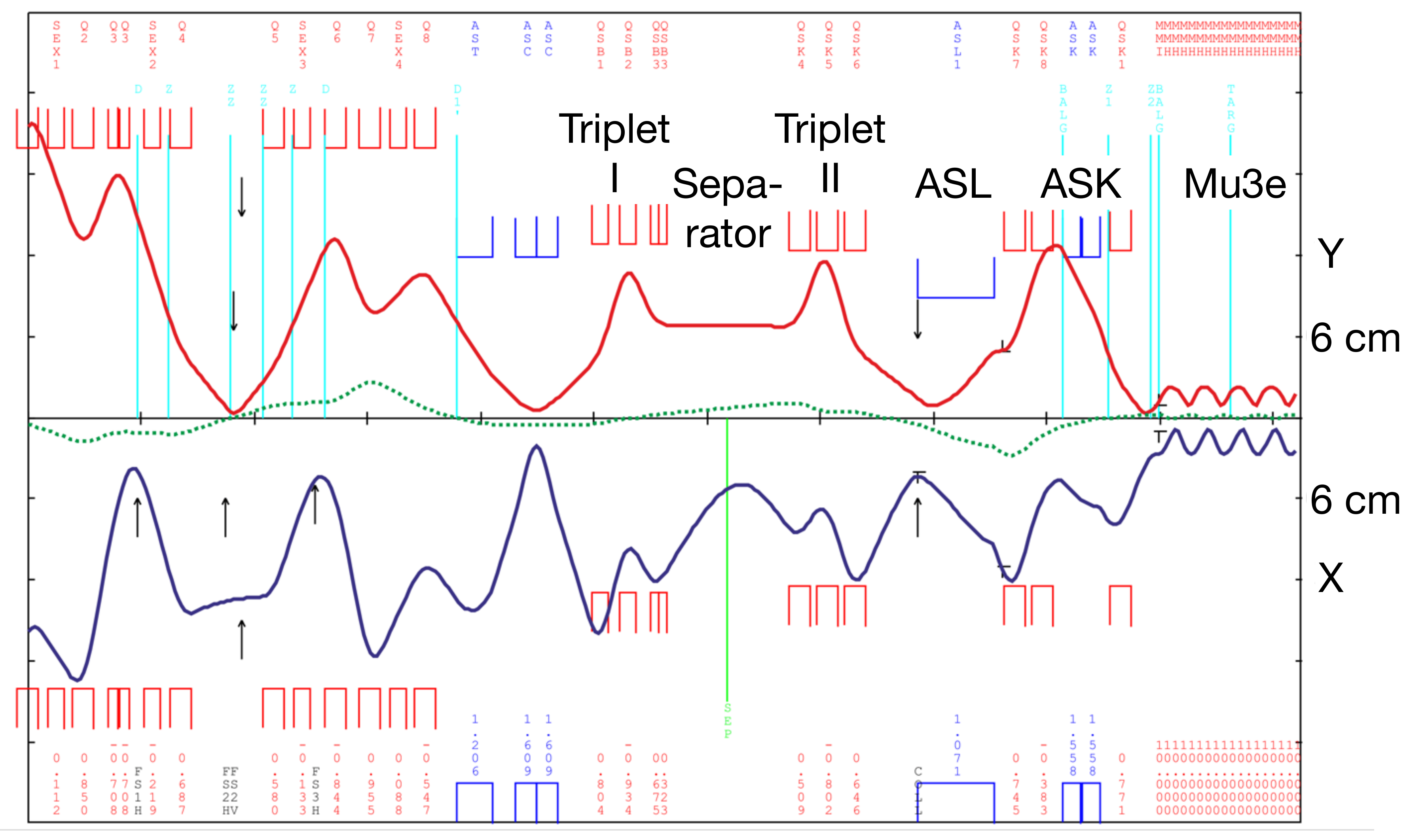} 
   \caption{Optical Model of the CMBL from the \textsc{graphic
       transport framework} program, showing 1st-order vertical and
     horizontal beam envelopes along the entire beam line from
     Target~E to the end of the Mu3e solenoid with some of the beam
     elements labelled (note the horizontal scale unit is \SI{2}{m},
     whereas the vertical is \SI{6}{cm}).  The dotted line shows the
     dispersion trajectory for a 1\% higher central momentum.}
   \label{fig:beam_optics}
\end{figure}

\begin{sloppypar}
The initial optical design of the CMBL was modelled using the beam
optics matrix code programs \textsc{graphic transport framework}
\cite{TRAN} and \textsc{graphic turtle framework} \cite{TURT}, while
the detailed modelling was undertaken using the newer \textsc{geant4}
based simulation software \textsc{g4beamline} (\textsc{g4bl})
\cite{g4bl}.  The 1st-order optical design showing the vertical and
horizontal beam envelopes from Target~E to the downstream end of the
Mu3e detector are shown in \autoref{fig:beam_optics}.
\end{sloppypar}

\begin{figure*}[t]
   \centering
   \includegraphics[height=5cm]{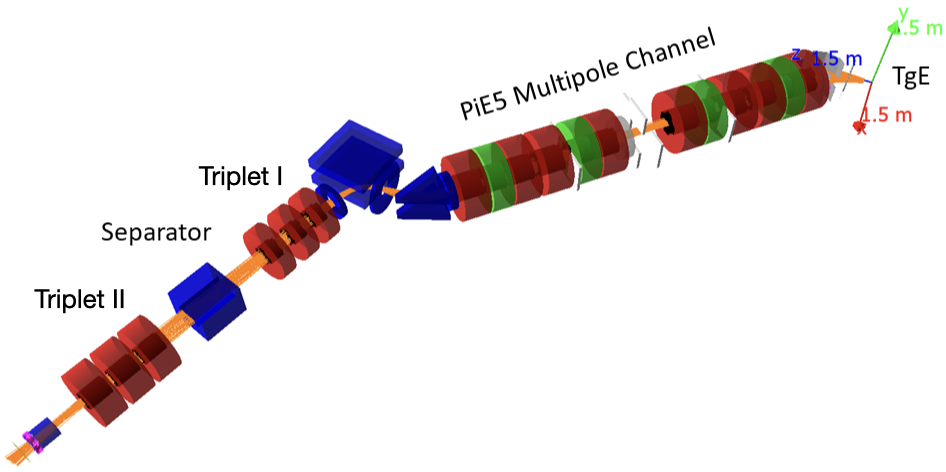}
   \includegraphics[height=5cm]{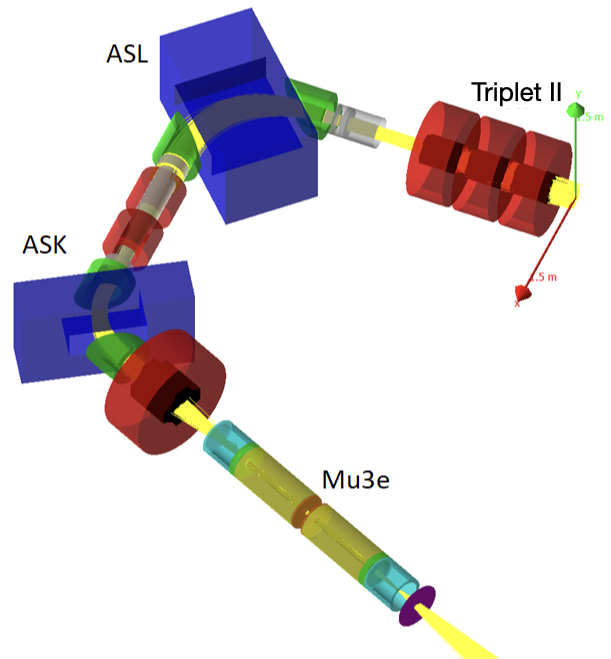}
   \caption{Shows the graphical outputs of the \textsc{g4bl}
     simulations with some of the beam elements labelled.  (Left) --
     simulation of the full beam line from Target~E up to the
     intermediate collimator system.  (Right) -- shows the shorter
     version of the simulation from Triplet II past the intermediate
     collimator system to the end of the Mu3e Detector solenoid.}
   \label{fig:gui_long_and_short}
\end{figure*}

The design includes the elements of the backward (165$^\circ$)
extraction channel $\pi$E5 from Target~E up to the ASC dipole magnet,
the background cleaning-stage including triplet I, the Wien-filter
(SEP41), triplet II and the collimator system, used to eliminate the
beam-related background.  The final injection stage is based on a very
compact ``split triplet'' layout which starts after the 90$^\circ$
dipole ASL41. The  ``split triplet'' consists of the quadrupole doublet QSO41/42 and quadrupole singlet QSM41.
In combination with the vertical edge-focusing of the ASK41
65$^\circ$ dipole magnet they serve the same purpose as a total of six
quadrupoles that would be needed in a more standard beamline configuration.  This allows sufficient space to place the \SI{3.2}{m}
long Mu3e detector in the front area without compromising the optics
and physics goals of the experiment.

Based on the \textsc{graphic transport} model, two \textsc{g4bl}
models were constructed, one including the full $\pi$E5 channel and
Target~E, simulating the whole pion production process by protons in
the primary target, followed by surface muon production and transport
to the intermediate collimator.  The second shorter version starts
from Triplet II, just upstream of the intermediate focus at the
collimator system, where measured phase space parameters determine the
initial beam used for the simulation - see
\autoref{fig:gui_long_and_short}.  The shorter version predictions were used
as a direct comparison to the CMBL commissioning measurements
described in the next section.

\section{CMBL Commissioning Steps}
\begin{figure}
   \centering
   \includegraphics[width=\columnwidth]{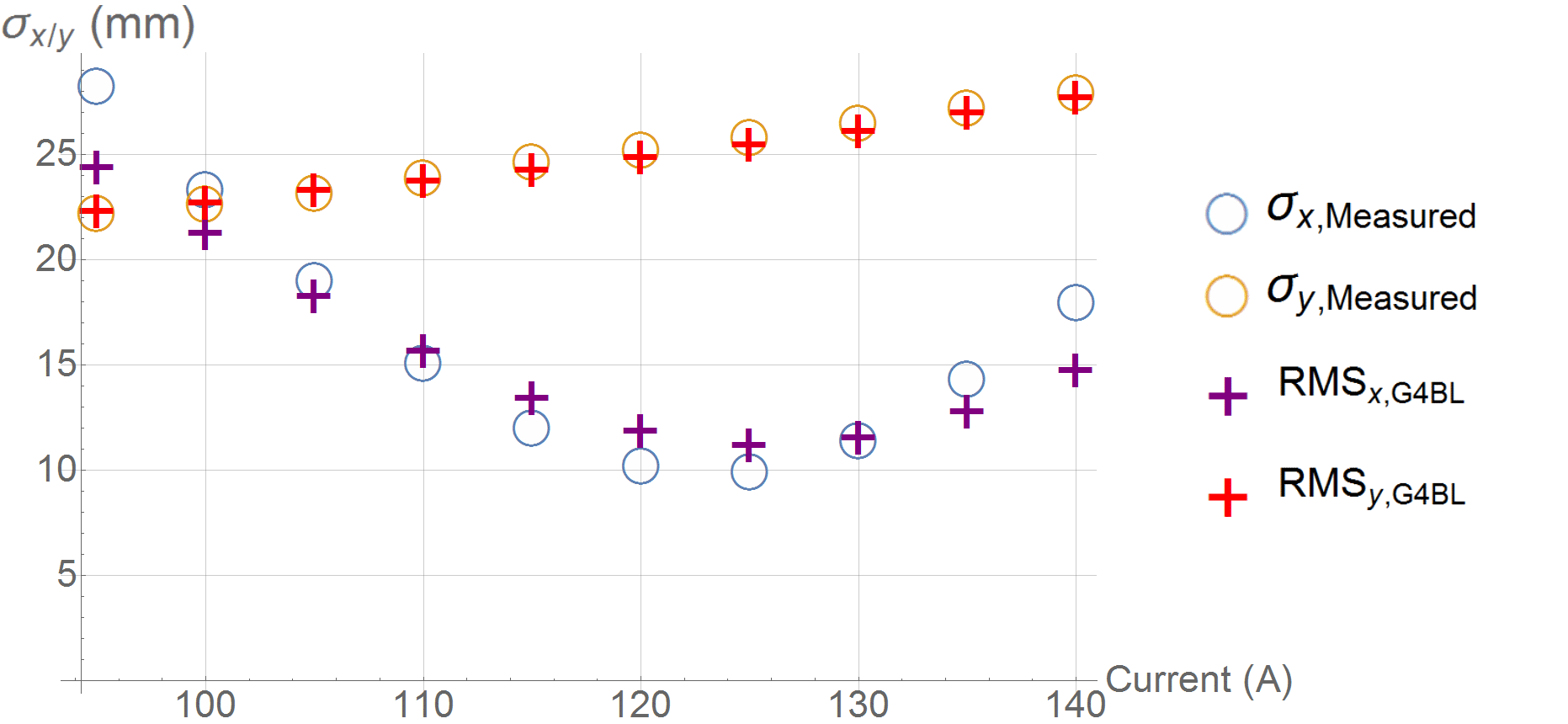}
   \caption{Simulated and measured 2-D beam sizes at the Mu3e solenoid injection point, showing good agreement for a wide range of currents applied to the last quadrupole QSM41. }
   \label{fig:sim_meas_beam_solenoid}
\end{figure}

Initial commissioning of the CMBL beam layout was undertaken in two
4-week beam periods in November and December 2014 and May 2015, using
mostly existing elements.  \autoref{fig:sim_meas_beam_solenoid}
shows the good agreement between predicted and measured beam sizes at
the injection point to the Mu3e solenoid, based on a 1st-order
transverse phase space reconstruction.  The validated \textsc{g4bl}
model was then used, and identified the ASL and ASK dipole apertures
as the main limitations for the transmission to the final focus.
Consequently, increased pole-gaps and modified vacuum chambers for
both dipole magnets allowed for an expected enhanced transmission of
18\%, which was proven in the following 2016
measurements~\cite{Ber17}.

\begin{figure}
   \centering
   \includegraphics[width=\columnwidth]{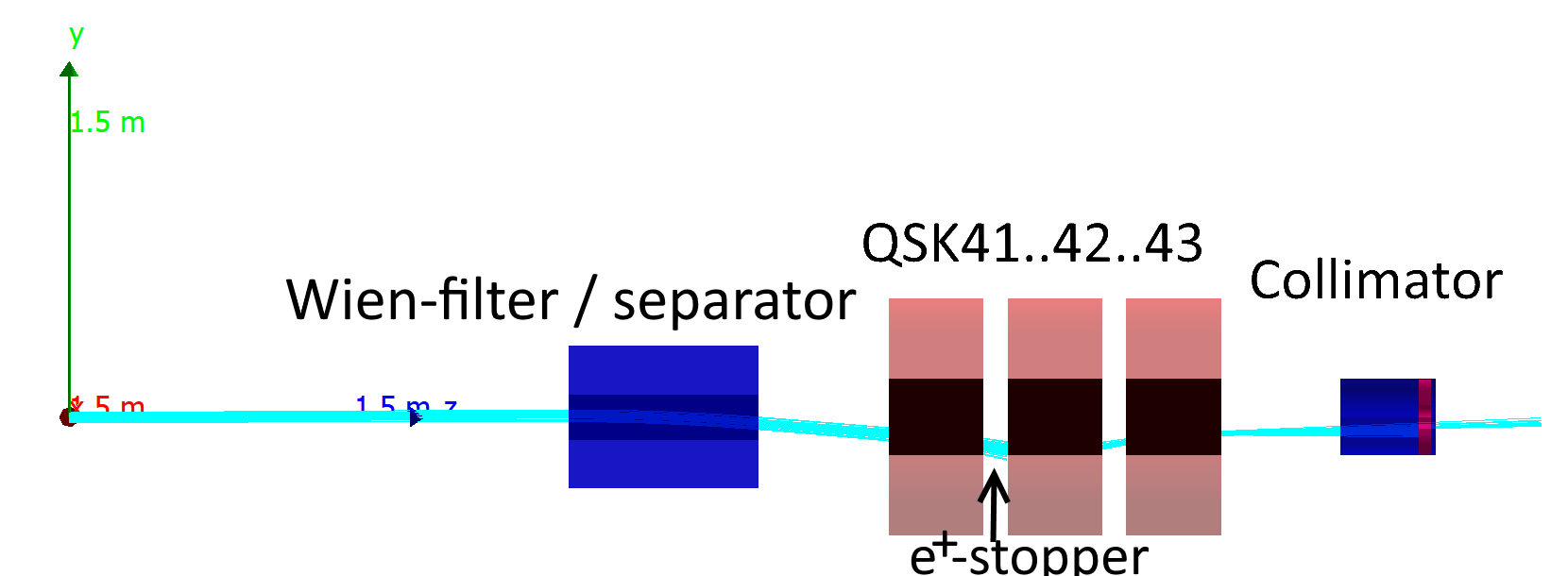} 
   \caption{Demonstration of beam positron contamination vertically separated post Wien-filter being swept-back into the acceptance at the collimator by QSK42/43.}
\label{fig:sepcontam}
\end{figure}

In 2017, the commissioning emphasis was placed on confirmation of increased muon yield using a \SI{60}{mm} long production Target~E instead of the usual \SI{40}{mm} version. 
The expectation of only an $\sim$30\% increase in muon yield (surface phenomenon) with a full 50\% increase in beam positron contamination (bulk phenomenon) for the 165$^\circ$ backward extraction was confirmed.
Furthermore, the expected impact on the experiment from an increased beam positron background was also studied and a differential measurement technique developed to distinguish Michel positrons from beam positrons at the final focus~\cite{Hod18}. 
These measurements showed that for the \SI{60}{mm} Target~E a beam-$e^+$/$\mu^+$-ratio = 10.1 was measured, with no Wien-filter in operation, whereas for a \SI{40}{mm} target the ratio was $\sim$7.
However, with the Wien-filter on, an unacceptably high number of beam positrons, seen as a vertically displaced spot, were measured. 
On investigation it was found that the off-centre, vertically displaced (by the Wien-filter) beam positrons entering triplet II are partially swept-back into the acceptance of the downstream collimator, as demonstrated in \autoref{fig:sepcontam}. 
The situation was quickly and temporarily solved by placing a lead $e^+$-stopper between QSK41/42 reducing the contamination by a factor of 15, with a 10\% loss of muons. 
The final solution is the modification of the Wien-filter, which was upgraded in 2019 to have a symmetric electric field with double the present voltage of \SI{200}{kV}. 
While not yet experimentally confirmed, this is expected to reduce any beam positron contamination by 3-orders of magnitude.  

Finally, using the measured contamination rate, the impact of this on
the experiment's sensitivity to combinatorial Michel and beam positron
events mimicking a 3-particle signature via Bhabha scattering was
investigated \cite{Hod18} and found that only muon decay-in-flight
events have a chance of coming close to the reconstructed muon mass
region, though occurring at a rate twelve orders of magnitude lower
than the most dominant background (Bhabha scattering with overlapping
Michel decays).

\begin{figure*}[p]
   \centering
   \includegraphics[width=0.8\columnwidth]{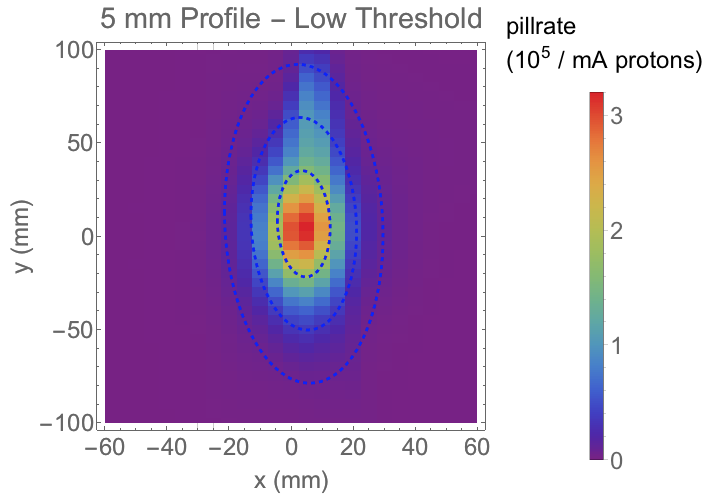}
   \includegraphics[width=0.8\columnwidth]{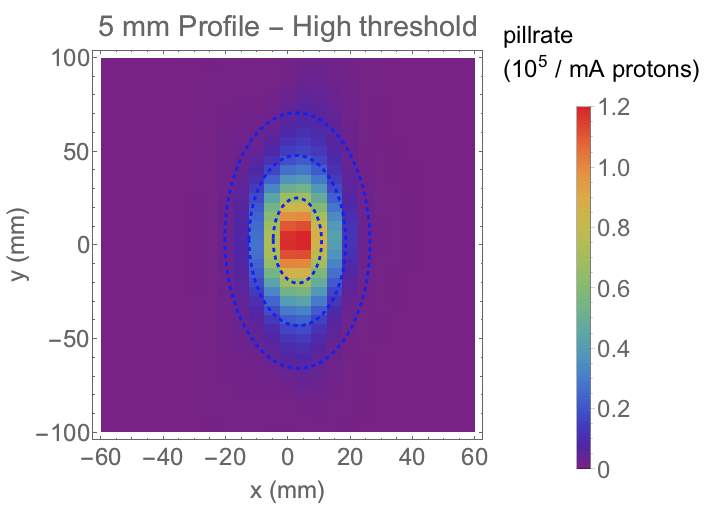}
   \caption{Measured beam spot at the injection point to the Mu3e solenoid triggering on either a low (left: muons + Michels + beam positrons) or high  (right: muons only) threshold. 
   A 2D Gaussian fit to the muon data yields $\sigma_x$ = \SI{8}{mm} and  $\sigma_y$ = \SI{23}{mm} with a total rate of $1.1 \times 10^8$~$\mu^+$/s at a proton current of \SI{2.4}{mA} for a \SI{40}{mm} long Target~E. 
   The vertical beam positron tail in the low threshold profile (top-part) is without the $e^+$-stopper in triplet II and will be totally removed with the upgraded Wien-filter.}
   \label{fig:final:focus:scans}
\end{figure*}

During the shutdown 2017/18 all magnet power supplies for $\pi$E5 were
replaced with digitally controlled ones.  The better stabilisation of
magnet currents contributed to a further increased transmission during
the 2018 commissioning run.  Optimisation at the injection point of
the Mu3e solenoid yielded a final rate of $1.1 \times 10^8$~$\mu^+$/s,
normalised to the expected future proton beam current of \SI{2.4}{mA}
for a \SI{40}{mm} long Target~E, with profile widths of $\sigma_x$ =
\SI{8}{mm}, $\sigma_y$ = \SI{23}{mm}, and a momentum-byte of 8\%.  The measured high (muons
only) and low threshold (muons + Michels) profiles are shown in
\autoref{fig:final:focus:scans}, these \SI{5}{mm} raster scan
profiles were measured with a 2D automated pill scintillator scanner
system with each profile consisting of $\sim$1025 single measurements.
The beam intensity is extracted from a 2D Gaussian fit to the
profiles.

The 2018 measurements therefore successfully conclude the beam commissioning up to the injection point of the Mu3e solenoid. 
The final commissioning to the centre of the Mu3e detector will be undertaken when the magnet is placed in the area.

\begin{figure*}[p]
   \centering
   \includegraphics[width=0.9\textwidth]{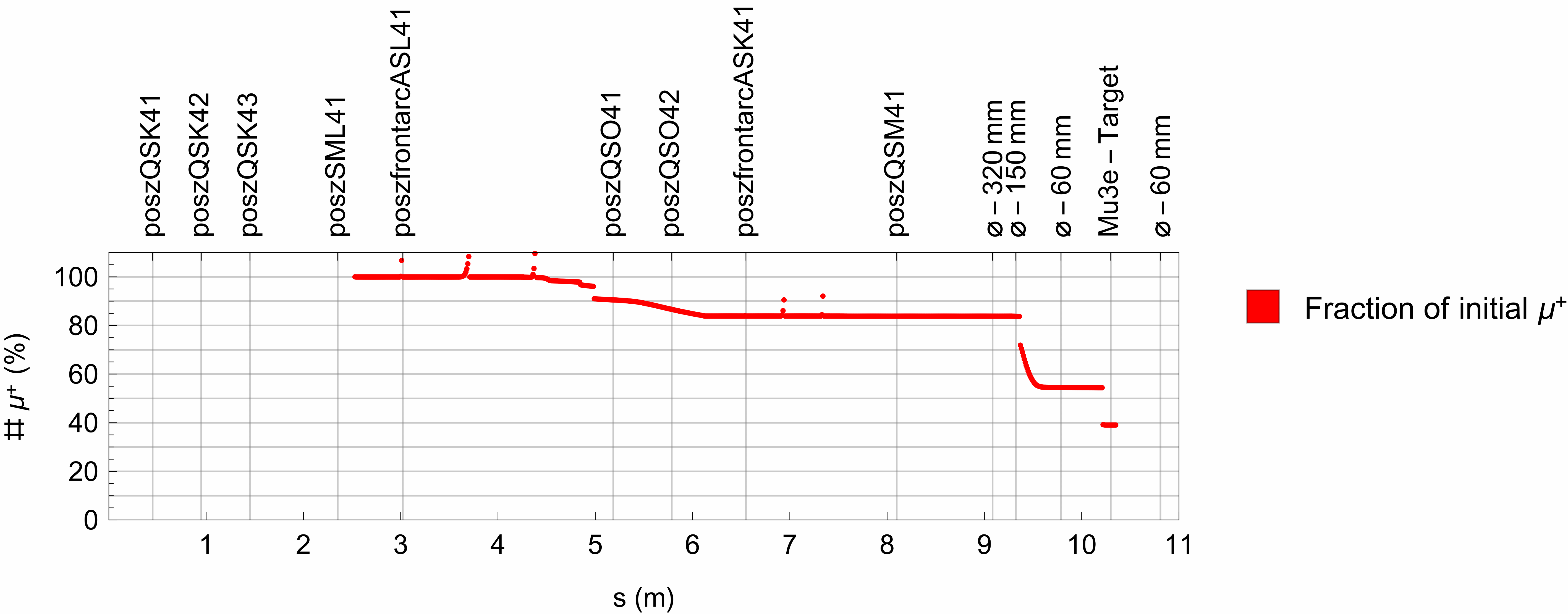}
   \caption{Beam losses along the Mu3e Compact Muon Beam Line (CMBL) starting from the intermediate collimator system to the centre of the Mu3e magnet. 
   In front of the Mu3e target a narrowing of the beam pipe down to \SI{40}{mm} diameter takes place. }
   \label{fig:beam:losses}
\end{figure*}

\begin{figure*}[p]
   \centering
   \includegraphics[width=0.9\textwidth]{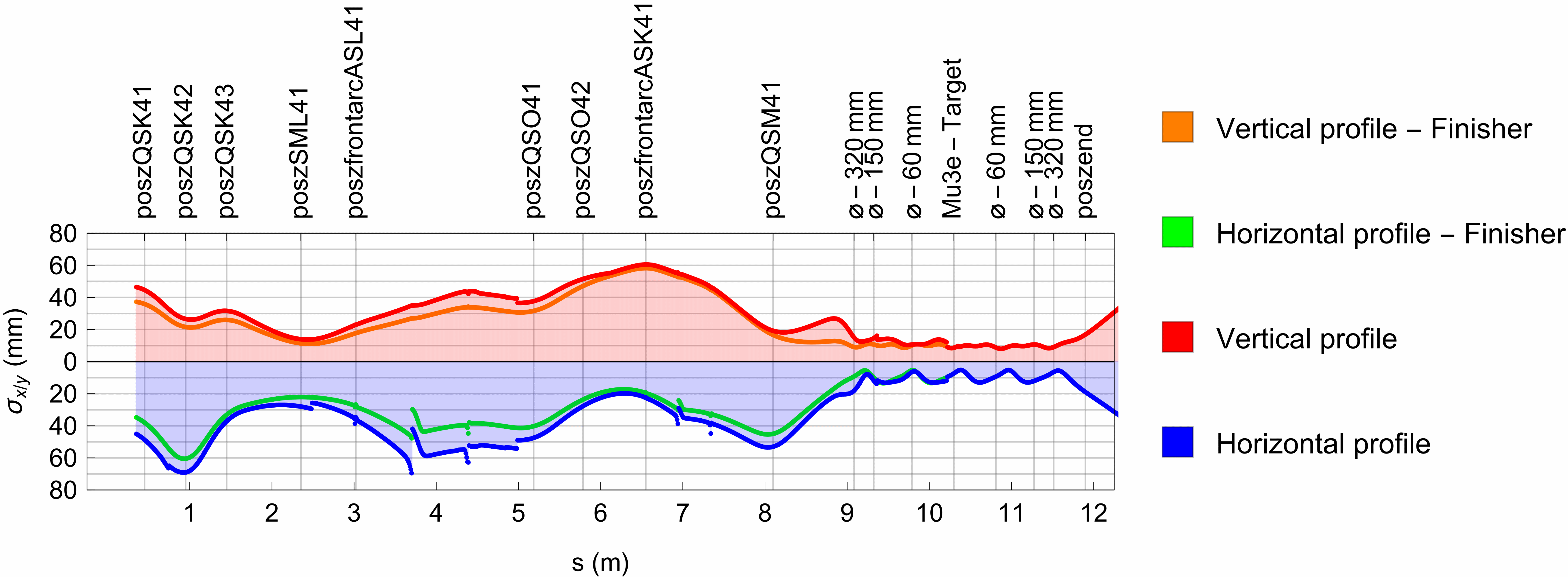}
   \caption{Horizontal and vertical beam envelopes for `all' particles started in the simulation or only for those that reach the centre of the solenoid (`finishers'). Due to the losses along the beamline on apertures and other restrictions the two envelopes converge at the center of the Mu3e target.}
   \label{fig:beam:envelopes}
\end{figure*}

\section{Expected Muon Rate and Distribution on the Mu3e Stopping Target}

As described in the previous section the final optimisation of the CMBL resulted in a surface muon rate at the injection point of the Mu3e magnet of  $1.1 \times 10^8$~$\mu^+$/s, normalised to the expected future proton beam current  of \SI{2.4}{mA} for a \SI{40}{mm} long Target~E, with profile widths of $\sigma_x$ = \SI{8}{mm}, $\sigma_y$ = \SI{23}{mm}, and a momentum-byte of 8\%. 

The coupling to the central detector region inside the solenoid magnet is planned to be with a custom bellows system (see \autoref{fig:beamline_detector}) reducing step-wise the aperture to an inner diameter of \SI{60}{mm} for the inner vacuum-pipe. 
This will contain a \SI{600}{\mu m} thick Mylar (biaxially-oriented polyethylene terephthalate) moderator located at an intermediate focus point some few hundred millimetres in front of the target and will end with a \SI{35}{\mu m}  Mylar vacuum window, placed just in front of the Mu3e target, where the aperture narrows down to 40 mm diameter due to the support structure of the inner pixel layers. 
A double-cone Mylar target of radius \SI{19}{mm}, length \SI{100}{mm} and total thickness of \SI{150}{\mu m} (see \autoref{sec:Target}) is located close to the vacuum window at the centre of the solenoid. 
The warm bore of the solenoid is filled with helium gas at atmospheric pressure to reduce multiple scattering. 
Furthermore, a \SI{20}{mm} thick lead collimator system will be introduced shortly after the moderator to protect the inner pixel layers from hits by the muon beam as well as from particles outside of the target acceptance. 

\begin{figure}[]
   \centering
   \includegraphics[width=\columnwidth]{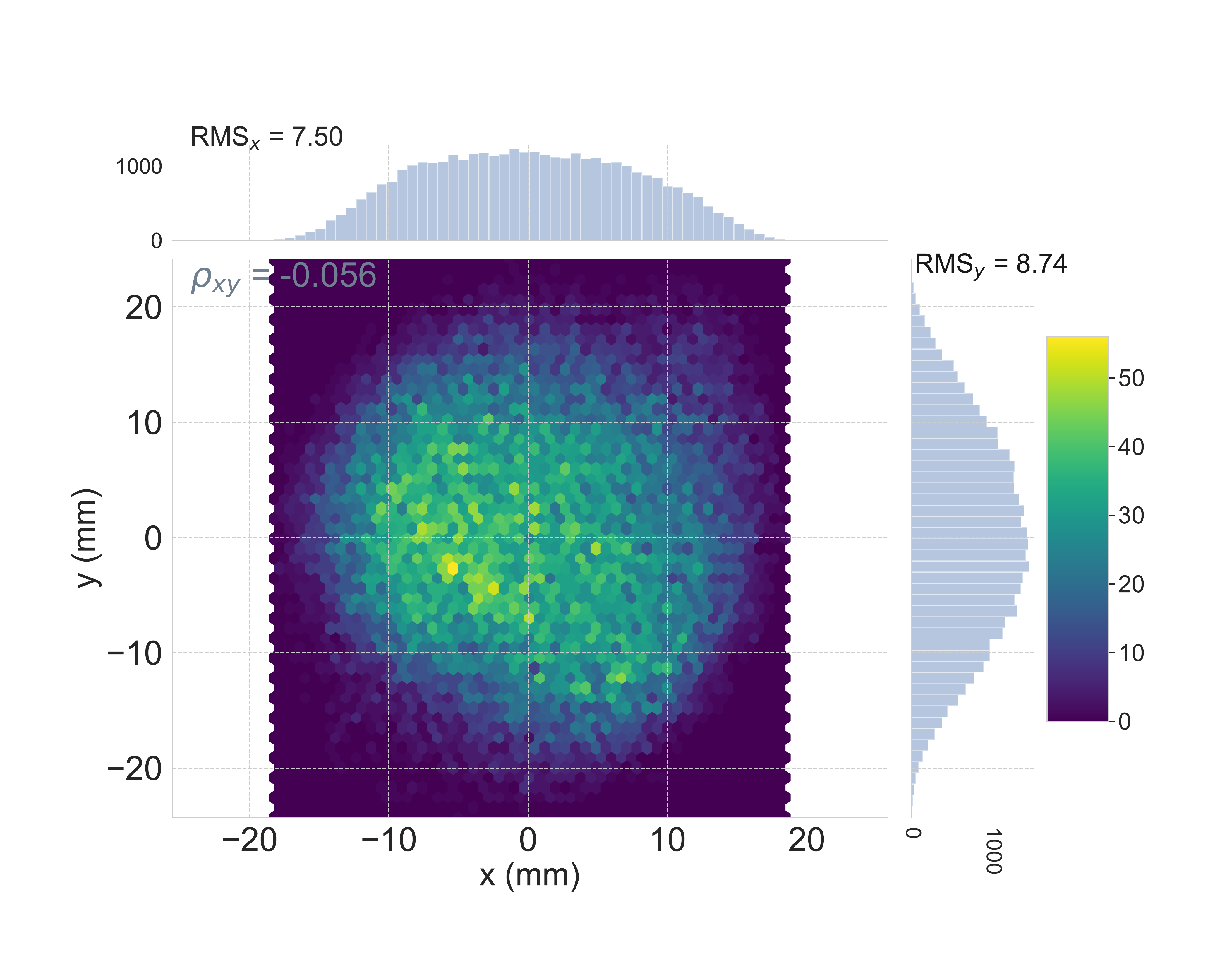}
   \caption{Estimated beam profile at the Mu3e target position.}
   \label{fig:target:beamspot}
\end{figure}
\begin{sloppypar}
Estimates for the final muon stopping rate on the target are based on the re-measured 1-$\sigma$ beam emittances at the intermediate collimator system in 2018, corresponding to $\epsilon_x$ = \SI{950}{\pi \cdot mm \cdot mrad}, $\epsilon_y$ = \SI{490}{\pi \cdot mm \cdot mrad} and the \textsc{g4bl} simulation. The beam losses along the beam line can be seen in \autoref{fig:beam:losses} and the corresponding beam envelope sizes in \autoref{fig:beam:envelopes}.
\end{sloppypar}

Even though the muon beam intensity at injection into the solenoid achieves the commissioning goal, it is the inner silicon detector diameters and the associated beam pipe size that determine the stopping target diameter, which has been maximised to a radius of \SI{19}{mm}. 
These conditions are a compromise between stopping rate, occupancy and vertex resolution. 

The main losses are associated with the transition to the initial
diameter of the beam pipe, and the final narrowing to a \SI{40}{mm}
diameter at its end.  The final beam-spot at the target is shown in
\autoref{fig:target:beamspot}.  The beam intensity on the target is
expected to be $\sim$$5 - 6 \times 10^7$~$\mu^+$/s at \SI{2.4}{mA}
proton current for the current \SI{40}{mm} long production Target~E.
The final muon rate can further be enhanced by the use of the
\SI{60}{mm} production target, or the recently tested 40-mm long
slanted target.  Both of these targets lead to a further $\sim$30-40\%
enhancement, so yielding muon rates on the Mu3e target of about
$\sim$$7 - 8 \times 10^7$~$\mu^+$/s at \SI{2.4}{mA} proton current.
Further enhancements are still under study.


\chapter{Magnet}
\label{sec:Magnet}

\chapterresponsible{Andre}

The magnet for the Mu3e experiment has to provide a homogeneous solenoidal
magnetic field of $B=\SI{1}{T}$ for the precise momentum determination of the muon decay
products. 
Field inhomogeneities along the beam line are required to stay below $\num{e-3}$ within $\pm\SI{60}{cm}$ around the center of the magnet. 
The magnet also serves as beam optical element for guiding the muon beam to the target. 
To further improve the field homogeneity, and for matching the magnetic field of the last beam elements of the compact muon beam line, compensating coils are included on either side of the magnet.

\begin{sloppypar}
The basic parameters of the superconducting solenoid magnet are given in \autoref{tab:MagnetRequirements}.
The outer dimensions also include an iron shield, reducing stray fields
to less than $\SI{5}{\milli\tesla}$ at a distance of \SI{1}{m}.
This increases the overall weight of the magnet to 31~tons, 27 of which are due to the iron shielding.
\end{sloppypar}

The long term stability of the magnetic field should be $\Delta B/B\leq\num{e-4}$ over each 100 days data-taking period. 
This is achieved with state of the art power supplies and will permanently be monitored by NMR and Hall probes inside the apparatus. 
The NMR system and hall probes will also be used to map the field.
The goal is to measure and describe the field distribution with a precision better than \num{2.0e-4}.
This measurement uncertainty will allow us to provide accurate field distributions for the track reconstruction of the positrons and electrons that requires knowledge of the magnetic field at the $\num{e-3}$ level in order to not impact the momentum resolution.

The tight requirements on the dimensions of the magnet come from the space
constraints of the $\pi$E5 area as described in the next chapter.
In this respect a good compromise had to be found as in particular the total length of the magnet is a critical parameter impacting the specified homogeneity of the field in the central region.

In addition, the magnet also acts as a container for the helium gas used for cooling as described in \autoref{sec:Cooling}.
For this reason, the warm bore is designed with helium-tightness in mind and is sealed off on both ends by removable flanges. 

A superconducting magnet design with a closed cooling system was determined to 
be the most stable and economic solution.
The coils made from niobium-titanium superconductor will operate at nominally
\SI{4}{\kelvin} and be cooled by four Gifford McMahon two-stage
cryocoolers, each delivering \SI{1.5}{\watt} cooling power at their second
stage.
The cool-down time for the system is about 10~days
with liquid nitrogen pre-cooling and the  
ramp up time to \SI{1}{T} will be less than 2~hours.

\begin{table}[tb!]
	\centering
		\begin{tabular}{lr}
			\toprule
			\sc Magnet parameter 											& \sc Value \\ 
			\midrule
			nominal field											& $\SI{1.0}{\tesla}$ \\
			warm bore diameter 												& $\SI{1.0}{\m}$ \\
			warm bore length      										& $\SI{2.7}{\m}$ \\
			field inhomogeneity $\Delta B/B$					& $\leq\num{e-3}$ \\
			field stability $\Delta B/B$ (100 days)  	& $\leq\num{e-4}$ \\
			field measurement accuracy $\Delta B/B$						& $\leq\num{2.0e-4}$ \\
			outer dimensions: length				    			& $\leq \SI{3.2}{\m}$\\
						\hspace{2.8cm}	width					  		& $\leq \SI{2.0}{\m}$\\
						\hspace{2.8cm}	height          		& $\leq \SI{3.5}{\m}$\\
			\bottomrule
		\end{tabular}
	\caption{Specifications for the Mu3e magnet. 
	The field inhomogeneity requirement concerns the central $\pm\SI{60}{cm}$ along the magnet axis.}
	\label{tab:MagnetRequirements}
\end{table}

\begin{figure}[tb!]
   \centering
   \includegraphics[width=0.49\textwidth]{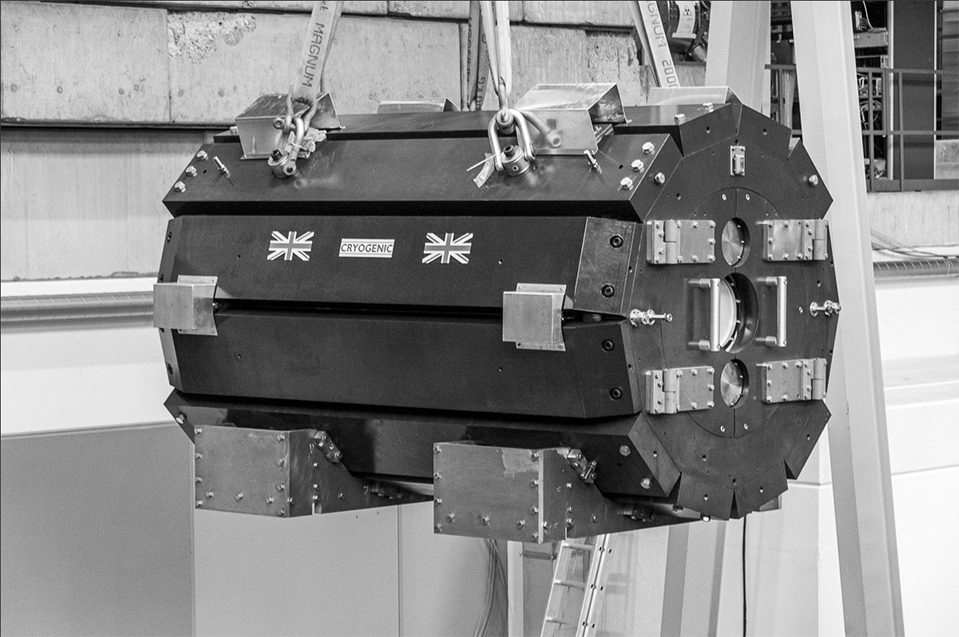} 
   \caption{Picture of the delivery of the 31-ton Mu3e magnet to PSI's experimental hall where the Mu3e experiment will take place.}
   \label{fig:magnet_delivery}
\end{figure}

Cryogenic Ltd.\footnotemark
was tasked to design and produce the Mu3e solenoid magnet.
The picture shown in \autoref{fig:magnet_delivery} depicts the delivery of the magnet to PSI's experimental hall in July 2020 after initial testing of the magnet at the company showed excellent performance.
\footnotetext{Cryogenic Ltd., Acton Park Industrial Estate, The Vale, London~W3~7QE, United Kingdom}


\chapter{Area Layout, Infrastructure \& Beam Line Connection}
\label{sec:Area}

\chapterresponsible{Peter-Raymond}

\nobalance

\label{sec:areaLayoutBeamLine} 

\begin{figure}[tb!]
   \centering
   \includegraphics[width=\columnwidth]{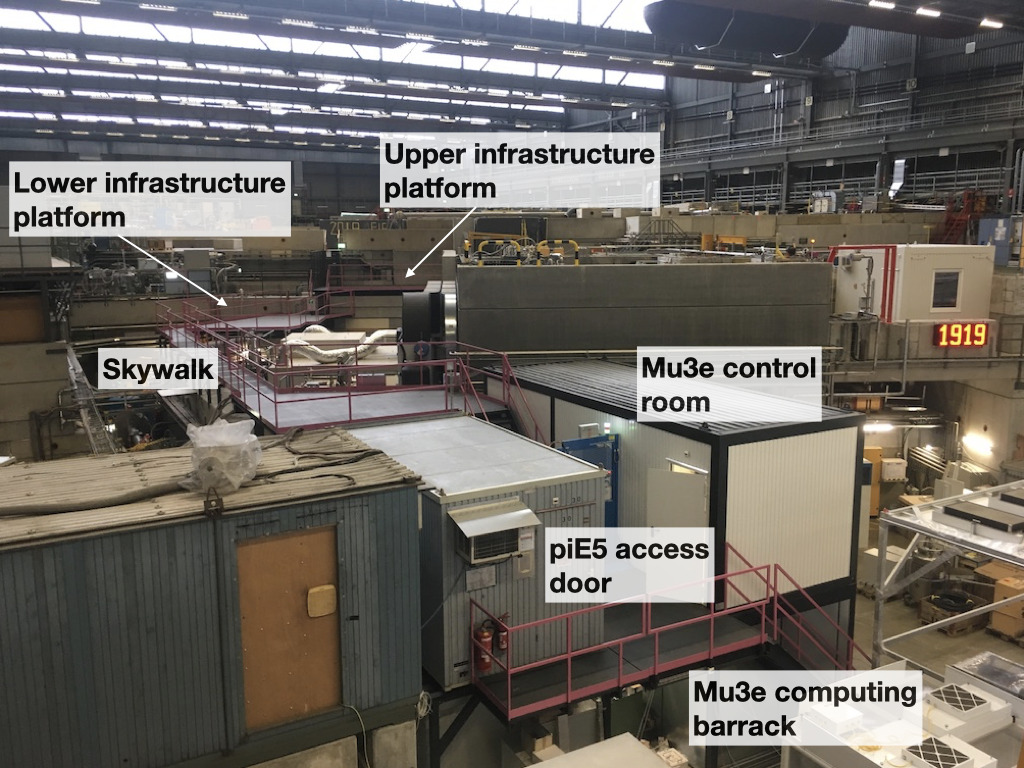}
   \caption{The new rear access to the $\pi$E5 Area with the Skywalk and the two new infrastructure platforms for the Mu3e experiment. 
   In the front, the Mu3e control room and computing barrack are located.}
   \label{fig:area_infra}
\end{figure}

\begin{sloppypar}
Due to the spatial restrictions in the $\pi$E5 front area and the
substantial infrastructure needs of the experiment, an optimised area
layout is necessary.  Upgrades were needed to both the electrical
installation and cooling-water and, due to safety requirements, an
additional access route to the front area had to be added.
\autoref{fig:area_infra} shows the overview of the new rear access
to the $\pi$E5 Area via the `skywalk' with its two new infrastructure
platforms and the Mu3e control room and computing farm barrack.  The
experimental area in the front part of $\pi$E5 is located below the
two infrastructure platforms and will have a stairway added as a
safety requirement, once the large magnet is in place, leading from
the lower platform into the experimental area.
\end{sloppypar}

The upper infrastructure platform, above the beam entrance wall, is
constructed to be removable in order to grant service access to the
$\pi$E5 channel during accelerator shutdown periods, if required.
This platform is closest to the magnet and detector and will house the
cooling elements such as the compressors for the cryogenic cold-heads
as well as the helium and water cooling circuits for the Mu3e
detector.  The lower, larger platform will not be removable and will
carry the magnet power supplies, quench detection system and
electronics as well as the power-control circuitry associated with
both magnet and detectors.

Also seen in \autoref{fig:area_infra} are the two new $\pi$E5
barracks located on top of each other.  The upper barrack will serve
as the Mu3e experiment's control room, while the lower barrack will
house the filter farm responsible for the readout of the detector (see
\autoref{sec:DAQ}).

\begin{figure}[tb!]
   \centering
   \includegraphics[width=\columnwidth]{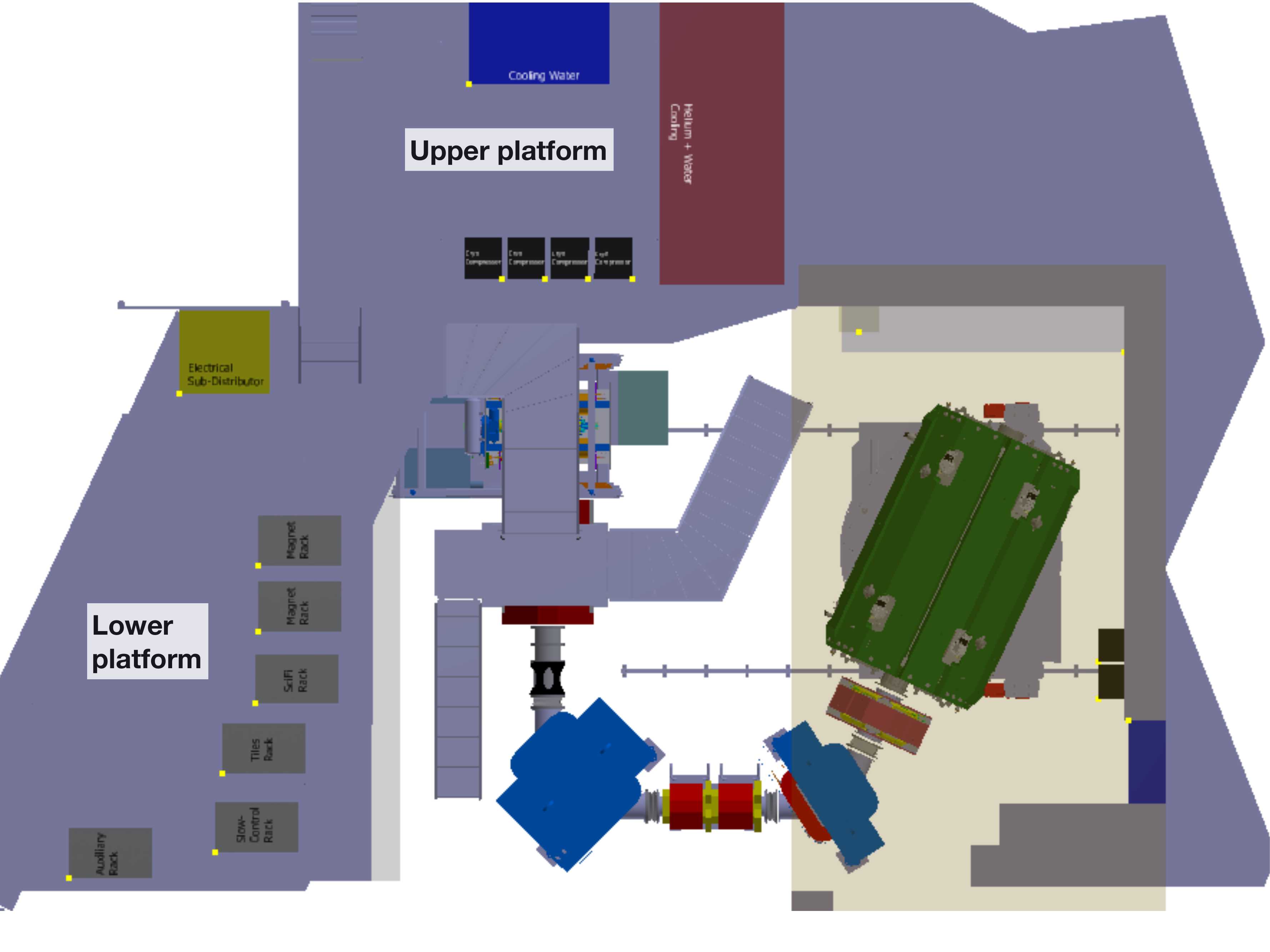}
   \caption{Top view of the $\pi$E5 experimental area showing the
     completed installation.  Also visible are the two new
     infrastructure platforms located on the shielding blocks above
     the area and the stairs leading down to the experiment.  The
     transparent beige-area marks the roof underneath which the Mu3e
     magnet has to be installed.}
   \label{fig:area_front}
\end{figure}

\begin{figure}[t!]
   \centering
   \includegraphics[width=\columnwidth]{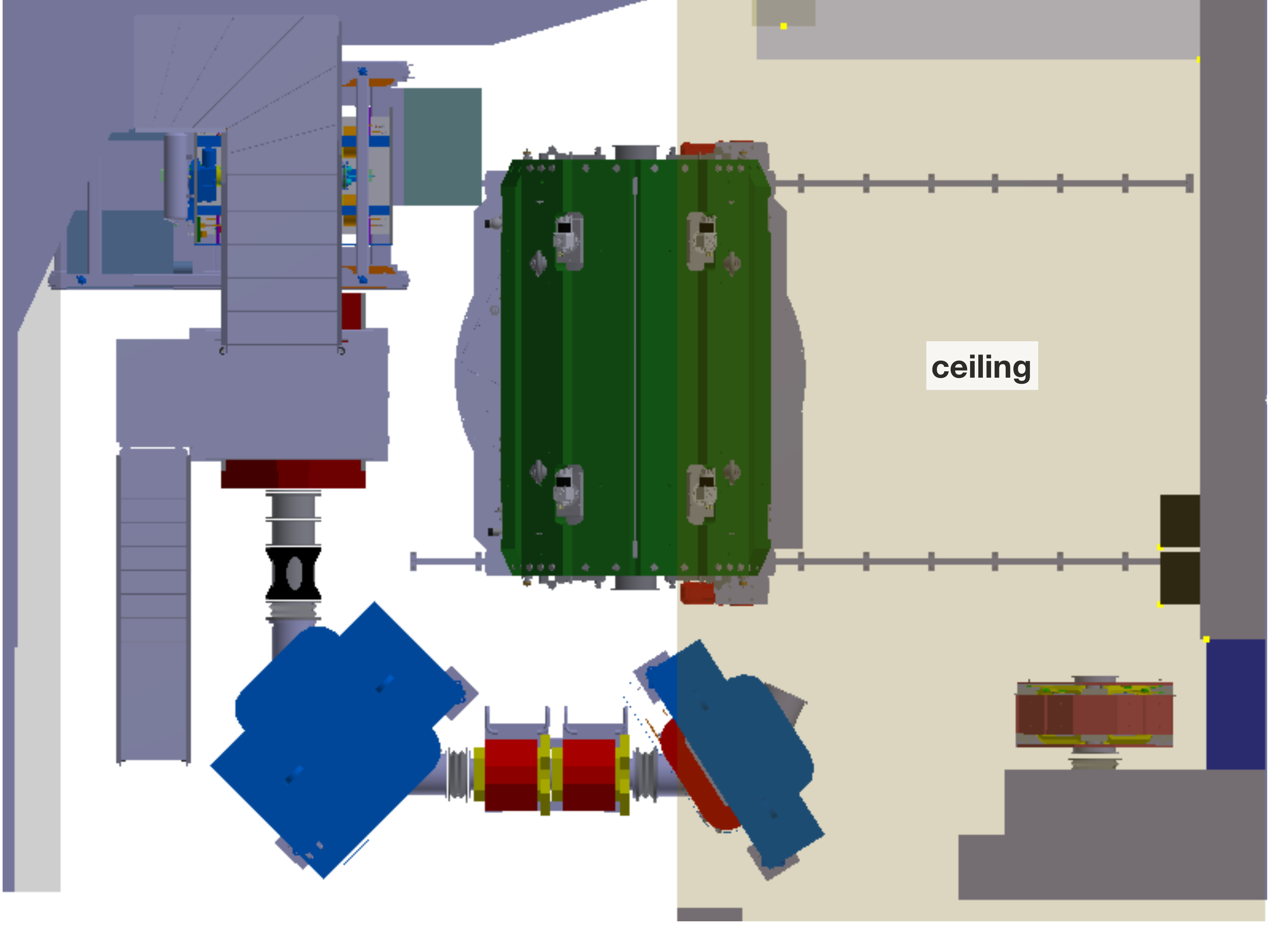}
   \caption{Position of the magnet when lowered into the experimental
     area onto its rail system.  The rail system allows to move the
     30-ton magnet underneath the ceiling and turn it in line with the
     rest of the beam elements.}
   \label{fig:area_magnet_insertion}
\end{figure}

\begin{figure}[t!]
   \centering
   \includegraphics[width=\columnwidth]{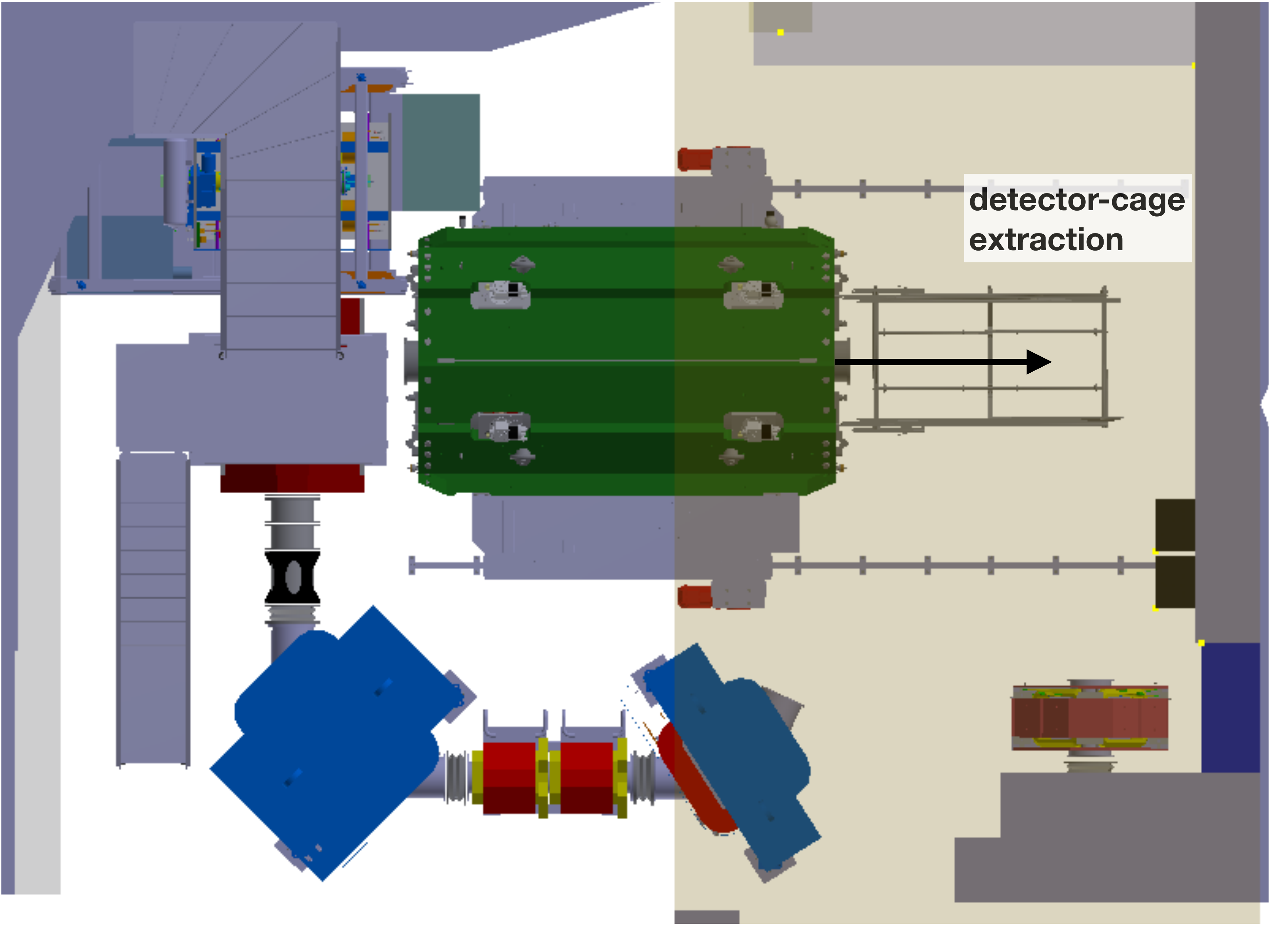}
   \caption{Maintenance position of the magnet on the rail system used
     to extract the detector-cage onto its transport unit. }
   \label{fig:detector_extraction}
\end{figure}

Due to the limited space in the front part of the $\pi$E5 area, as can
be seen in \autoref{fig:area_front}, as well as the fact that the
Mu3e magnet is located underneath the roof formed by the $\pi$E3 area
above, a rail system is required to move the Mu3e magnet
from a position where it can be lowered down into the experimental
area by crane -- shown in \autoref{fig:area_magnet_insertion} -- to
its final position underneath the roof shown in
\autoref{fig:area_front}.  The crane operation will be a challenging
one and extra degrees of rotational freedom included in the rail
system are needed to allow for such a movement of the 30-ton magnet to
its final position under the roof.  In addition, a small crane is
needed to move the last quadrupole QSM41 away from its position along
the beamline in order to allow the free movement of the magnet.

\autoref{fig:detector_extraction} shows the magnet in its
maintenance position.  This position allows the Mu3e solenoid to be
rotated in such a way that the full detector-cage can be extracted
onto its transport support structure for repairs, maintenance or
transportation.  A detailed description on how the detector can be
extracted onto the support structure can be found in
\autoref{sec:MechanicalIntegration}.

\begin{figure}
   \centering
   \includegraphics[width=\columnwidth]{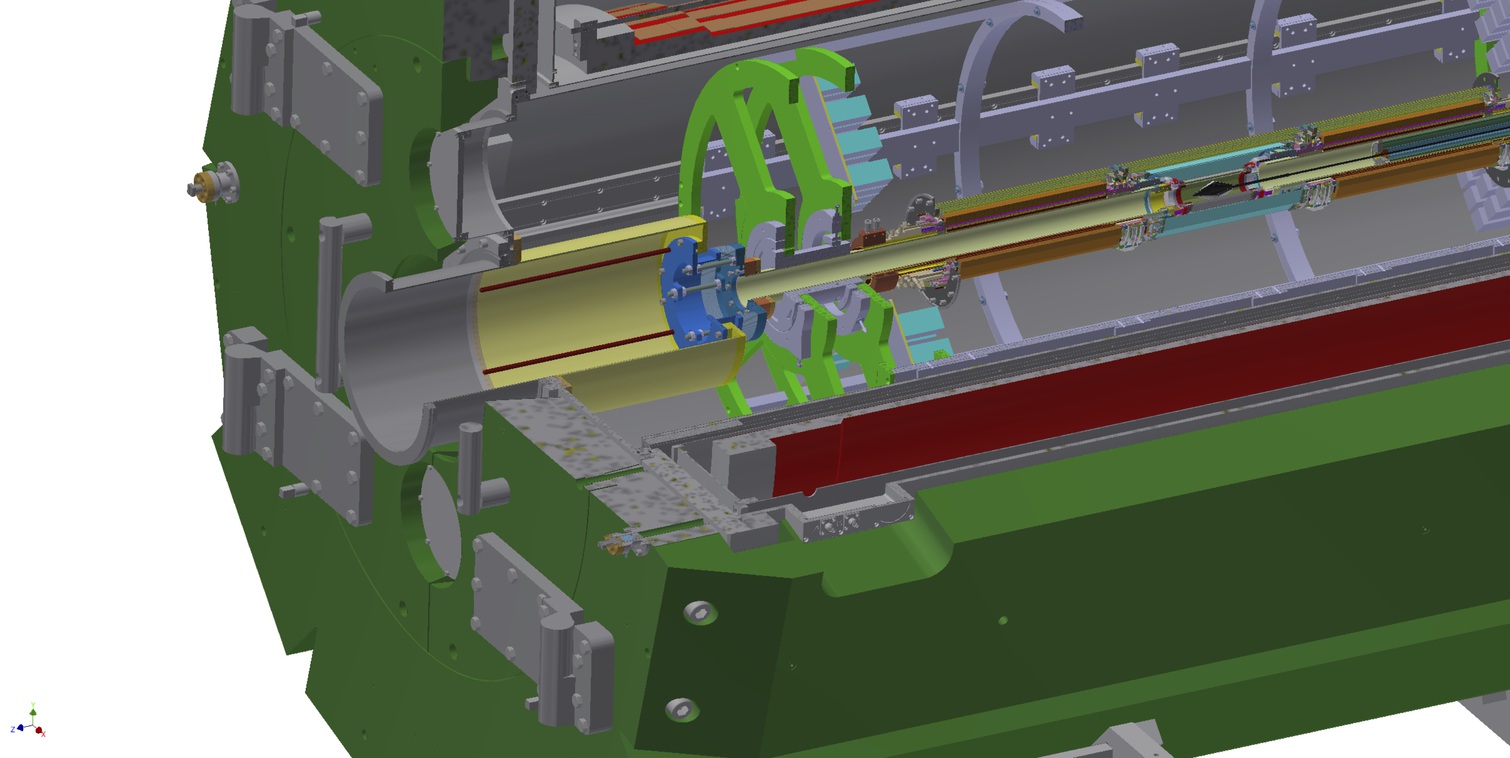}
   \caption{Connection of the Mu3e detector to the muon beam line is achieved through a custom bellows assembly inside the solenoid.}
   \label{fig:beamline_detector}
\end{figure}

Finally, the detailed coupling mechanism of the beam line to the solenoid magnet is described. 
The components are shown in \autoref{fig:beamline_detector}. 
The standard ISO-320-K beam line vacuum tube, with its upstream bellows connection protrudes into the magnet allowing a maximum acceptance of the converging beam envelopes before entering a custom 150-mm diameter intermediate bellows connection to the final 60-mm diameter beam tube of the Mu3e detector. 
The mounting sequence is as follows: 
In a first step the detector system is mounted with its cage inside the magnet bore and fixed in position.
Subsequently, the internal beam line elements are mounted onto the inside end of the He-tight flange of the magnet bore, which is then bolted onto the cryostat.
In order to achieve a vacuum tight connection between the custom bellows assembly and the Mu3e detector cage beam-flange, the final screws are tightened from the inside of the ISO-320-K vacuum tube, so pressing on the O-ring seal.
As a last step, internal tensioning supports for the bellows are mounted and securely fixed in place to prevent the bellows from collapsing when evacuated.


\chapter{Stopping Target}
\label{sec:Target}

\chapterresponsible{Malte}

\nobalance

\begin{figure}[b!]
	\centering
  	\includegraphics[width=0.4\textwidth]{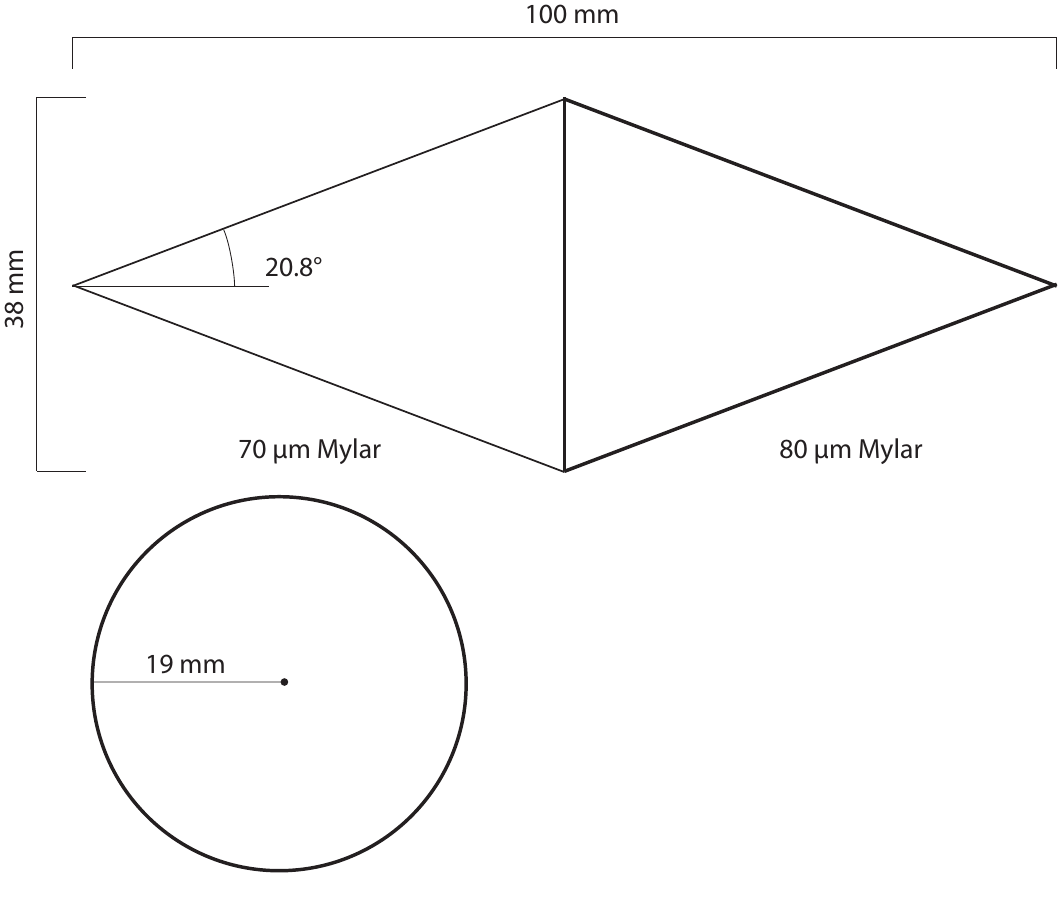}
	\caption{Dimensions of the baseline design target. The muon beam enters from the left.
		Note that the material thickness is not to scale.}
	\label{fig:Target}
\end{figure}

The main challenge for the design of the stopping target is to
optimise the stopping power, while also minimising the total amount of
material in order to reduce both backgrounds and the impact on the
track measurement.  Therefore the stopping target should contain just
enough material in the beam direction to stop most of the muons, which
is facilitated by a moderator in the final part of the beam line, but
should be as thin as possible to minimise the material in the flight
direction of decay electrons entering the detector acceptance.  Usage of a
low-$Z$ material is advantageous as photon conversion and large-angle
Coulomb scattering are suppressed.  In addition, the decay vertices
should be spread out as wide as possible in order to reduce accidental
coincidences of track vertices and to produce a more or less even
occupancy in the innermost detector layer.

\section{Baseline Design}
\label{sec:TargetBaselineDesign}

\begin{figure}[b!]
	\centering
		\includegraphics[width=0.35\textwidth]{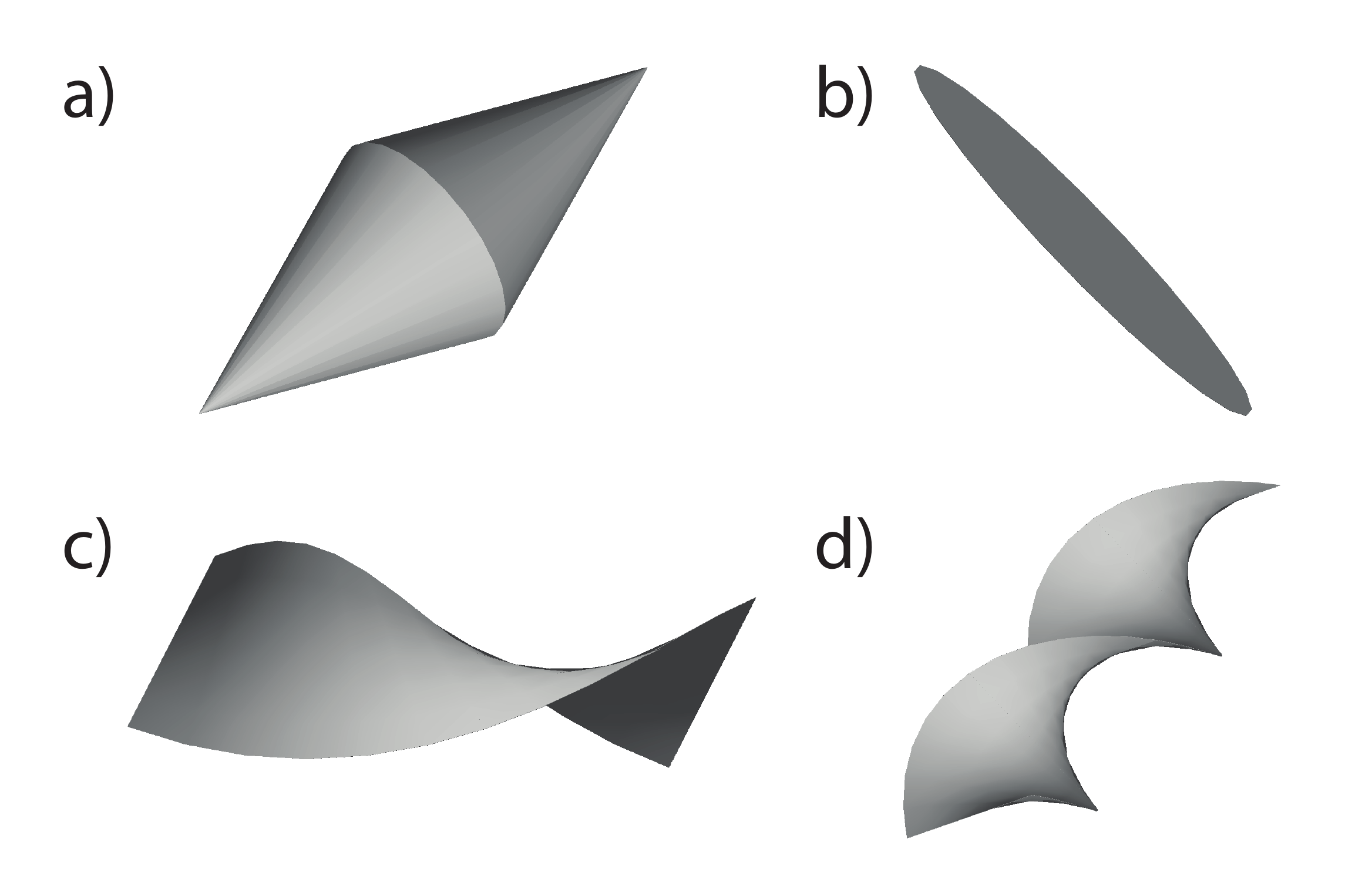}
	\caption{Target shapes studied. a) Is the default hollow double cone, b) a
	simple plane, c) a single-turn garland and d) a double-turn garland. For the
	chiral shapes c) and d), both senses of rotation were tried.}
	\label{fig:TargetShapes}
\end{figure}

\begin{figure}[b!]
	\centering
		\includegraphics[width=0.42\textwidth]{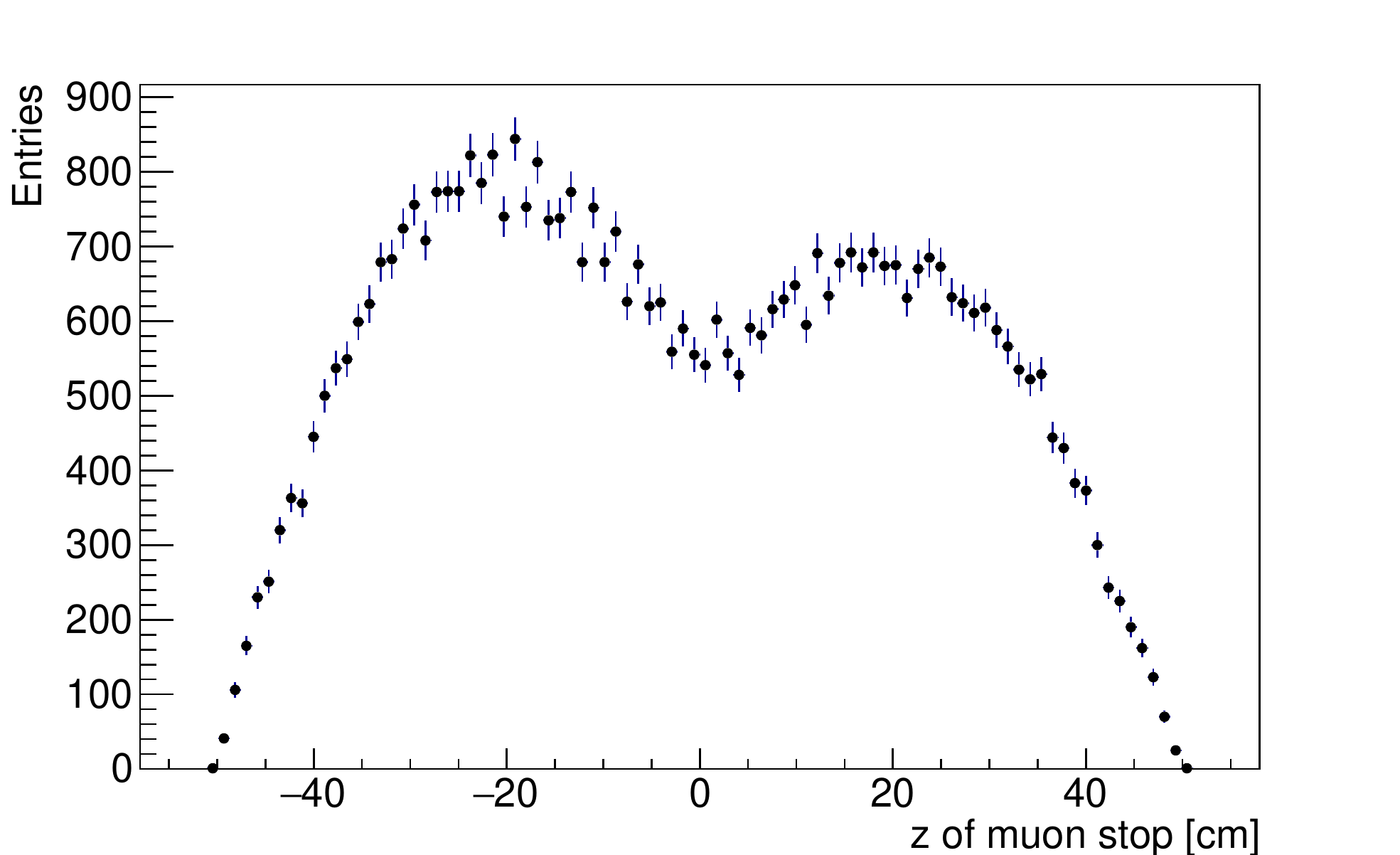}
	\caption{Simulated stopping distribution along the beam ($z$) direction for the
	baseline target.}
	\label{fig:target-zstop}
\end{figure}

These requirements can be met by a hollow double cone target \`a la SINDRUM
\cite{Bellgardt:1987du, bertl2008}. In our baseline design (see
\autoref{fig:Target}), the target is made from $\SI{70}{\micro \meter}$ of
Mylar in the front part and $\SI{80}{\micro \meter}$ Mylar in the back
part, with a total length of $\SI{100}{mm}$ and a radius of $\SI{19}{mm}$. 
This leads to an incline of \ang{20.8} of the target surface with regards to the
beam direction.
The projected thickness is thus $\SI{197}{\micro \meter}$ for the front and
$\SI{225}{\micro \meter}$ for the back part, giving a total of $\SI{422}{\micro \meter}$
of Mylar corresponding to 0.15\% of a radiation length.
The mass of the Mylar in the target is \SI{0.671}{g}.
The total area of the target is \SI{6386}{mm^2}.

\begin{figure*}[t!]
	\centering
		\includegraphics[width=0.98\textwidth]{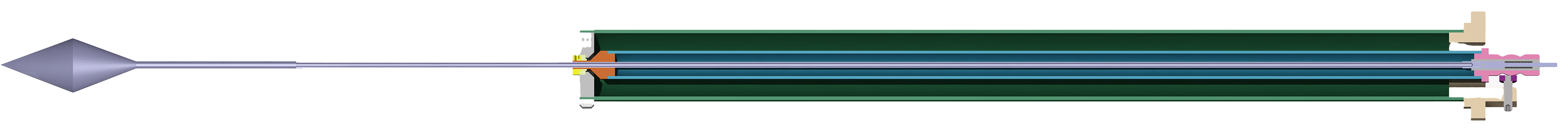}
	\caption{Cross section of target support and alignment mechanism. Muons hit the target from the left. The stopping target is mounted on a thin carbon tube which is steered and fixed in the support structure. The rear end of the support structure consists of an alignment mechanism to adjust the position of the target.}
	\label{fig:target-support}
\end{figure*}

We have studied the stopping power and material budget for a variety of target
shapes (see \autoref{fig:TargetShapes}) 
and found that for the given beam parameters and geometrical constraints,
the double cone offers the highest stopping fraction with the least material.
The simulation was performed with Mylar as the target material, a previous study
using aluminium however gave very similar results. 
The stopping distribution along the beam axis for the baseline target is shown 
in \autoref{fig:target-zstop}; about \SI{95.5}{\percent} of the muons reaching the
target are stopped, the reminder ends up in the downstream beam pipe, the downstream 
support of the first pixel layer and the sensors of the first pixel layer.

\section{Production}
\label{sec:TargetProduction}

\begin{figure}
	\centering
		\includegraphics[width=0.49\textwidth]{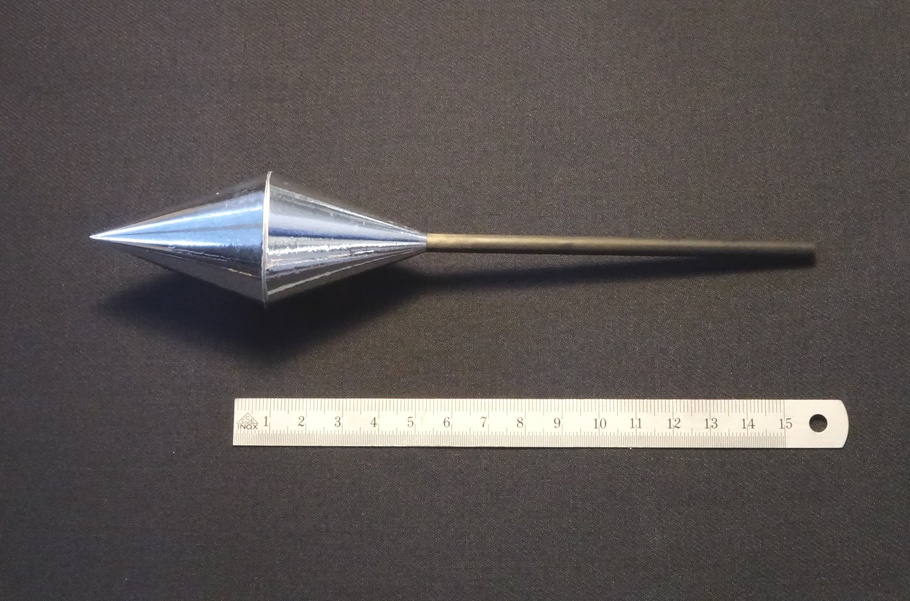}
	\caption{Hollow double-cone muon stopping target made of aluminised Mylar foil.}
	\label{fig:target-manufactured-final}
\end{figure}

At PSI, a manufacturing procedure was developed and a complete 
target was produced, see \autoref{fig:target-manufactured-final}.
Each single hollow cone of the double cone structure is manufactured 
separately and is a sandwich structure consisting of 2 or 3 rolled up thin Mylar 
foils glued together with epoxy glue. 
The thickness of the individual Mylar foils and the combination of several foils 
are chosen to match best with the desired final thickness. Finally, the two individual cones are glued together to build up the hollow 
double cone structure.

The inner and the outer foil in each sandwiched stack is aluminium coated and the orientation of the aluminium layers is such that the inner and outer surface of the cones features an aluminium layer. The conductive surfaces, in combination with the mounting on a conductive carbon tube avoid a possible charging up of the target due to the high stopping rate of positive muons.

\section{Support}
\label{sec:TargetSupport}

\begin{figure}
	\centering
		\includegraphics[width=0.45\textwidth]{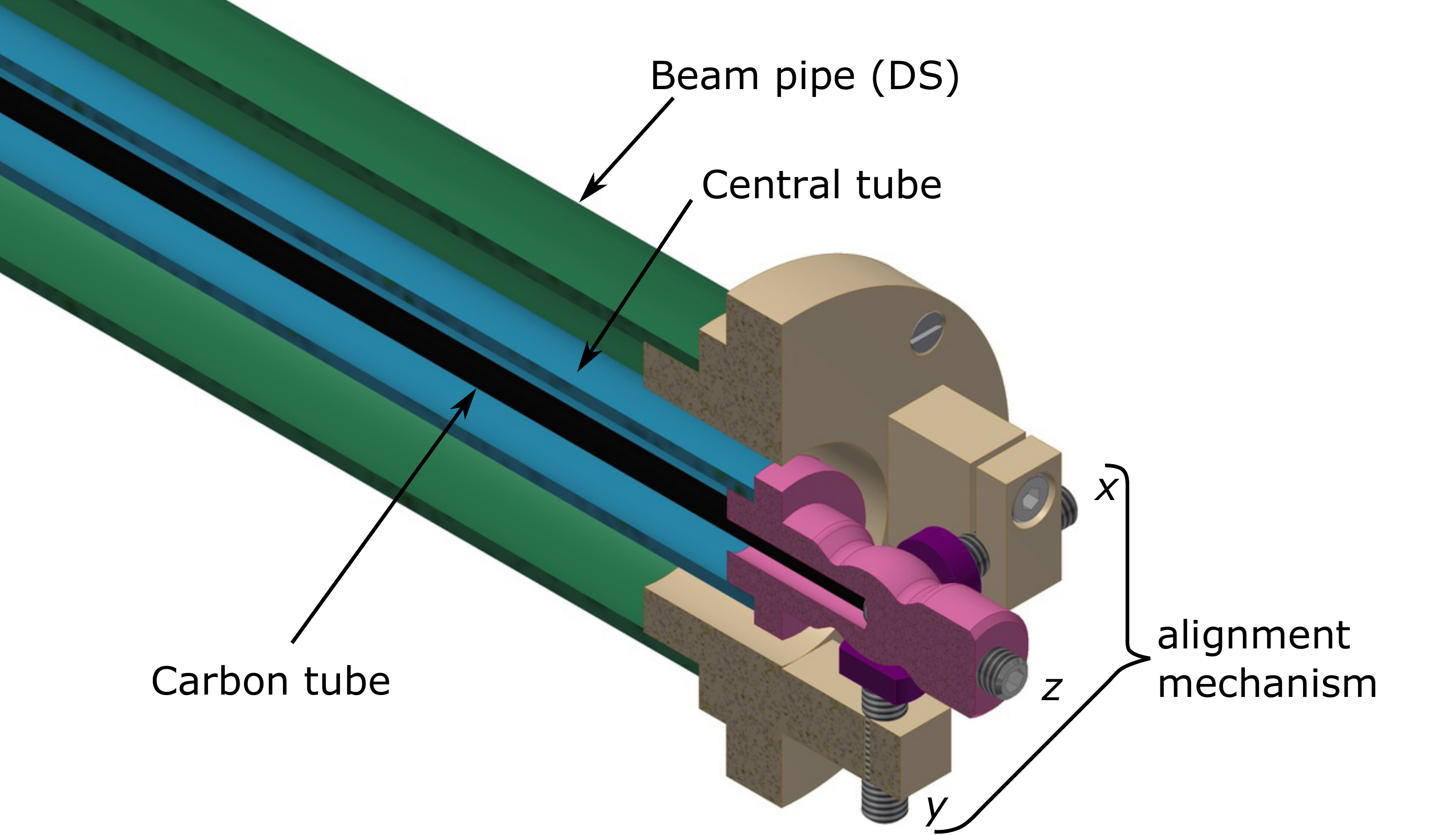}
	\caption{Cross section of alignment mechanism. The setup is spring-loaded towards the two screws and allows an adjustment of the target position in 3 coordinates. The direction of the spring is in the bisecting line with respect to the two screws and is in this view therefore hidden by the holder.}
	\label{fig:target-adaptation-mechanism}
\end{figure}

The double cone structure will be glued on a carbon tube which will be
fixed in a dedicated support structure with an alignment mechanism.
\autoref{fig:target-support} shows a cross section of the complete
target system consisting of stopping target, carbon tube and support,
while \autoref{fig:target-adaptation-mechanism} shows an enlarged view
of the rear end of the support structure consisting of the alignment
mechanism. The target support structure will be placed on the
downstream side of the experiment in order not to disturb the incident
muon beam.

The carbon tube has an
inner diameter of \SI{5}{mm} and will be glued on the tip of the
downstream cone of the stopping target. Along the first \SI{10}{cm}
downstream of the target the original wall thickness of \SI{0.5}{mm}
of the carbon tube is reduced to $\sim$\SI{0.125}{mm} by means of
centerless-grinding to reduce the material budget in the
central region of the detector.

To avoid possible vibrations of the target due to a long lever arm the
carbon tube is not only rigidly fixed in the alignment mechanism,
but also guided in a joint at the front end of the structure close to
the target itself.

The alignment mechanism allows an adjustment of the
target position in all 3 coordinates. To ensure sufficient clearance
between the target and the innermost layer of the silicon detectors,
the range of movement for the target is limited to
$\pm~\SI{2}{mm}$ in x- and y-directions, and $\pm~\SI{4}{mm}$ in
z-direction.  This is achieved with a limited range for the adjustment
screw at the rear end of the support structure, in conjunction with
the transformation ratio due to the different lengths of carbon tube
and support structure.

The central tube of the support structure
hosting the carbon tube and connected to the holder at the end is spring-loaded
towards the adjustment screws to allow for a hysteresis-free
adjustment of the target.


\chapter{Pixel Tracker}
\label{sec:Pixel}

The Mu3e pixel tracker provides precision hit information for the track
reconstruction of the electrons produced in muon decays.
Achieving the best possible vertex and momentum resolution measurements for these electrons is of
key importance to the success of the experiment.
Due to the dominance of multiple scattering, a rigorous minimisation of the material in the active region of the tracking detector is critical.
For this reason, the tracker relies on High-Voltage Monolithic Active Pixel Sensors (HV-MAPS)~\cite{Peric:2007zz}, thinned to $\SI{50}{\micro\meter}$ and mounted on a low mass service flexible printed circuit (``flex''). The detector is operated inside a dry helium atmosphere and cooled by helium gas flow to further reduce multiple scattering.

\section{Overview of the Pixel Tracker}
\begin{figure}[bht]
        \centering
                \includegraphics[width=\columnwidth]{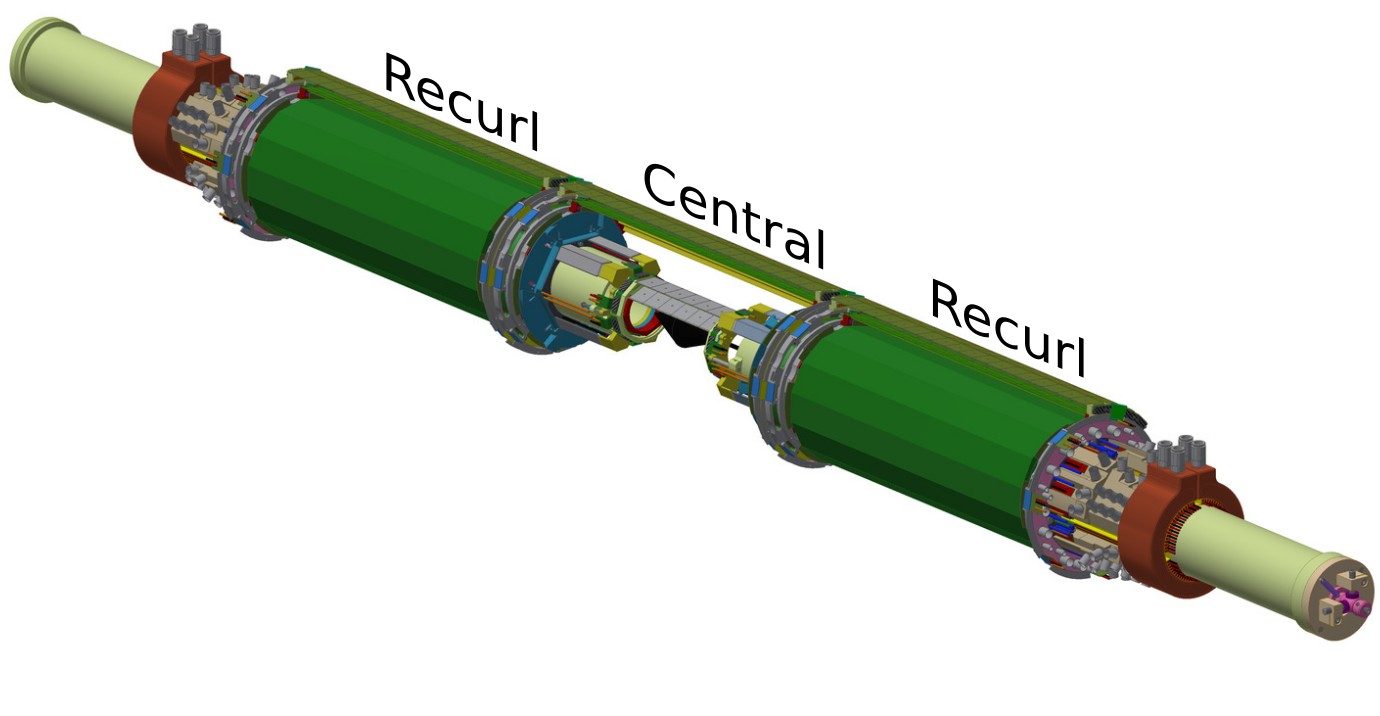}
        \caption{The Mu3e pixel tracker with the central pixel tracker in the
          middle and the two recurl stations down- and upstream. Some modules in the
          central pixel tracker have been removed for visibility.}
        \label{fig:PixelTracker}
\end{figure}

The Mu3e pixel tracker consists of three parts,
the central pixel tracker and two recurl stations, see
\autoref{fig:PixelTracker}.
Pixel layers in the central tracker provide the main hits used for the reconstruction of tracks and of the decay vertex associated with multiple tracks.
The hits detected in the recurl stations allowing the reconstruction of tracks with higher
purity and improved momentum resolution.

Throughout the pixel tracker all \mupix sensors have the same
dimensions, with an active area of $20 \times \SI{20}{\mm\squared}$. A small non-active area of the sensor chip houses peripheral digital and analogue circuitry, enlarging the chip in one dimension to about \SI{23}{\mm}. The chips are mounted on High Density Interconnect (HDI) circuits, which
incorporate both signal and power lines as aluminium traces on thin polyimide substrates.
The HDIs provide power and bias voltage, and transmit control signals and data. The data over, up to, 3 differential lines per chip at a bandwidth of $\SI{1.25}{Gbit/s}$ per line.

The \mupix chips are bonded to the HDI using \emph{Single-point Tape Automated
Bonding} (SpTAB) without the need for additional bonding material \cite{Oinonen:2005rm}.
Pixel modules are constructed from self-supporting sensor-HDI-polyimide ladders. These host between 6 and 18 sensors,
and represent a total radiation length of approximately $X/X_0=0.115\%$ per layer.

\begin{figure*}[tb!]
        \centering
                \includegraphics[width=0.85\textwidth]{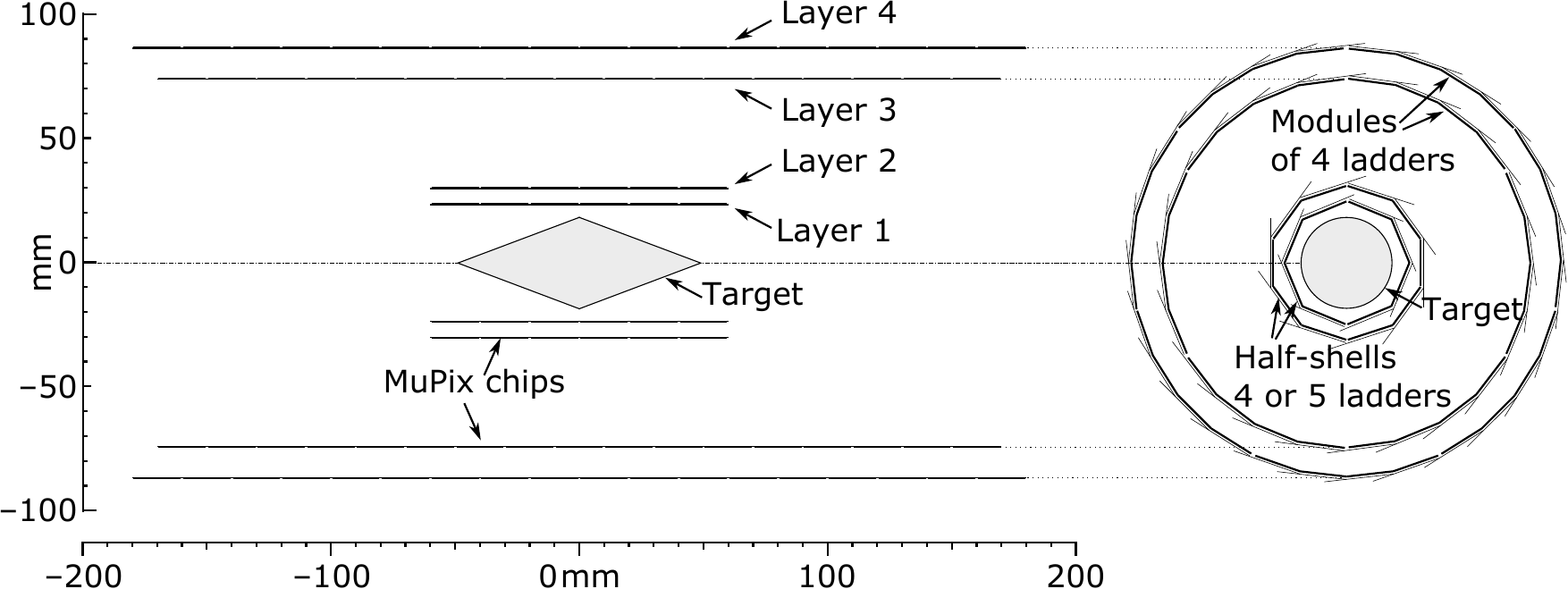}
        \caption{Geometry of the central pixel tracker including the target.}
        \label{fig:SketchPixelGeometry}
\end{figure*}

\subsection{Pixel Tracker Layout}
The central pixel tracker has four layers of \mupix sensors: two inner layers (layer~1 and~2) at small radii and two outer layers (layers~3 and~4 ) at larger radii. The inner and the outer layers are both arranged as double layers, pairs that work together to provide a track trajectory.
The layout of the central pixel tracker is shown in \autoref{fig:SketchPixelGeometry}
and the corresponding geometrical design parameters are listed in
\autoref{tab:TrackerSpecifications}. The recurl stations have only two pixel layers (layers~3 and~4) which are identical in design to the outer layers in the central tracker.

\begin{figure}[bht]
        \centering
                \includegraphics[width=\columnwidth]{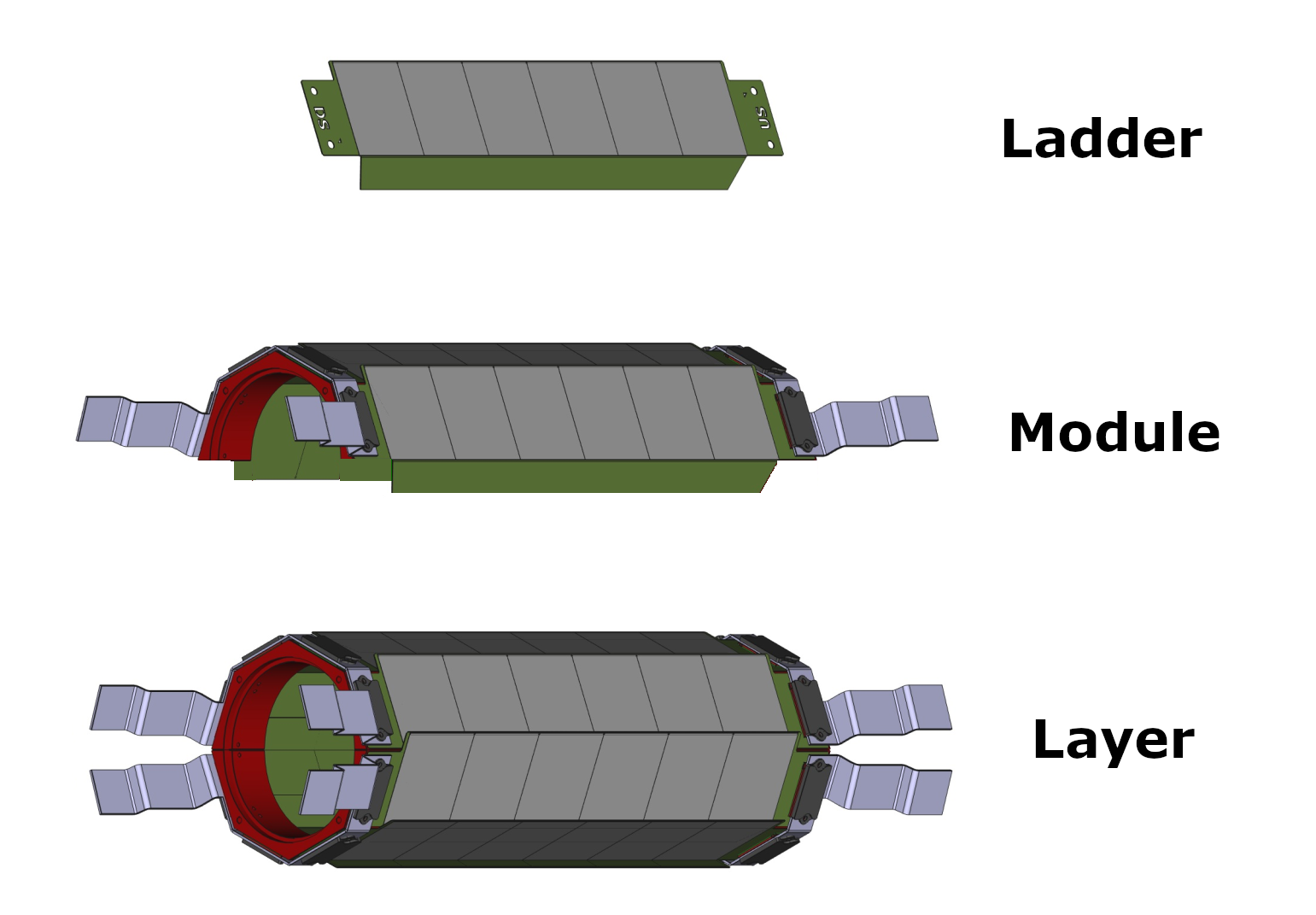}
        \caption{3D-model reproduction of the Tracker Layer 1 assembly, from single ladder to module to full layer.}
        \label{fig:PixelLayerConstruction}
\end{figure}

Each tracking layer is composed of mechanically robust modules, which in turn
integrate 4 or 5 of the more fragile sub-modules (ladders). Ladders represent the smallest mechanical unit in the tracker. A 3D-model representing the mechanical integration of these components for Layer~1 can be found in \autoref{fig:PixelLayerConstruction}. The same principle applies to all layers.

\begin{table*}[bth]
  \centering
  \small
  \begin{tabular}{lrrrr}
    \toprule
    layer                               &  1  &  2  &  3   &  4   \\    \midrule
    number of modules                   & 2        & 2       & 6        & 7        \\      
    number of ladders                   & 8        & 10       & 24       & 28       \\      
    number of \mupix sensors per ladder & 6        & 6        & 17       &18      \\
    instrumented length [$\SI{}{\mm}$]  & 124.7  & 124.7  & 351.9  & 372.6  \\
    minimum radius [$\SI{}{\mm}$]       & 23.3   & 29.8     & 73.9      & 86.3  \\
    \bottomrule
  \end{tabular}
  \caption{Pixel tracker geometry parameters of the central barrel.
   The radius is defined as the nearest distance of \mupix sensor w/o polyimide
   support to the symmetry axis (beam line).}
  \label{tab:TrackerSpecifications}
\end{table*}

The inner tracking layers, 1 and 2, are of equal length, $\SI{12}{\cm}$, hosting 6 chips per ladder. These provide the vertexing in Mu3e.
The inner layers have full overlap in z with the muon stopping target, which has a length of $\SI{10}{\cm}$. The outer and recurl pixel tracker modules are significantly longer and provide
a larger acceptance for downstream and upstream going particles. The outer and recurl layers are critical for selection of high-quality tracks and for the momentum resolution in Mu3e.
The outer layers instrument a region with a length of $\SI{34}{\cm}$
(layer 3) and $\SI{36}{\cm}$ (layer 4), corresponding to 17 and 18 \mupix chips, respectively.

\begin{figure}[thb]
        \centering
                \includegraphics[width=\columnwidth]{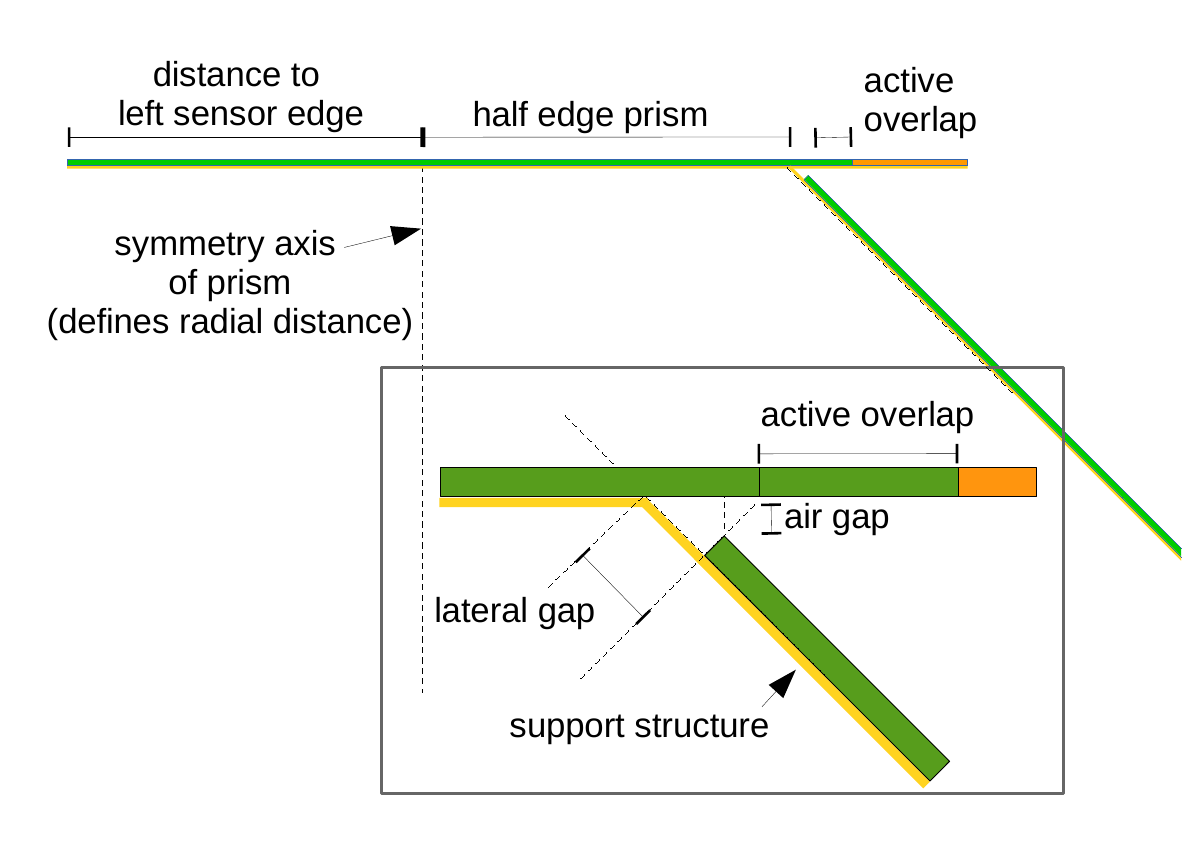}
        \caption{Design of the \mupix ladder overlap region of pixel layer~1. 
        The green region indicates the active part of the sensor, the
        orange part the inactive periphery and the yellow part the 
        polyimide support structure. 
        The lower edge shows a zoom into the overlap region.
        Note that the sensor thicknesses are not to scale.}
        \label{fig:OverlapRegionLayer1}
\end{figure}
The \mupix ladders are mounted with a small overlap, in the tangential direction, of the active area
with the adjacent ladder, see 
\autoref{fig:OverlapRegionLayer1}. 
The lateral overlap is $\SI{0.5}{\mm}$, which ensures high acceptance
for low momentum tracks and also helps with the alignment of the pixel tracker.
There is a small physical clearance, along the radial direction, between overlapping 
sensors of $\approx\SI{200}{\micro\meter}$. 

\subsection{Signal path}
\label{sec:signalPath}
\begin{figure*}[th!]
        \centering
                \includegraphics[width=0.85\textwidth]{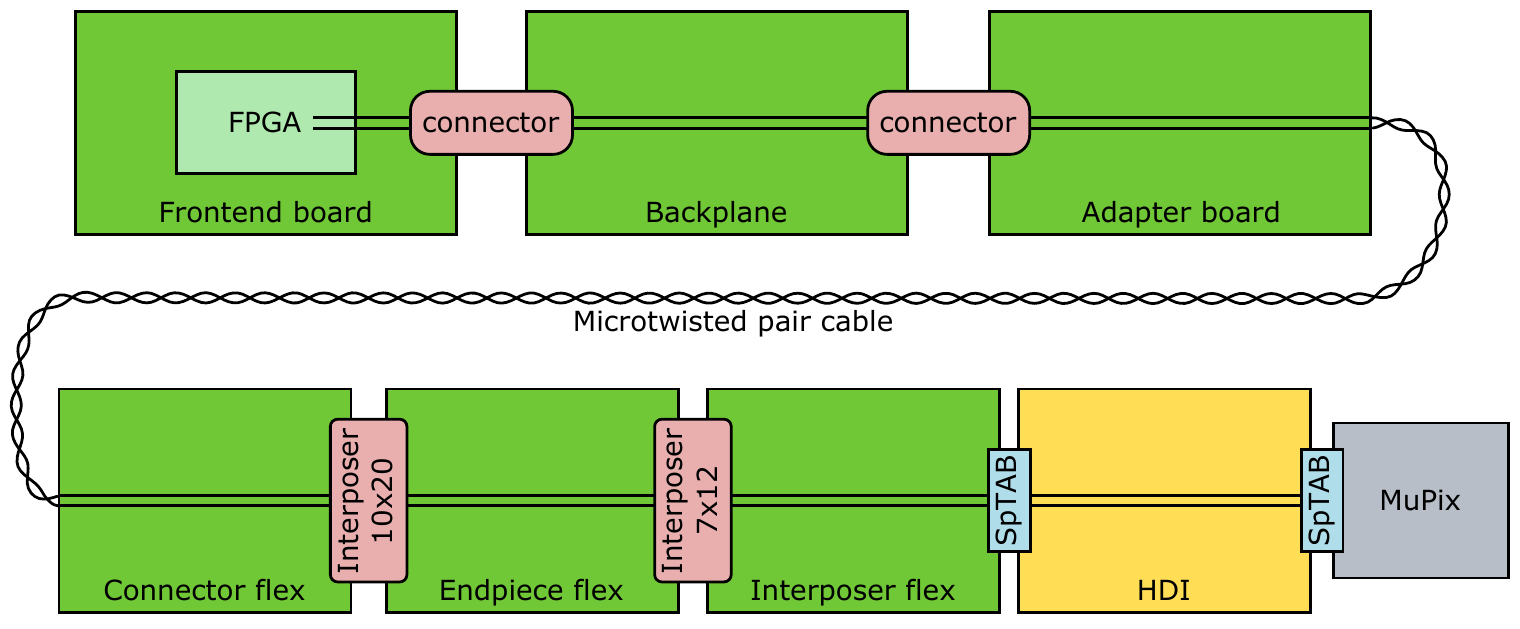}
        \caption{Signal path between \mupix chip and FPGA for a differential readout line. The parts on the top are located on the SSW, the ones on the bottom in the tracker barrels.}
        \label{fig:pixelSignalConnectFEB}
\end{figure*}

The signal connection between the front-end FPGA board, located on the service support wheels (SSW, \autoref{sec:mechintSupplyAndRouting}), and the \mupix chips is purely electric and differential with impedance-controlled lines. 
\begin{sloppypar}
A schematic path of a differential signal is shown in \autoref{fig:pixelSignalConnectFEB}. The FPGA board is plugged into a back-plane where basic routing is performed. The distance to the detector (about \SI{1}{\metre}) is bridged with micro-twisted pair cables, each consisting of two copper wires with \SI{127}{\micro\metre} diameter, insulated with \SI{25}{\micro\metre} polyimide and coated together with a polyamide enamel. The differential impedance of this transmission line is $Z_\text{diff}\approx\SI{90}{\ohm}$. 50 such pairs are combined to a flexible bundle with a diameter of less than \SI{2}{\mm}. At both ends, the wires are soldered onto small PCBs, plugged into zero-insert-force (ZIF) connectors. On the detector end, the signals are routed on flexible PCBs to the HDI (see \autoref{sec:HDI}). The connections between the components use industry-standard parts (back-plane connectors, gold-ball/gold-spring array interposers) and SpTA-bonding, as shown in the figure.
\end{sloppypar}

\section{Pixel Tracker Modules}
\label{sec:PixelMechanics} 

The pixel tracker modules of all layers have a very similar design. 
They consist of either  four or five instrumented ladders mounted to a polyetherimide (PEI)
endpiece at the upstream and downstream ends.
The ladders host between 6 and 18 \mupix chips glued and electrically connected to a single HDI circuit.
For the inner two layers, self-supporting half-shells define a module, with each 
half shells comprising four (layer~1) or five (layer~2) short ladders with
six \mupix sensors. 

For the outer two layers, a single module is an arc-segment, corresponding to either 1/6th (layer~3) or 1/7th (layer~4) of a full cylinder. Outer layer modules comprise four ladders with either 17 (layer~3) or 18 (layer~4)  \mupix sensors. 

\begin{figure}[tb]
        \centering
                \includegraphics[width=\columnwidth]{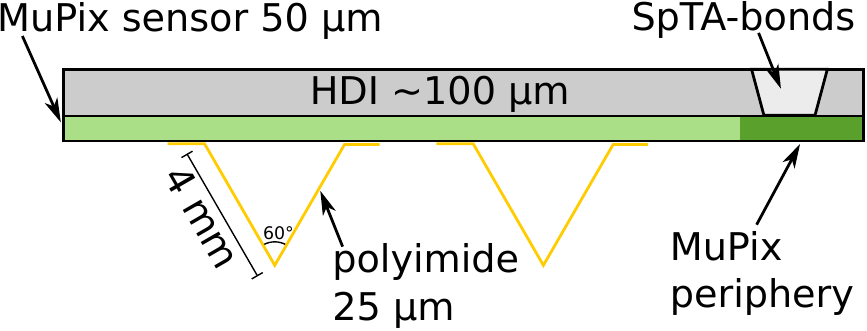}
        \caption{Cross section of an outer layer ladder. From top to bottom: HDI, \mupix
          sensor, polyimide support structure. Not to scale.}
        \label{fig:outer_layer_cross_section}
\end{figure}

The longer outer layer ladders require additional reinforcement to achieve a 
mechanical stability comparable to the shorter ladders of the inner layers. 
For this purpose, two polyimide strips folded into a v-shape (yellow structure in
\autoref{fig:outer_layer_cross_section}) are glued to each ladder on the inner side. The obtained v-channels also serve as high flux helium cooling channels.

\begin{figure}[tb]
        \centering
                \includegraphics[width=\columnwidth]{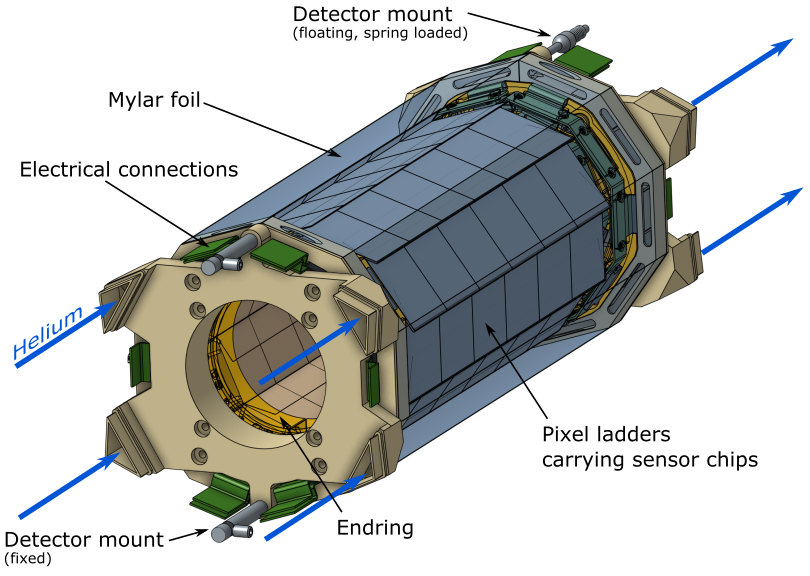}
        \caption{Schematic view of the vertex pixel barrel. Each of the 8+10 ladders carry 6 sensor chips. Endrings (split in halves) provide the mechanical support to the ladders.}
        \label{fig:InnerTracker}
\end{figure}

\subsection{Inner Layer Modules}
Modules for layers 1 and 2 are constructed by mounting four or five ladders to upstream and downstream half-shell endpieces.
The half shells are strengthened with a $\SI{25}{\micro\meter}$ polyimide foil, glued to the \mupix ladders on the inward facing side. The foil also restricts helium from flowing through the gaps between   ladders. After mounting and gluing, the half module represents a mechanically robust structure.
To construct the full layers 1 and 2, half-shell modules are mounted on
two endrings, see \autoref{fig:InnerTracker}.

\begin{figure}[tb]
        \centering
                \includegraphics[width=\columnwidth]{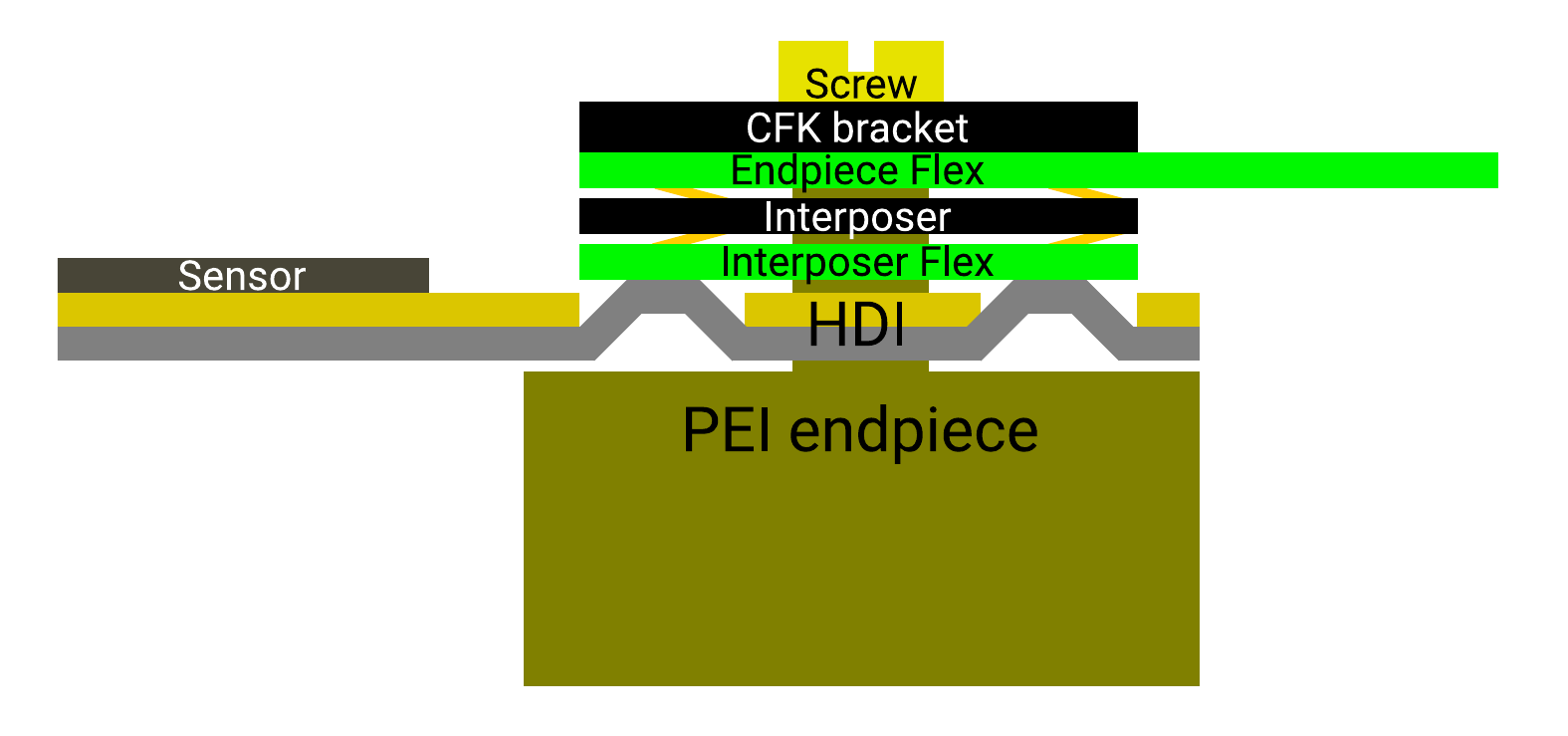}
        \caption{Schematic representation of the connections within an interposer at the end of an HDI flex.}
        \label{fig:interposer_connections}
\end{figure}

The electrical connection to the outside is made through multilayer copper-polyimide 
interposer flexible printed circuit which are SpTA-bonded to the HDI just outside the
 active region at the position of the endpieces (see \autoref{fig:interposer_connections}).
The interposer flex is connected to an andpiece flex via the interposer, which provides a $7\times 12$ micro grid array of gold-spring contacts.

\subsection{Outer Layer Modules}
\label{sec:outer_pixel_layers}
\begin{figure}[tb]
        \centering
                \includegraphics[width=\columnwidth]{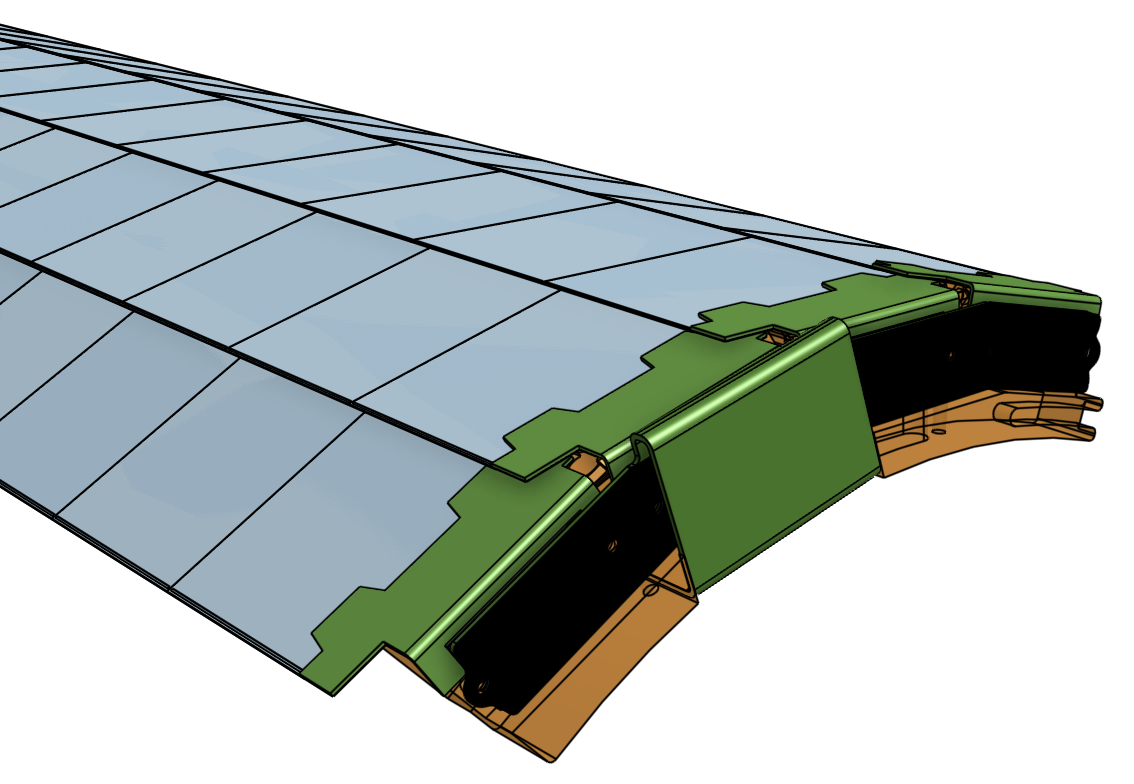}
        \caption{Schematic representation of a layer~4 module, integrating four long ladders
        with 18 \mupix sensors each. The picture shows one end, including the holding
        endpiece which also provides the electrical connections. An exploded view can be found in \autoref{fig:endpiece}.
}
        \label{fig:pixel_module}
\end{figure}

The thirteen outer tracking modules in the central detector, see \autoref{fig:pixel_module}, 
have a modular structure. 
Each module comprises four \mupix ladders which 
are glued to upstream and downstream module endpieces. As with the inner layers, the HDI circuit is SpTA-bonded to a multilayer copper-polyimide interposer flex circuit that connects to the $7\times 12$ micro grid interposer plate. Connections from four ladders are combined on the endpiece flex circuit. The final connection from a  module to the outside world is made through a further $10\times 20$ interposer plate, combining gold-spring and ball-grid array contacts, to which the module connects when mounted on its endrings. Layer 3 and 4 modules are assembled into full cylinders by mounting to a PEI endring. The design foresees a swing-in mechanism for installation, where modules are located by a dowel pin on each endring and fixed by two screws on the endpieces at either end.
Modules for the recurl stations are identical to the outer layer modules in the central region.

\subsection{Pixel Ladder Design}
\label{sec:ladder_designs}
The \mupix ladders integrate and support the 
pixel sensors. They have a compound structure optimised for a minimal
material budget.
The material composition for the inner and outer \mupix ladders is listed
in \autoref{tab:material_ladder} and amounts to
a radiation length of approximately $X/X_0=0.115\%$ per layer.

\begin{table*}
  \centering
  \small
  \begin{tabular}{lrrrr}
    \toprule
     & \multicolumn{2}{r}{Layer 1-2} & \multicolumn{2}{r}{Layer 3-4}   \\ 
     & thickness [$\SI{}{\micro\meter}$] & $X/X_0$
     & thickness [$\SI{}{\micro\meter}$] & $X/X_0$   \\    \midrule
    \mupix Si   & $45$  & $\SI{0.48e-3}{}$   & $45$  & $\SI{0.48e-3}{}$   \\ 
    \mupix Al   & $5$   & $\SI{0.06e-3}{}$   & $5$   & $\SI{0.06e-3}{}$   \\
    HDI polyimide \& glue  & $45$   &  $\SI{0.18e-3}{}$   & $45$   &  $\SI{0.18e-3}{}$   \\
    HDI Al      & $28$   &   $\SI{0.31e-3}{}$  & $28$   &  $\SI{0.31e-3}{}$  \\
    polyimide support   & $25$ & $\SI{0.09e-3}{}$   & $\approx 30$ & $\SI{0.10e-3}{}$   \\
    adhesives        & $10$   & $\SI{0.03e-3}{}$   & $10$   & $\SI{0.03e-3}{}$   \\   \midrule
    total   & $158$ & $\SI{1.15e-3}{}$  & $163$ & $\SI{1.16e-3}{}$  \\ 
    \bottomrule
  \end{tabular}
  \caption{Material budget of a \mupix ladder. The thicknesses and radiation
    length are given as an average over the $\SI{23}{mm}$ width of the ladder.}
  \label{tab:material_ladder}
\end{table*}

The mechanical stability of the outer \mupix ladders is mainly determined by the
two v-fold channels on the inner side which also serve as cooling channels. 
The inner layers do not have v-folds and are supported by the polyimide
support structure, see \autoref{fig:OverlapRegionLayer1}.

Every ladder is electrically divided into two halves and \mupix sensors are read out
from both ends of the ladder, i.e. three sensors per half ladder for the inner
layers and eight or nine sensors per half for the outer layers.
The components of the \mupix ladders and modules are described in the following in more detail.

\subsection{Sensors}
The \mupix sensors are monolithic pixel sensors in HV-CMOS~\cite{Peric:2007zz} technology. A full discussion of the functionality of the \mupix sensors as well as detailed  performance results can be found in \autoref{sec:hvmaps}. For the purpose of this section we discuss geometric properties and aspects relevant to the physical  connectivity between the sensors and the outside world.
Each sensor has a sensitive area of $20.48 \times \SI{20.00}{mm\squared}$ 
equipped with pixels of size $80 \times \SI{80}{\micro\meter\squared}$,
 corresponding to $256 \times 250$ pixels. 
 The overall dimensions of each chip are $20.66 \times \SI{23.18}{mm\squared}$, where the additional non-sensitive area hosts a comparator and memory cells for each pixel, as well as voltage regulation and digital logic circuits. All \mupix sensors will be thinned to a thickness of $\SI{50}{\micro\meter}$. The \mupix sensors can send data over up to
three serial links, each running at a data rate of $\SI{1.250}{Gbit/s}$.
The sensors require an operating voltage of $\SI{1.8}{\volt}$, 
a sensor bias voltage of up to $\SI{-100}{\volt}$, ground potential, and differential
signal traces for the readout, clock, slow control and monitoring.
SpTA-Bond pads of size $200\times\SI{100}{\micro\meter\squared}$
provide all electrical connections.
All pads are arranged in one row in the inactive peripheral area of the chip.

\subsection{High Density Interconnects}
\label{sec:HDI}
\begin{figure}
        \centering
                \includegraphics[width=0.8\columnwidth]{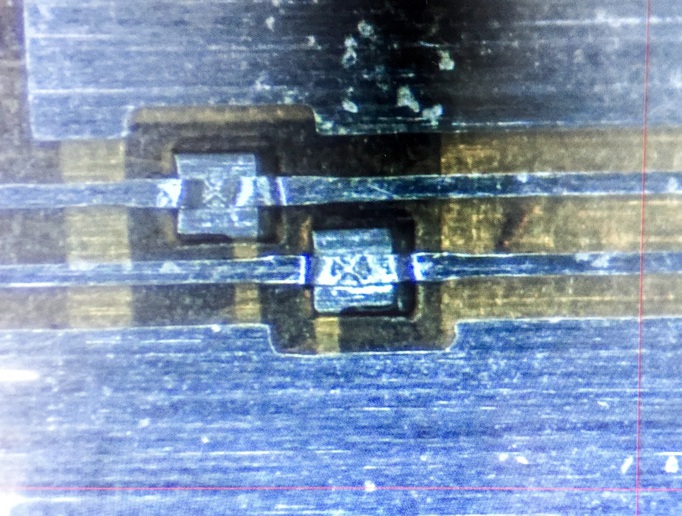}
        \caption{Photograph of two SpTA-bonds on a test flexprint produced by
          LTU Ltd~\cite{ref:LTU}.
        } 
        \label{fig:SpTABl}
\end{figure}

The High Density Interconnects (HDI) provide all electrical connections for the
\mupix sensors which are directly glued onto the HDI after which electrical connections are made using SpTA-bonding.
In order to achieve the target material budget of 
one per mille of a radiation length per layer, the HDIs 
have to be very thin and must not contain any high $Z$ materials.
The HDIs are produced by LTU Ltd.~(Ukraine)~\cite{ref:LTU}, who offer thin aluminium/polyimide technology as well as preparing the HDI  for SpTA-bonding. 
With the latter, aluminium traces are directly connected through vias either 
to chip pads or to 
other aluminium layers, see \autoref{fig:SpTABl} for an image of such bonds. This technique avoids the use of fragile wires and
also saves material. 
Tests with prototypes circuits have shown good
results~\cite{Noehte2016}.

\begin{figure}
        \centering
                \includegraphics[width=0.35\columnwidth]{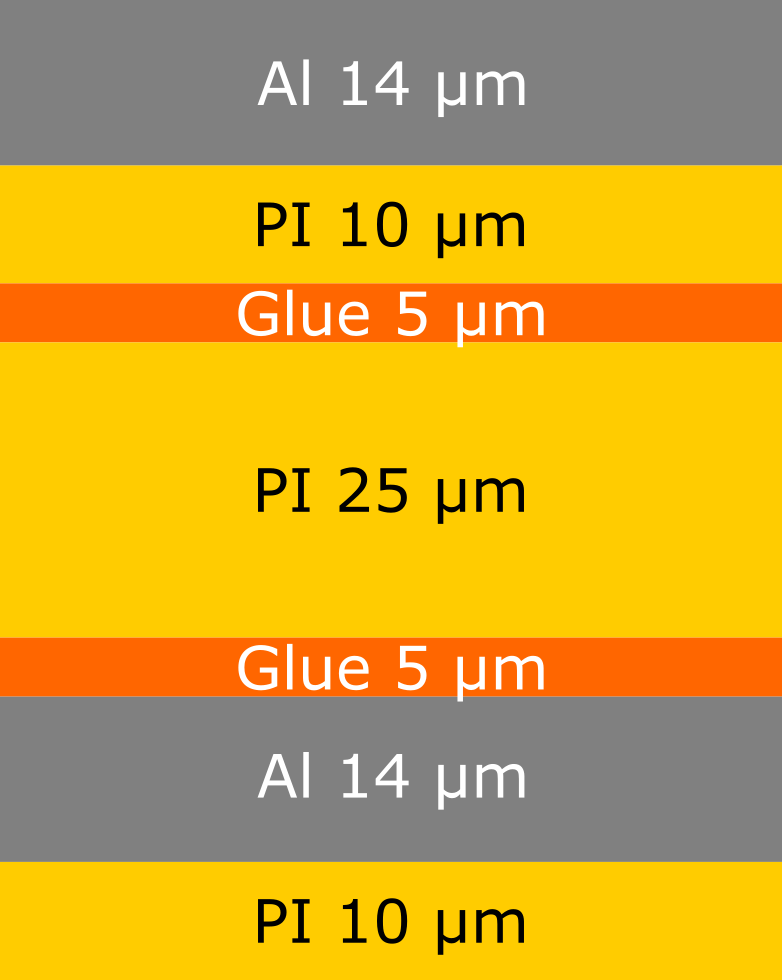}
        \caption{Stack chosen for the LTU produced 2-layer HDI circuits. PI=polyimide, Al=aluminium.
        } 
        \label{fig:Stack_Al}
\end{figure}

\begin{table}
  \centering
  \small
  \begin{tabular}{lrr}
    \toprule
    Material & Thickness [$\SI{}{\micro\meter}$] & $X/X_0$   \\    \midrule
    upper Al layer         & $14$  & $\SI{1.57e-4}{}$  \\
    insulator (PI)          & $35$  & $\SI{1.22e-4}{}$   \\
    glue                   & $10$  & $\SI{0.25e-4}{}$   \\
    lower Al layer         & $14$  & $\SI{1.57e-4}{}$   \\
    lower PI shield        & $10$  & $\SI{0.35e-4}{}$   \\   \midrule
    total                  & $83$  & $\SI{4.96e-4}{}$   \\ 
    \bottomrule
  \end{tabular}
  \caption{Material composition of the HDI for a $Z_\text{diff}=\SI{100}{\ohm}$ prototype.}
  \label{tab:hdi}
\end{table}

The performance of all electrical lines on the HDI is critical to the successful performance of the \mupix ladders. 
The traces for power and ground have to be large enough to provide 
the required power of up to 
$\SI{30}{W}$ on the longest \mupix ladders.
On the other hand, all traces should be as small as possible to fit them in 
the two aluminium layers available within the material budget.
All fast signals (serial links, clocks, resets, etc.) are implemented using the
LVDS standard to minimise cross-talk.
The differential impedances of all fast differential transmission lines 
are designed to match the specification.
Differential bus terminations are foreseen on the last chip in a row, i.e. at the
centre of the HDI.
All fast differential transmission lines are laid out underneath 
wide ground and VDD potential traces which serve as shielding and 
define the impedance~\cite{Noehte2019}.
With a minimum possible width of the aluminium traces of ~$\SI{63}{\micro\meter}$ (LTU),
the distance between signal and shielding layer is $\SI{45}{\micro\meter}$ with polyimide as the insulator.
The thickness of the insulator and  aluminium layers and the outer shielding 
define the total thickness and thus the material budget of the HDI.
The main parameters of the HDIs are listed in \Autoref{tab:hdi,tab:spec_ladder} and a schematic stack is shown in \autoref{fig:Stack_Al}.

\begin{table*}
  \centering
  \small
  \begin{tabular}{lccc}
    \toprule
                                            & layer 1\&2    & layer 3   & layer 4  \\    \midrule
    HDI length [$\SI{}{mm}$]                & $140$         & $360$ & $380$  \\
    instrumented area [$\SI{}{mm\squared}$] & $120 \times 20$ & $340 \times 20$ & $360 \times 20$  \\ 
    number of \mupix chips                  & 6             &  17       & 18 \\ \hline  
    \multicolumn{4}{c}{\sl the following numbers refer to ladder halves}  \\
    number of traces:             &          &   &   \\ 
    ~~bias (HV)        (BIAS)               & 1             & 1         & 1 \\
    ~~$\SI{1.8}{V}$  (VDD)                  & 1             & 1         & 1 \\
    ~~ground    (GND)                       & 1             & 1         & 1 \\  
    number of differential pairs:      &          &   &   \\ 
    ~~data     (SOUT)                       & $3\times 3$        & $8\times 1$, $9\times 1$& $9\times 1$ \\
    ~~clock bus   (CLK)                     & 1             & 1         & 1 \\
    ~~reset bus  (SynRes)                   & 1             & 1         & 1 \\
    ~~slow control bus (SIN)                & 1             & 1         & 1 \\ \bottomrule
    \end{tabular}
    \caption{Specification of the HDIs for inner and outer layers. All lines have a shared bus topology except for the fast data lines (SOUT), which are point-to-point. Note that the numbers of electrical lines refer to each half of the HDI. } 
  \label{tab:spec_ladder}
\end{table*}

Space constraints on the HDI have motivated the use of a minimal set of
traces for power, control and readout.
Differential buses are used for slow control, clock and reset. 
Global power and ground lines are foreseen. 
Voltage gradients between sensors due to path length dependent ohmic
losses are minimised by design.
The remaining voltage differences are at the level of a few tens of millivolts
and significantly smaller than the operation range of the \mupix which is $1.8-\SI{2.0}{V}$.
The supply of the correct operation voltage is ensured by sense lines on the HDI.

\begin{figure}
        \centering
                \includegraphics[width=\columnwidth]{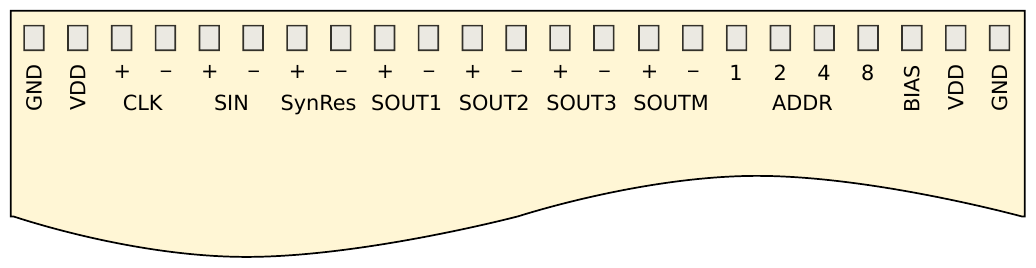}
        \caption{Conceptual \mupix pad layout on the HDI. Depending on location, either SOUT1 to SOUT3 or SOUTM is connected to accommodate for different data rate needs (vertex or recurl layers, respectively). Power and ground have multiple pads to reduce effects of voltage drop.}
        \label{fig:mupix_pads}
\end{figure}

Every pixel sensor is electrically connected by only 21 pads, 
see \autoref{fig:mupix_pads}.
Four address bits, selected by SpTA-bonding pads to ground or the supply voltage bus are used to set the chip address for the uplink communication bus.
The SpTA-bond pads have a relatively large size of 
$200\times\SI{100}{\micro\meter\squared}$ to fulfil the specification requirement 
for SpTA-bonding and to ensure a high production yield.

\begin{figure}[thb]
        \centering
                \includegraphics[width=0.5\columnwidth]{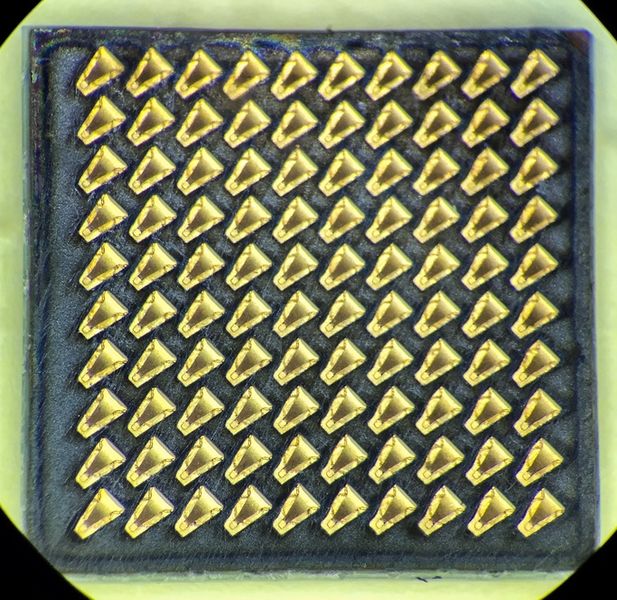}
        \caption{Interposer ZA8 from Samtec, version with $10\times 10$ connections. The pins have a pitch of $\SI{0.8}{mm}$}
        \label{fig:interposer}
\end{figure}

The \mupix ladders will be electrically connected to the endrings by interposers.
The chosen interposer from Samtec (ZA8H)\cite{interposer_samtec} is
a type of flat
connector allowing for high speed electrical signal transmission up to $\SI{30}{GHz}$,
see \autoref{fig:interposer}, with a high density of connections and a total thickness (compressed) of \SI{0.3}{\mm}.

\begin{figure}[bt]
        \centering
                \includegraphics[width=0.7\columnwidth]{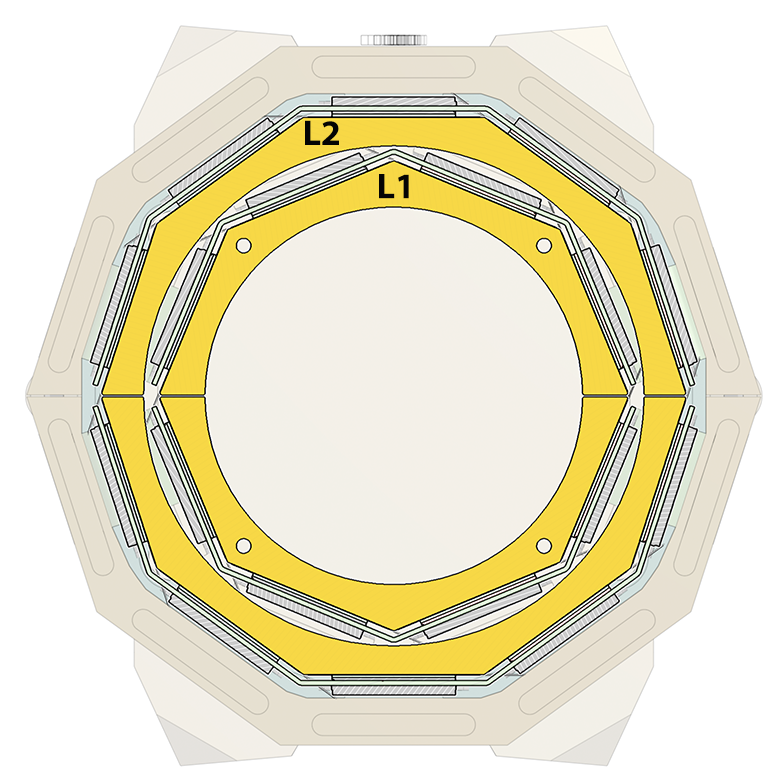}
        \caption{Holding endpieces and endrings of the inner layers with the octagonal (layer~1)
          and decagonal (layer~2) geometry, shown in yellow.
              }
        \label{fig:InnerEndPiece}
\end{figure}

\subsection{Module endpieces}
The endpieces of the inner layers consist of half-arcs made from
PEI\footnote{We use this insulator material to mitigate the risks of eddy currents in case of magnet quenches.}
to which the polyimide support structure and the \mupix ladders are glued.
The endpieces of layer~1 (layer~2) have a four (five)-fold segmentation,
 see \autoref{fig:InnerEndPiece}.

\begin{figure*}[tb]
    \centering
        \includegraphics[width=0.6\textwidth]{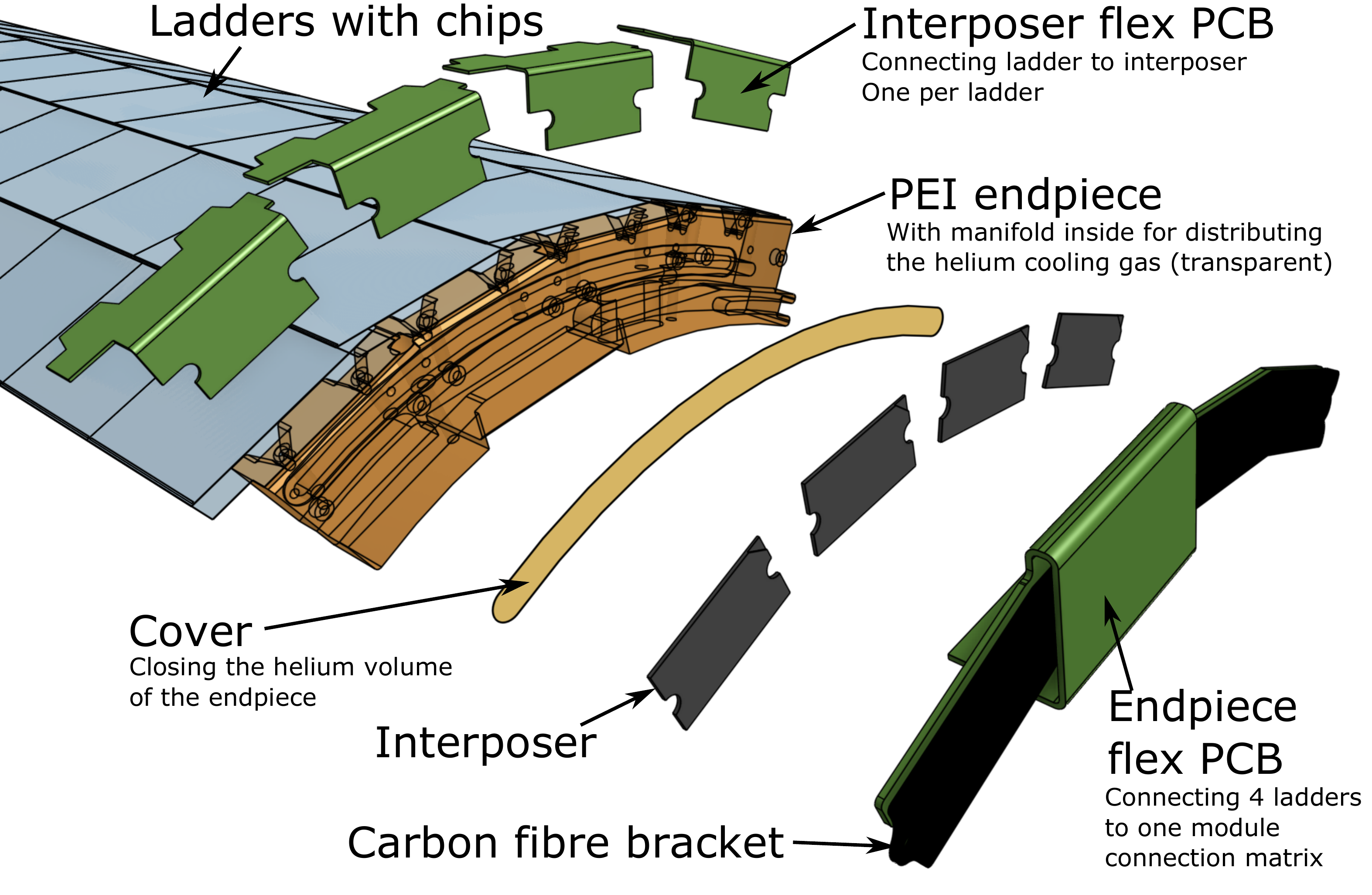}
    \caption{Exploded view of the outer layer module assembly. The end piece region provides a manifold for distributing the helium cooling gas and holds the flexible circuits to connect ladders to a single electrical connection matrix on the bottom.}
    \label{fig:endpiece}
\end{figure*}

The endpieces of the outer layers have a fourfold segmentation and include an internal open volume to distribute helium gas for the cooling-system, see \autoref{fig:endpiece}. To realise the internal volume for helium distribution the endpieces are manufactured out of a main part with a thin lid, glued on, to seal the open volume after machining.  
The endpieces also accommodate the interposer connectors for power, control and
data transmission.

\subsection{\mupix Ladder Integration and Chip Bonding}
During fabrication of the \mupix ladders, chips are placed accurately on the HDI by
use of fiducial marks on the chips and cut-outs on the HDI. 
The chips are then glued using an epoxy (Araldite 2011), applied in small dots of \SI{10}{\micro\metre} thickness, and the positions are
checked again. The flex circuit for connecting to the interposer is placed and glued in a similar manner. After curing, all connections between the HDI and each chip, and between the HDI and the interposer are SpTA-bonded (any vias
on the HDI are bonded beforehand by the manufacturer). Once all the connections are
in place, the unit is electrically fully functional. This allows for the
comprehensive quality testing of a \mupix ladder before they are assembled into
modules.

\section{Pixel Tracker Global Mechanics}
Pixel tracker inner and outer layer modules are integrated into the full cylindrical tracking layers by mounting the modules to the inner or outer layer pixel endrings. The latter in turn are connected to the up- and downstream beam pipes.
Like the module endpieces these are manufactured out of PEI. 
For the inner layers the endrings have gas inlets and outlets to provide the helium flow between layers 1 and 2.

\begin{figure}[tb]
        \centering
                \includegraphics[width=\columnwidth]{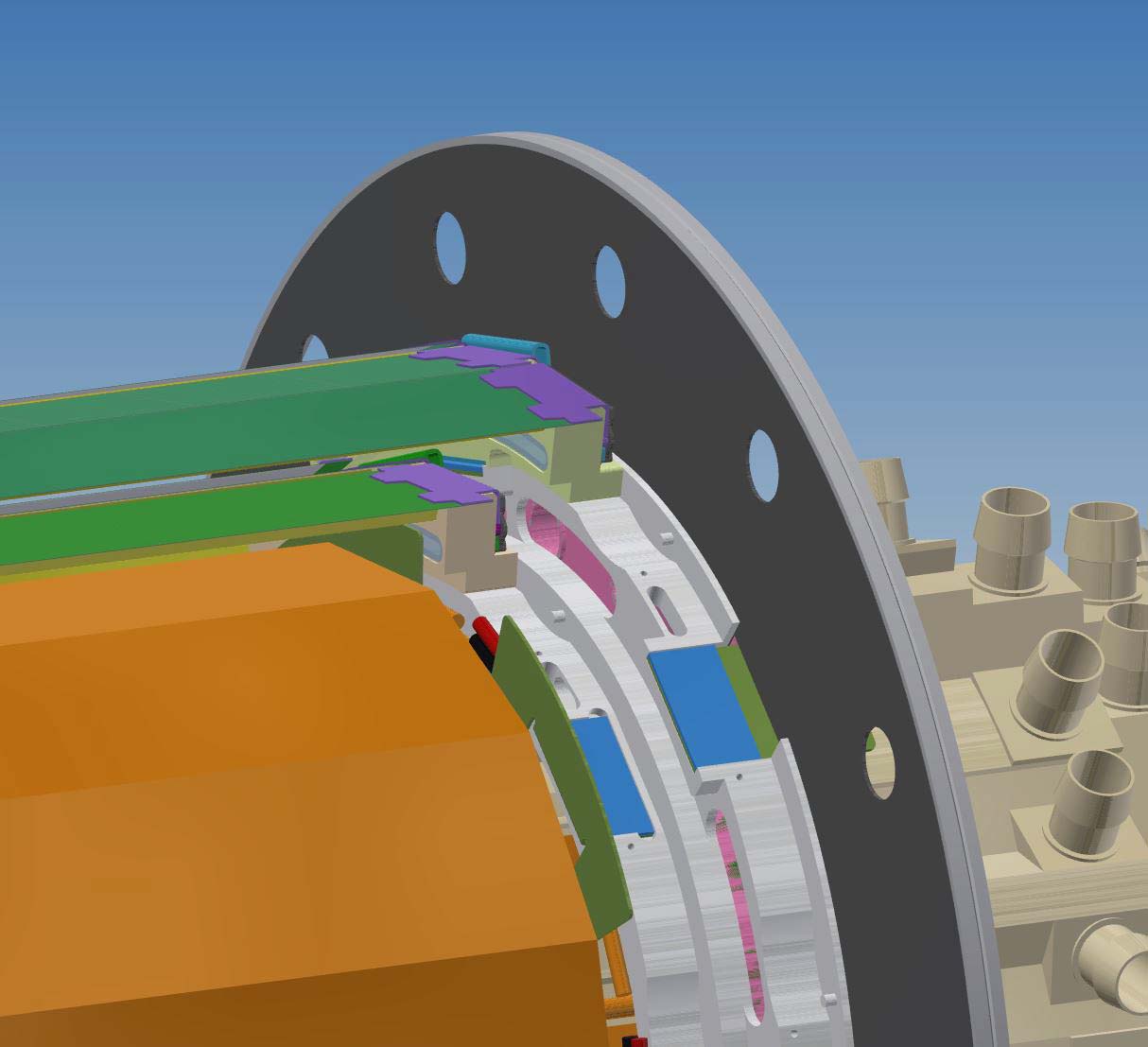}
        \caption{Endring situation shown with modules inserted in layers 3 and 4.}
        \label{fig:endring}
\end{figure}

A drawing of an outer layer endring equipped with a layer~3 module is
shown in \autoref{fig:endring}.
The outer endring supports six modules of layer~3 and seven modules of layer~4.
The endrings have dowel pins for every module to guide the module when it is rotated into position, ensuring no accidental collision happens with already mounted modules during installation.  
The final mechanical connection is done with screws. This also secures the contact through the endring interposer   
which provides the further electrical connection from the module, via the front-end flexprints,
to the front-end boards, from where the steering and control of the \mupix sensors is handled and where the signal processing is done.

The outer layer endrings provide conduits for the helium gas flow between layers 3 and 4 and to the module endpieces for the gas flow in to the v-fold channels.
At the upstream end, the inner and outer endrings are rigidly connected to the
beam pipe. 
At the downstream end, the endrings are supported by bearings and connected via a small spring tension such that the downstream endrings can move along the beam direction to accommodate thermal expansion of the ladders.

\section{Pixel Tracker construction and Quality Control}
\begin{figure*}[tb!]
        \centering
                \includegraphics[width=0.9\textwidth]{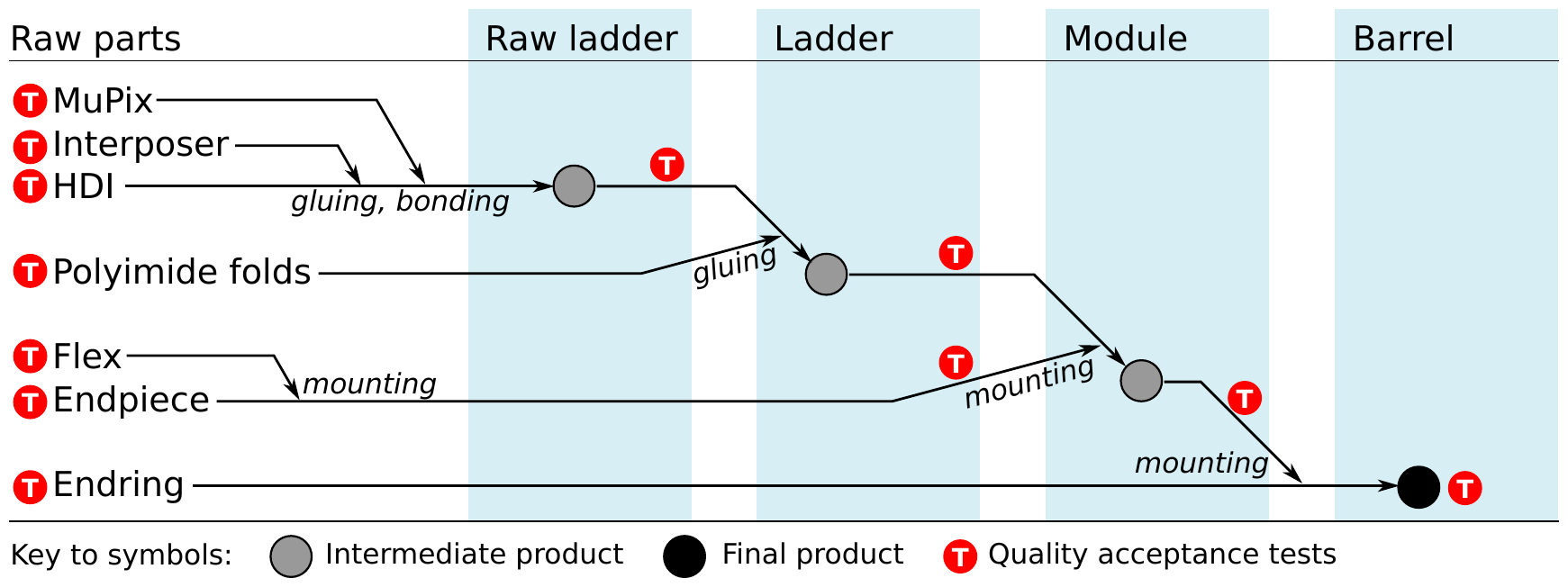}
        \caption{Module manufacturing workflow and quality points. Only main steps are shown.}
        \label{fig:pxmodprodworkflow}
\end{figure*}
\begin{sloppypar}
The production workflow of the pixel detector parts consists of manufacturing 
steps and quality control points, shown in \autoref{fig:pxmodprodworkflow}. 
The manufacturing steps make use of custom-made tooling for careful picking and 
accurate placing of parts. 
\end{sloppypar}
To protect parts from damage and contamination, manufacturing will take place in 
controlled environments, e.g.~cleanrooms of suitable levels, and standard ESD 
protection procedures will be in place. 
Polyimide expands when exposed to humidity.
All manufacturing steps crucial to defining tolerances will be carried out in
environments with strict temperature and humidity control and material will be stored
therein for proper equilibration prior to use.
At any other times during the production, storage or transport, components will be mounted in such a way to ensure thermal or humidity induced expansion can be accommodated safely.
Raw parts are either obtained from suppliers (e.g.~\mupix, HDI, interposer, etc.) 
or made in-house using custom tooling (e.g.~polyimide folds) or CNC machines (e.g.~endpieces).

Quality control takes place before and after every manufacturing step. 
Tests include (but are not limited to): visual inspection, dimension control, 
electrical testing, and gas leak testing. All components and their test results are tracked and documented in a production database. 
Raw parts will be acceptance tested upon receipt. 
In case of the \mupix chips, electrical testing will take place on the wafer, and 
on single die after dicing, using appropriate probe cards. 
Thanks to the modular design of the process, full electrical testing of all 
intermediate products is possible and foreseen. 
This includes the possibility to check sensor response using lasers or 
lab-grade radioactive sources.
In addition to checks of the electrical functionality at all stages of production, \mupix ladders will undergo a long term burn-in test.

\subsection{Inner Pixel Layers: Ladder and Module Production}
The full inner pixel production and assembly takes place at Heidelberg. 
The small nature of this detector part (18 ladders with 6 chips per ladder) makes a manual procedure a cost-effective choice. 

\begin{figure}[ht]
	\centering
	\includegraphics[width=\columnwidth]{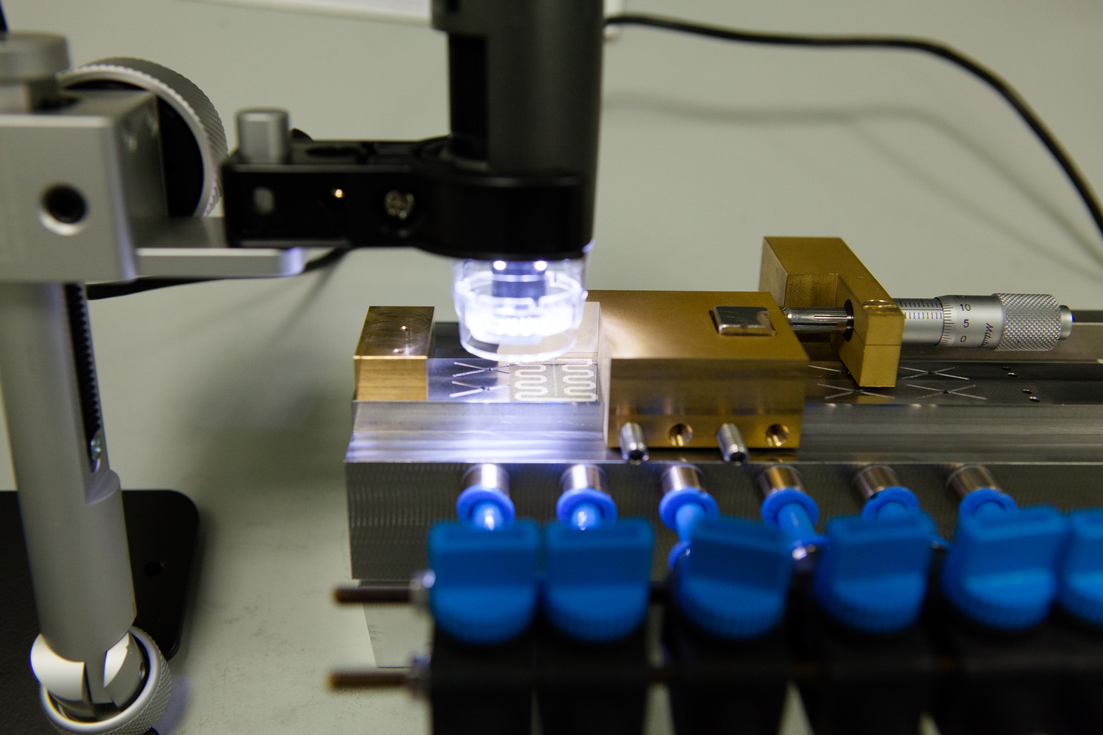}
	\caption{Assembly tool for the inner tracking ladders. Brass sliding block in the middle is guiding a prototype chip into position. Brass stop edge to the left. Micrometer screw to the right. Microscope to monitor position at the top.}
	\label{fig:InnerTrackerPlacementTool}
\end{figure}

Chips are positioned relative to each other and to the interposer flexes on a custom jig. The interposer flexes define the position of a ladder on the endrings. The positioning is done by moving the chips with a sliding block and fixating each chip at the desired position by vacuum (\autoref{fig:InnerTrackerPlacementTool}). 
While the position of the first chip is defined by a stop edge, following chips are placed using a micrometer screw and by monitoring the chip-to-chip gap with a microscope\footnote{Dino-Lite AM4515T8-EDGE, resolution of  \SI{1.5}{\micro\metre}}.

\begin{figure}
	\centering
	\includegraphics[width=\columnwidth]{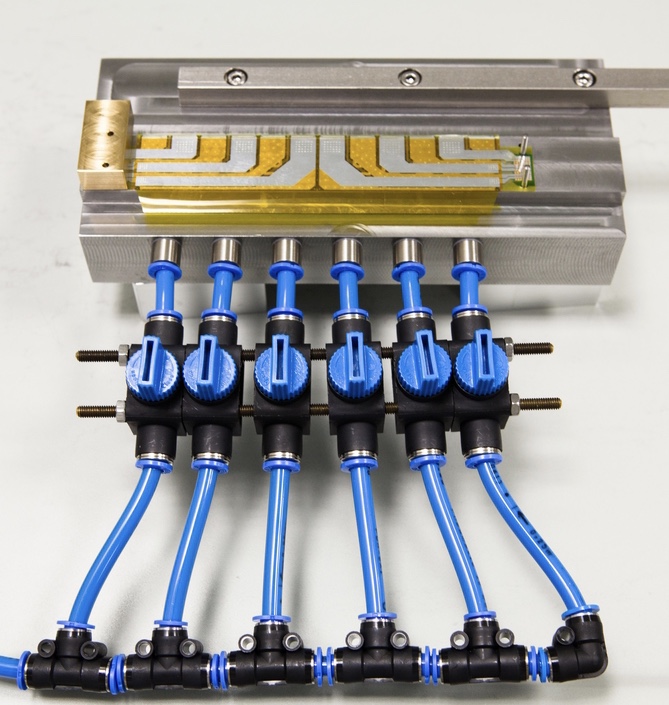}
	\caption{Prototype ladder for the inner pixel layers after gluing on mounting tool.} 
	\label{fig:InnerPixelPrototypeLadder}
\end{figure}

Epoxy (Araldite 2011)  is applied to the chips and the interposer flexes manually in small dots.
The HDI is aligned to the chips and flexes by fiducial marks on both parts under the microscope.
Weights ensure flatness and a uniform distribution of glue. A finished prototype ladder on the jig is shown in \autoref{fig:InnerPixelPrototypeLadder}.
Prototype construction has demonstrated a placement precision of  $\sigma<\SI{5}{\micro\metre}$ and an average glue thickness of \SI{5\pm4}{\micro\metre}.

After curing, connections between the chips and the HDI and between the interposer flexes and the HDI are made using SpTA-bonding.
From this point on, the ladder is electrically fully functional.
Each ladder undergoes a basic functionality test including powering, configuration and the readout of each \mupix chip. 

\begin{figure}
 	\centering
 	\includegraphics[width=\columnwidth]{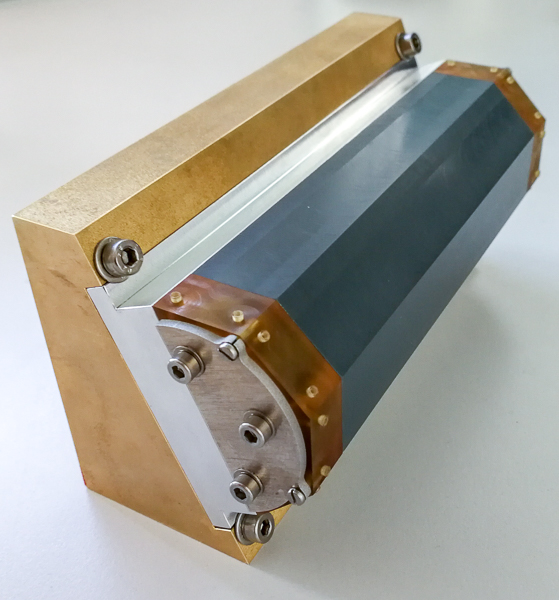}
 	\caption{Assembly tool for layer~2 modules. Tilting of the full tool and sliding of the grey block allows to bring every facet into the horizontal position.}
 	\label{fig:L2assyTool}
 \end{figure}

 Ladders that pass all QA checks are mounted into half-shells on custom assembly tools (\autoref{fig:L2assyTool}). 
 These tools, for layer 1 and layer 2, accommodate the module endpieces and are designed such that each facet can be brought into the horizontal position for ladder placement.
 The ladders are glued consecutively to the polyimide flap of the previously positioned ladder.
 Again, weights ensure flatness and a uniform glue distribution.
 At the same time, the ladders are attached to the PEI endpieces by clamping them to a stack comprising the end of the ladder, the interposer and the endpiece flex held by a carbon fibre bracket. 
 
\subsection{Outer Pixel Layers: Ladder and Module Production}
\begin{figure}[tb]
        \centering
                \includegraphics[width=\columnwidth]{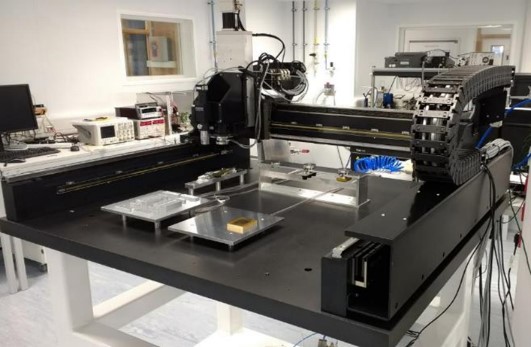}
                \includegraphics[width=\columnwidth]{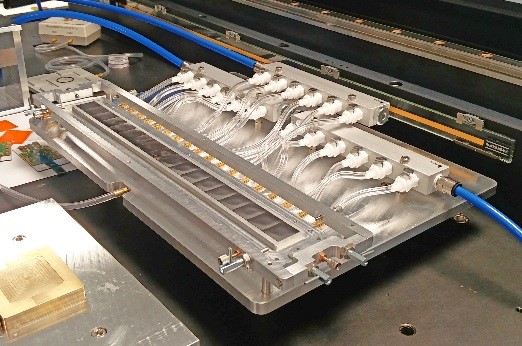}
        \caption{Robotic gantry (upper figure) for placement of 17 or 18 \mupix chips on the vacuum jig (lower figure).
}
        \label{fig:chip_placement}
\end{figure}

\begin{figure}[tb]
        \centering
                \includegraphics[width=\columnwidth]{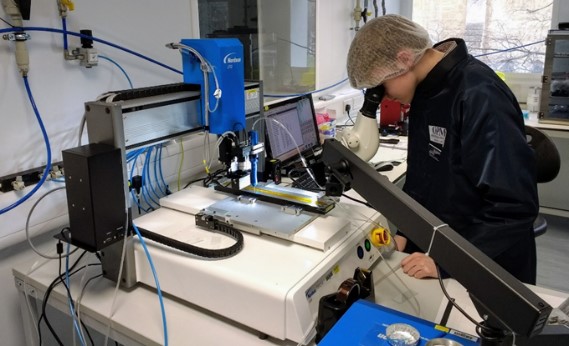}
                \includegraphics[width=\columnwidth]{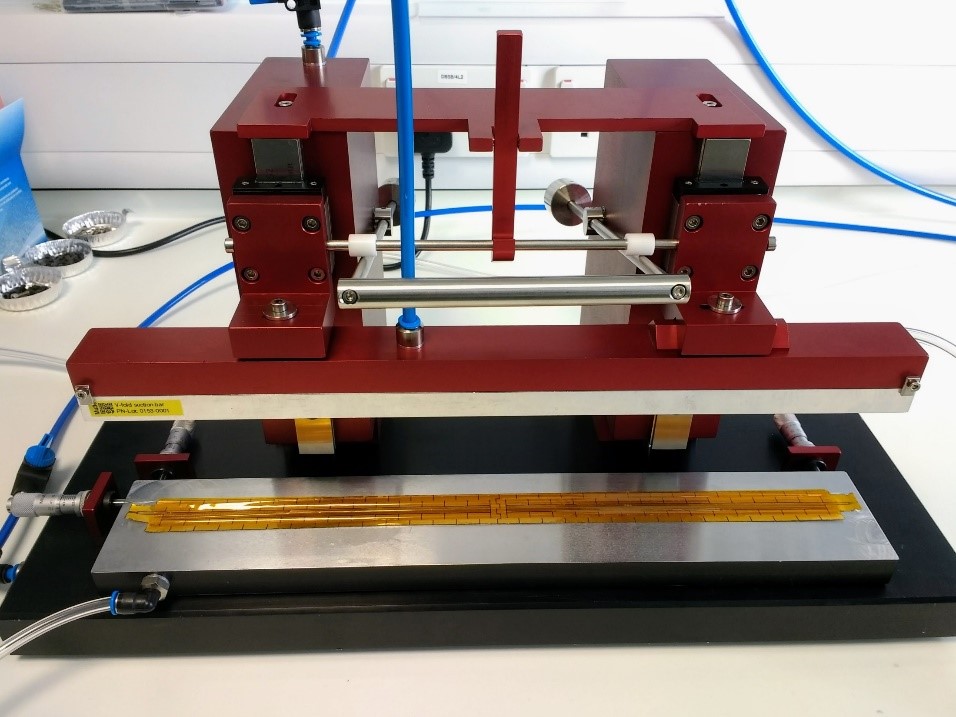}
        \caption{Glue dispensing robot (upper figure) and custom tooling for the gluing of v-channel reinforcements (lower figure).
}
        \label{fig:ladder_assembly}
\end{figure}

Ladder assembly for layers 3 and 4 of the \mupix tracker 
takes place at the Oxford Physics Microstructure Detector (OPMD) Laboratory. To make a ladder, 18 (17) chips are positioned on a vacuum jig using a 4-axis gantry positioning system, integrated with vision and electro-valve controls, and custom built tooling (see \autoref{fig:chip_placement}). A positioning accuracy within 10 microns (see \autoref{fig:ladder_assembly}) is achieved.
After this the interposer flex circuit is added, located by the jig, glue is deposited by a commercial machine vision guided liquid dispensing robot and the HDI is glued to the chips using a counter-jig. 
Connections between the sensor chips and the HDI circuits are made using SpTA-bonding. The completed assembly is reinforced with two V shaped, folded polyimide support structures glued to each ladder. The liquid dispensing robot is used to accurately apply the required epoxy to achieve 5 micron thick glue layers to adhere sensor chips to the flexprints and polyimide V-folds to the ladders. 

The polyimide V-folds are repeatably aligned and joined to the ladders using a custom jig with linear rails and micrometer adjusters. Semi-automated non-contact metrology of components and completed ladders is performed with an optical probe on a coordinate measuring machine.

After testing, \mupix ladders are shipped to Liverpool, for the assembly into modules. Upstream and downstream module endpieces are mounted to a custom jig that defines the overall length of a module. Ladders are positioned and glued onto the endpiece. Glue is applied manually to the surface of the endpiece and inside the v-shaped cut-out in the endpiece, as well as to the underside of the end of each ladder and the outside of the polyimide v-channels. Weights are used to ensure flatness and a uniform distribution of glue. After four ladders are assembled into a module, the v-channels on the inward facing side of the module are sealed with additional adhesive and electrical connections are made by fixing a stack of the ends of the four ladders, four interposers and the endpiece flex with a single carbon fibre bracket. 
Modules are checked for gas flow and leaks and for electrical conformity.

\section{Prototyping and System Tests} \label{sec:PxTMM}
A programme of manufacturing thermo-mechanical prototype modules for both the inner and outer layers of the \mupix tracker has been used to develop and commission the assembly tooling and processes. At the same time the built modules are intended to provide a testbed, called the \emph{thermo-mechanical mockup}~(TMM), to develop and demonstrate the helium cooling concept for the \mupix tracker. Modules for the TMM provide a close match to the final detector in terms of their mass and materials used and provide the means to dissipate heat loads, matching those in the real detector into the structure. Circuitry to monitor temperatures is also incorporated. Modules are built out two types of ladders: Silicon heater ladders and Tape heater ladders.\\

\begin{figure}
    \centering
    \includegraphics[width=\columnwidth]{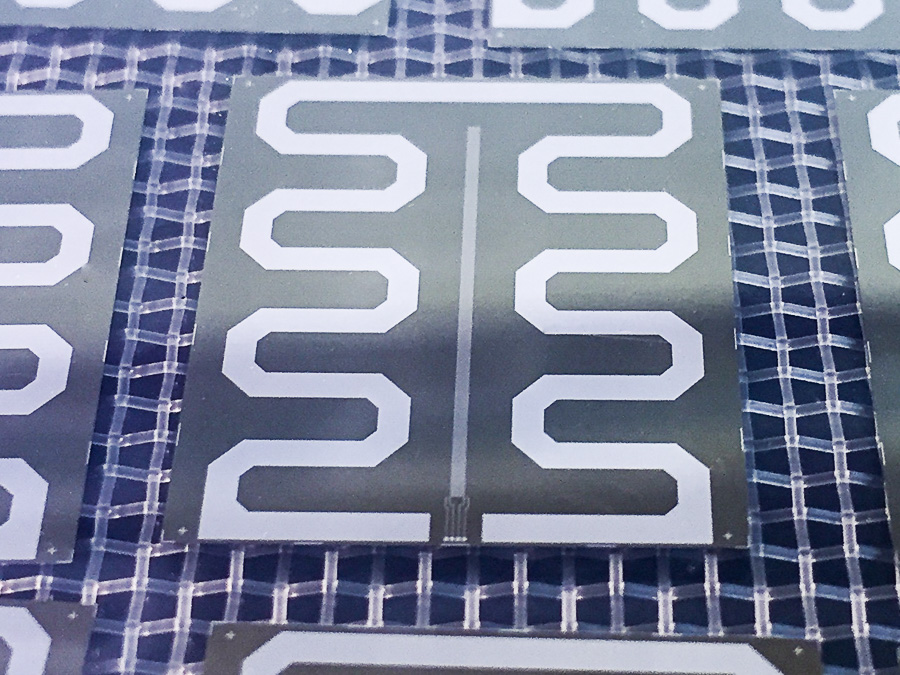}
    \caption{Silicon heater chip. The large meander for heating the chip and a narrow meander used as an RTD can both be seen. Contact pads are arranged on the bottom edge corresponding to the final chip connection locations.}
    \label{fig:siheaterchip}
\end{figure}

\begin{figure}
    \centering
    \includegraphics[width=\columnwidth]{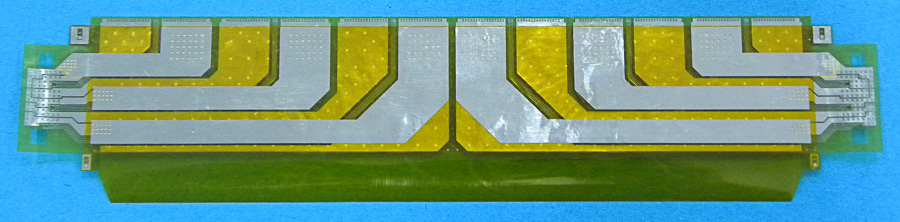}
    \caption{HDI for silicon heaters, layer~1 and~2. Six silicon heaters can be mounted on the back side.}
    \label{fig:siheaterHDIL1}
\end{figure}

\paragraph{Silicon heater ladders.}
These ladders closely match the material stack of the final detector. Silicon heater chips  (\autoref{fig:siheaterchip}) have been manufactured at the Max-Planck Halbleiterlabor in Munich using sputtered aluminium on silicon without a passivation layer. A meander with $R=\SI{3.24}{\ohm}$ allows heat to be generated in the chip in the range of 1 to \SI{1.6}{\watt} with similar voltages as for \mupix chips. An additional meander with $R \approx \SI{1000}{\ohm}$ is used as a resistance temperature detector (RTD) to measure the temperature in situ. The chips are thinned to  \SI{50}{\micro\metre} thickness. Ladders are fabricated using adapted versions of the HDI with the same stack as foreseen for the detector (\autoref{fig:siheaterHDIL1}). Connections are made with SpTA-bonding. The manufacturing steps needed and tooling used are the same as for detector fabrication, providing the ideal test bed to develop, commission and qualify the tooling and processes  for the final detector production.\\

\begin{figure}[tb!]
    \centering
    \includegraphics[width=\columnwidth]{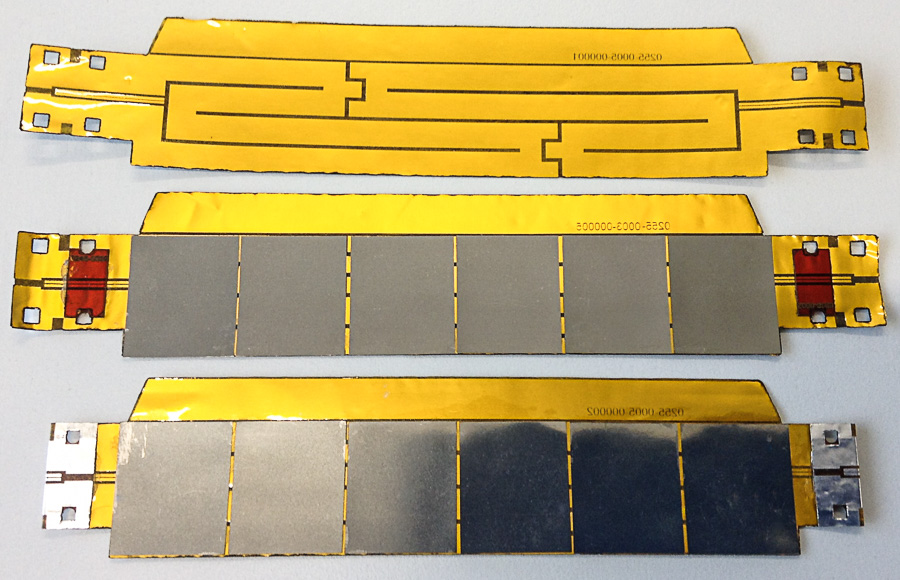}
    \caption{Tape heaters for layer 1 and 2. Top to bottom: bare heater with meander, stiffener attached to match final dimensions, dummy chips glued on. Large contact pad pair on both ends used for powering. Chip size $\SI[parse-numbers = false]{20 \times 23}{\mm\squared}$.}
    \label{fig:tapeheater}
\end{figure}

\paragraph{Tape heater ladders.}
These ladders are simpler objects, based around an aluminium-polyimide laminate (\autoref{fig:tapeheater}) resistive heating circuit that has the same shape as the HDI plus interposer flex assembly in the final detector. Laser cutting and etching are used to manufacture the tape heater flexes in sizes corresponding to inner ladders ($R\approx\SI{0.5}{\ohm}$) and outer ladders ($R\approx\SI{3.7}{\ohm}$). To create a more realistic mechanical model and material budget, \SI{50}{\micro\metre} thick stainless steel dummy chips can be attached if needed for specific test purposes. This more cost-effective option is used for simpler manufacturing tests and to instrument most of the full TMM.

\paragraph{} The TMM is assembled by assembling the silicon heater and tape heater ladders into modules and barrels using the same mechanical components as are use in the final detector. With the heating capabilities and all the cooling facilities in place, realistic measurements of the cooling and mechanical stability will be possible. Tests stands for intermediate and final assemblies have been developed in preparation for the final testing of detector assemblies. All manufacturing steps are taking place at the locations foreseen for detector fabrication. 

\begin{figure}[tb!]
    \centering
    \includegraphics[width=\columnwidth]{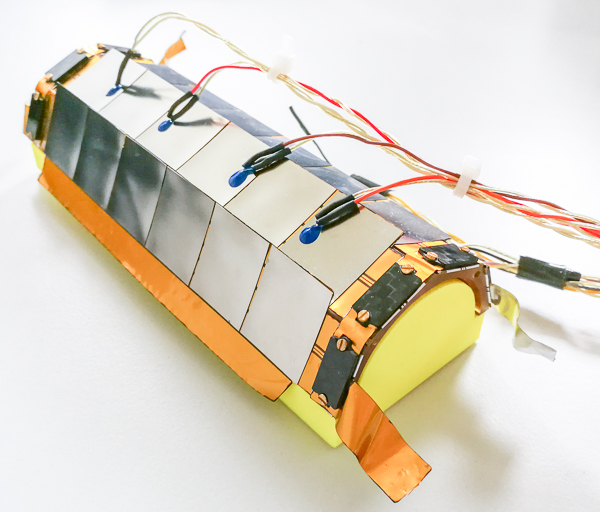}
    \caption{Layer~1 half shell made with tape heaters and stainless steel dummy chips, placed on a handling block (yellow). RTDs attached with conductive glue for temperature profiling in cooling tests.}
    \label{fig:L1halfsheelTapeheater}
\end{figure}

An example of ladders manufactured for the TMM is shown in \autoref{fig:L1halfsheelTapeheater}.

\section{Pixel Tracker Cooling}
\label{sec:tracker_cooling}

\begin{sloppypar}
The full pixel detector will dissipate about $\SI{4.55}{kW}$ of heat\footnote{Throughout this section, heat from chips and losses in conductors inside the HDI are taken into account, summing up to the heat density used in the scenarios. } in a \emph{conservative scenario} assuming $\SI{400}{mW/\cm\squared}$. The latest chip versions have shown a heat dissipation below $\SI{250}{mW/\cm\squared}$. This heat load is used for our most  \emph{realistic scenario}. \autoref{tab:pixel_heatload} shows expected heat load in each layer of the tracker under these two scenarios. The cooling system must keep the maximum temperature, anywhere in the pixel detector, safely below \SI{70}{\celsius}, given by the glass-transition temperature of the adhesives used for construction.
\end{sloppypar}

\begin{table*}[t!]
    \centering
    \footnotesize
    \begin{center}
        \begin{tabular}{lrrr}
            \toprule
            Detector Part & Area & $\SI{250}{\mW\per\cm\squared}$ & $\SI{400}{\mW\per\cm\squared}$ \\
              & [$\SI{}{\cm\squared}$] & [$\SI{}{\W}$] & [$\SI{}{\W}$] \\
            \midrule
            layer 1  & $ 192$ & $ 48$ & $ 77$ \\
            layer 2  & $ 240$ & $ 60$ & $ 96$ \\
            layer 3  & $1632$ & $408$ & $652$ \\
            layer 4  & $2016$ & $504$ & $807$ \\
            Recurl Station  ($2\times$) & $3648$& $912$ & $1459$	\\ \midrule
            total & $11376$ & $2844$ & $4550$	\\ \midrule
            \bottomrule
        \end{tabular}
        \caption{Heat dissipation of the pixel detector for a power consumption of
        $\SI{250}{\mW\per\cm\squared}$ (realistic scenario) and
        $\SI{400}{\mW\per\cm\squared}$ (conservative scenario).}
        \label{tab:pixel_heatload}
    \end{center}
\end{table*}

\begin{figure*}[bht]
    \centering
    \begin{subfigure}[c]{\textwidth}
        \includegraphics[width=\textwidth]{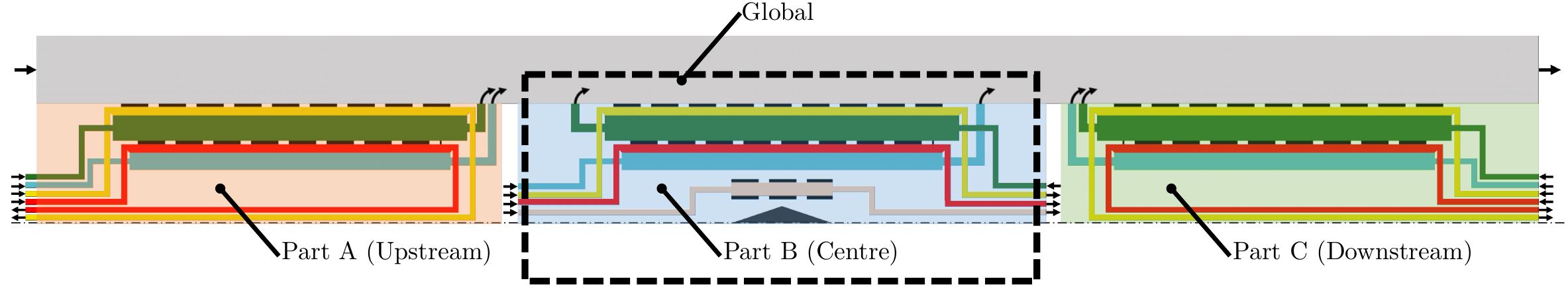}
        \caption{Schematic of the helium flows in the full pixel detector}
        \label{fig:cooling_he_pixel-schematic_full}
    \end{subfigure}
    \begin{subfigure}[c]{\textwidth}
        \bigskip
        \includegraphics[width=\textwidth]{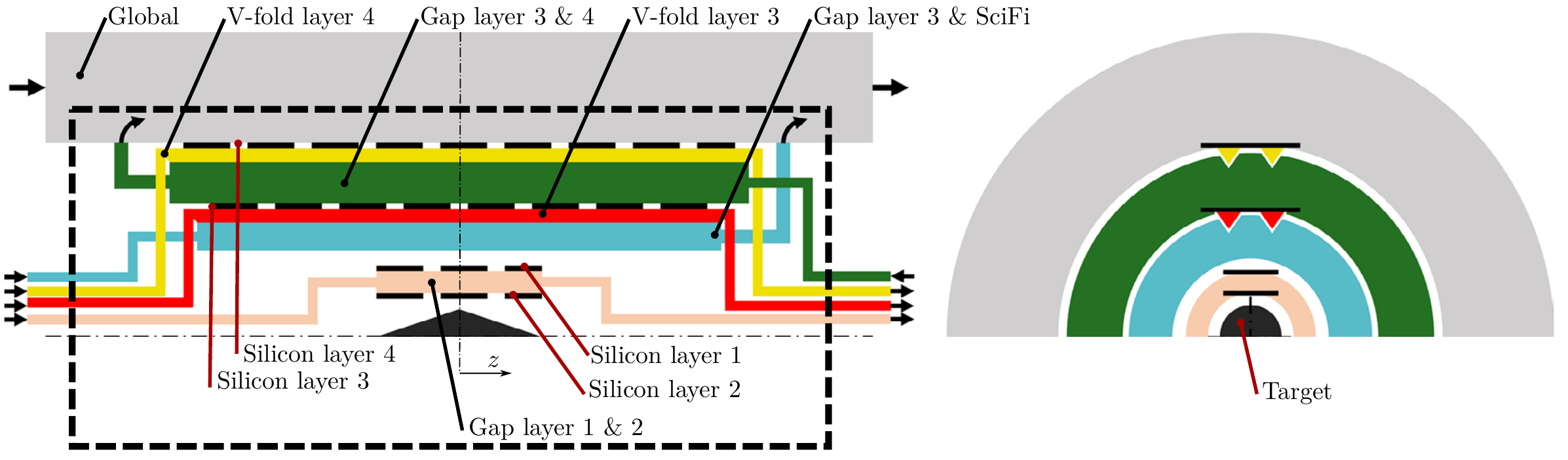}
        \caption{Central detector cooling detail}
        \label{fig:cooling_he_pixel-schematic_detail}
    \end{subfigure}
    \caption{Sketch of the helium cooling system for the pixel detector. (\subref{fig:cooling_he_pixel-schematic_full}) shows all volumes with its flow directions in a cut view. The system is cylindrically symmetric around the long dashed-dotted line. Some volumes vent into the global flow inside the experiment, indicated by bent arrows. Every circuit is individually controlled for flow and pressure inside the detector volume. (\subref{fig:cooling_he_pixel-schematic_detail}) shows a cut in the transverse direction as well. The triangles (in red and yellow) indicate the V-fold channels, which exist in pairs for every ladder.}
    \label{fig:cooling_he_pixel-schematic}
\end{figure*}

\begin{table*}[h]
    \centering
    \footnotesize
    \begin{tabular}{clcccccc}
        \toprule
        No. &  Description                         & \# & \multicolumn{2}{c}{Inlet} & & \multicolumn{2}{c}{Outlet} \\ \cmidrule{4-5}\cmidrule{7-8}
            &                                      &    & $\dot{m}$ & $\Delta p$ & $v$ & $\dot{m}$ & $\Delta p$ \\
            &          &    & \si{\gram\per\second} & \si{\milli\bar} & \si{\metre\per\second} & \si{\gram\per\second} & \si{\milli\bar} \\
        \midrule
        1 & Gap flow vertex detector            &  1 &  2.0 & +40 & 10 & 2.0 & -40 \\
        2 & Gap flow b/w SciFi and L3           &  1 &  6.9 & +25 & 10 &   0 &   0 \\
        3 & Gap flow b/w SciTile and L3         &  2 &  5.7 & +28 & 10 &   0 &   0 \\
        4 & Gap flow b/w L3 and L4              &  3 &  7.6 & +25 & 10 &   0 &   0 \\
        5 & Flow in V-folds L3                  &  3 &  1.3 & +90 & 20 & 1.3 & -90 \\
        6 & Flow in V-folds L4                  &  3 &  1.5 & +80 & 20 & 1.5 & -80 \\
        7 & Global flow, $D\approx\SI{300}{\mm}$ & 1 &  4   & +0.04 & var. & 45 & -0.04 \\
        \midrule
          & Total                               & 14 &  56  &    &       & 56 & \\
        \bottomrule
    \end{tabular}
    \caption{List of helium circuits inside the experiment. Pressures are given relative to ambient in the experiment and were obtained from CFD simulations. Circuits with outlet flows and pressures of 0 vent into the main volume, collected in the global flow outlet. The total flow corresponds to about \SI{20}{\metre\cubed\per\minute} under standard conditions. Column \# gives number of identical circuit copies in the detector.}
    \label{tab:hecoolantcircuits}
\end{table*}

We use gaseous helium at ambient conditions\footnote{This means temperatures above \SI{0}{\celsius} and the absolute pressure around \SI{1}{\bar}.} as coolant. The helium is distributed in separate circuits, serving different parts of the detector separately. The concept is shown in \autoref{fig:cooling_he_pixel-schematic} and the different helium circuits are listed in \autoref{tab:hecoolantcircuits}. The flow in the global circuit increases along $z$ because of other circuits directly venting into the global flow. The global flow is constrained by a thin mylar foil (thickness \SI{5}{\micro\metre}) surrounding the full pixel detector in a conical shape that keeps the helium velocity near constant along z (see \autoref{fig:InnerTracker}.

\begin{figure*}
    \centering
        \includegraphics[width=.49\textwidth]{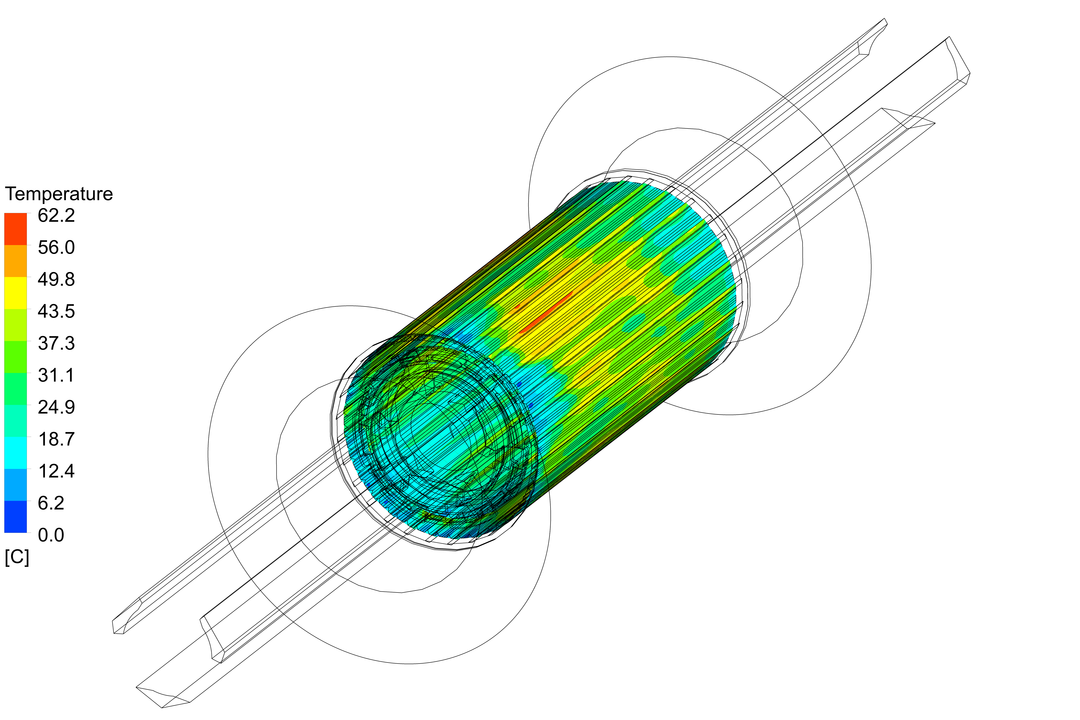}
        \includegraphics[width=.49\textwidth]{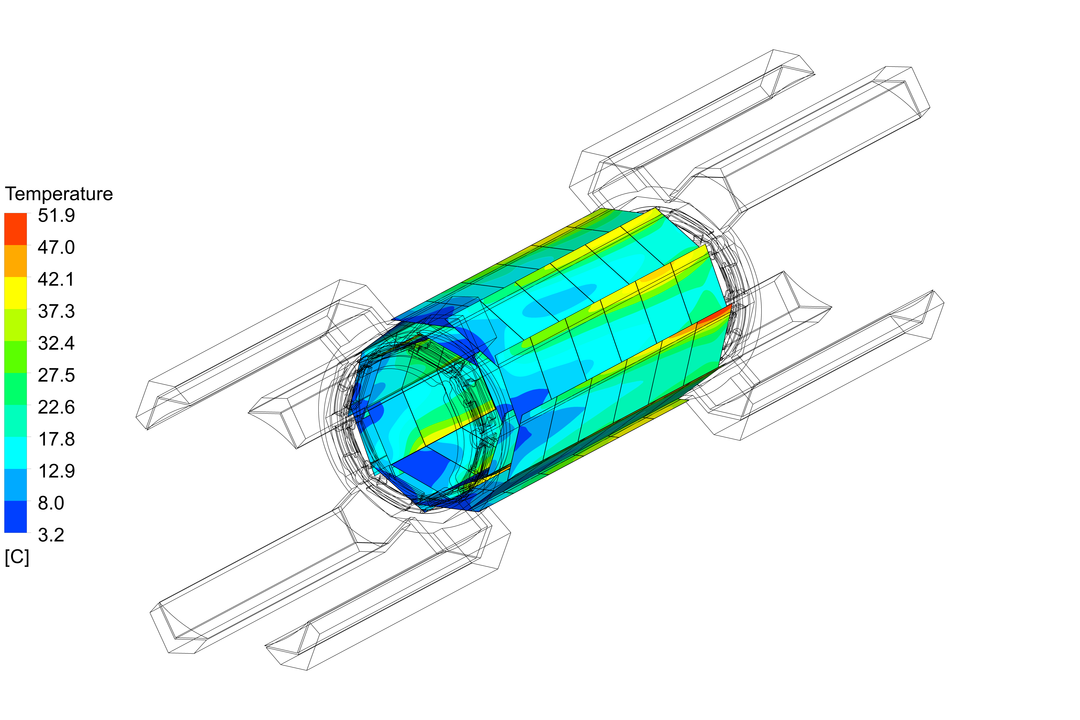}
    \caption{Simulated $\Delta T$ distribution of the silicon in the tracking detector for a power dissipation of $\SI{400}{mW/cm^2}$ (non-uniform distribution, more heat power at periphery edge). Inlet gas temperature is $T=\SI{0}{\degreeCelsius}$ Left: full barrel. Right: vertex barrel inside the full barrel.}
    \label{fig:pixel_cooling_simulated}
\end{figure*}

The system described is the result of a process of optimisations through simulation studies using computational fluid dynamics (CFD)\footnote{Autodesk\textsuperscript{\textregistered} and ANSYS~CFX\textsuperscript{\textregistered} CFD software were used.} and verifications in the laboratory~\cite{Zimmermann2012, Huxold2014, Ng2015, Herkert2015, Tormann2018, Deflorin2019}. The models used in both simulation and laboratory measurements progressed in detail and the final mock-up models described in \autoref{sec:PxTMM} match the final detector to a great extent in shape, materials and heat-density.
\begin{sloppypar}
The heat-load density distribution used in the simulations take care of the uneven heat density on the pixel chip. Half of the power dissipation on the chip is expected to be located on the periphery, the remaining half within the pixel matrix, equivalent to \SI{200}{\milli\watt\per\cm\squared} and \SI{1730}{\milli\watt\per\cm\squared} respectively, for an averaged \SI{400}{\milli\watt\per\cm\squared} in the conservative scenario. Simulation results for the pixel tracker are shown in \autoref{fig:pixel_cooling_simulated}, confirming a safe $\Delta T$ even in the conservative scenario. In simulations with the realistic scenario the obtained $\Delta T$ values have been found to scale down linearly with the reduced power dissipation and are therefore not shown.
\end{sloppypar}

\begin{figure*}[hbt]
    \centering
    \begin{subfigure}[c]{\textwidth}
        \centering
        \includegraphics[width=.427\textwidth]{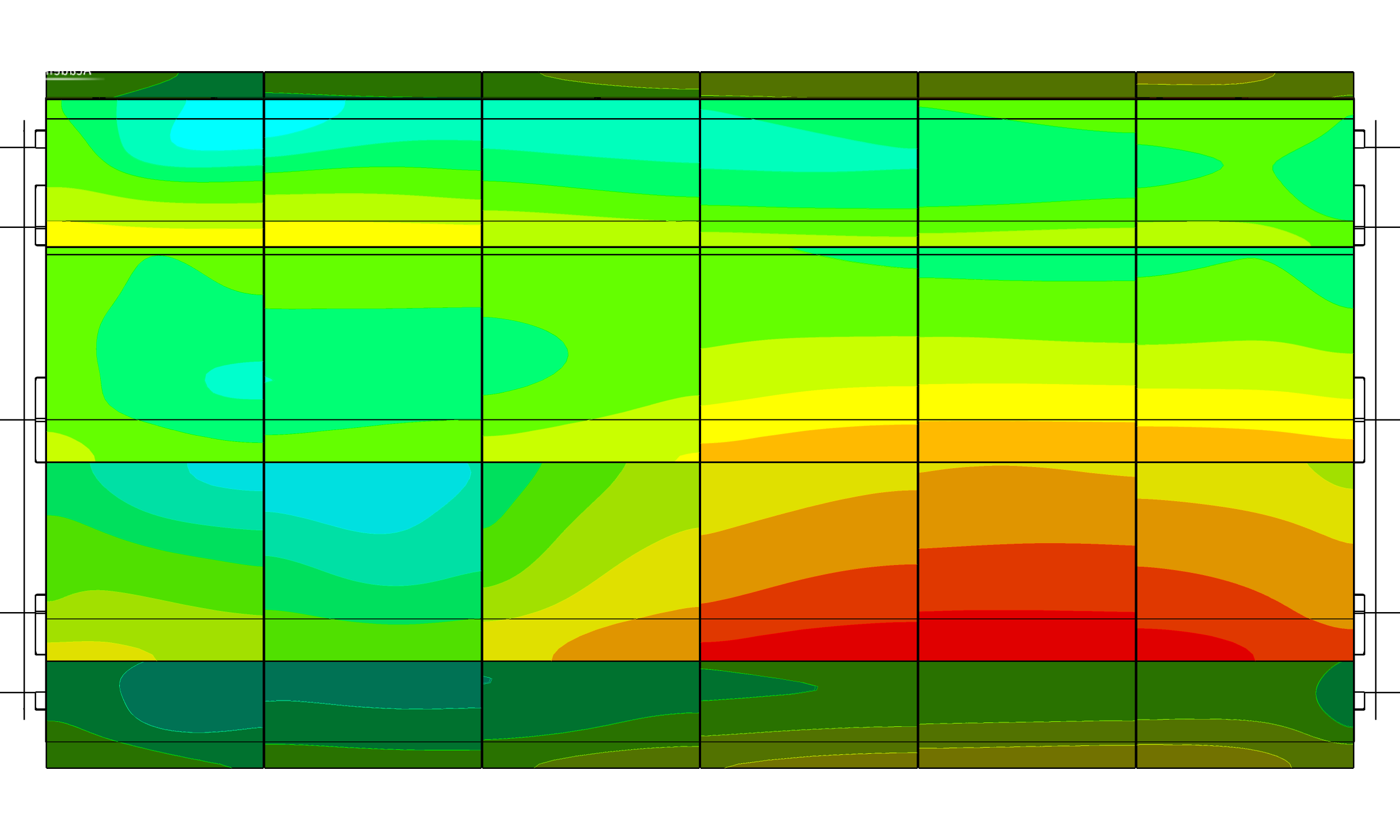}
        \includegraphics[width=.05\textwidth]{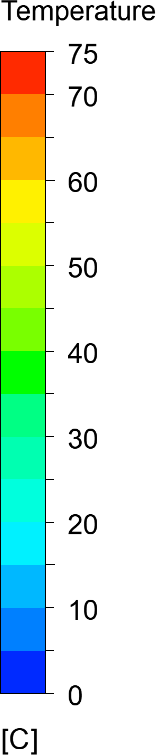}
	\caption{Simulated temperature on the outer layer of the mock-up.}
        \label{fig:L12_T_CFD}
    \end{subfigure}
    \begin{subfigure}[c]{\textwidth}
        \bigskip
        \centering
        \includegraphics[width=.58\textwidth]{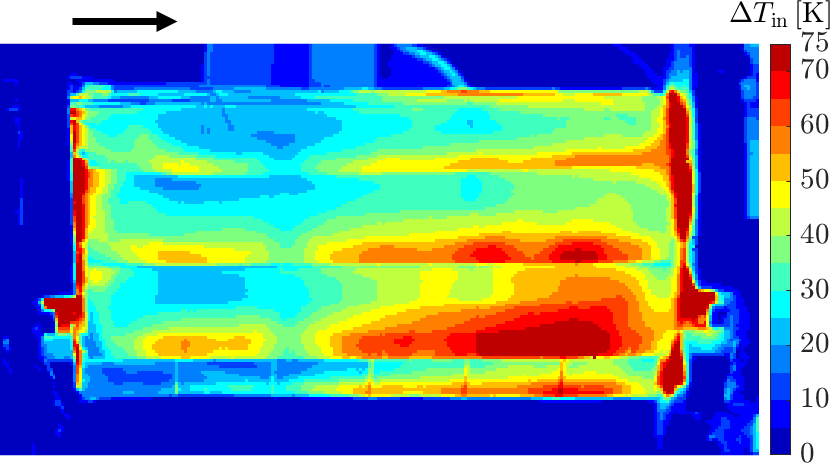}
	\caption{Measured temperature on the outer layer of the mock-up using an infrared camera.}
        \label{fig:L12_T_Meas}
    \end{subfigure}
    \caption{Temperature obtained by measurement and CFD-simulation. Angle of view of simulation has been carefully matched to the camera view. Cold helium enters from the left. Hot zones on the right of (\subref{fig:L12_T_Meas}) are cable connections from the setup not present in the final detector.}
    \label{fig:L12_Temp}
\end{figure*}

The experimental cooling tests were performed inside a cylindrical closed volume with a diameter of \SI{22}{cm} and a length of approximately \SI{1}{m}. Helium was initially provided by compressed gas bottles, limiting measurements to a few minutes. This was overcome by using a miniature turbo compressor (described in \autoref{sec:Cooling}) allowing for  helium recirculation and hence continuous operation. The agreement between simulation and mock-up measurements are good, see \autoref{fig:L12_Temp} for an example comparison for the vertex detector~\cite{MeierAeschbacher:2020ldo}.

Vibrations induced by the helium flows must not damage the structures or have a substantial impact on the hit resolution. Such vibrations were studied using a setup based on a Michelson interferometer pointing to reflective surfaces on a realistic mock-up. For velocities up to \SI{20}{\metre\per\second}, average amplitudes of $\SI{2}{\micro\meter}$ were observed, with peaks of \SI{10}{\micro\meter}~\cite{Austermuehl2015,Herkert2015,Henkelmann2015}. This is well below the single hit resolution of the pixel sensors. An excitation spectrum using a speaker showed resonances between \SI{50}{Hz} to \SI{1000}{Hz} with no major peaks. No damage to the test structures has been observed during these studies.


\chapter{\mupix Pixel Sensor}
\label{sec:Mupix}

The very challenging requirements on the allowed amount of material in the tracking layers
can only be fulfilled with a monolithic silicon pixel technology.
Monolithic sensors efficiently integrate sensor and readout in the same
device, thereby greatly reducing the detector material in comparison to
classical hybrid pixel module designs, which require additional readout chips and
interconnects (bonds). 
For the Mu3e pixel detector, 
High-Voltage Monolithic Active Pixel Sensors (HV-MAPS)~\cite{Peric:2007zz}
were chosen.
They are produced in a commercial $\SI{180}{\nano\meter}$ HV-CMOS process~\cite{TSI}
and can be thinned to $\SI{50}{\micro\meter}$ to reduce material~\cite{Augustin:2016iff}.

For Mu3e an experiment-specific HV-MAPS, the \mupix, has been developed.
All \mupix sensors in the pixel tracker are the same 
size, each instrumenting an (active) area of about $20 \times \SI{20}{\mm\squared}$.
The main parts of the digital electronics are located in the chip
periphery, a region about \SI{3}{\mm} wide on one side of the sensor.
The periphery also integrates dedicated pads for SpTA-bonding \cite{Oinonen:2005rm}
(see~\autoref{sec:Pixel}), and additional pads for testing. 
The number of electrical lines to operate the sensor is kept to a
minimum in order to reduce the number of interconnects and to ease
routing. All electrical connections for signal, control and
monitoring are differential and run at high speed. Additional
connections are provided for power, ground, bias-voltage, and for
passive temperature monitoring using a diode.

All \mupix sensors will be operated synchronous to the Mu3e system clock with $\approx \SI{1}{\nano\second}$ precision. This alignment is achieved by means of a synchronous reset command.
Hit timestamps are derived from an internal phase-locked loop (PLL) running at a nominal frequency of \SI{625}{\MHz}.
Data are sent over up to three configurable serial links,
each providing a bandwidth of \SI{1.25}{Gbit/s} using an 8~bit/10~bit encoding protocol.

Operating temperature, and therefore the power consumption of the
\mupix sensors, is critical for the tracking detector. We have tested
and qualified HV-MAPS for temperatures up to
$\SI{100}{\degreeCelsius}$ but define a maximum temperature of
$T_{max}=\SI{70}{\degreeCelsius}$ to stay in the specified range for
the adhesives used in the tracking detector. The minimum temperature
is defined by the $\SI{0}{\degreeCelsius}$ icing limit
(\autoref{sec:tracker_cooling}). The power consumption of the pixel
tracker per unit area must not exceed the maximum cooling capacity of
the helium gas cooling system, $P_{max}$
$=\SI{400}{\milli\watt/cm^2}$, see ~\autoref{sec:Cooling}. Taking
into account electrical losses on the HDI and power cables, the \mupix
sensor must therefore be operated below the power consumption limit of
$\SI{350}{mW/cm^2}$. The main requirements for the pixel sensor are
summarised in~\autoref{tab:SensorRequirements}.

\label{sec:SensorRequirements}
\begin{table}
        \centering
        \small
                \begin{tabular}{lr}
                        \toprule
                         sensor dimensions [$\SI{}{\mm\squared}$]& $\leq 21\times23$\\
                         sensor size (active) [$\SI{}{\mm\squared}$]& $\approx 20 \times20$\\
                         thickness [$\SI{}{\micro\meter}$]& $\leq 50$\\
                         spatial resolution $\SI{}{\micro\meter}$& ${\leq30}$\\
                         time resolution [$\SI{}{\nano\second}$]& $\leq20$\\
                         hit efficiency [$\SI{}{\percent}$]& $\geq99$\\
                         \#LVDS links (inner layers) & 1 (3)\\
                         bandwidth per link [$\SI{}{Gbit/s}$] & $\geq 1.25$ \\
                         power density of sensors [$\SI{}{\milli\watt\per\cm\squared}$] & ${\leq350}$\\
                         operation temperature range [$\SI{}{\degreeCelsius}$] & 0 to 70\\
                        \bottomrule
                \end{tabular}
        \caption{Main requirements of the Mu3e pixel sensor.}
        \label{tab:SensorRequirements}
\end{table}

\begin{sloppypar}
After introducing the HV-MAPS concept, an overview of the \mupix
R\&D and the characterised prototypes is given. The final \mupix
design is presented in~\autoref{sec:mupix_design}. The main results
obtained by prototypes are discussed in~\autoref{sec:mupix_results}
including first characterisation results from the final \mupix{10}
prototype.
\end{sloppypar}

\section{HV-MAPS}

\label{sec:hvmaps}

\begin{figure}[hb]
        \centering      \includegraphics[width=0.48\textwidth]{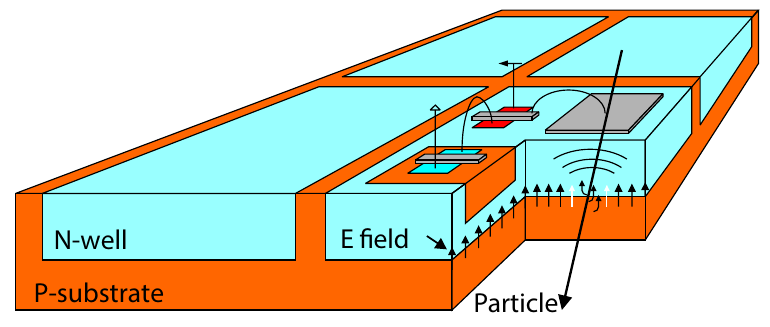}
        \caption{Sketch of the HV-MAPS detector design from \protect\cite{Peric:2007zz}.}
        \label{fig:cmos_sketch}
\end{figure}

HV-MAPS collect ionisation charge mainly via drift and therefore
provide time resolutions of a few nanoseconds, in contrast to standard
MAPS~\cite{Turchetta:2001dy,Baudot:2009dta,Baudot2013,Dorokhov:2011zzb,Senyukov:2013se,Valin:2012zz}
which collect ionisation charge mainly by diffusion with a
typical timescale of several hundreds of nanoseconds.
In standard MAPS, the in-cell
electronics is implemented outside the n-well which serves as charge
collecting diode (sometimes referred to as ``small fill factor'' design).
In HV-MAPS, instead, the pixel amplifier electronics is 
implemented inside the deep n-well, see~\autoref{fig:cmos_sketch}.
By reverse biasing the charge collecting diode with high voltage ($\ge \SI{60}{V}$) the substrate of the pixel cell is depleted.
Using the $\SI{180}{\nm}$ HV-CMOS process developed by IBM (or variants), this HV-MAPS concept was first proposed in~\cite{Peric:2007zz} and has been successfully tested with several prototypes since~\cite{Peric2010504,Peric2010,Augustin:2015mqa,Augustin:2016iff,Augustin:2018Elba,Augustin:2020tnk}.
In recent years, the concept of depleting MAPS was
successfully applied also on other technologies,
see e.g.~\cite{Fernandez-Perez:2016tmr,Obermann:2017nlq}.

At maximum high-voltage, the size of the depletion zone is below
$\approx 30-\SI{40}{\micro\meter}$ for low-ohmic wafer substrates
$\le \SI{200}{\ohm\cm}$, as confirmed by edge-TCT
characterisation~\cite{Cavallaro:2016gmx}.
As the stack containing the electronics and metal layers is only
$\approx \SI{16}{\micro\meter}$ for the technology used,
it is possible to remove substantial parts of the substrate that do
not contribute to the charge generation process.
Therefore, HV-MAPS can be produced with a
thickness of $\SI{50}{\micro\meter}$, corresponding to about $X/X_0=
0.054\%$.
Depending on the choice of the substrate, up to about $3000$ primary
electrons are expected for minimum ionising particles with trajectories
perpendicular to the plane of the substrate.

\begin{figure}[h!]
  \centering
  \begin{subfigure}[c]{0.4\textwidth}
    \includegraphics[width=1.0\textwidth]{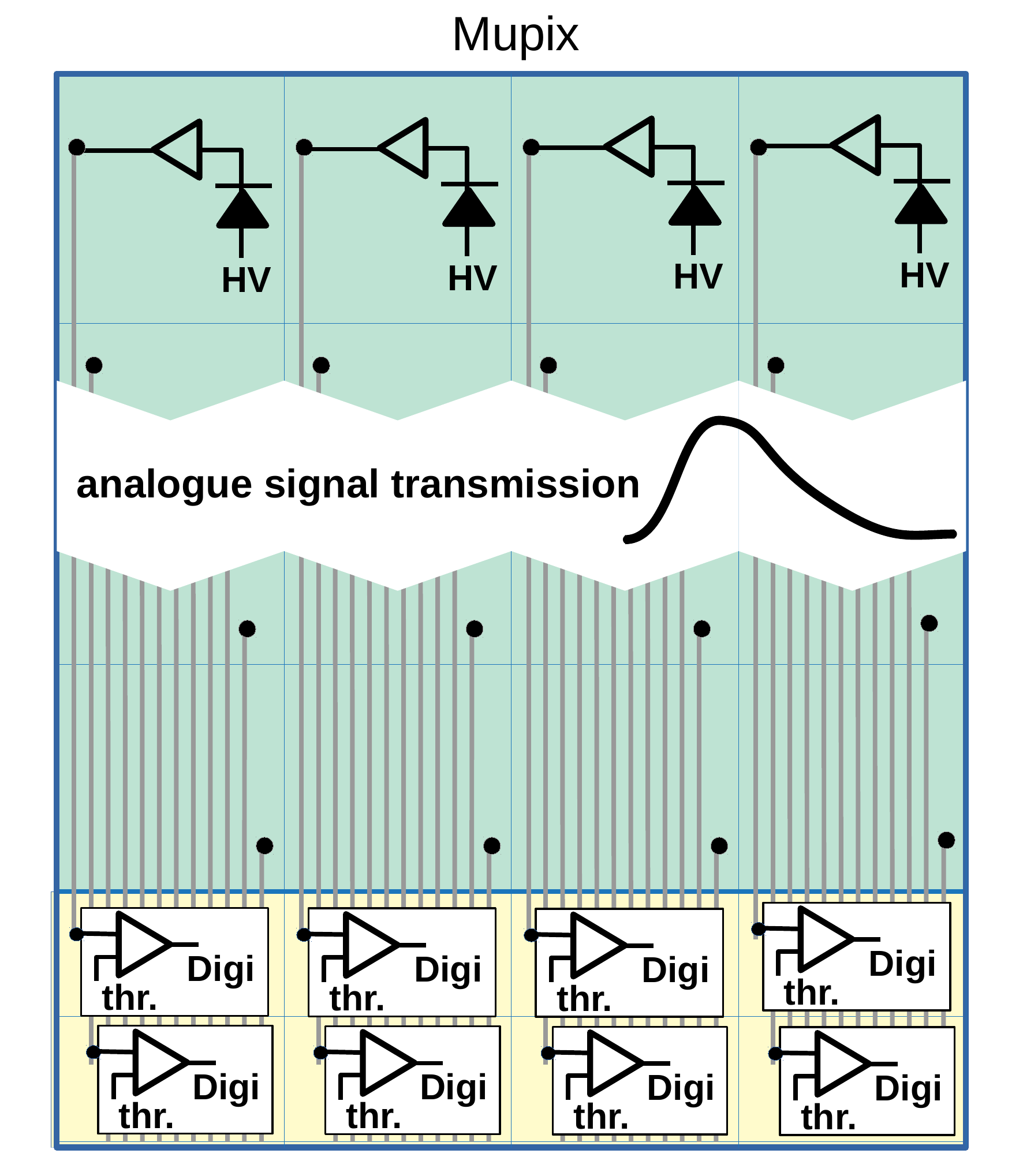}
  \end{subfigure}
        \caption{Sketch of \mupix readout concept with
          analogue cell readout.
          The triangles in the active matrix (green) represent the amplifiers. Source followers
          transmit the signals (vertical lines) to the comparators at the periphery.}
        \label{fig:mupix_readout_concept}
\end{figure}

Specific for all \mupix designs is the spatial
separation of charge-sensitive amplifiers (in the active pixel matrix)
from the comparators (in the chip periphery with the readout
circuitry). Each pixel cell implements a source follower which drives
the analogue signal to the periphery,
see~\autoref{fig:mupix_readout_concept}.

\begin{sloppypar}
The \mupix readout circuitry provides zero suppression and generates
timestamps to enable time-matching of hits from different pixel
layers. For \mupix{8} and later generations, a second timestamp is
generated to provide time-over-threshold (ToT) information. Offline,
the ToT information can be used for time-walk corrections, better noise
suppression, and for improving the spatial resolution by means of
charge sharing.
\end{sloppypar}

After hit detection and digitisation, an internal state
machine collects all hits using an address prioritisation scheme.
Data are sent via up to three serial links at a nominal data rate of
$\SI{1.25}{Gbit/s}$ each using a simple protocol with time frames and
8~bit/10~bit encoding.
Alternatively, in the multiplexed mode it is possible to use only one serial link.

\subsection{HV-CMOS Process and Manufacturers}
The $\SI{180}{\nm}$ H18 HV-CMOS process was selected based on the high
level of achievable integration, and the positive results from
prototypes for high rate capability, timing resolution and efficiency.
The breakdown voltage of the IBM HV-CMOS process is \SI{60}{V} or
greater, depending on the design rules. The maximum available reticle
size depends on the manufacturer, and is slightly larger than the
envisaged chip size of about $20 \times 23 \SI{}{\mm^2}$ for all
foundries.

The HV-CMOS process was originally developed for the automotive industry 
and thus offers long-term availability as well as specifications covering a wide 
range of operating conditions.
Although the HV-CMOS production costs are higher than for standard CMOS 
processes, they are significantly lower than for hybrid
silicon sensors, thus making the large Mu3e pixel detector with an instrumented area of about $\SI{1}{\meter\squared}$ affordable.

The original $\SI{180}{nm}$ HV-CMOS process from IBM \cite{IBM7HV} was offered by
ams AG~\footnote{ams AG, Austria, {\tt http://www.ams.com}} until 2015 and was used for the production of several
prototypes including \mupix{7}.
In 2017, ams AG changed to a new in-house developed process which was announced
to be similar to the original H18 process from IBM and used for the production of the \mupix{8} and \mupix{9} prototypes.

The $\SI{180}{nm}$ HV-CMOS process from IBM is also offered by
Global\-Foundries Inc.\footnote{Global\-Foundries Inc., USA, {\tt
    https://www.global\-foun\-dries\-.com}} and TSI
Semiconductors~\footnote{TSI Semiconductors, USA, {\tt
    http://www.tsisemi.com}}. In contrast to Global\-Foundries, TSI
also offers chip production with non standard substrates. This allows
for higher resistivity substrates and thus higher charge collection
signals. In addition, TSI provides seven metal layers instead of six
(ams AG). This feature is crucial for the reduction of cross-talk
(see the next section). Also for cost reasons, we have chosen the
$\SI{180}{nm}$ HV-CMOS process from TSI as baseline for the production
of the \mupix sensor. Since 2018, several HV-MAPS have been
successfully produced at TSI.

\subsection{Limitations and Design Challenges}

\begin{figure}[t!]
  \centering
  \includegraphics[width=0.43\textwidth]{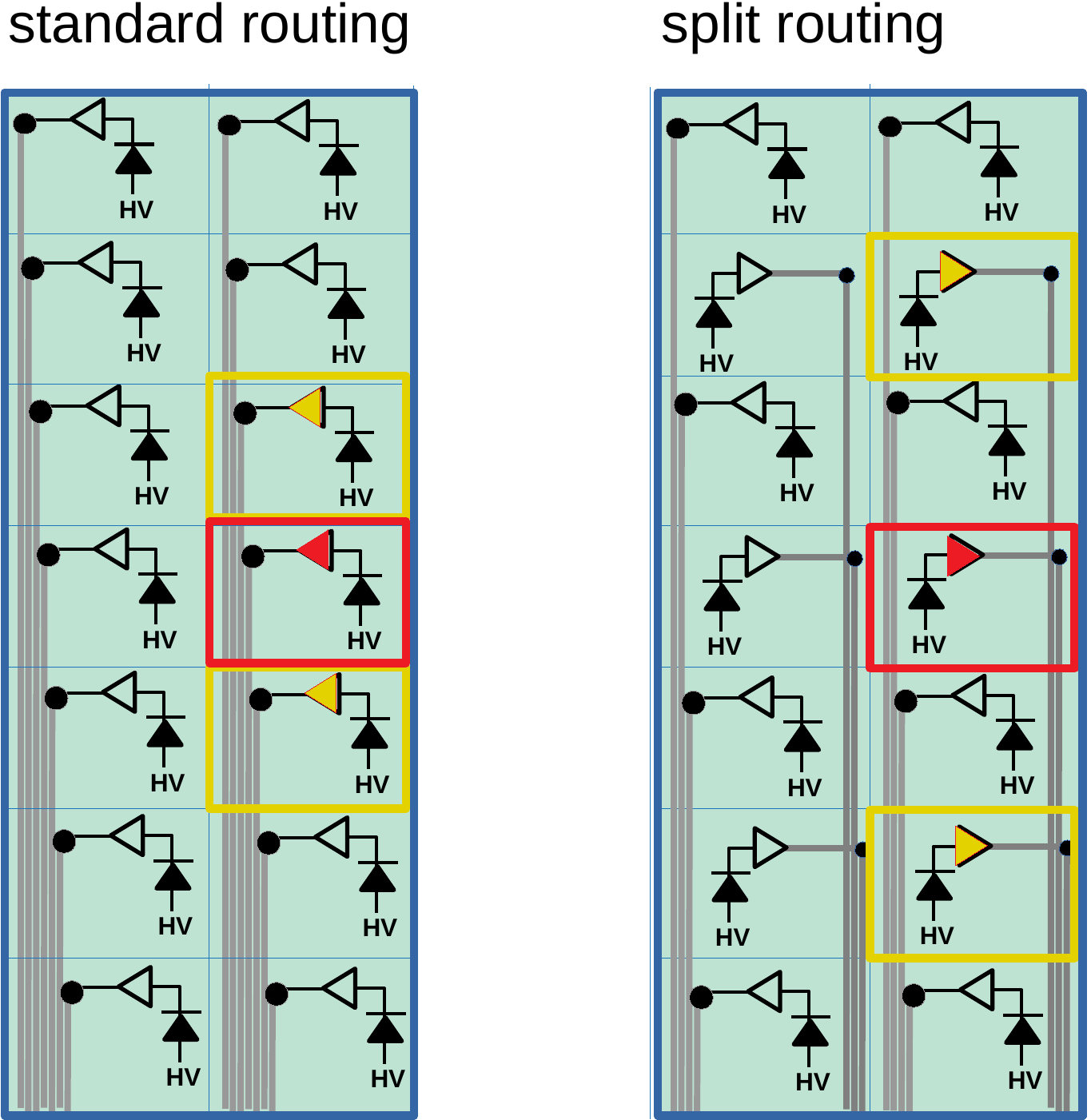}
  \caption{Schematics of two pixel matrices with different signal line routing schemes:
    conventional (left) and split routing (right). Large capacitive couplings
    between adjacent signal lines may lead to triplet hit patterns (red and
    yellow boxes). For the final \mupix design, a split routing scheme is used. 
        }
        \label{fig:split_readout}
\end{figure}

\begin{sloppypar}
The high bias voltage of the HV-MAPS concept requires a careful design
of the pixel cell geometry. TCAD simulation~\cite{TCADsim,TCADdevice}
is mandatory for the design to avoid large field gradients which lead
to early breakdown. The large fill factor of the HV-MAPS cell design
implies relatively large pixel capacitances, thus increasing noise,
compared to the more standard ``low fill factor designs''. As a general
design principle, pixel capacitances should be minimal since they lead
to signal deformation and increase time-walk effects, thus
compromising the maximum achievable time resolution. Large pixel
capacitances can be partially compensated by amplification stages with
larger gain. However, this goes along with a high power
consumption and also increases the risk of electronic cross-talk.
\end{sloppypar}

A limitation of the \mupix concept is related to the long analogue
readout lines from the pixel cell to the periphery,
see~\autoref{fig:mupix_readout_concept}. These long interconnects (up
to \SI{2}{\cm}) have large capacitive couplings with the neighbouring
lines and are prone to cross-talk, strongly dependant on the spacing
of the readout lines -- and thus on the number of available metal layers
for routing. To reduce cross-talk, several mitigation strategies have
been studied. With \mupix prototypes, special multi-layer routing
schemes which reduce the coupling between readout lines were
implemented and tested. In addition, dedicated routing topologies --
one of the simplest being split readout of even and odd rows,
see~\autoref{fig:split_readout} -- can be used to differentiate
between charge sharing among neighbouring pixels and cross-talk between
adjacent signal lines. Furthermore an approach with a two-stage
amplifier was studied \cite{Augustin2014} but not further considered.
Detailed cross-talk measurements have been performed with the
\mupix{8} prototype and are discussed in more detail
in~\autoref{sec:mupix_results}.

\section{\mupix Prototypes}
\label{sec:prototypes}

\begin{figure}[htb]
        \centering      \includegraphics[width=0.48\textwidth]{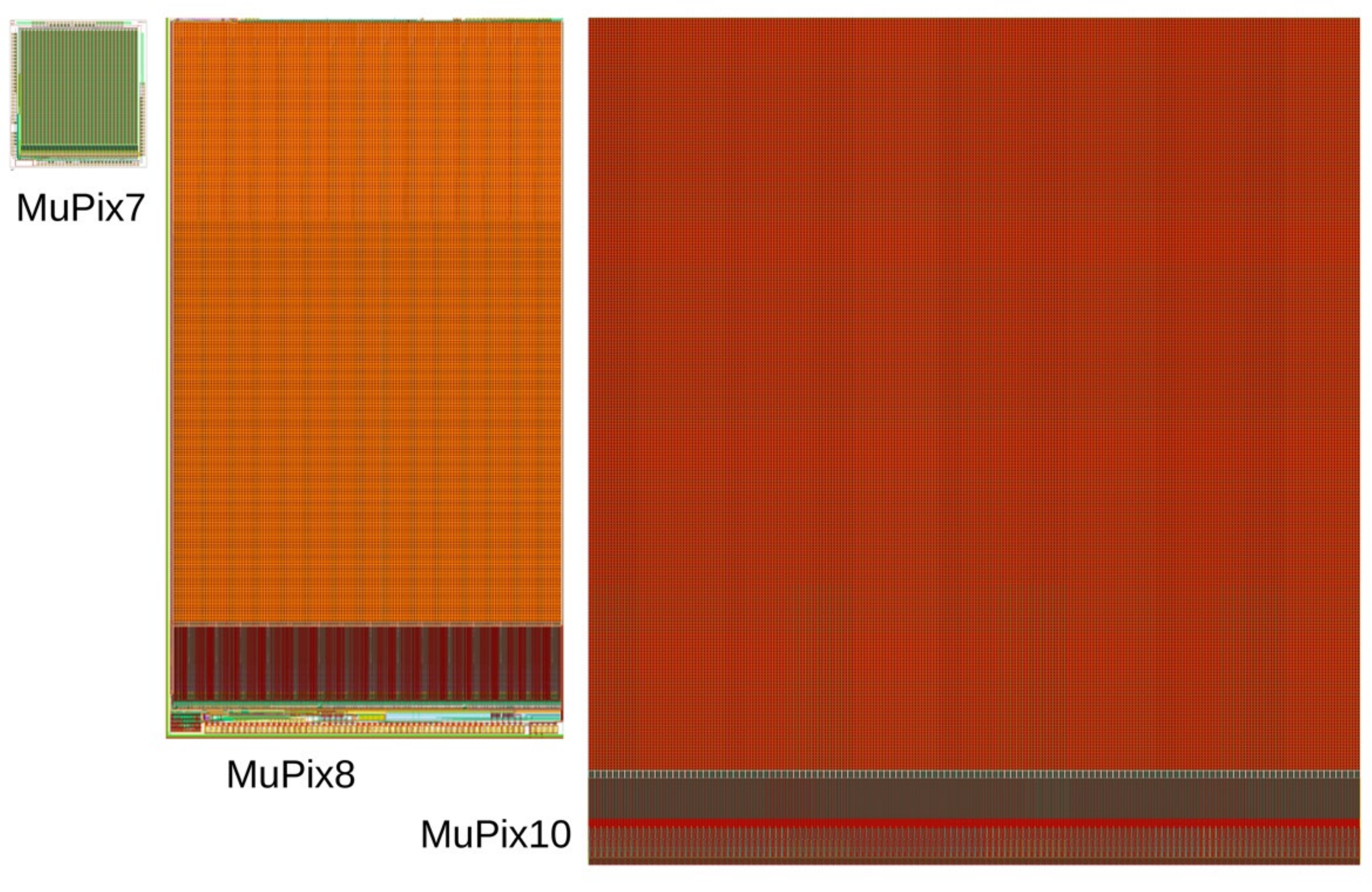}
        \caption{Layouts and size comparison of selected \mupix prototypes.
          The main area is the active pixel matrix, the stripe at the bottom contains the digital electronics (periphery).
          For reference, the size of \mupix{10}
        is $20.66\times\SI{23.18}{\milli\meter\squared}$.
        }
        \label{fig:mupix_comparison}
\end{figure}

Nine prototypes were produced so
far in preparation of the Mu3e experiment, the latest being \mupix{10}.
The layout of selected prototypes is shown in~\autoref{fig:mupix_comparison}.
Their specification and measured performance parameters are listed in~\autoref{tab:Mupix_Overview}, compared
with the main Mu3e requirements.

\begin{table*}
        \centering
        \small
        \begin{center}
                \begin{tabular}{lrrrrr}
                        \toprule
                                             & Requirements  & \mupix{7}        & \mupix{8}       & \mupix{10} \\ \midrule
  pixel size [$\SI{}{\micro\meter\squared}$] & $80\times80$  & $103\times80$  & $81\times80$   & $80\times80$    \\
  sensor size [$\SI{}{\mm\squared}$]         & $20\times23$  & $3.8\times 4.1$& $10.7\times 19.5$   & $20.66\times23.18$   \\
  active area [$\SI{}{\mm\squared}$]         & $20\times20$  & $3.2\times 3.2$& $10.3\times 16.0$   & $20.48\times20.00$    \\
  active area [$\SI{}{\mm\squared}$]         & $400$         & $10.6$         & $166$              & $410$    \\
  sensor thinned to thickness [$\SI{}{\micro\meter}$] & $50$          & $50$, $63$, $75$
   & $63$, $100$ &  $50$, $100$     \\
  LVDS links                                  & $3+1$           & $1$            & $3+1$               & $3+1$                     \\
  maximum bandwidth$^\mathsection$  [$\SI{}{Gbit/s}$]                & $3 \times 1.6$       & $1 \times 1.6$        & $3 \times 1.6$           & $3 \times 1.6$                  \\ 
  timestamp clock  [$\SI{}{MHz}$]            &  $\geq50$     &    $62.5$     &  $125$             & $625$   \\ \midrule
  RMS of spatial resolution [$\SI{}{\micro\meter}$]  & ${\leq30}$    &  $\leq30$ & $\leq30$       & $\leq30$   \\
  power consumption [$\SI{}{\milli\watt\per\cm\squared}$]  & $\leq350$  &
  $\approx 300^\dagger$ &  $250-300$ &  $\approx 200$    \\
  time resolution per pixel [$\SI{}{\nano\second}$]    & $\leq20$      & $\approx 14$  &
        $\approx 13~(6{^*})$      &   not meas.$^\ddagger$       \\
  efficiency at $\SI{20}{Hz/pix}$ noise [$\SI{}{\percent}$]             & $\geq99$    & $99.9$ &   $99.9$   & $99.9$         \\
 noise rate at $\SI{99}{\percent}$ efficiency [$\SI{}{\hertz\per pix.}$]     &  $\leq20$  &$<10$    &$<1$     &$<1$   \\ \midrule
  amplifier type                          & no spec.     & PMOS            & PMOS             & PMOS                     \\
  amplifier stages                         & no spec.     & 2            & 1                  & 1                     \\
  timestamp representation                 &   no spec.    &     8 bit   &     10 bit         &    11 bit  \\ 
  ToT representation                       &   no spec.    &     -          &     6 bit        &      5 bit  \\ 
  ring transistors (irradiation tolerant)    &   no spec.    &     no         &       yes        &  yes   \\
  approx. substrate resistivity$^\mathparagraph$ [$\SI{}{\ohm\cm}$]    &  no spec.     & $\approx 20$  &
  $\approx 20, 80, 200$   & $\approx 200$  \\
\bottomrule
                \end{tabular}
        \caption{Mu3e pixel sensor specification and performance
          parameters achieved for selected \mupix prototypes.
          Notes:  $^\mathsection$The nominal bandwidth is only
          $\SI{1.25}{Gbit/s}$ per serial link but a higher value was specified
          as a safety margin.
          $^\dagger$The operation points of the DAC values were set according to
          the given power consumption.
          $^*$The time resolutions given in brackets refer to 
          offline time-walk corrected values.
          $^\ddagger$The time resolution of \mupix{10} has not been measured yet;
          time resolutions of $\approx \SI{10}{\nano\second}$ w/o and
          $\approx \SI{5}{\nano\second}$ with offline time-walk correction are expected. $^\mathparagraph$The given resistivities are only approximate and are specified in the ranges $10-\SI{20}{\ohm\cm}$ (20), $50-\SI{100}{\ohm\cm}$ (80) and $200-\SI{400}{\ohm\cm}$ (200).
      }
        \label{tab:Mupix_Overview}
        \end{center}
\end{table*}

The \mupix{7} sensor~\cite{Augustin:2018ppf} was the first prototype which included
all main functionalities required for the Mu3e experiment:
 a fully integrated readout state machine, high
speed clock generation circuits (PLL) and a fast serial output link
running at up to $\SI{1.6}{Gbit/s}$, able to drive signals over
$\SI{2}{\meter}$.
The \mupix{7} sensor has an active area of about $3.2\times
\SI{3.2}{\milli\meter\squared}$.

The \mupix{8} sensor~\cite{Augustin:2018Elba,Schoning:2020zed} was the first large scale sensor produced in an
engineering run and was produced with different p-substrate resistivity.
Unlike previous prototypes, \mupix{8} was produced in the new AH18 process by ams AG.
\mupix{8} features a more radiation tolerant design\footnote{The HV-CMOS process has been qualified for
fluences of up to $2 \cdot 10^{15}$~(1 MeV)~neq in the context of the ATLAS high
luminosity upgrade~\cite{Augustin:2017guc,Kiehn:2019wpe,phd_thesis_herkert,Schoning:2020zed}.}
and is prepared for high data
rates by implementing three additional serial links.
Furthermore, a second time stamp was added to measure ToT,
thereby enabling the offline correction of time-walk effects due to
pulse height variations.
The main purpose of the \mupix{8} prototype was the study of topics related to the size of the sensor: the power network design and potential cross-talk
between the long analogue readout lines.

Special system aspects relevant for the construction of the pixel tracker
were addressed with the dedicated small area \mupix{9} sensor which
features 
fast differential control inputs, a new chip configuration scheme and a
shunt-Low Drop Out (LDO) regulator to study serial powering.

\mupix{10} is the final prototype, prepared for the construction of
pixel modules.
The active area is $20.48\times \SI{20.00}{\milli\meter\squared}$. The
pad layout is compatible with the spatial requirements of the pixel modules and the HDI design rules.
The sensor was produced at TSI in an engineering run and delivered in March 2020.

In the following, the final design of the \mupix sensor -- as it will be used in the Mu3e experiment -- is described.

\begin{figure}[tb!]
        \centering
                \includegraphics[width=0.45\textwidth]{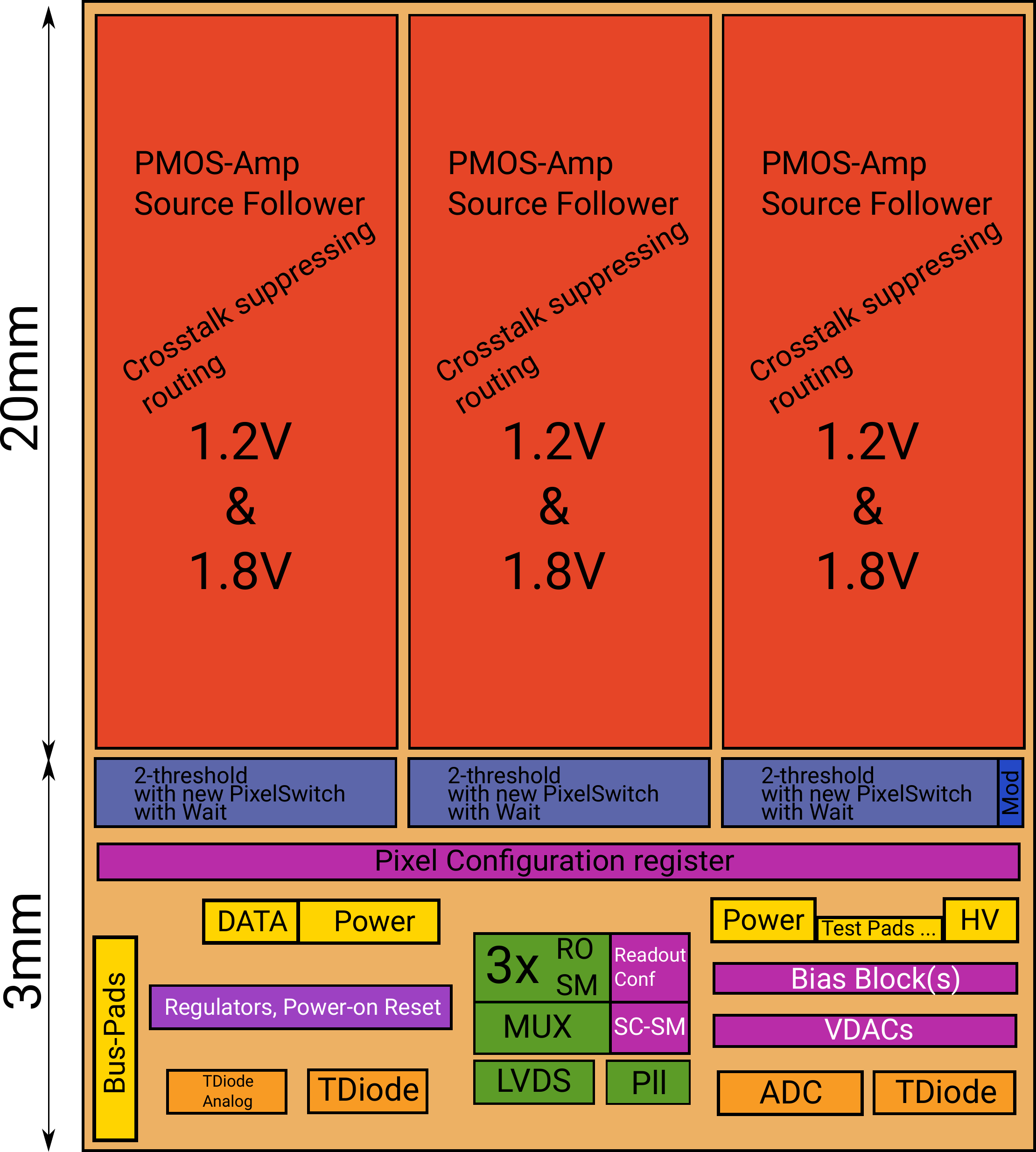}
\caption{\mupix{10} block diagram (not to scale).}
        \label{fig:mupix10_blocks}
\end{figure}

\begin{figure}[tb!]
  \centering    \includegraphics[width=0.45\textwidth]{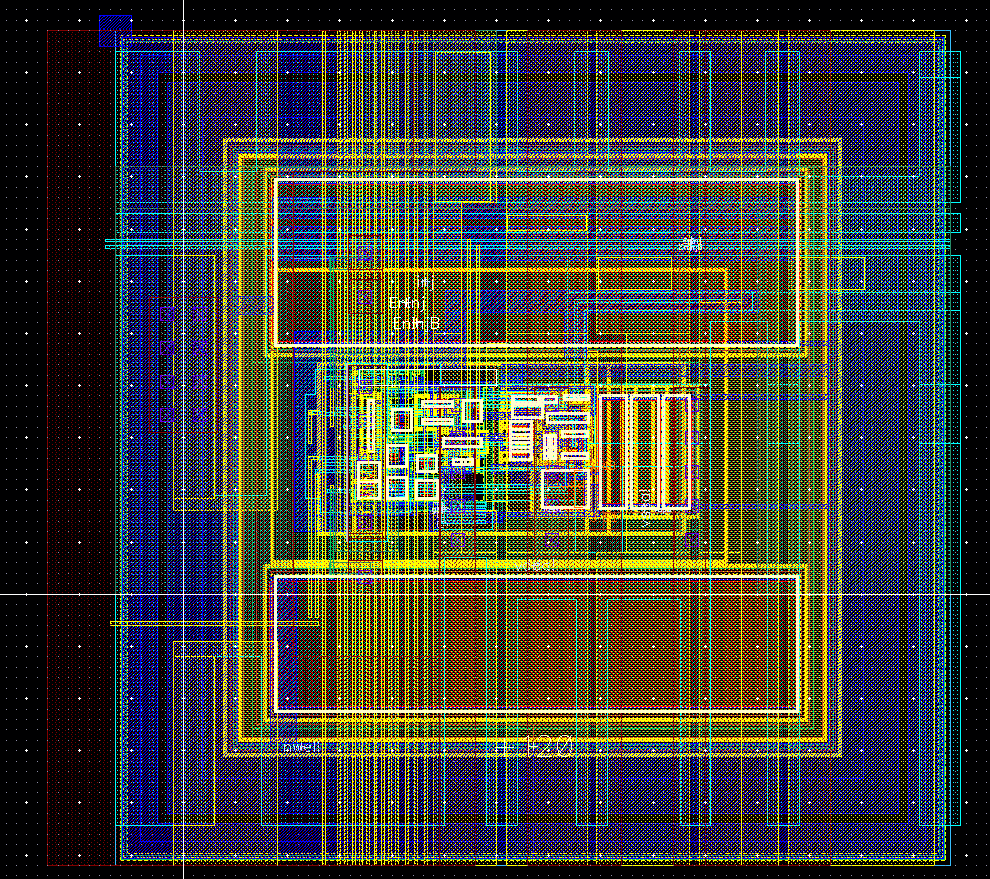}
  \caption{Layout of the in-pixel circuitry (amplifier and source follower) in
  the \mupix{10}. The pixel size is $80\times\SI{80}{\micro\meter\squared}$.}
        \label{fig:layout_mupix_pixel_amp}
\end{figure}

\begin{figure*}
        \centering
                \includegraphics[width=0.8\textwidth]{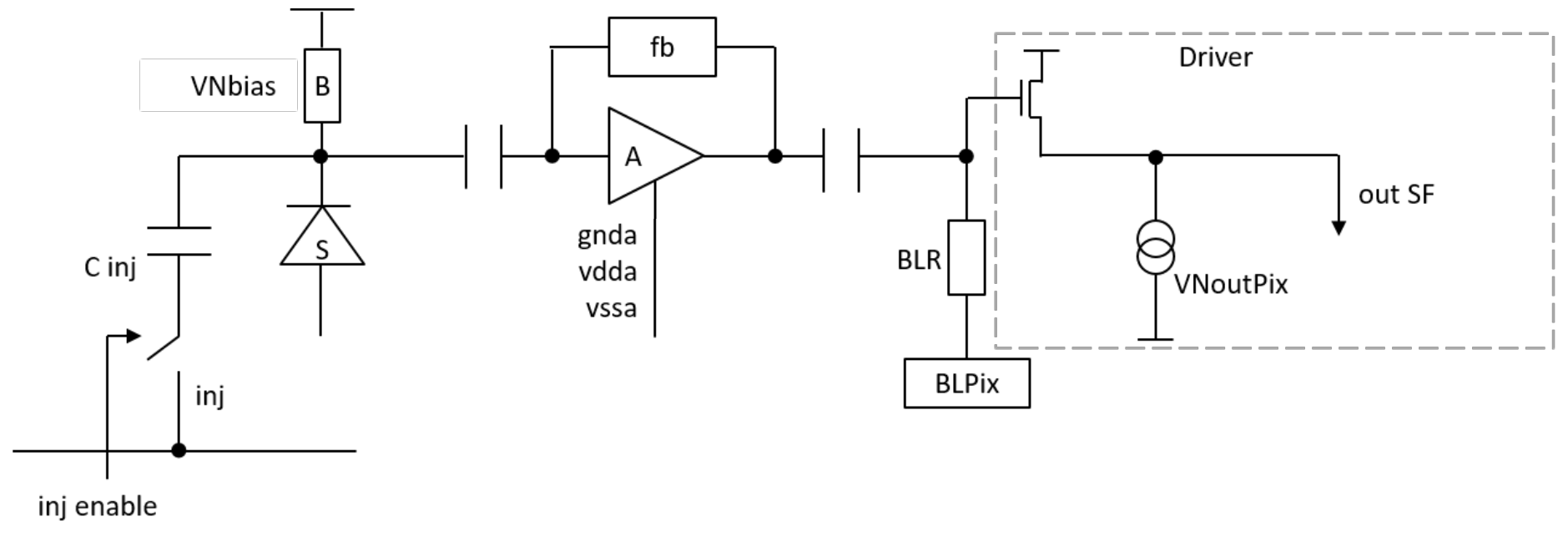}
        \caption{Schematic of the pixel cell analogue electronics in
                the \mupix chips. The main components are from left to right:
                charge injection (C Inj), the charge collecting diode (S), the
                n-well bias restoration circuit (B), the amplifier (A) with the feedback line
                (fb), the baseline circuit with base line restoration (BLR)
                and the source follower (out SF). See text for further detail.}
        \label{fig:PixelElectronics}
\end{figure*}

\section{\mupix Final Design}
\label{sec:mupix_design}
The pixel cell has a size of $80\times\SI{80}{\micro\meter\squared}$,
and the pixel matrix is divided into three sub-matrices (A-C),
consisting of 42+43+43 double-columns. In total there are 256 pixel
columns and 250 pixel rows. Readout of hits in the three sub-matrices
are handled by state machines. Depending on the readout mode, hits in
the three sub-matrices are sent to corresponding serial links
SOUT1-SOUT3 (\textit{high bandwidth mode}) or to a common readout link
SOUTX (\textit{multiplexed mode}). Hits read from a pixel column are
collected at the end of the column using address prioritisation. This
implies that the hits are not in chronological order.

A block diagram of the \mupix{10} layout is shown in~\autoref{fig:mupix10_blocks} and the main function blocks are detailed in
the following.

\subsection{Pixel Cell Electronics}

The layout of the pixel cell housing the amplification circuitry is shown in~\autoref{fig:layout_mupix_pixel_amp} for \mupix{10}.
Each pixel consists of the sensor diode, a charge-sensitive amplifier and a
source follower to drive the signal to the chip periphery, see~\autoref{fig:PixelElectronics}.
Every pixel has a capacitor allowing to inject test charges. 
As baseline the implementation of a PMOS-based amplifier with source follower
is chosen.
Pulse shaping is adjustable via bias currents with
typical shaping times of ${\cal O}(\SI{1}{\micro s})$.
The size of the charge collecting diode was optimised using TCAD simulation, to ensure
a homogeneous electrical field, for a substrate resistivity
of $\SI{200}{\ohm\cm}$ and a depletion voltage of $\SI{-60}{\volt}$. 
The guard ring was optimised with a design goal of
$\SI{-120}{\volt}$ for the breakdown voltage.

\subsection{Readout Buffer Cell}
\label{sec:mupix_digital}
For all \mupix designs the digital electronics was placed at the chip
periphery. This design decision was motivated
by the goal to reduce cross-talk between the quickly switching digital signals
and the sensitive analogue circuits.
The Readout Buffer Cell occupies about
$160 \times \SI{4.2}{\micro\meter\squared}$, corresponding to about 10\% of the
active cell, and covers two pixel columns in width
(double column routing).
The main functionalities of the Readout Buffer Cell are described in the
following.

Comparators convert the analogue signal into an arrival time signal. 
The common threshold for the comparators is set globally.
Individual 3-bit digital-to-analogue converters (DAC) allow for fine-tuning the
threshold for each pixel.
This feature can be used to ensure an uniform signal response or noise suppression over the
pixel matrix. A fourth enable bit can be used to mask out noisy pixels.
For test purposes, the comparator output of pixels can be monitored via a 
dedicated output line (hitbus signal).

A hit is defined by the rising edge of the comparator output.
For timestamp generation the output is sampled with an adjustable frequency derived from the internal clock.

\begin{figure}[tb!]
  \centering
      \includegraphics[width=0.49\textwidth]{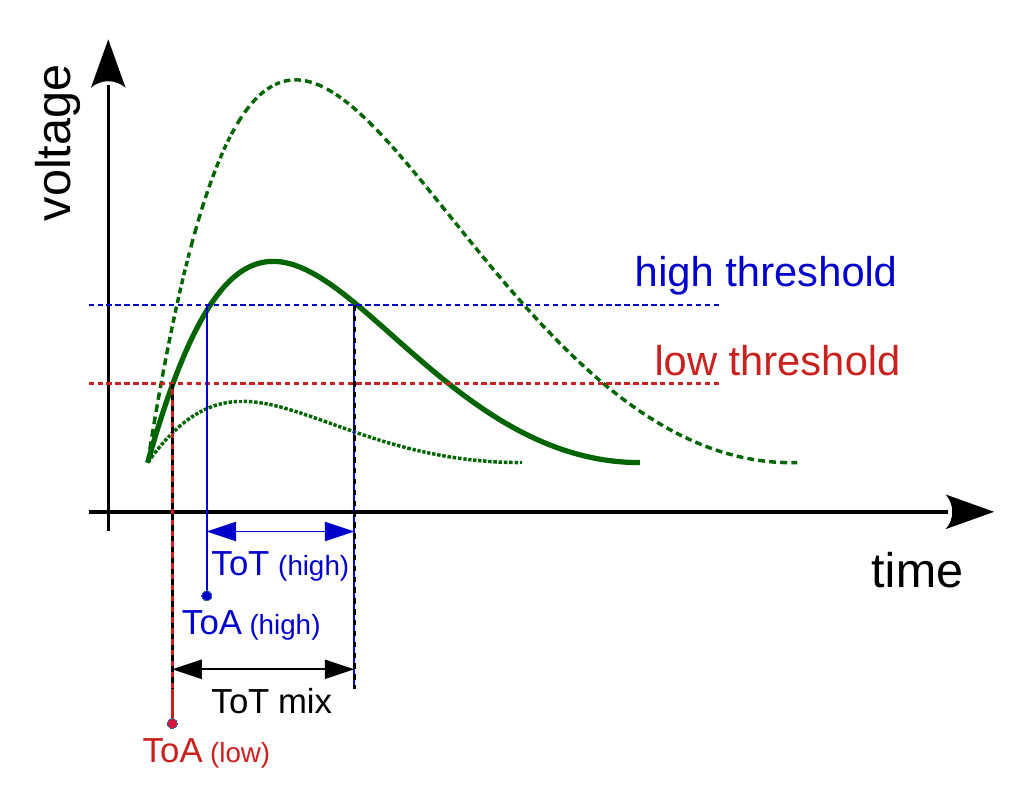}
      \caption{2-comparator threshold methods as implemented in \mupix{10}.
        In the mixed mode the ToA is determined by the low threshold, time-of-fall by the high threshold.
        The ToT information is correlated with pulse height and can be
	used to correct for time-walk.
        See text for further detail. 
}
	\label{fig:Mupix_TDC}
\end{figure}

For the latest \mupix prototypes, a second comparator was
added to each Readout Buffer Cell.
The 2-comparator threshold scheme allows the implementation of a very low threshold (close to noise) for
measuring the time-of-arrival (ToA) of the rising edge with little time-walk, and a high
threshold (well above noise) for the generation of the hit flag, see also~\autoref{fig:Mupix_TDC}.
This scheme proved successful with the \mupix{8}
prototype, for which an improvement of the time resolution was demonstrated~\cite{master_thesis_hammerich}.
In \mupix{10} it is possible to use only one comparator or a mixed mode where the ToA is defined by the lower threshold and the time-of-fall by the higher threshold. The mixed mode features time-walk mitigation and, additionally, provides robust ToT information for a residual time-walk correction.

In \mupix{10}, the timestamp is represented by 11~bits,
and the ToT by 5~bits. Both measurements are sampled with 
adjustable frequencies derived from the internal clock. All counters are
implemented as Gray counters and ToA and ToT are stored in floating capacitors\footnote{Actually, the time of the falling edge is stored from which the ToT is calculated at a later stage}.
\mupix{10} can be operated in different modes; with only one comparator
or with two comparators where the ToT measurement can be taken either from the lower or
 higher threshold (configurable). 

\begin{figure*}
	\centering
		\includegraphics[width=0.7\textwidth]{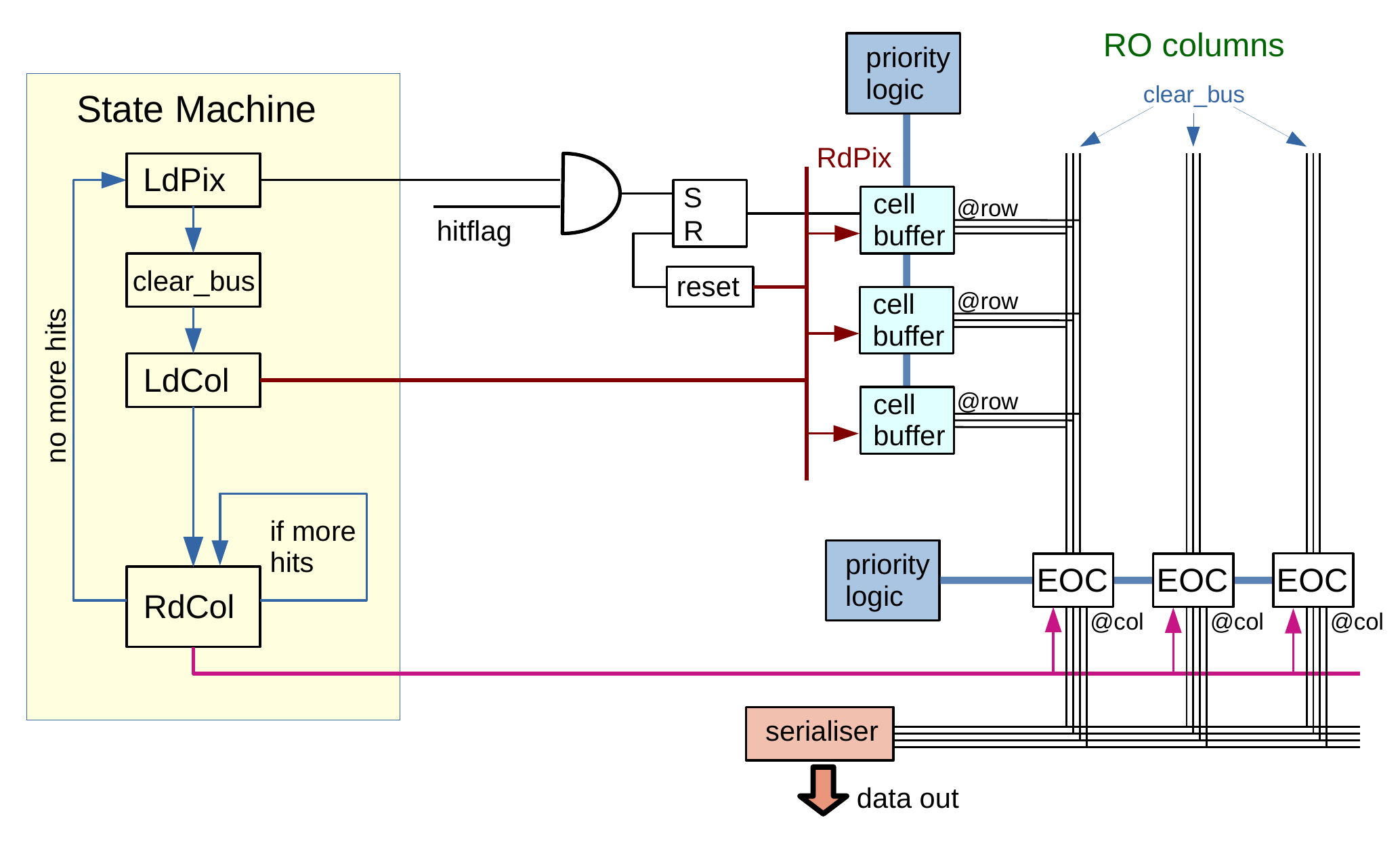}
	\caption{Schematic of the readout state machine and the column drain
		readout as implemented in \mupix{10}.
	See text for a full description.}
	\label{fig:sketch_RO_Mupix10}
\end{figure*}

In \mupix{10}, the hit-flag is generated by a hit-delay circuit with a
programmable timer in order to avoid huge sampling times for signals
with large ToT values. The purpose of the hit-delay circuit is
two-fold: first, it makes the hit-flag generation independent from the
pulse-height and helps to keep the chronological order of the
hits; second, it reduces dead-time by cutting out long sampling times.
The impact on ToT-based time-walk corrections is minor since time-walk
effects are small for large signals.

\subsection{Pixel Routing}
\label{sec:pixel_routing}
The routing for MuPix10 is slightly more complex due to
the double column RO scheme, i.e. two pixel columns are read out by one column of hit buffers.
In addition, a special routing scheme has been implemented by fully exploiting two metal
layers with the goals to A) reduce signal line cross-talk, B) enabling
the identification of signal line cross-talk and C) reduce 
row dependent time delays.
In comparison to a conventional comb-like routing scheme,
cross-talk is reduced by implementing a scheme where neighbouring lines are at their closest proximity 
at only for $1/4$ of the total column length. This routing scheme is fully described
in \cite{Mu3eNote:Mupix10,Augustin:2020pkv}.
The expected cross-talk rate is discussed in~\autoref{mupix:noise_crosstalk}.

The identification of remaining cross-talk is achieved by the addressing choice:
neighbouring signal lines are not connected to neighbouring
pixels, see~\autoref{fig:split_readout}.
Physical space and address space are consequently defined differently. Cross-talk is expected to
show (triplet) clusters in the address space whereas charge
sharing among pixels cells 
creates clusters in physical space. 

Furthermore, row dependent time delays are caused by signal line differences in the capacitive coupling
and ohmic losses. In \mupix{10} only four different lengths of the signal
lines are routed. Therefore, the time distribution of hits is expected
to show 4 discrete delays only.

\subsection{State machine}

Finally, the internal state machine reads out the all the hit information (address,
timestamps and ToT-values) from the Readout Buffer Cells.
We now describe the readout logic as
implemented in the chip internal state machine.

The readout scheme follows a standard column drain architecture, see~\autoref{fig:sketch_RO_Mupix10}.
A readout cycle starts with the issuing of the {\tt LdPix} signal. The hit flags, generated by the hit delay circuit, are then stored in a second register/latch.
Hits arriving after {\tt LdPix} will only be considered in the next readout
cycle, thus preventing race conditions due to hot pixels.
In the next step, the column buses are cleared.
Then, the {\tt LdCol} signal is sent to all columns.
By exploiting priority logic the {\tt RdPix} signal is generated for the 
highest priority pixel containing a hit. {\tt RdPix} reads the corresponding buffer cell
and drives timestamp information and the row address on the column bus.
These signals are registered at the end of column (EOC).
The corresponding pixel is reset (initiated by {\tt RdPix})
and is ready to accept the next hit.

If at least one column (or rather EOC) contains a hit, the {\tt RdCol} signal is 
issued, upon which the column with the highest priority drives the EOC data
(containing timestamp information and row address) and the column
address to the serialiser, and then resets itself.
This is repeated, until all columns are empty. Then the next hit in each column
is loaded to the column periphery, and so on.
The speed of the state machine is adjustable.

\subsection{Clocking and PLL}

The \mupix is synchronised to an external clock and generates 
internal clocks for hit sampling, the state machine and the serialiser.
It contains a tunable voltage controlled oscillator (VCO), running at up to
\SI{800}{\MHz}, and a PLL to keep the VCO in phase with respect to the 
reference clock.
The various stages of the serialiser, the readout state machine, and the 
timestamp counter all run at frequencies which are adjustable integer 
divisions of the fast base clock.
\mupix prototypes have been operated at a sampling frequency of
$\SI{62.5}{MHz}$ (\mupix{7}) and $\SI{125}{MHz}$ (\mupix{8});
a sampling frequency of up to $\SI{625}{MHz}$ was also successfully tested.

\subsection{Serial Links and Data Output}

After serialisation the data are sent out via fast serial links, which are
clocked with the internal fast clock ($\SI{625}{MHz}$) which corresponds to 5 times the
reference clock frequency ($\SI{125}{MHz}$). Data are sent at both, the rising and
falling edge of the clock, corresponding to a bandwidth of $\SI{1.25}{Gbit/s}$
per link.
Pre-emphasis of the signal can be controlled via bias currents to reduce rise
and fall times. Each LVDS links consumes about $\SI{15}{\milli\watt}$ of power.

Data are packed in 32 bit words which include the hit address (8 bit for
column and row address each) and Gray-encoded timestamp information (11 bits
for time-of-arrival and 5 bits for ToT).
Data on the serial link are 8bit/10bit encoded, using the standard IBM 
encoding \cite{Widmer1983, franaszek1984byte}. 
Operating at \SI{1.25}{GBit/s}, this leaves up to \SI{1}{GBit/s} for user data.
Comma words are used to define event frames and ease synchronisation of the
data. The data protocol~\cite{Mu3eNote:Mupix10} also foresees the sending of slow
control data.

In the high bandwidth mode three serial links are operated in parallel,
yielding a nominal data rate of \SI{3.75}{GBit/s} with a total payload of \SI{3}{GBit/s}.
In the multiplexed mode the state machine is alternately reading hits from
the three sub-matrices which are sent to a common serial link.
The configuration of the RO-mode is done by control commands.

\subsection{Powering and Configuration}

The final \mupix will work with only one supply voltage of $1.8-\SI{2.0}{V}$
(at chip) from which the internal VSSA$=\SI{1.2}{\volt}$ is generated using
an LDO which can also be bypassed by extra pads.
A power up reset, with a duration not exceeding $\SI{1}{ms}$, is
implemented to ensure that the chip starts up in a stable
configuration with only minimal power consumption (\textit{standby
mode}). Only the slow-control block is directly connected to the
supply voltage line, all other bias blocks are by default disabled
after power reset and need to be switched on by control commands. The
standby mode should allow for basic communication with the chip
without dedicated cooling measures.

All capacitors required for power reset, power regulators and noise decoupling are
implemented in-chip. The chip is designed such that no external de-coupling
capacitors and pull-ups or pull-downs are needed.

Control and configuration of the chip is possible over a differential serial
link (SIN). 64 bit long commands are used to fill the shift registers.
The decoding of the commands is handled by a slow control state machine.
The shift registers \textit{Bias}, \textit{Config} and \textit{VDAC} are responsible for 
the global chip configuration. The shift registers \textit{Col}, \textit{TDAC} and \textit{Test} are responsible for the configuration
of the test infrastructure (injection, \textit{ampout}, \textit{hitbus}), as well as the pixel tuning.

\subsection{Monitoring}

Monitoring information is primarily sent via the LVDS links. An ADC
is implemented to measure several internal voltages, such as
thresholds, baselines and regulated power, as well as temperatures
using thermo-circuits~\cite{Mu3eNote:Mupix10}. The ADC readout can be
configured via special control commands - for example, it is possible
to read out all slow control parameters round robin. The ADC measured
voltages can also be tapped by additional slow-control outputs. The
ADC is in the same power block as the slow-control and always
operational.

\begin{table*}
        \centering
        \small
                \begin{tabular}{lrrrrr}
                        \toprule
                         Pad & \# pads & IN/OUT & type & routed & description\\ \hline
                         GNDD  & $3$ & IN & mand. & GND & digital ground\\
                         GNDA  & $3$ & IN & mand. & GND & analog ground \\ 
                         VDDD & $5$ & IN & mand. & VDD & power supply\\ 
                         VDDA & $5$ & IN & mand. & VDD & power supply\\ \midrule

                         VOUTS & $2$ & OUT & select & - & regulated voltage out \\
                         VSSA  & $2$ & IN & mand. & - & regulated  voltage in \\
                         BIAS & $2$ & IN & mand. & BIAS & HV bias voltage (bus) \\ \midrule
                         CLK  & $2$ & IN & diff.  & CLK & $\SI{125}{MHz}$ system clock \\
                         SIN  & $2$ & IN & diff.  & SIN &  control bus\\
                         SOUTX  & $2$ & OUT & diff. & SOUTX & multiplexed   LVDS data output\\
                         SOUT1  & $2$ & OUT & diff. & SOUT1 &  LVDS data sub-matrix~1 \\
                         SOUT2  & $2$ & OUT & diff. & SOUT2 &  LVDS data sub-matrix~2 \\
                         SOUT3  & $2$ & OUT & diff. & SOUT3 &  LVDS data sub-matrix~3\\ \midrule
                       \bottomrule
                \end{tabular}
        \caption{List of mandatory \mupix{10} inputs and outputs.
        Two powering scheme for VSSA are bondable:
          \textit{regulated} by shorting VOUTS and VSSA or 
          \textit{bypassed} by supplying VSSA externally.}
        \label{tab:mupix_pinout}
\end{table*}

An additional analogue temperature measurement is implemented to allow
measurements if the chip is not powered or configured. This circuit
is optimised to be maximally sensitive the temperature range of
$0-\SI{100}{\celsius}$ with $\Delta V \approx \SI{300}{\milli\volt}$.
This diode is externally connected to power and floating if the chip
is not powered.

\subsection{I/O and Pad Layout}

\begin{sloppypar}
For the construction of \mupix modules the number of inputs and outputs
required to operate the chip must be kept at a minimum.
In total only 9 (13) lines have to be connected 
to operate the sensor in the multiplexed (high bandwidth) RO mode,
assuming that common ground (GNDD=GNDA) and power (VDDD=VDDA) are used for the digital and
analogue domains.
The list of mandatory connections is shown in \autoref{tab:mupix_pinout}. 
Multiple pads are implemented for power and ground to reduce ohmic losses.
The usage of a power regulator to generate VSSA internally is hardware
configurable\footnote{Actually, two parallel regulators are implemented for
distributing the power dissipation over a larger area at the
periphery.}.
All differential lines are ESD protected.
For all mandatory connections two different types of pads are implemented:
standard wedge bonds for chip tests and SpTA-bonds for pixel module production.
Additional standard pads are implemented for chip characterisation
studies. They enable monitoring of internal signals, thresholds and voltages.
The pad size for wedge bonds is $76 \times \SI{146}{\micro\meter\squared}$.
The pads for SpTA-bonds have a size of $200 \times \SI{100}{\micro\meter\squared}$ and
fulfil the specification for bonding the aluminium-polyimide HDIs produced by LTU.
\end{sloppypar}

\section{Performance of \mupix Prototypes}
\label{sec:mupix_results}

\begin{figure}
  \begin{subfigure}[l]{0.22\textwidth}
    \includegraphics[width=1.05\textwidth]{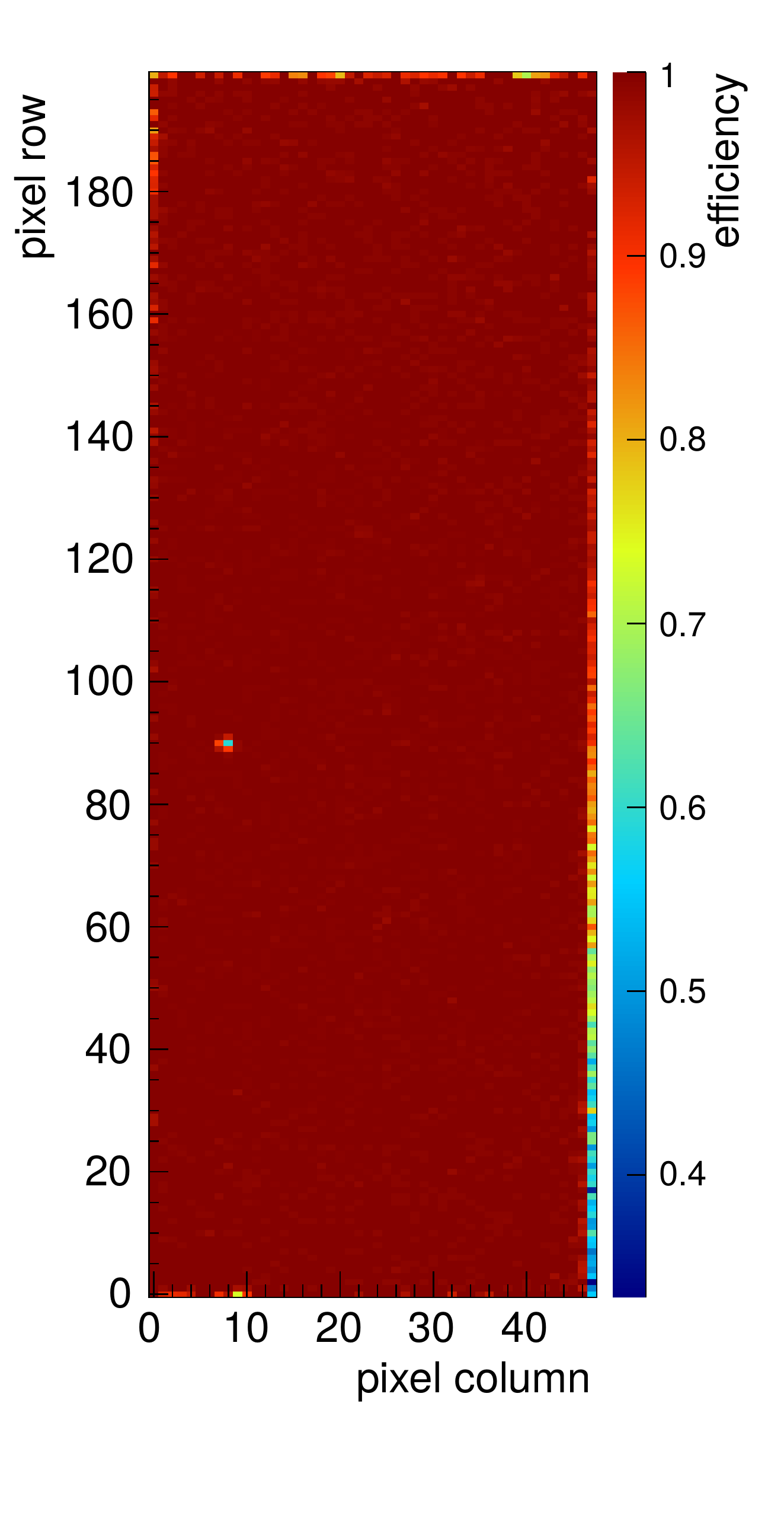}
    \subcaption{Efficiency}
  \end{subfigure}
  \begin{subfigure}[r]{0.22\textwidth}
    \includegraphics[width=1.05\textwidth]{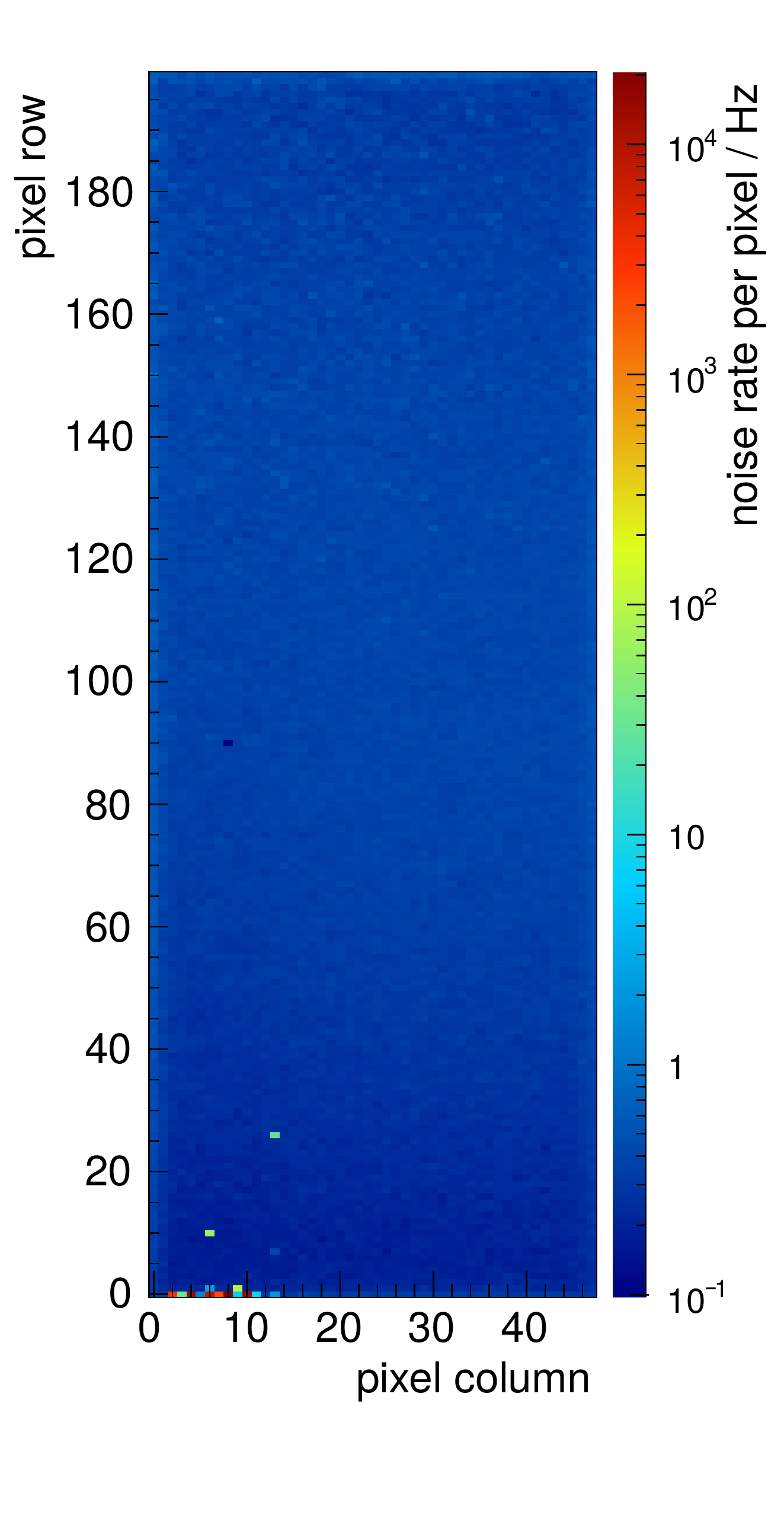}
    \subcaption{Noise}
  \end{subfigure}
  \caption{ Efficiency and noise maps for the \mupix8 sensor 084-2-03
    ($\SI{80}{\ohm\cm}$) at a threshold of $\SI{56}{\milli\volt}$ and
    a bias voltage of $\SI{-60}{\volt}$. Note that the noise was
    measured during beam and that the noise rate also includes
    non-reconstructed beam particles. Plot
    from~\cite{phd_thesis_huth}.}
        \label{fig:mupix8_eff_noise}
\end{figure}

As of 2020 nine \mupix prototypes have been studied by exploiting different
techniques for chip/sensor characterisation:
injection pulses, 
LEDs, laser diodes, X-rays, radioactive sources and test beam campaigns. 
The main results obtained from the latest \mupix prototypes
are presented here. Emphasis is given to results obtained from \mupix{8} which was thoroughly studied and has a design very similar to the final \mupix. 


\subsection{Single Hit Efficiencies}

Single hit efficiencies of \mupix sensors were determined 
in test beam campaigns at CERN (Geneva), DESY (Hamburg), MA\-MI (Mainz)
and PSI (Villigen). Studies were performed as a function of
various DAC and HV settings for different comparator thresholds and powering schemes. 
Beam telescopes were used for the reconstruction of reference tracks and the measurement of 
single hit efficiencies.
The efficiency map and noise map for a \mupix{8} sensor produced with an $\SI{80}{\ohm\cm}$ substrate is shown in~\autoref{fig:mupix8_eff_noise}.

\begin{figure}
        \centering      \includegraphics[width=0.48\textwidth]{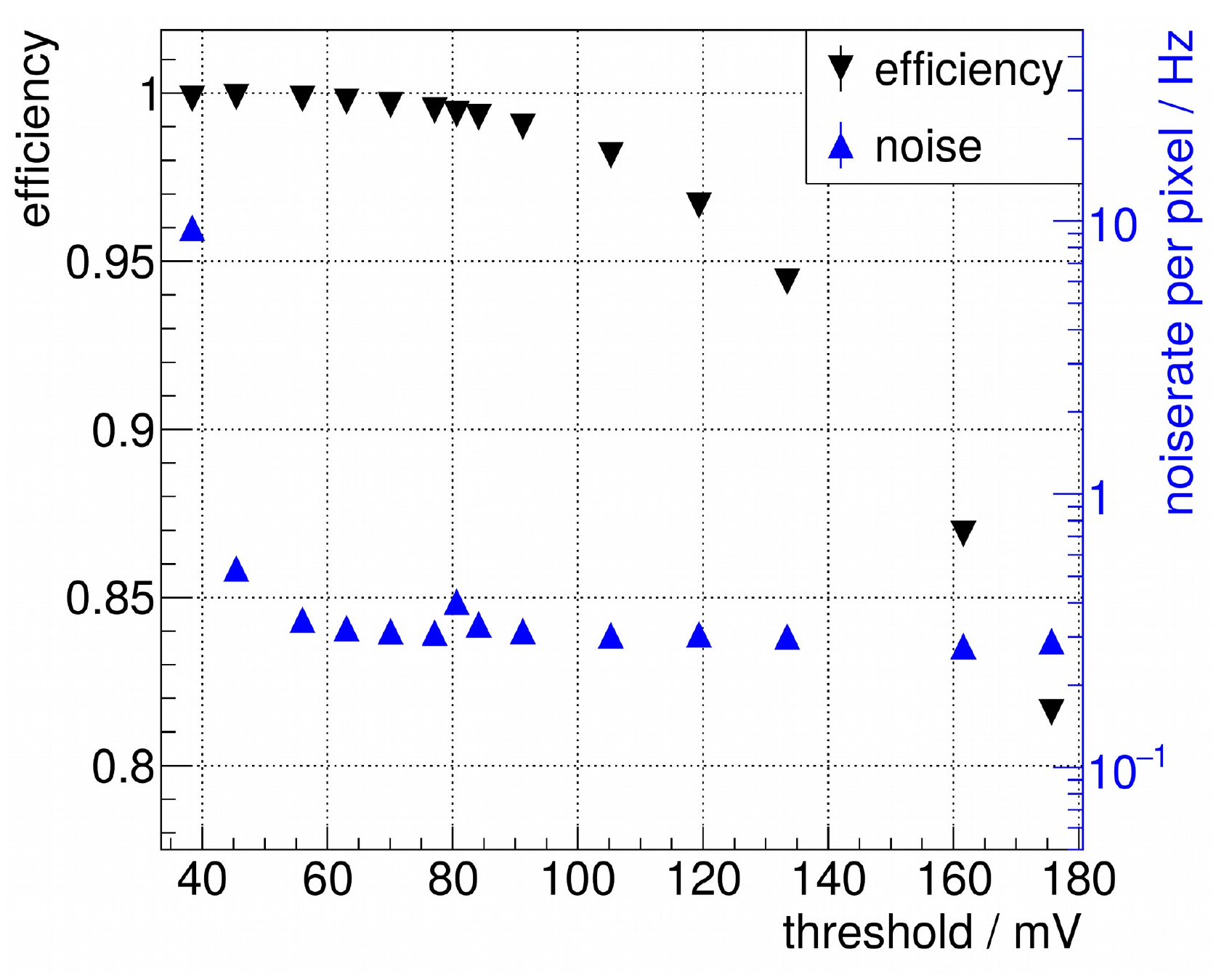}
        \caption{Hit efficiency and noise as a function of the charge threshold 
        for the \mupix{8} sensor 084-2-03 ($\SI{80}{\ohm\cm}$, thickness $\SI{62}{\mu\m}$) as measured for $\SI{4}{GeV}$ electrons for a beam
        inclination angle of $\ang{0}$.
        The bias voltage was set to $\SI{-60}{\volt}$ and the pixel cells were
        untuned. Plot from~\cite{phd_thesis_huth}.
         }
        \label{fig:mupix8_eff_unmasked}
\end{figure}

In~\autoref{fig:mupix8_eff_unmasked} the single hit efficiency of the same
\mupix{8} sensor is shown as a function of the threshold.
This sensor was operated at a bias voltage of $\SI{-60}{\volt}$ and
noisy pixels have not been masked.
The Mu3e efficiency and noise requirements are fulfilled in a large threshold
range of about $40$-$\SI{90}{\milli\volt}$, corresponding to about $650$-$\SI{1450}{electrons}$. This high efficiency range can be further extended by
tuning the individual pixel thresholds, by masking noisy pixel and by
increasing the bias voltage.
This measurement confirmed the expected increase of the depletion area, and the resulting hit efficiency,
by using a higher resistivity substrate than the standard $10-\SI{20}{\ohm\cm}$.
The increase of the depletion region is also supported by TCAD simulations~\cite{Buckland_TCAD} and HV-CMOS characterisation studies including 
Edge-TCT measurements~\cite{Cavallaro:2016gmx}.
For substrate resistivities of $\approx \SI{200}{\ohm\cm}$ an even wider plateau of high efficiency has been obtained~\cite{Augustin:2019qiv}.

\subsection{Noise and Cross-Talk}
\label{mupix:noise_crosstalk}

For optimised DAC settings a noise of about 90 electrons was measured for \mupix{8} using a threshold scan.
The source is mainly thermal noise from the capacitances of the diode and the amplifier input transistors.
The noise figure has to be compared to the expected number of primary
electrons which strongly depends on the substrate resistivity.
For the envisaged substrate of $\approx \SI{200}{\ohm\cm}$ and approx. $\approx \SI{30}{\micro\meter}$ depletion
more than $3000$ primary electrons are expected in the experiment.

\begin{figure}[htb!]
        \centering
                \includegraphics[width=0.48\textwidth]{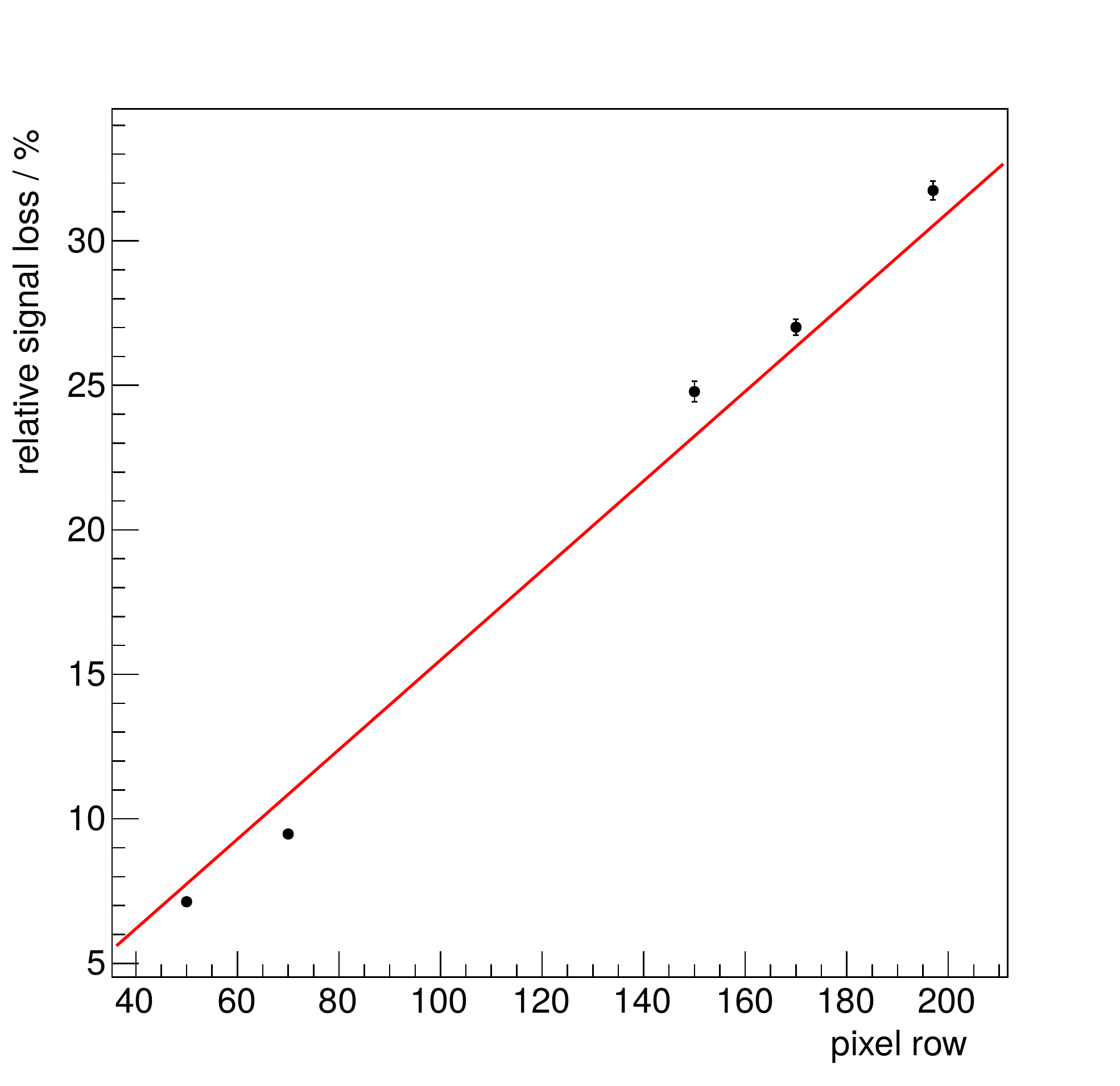}
        \caption{Relative amplitude loss of an injected
          signal in \mupix{8} as a function of the row number due to the capacitive couplings
          of the readout lines (full points).
          The red line corresponds to a proportionality constant of
          $\SI{0.155}{\%}$ per pixel row. Plot based on \cite{master_thesis_hammerich}.
        }
        \label{fig:MuPix8_capacitive_coupling}
\end{figure}

\begin{figure}[htb!]
        \centering
                \includegraphics[width=0.48\textwidth]{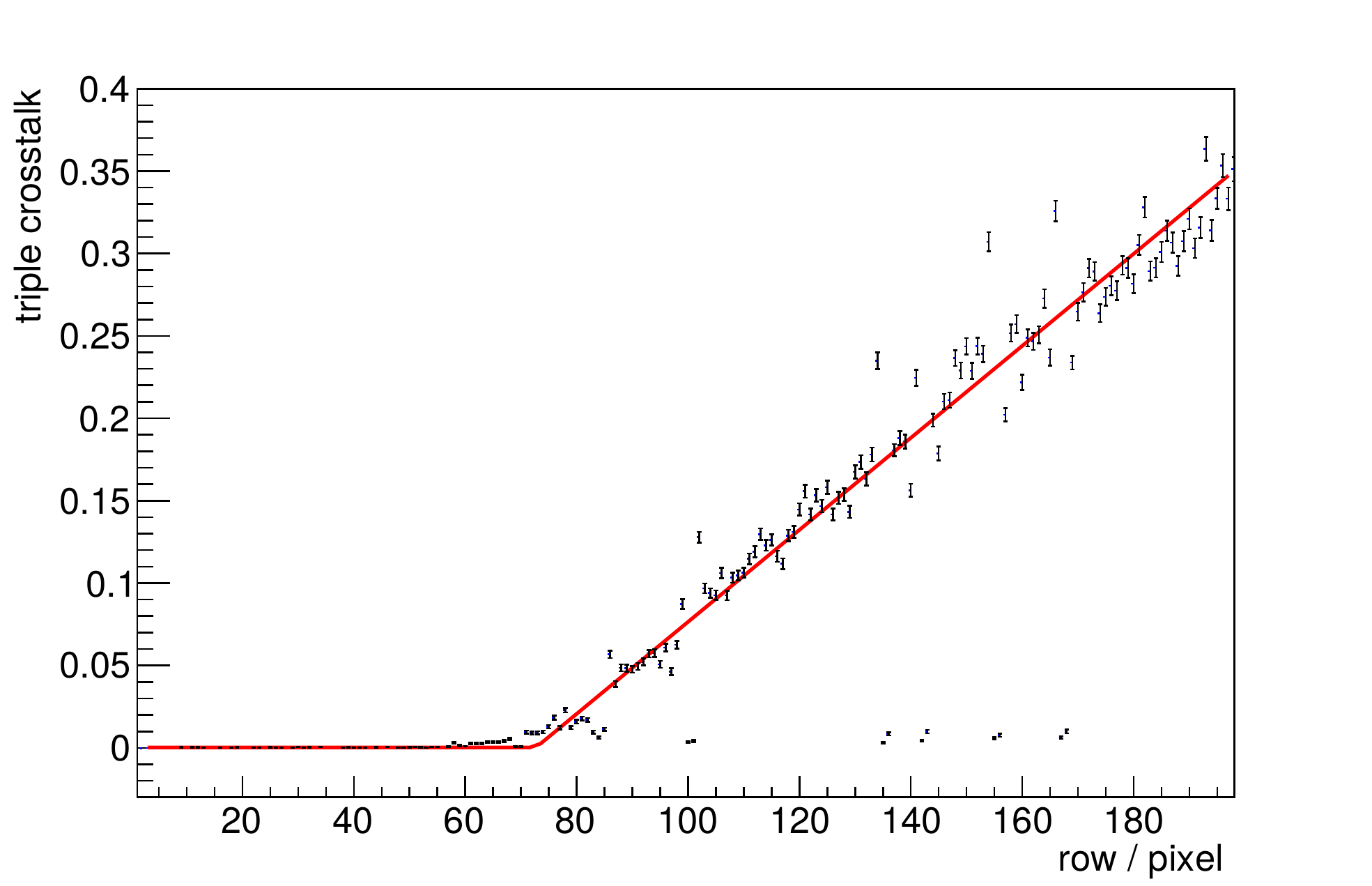}
        \caption{Triplet pattern probability due to cross-talk as a function of the
          row number in \mupix{8}.
          The red solid line shows a fit to the data. For more detail see~\cite{phd_thesis_huth}.}
        \label{fig:MuPix8_triple_row_function}
\end{figure}

Another source of noise is cross-talk which is particularly
dangerous in mixed signal designs where frequently switching signals in the digital circuitry
induce noise in the analogue section. 
Various tests have been performed and no cross-talk from
the digital section was detected for reasonable hit thresholds, even when the \mupix prototypes were operated at 
very high readout rates ($>\SI{1}{Mhits/s}$), thus confirming the \mupix design. 

Cross-talk between pixel cells was studied by analysing hit correlations.
Hit correlations are naturally expected from charge sharing if tracks
create ionisation charges in the vicinity of two pixels inside a cone of
about $\SI{3}{\micro\meter}$ \cite{phd_thesis_huth}. 
A clear correlation between the position of the charge deposition and 
charge sharing was seen in test-beam measurements 
but no significant cross-talk between pixels could be measured.

Significant cross-talk, however, was observed in \mupix{8} between the long analogue
readout lines
connecting the pixel cells with the comparators in the periphery, see~\autoref{fig:mupix_readout_concept}.
\autoref{fig:MuPix8_capacitive_coupling} shows the signal measured at
the comparator inputs of adjacent pixels in the same column after injecting a
pulse to the middle pixel. 
In \mupix{8} a conventional comb-like routing scheme was implemented where the length of RO lines scales linearly with the row number.
This scheme allows a detailed study of the cross-talk probability as a function of the row number, and thus the length of the RO line.
The capacitive coupling has been derived from the amplitude ratios of injected to measured signal,
and was found to be proportional to the length of the readout line, see \autoref{fig:MuPix8_capacitive_coupling},
with a signal loss of roughly 0.155 \% per pixel row.
Small deviations from linear behaviour are expected and due to non-linear routing effects, e.g. change of metal layers.

The capacitive coupling between RO lines leads to a specific triplet pattern, see discussion of \autoref{fig:split_readout}.
The frequency of this cross-talk has been derived from test-beam data as a function of the row number
and is shown in~\autoref{fig:MuPix8_triple_row_function}. The triplet
pattern probability above row number $\approx 70$ shows a linear
increase with the length of the readout line. For the highest row
numbers, corresponding to a signal line length of $\SI{1.6}{cm}$, the
probability is approx. 35\% that a triplet pattern fires.

From the \mupix{8} characterisation results the capacitive coupling between RO
lines is estimated for \mupix{10} to be $\approx 13$\%, considering the improved routing scheme (see \autoref{sec:pixel_routing}) and the 20\% increase of the signal line density.
For most hits, the amplitude of the cross talk signal is expected to be small enough to be below detection threshold.
If the cross-talk is above the hit threshold, special easy-to-identify patterns will emerge due to the \mupix{10} routing scheme.

\subsection{Time Resolution}

Several effects contribute
to the timing of hits in a monolithic sensor: pixel-to-pixel
variations in the amplifier response, signal routings of different length,
effects due to variations of the signal amplitude (time-walk), discretisation due to
the timestamp sampling and jitter due to noise.
The time resolution was studied in detail for the \mupix{7} and~8 designs.

For \mupix{7}, which has a special $3 \times 3$ diode structure in the pixel cell, small time variations
depending on the spatial position within the cell
have been measured~\cite{Augustin:2018ppf} with the high resolution EUDET
telescope at DESY.
These variations can be explained by inhomogeneities in the charge collecting field and
were found to be about $\SI{1.5}{ns}$, much smaller than the measured time resolutions of \mupix{7}
of about $\SI{14}{ns}$ which is dominated by time-walk effects.

\begin{figure}[htb!]
        \centering
                \includegraphics[width=0.48\textwidth]{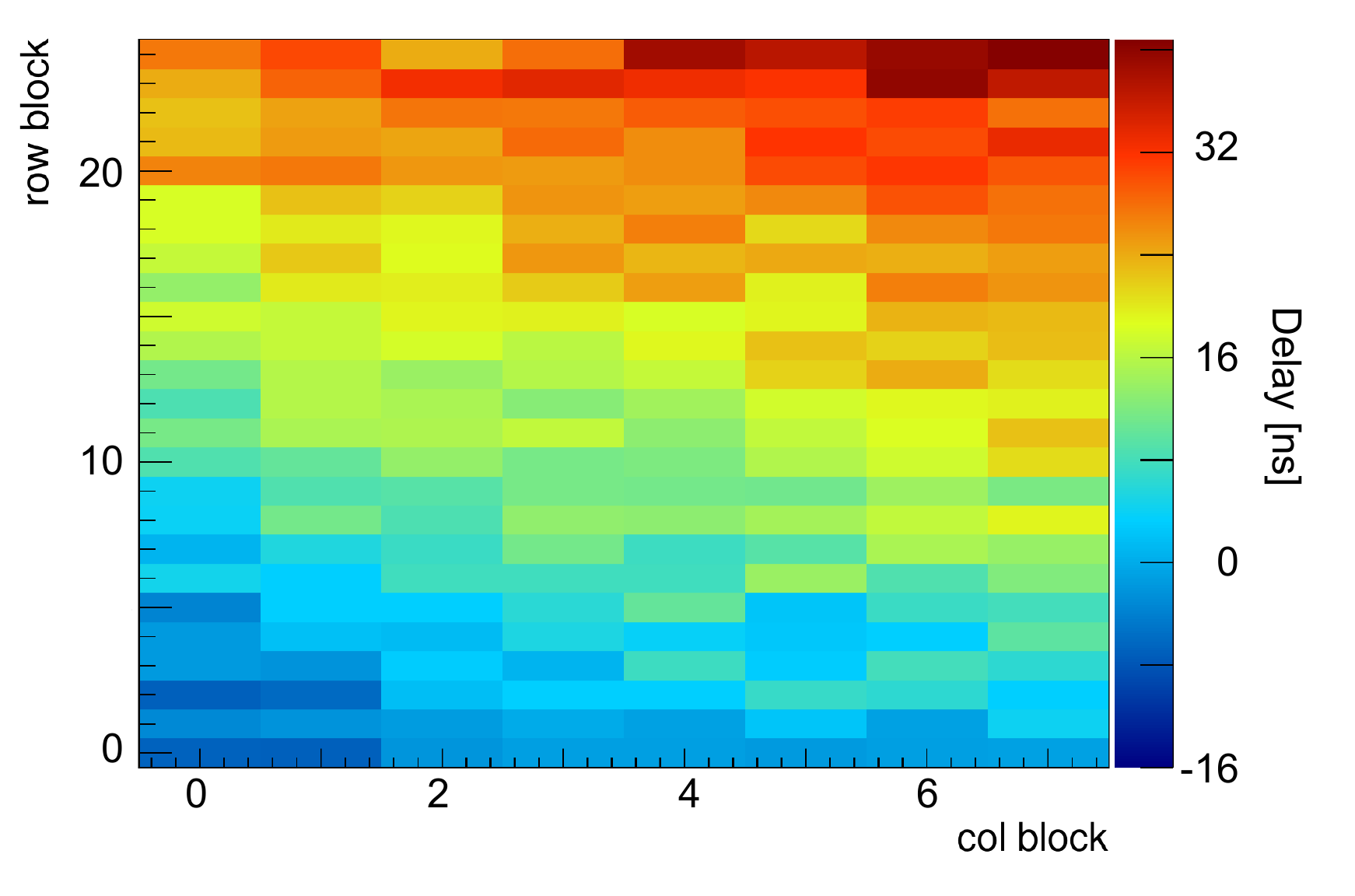}
        \caption{
                  Hit delay distribution of the first 48 columns of \mupix{8} derived using
a $^{90}$Sr source. Plotted is the discriminator output delay  with respect to a time reference
as a function of the column and raw number in units of a time sample of $\SI{8}{ns}$.
          The hit delays are determined for
          small areas of size $6 \times 8$ pixel cells.
        Plot from \cite{master_thesis_hammerich}}
        \label{fig:MuPix8_delay_blocks}
\end{figure}

\begin{figure}[htb!]
        \centering
                \includegraphics[width=0.46\textwidth]{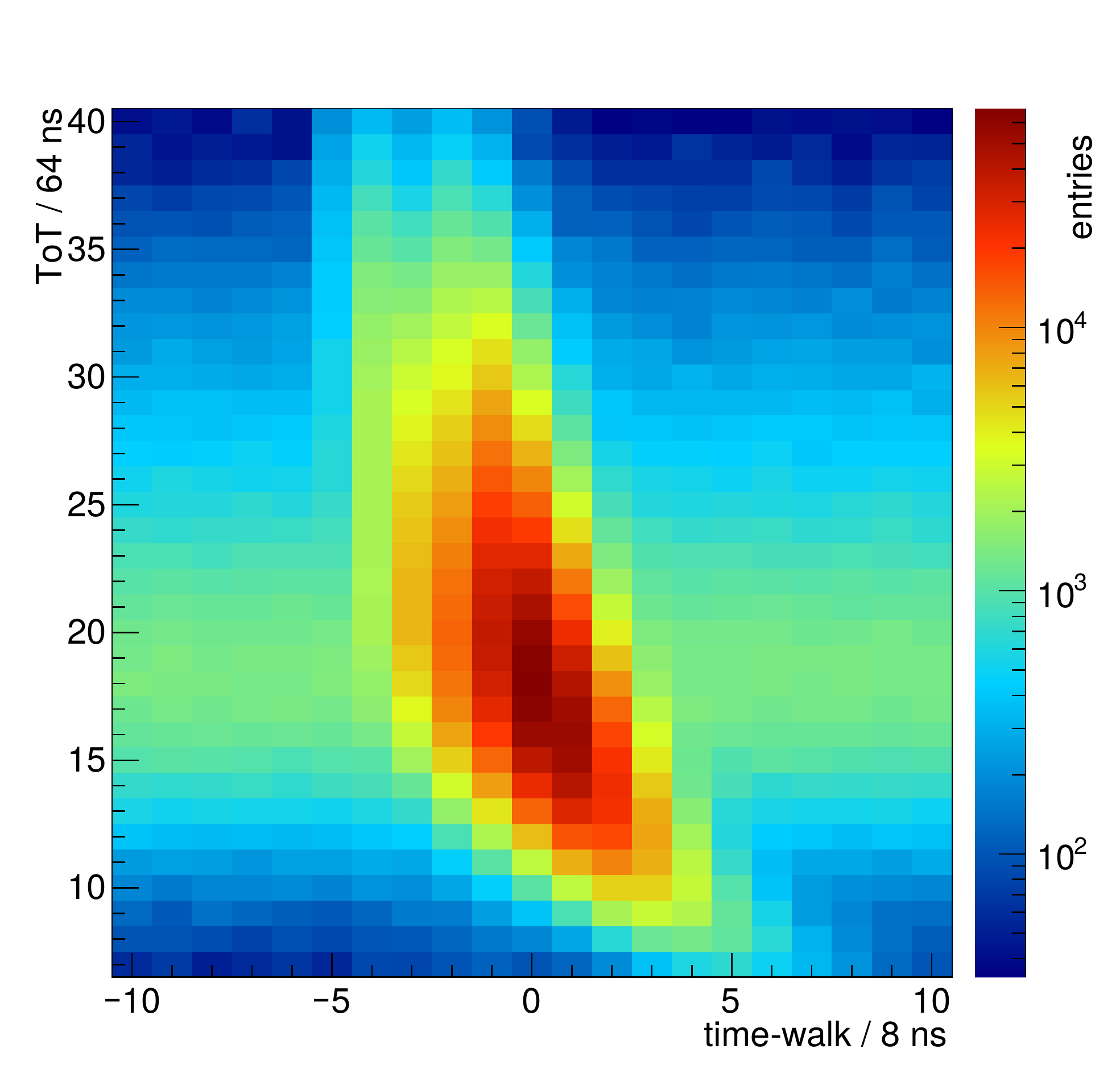}
                \caption{
                  Correlation between ToT and time-walk (ToA minus
                  scintillator reference after correcting for individual pixel delays) for \mupix{8}
                  using a $\SI{4}{GeV}$ electron beam at DESY.
        Plot from \cite{master_thesis_hammerich}.}
        \label{fig:MuPix8_tot_correlation}
\end{figure}

Hit delays over the matrix were studied in detail with the first large scale sensor, \mupix{8}.
Hit delay variations were found to be significant, see~\autoref{fig:MuPix8_delay_blocks}.
A strong position dependence is measured which is more
pronounced as a function of the row number.
The time difference between the lower left edge and the upper right edge is more than
$\SI{50}{ns}$ as measured with a $^{90}$Sr source.
For the hit delay and its spatial variation a strong dependence on the DAC settings and the supply
voltages was observed.
For \mupix{10} the hit delay variation could be reduced with the new routing
scheme of the readout lines where all traces in one out of four blocks have
identical lengths~\cite{Augustin:2020pkv}.
In addition the power net was improved such that the total delay variation is reduced by about a factor $2$.
Time variations over the sensor can be corrected either offline or in the Mu3e filter farm.
Therefore, they are not relevant for the ultimate time resolution achievable with the pixel detector system.

Next, the impact of time-walk on the time resolution is
discussed. \autoref{fig:MuPix8_tot_correlation} shows 
the measured correlation between ToT and time-walk for \mupix{8}, which was
determined with respect to a scintillator reference using a beam of
$\SI{4}{GeV}$ electrons. Here, the timewalk is plotted for all pixels
after applying a delay correction dependent on pixel position (see
above). A mean shift of the time-walk of about $\SI{25}{ns}$ is seen
between large signals (large ToT) and small signals (small ToT) which
can be corrected for on average by using the measured ToT information.

After correcting hit delays,
overall a time resolution of $\approx \SI{8}{ns}$ is obtained.
The ToT information can be used to further improve the time resolution
and the results are shown in \autoref{tab:mupix8_time_resol}. Time
resolutions before and after time-walk correction are given here for row numbers $<18$, where signal losses due to the capacitive coupling between RO lines are small.
The numbers are given for the standard method and the 2-comparator threshold method where the first threshold is $\SI{15}{\milli\volt}$ below the second threshold. For both methods a significant improvement of the time resolution is achieved by applying time-walk corrections.
In contrast, the improvement of the time resolution by using two thresholds is small, both with and without the time-walk correction.

\begin{table}
        \centering
        \small
                \begin{tabular}{lrr}
                        \toprule
                          & \multicolumn{2}{c}{time-walk correction}   \\ 
                          & w/o & with   \\ \midrule
                        $\sigma_t$(1 comparator)& $\SI{8.4}{\nano\second}$ & $\SI{6.6}{\nano\second}$  \\
                        $\sigma_t$(2 comparators)& $\SI{7.8}{\nano\second}$ & $\SI{6.2}{\nano\second}$  \\
                        \bottomrule
                \end{tabular}
        \caption{\mupix{8} time resolutions obtained with a $^{90}$Sr source using the 1-comparator and the 2-comparator threshold method before and after applying a time-walk correction for 1-hit clusters.
        For the correction of the hit delay variations a row and column number dependent method is used. The average time resolution is given for all pixels with row number $<18$.
        The sampling frequency used for time measurement is $\SI{125}{MHz}$.
        Numbers taken from~\cite{Mu3eNote:Note53}}
        \label{tab:mupix8_time_resol}
\end{table}

\begin{figure}
        \centering
                \includegraphics[width=0.48\textwidth]{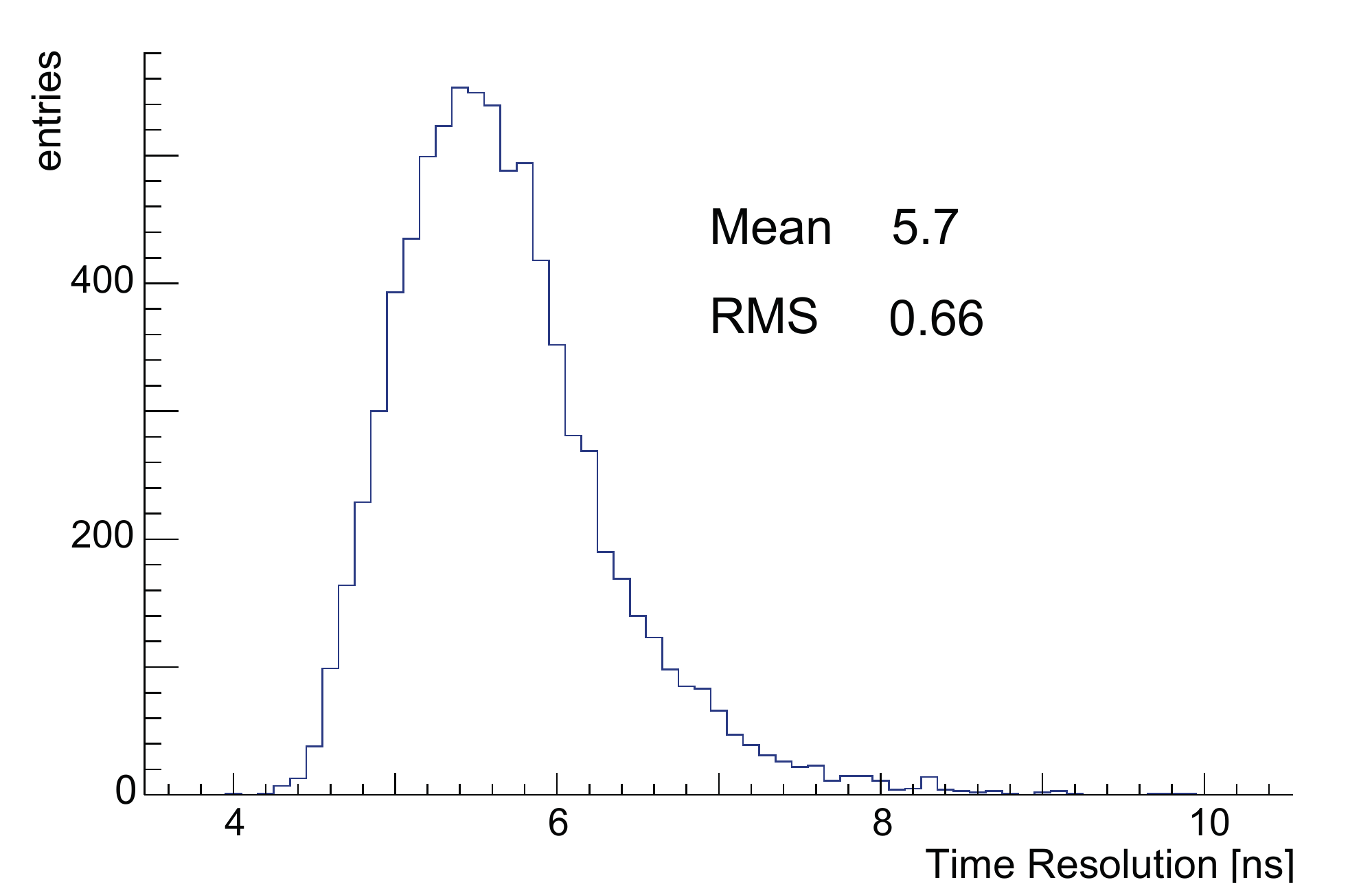}
        \caption{Distribution of the measured \protect\mupix{8} time resolutions of individual pixels obtained with the 2-comparator threshold method and after time-walk correction using a $^{90}$Sr source. Plot taken from~\protect\cite{Mu3eNote:Note53}.}
        \label{fig:MuPix8_time_resolution_best}
\end{figure}

\begin{figure}
        \centering
                \includegraphics[width=0.48\textwidth]{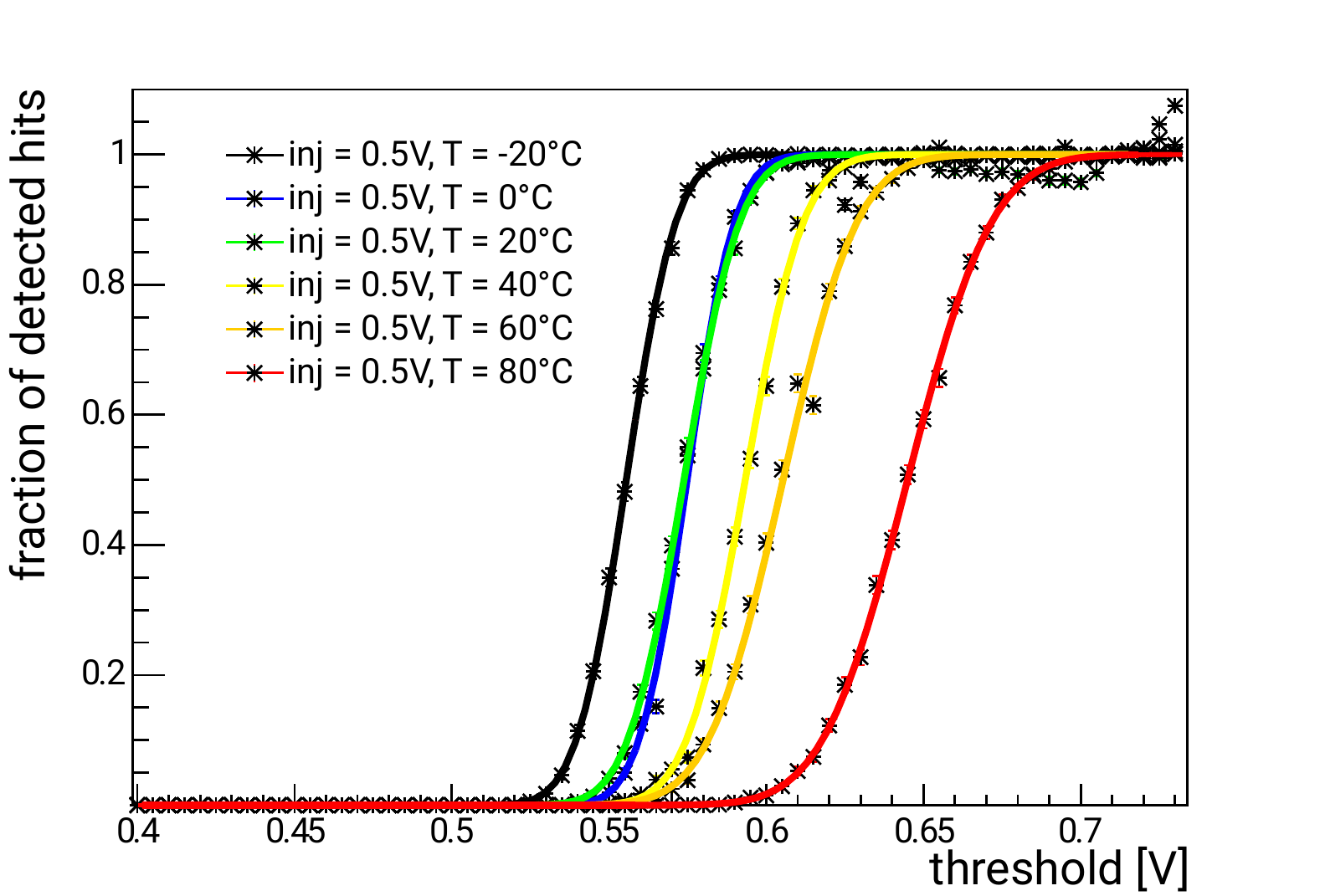}
        \caption{\mupix{7} response to a constant injected charge as a function 
        of the comparator threshold
          (``s-curve'') for different temperatures ranging from
          $-20$ to $+\SI{80}{\degreeCelsius}$~\cite{Immig2016}. 
          Note that due to the 2-stage amplifier in \mupix{7} high voltages correspond
          to low signals (thresholds). For this measurement the baseline is at about $\SI{0.8}{V}$.
        }
        \label{fig:scurves}
\end{figure}

The time resolutions listed in \autoref{tab:mupix8_time_resol} are still affected by pixel-to-pixel variations of the hit delay parameters. These variations can be accounted for by measuring the time resolutions for individual pixels. The resulting time resolution using the 2-comparator threshold method and after applying a time-walk correction is shown for all pixels in \autoref{fig:MuPix8_time_resolution_best}. An average time resolution of about $\SI{5.7}{\nano\second}$ is obtained. The spread of time resolutions is small but there is a trend that the best time resolutions are obtained for pixels with a small row number. Taking into account the sampling frequency of $\SI{125}{MHz}$, the intrinsic time resolution is estimated to be $\SI{5.2}{\nano\second}$ on average.

To conclude, the \mupix{8} sensor with a substrate resistivity of $\ge \SI{80}{\ohm\cm}$ fulfils the time resolution requirement of $\SI{20}{\nano\second}$ even without applying corrections. The excellent time resolution is beneficial for the Mu3e experiment and can be used to reduce combinatorics in the online reconstruction of tracks (in the filter farm) and offline.

\subsection{Temperature Dependence}

The temperature dependence of the \mupix design has been investigated for
prototypes in the temperature range of $T_{ambient}=-20$ to $+\SI{80}{\degreeCelsius}$ using a climate chamber. 
The actual sensor temperature was not monitored in these measurements but known from infrared measurements to be
typically $25$-$\SI{30}{\degreeCelsius}$ higher.
The temperature dependence was studied for both the analogue and the digital part.
As expected the hit noise was found to increase by about a factor~$2$
in the temperature range from $T=0-\SI{80}{\degreeCelsius}$~\cite{Immig2016}. 
At the same time the amplified signal was found to significantly decrease, see~\autoref{fig:scurves}.
The \mupix prototypes were fully operational in the targeted operation range. 
The PLL was successfully locking at all temperatures and no significant
increase of clock jitter was measured, even at the highest temperatures.
However, in order to keep the signal-to-noise ratio high, the \mupix temperature should 
not exceed $T=60-\SI{70}{\degreeCelsius}$.

\subsection{Power Consumption}

\begin{figure*}
    \centering
      \includegraphics[width=0.95\textwidth]{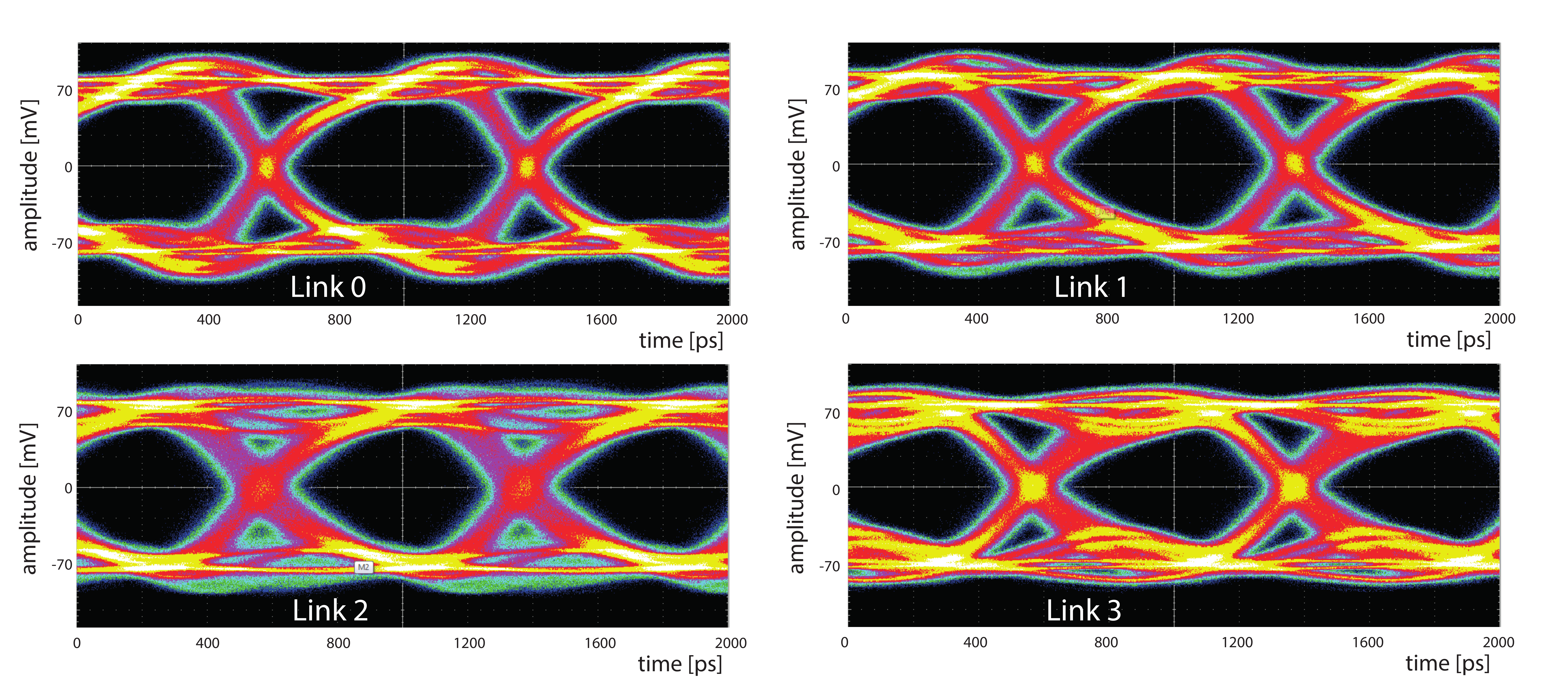}
                \caption{\mupix{8} serial data outputs at \SI{1.25}{Gbit/s}. The
                  links 0, 1, and 2 refer to the pixel matrices A, B and C.
                  Link 3 is switchable.
                            Plot modified from~\cite{phd_thesis_dittmeier}.
                }
        \label{fig:LinkEyeDiagram}
\end{figure*}

The power consumption of \mupix sensors, and thus their heat dissipation, depends strongly
on the amplifier and other DAC settings. Studies of prototypes have shown
that high performance operation points can be found if the power consumption
normalised to the active area is $\approx \SI{200}{mW/cm^2}$ (see \autoref{tab:Mupix_Overview}). But \mupix prototypes can also be operated at significantly higher power consumption, for example when aiming for better time resolutions.
For the final \mupix sensor the total power consumption will depend on the readout mode and the powering scheme.
If three LVDS links are used (high bandwidth mode for inner vertex layers)
the total power consumption is about
$\SI{30}{mW}$ higher compared to the standard operation with one multiplexed link\footnote{Dis-/Enabling of the LVDS links has not been implemented in existing prototypes but will be implemented in the final \mupix sensor.}.
Furthermore, the VSSA voltage can be externally provided or generated in-chip with a
regulator. The latter is the standard operation for Mu3e and adds another $\approx \SI{100}{mW}$, corresponding to a relative power increase of 10\%.
First characterisation measurements of \mupix{10} are consistent with previous
results obtained from \mupix{7} and \mupix{8}, and suggest that the final
sensor can be safely operated with all LVDS links running with a power consumption of $< \SI{250}{mW/cm^2}$, well below the critical cooling limit of $P_{crit} = \SI{350}{mW/cm^2}$.

\subsection{LVDS Links}
\label{sec:LVDSLink}

The quality of the serial data links has been extensively studied for the \mupix{7}
and \mupix{8} prototypes using bit error rate measurements and eye diagrams. 
For tests the signal was transferred over SCSI-2 twisted pair cables up to \SI{2}{m} long, from the sensor to an Altera Stratix IV FPGA development board.
In this \SI{10}{\hour} long test, no errors were found, giving an upper limit on 
the bit error rate $BER \le \num {5e-14}$ at 90\% confidence level.

\autoref{fig:LinkEyeDiagram} shows the eye diagrams of the four LVDS links
of \mupix{8} at \SI{1.25}{Gbit/s} without pre-emphasis.
A degradation of the eye opening from link~0 (width = $\SI{600}{ps}$,
height = $\SI{117}{mV}$) to link~3 (width = $\SI{565}{ps}$, height = $\SI{78}{mV}$) is
visible. Moreover, the jitter of link~2 was found more than 25\% larger than that of
the other links.
This degradation of the signal quality from link~0 to link~3 is probably caused by an in-chip drop of the voltage in \mupix{8}.

\subsection{Irradiation Effects and High Rate Tests}

High particle rates increase the readout dead-time and can also lead to irradiation damage. 
For Mu3e, bulk damage of the \mupix sensor due to non-ionising radiation
(fluence $<10^{13}$~(1 MeV)~neq for Phase~I) is considered to be negligible.
The situation might be different for ionising radiation which can lead to oxide damage in transistors.
In Mu3e, the ionising dose is expected to be highest in the inner-most vertex layer; not only from Michel decays electrons 
but also from scattered muons that miss the target and stop on the innermost vertex layer where they deposit all their kinetic energy.

The impact of large irradiation doses to the \mupix sensor was studied
in test beam campaigns at MAMI and PSI. For several prototypes,
temporary damage was measured in the $\SI{855}{MeV}$ electron beam at
MAMI if the beam intensity exceeded $\approx
\SI{1e6}{e/mm\squared/\second}$ and if the sensors were operated in
the avalanche region which starts at about $V_{HV}>\SI{60}{V}$. The
temporary damage caused a loss of the signal detection efficiency and
lead to an ``after-glowing'' (noise) of the irradiated region with a
time constant of minutes to hours. Similar effects were also observed
at PSI in a pion beam-line. These effects were only seen at rates
which exceed the Mu3e Phase~I conditions by several orders of
magnitude. A dedicated campaign where \mupix sensors are irradiated
 with a strong $^{90}$Sr source for long periods is ongoing.

\begin{figure}
        \centering
                \includegraphics[width=0.49\textwidth]{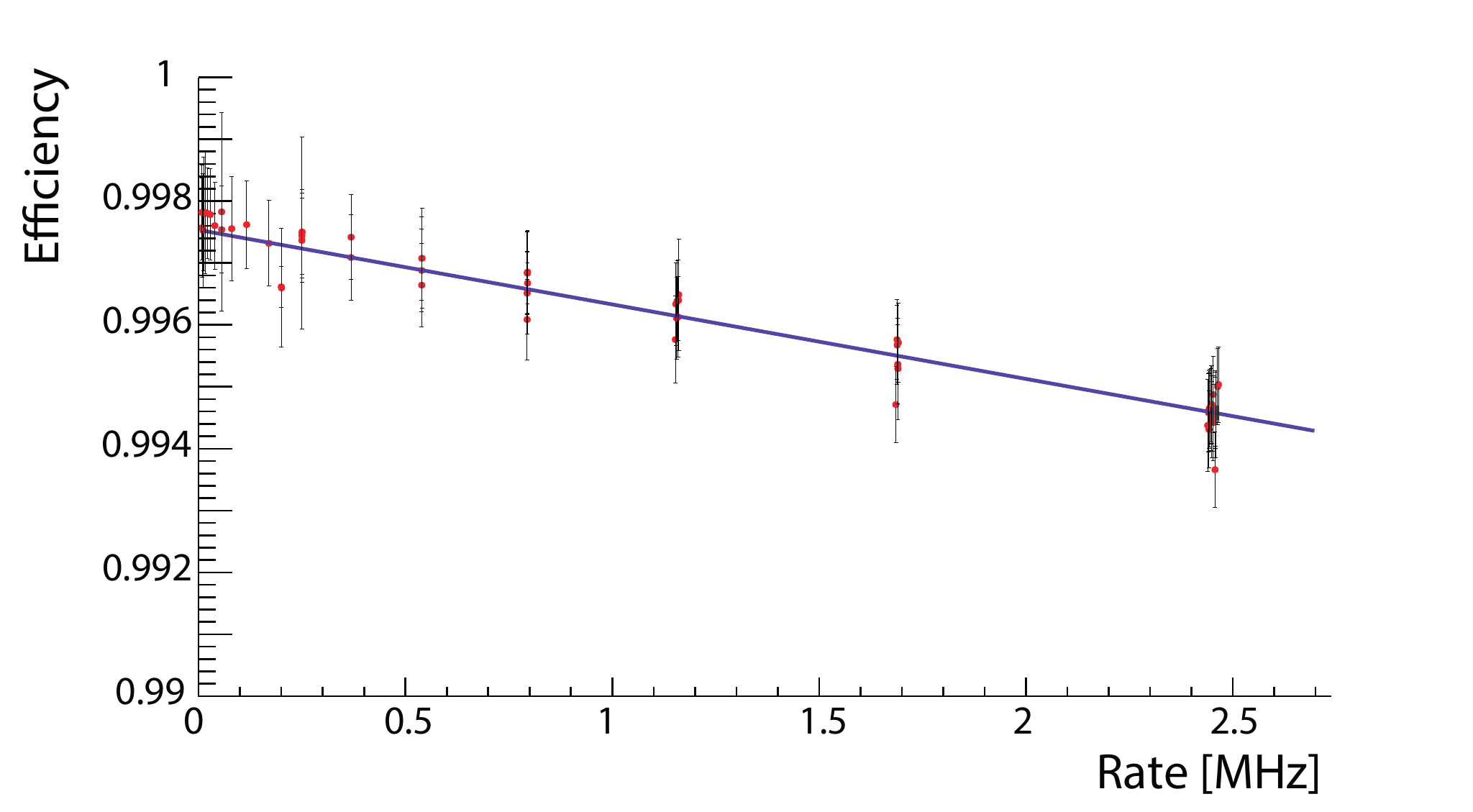}
        \caption{Rate dependence of the \mupix{7} hit detection efficiency measured with the
          high intensity electron beam at MAMI at a beam energy of $\SI{855}{MeV}$,
                                        the beam spot size of about ~$\sigma
                                        \approx \SI{0.5}{mm}$ was significantly
                                        smaller than the chip size.}
        \label{fig:Mupix7_Efficiency_MAMI}
\end{figure}

The impact of high particle rates on the hit detection efficiency due to
dead-time in the pixel readout cell is shown in \autoref{fig:Mupix7_Efficiency_MAMI}.
The efficiency was measured using an extremely focused
electron beam at MAMI with local beam 
intensities of up to $\SI{2.5e6}{e/mm\squared/\second}$. A degradation of
only 3 permil was measured at rates which exceed the expected rates in
Mu3e. The degradation is caused by dead-time in the pixel readout cell.

\subsection{Mupix10 results}
\label{sec:mupix10_results}
\begin{figure}[htb!]
        \centering
                \includegraphics[width=0.48\textwidth]{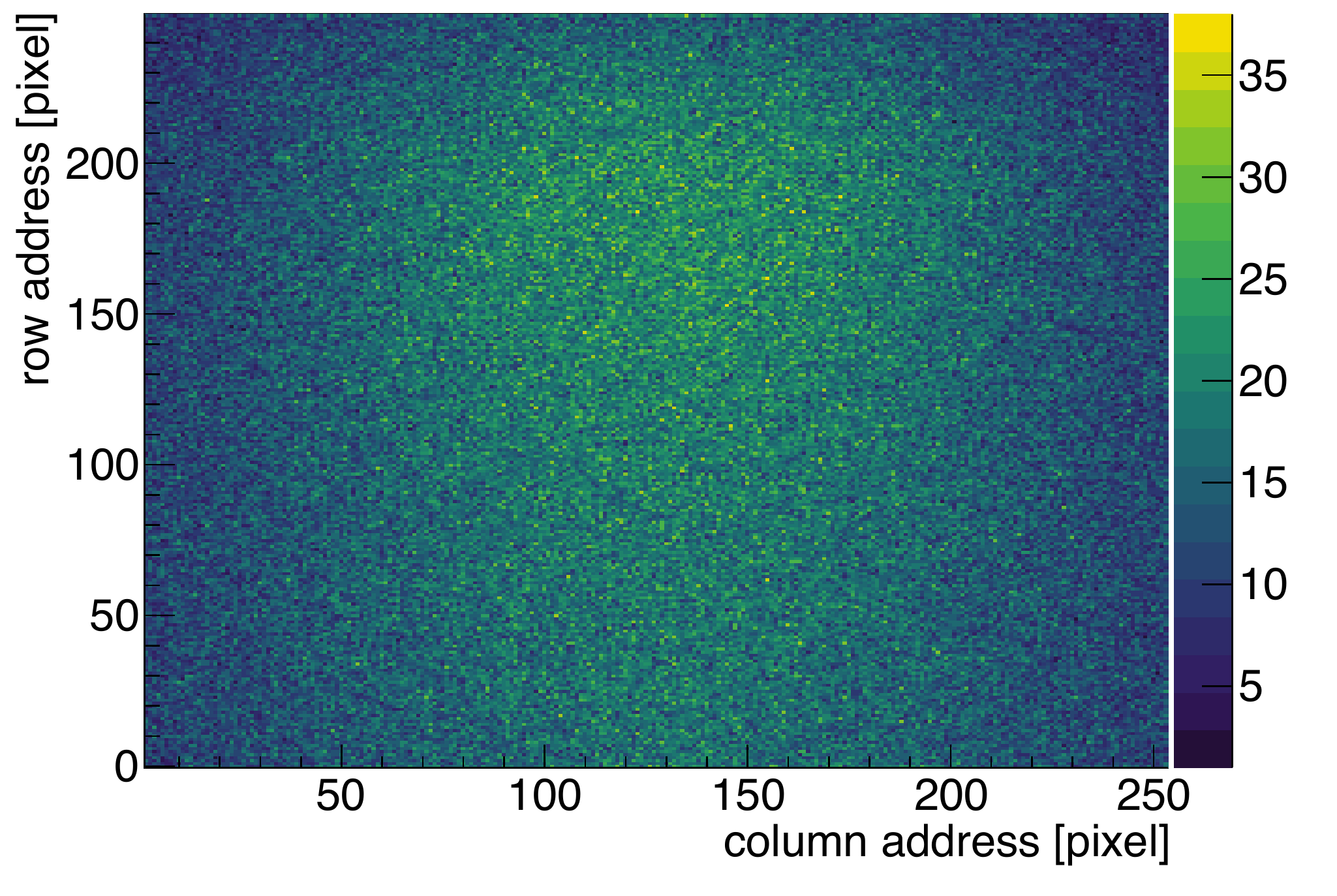}
        \caption{\mupix{10} hit map in a beam of $\SI{3}{GeV/c}$ electrons measured at DESY. The active area is $\SI{4.1}{\centi\meter\squared}$.}
        \label{fig:MuPix10_hitmap}
\end{figure}
The \mupix{10} prototype was delivered by TSI in May 2020, manufactured on \SI{200}{\ohm\cm} substrate. This prototype could be successfully brought to operation and is currently being characterised in the lab and in test beam campaigns at DESY and PSI. At time of publication the characterisation studies of \mupix10 have just started and preliminary results are presented in the following.
\autoref{fig:MuPix10_hitmap} shows the \mupix{10} hit map as measured at DESY in a beam with $\SI{3}{GeV/c}$ electrons without any tuning of the sensor. All $256\times 250$ pixels are operational and no noisy (hot) pixels are visible.
First results indicate that the efficiency and noise performance of \mupix{10} is similar to \mupix{8}.

\begin{figure}
        \centering
        \includegraphics[width=0.48\textwidth]{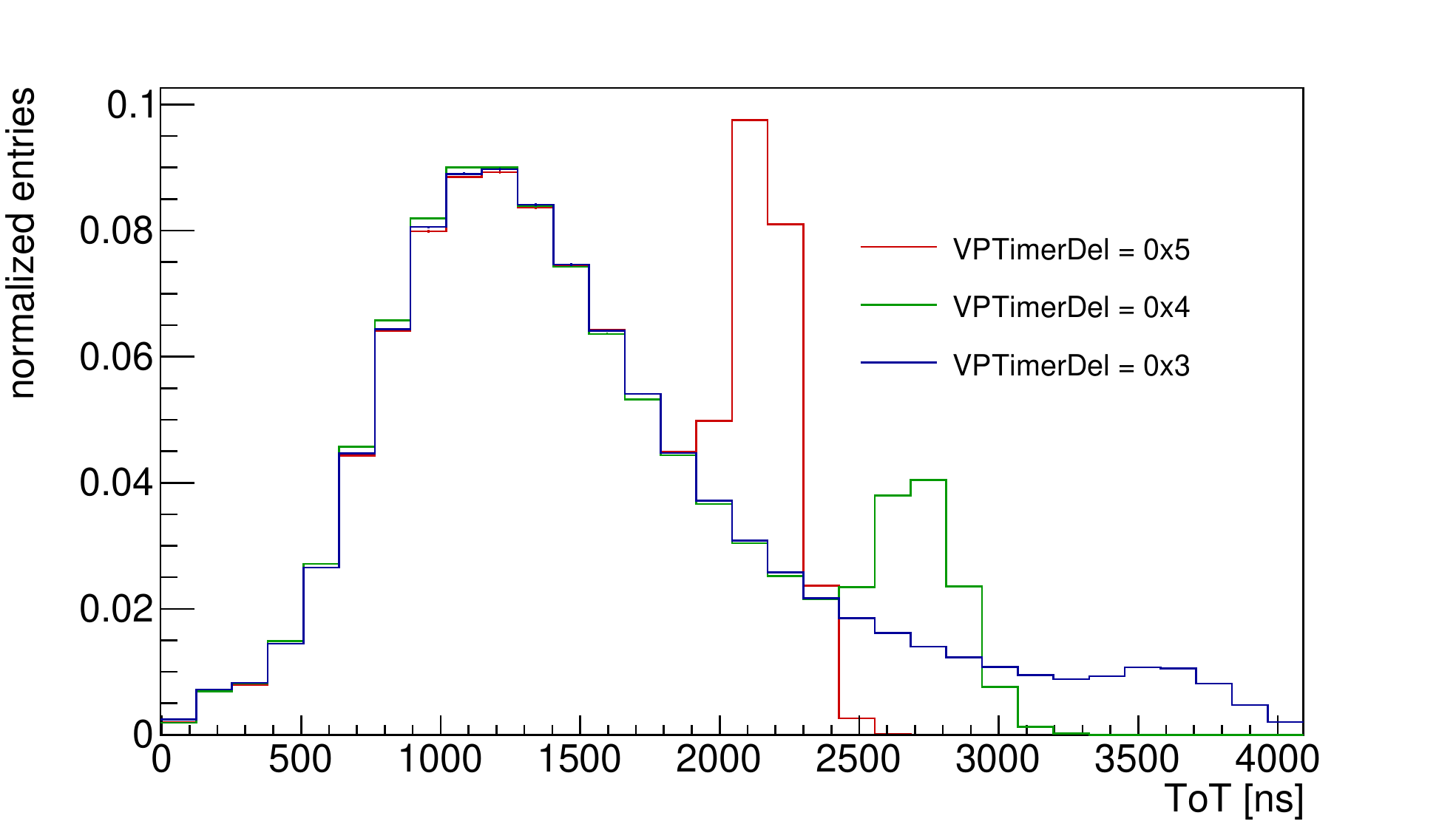}
        \caption{\mupix{10} ToT spectra for $\SI{350}{MeV/c}$ pions (PSI) as measured for
for different settings of the readout timer. The first peak in each distribution is the ionisation peak (Landau), the second peak the timer end which slightly varies for different pixels and hits.}
        \label{fig:MuPix10_tot_spectra}
\end{figure}
A new feature of the \mupix{10} prototype is the adjustable timer delay readout. \autoref{fig:MuPix10_tot_spectra} shows the ToT distribution of $\SI{350}{MeV/c}$ pions as measured for different timer delays. The peak at the end of each ToT distribution is due to the readout timer delay which slightly varies for different pixels and hits.
This new feature limits the readout time (and therefore dead-time) without affecting
the measurement of small or medium signal amplitudes where time-walk corrections are important.
\begin{figure}
        \centering
                \includegraphics[width=0.48\textwidth]{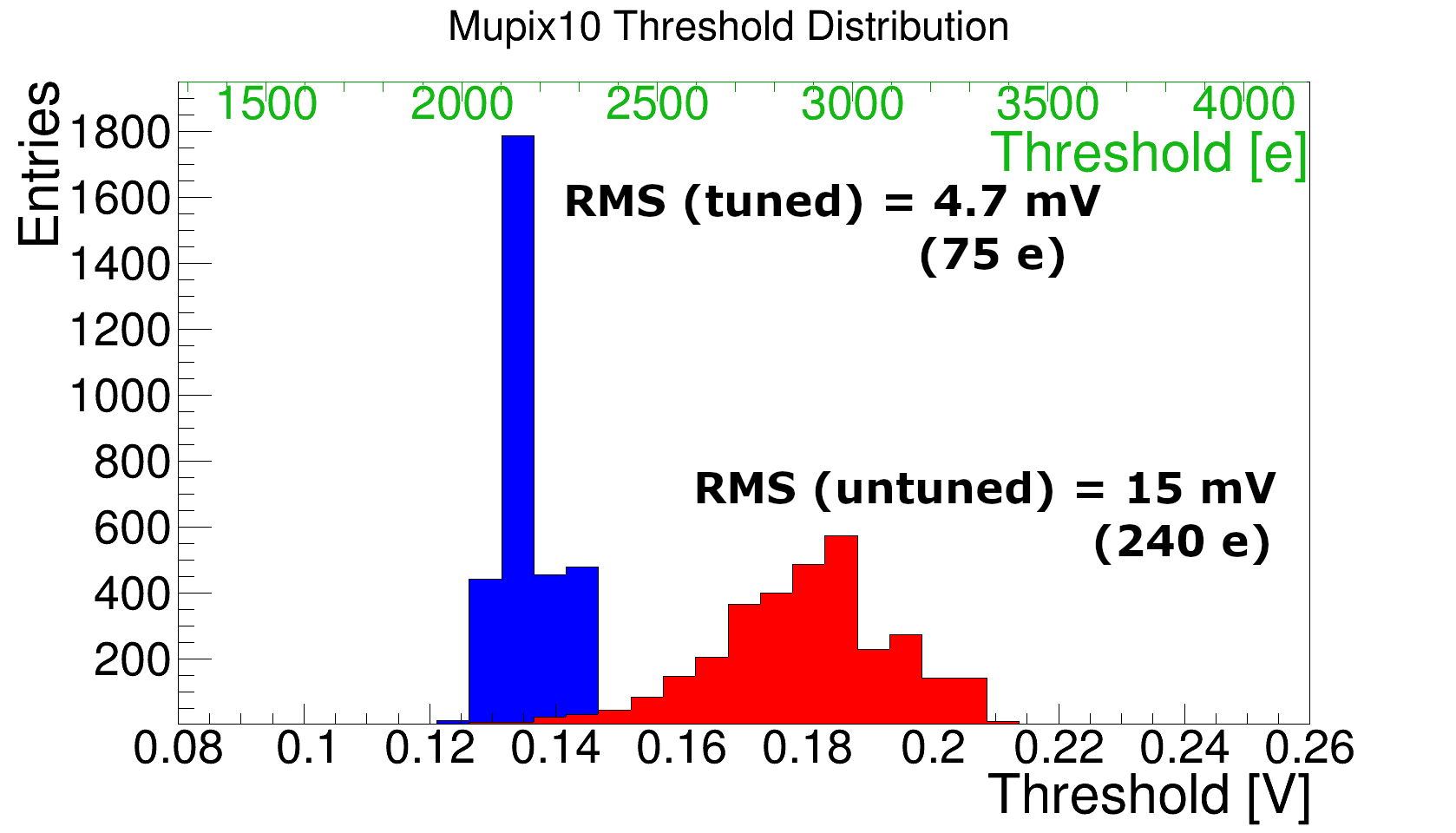}
                \caption{Distribution of pixel comparator thresholds
                  corresponding to constant injected charges obtained
                  in threshold scans (s-curves).
        The histogram is shown before and after calibrating the comparator TDAC.
        Tuning enables the dispersion of the threshold to be improved from
        $\SI{15}{\milli\volt}$ ($\approx 240$ electrons) to $\SI{4.7}{\milli\volt}$ ($\approx 75$ electrons).
}
        \label{fig:MuPix10_threshold_tuning}
\end{figure}
Finally, \autoref{fig:MuPix10_threshold_tuning} shows the threshold distribution of injected charges for a block of pixels before and after calibration. The spread of the pixel thresholds can be significantly reduced from $\SI{15}{mV}$ (corresponding to $\approx 240$ electrons) to $\SI{4.7}{mV}$ (corresponding to $\approx 75$ electrons).
These numbers can be compared with the expected signal amplitude of about
$\SI{220}{mV}$ signal for a MIP ($\approx 3600$ electrons) and a measured
noise of about $\SI{5.3}{mV}$ (corresponding to $\approx 85$ electrons) after
TDAC tuning.

Further measurements of the \mupix{10} sensor indicate a similar
or even better performance than the one obtained for \mupix{8}.
All preliminary results obtained so far suggest production readiness.
The final sensor, the \mupix{11}, will include only  minor
improvements, most importantly a fix in one of the configuration lines. It 
will be produced in 2021 for constructing the Mu3e pixel tracker.


\chapter{MuTRiG}
\label{sec:Mutrig}

\chapterresponsible{Wei}

A common Application Specific Integrated Circuit (ASIC) has been
developed for both the fibre and tile detectors in Mu3e, capable of
operating with the rather different conditions of the two systems.

\section{Introduction}

\begin{sloppypar}
\mutrig (\underline{Mu}on \underline{T}iming \underline{R}esolver
\underline{i}ncluding \underline{G}igabit-link) is a 32 channel,
mixed-signal Silicon photo-multiplier (SiPM) readout ASIC designed and
fabricated in UMC \SI{180}{\nano\metre} CMOS technology. It has been
developed to read out the fibre and tile detectors in Mu3e, and is
designed to achieve the required timing resolution for both systems
while keeping up with the high event rate in the scintillating
fibre detector.
\end{sloppypar}

\begin{figure}[b!]
	\centering
	\includegraphics[width=0.48\textwidth]{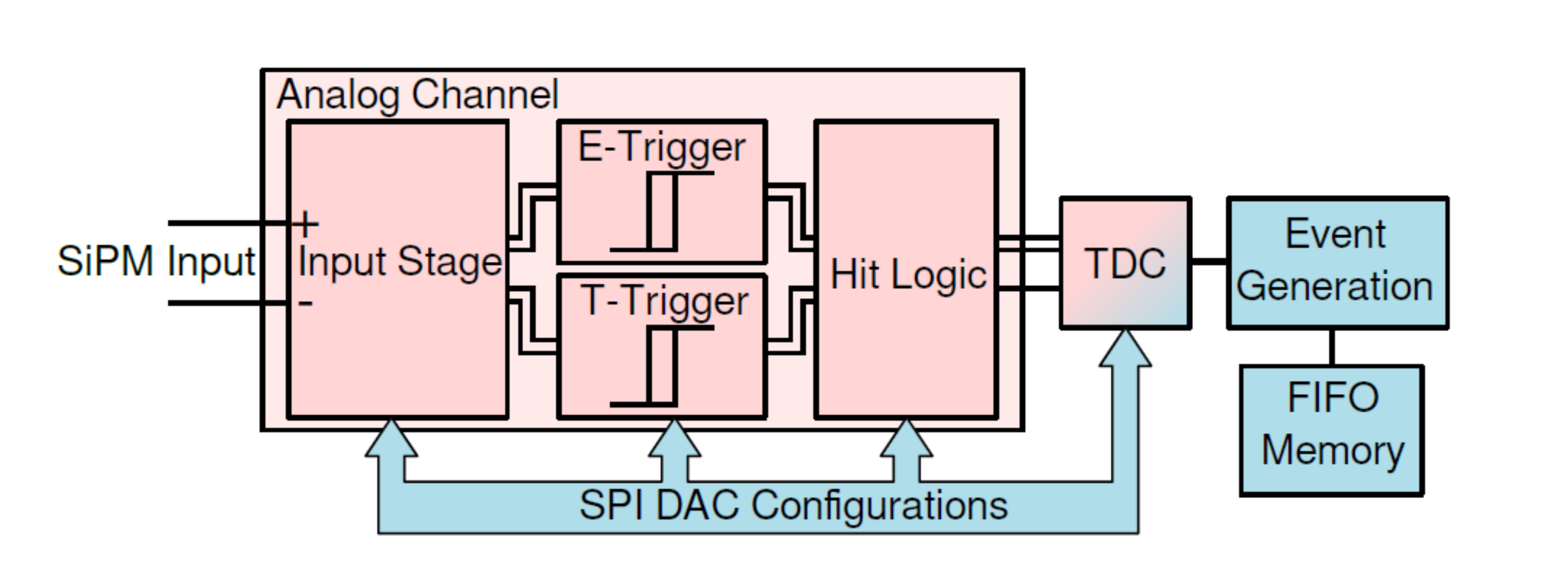}
	\caption{Diagram of a \mutrig channel. After taking signal from
          SiPM by the input stage, separate signals are provided to the
          T-Trigger and E-Trigger branches for time and energy
          discrimination respectively.  The discrimination signals are
          encoded in the hit logic module to generate the combined hit
          signal, and then converted to digital time stamps after the
          TDC module. The signal is then buffered in the on-chip memories
          before being transferred out of the chip.  The analogue front-end,
          TDC and digital modules are configured using a Serial
          Peripheral Interface (SPI) interface.}
	\label{fig:Diagram_of_a_MuTRiG_channel}
\end{figure}

\begin{figure}[b!]
	\centering
	\includegraphics[width=0.48\textwidth]{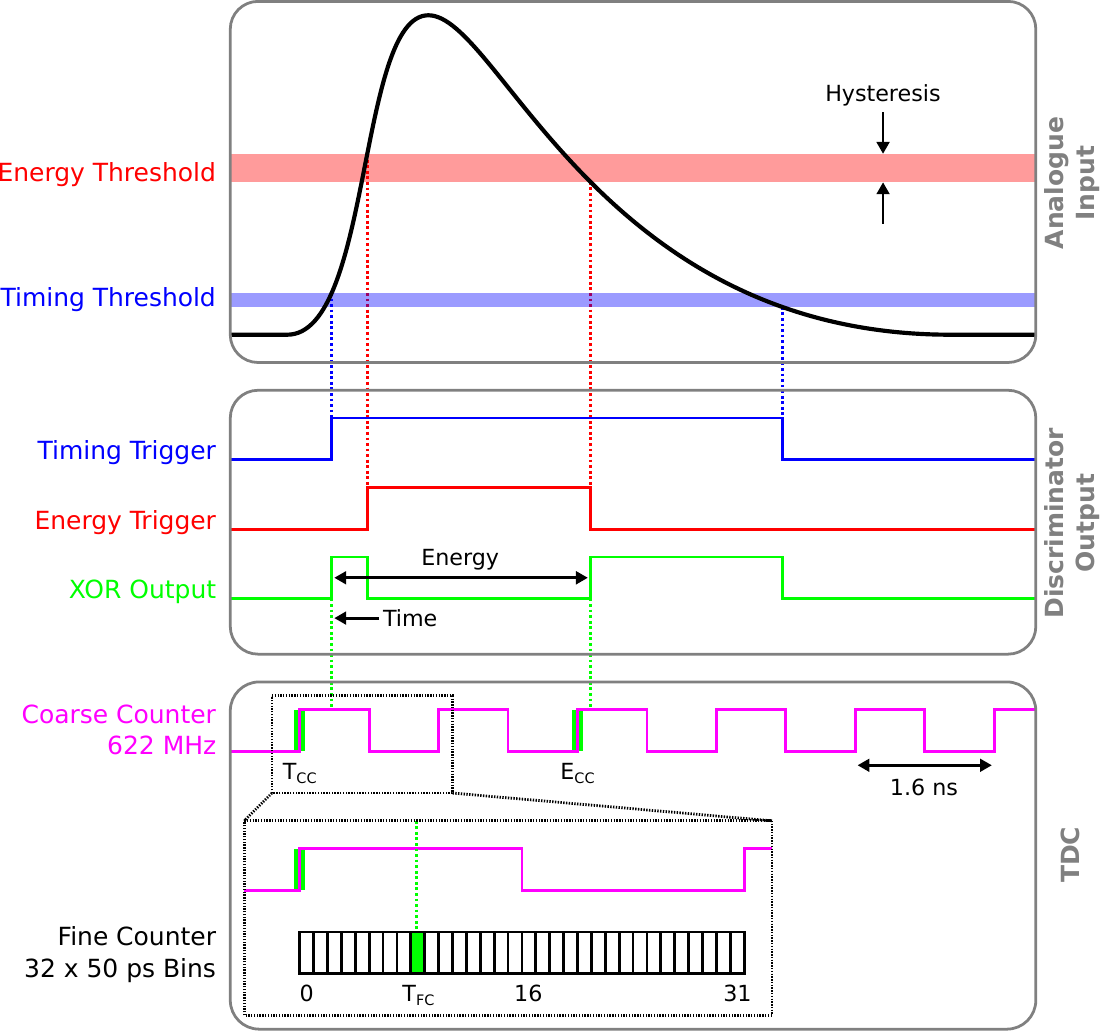}
	\caption{Sketch of the functionality of the \mutrig chip. The
          time and energy information of the analogue input signal is
          obtained via two discriminator units. The discriminator
          output is processed by a TDC with a \SI{625}{\MHz} coarse
          counter and a fine counter with a bin size of \SI{50}{\ps}.}
	\label{fig:STiC_Discri}
\end{figure}

\section{ASIC Description}

\mutrig is an evolutional development from the STiCv3.1 chip~\cite{STiC3} developed
at the Kirchhoff Institute in Heidelberg for medical applications of SiPMs
(EndoTOFPET-US~\cite{proposal186novel}).

\begin{sloppypar}
The analogue processing building blocks of \mutrig inherit from the
STiCv3.1 chip, whose satisfactory performance has been validated in
several testing conditions. However, the STiCv3.1 chip is only capable
of transferring $\sim$\SI{50}{\kilo\hertz} per channel through the
\SI{160}{Mbit\per\second} data link, which is too slow for the Mu3e
timing detectors, especially for the fibre detector which is required
to handle \SI{1}{\mega\hertz}/channel event rate to achieve 100\% data
acquisition efficiency. The \mutrig chip extends the excellent timing
performance of the STiCv3.1 chip with a newly developed fast digital
readout for high rate applications.  The analogue timing jitter of the common frontend
is expected to be around 15 ps.
\end{sloppypar}

\autoref{fig:Diagram_of_a_MuTRiG_channel} shows the channel diagram of
the \mutrig chip and \autoref{fig:STiC_Discri} shows the sketch of the
chip functionality.  (More details can be found in~\cite{Chen2018}.)

\begin{figure*}[t!]
	\centering
	\includegraphics[width=0.95\textwidth]{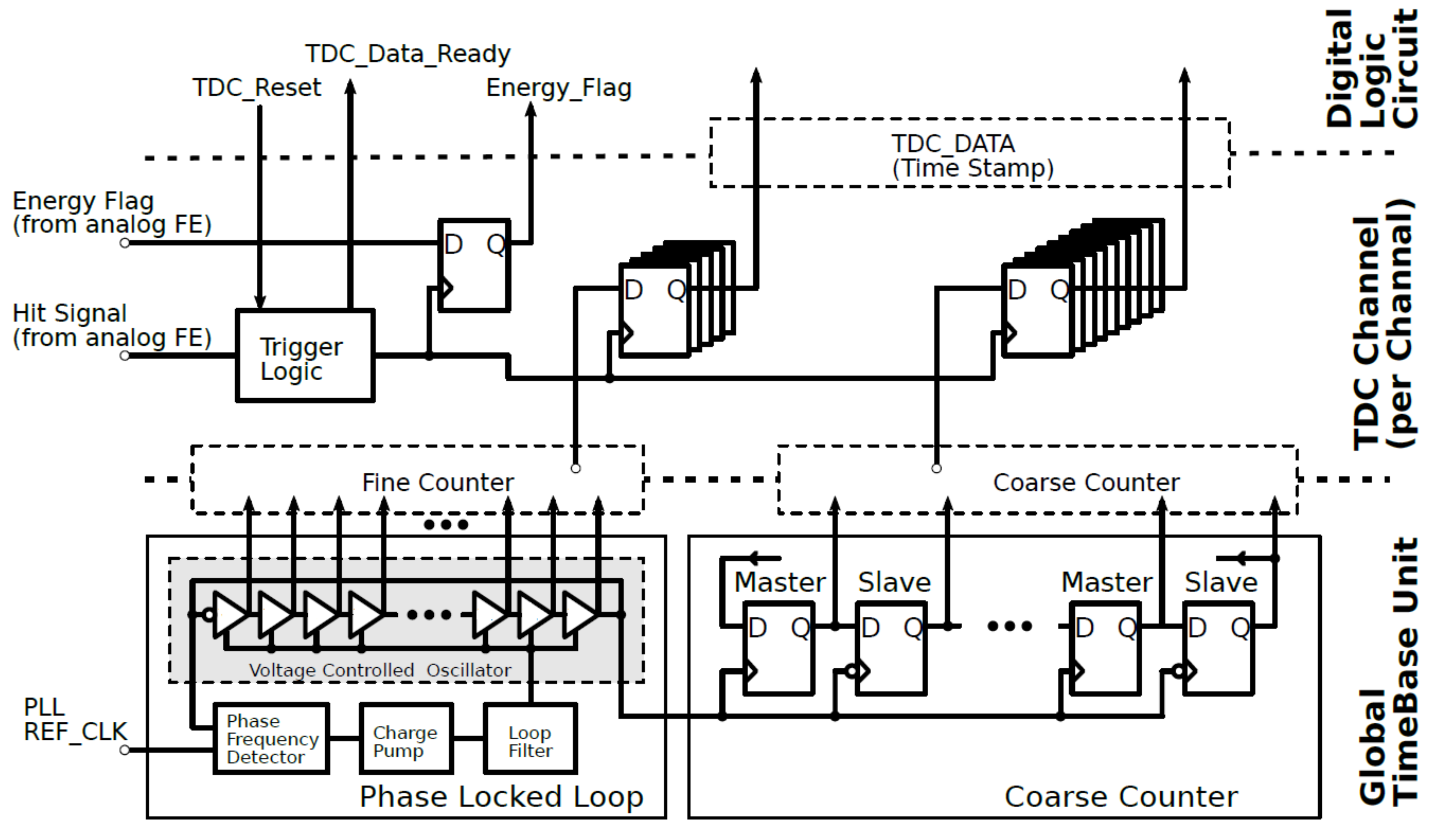}
	\caption{Schematic of the \mutrig TDC.}
	\label{fig:TDC_MuTRiG}
\end{figure*}

\begin{figure}
	\centering
	\includegraphics[width=0.48\textwidth]{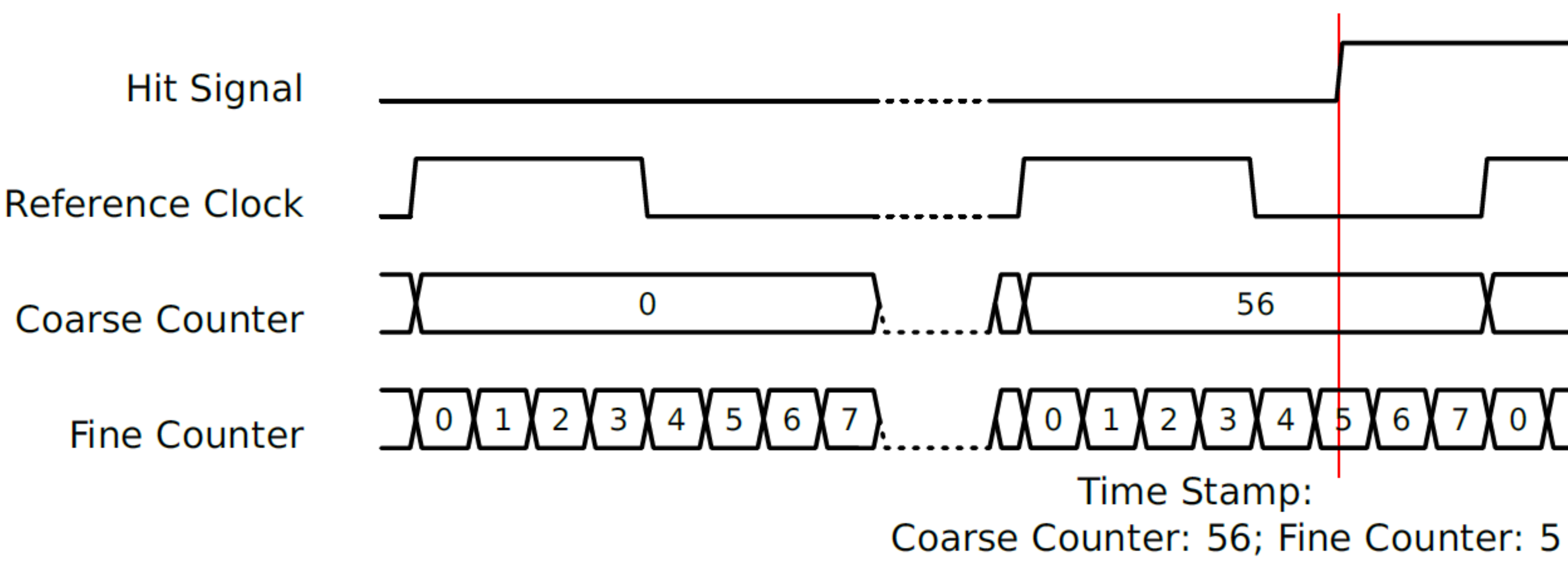}
	\caption{Working principle of the TDC, showing the fine and
          coarse counters, reference clock and example arrive of a hit
          signal.}
	\label{fig:TDC_principle}
\end{figure}

The good timing resolution of \mutrig derives from its differential
analogue front-end and the \SI{50}{\pico\second} binning
time-to-digital converter (TDC), which were inherited from the
STiCv3.1 chip. The working principle of a TDC is shown in
\autoref{fig:TDC_principle}. At the arrival of a hit signal over
threshold, the TDC module samples the state of a
\textit{coarse-counter}, which is incremented at \SI{625}{\MHz} by a
reference clock. A \textit{fine counter} with \SI{50}{\ps} bins is
then used to make a more precise measurement of the hit time within
the \SI{1.6}{\ns} coarse counter bin. The coarse and fine counter
values are then recorded as the time stamp of the hit signal. The time
the signal drops back below threshold is similarly recorded. The
\textit{Global TimeBase Unit} provides common coarse and fine counter
values to all the channels for time stamping, as shown in
\autoref{fig:TDC_MuTRiG}. The TDC requires $\sim$\SI{30}{\nano\second}
to reset after a hit, which corresponds to a maximum occupancy of around 30M Event per second per channel.

In order to fulfil the high rate data readout, a double data rate
serialiser and a customised low-voltage differential signalling (LVDS)
transmitter were developed to establish a gigabit data link with the
data acquisition system (DAQ) for data transmission. The event data
from all the channels are buffered and sent out in frames via the 1.25
Gbps LVDS serial data link. In order to increase the event rate
capability of the \mutrig chip, the output event structure can be
switched from the standard 48 bits, containing both the time stamps a
hit signal passes above and back below threshold, to a short event
structure of 27 bits, containing only the first of these times and a 1
bit energy flag of the hit.

\begin{table}[b!]
	\begin{tabular}{l r r}
		\toprule
		& STiCv3.1 & \mutrig\\
		\midrule
		number of channels & 64 & 32 \\
		LVDS speed [Mbit/s] & 160 & 1250\\
		8b/10b encoding & yes & yes\\
		event size [bit] & & \\
		~~\textit{standard event} & 48 & 47 \\
		~~\textit{short event} & - & 27 \\
		event rate / chip [MHz] & & \\
		~~\textit{standard event} & $\sim$2.6 & $\sim$20\\
		~~\textit{short event} & - & $\sim$38\\
		event rate / channel [kHz] & &  \\
		~~\textit{standard event} & $\sim$40 & $\sim$650\\
		~~\textit{short event} & - & $\sim$1200\\
		power per channel [mW] & 35 & 35 \\
		size [mm x mm] & 5x5 & 5x5\\
		number of PLLs & 2 & 1 \\
		\bottomrule
	\end{tabular}
	\caption{Comparison of STiCv3.1 and \mutrig.}
	\label{tab:FibrereadoutSTiCComparision}
\end{table}

A few more new functionalities were implemented in the digital logic
circuit of the \mutrig chip for convenient and reliable operation of
the chip.  \autoref{tab:FibrereadoutSTiCComparision} shows a summary
of the major differences in event and data handling capabilities
of the STiCv3.1 and \mutrig chips.

\begin{figure*}[t]
	\centering
	\includegraphics[width=0.98\textwidth]{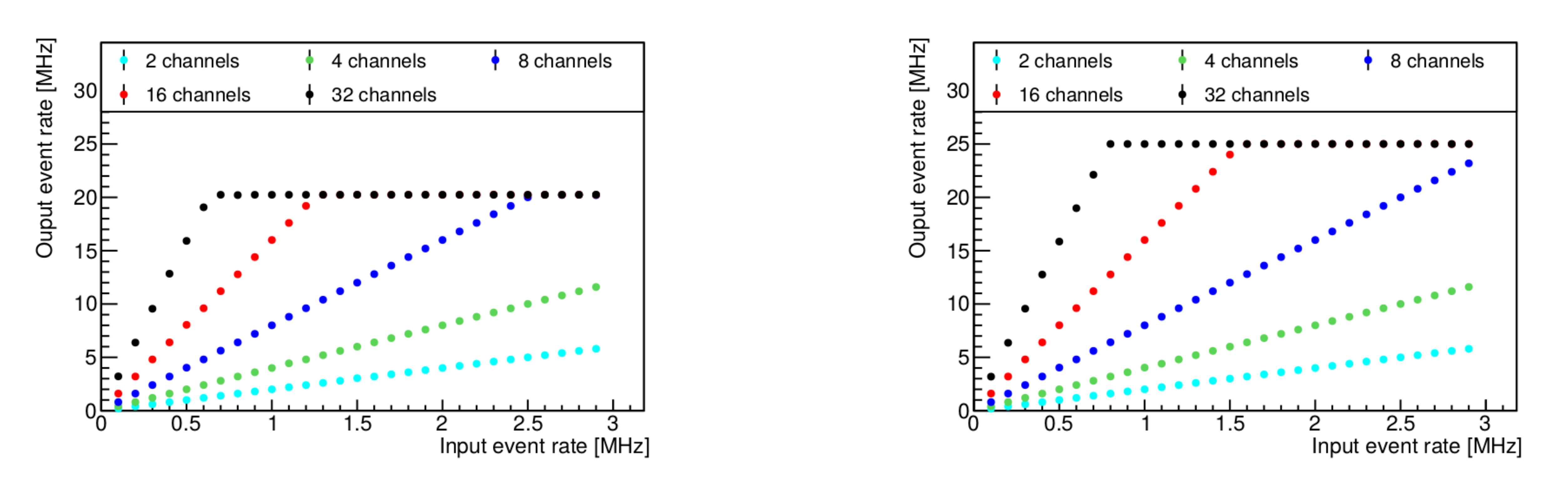}
	\caption{Event rate measurements for the standard output event
          structure (left) and short output event structure (right).}
	\label{fig:rate_limit}
\end{figure*}

\section{Characterisation Measurement}
\subsection{Rate Limitation Measurement}

The event rate limit of the chip is measured by injecting test pulses
to multiple channels and measuring the output event rate for a serial
data link bit rate of \SI{1.25}Gbps. Results are shown in
\autoref{fig:rate_limit}.  For the standard event structure
configuration of 48 bits, the output event rate is limited to
\SI{20.24}{\mega\hertz} (on average \SI{632}{\kilo\hertz}/channel) by
the bit rate of the serial data link. The maximum event rate for the
27 bits short event configuration is \SI{25}{\mega\hertz}
(\SI{781}{\kilo\hertz}/channel), 1/5th of the system clock
frequency(\SI{125}{\mega\hertz}).  The event rate limit of both configurations
fulfills the maximum event rate requirements of the experiment.

\begin{figure}[tb!]
	\centering
	\includegraphics[width=0.48\textwidth]{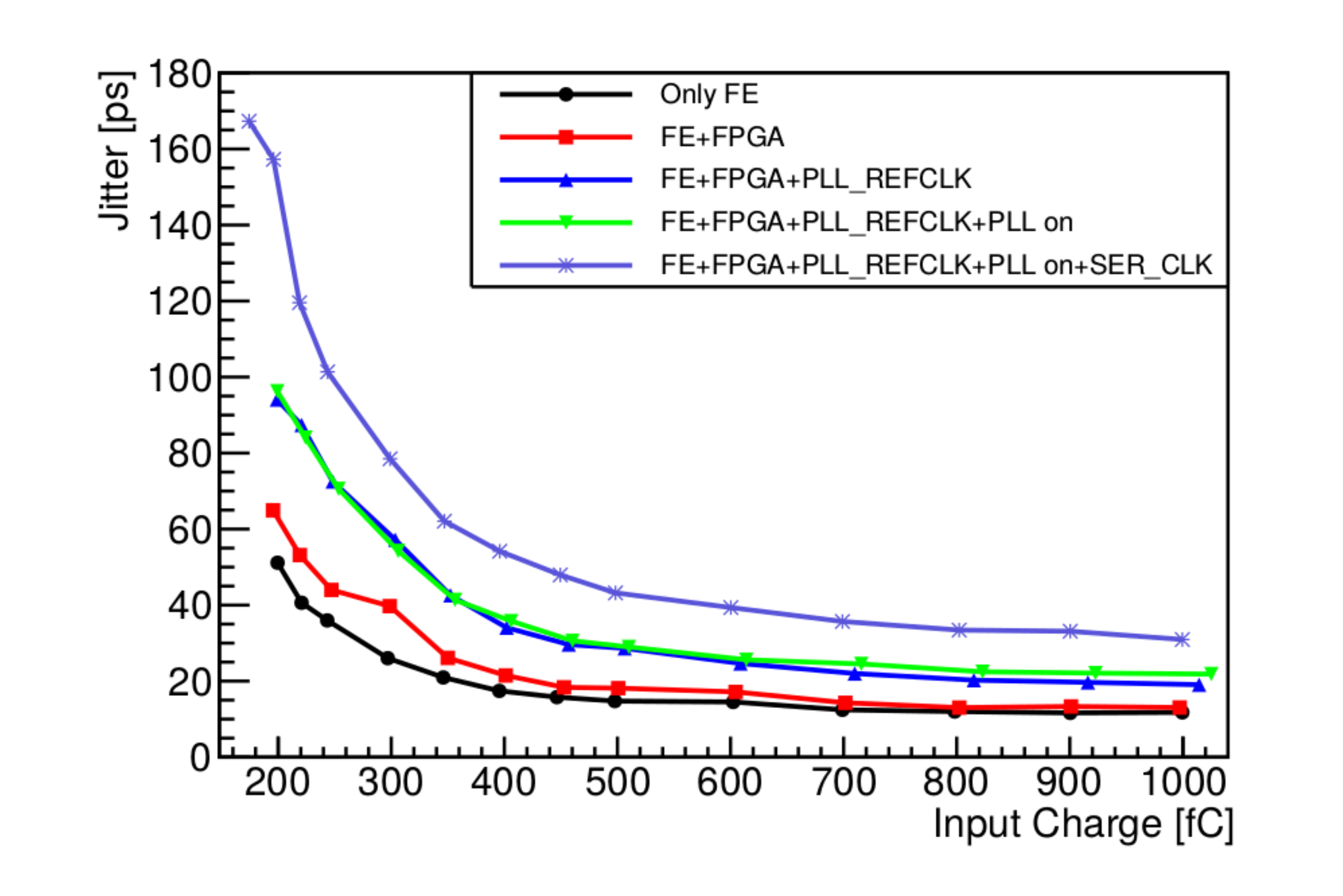}
	\caption{\mutrig front-end jitter measurement by injecting
          charge over a \SI{33}{\pico\farad} capacitor.}
	\label{fig:front_end_jitter}
\end{figure}

\begin{figure}[]
	\centering
	\includegraphics[width=0.48\textwidth]{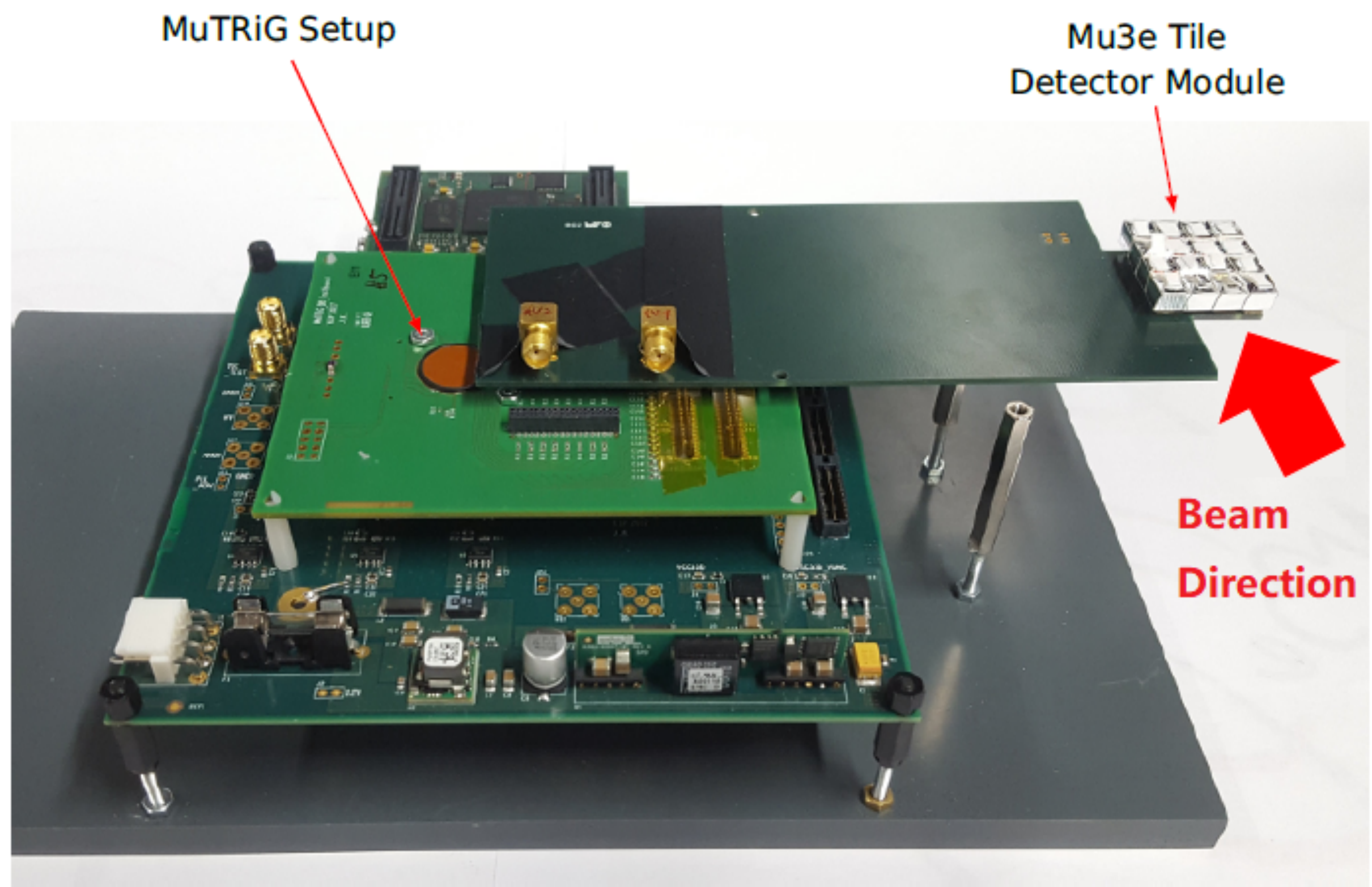}
	\caption{The \mutrig and Mu3e Tile Detector test beam setup.}
	\label{fig:testbeam_setup_mutrig}
\end{figure}

\begin{figure}[]
	\centering
	\includegraphics[width=0.48\textwidth]{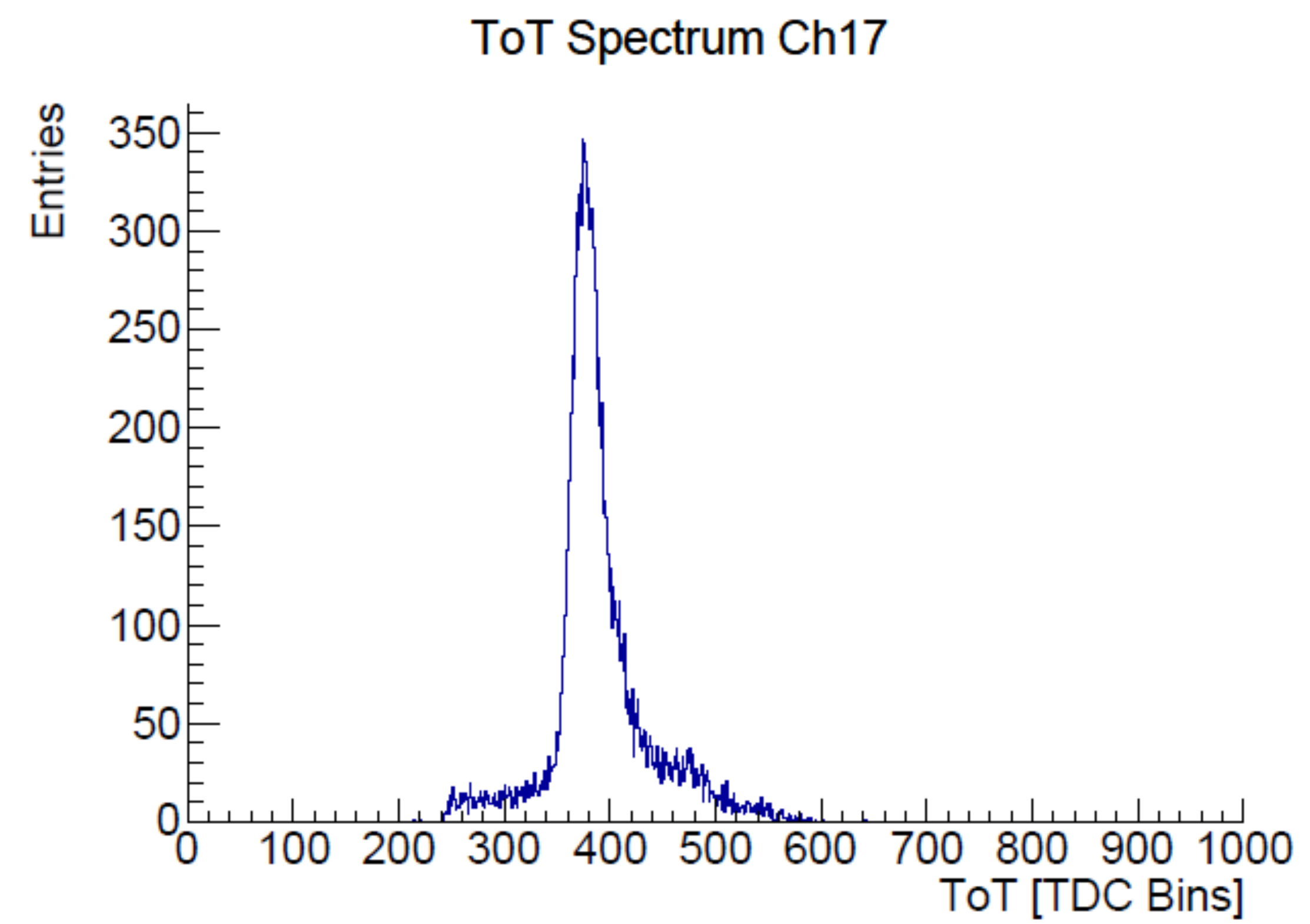}
	\caption{The time-over-threshold (ToT) of minimum-ionising electrons recorded on channel 17.}
	\label{fig:testbeam_E}
\end{figure}

\begin{figure}[]
	\centering
	\includegraphics[width=0.48\textwidth]{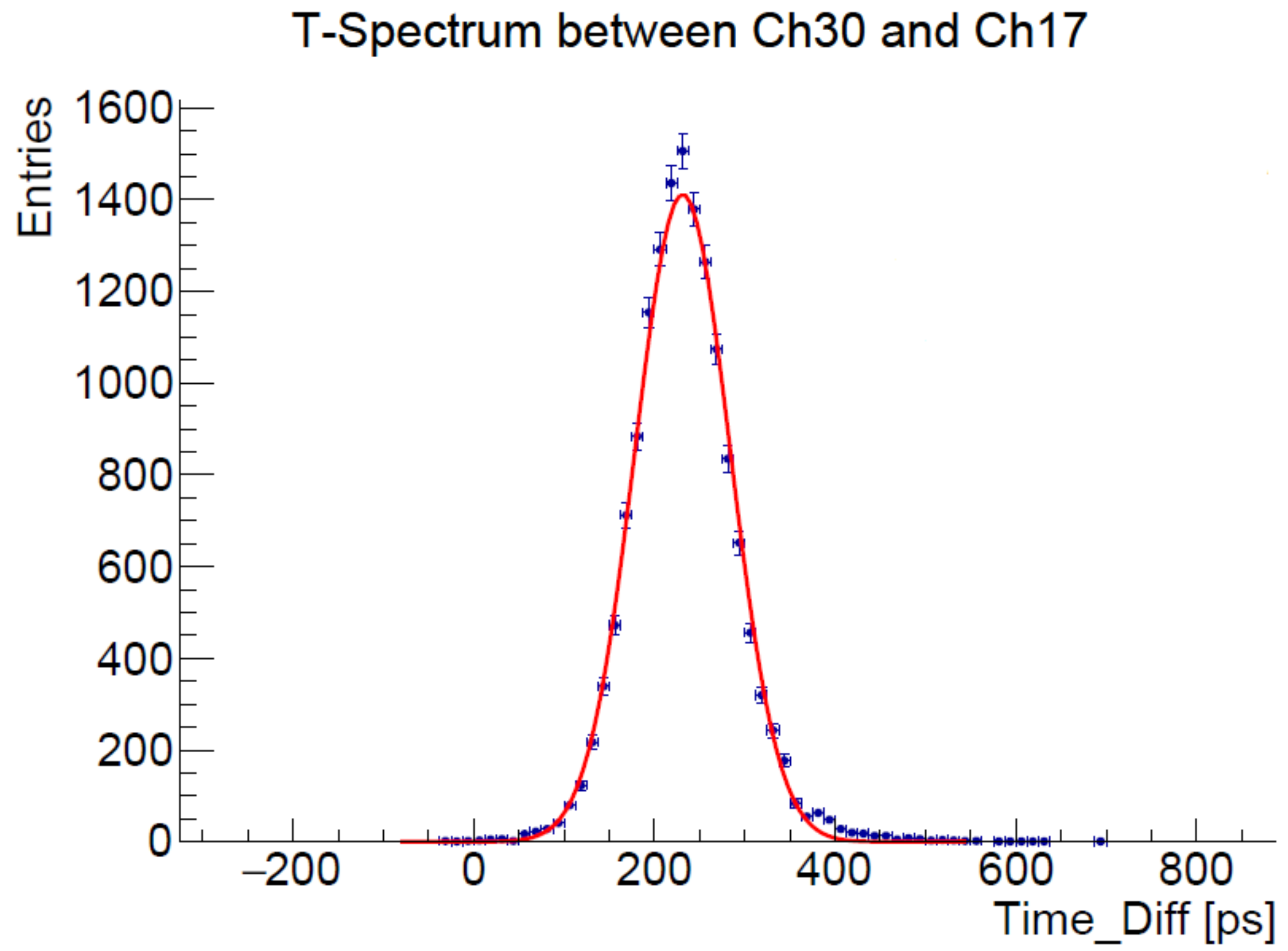}
	\caption{Coincidence timing spectrum between channel 30 and 17.}
	\label{fig:testbeam_CTR}
\end{figure}

\subsection{Jitter Measurement}
The jitter found in just the front-end, and in a full channel
(front-end, TDC and the digital part of the chip), have been measured.
The front-end jitter was measured by charge injection over a
\SI{33}{\pico\farad} capacitor. The time difference between the marker
signal from the arbitrary waveform generator and the \mutrig timing
trigger signal was then measured using a high bandwidth oscilloscope.
The front-end jitter in five different cases is shown in
\autoref{fig:front_end_jitter}. The jitter on a full channel was
measured with input charges of \SI{1}pC and an optimised time
threshold.

\subsection{Test-beam Result}

In order to verify the functionality and the timing performance of the
\mutrig chip under realistic experimental conditions, the \mutrig chip
was tested with the Mu3e Tile detector prototype in an electron test
beam campaign at DESY (Feb. 2018). The setup, shown in
\autoref{fig:testbeam_setup_mutrig}, was the same as a tile detector
submodule: 16 scintillator tiles arranged in a 4 by 4 matrix and read
out by SiPM photon detectors. Example time-over-threshold spectra and
coincidence time resolutions are given in
\Autoref{fig:testbeam_E,fig:testbeam_CTR}. Excellent
channel-to-channel coincidence timing resolutions of $<$\SI{50}{\pico\second} were
been obtained over a large chip configuration parameter range,
confirming the performance and functionality of the chip.


\chapter{The Fibre Detector}
\label{sec:Fibre}

\chapterresponsible{Sandro}

\nobalance

To suppress all forms of combinatorial background from tracks with different timing,
a very thin detector with good spatial and very good timing resolution, very high efficiency,
and high rate capability
is required in the central region of the Mu3e apparatus.
For this, a thin Scintillating Fibre (SciFi) detector
with a time resolution of \SI{250}{ps}, an efficiency in excess of \SI{95}{\percent},
a spatial resolution around \SI{100}{\micro m},
and a thickness of $X/X_0 < \SI{0.2}{\percent}$
has been developed.
This chapter describes this detector and also shows how well it copes with the expectations.

\begin{figure}[!b]
   \vspace*{-6mm}
   \centering
   \includegraphics[width=0.49\textwidth]{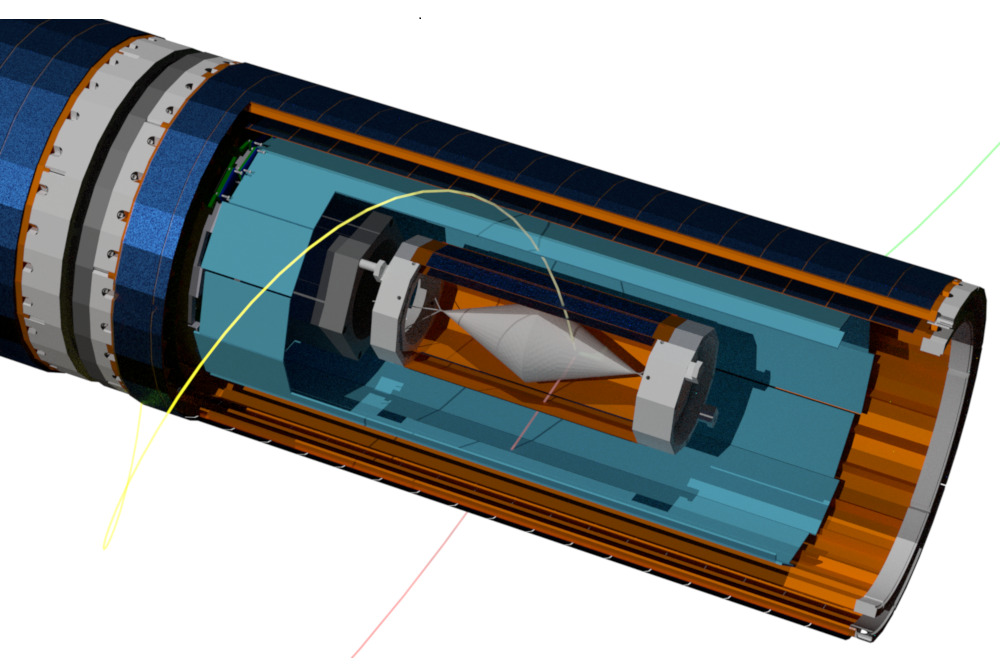}
   \caption{Open view of the central part of the Mu3e detector.
The SciFi ribbons are depicted in light blue.}
   \label{fig:scifimu3e}
\end{figure}

\autoref{fig:scifimu3e} shows the SciFi detector inside the Mu3e experiment.
In particular, the space constraints in the central part of the Mu3e experiment
impose a very compact design on this sub-detector.
In addition to timing, the SciFi detector helps resolve the direction of rotation (i.e., the charge)
of the recurling tracks in the central region of the Mu3e detector by time of flight measurements.

\begin{sloppypar}
The SciFi detector is roughly cylindrical in shape,
with a radius of \SI{61}{mm} and a length of about \SI{300}{mm} (\SI{280}{mm} in the Mu3e acceptance region).
It is composed of 12 SciFi ribbons, each \SI{300}{mm} long and \SI{32.5}{mm} 
wide\footnote{This particular value is set by the size of the photo-sensor:
the radius of a circle inscribed inside a regular dodecagon with side \SI{32.5}{mm},
i.e., the size of the photo-sensor, is indeed \SI{61}{mm}.}.
The width of the ribbons matches the size of the photo-sensor (see below).
The detector is located \SI{5}{\mm} below the outer double-layer silicon pixel detector.
\end{sloppypar}

\begin{figure}[!t]
   \centering
   \includegraphics[width=0.40\textwidth]{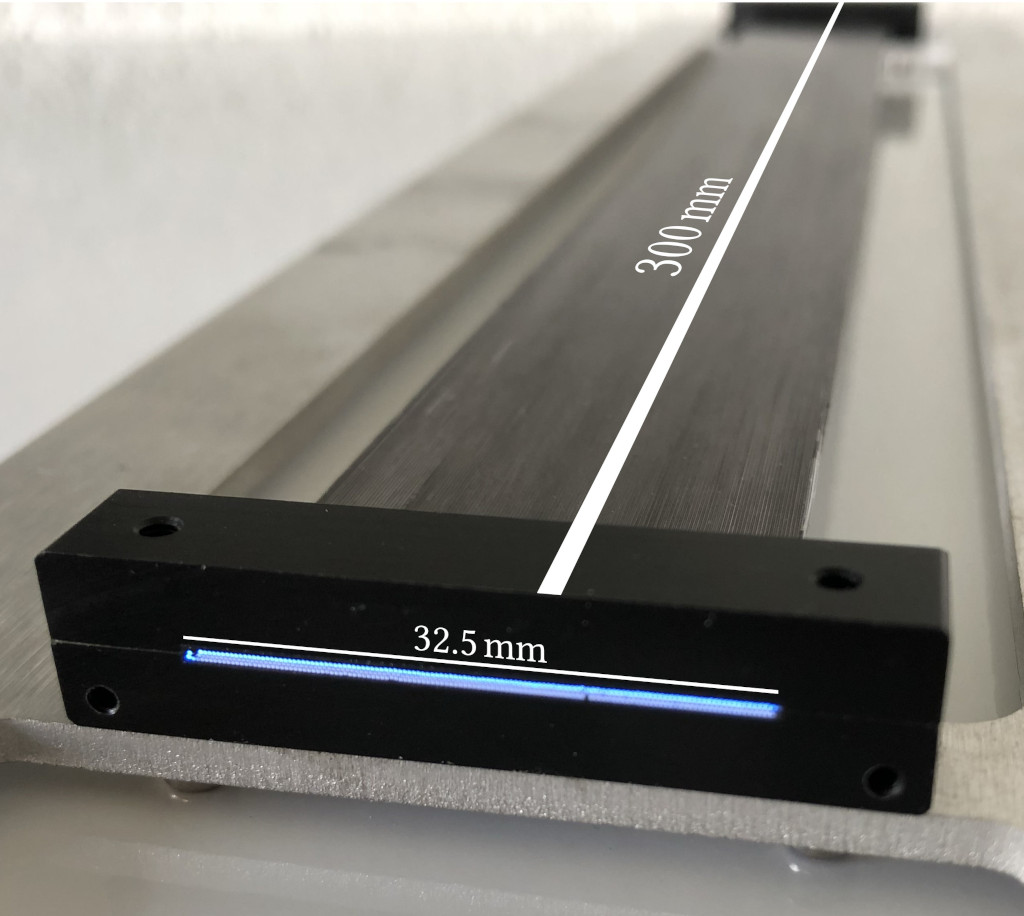}
   \caption{Full size SciFi ribbon prototype with preliminary holding structure.
The SciFi ribbon is formed by staggering three layers of round scintillating fibres.}
   \label{fig:scifi}
\end{figure}

A SciFi ribbon consists of three layers of scintillating fibres that are staggered in order to assure continuous coverage
and high detection efficiency.
\autoref{fig:scifi} shows a full size SciFi ribbon prototype.
\SI{250}{\micro m} diameter round multiclad fibres from Kuraray, type SCSF-78MJ, were selected.
Both ends of the SciFi ribbons are coupled to silicon photomultiplier (SiPM) arrays.
After careful evaluation the 128-channel Hamamatsu S13552-HRQ SiPM array, that is also being used in the LHCb experiment, was selected.
The SiPM arrays are read out with a dedicated mixed-mode ASIC, the \mutrig
(\autoref{sec:Mutrig}).

\begin{figure}[b!]
   \centering
   \includegraphics[width=0.49\textwidth]{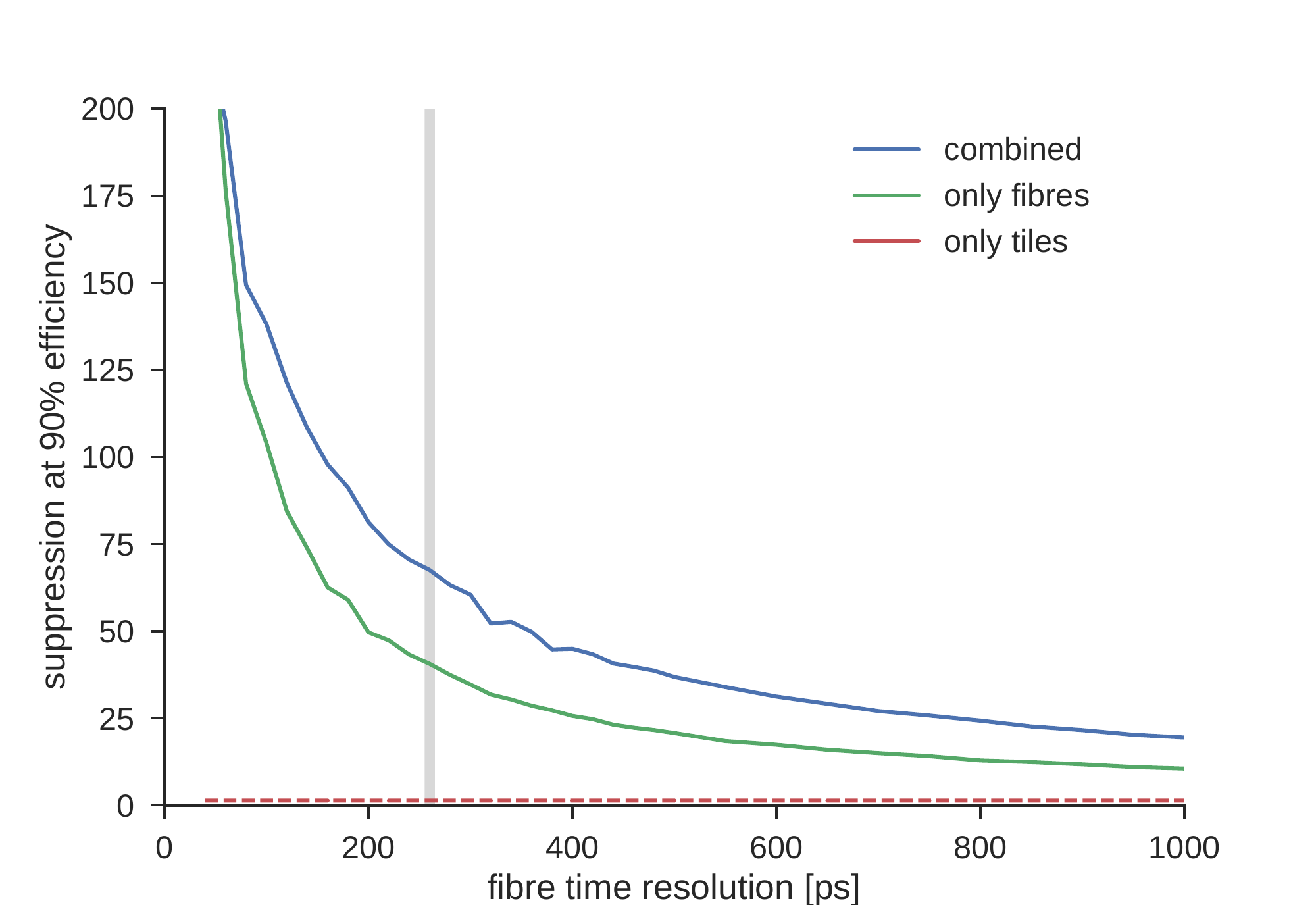}
   \caption{Suppression of Bhabha $e^+ / e^-$ pairs plus Michel $e^+$ accidental background
as a function of fibre detector time resolution if only the fibre detector (green) is used
or both timing detectors (blue) are used.
A time resolution of \SI{60}{\pico\second} for the tile detector and
a working point with a \SI{90}{\percent} overall signal efficiency are assumed in this simulation.
The vertical line (in grey) corresponds to a \SI{250}{ps} time resolution for the fibre detector. The tile detector alone has no suppression power. }
   \label{fig:FibresTimingSuppression}
\end{figure}

By far the largest source of background to the \mte search comes from
the accidental combination of positron tracks from muon decays,
in which two muons decay very closely in space, such that the decay vertices cannot be resolved,
with at least one decay positron undergoing Bhabha scattering and ejecting an electron from the target,
thus mimicking the topology of a single three-prong decay.
Such backgrounds can be efficiently suppressed by timing.
\autoref{fig:FibresTimingSuppression} shows the background suppression
power of the SciFi detector as a function of the detector time resolution.
Exploiting the fibre detector alone,
in a scenario with a time resolution of \SI{250}{ps} and a \SI{90}{\percent} overall efficiency,
leads to a suppression of the accidental background 
of $\mathcal{O}{(2.4 \cdot 10^{-2})}$.
Combining the fibre and tile (see \autoref{sec:Tiles}) timing detectors
the background is further suppressed to $\mathcal{O}{(1.4 \cdot 10^{-2})}$.
For this study, we simulated a Bhabha electron/positron pair plus a Michel positron
emerging from the same vertex
and distributed in a \SI{50}{ns} time window,
assuming a beam intensity of $10^8$ stopping $\mu^+$ per second.
The three outgoing tracks are required to pass the selection criteria
described in \autoref{sec:SensitivityStudy}.

\begin{figure}[b!]
   \centering
   \includegraphics[width=0.37\textwidth]{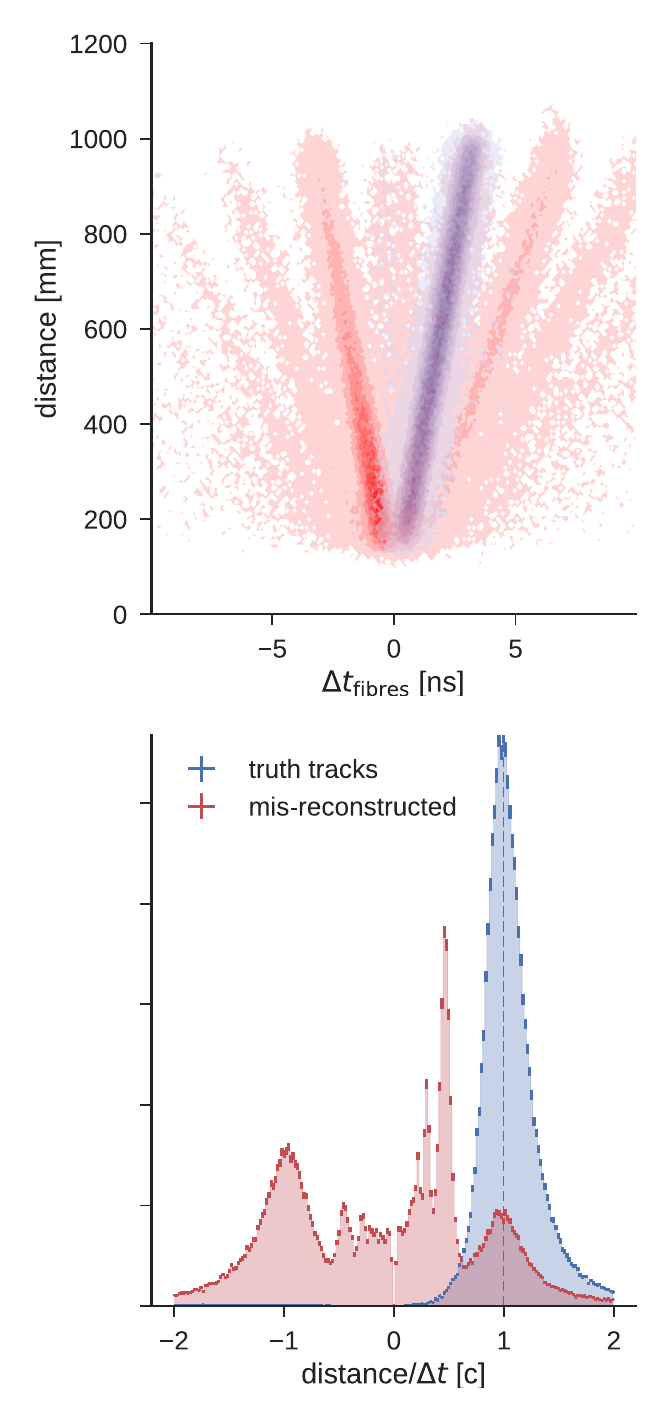}
   \caption{(top) Correlation between the time difference between two consecutive crossings of the fibre detector
and the length of the trajectory of a recurling track.
The different branches correspond to the combination of different track segments.
The correctly reconstructed tracks with the correct charge assignment are shown in blue,
while tracks with wrong charge assignment and/or mis-reconstructed tracks are shown in red.
(bottom) Speed $v=\text{track length}/\Delta t \times c$ of recurling tracks.
The different branches in the top plot correspond to the peaks in the bottom spectrum.
Track candidates with $\Delta t<0$ ($v<0$) have a wrong charge assignment.
}
   \label{fig:FibresChargeId}
\end{figure}

\begin{sloppypar}
\autoref{fig:FibresChargeId} shows the time difference (time of flight)
between two consecutive SciFi detector crossings of recurling track candidates. 
The correlation between the time difference and the reconstructed trajectory length
allows one to determine the sense of rotation of the track (and thus the charge)
and/or to reject mis-reconstructed tracks with confused recurling track segments.
\end{sloppypar}

\section{Scintillating Fibre Ribbons}
\label{sec:FibreRibbons}

Three considerations determine the SciFi detector location.
Firstly, no material should be placed outside of the fourth silicon pixel layer,
where the main momentum measurement is performed.
Secondly, it has to be in close proximity to a pixel layer,
as the track finding algorithm accounts for multiple Coulomb scattering only in the tracking layers.
And thirdly, with a larger radius the SciFi detector occupancy is reduced along with the resulting detector pile-up.
The best performance is obtained with the SciFi detector positioned just inside the third silicon pixel layer.

\begin{figure}[t!]
   \centering
   \includegraphics[width=0.48\textwidth]{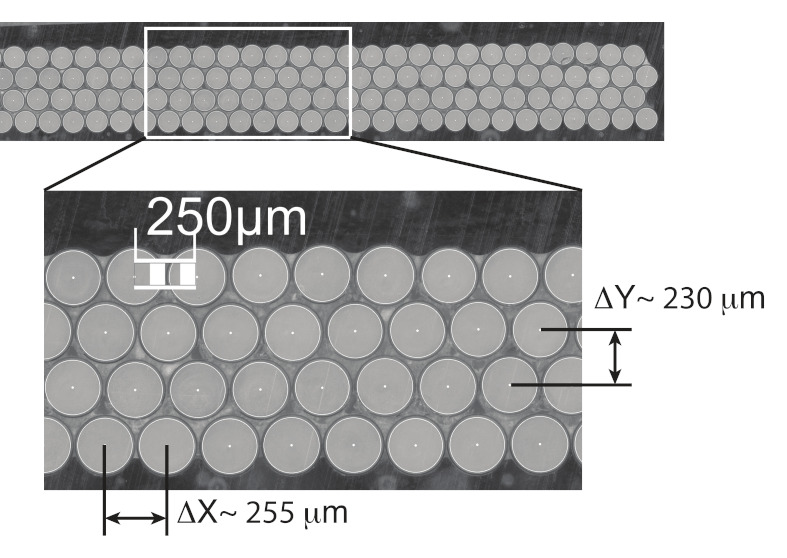}
   \caption{Front view of a SciFi ribbon prototype.
   A very good uniformity can be achieved by this ribbon construction technique.
Note that the photograph shows a four-layer SciFi ribbon,
while in Mu3e three-layer ribbons are used. The multi-cladding makes the fibres to appear smaller due to restricted light propagation.}
   \label{fig:fibres:ribbon:front}
\end{figure}

Each SciFi ribbon is formed by staggering three layers of \SI{250}{\micro m} diameter round fibres
(there are 128~fibres in a layer) with a length of \SI{300}{mm}.
Polytec EP 601-Black epoxy is used for the assembly of the final SciFi ribbons.
This two component, low viscosity, black-coloured adhesive was chosen for its excellent handling properties. 
Using a titanium dioxide loaded adhesive hasn't been an option due to the high $Z$ of titanium.
\autoref{fig:fibres:ribbon:front} shows the cross-section of a fibre ribbon prototype.
As can be observed, the fibres in a layer are separated by $\sim \SI{255}{\micro m}$ centre to centre
with a very good uniformity
and the separation between the layers is $\sim \SI{230}{\micro m}$,
which gives an overall thickness of  approximately  \SI{700}{\micro m} for a three-layer ribbon.

\subsection{Scintillating Fibres}
\label{sec:Fibres}

\begin{table}[b!]
   \centering
   \begin{tabular}{l l}
      \toprule
                characteristic &  value \\
                \midrule
      cross-section                        & round \\
      emission peak [nm]              & 450   \\
      decay time [ns]                    & 2.8    \\
      attenuation length [m]          & >4.0  \\
      light yield     [ph/MeV]          & n/a ({\it high})   \\
      trapping efficiency [\%]        & 5.4   \\
      cladding thickness [\%]        & 3 / 3   \\
      core                                   & Polystyrene (PS) \\
      inner cladding                     & Acrylic (PMMA)   \\
      outer cladding                     & Fluor-acrylic (FP) \\
      refractive index                    & 1.59/1.49/1.42  \\
      density [g/cm$^3$]             & 1.05/1.19/1.43  \\
      \bottomrule
   \end{tabular}
   \caption[Scintillating fibre properties.]{Properties of the \SI{250}{\micro m} diameter
round multi-clad Kuraray SCSF-78MJ scintillating fibres as quoted by the manufacturer.}
   \label{tab:FibresScintillators}
\end{table}

The constraints on the material budget, the occupancy, and position resolution
require the use of the thinnest available scintillating fibres. 
In extensive measurement campaigns, a detailed comparison was undertaken of different types
of \SI{250}{\micro m} diameter round scintillating fibres produced by Kuraray (SCSF-78, SCSF-81 and NOL-11)
and Saint-Gobain (BCF-12), as well as square cross-section fibres by Saint-Gobain (BCF-12).
Scintillating fibre ribbon prototypes  coupled to SiPM arrays 
have been tested in test beams at the CERN PS (T9 beamline) and PSI ($\pi$M1 beamline)
and with $^{90}$Sr sources.
The detailed results of these studies are reported in~\cite{gredig:phd,rutar:phd,damyanova:phd,Corrodi2018}.
Based on their performance with respect to light yield and time resolution,
round double-clad SCSF-78MJ fibres from Kuraray were chosen.
\autoref{tab:FibresScintillators} summarises the characteristics of this fibre type.
Novel NOL fibres, based on Nanostructured Organosilicon Luminophores,
give the best performance, but will only become commercially available in the years to come
and will be considered for future SciFi detector upgrades.

\subsection{Number of SciFi Layers}
\label{sec:NumberOfLayers}

A critical point of optimization is the number of staggered fibre layers.
More layers lead to an improved timing resolution and a higher detection efficiency but
reduces the momentum resolution of the pixel tracker due to multiple Coulomb scattering.
Since the particles cross the SciFi ribbons at an angle,
more layers lead also to a larger cluster size
(i.e., the number of channels in the SiPM array excited by the scintillating light)
and therefore to a larger occupancy.

Using the physical characteristics of the SciFi ribbons
 extensive simulation studies were performed on the impact of this sub-detector
on the momentum resolution, efficiency and track reconstruction
(details on the complete detector simulation, reconstruction algorithm and event selection
can be found in \Autoref{sec:Simulation,sec:Reconstruction,sec:SensitivityStudy}).

\begin{figure}[b!]
  \includegraphics[width=0.48\textwidth]{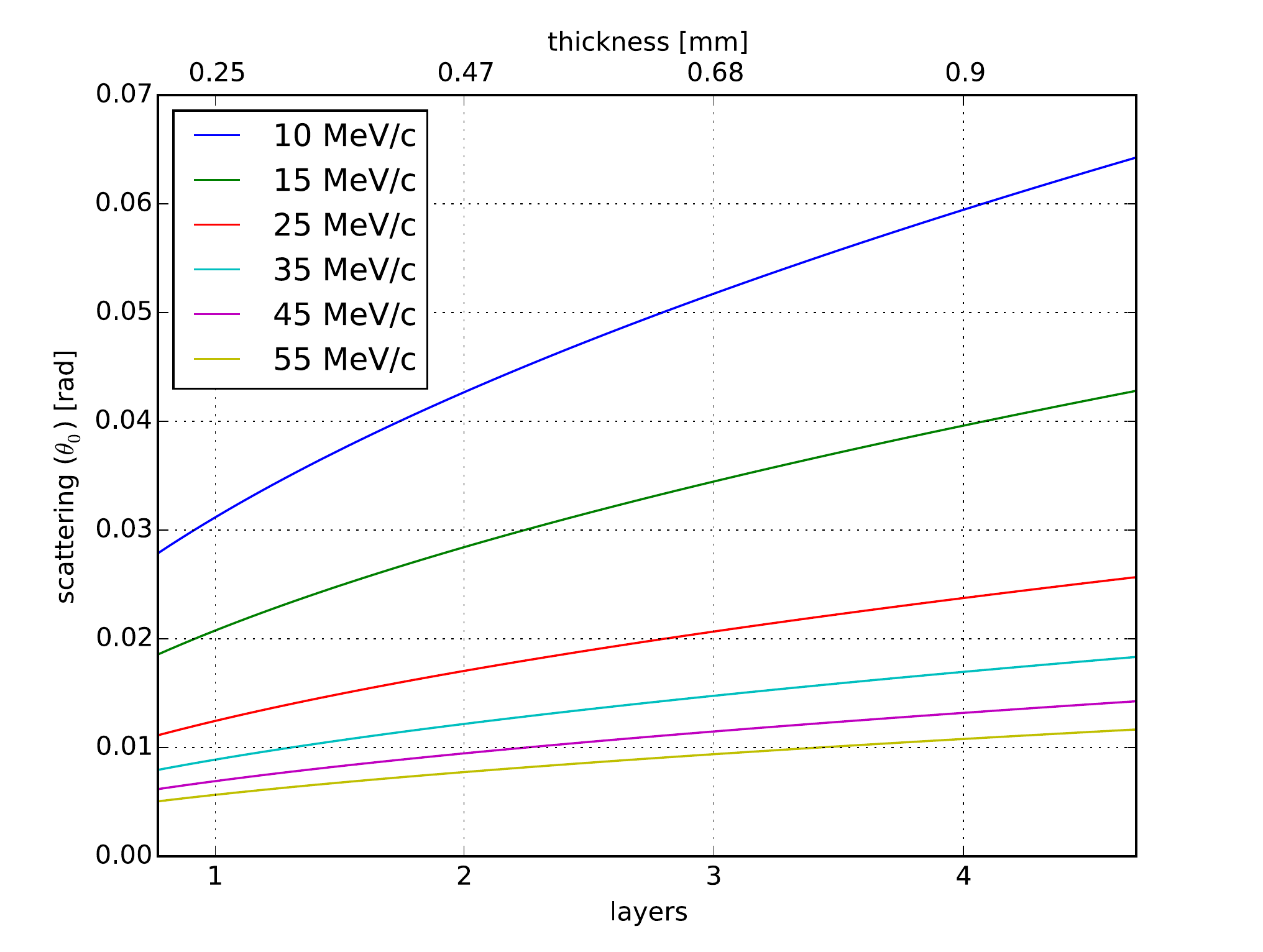}
  \caption{Multiple Scattering $\theta_0$ depending on electron/positron 
momentum and fibre ribbon thickness.}
  \label{fig:FibresScattering}
\end{figure}

\begin{figure}[t!]
  \includegraphics[width=0.46\textwidth]{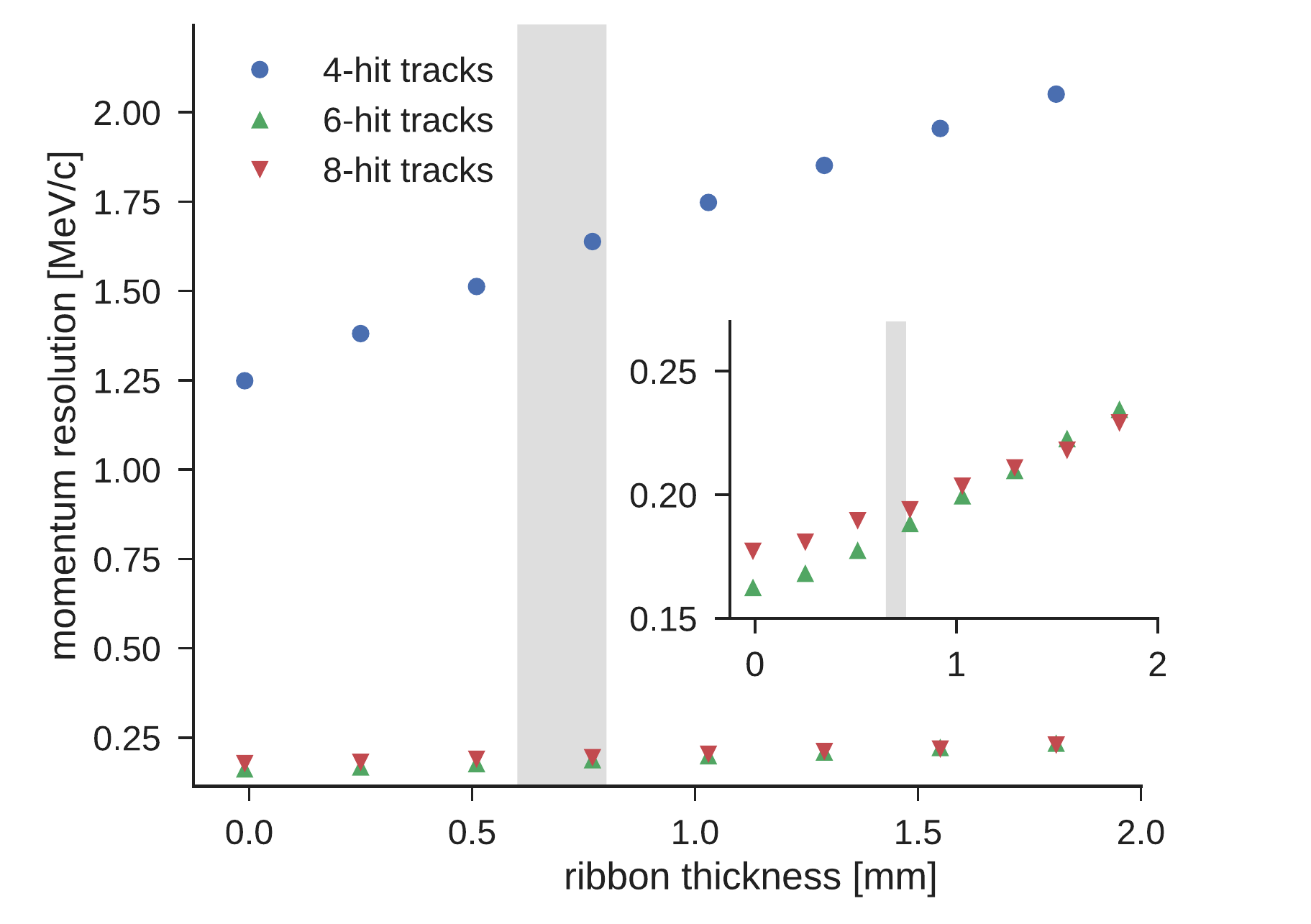}
  \caption{Momentum resolution for short (outgoing only) and long (outgoing and recurling) tracks
as a function of fibre ribbon thickness using simulated Michel decays.
The highlighted region corresponds to a three-layer SciFi ribbon thickness of $\sim \SI{0.7}{mm}$.
The momentum resolution of long (6- and 8-hit) tracks is improved over short (4-hit) tracks due to recurling
(more measured points).
}
  \label{fig:FibreLayerMomentumRes}
\end{figure}

\begin{figure}[t!]
   \includegraphics[width=0.46\textwidth]{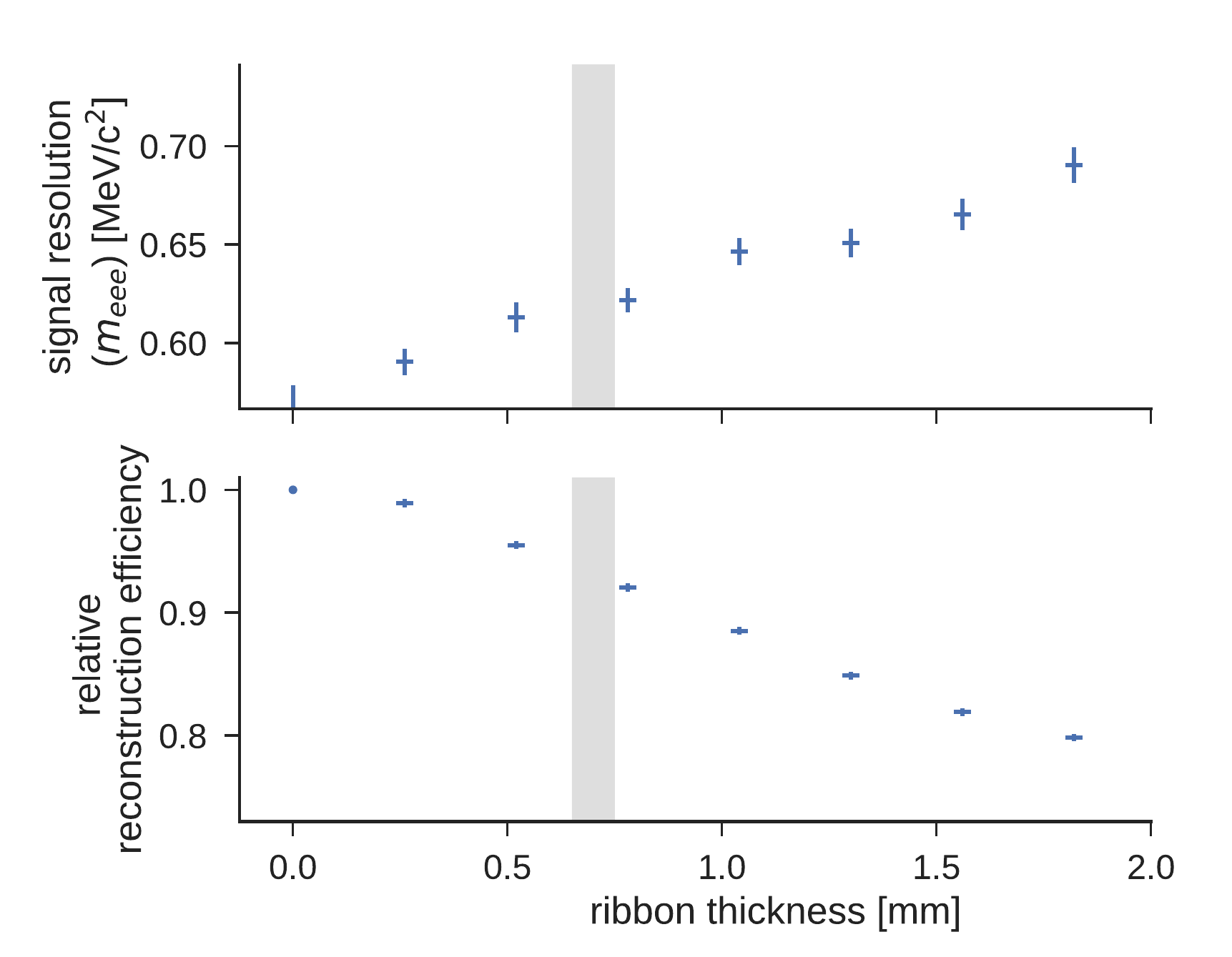}
   \caption{(top) Signal resolution in terms of the invariant mass of the three tracks of a candidate $m_{\text{eee}}$ decay
and (bottom) loss in reconstruction efficiency as a function of the fibre ribbon thickness. 
The highlighted region corresponds to a three-layer SciFi ribbon thickness of $\sim \SI{0.7}{mm}$.
}
  \label{fig:FibreLayerSignalRes}
\end{figure}

The amount of multiple Coulomb scattering generated by the fibre detector is shown in \autoref{fig:FibresScattering}.
Note that a ribbon of three layers of \SI{250}{\micro m} round fibres corresponds
to $X/X_0 \approx \SI{0.2}{\percent}$.
Multiple Coulomb scattering affects the momentum resolution (\autoref{fig:FibreLayerMomentumRes})
and thus the \mte signal invariant mass resolution
and reduces the overall reconstruction efficiency (\autoref{fig:FibreLayerSignalRes}).

As a compromise between these constraints,
ribbons consisting of three staggered layers of \SI{250}{\micro m}  diameter round fibres are chosen.
With a thinner detector it would be challenging to fulfill the efficiency requirements
and the time resolution would not be sufficient to effectively reject accidental backgrounds,
reliably determine the sense of rotation of tracks and reject misreconstructed track candidates.

\section{Silicon Photomultiplier Arrays}

The light produced in the scintillating fibres is detected in SiPM arrays at both fibre ends.
Acquiring the signals on both sides increases the time resolution
(two time measurements instead of one),
helps to distinguish between noise and signal and increases the detection efficiency of 
the whole system (because of the noise rejection).
Moreover, by taking the mean time of the two time measurements,
the timing measurements is made independent of the hit position
(assuming that light propagates at the same speed to both fibre ends)
and thus no position correction is necessary.

\begin{figure}[b!]
   \centering
   \includegraphics[width=0.48\textwidth]{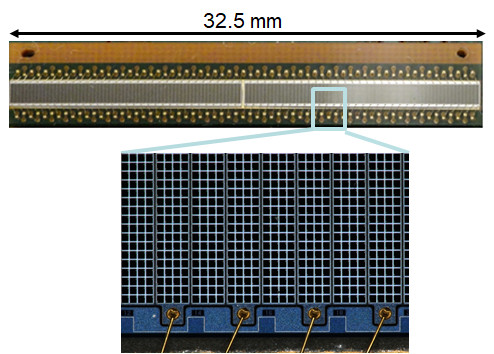}
   \caption[Hamamatsu S13552-HRQ SiPM column array.]
     {Picture of a Hamamatsu S13552-HRQ SiPM column array
     including a close view showing the pixel structure of the sensor.}
   \label{fig:sipmArray}
\end{figure}

\begin{sloppypar}
The Mu3e fibre detector is read out with Hamamatsu S13552-HRQ SiPM arrays, 
with a high quenching resistance.
The segmentation of the sensor is obtained by arranging the individual SiPM pixels into independent
readout columns (channels).
Each channel consists of 104 pixels, each measuring \SI{57.5 x 62.5}{\micro m},
arranged in a $4 \times 26$ grid.
The sensitive area of one channel is therefore \SI{230 x 1625}{\micro m}.
The pixels are separated by trenches of the fifth generation Hamamatsu low-crosstalk development (LCT5).
A \SI{20}{\micro m} gap separates the array's columns,
resulting in a \SI{250}{\micro m} pitch.
Each sensor comprises 64 such channels, which share a common cathode.
Two sensors, separated by a gap of \SI{220}{\micro m},
form the 128 channel device shown in \autoref{fig:sipmArray}.
The overall current consumption of one array is expected to be below \SI{1}{mA} even for heavily irradiated sensors.
The sensors are delivered wire-bonded on a PCB with solder pads on the backside.
The sensors are covered with a \SI{105}{\micro m} thick protective layer of epoxy resin.
\autoref{tab:LHCbArrays} summaries the most important features of the sensor.
\end{sloppypar}

\begin{table}[t!]
    \centering
        \begin{tabular}{l l}
                \toprule
                characteristic &  value \\
                \midrule
                breakdown voltage                  & \SI{52.5}{V} \\
                variation per sensor         & \SI{+-250}{\milli\volt} \\
                variation between sensors    & \SI{+-500}{\milli\volt} \\
                temperature coefficient      & \SI{53.7}{\milli\volt\per\kelvin} \\
                gain                         & \num{3.8e6} \\
                direct crosstalk             & \SI{3}{\percent} \\
                delayed crosstalk            & \SI{2.5}{\percent} \\
                after-pulse                  & \SI{0}{\percent} \\
                peak PDE               & \SI{48}{\percent} \\
                max PDE wavelength     & \SI{450}{\nano\meter} \\
                mean quench resistance $R_Q$ & \SI{490}{\kilo\ohm} at \SI{25}{\celsius} \\
                recovery time $\tau_{\text{recovery}}$ & \SI{68.9+-2.1}{\nano\second} \\
                short component $\tau_{\text{short}}$  & $<\SI{1}{\nano\second}$ \\
                long component $\tau_{\text{long}}$  & \SI{50.1+-4.1}{\nano\second} \\
                \bottomrule
        \end{tabular}
   \caption[Characteristics of SiPM column arrays.]
   {SiPM array (model S13552-HRQ) characteristics at $\Delta V = V_\text{op} - V_\text{breakdown} = \SI{3.5}{V}$
and $T=\SI{25}{\celsius}$ from \cite{LPHE-2017-001}.}
   \label{tab:LHCbArrays}
\end{table}

This sensor was developed for the LHCb experiment
and matches the requirements of the Mu3e fibre detector.
The photon detection efficiency (PDE\footnote{With contributions from quantum 
efficiency and geometrical fill factors.}) of up to \SI{50}{\percent},
single photon detection capabilities and very fast intrinsic time response 
(single photon jitter of approximately \SI{200}{ps}) are the key 
features for the use in the Mu3e fibre detector.
The SiPM arrays are read out with a dedicated mixed-mode ASIC, the \mutrig
(see \autoref{sec:Mutrig}).
The high gain ($> 10^6$) allows for the use of the \mutrig without any pre-amplification.
Typical dark-rates are around \SI{100}{kHz} at room temperature per SiPM array channel for unirradiated sensors.
In contrast to LHCb, where the SiPM arrays are operated around \SI{-40}{\celsius},
the Mu3e sensors are being operated at a temperature of $\sim\SI{0}{\celsius}$,
but in a less intense radiation field.
The moderate cooling of the detector is required
to further reduce the dark count rate and mitigate the radiation damage effects.

\begin{figure}[t!]
   \includegraphics[width=0.48\textwidth]{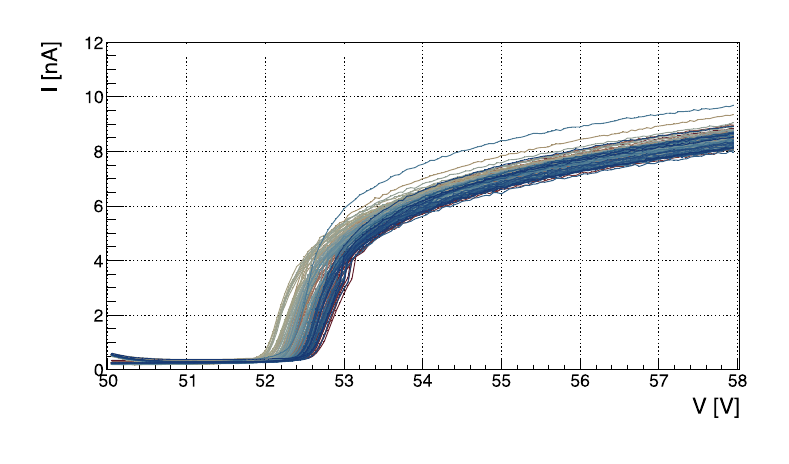}
   \caption{I-V curves for each channel of the SiPM array.
   All breakdown voltages are comprised within $\pm \SI{0.25}{V}$ of the central value of \SI{52.5}{V}.}
  \label{fig:SiPMIVcurves}
\end{figure}

\autoref{fig:SiPMIVcurves} shows the I-V curves for one SiPM array for each channel of the sensor.
All breakdown voltages are comprised within $\pm \SI{0.25}{V}$ of the central value of \SI{52.5}{V}.
The best performance is obtained for an operational voltage ($V_\text{op}$)
\SI{3.5}{V} above the breakdown voltage ($V_\text{breakdown}$),
but the sensor can also be operated at higher voltages for an increased gain.
Since all channels share a common cathode, the sensor is usually operated at a common voltage
for all channels.
The performance of the photo-detector can be further improved by adjusting $V_\text{op}$
individually for each channel.
The \mutrig readout ASIC allows for the fine tuning of the bias voltage
around a common value for each individual channel of the sensor.

\begin{figure}
   \centering
   \includegraphics[width=0.48\textwidth]{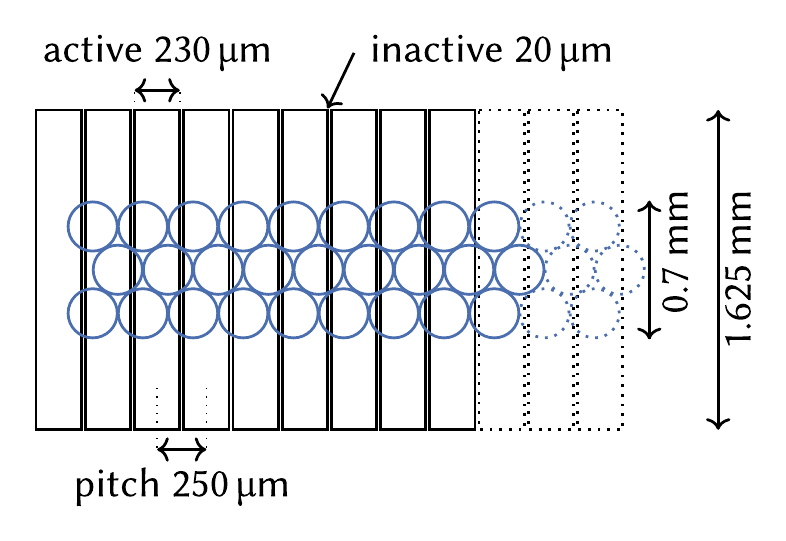}
\caption{Mapping of the SciFi ribbon on the SiPM array.
No one to one matching is possible between the fibres and the SiPM columns.}
   \label{fig:FibreMapping}
\end{figure}

The fibre ribbons are coupled directly to the surface of the SiPMs on both sides.
\autoref{fig:FibreMapping} shows the mapping of the SciFi ribbon on the SiPM array.
As can be seen, no one-to-one matching is possible between the fibres and the SiPM columns
because of the staggering of the fibres.
To ease detector assembly and maintainability, the coupling is realised by
only mechanical pressure without the use of optical interfaces.

\section{SciFi Readout Electronics}

\begin{figure}
   \centering
   \includegraphics[width=0.46\textwidth]{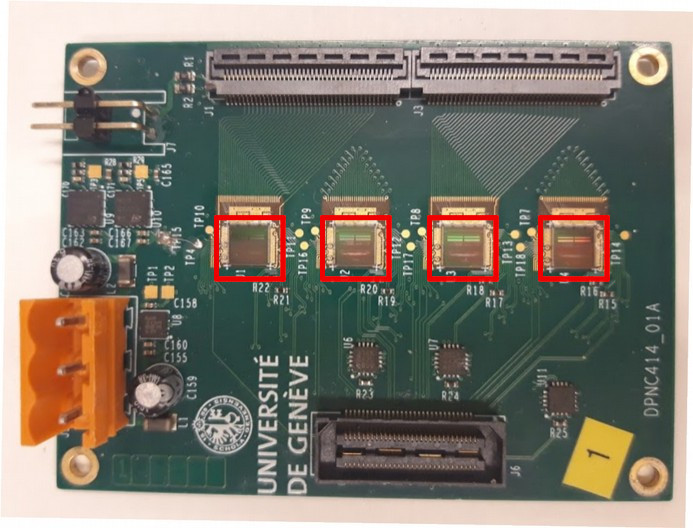}
\caption{First version of the SciFi module board hosting 4 \mutrig ASICs
(outlined in the red boxes) wired bonded directly on the board.}
   \label{fig:scifiBoard}
\end{figure}

The Mu3e scintillating fibre detector requires the digitization of the crossing time
information at a single photon level.
That leads to very high rates per SiPM channel
coming from the particles crossing the SciFi ribbons ($\sim \SI{200}{kHz}$ signal rate)
and the dark noise ($\sim \SI{1}{MHz}$ for irradiated sensors).
The latter is reduced by clustering during the real-time processing of the data.

For the readout of the 3072 SiPM channels we use the mixed-mode \mutrig ASIC
with \SI{50}{ps} TDC time binning (see \autoref{sec:Mutrig} for a detailed description of the ASIC).
Each ASIC comprises 32 fully differential input analogue channels,
therefore four \mutrig ASICs are required for the readout of one SiPM array.
Although the ASIC has a fully differential input, single ended signals are used,
because the SiPM array channels share a common cathode. 
When operated with the SiPM arrays, the signal is compared to two thresholds: a low one for timing
(a time stamp is generated) and a high one for hit selection (single flag).


The analog signals from each SiPM array (128~channels) are digitised by one SciFi module board (SMB)
hosting four \mutrig ASICs.
\autoref{fig:scifiBoard} shows the first version of the SMB.
The space limitations in the Mu3e setup require a very compact design of the board.
The ASICs are wire bonded directly to the board.
The final version of the board will measure \SI{26 x 50}{mm}
and is currently under development.
The electrical connection between the SiPM sensors and the readout electronics is realised through
flex-print circuits.
A 128-channel SiPM array is soldered to a support PCB with an embedded flex-print,
which continues to a second PCB hosting the \mutrig ASICs.
In addition to the \mutrig ASICs, the SMB hosts the clock and reset distribution circuits,
components for the control of the \mutrig, LDO voltage regulators for power distribution, and temperature probes.
In total 24 such SMBs are needed, one per SiPM array.
Finally all SMBs are connected to front-end FPGA boards (see \autoref{sec:FrontEndFPGABoards}) via micro twisted-pair cables.

\subsection{Power requirements}
\label{sec:PowerMUTRIG}
The power requirements of the \mutrig ASICs are given in \autoref{sec:Mutrig}.
The powering of one SMB requires \SI{2}{V} at \SI{2.5}{A}, \SI{3.5}{V} at \SI{0.1}{A},
and a bias line (around 55 to \SI{57}{V}) for the SiPM array.
Each SMB generates around \SI{5}{W} of thermal output,
which has to be cooled.

\section{SciFi Detector Performance}

\begin{figure}[t!]
   \centering
   \includegraphics[width=0.48\textwidth]{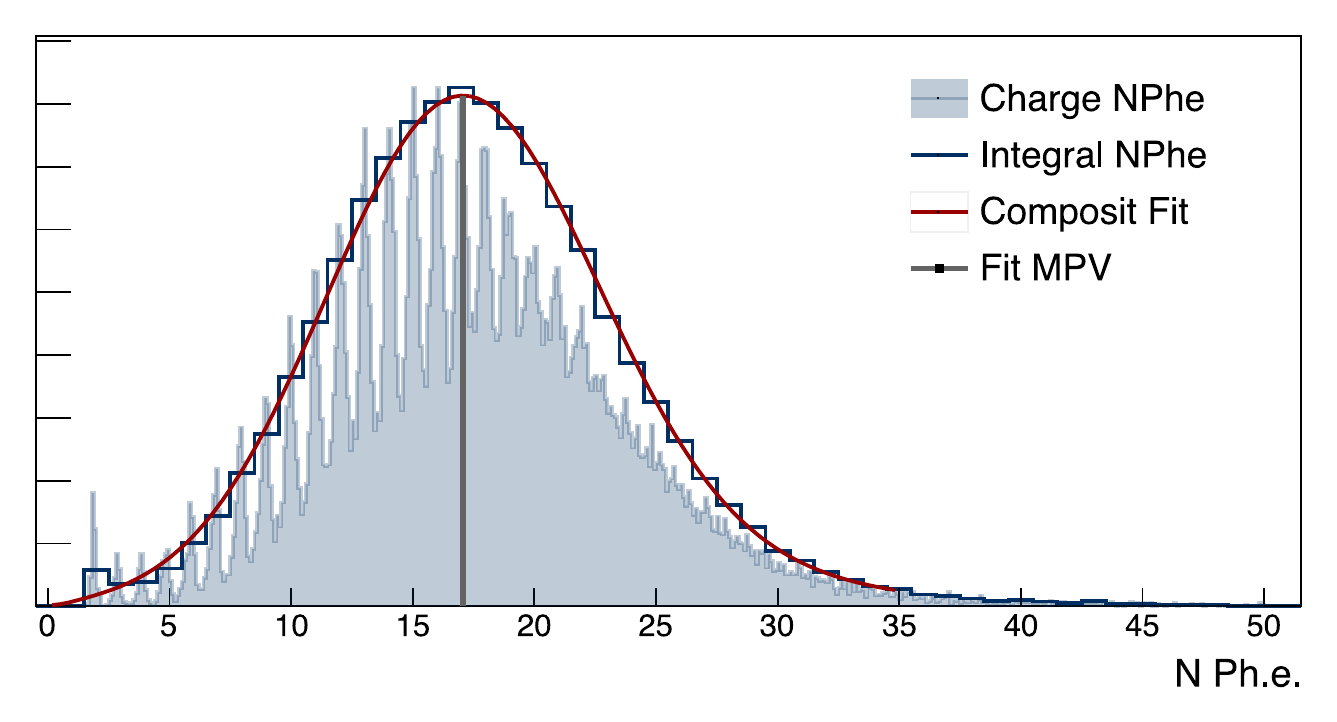}
   \caption[Spectrum of charge and number of photons.]
{Light yield of a cluster (see text) for a m.i.p. crossing a three-layer SCSF-78MJ fibre ribbon
prepared with clear epoxy.
The integral NPhe is obtained by integrating the charge in a region of $\pm 0.5$ ph.e. around each peak (integer).
A convolution of a Gaussian and of a Landau is used to fit the data
and the MPV of the spectrum is marked with the vertical line.}
   \label{fig:clCharge}
\end{figure}

\autoref{fig:clCharge} shows the light yield in a {\it cluster} excited by a minimum-ionizing particle
crossing a three-layer SCSF-78MJ fibre ribbon prepared with clear epoxy.
A cluster is defined as the sum of all consecutive SiPM channels
with an amplitude larger than a specific threshold (in this case 0.5 photo-electrons)
and a cluster multiplicity of at least two adjacent SiPM channels above the same threshold.
The number of photo-electrons (ph.e.) is defined by the charge sum of all channels in a cluster
at one side of the SciFi ribbon matched to a crossing track.
The light yield is measured with respect to the centre of the fibre ribbon (i.e., \SI{150}{mm} from the edge).
A convolution of a Gaussian and of a Landau distribution is used to fit the data.
The fit provides also the most probable value (MPV) for the number of detected ph.e.,
which is of about 17 for this configuration.
This ph.e.\ spectrum, however, is not accessible in the experiment
since the \mutrig provides only the timing information and no charge information.
Test-beam data were recorded using a fast pre-amplifier and readout digitizing electronics
based on the DRS4 ASIC.
The recorded waveforms were then processed using timing algorithms close to the \mutrig functioning
(i.e., 0.5 ph.e.\ low threshold leading edge discriminator).

\begin{figure}[t!]
   \centering
   \includegraphics[width=0.44\textwidth]{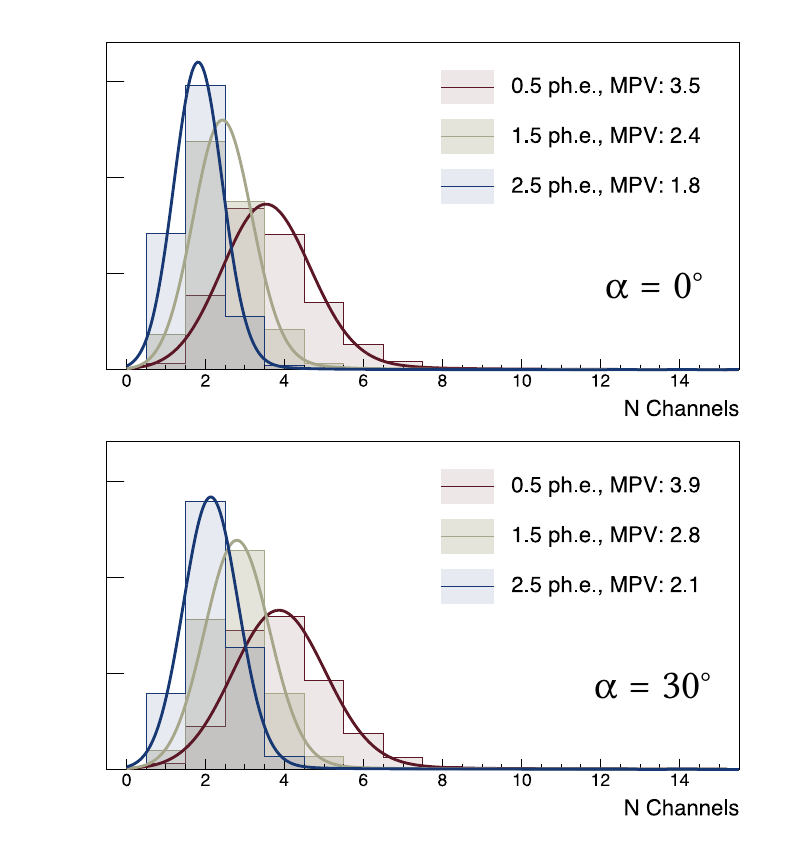}
   \caption[SciFi cluster size.]
{Cluster size for a particle crossing the ribbon at two different angles and different thresholds.
Electrons from a radioactive $^{90}{\rm Sr}$ source are used for this measurement.
An angle of $\alpha = \SI{0}{\degree}$ describes a perpendicular crossing.}
   \label{fig:clSize}
\end{figure}

The cluster size distribution for the same SciFi ribbon is shown in \autoref{fig:clSize}.
Typical cluster sizes are around 3.5 for a threshold of 0.5 ph.e.,
for a particle crossing the ribbon at \SI{0}{\degree} (i.e., perpendicularly to the ribbon).
The cluster size can be reduced by increasing the detection threshold to e.g., 1.5 ph.e.\ or higher.
The figure shows also cluster sizes for particles crossing the ribbon at an angle of \SI{30}{\degree},
which is close to the mean crossing angle in Mu3e of \SI{25}{\degree}\footnote{Due to curling tracks in the magnetic field}.
A larger crossing angle increases the average cluster size.

\begin{figure}
   \centering
   \includegraphics[width=0.48\textwidth]{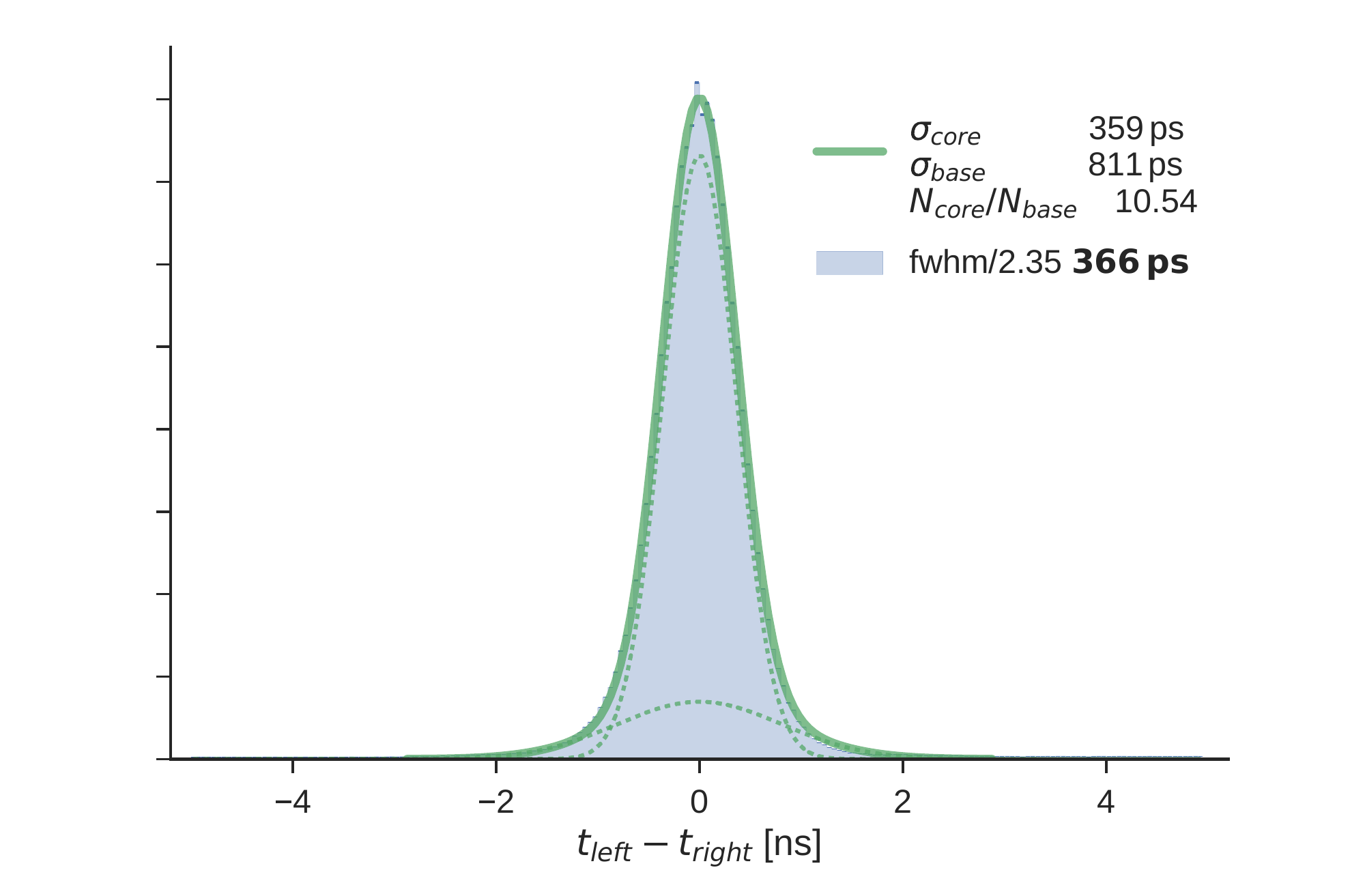}
   \caption[Time Resolution of fibre prototype extracted with \mutrig.]
{Time resolution of a 4 layer SCSF-78MJ SciFi ribbon extracted from clusters with at least 2 active columns.
No channel by channel time offset correction has been applied.}
 \label{fig:mutrig:deltaT}
\end{figure}

The detection efficiency of the SciFi detector depends on the applied thresholds, minimal cluster multiplicity
and the requirement of time matched clusters at both SciFi ribbon ends.
For the selected working point,
which requires a threshold of 0.5 ph.e., with a minimal cluster multiplicity of two
and a $5~\sigma$ timing cut on the matched clusters,
where $\sigma$ is the intrinsic time resolution of the SciFi detector,
the detection efficiency is around \SI{95}{\percent}.
Without the timing cut, the detection efficiency increases close to \SI{100}{\percent}.
It should be noted that the cluster matching and the timing cut can only be applied
in the offline analysis of the SciFi data
and can be tuned to optimise the detection efficiency.

Finally, an example of the timing performance of the SciFi detector is shown in \autoref{fig:mutrig:deltaT}.
This measurement has been performed with the \mutrig evaluation board,
see \autoref{fig:testbeam_setup_mutrig},
 using a four-layer SciFi ribbon with a $^{90}{\rm Sr}$ source
requiring a minimal cluster multiplicity of two neighbouring channels with an amplitude of at least 0.5 ph.e.
Similar results have also been obtained with the analogue electronics (DRS4-based DAQ) mentioned above
and particle beams~\cite{damyanova:phd}.
The spread of the time difference distribution from the two ribbon sides $\sigma(t_{\text{left}} - t_{\text{right}})$
corresponds to twice the intrinsic detector resolution (mean time).
For example, the FWHM/2.35 of the distribution obtained in this measurement is \SI{366}{ps}
implying a resolution on the mean time around \SI{200}{ps}.
For a three-layer ribbon as used in Mu3e, the time resolution is slightly worse, at around \SI{250}{ps}.

\section{SciFi Detector Mechanics}
\label{sec:FibreMechanicsAndCooling}

\begin{figure}
   \centering
   \includegraphics[width=0.48\textwidth]{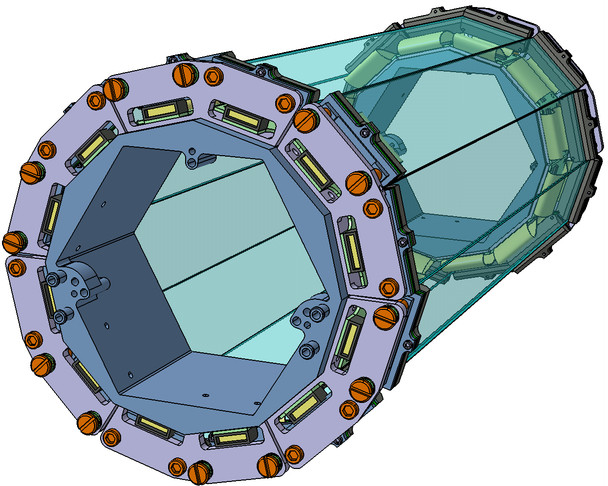}
   \caption[SciFi detector design.]
{Overall structure of the scintillating fibre detector.}
   \label{fig:scifi:full}
\end{figure}

\begin{figure}
   \centering
   \includegraphics[width=0.48\textwidth]{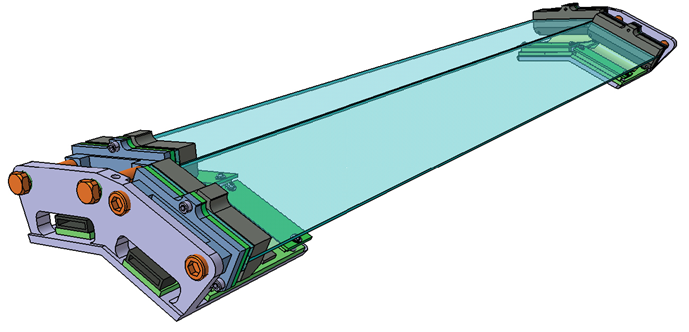}
   \caption{A fibre module consists of two SciFi ribbons with the associated support structure.
The ribbons are staggered longitudinally to minimise dead spaces between the ribbons
and are spring loaded alternately on opposite sides of the structure.}
   \label{fig:fibresmodule}
\end{figure}

\autoref{fig:scifi:full} shows the overall structure of the SciFi detector.
The detector is composed of 12 SciFi ribbons, \SI{300}{mm} long and \SI{32.5}{mm} wide.
The ribbons are staggered longitudinally by about \SI{10}{mm} (\autoref{fig:fibresmodule})
in order to minimise dead spaces between the ribbons
and to provide sufficient space for the spring loading of the ribbons.
To avoid sagging and to compensate for the thermal expansion 
the ribbons are spring loaded on one side of the structure
(6 ribbons on one side and the other 6 on the other side).

\begin{figure}
   \centering
   \includegraphics[width=0.48\textwidth]{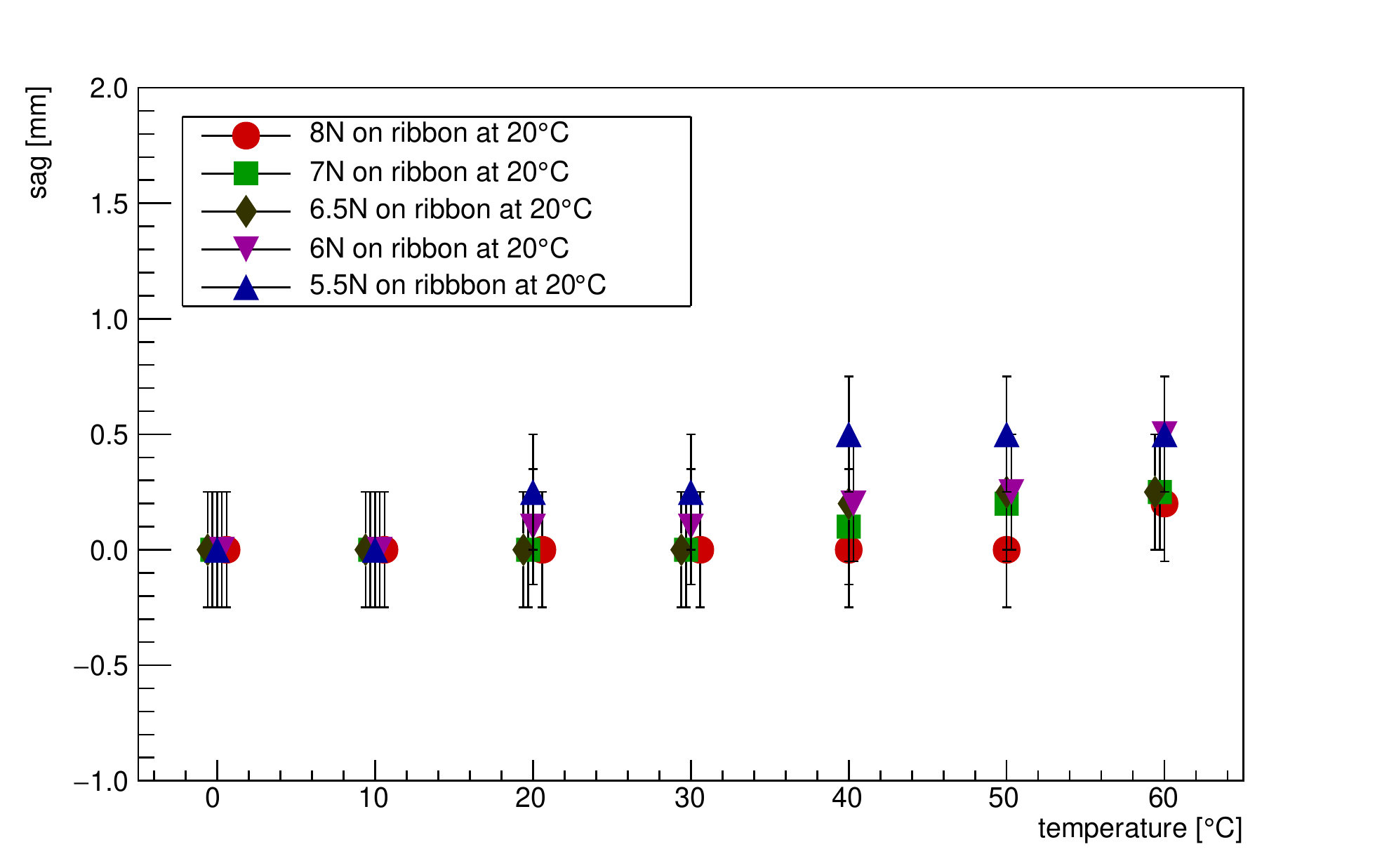}
   \caption{Sag of the SciFi ribbon as a function of the temperature for different values
of the applied spring tension.}
   \label{fig:scifiSag}
\end{figure}

A detailed study to determine the effects of the thermal expansion and sagging
has been performed.
A thermal expansion coefficient for the \SI{300}{mm} long SciFi ribbon of
$(65 \pm 16) \cdot 10^{-6} / {\rm K}$ has been measured.
Therefore, for a \SI{50}{\celsius} thermal excursion,
an elongation of the ribbons of around \SI{1}{mm} is expected.
This elongation effect can be compensated by spring-loading the ribbons as mentioned above.
\autoref{fig:scifiSag} shows the sag of
a three-layer \SI{300}{mm} long and \SI{32.5}{mm} wide SciFi ribbon as a function of the temperature
for different values of the applied tension.
\autoref{fig:scifiSag} also shows that a tension of \SI{8}{N} is required to prevent sagging
over the whole temperature interval
and to guarantee the correct positioning of the detector.

To ease the sub-detector installation, the SciFi ribbons are assembled in modules.
Each module consists of two SciFi ribbons, as shown in \autoref{fig:fibresmodule}.

\begin{figure}[]
   \centering
   \includegraphics[width=0.4\textwidth]{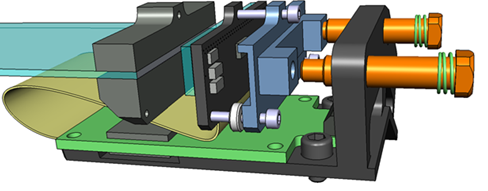}
   \caption[SciFi detector design top view.]
{Expanded view of the SciFi support structure,
showing all the elements of the detector:
SciFi ribbon, SiPM sensor, SciFi front end board
and the L-shaped support structure.}
   \label{fig:SciFiExpanded}
\end{figure}

The SciFi ribbons are coupled to the SiPM arrays by simple mechanical pressure
(no grease or other optical interface).
Each SiPM sensor is connected to a front-end digitizing board via a flex-print circuit.
\autoref{fig:SciFiExpanded} shows an expanded view of the assembly structure:
the SciFi ribbons are attached to the SiPM arrays, which in turn are supported by stiffeners
fixed to L-shaped supports,
where the assembly is also spring loaded.
The same L-shaped supports are also used to mount the SciFi front-end boards.

\begin{figure}[]
   \centering
   \includegraphics[width=0.4\textwidth]{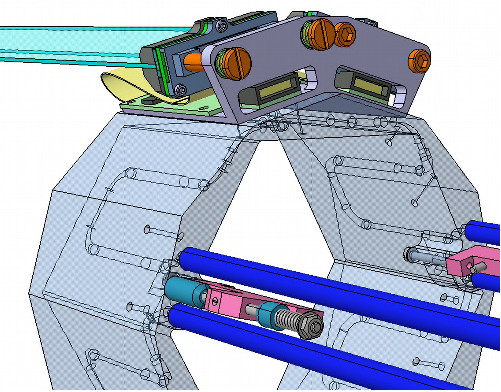}
   \caption[SciFi detector design side view.]
{Support structure of the SciFi detectors, which serves also as cold mass to cool away the heat
generated by the SciFi module board and the cooling of the SiPM arrays.}
   \label{fig:SciFiSupport}
\end{figure}

The L-shaped supports are fixed to a hollow dodecagonal prism as shown in \autoref{fig:SciFiSupport},
\SI{50}{mm} tall with an outer diameter of \SI{100}{mm},
which also provides the necessary cooling for the front-end electronics.
Two such cooling structures are attached to the beam pipe on each side of the Mu3e detector
and connected to the pipes of the Mu3e cooling system.
This cooling structure is created by 3D printing in aluminium
with embedded piping for the circulation of the coolant.
Each \mutrig ASIC generates about \SI{1}{W} of thermal output,
therefore around \SI{50}{W} has to be cooled away on each side of the SciFi detector.
Since the SiPM arrays are in thermal contact with the L-shaped supports, they are cooled
by the same cooling structure.
The goal is to cool the SiPM arrays down to \SI{0}{\celsius}.


\chapter{Tile Detector}
\label{sec:Tiles}

\nobalance

\chapterresponsible{Y.Munwes}

The tile detector aims at providing the  most precise timing information of the 
particle tracks possible.
As it is located at the very end of recurling particle trajectories, there are no constraints on the amount of detector material; the placement inside the recurl pixel detectors however implies very tight spatial constraints.
The detector consists of plastic scintillator segmented into small tiles.
Each tile is read out with a silicon photomultiplier (SiPM) directly attached to the scintillator.
The main goal of the tile detector is to achieve a time resolution of better than $\SI{100}{ps}$ and a detection efficiency close to $100\%$ in order to efficiently identify coincident signals of electron triplets and suppress accidental background.

\section{Detector Design}
\label{sec:TileDetectorDesign}

The tile detector is subdivided into two identical stations -- one in each recurl station.
Each tile detector segment has the shape of a hollow cylinder enclosing the beam pipe.
The length of a segment is \SI{34.2}{\cm} along the beam direction ($z$ direction) including the endrings, while the outer radius is \SI{6.4}{\cm}, which is limited by the surrounding pixel sensor layers.
The detector in each recurl station is segmented into 52 tiles in $z$ direction and 56 tiles along the azimuthal angle ($\phi$ direction).
This is the highest feasible channel density, considering the space requirements for the readout electronics.
The high granularity is essential in order to achieve a low occupancy as well as a high time resolution.

\begin{figure}[tbh]
        \centering
                \includegraphics[width=0.48\columnwidth]{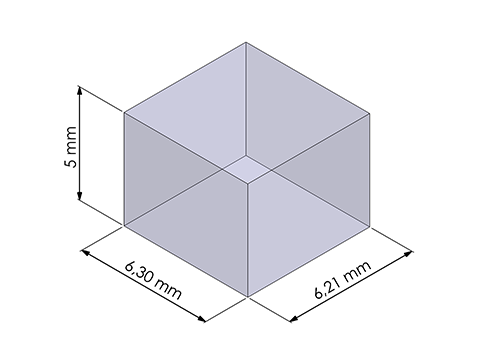}
                \includegraphics[width=0.48\columnwidth]{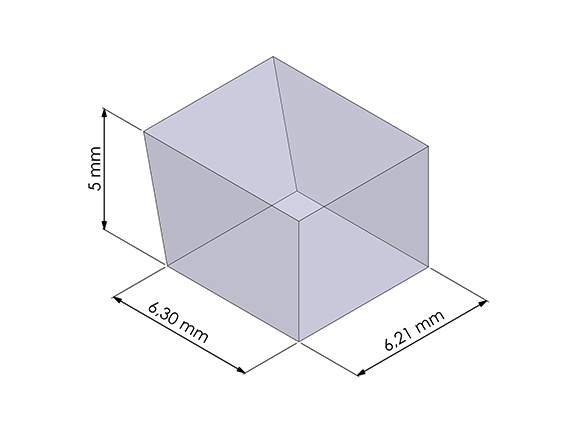}
        \caption{Tile scintillator geometry: (left) central tile, (right) edge tile.} 
        \label{fig:Tiles_tile}
\end{figure}

\begin{figure*}[!bht]
	\begin{subfigure}[t]{0.48\textwidth}
        \centering
        \includegraphics[width=0.8\linewidth]{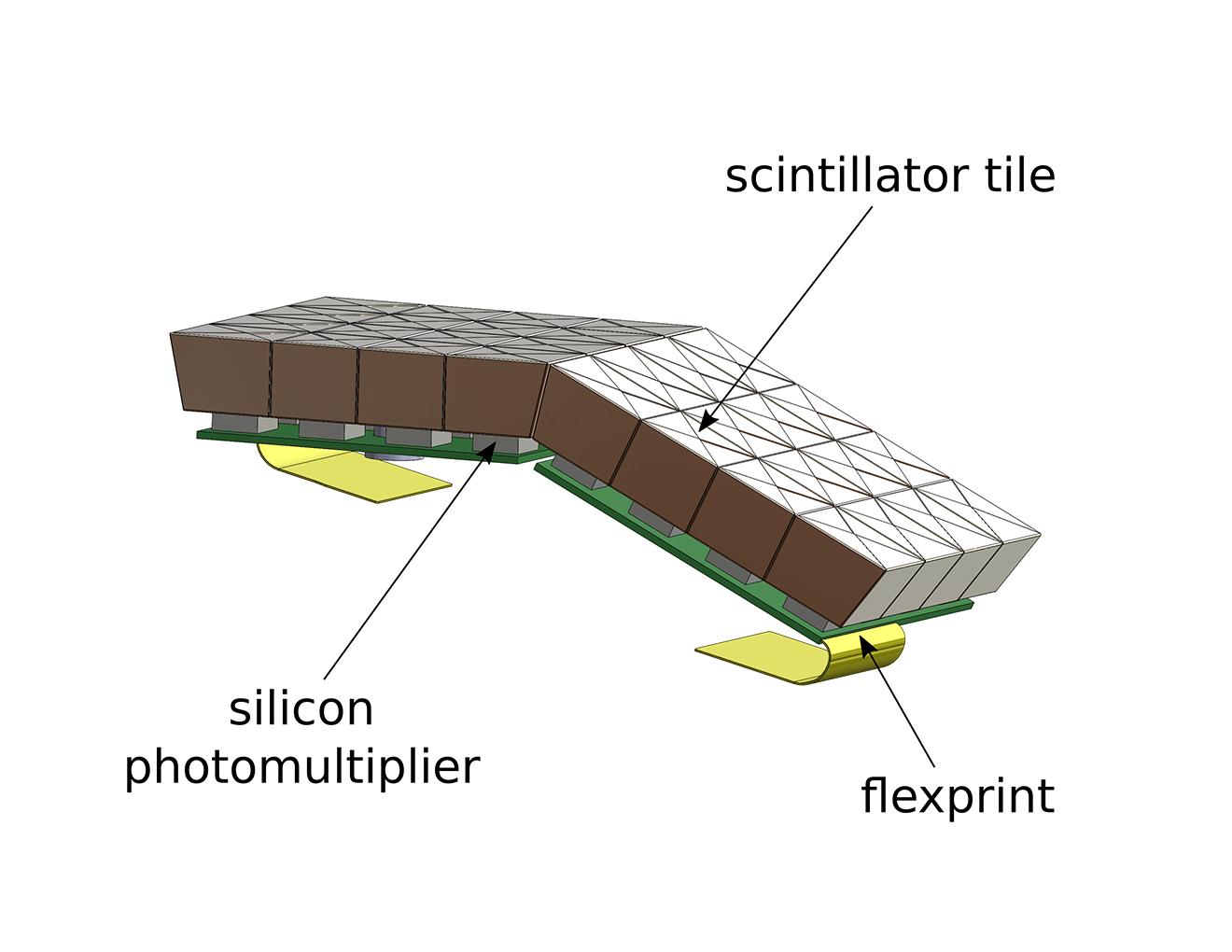}
        \subcaption{The tile detector base unit consisting of 32 scintillator \newline and SiPM channels. The sensors are mounted on a flex-rigid \newline PCB. CAD rendering.}
        \label{fig:Tiles_CAD_submodule}
     \end{subfigure}
     \begin{subfigure}[t]{0.48\textwidth}
        \centering
        \includegraphics[width=0.8\linewidth]{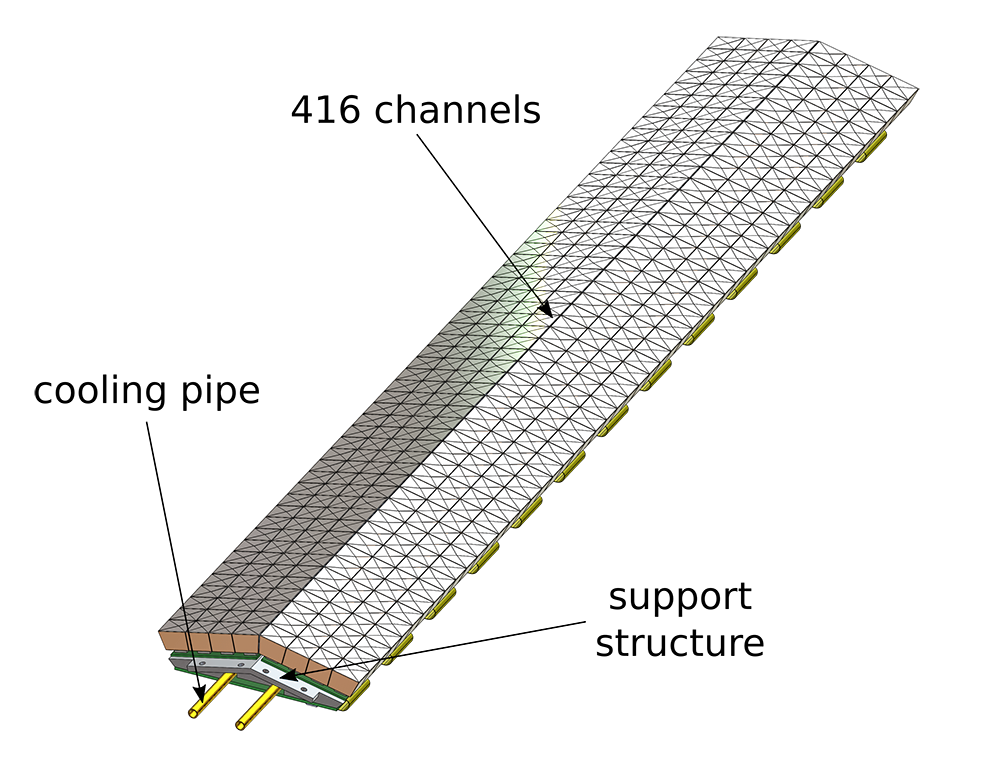}
        \caption{Module (416 channels) of the tile detector consisting of 13 base units, which are mounted on a support structure. A copper pipe for cooling liquid is placed inside the support structure to cool the readout chips and SiPMs. CAD rendering.}
		\label{fig:Tiles_CAD_module}
	\end{subfigure}
	\newline
	\begin{subfigure}[t]{0.48\textwidth}
        \centering
        \includegraphics[width=0.8\linewidth]{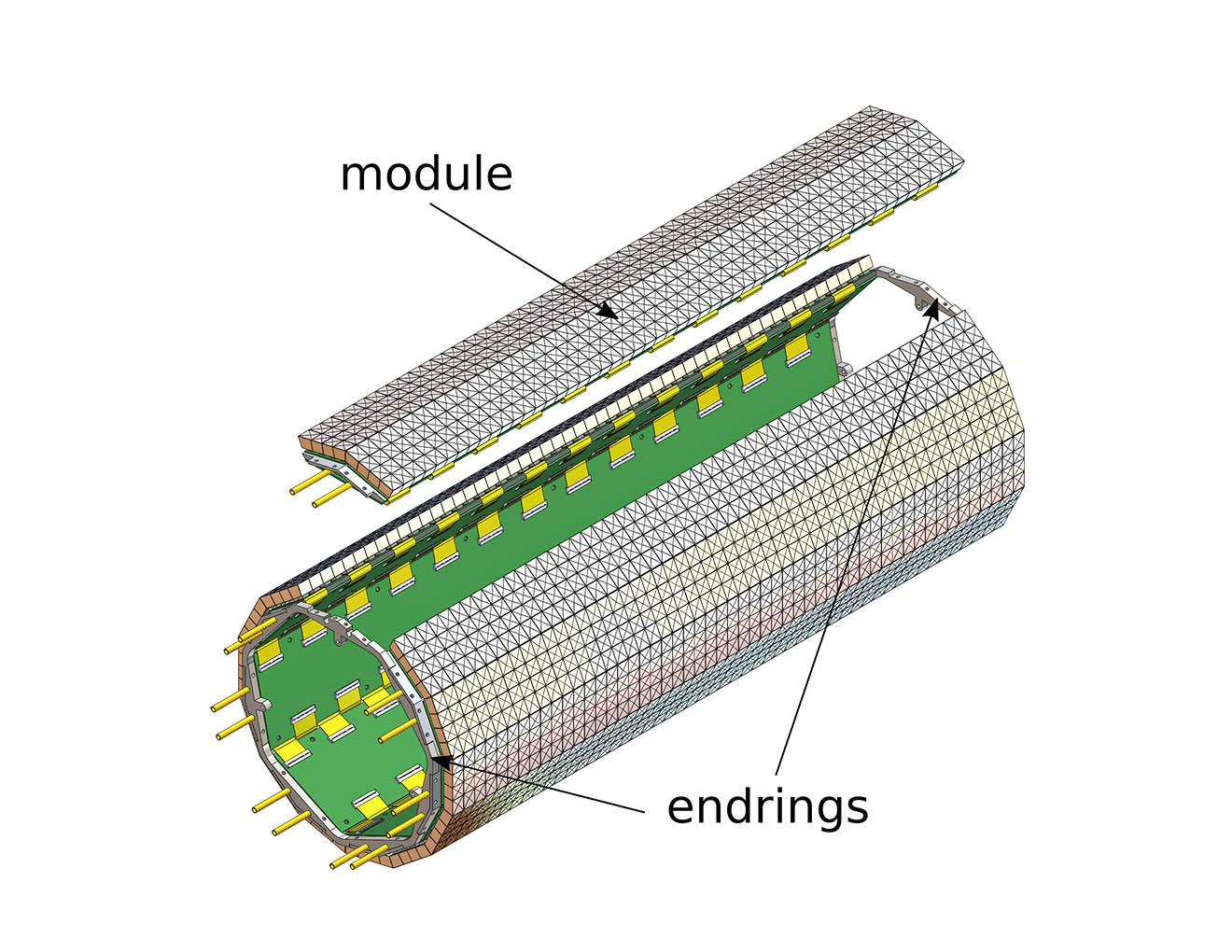}
        \caption{Full tile detector (CAD rendering, exploded view).}
		\label{fig:Tiles_CAD_explode}
	\end{subfigure}
	\begin{subfigure}[t]{0.48\textwidth}
        \centering
        \includegraphics[width=0.8\linewidth]{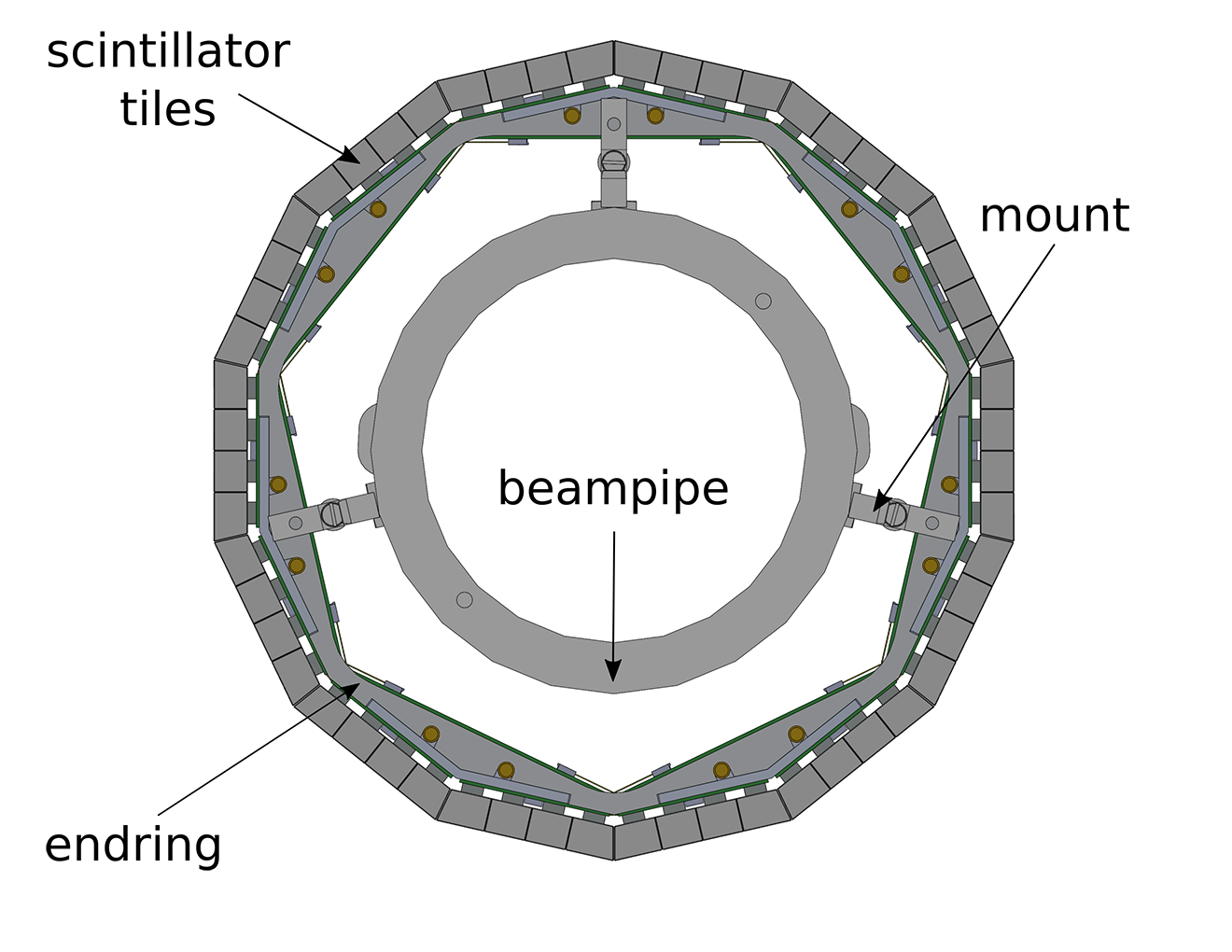}
        \caption{Full tile detector (front view). The detector modules are mounted on two endrings connected to the beam pipe. CAD rendering.}
		\label{fig:Tiles_CAD_front}
	\end{subfigure}
        \caption{CAD rendered views of the tile detector.}

\end{figure*}

The technical design of the tile detector is based on a modular concept, i.e.~the detector is composed of small independent detector units.
The base unit of the tile detector, referred to as \textit{sub-module}, is shown in \autoref{fig:Tiles_CAD_submodule}.
It consists of 32 channels arranged in two $4\times4$ arrays. The tiles are made out of Eljen technology EJ-228 plastic scintillator and have a size of \SI[parse-numbers = false]{6.3\times6.2\times5.0}{\mm\cubed}, see \autoref{fig:Tiles_tile}. The edges of the two outer rows of an array are bevelled by \SI{25.7}{\degree}, which allows for seven base units to be arranged approximately in a circle. 
\begin{figure}[tbh]
	\centering
	\includegraphics[width=0.48\textwidth]{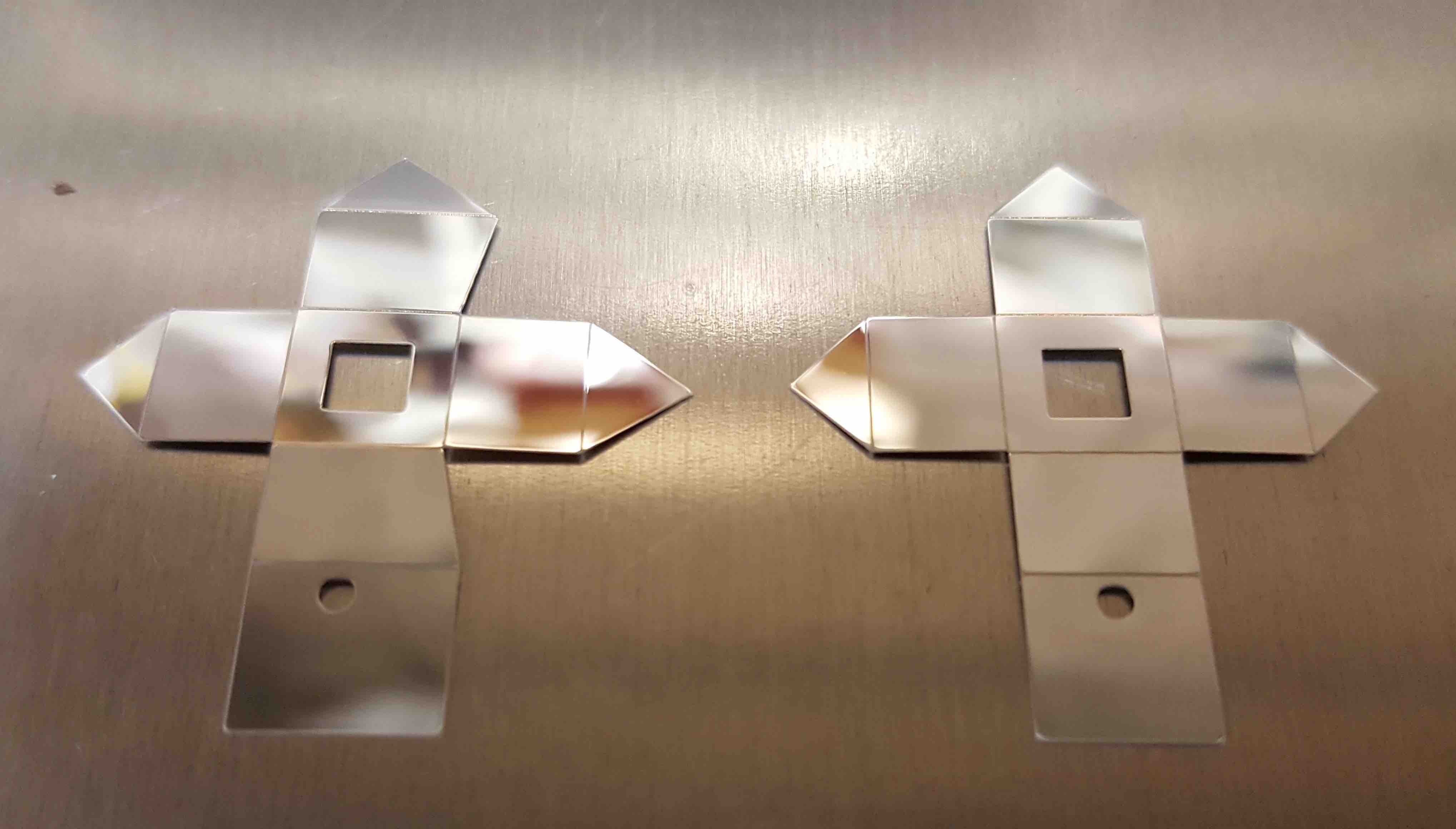}
	\caption{Individual ESR reflective foils for two types of scintillator tiles:(left) edge tile, (right) central tile.}
	\label{fig:Tiels_foils}
\end{figure}

The individual tiles are wrapped with Enhanced Specular Reflector (ESR) foil. In order to increase the light yield and optically isolate the channel, the foil is designed to cover the entire tile except for an opening window of the size of the SiPM surface, as can be seen in \autoref{fig:Tiels_foils}.
Every tile is read out by a \SI[parse-numbers = false]{3 \times 3}{\mm\squared} SiPM with \num{3600} pixels, which is glued to the bottom \SI[parse-numbers = false]{6.3 \times 6.2}{\mm\squared} side of the tile.
The SiPMs are soldered to a printed circuit board (PCB), which is connected via a 
flexible PCB (flexprint) to one of the ASICs on the readout board, the Tile Module Board (TMB).

A tile  \textit{module} is comprised of  \num{13} sub-modules, and contains a total of  \num{416} channels.  A CAD rendering of such a module is shown in~\autoref{fig:Tiles_CAD_module}. The sub-modules are mounted on a water-cooled aluminium support structure and are read out by \num{13} \mutrig ASICs assembled on one TMB, which collects the analog signals of the SiPMs and forwards the digitised signals to the front-end FPGAs. The subsequent data transmission is discussed in \autoref{sec:DAQ}. The heat of the readout chips is dissipated via liquid cooling through a copper tube, with an outer diameter of \SI{2.5}{\mm} and an inner diameter of \SI{2.0}{\mm}, which is placed in a U-shaped groove on the bottom side of the support structure.

\autoref{fig:Tiles_CAD_explode} shows an exploded view of a full tile detector recurl station, which consists of seven modules. The modules are assembled on two endrings, which in turn are mounted on the beam pipe, as shown in \autoref{fig:Tiles_CAD_front}.

\begin{figure}[tb!]
	\centering
	\includegraphics[width=0.48\textwidth]{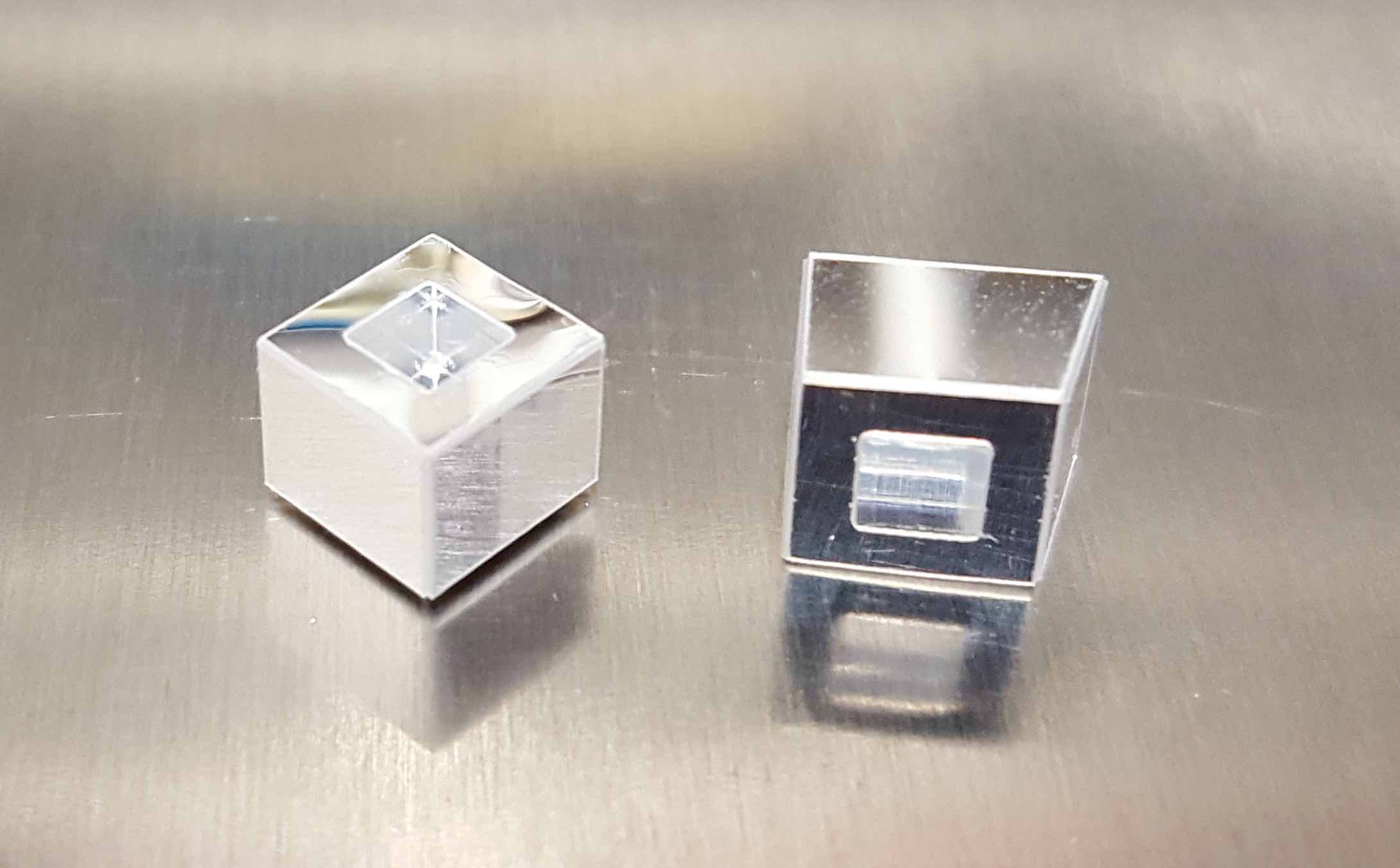}
	\caption{Individual tiles wrapped with ESR reflective foil.}
	\label{fig:Tiles_wrapp}
\end{figure}

Based on previous studies~\cite{Eckert2015}, the best timing resolution is achieved with the plastic scintillator BC418 (equivalent to EJ-228), which has both a high light yield and a fast response time, and therefore is chosen as the baseline material for the tile detector.  This scintillator has a nominal light output of about \num{10200} photons per MeV, a rise time of \SI{0.5}{\ns} and a decay time of \SI{1.4}{\ns}.
The emission spectrum of the scintillator peaks at a wavelength of \SI{391}{\nm}, which roughly matches the maximum spectral sensitivity of the SiPM.
This allows the direct read-out of the scintillation light without the need of an additional wavelength shifter.

\begin{figure}[tb]
	\centering
	\includegraphics[width=0.32\textwidth]{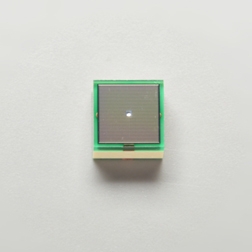}
	\caption{Hamamatsu MPPC S13360-3050VE.}
	\label{fig:MPPC_S13360-3050VE}
\end{figure}

Different SiPM types have been compared in simulation studies in order to find the best suited device for the tile detector.
Based on the simulation studies, a \SI[parse-numbers = false]{3 \times 3}{\mm\squared} SiPM with \SI{50}{\um\square} pixel size is chosen as the baseline photo-sensor.
A respective SiPM from Hamamatsu (MPPC S13360-3050VE, see \autoref{fig:MPPC_S13360-3050VE}) has been successfully tested in the tile detector technical prototype (see \autoref{sec:Tiles_prototype}).

\section{SiPM Radiation Hardness}

Ionising radiation can have a large impact on the SiPM characteristics and performance.
The most prominent effect caused by irradiation is a strong increase in the SiPM dark-rate.
Furthermore, there are several studies (e.g.\ \cite{Qiang:2012zh,SanchezMajos:2009zza}), which have observed a slight decrease in the detection efficiency after exposure of the SiPM to radiation.
A possible explanation for this effect is the progressively larger amount of pixels in a permanent off-state~\cite{SanchezMajos:2009zza}.
Both an increasing dark-rate and a reduced signal amplitude directly influence the time resolution of the sensor.
The exact amount of signal degradation caused by radiation depends on the particle energy and type, as well as the specific SiPM device.

During the data taking period of phase~I of the Mu3e experiment, the SiPMs will be exposed to a total radiation dose of about \SI{e10}{e^+ \per \square\mm}. 
So far, no conclusive experimental data of the SiPM signal degradation is available for the given irradiation dose, particle type and energy.
First studies of the radiation damage in SiPMs using a $^{90}$Sr source indicate that the degradation of time resolution during the Mu3e phase~I is of the order of a few percent.
A study with a collected dose of up to \SI{50}{\percent} of the full phase~I dose showed an increase of the dark rates. Due to the large signals we obtain in our setup, and the ability to set the threshold above the dark noise, the timing resolution was still better than \SI{100}{\pico\second}. It is foreseen to heat up the sensors during no-operation periods for a faster annealing.

\section{Tile Readout}
\label{sec:Tile_Readout}

\begin{figure}
	\centering
	\includegraphics[width=0.48\textwidth]{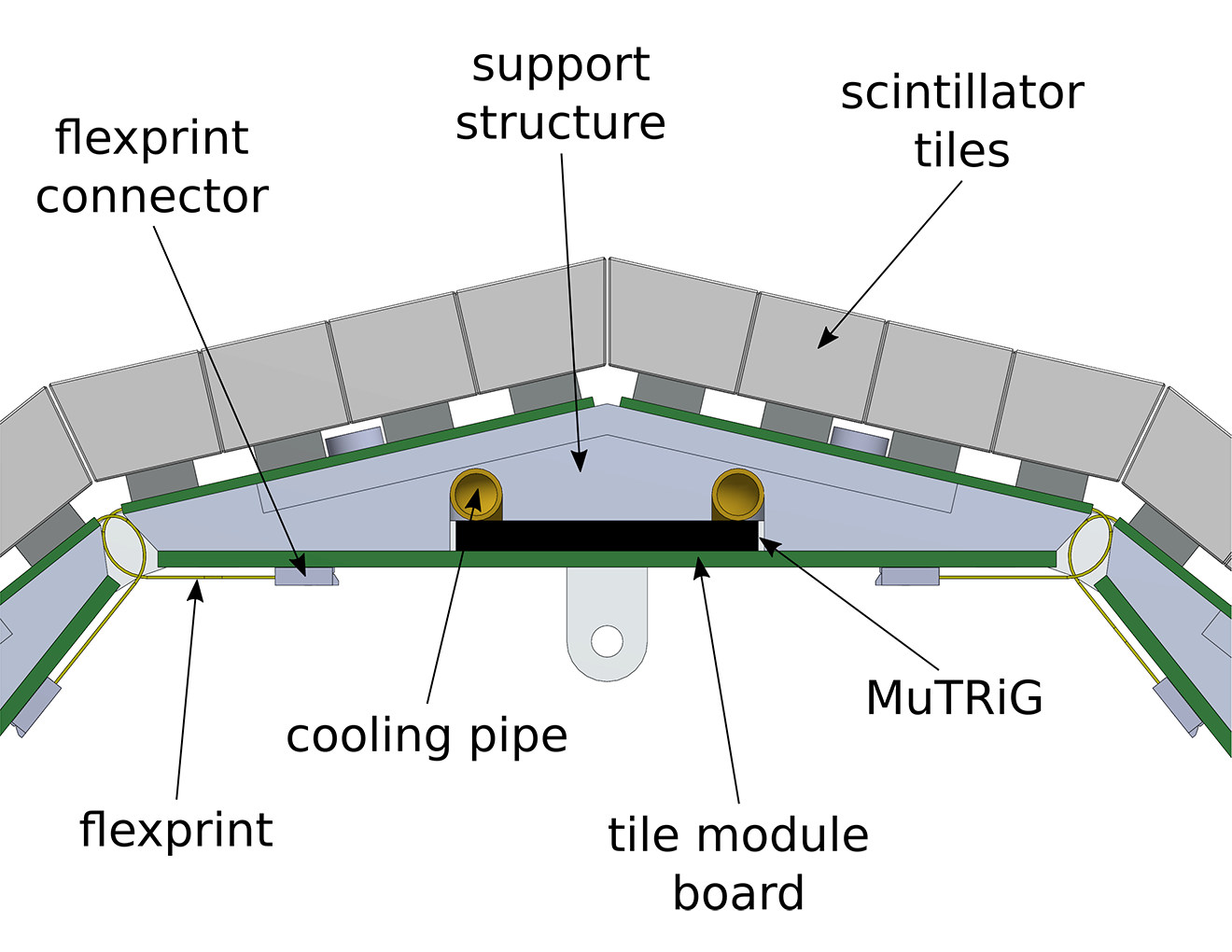}
	\caption{Tile detector with readout electronics. The tile module is divided into a PCB hosting the \mutrig chips and \num{2} $\times$ \num{13} PCBs hosting the SiPMs. The SiPM boards are connected to the \mutrig board via flex cables. The tile readout board is placed on the cooling structure, connecting the tile sub-modules to the front-end FPGA readout board.}
	\label{fig:Tiles_RO_Scheme_B}
\end{figure}

The tile detector will use the same \mutrig ASIC as the fibre detector, see \autoref{sec:Mutrig} for details.
The output signals of 32 tile SiPMs are connected via a flexible printed circuit board to a \mutrig chip. 
The arrangement of the SiPMs and the readout electronics around the cooling structure is shown in \autoref{fig:Tiles_RO_Scheme_B}.
The \mutrig will be operated in two-threshold mode (see \autoref{fig:STiC_Discri}), allowing for time-walk correction. 
The data are then forwarded via the TMB, mounted on the detector module, to the FPGA front-end boards
, see \autoref{sec:DAQ}.

\section{Assembly Tools and Productions steps}
As a first step, the SiPMs are sorted by breakdown voltage and preselected in groups of 32 SiPMs with a spread of the breakdown voltage smaller than 100~mV. This will allow the operation of each sub-module with the same operating High Voltage (HV). 

\begin{figure}[tb]
	\centering
	\includegraphics[trim={0 6cm 0 6cm},clip,width=0.48\textwidth]{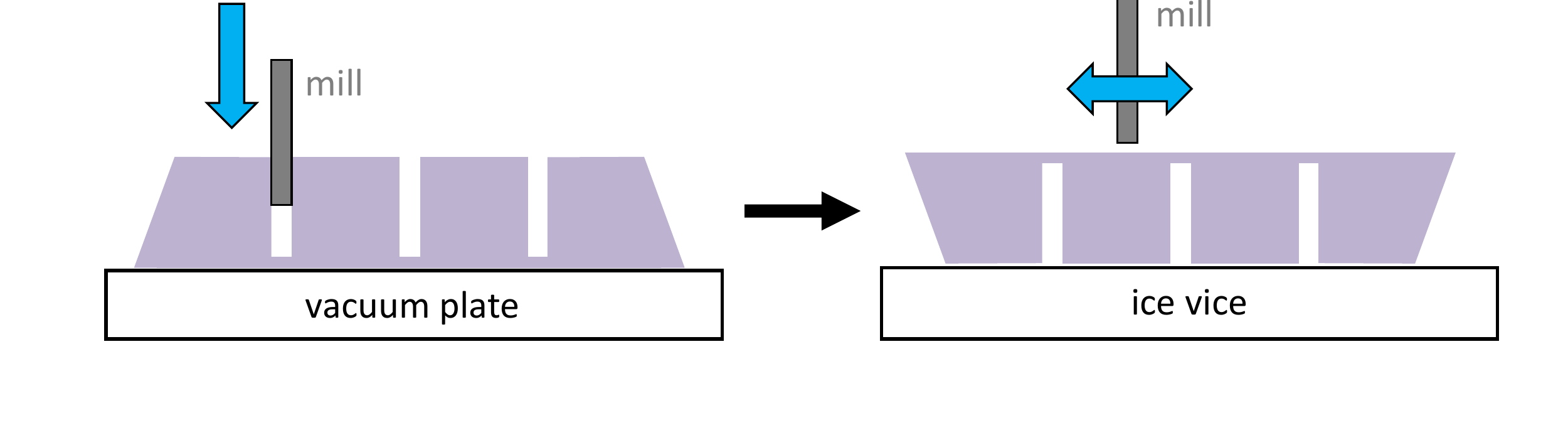}
	\caption{Tile milling procedure: (left) milling the matrix shape on a vacuum plate, (right) flip the matrix, freeze on the ice-vice, and mill from top.}
	\label{fig:Tiels_milling}
\end{figure}

The tiles are manufactured in the Kirchhoff-Institute for Physics in Heidelberg. The scintillator material is mounted on a vacuum plate, where the full matrix is milled from the top, only leaving a 0.5~mm base. The plate is flipped by 180 degrees on to an ice-vice, which freezes the matrix to mill off the base, as sketched in \autoref{fig:Tiels_milling}. Using this method, a relatively fast production rate with very high accuracy of several micrometers is achieved.

\begin{figure}[tb]
	\centering
	\includegraphics[width=0.48\textwidth]{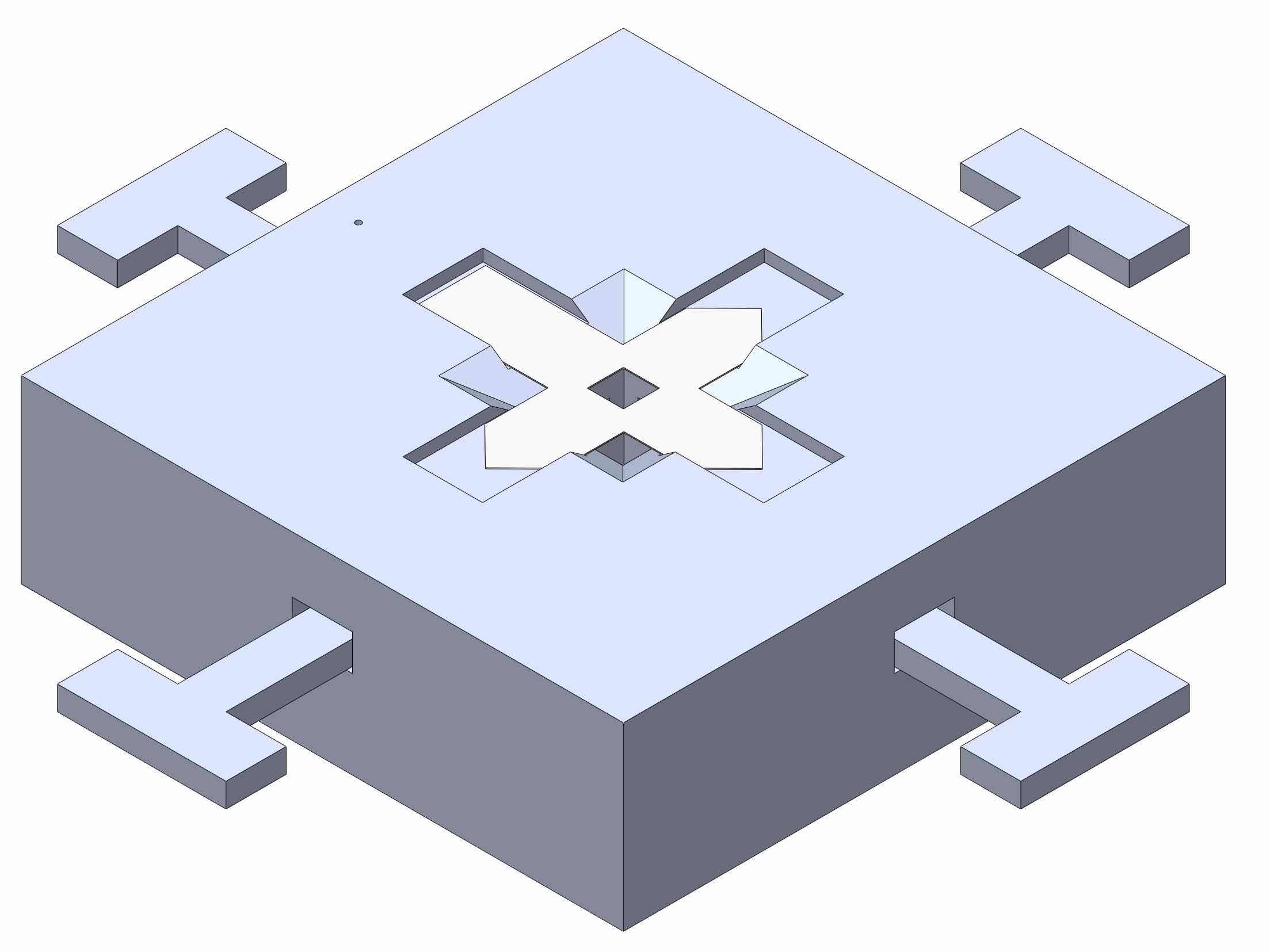}
	\caption{Sketch of the tile wrapping tool with a foil on top.}
	\label{fig:Tiels_wrappingTool}
\end{figure}

After cutting the tiles to the required shape, the tiles' length and width are measured using a digital micrometer before wrapping them with the reflective foil. In order to wrap the tiles, a semi-automatic tool was designed that allows for an easy wrapping of such small tile sizes. A sketch of the wrapping tool is presented in \autoref{fig:Tiels_wrappingTool}. The foil is placed into a dedicated groove on top of the tool; then the tile is placed onto the foil. By pushing the tile down into a customised funnel, the foil side walls are folded around the tile. Using the side rods of the tool, the wrapping is folded like an envelope and a small sticker is placed on top to close it. The resulting wrapped tiles are shown in \autoref{fig:Tiles_wrapp}.

\begin{figure}[tb]
    \centering
    \begin{minipage}{.48\linewidth}
            \begin{subfigure}[t]{.9\linewidth}
                \includegraphics[width=\textwidth]{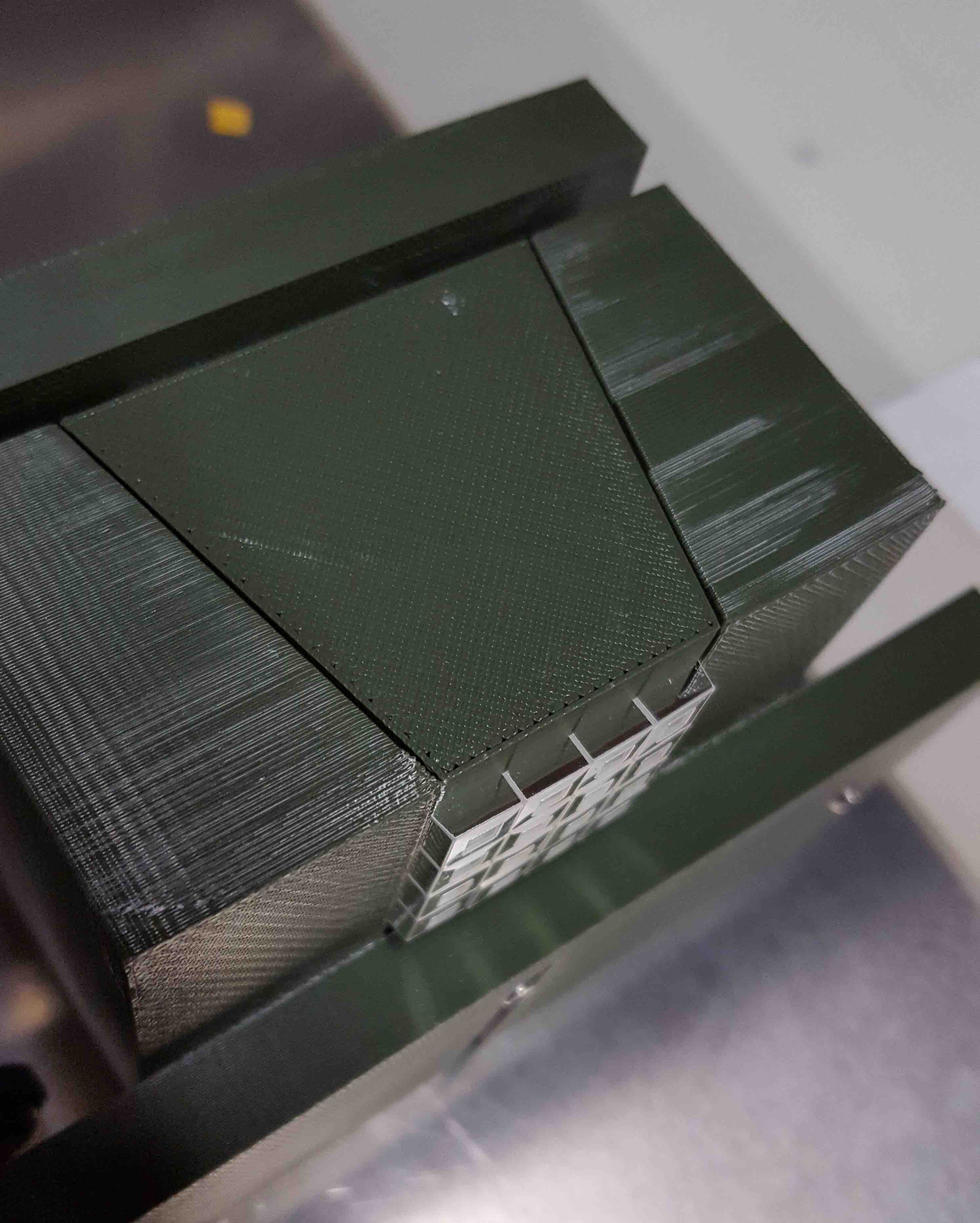}
                \caption{}
                \label{fig:Tiles_glueTest}
            \end{subfigure}
        \end{minipage}
    \begin{minipage}{.5\linewidth}
        \begin{subfigure}[t]{.95\linewidth}
            \includegraphics[width=\textwidth]{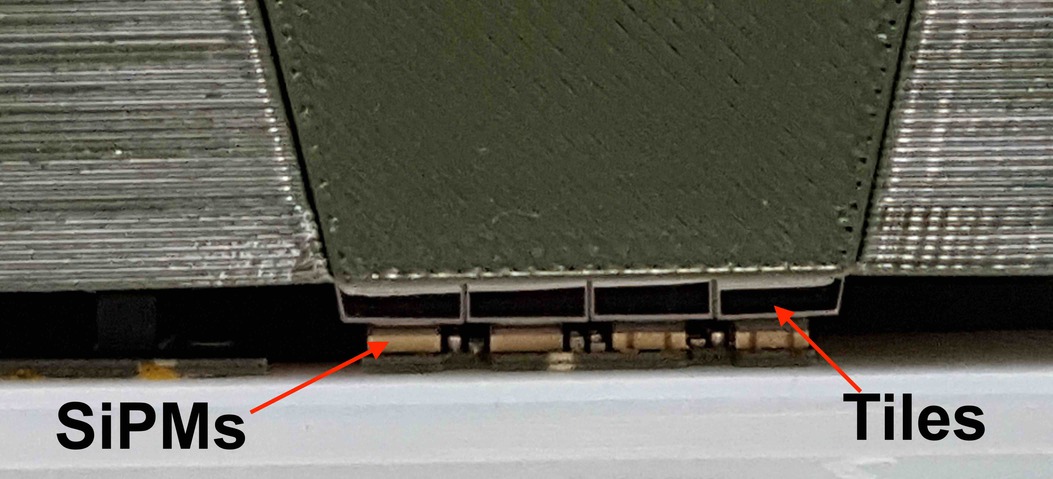}
            \caption{}
            \label{fig:Tiles_glueSiPM}
        \end{subfigure} \\
        \begin{subfigure}[b]{.95\linewidth}
            \includegraphics[width=\textwidth]{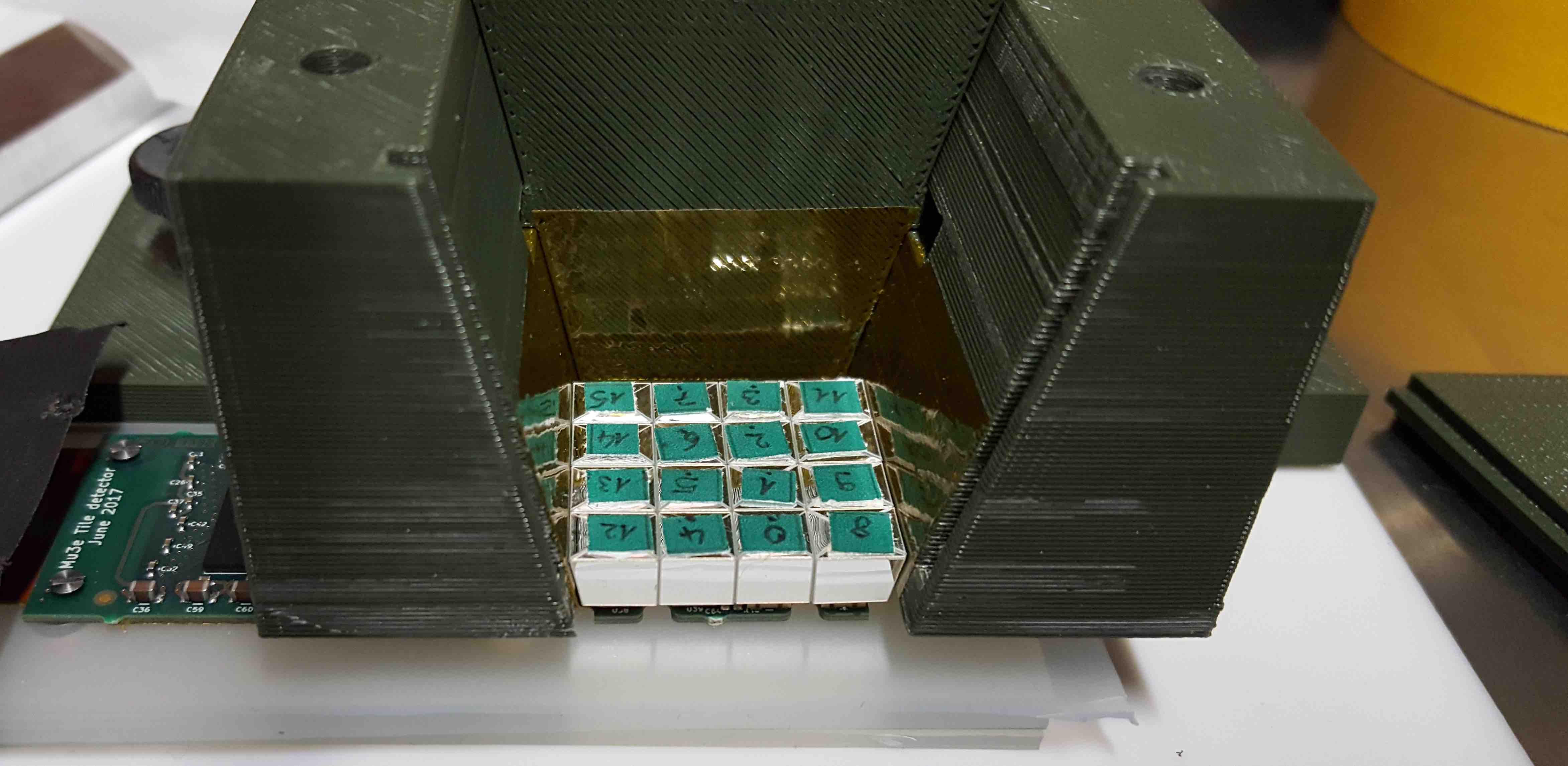}
            \caption{}
            \label{fig:Tiles_glued}
        \end{subfigure} 
    \end{minipage}
    \caption{Gluing tool design to glue a 16 channel matrix on one side of tile matrix board: (a) the tools with 16 tiles pressed before gluing, (b) tiles pressed on SiPMs during the gluing stage, (c) glued tile matrix after curing. }
\end{figure}

In the following step, the tiles must be glued to the SiPMs. This is done on matrix level in order to avoid tolerance issues. A gluing tool was designed with the emphasis of allowing a small degree of freedom with respect to the height of the individual tiles in order to compensate different SiPM heights due to soldering paste and tolerances of the SiPM manufacturing. The scintillator tiles are manually arranged inside the tool and are pressed from the back side and the top such that half of the tiles' height is outside of the tool as shown in \autoref{fig:Tiles_glueTest}. The matrix board is mounted on a pedestal and the glue is dispensed onto the SiPMs. At this stage, the tool is pressed onto the SiPMs as shown in \autoref{fig:Tiles_glueSiPM}, where the $x$-$y$ position is set using alignment pins. After a curing time of 24 hours, the outer wall of the gluing tool is taken out (see \autoref{fig:Tiles_glued}) and the gluing tool can be removed.

\section{Technical Prototype}
\label{sec:Tiles_prototype}
\begin{figure}[tb]
	\centering
	\includegraphics[width=0.48\textwidth]{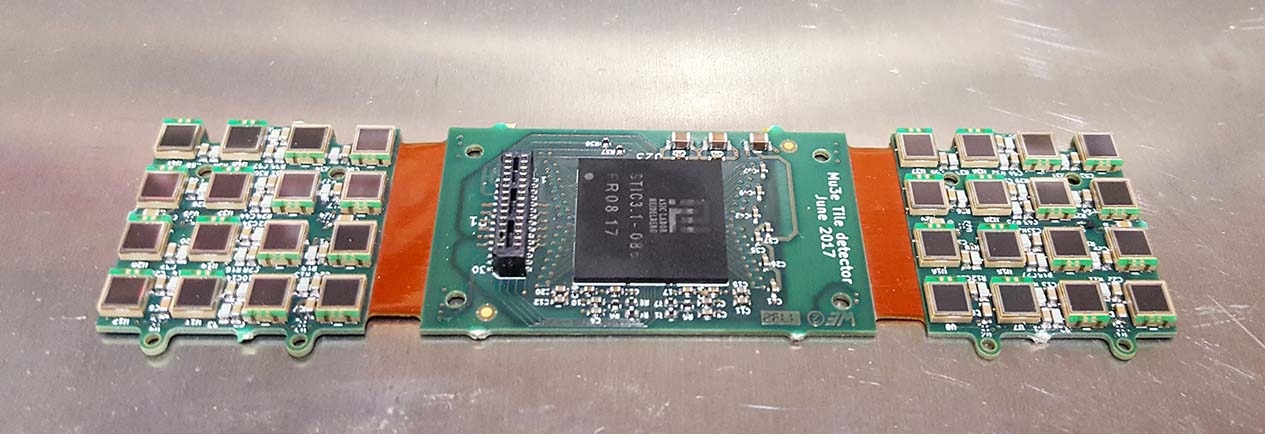}
	\caption{Tile matrix board assembled with SiPMs and a BGA packaged STiC 3.1 ASIC.}
	\label{fig:Tiels_FEB}
\end{figure}

A technical prototype of the tile detector has been developed and tested. The goal of this prototype was to evaluate the detector performance and cooling concept, develop production tools and finalise assembly procedures. This detector has a similar design to the one described in \autoref{sec:TileDetectorDesign}, with a few modifications in the sub-module layout that were done in a later stage based on the experience from this technical prototype.
For this prototype, the endrings, the cooling support structure and the tile matrix readout board were produced. At the time of production, the \mutrig ASIC was not available. Therefore, a BGA packaged STiC 3.1 was used, which has the same front-end as the \mutrig ASIC. In addition, a first version of the TMB, which allows the readout of a full module, was produced. 
In \autoref{fig:Tiels_FEB}, a tile matrix board assembled with SiPMs and a BGA-packaged STiC 3.1 ASIC is shown. 
In this design, eight digital temperature sensors were placed between the SiPMs and used for monitoring.

The scintillator material was manually cut in the workshop of the Kirchhoff-Institute for Physics in Heidelberg, with a tolerance of \SI{180}{\um} for two different tile geometries, as presented in \autoref{fig:Tiles_tile}. The tiles were individually wrapped with ESR reflective foil that was designed in a way to maximise the light yield while at the same time minimising optical cross-talk between the channels. The foils were cut to the desired shape using a laser cutter. For this prototype, an additional hole on the top side was added in order to monitor the gluing quality as shown in \autoref{fig:Tiels_foils}.

\begin{figure}[tb]
	\centering
	\includegraphics[width=0.48\textwidth]{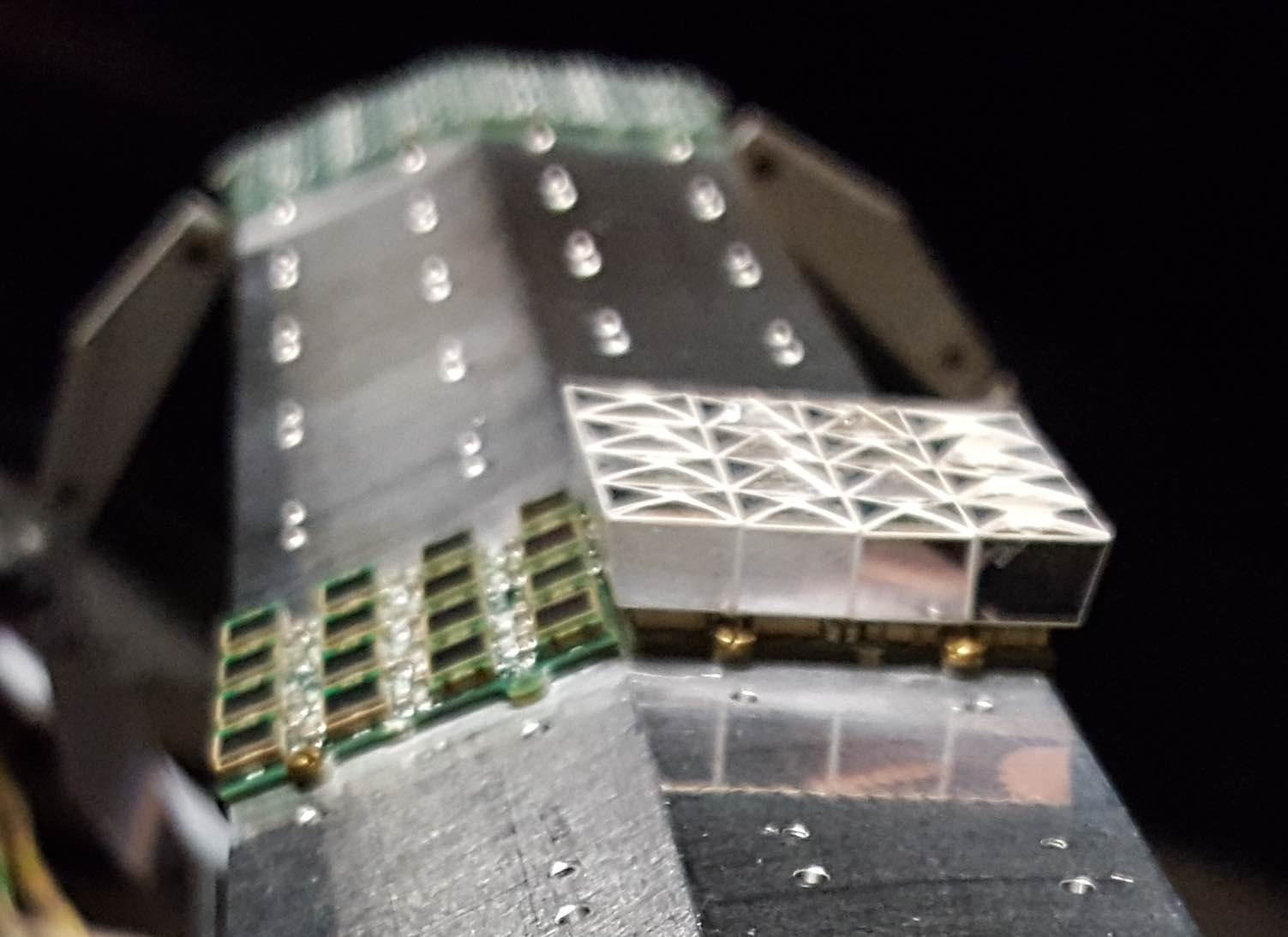}
	\caption{First half-assembled sub-module mounted on the cooling structure.}
	\label{fig:Tiels_submoduleAssemble}
\end{figure}

In total, three sub-modules consisting of 96 channels, were assembled and tested in test-beam conditions. In \autoref{fig:Tiels_submoduleAssemble}, the first half of a sub-module assembled on the cooling structure is presented. 

\subsection{Prototype Performance} 
\begin{figure}[tb]
	\centering
	\includegraphics[width=0.48\textwidth]{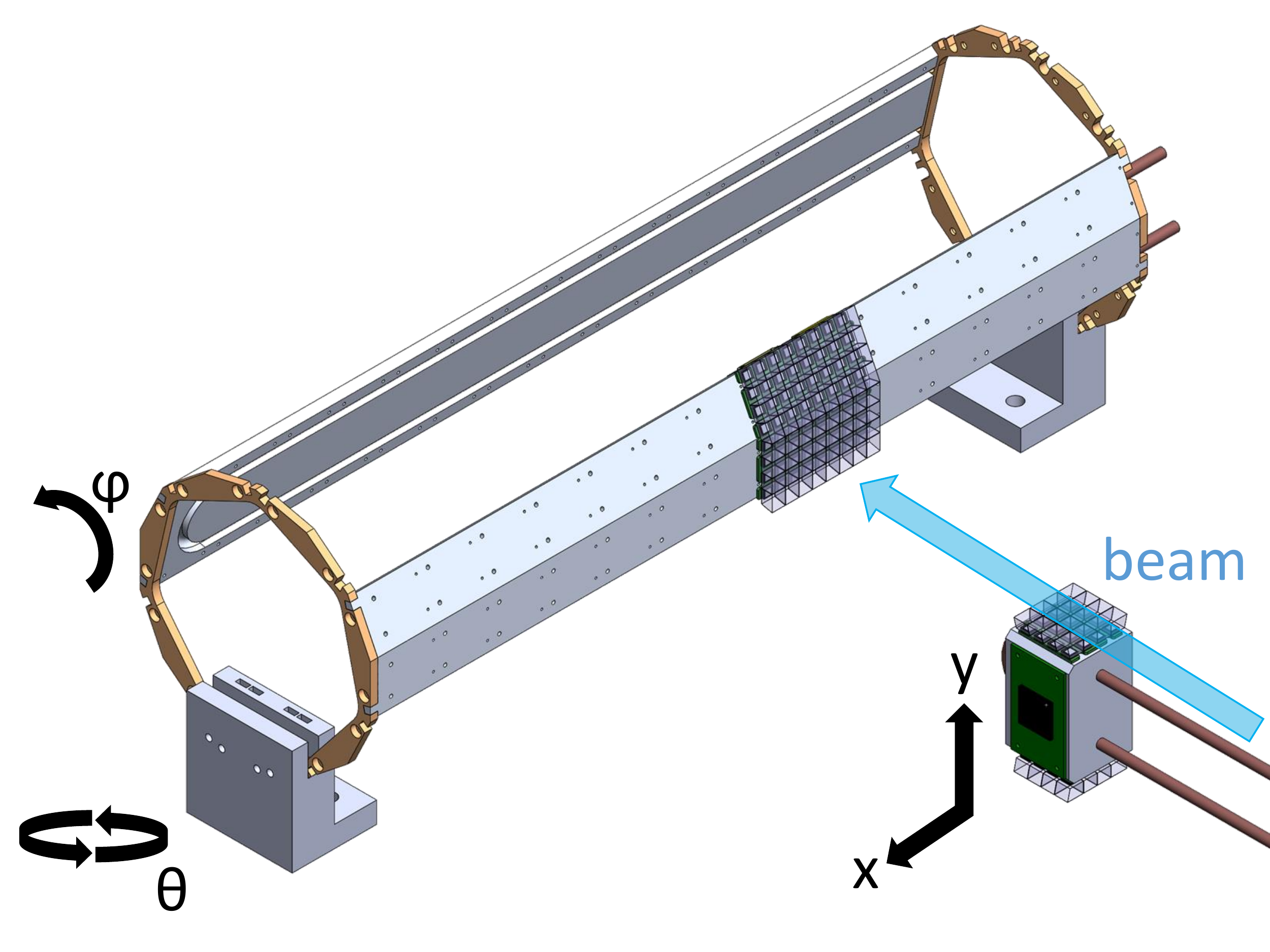}
	\caption{Schematic drawing of the test-beam setup at DESY, which includes three sub-modules.}
	\label{fig:Tiles_testbeamSetup}
\end{figure}

\begin{sloppypar}
In order to evaluate the detector performance, the timing and detection efficiency were measured with an electron beam at the DESY test-beam facility.
A schematic view of the test setup is shown in \autoref{fig:Tiles_testbeamSetup}. For the measurements, one sub-module array of $4\times4$ scintillator tiles was positioned in parallel to the beam and served as a reference, such that the incident particles traversed four tiles in a row for each electron event. The other two sub-modules were assembled on the cooling structures and used as devices under test (DUTs). The devices under test could be rotated in $\phi$  and $\theta$ with respect to the beam and were read out using the prototype TMB board. The reference detector and the prototype TMB board were connected with 50~cm cables to a test FPGA board, which merged the data from the three ASICs.
During the test-beam, both the reference matrix and all the channels of DUT$_0$ were calibrated, while for DUT$_1$ only a single row was optimised. 
\end{sloppypar}

\begin{figure}
	\centering
	\includegraphics[width=0.48\textwidth]{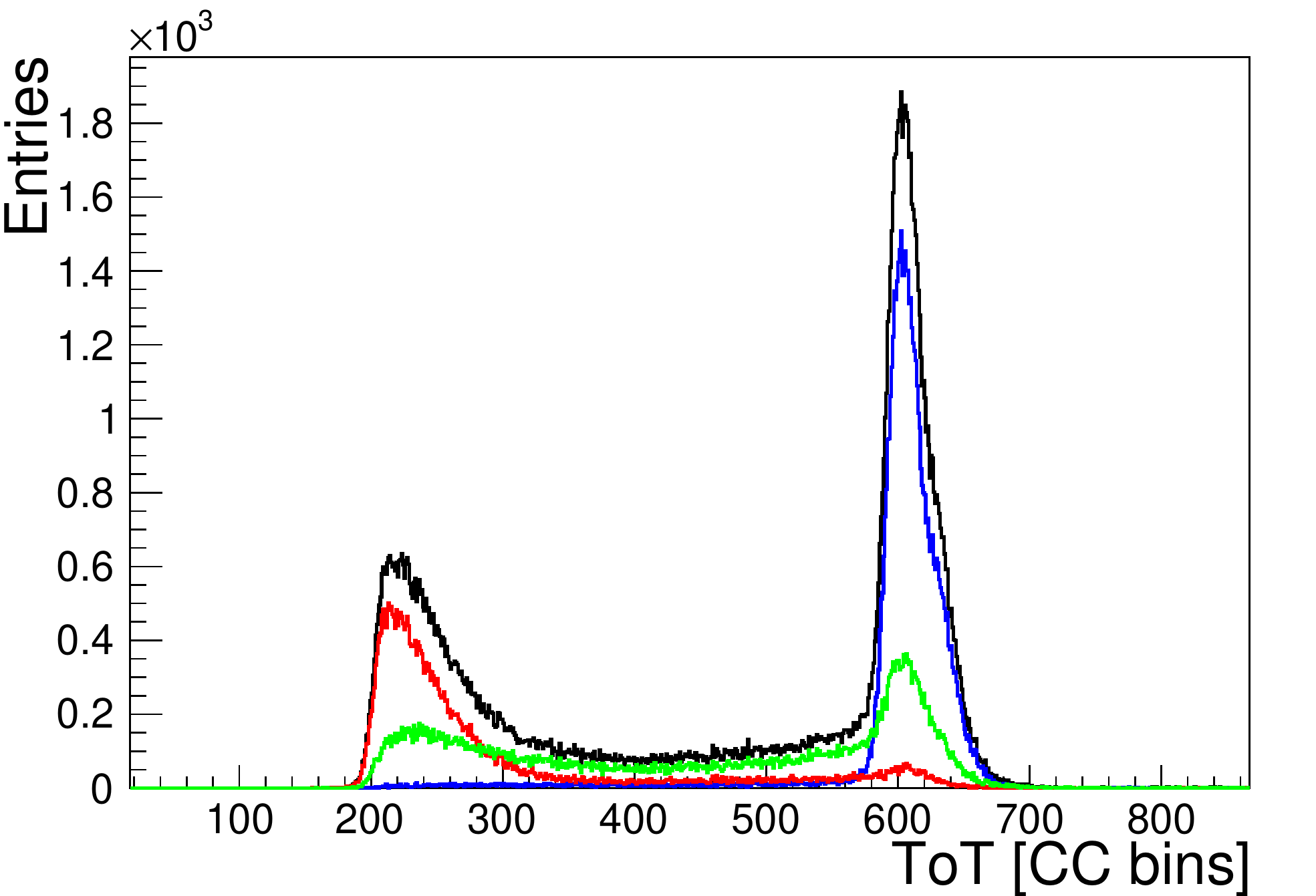}
	\caption{Energy deposition in a scintillator tile (black). The spectrum is composed of the Landau peak (blue), a plateau arising from edge effects (particles grazing the tile, green) and a peak from optical cross-talk (red).}
	\label{fig:Tiles_TypicalToT}
\end{figure}

\autoref{fig:Tiles_TypicalToT} shows a typical time-over-threshold (ToT) spectrum. Several distinct features are visible: The most prominent feature is the peak at a ToT of about 610~coarse counter (CC) bins, in the following referred to as Landau peak. This peak originates from electrons which fully traverse the tile. The second peak at a ToT of 210~CC~bins originates from cross-talk between neighbouring scintillator tiles. This can be shown by selecting hits where at least one direct neighbour in the rows above or below the selected tile has a large signal with an energy deposition in the Landau peak. The corresponding events are shown by the red curve in \autoref{fig:Tiles_TypicalToT}.
The green line refers to a plateau arising from edge effects, where particles pass only partially through the tile. The large plateau and gap between the two peaks indicate the excellent light collection and low optical cross-talk between tiles, which shows the benefit of the individual tile wrapping.  

\subsection{Detection Efficiency}

The detection efficiency is determined using electrons which traverse all four tiles of a certain row.
Such events are selected by requiring at least three hits in the row, two of which must be in the first and last tile.
The detection efficiency $\varepsilon$ is then given by the probability to detect a hit in the remaining channel of the row with an energy deposition above the cross-talk level.

\begin{figure}
	\centering
	\includegraphics[width=0.48\textwidth]{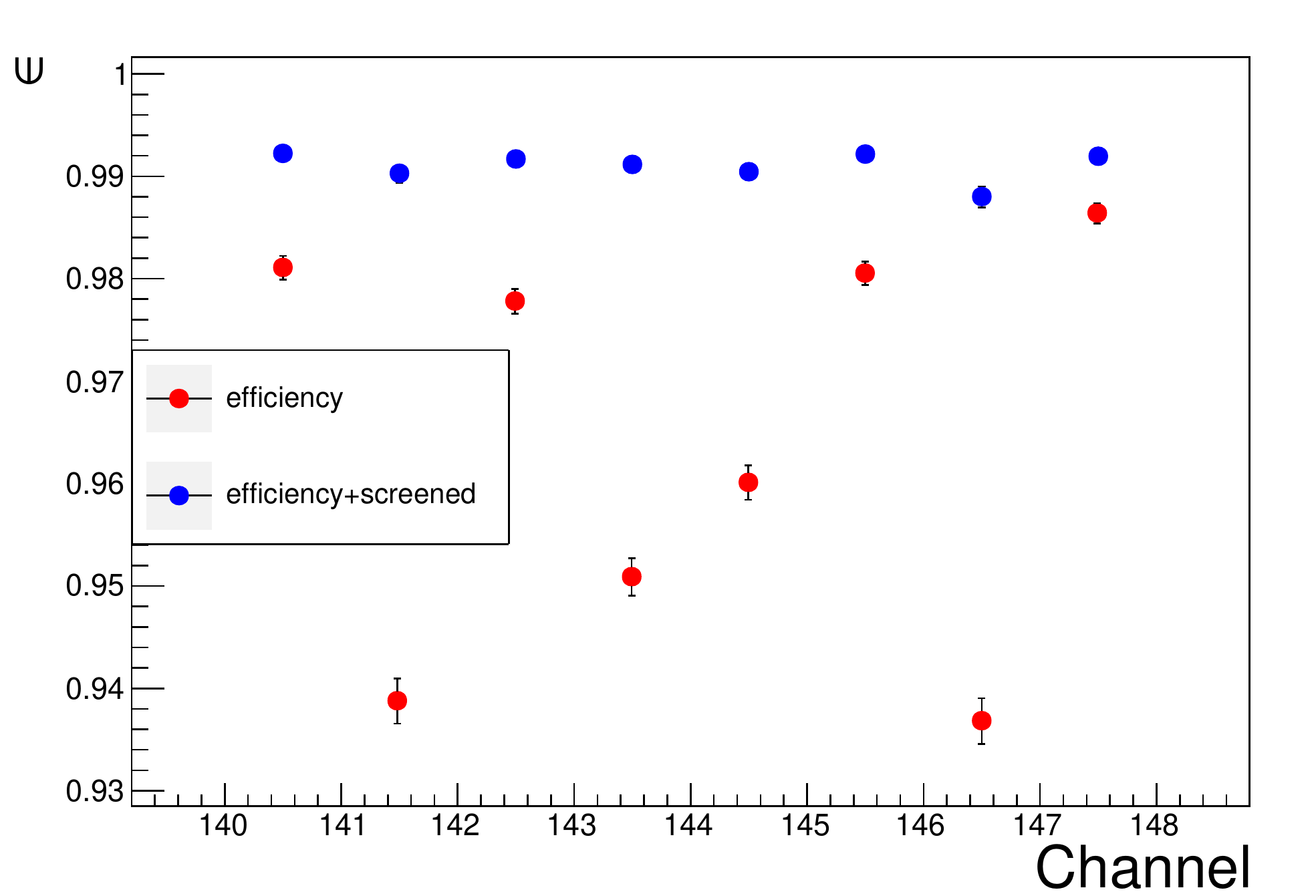}
	\caption{Efficiency calculated for the reference sub-module before and after the correction of the screening effect.}
	\label{fig:Tiles_efficiency}
\end{figure}

Due to the large light yield, which guarantees the signal to be well above the detection threshold, the efficiency is expected to be $\varepsilon \approx \SI{100}{\percent}$.
Experimentally, an efficiency between $\varepsilon = \SI{93.8}{\percent}$ and $\varepsilon = \SI{98.7}{\percent}$ is achieved, see~\autoref{fig:Tiles_efficiency}. In a small fraction of the events, a hit prior to the expected event was observed, which screens the expected hit due to a channel being busy, thus causing an inefficiency in the channel. Correcting for this screening effect leads to an efficiency above $\SI{99}{\percent}$. By changing the ASIC configuration, this busy time can be further reduced by a factor of three, which will reduce the screening effect and increase the efficiency. 
The remaining inefficiency can presumably be attributed to edge effects and misalignment of the tiles and inefficiency of data acquisition. For a better efficiency estimation, it is planned to repeat the measurement using a tracker.    

\subsection{Time Resolution}
The detector was optimised for timing measurements by fine-tuning the SiPM bias voltage and the timing thresholds. In order to evaluate the time resolution, a channel-to-channel time delay calibration was performed. These time delays are arising from different path lengths of the signal lines on the PCB and can vary up to 600~ps.
When measuring the timing using threshold discrimination, an additional time delay caused by time walk effects needs to be corrected.
For this correction, a tight ToT cut is applied on the reference channels in order to minimize time walk effects from the reference side.

In order to estimate the time resolution of a single channel, coincidence time distributions between at least three channels are used.
The channel time resolution can then be extracted by:
\begin{linenomath*}
\begin{equation}
\label{eq:resolution}
\begin{aligned}
\sigma^2_{i,3}=\sigma^2_{i}+\sigma^2_{3},  \quad i=1,2 \\
\sigma^2_{1,2}=\sigma^2_{1}+\sigma^2_{2} \\
\sigma=\sigma_{3}=\frac{1}{\sqrt{2}}\sqrt{\sigma^2_{3,1}+\sigma^2_{3,2}-\sigma^2_{1,2}}
\end{aligned}
\end{equation}
\end{linenomath*}

where $\sigma^2_{1,2}$, $\sigma^2_{1,3}$ and $\sigma^2_{2,3}$ are the three widths extracted from the coincidence time distribution between different pairs of channels.

\begin{figure}
	\centering
	\includegraphics[width=0.48\textwidth]{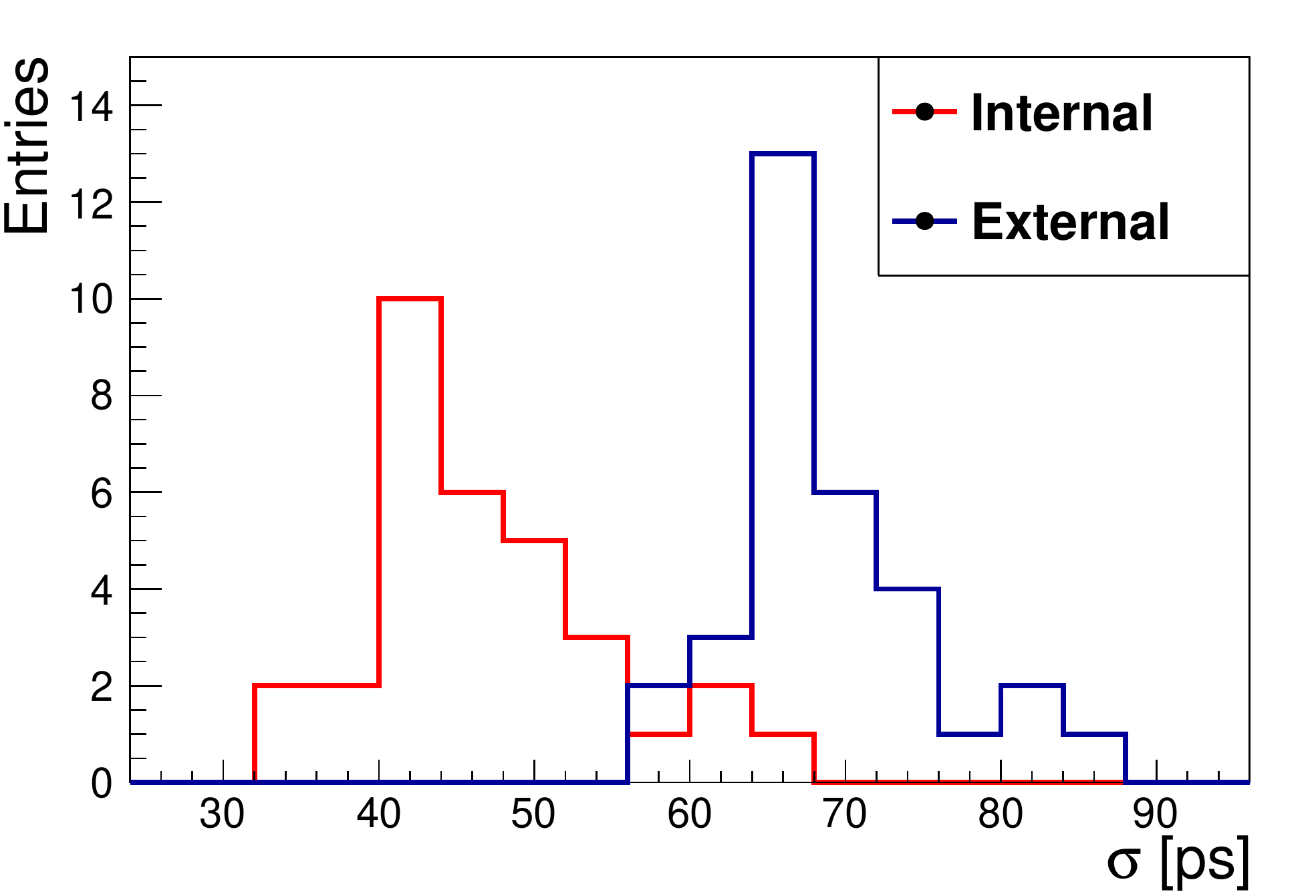}
	\caption{DUT channel resolution: (red) internal, (blue) external.}
	\label{fig:Tiles_InterResolution}
\end{figure}

The internal channel resolution, calculated with \autoref{eq:resolution} using channels in the same sub-module, is presented in \autoref{fig:Tiles_InterResolution} for the DUTs. For these results, runs with tracks parallel to the DUTs are used in order to have at least three channel hits for the same electron event within a sub-module matrix by requiring events in the Landau distribution.
A similar average resolution was measured both for the reference sub-module and for the two DUTs, where the average time resolution measured is 46.8~$\pm$~7.6~ps. However, when repeating the same calculation using channels from different sub-modules, an additional jitter between the sub-modules is observed.
The extra jitter between the reference sub-module and the DUTs of 45.5~$\pm$~3.2~ps leads to a worse time resolution as shown in \autoref{fig:Tiles_InterResolution} (blue). The main contribution to this arises from non-optimal design of the test board used for the read out of all sub-modules.  

\begin{figure}[hbt!]
    \centering
    \begin{minipage}{\linewidth}
            \begin{subfigure}[t]{\linewidth}
                \includegraphics[width=\textwidth]{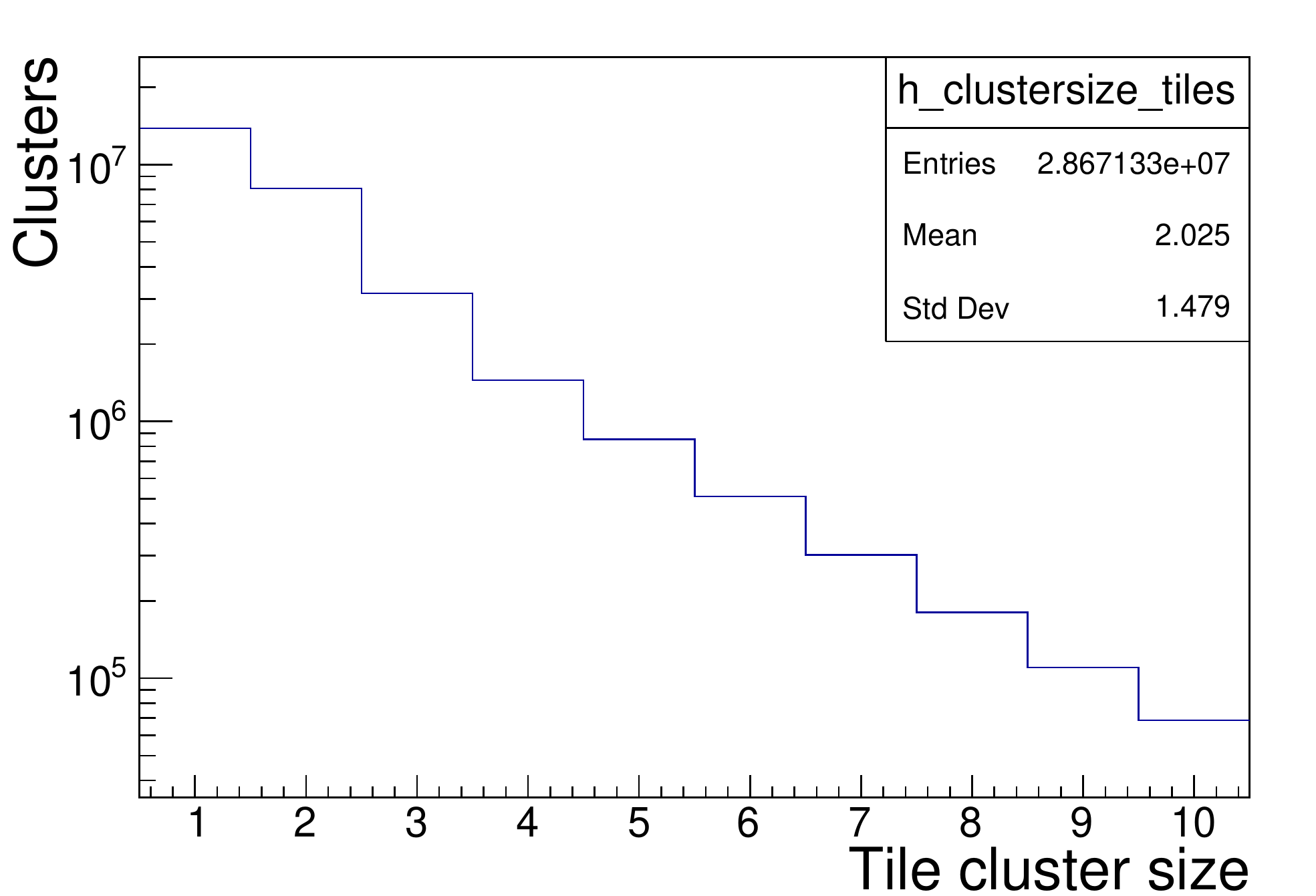}
                \caption{}
                \label{fig:Tiles_simuMulti}
            \end{subfigure}\\
        \end{minipage}
    \begin{minipage}{\linewidth}
        \begin{subfigure}[t]{\linewidth}
            \includegraphics[width=\textwidth]{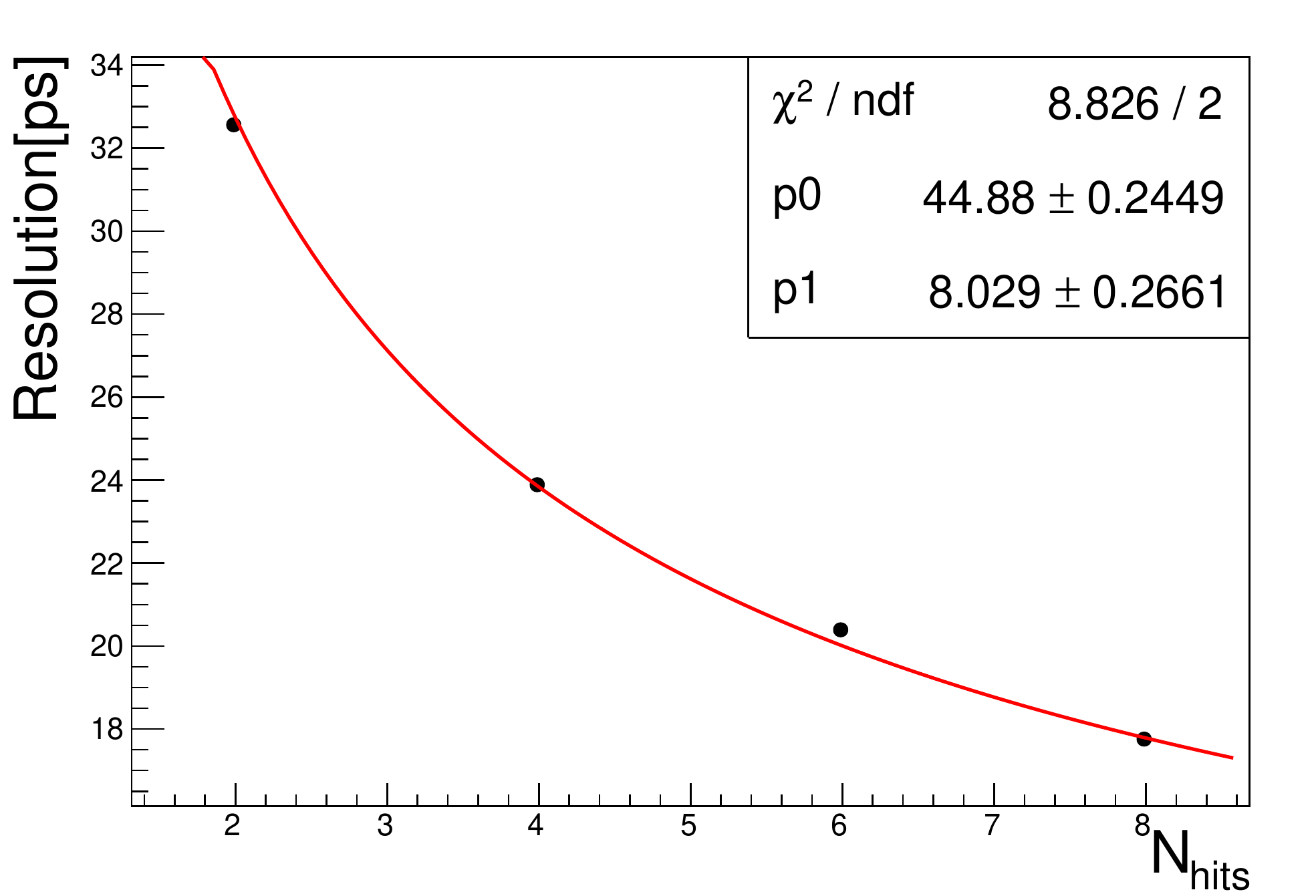}
            \caption{}
            \label{fig:Tiles_resVsHits}
        \end{subfigure}
    \end{minipage}
    \caption{Cluster size impact on time resolution: (a) Simulated phase~I cluster size per track. (b) Measured time resolution as a function of number of hits using the even-odd analysis.}
\end{figure}

The expected event multiplicity during phase~I of the experiment is presented in \autoref{fig:Tiles_simuMulti}. While the average cluster size is $\approx$2, also cluster sizes higher than 9 can be observed. 
In order to evaluate the time resolution as a function of cluster size, a run with beam parallel to the DUTs, where the electron can pass through up to four channels in each DUT, is used.

The time resolution is evaluated using an even-odd analysis. For a given electron track, all hits are grouped into 'odd' or 'even' based on their channel position and the time difference is defined by:
\begin{linenomath*}
\begin{equation}
\label{eq:evenOdd}
\Delta t_{even-odd}(N_{hits}) =\frac{1}{N_{hits}}\left\lbrace \sum_{i=1}^{N_{even}} t_{2i} -\sum_{i=1}^{N_{odd}} t_{2i-1} \right\rbrace
\end{equation}
\end{linenomath*}
where $N_{hits}$ is the cluster size. In order to avoid the additional jitter between the two DUTs, the sums in \autoref{eq:evenOdd} can be arranged such that the subtraction is only done within a sub-module, which leads to a requirement for an even total number of hits within each sub-module.
In \autoref{fig:Tiles_resVsHits}, the result for the even-odd analysis is shown. 
The resolution as a function of cluster size is extracted from \autoref{fig:Tiles_resVsHits} by fitting it with the following function:
\begin{linenomath*}
\begin{equation}
\label{eq:resHits}
\sigma_{t}(N_{hits})=\sigma^{single}_{t}/\sqrt{N_{hit}}\oplus \sigma_{t}^{const}
\end{equation}
\end{linenomath*}
where $\sigma_{t}(N_{hits})$ is the time resolution for events with $N_{hits}$, $\sigma^{single}_{t}$ is the time resolution of a single channel, and $\sigma_{t}^{const}$ is an additional jitter that can be caused by misalignment between the channels.
From the fit, a single channel resolution of $\approx45$~ps is measured, which is in agreement with the value extracted from the single channel measurements, see \autoref{fig:Tiles_InterResolution}. In addition, a small misalignment is also observed. Furthermore, it can be seen that a time resolution better than 20~ps can be reached for events with high multiplicities.

\begin{figure}
	\centering
	\includegraphics[width=0.48\textwidth]{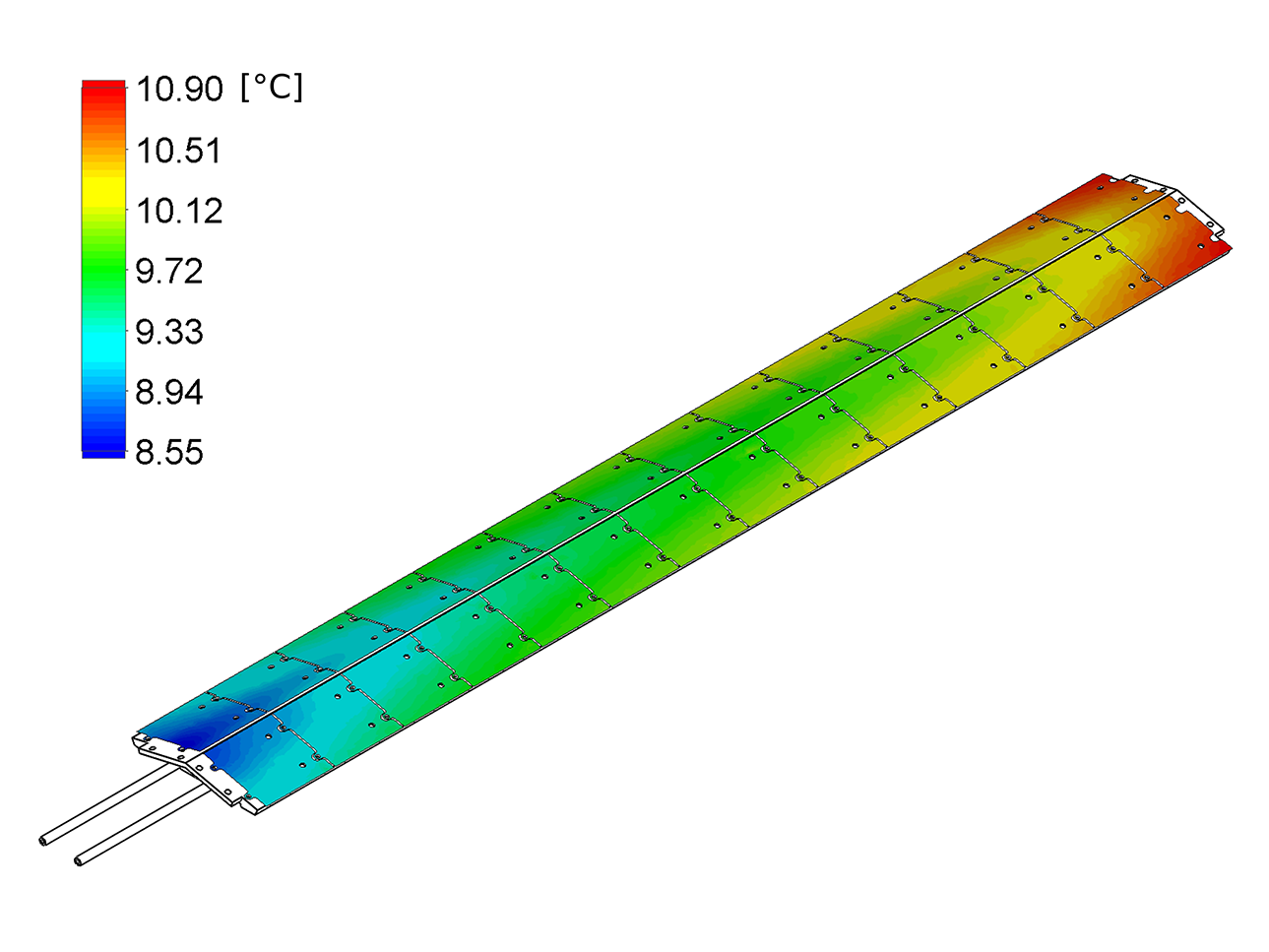}
	\caption{Simulated temperature of the SiPM PCBs. The temperature of the cooling water was set to $\SI{1}{\celsius}$ at a flow speed of $\SI{1}{\meter \per \second}$, while the environment temperature was set to $\SI{50}{\celsius}$.}
	\label{fig:thermalSimSiPM}
\end{figure}

\section{Cooling Simulation of the Tile Detector}
To study the feasibility of the cooling system, thermal simulations were performed using the CAD implementation of the technical prototype,  while in parallel, several measurements of the prototype in the laboratory environment were undertaken. After calibrating the simulation settings to the laboratory conditions, it was shown that the measurements can be reproduced in the simulation~\cite{KLINGENMEYER:2019162852}. The simulation was therefore modified to investigate the cooling performance of a full module operating at the \mutrig working power consumption of \SI{1.2}{W}. Furthermore, the temperature of the water was adjusted to $\SI{1}{\celsius}$ to be closer to the operating conditions foreseen for the tile detector within the experiment, while the environment temperature was increased to $\SI{50}{\celsius}$ in order to subject the system to a stress test. The temperature of the SiPMs and the \mutrig ASICs was investigated under these conditions. In \autoref{fig:thermalSimSiPM}, the temperature of the PCBs on which the SiPMs are assembled is examined. While the temperature on the single PCBs is uniform down to a few tenths of a degree, the temperature range across the full length of the module spans about $\SI{2}{\celsius}$. Considering the SiPM temperature coefficient $\Delta T_{Vop} = \SI{54}{\milli\volt \per \celsius}$, these differences can be compensated by adjusting the high voltage of the individual SiPMs. Overall, the temperature is clearly reduced compared to the environment temperature, demonstrating the influence of the cooling system. Furthermore, the maximum temperature of the ASICs can be extracted from the simulation as $\approx \SI{42}{\celsius}$. This is still well within the safe margin of operation.


\chapter{Cooling infrastructure}
\label{sec:Cooling}

\nobalance

\chapterresponsible{Frank}

\noindent
The detectors, their electronics, the power converters and the data acquisition systems are located inside the densely spaced Mu3e magnet. The heat they produce is transferred to the outside by forced convection cooling. Except for the pixel sensor chips, we are using water cooling everywhere. For the pixels, a novel gaseous helium cooling has been developed.

\section{Water cooling}
\label{sec:WaterCooling}
\begin{figure*}
  \centering
  \includegraphics[width=0.70\textwidth]{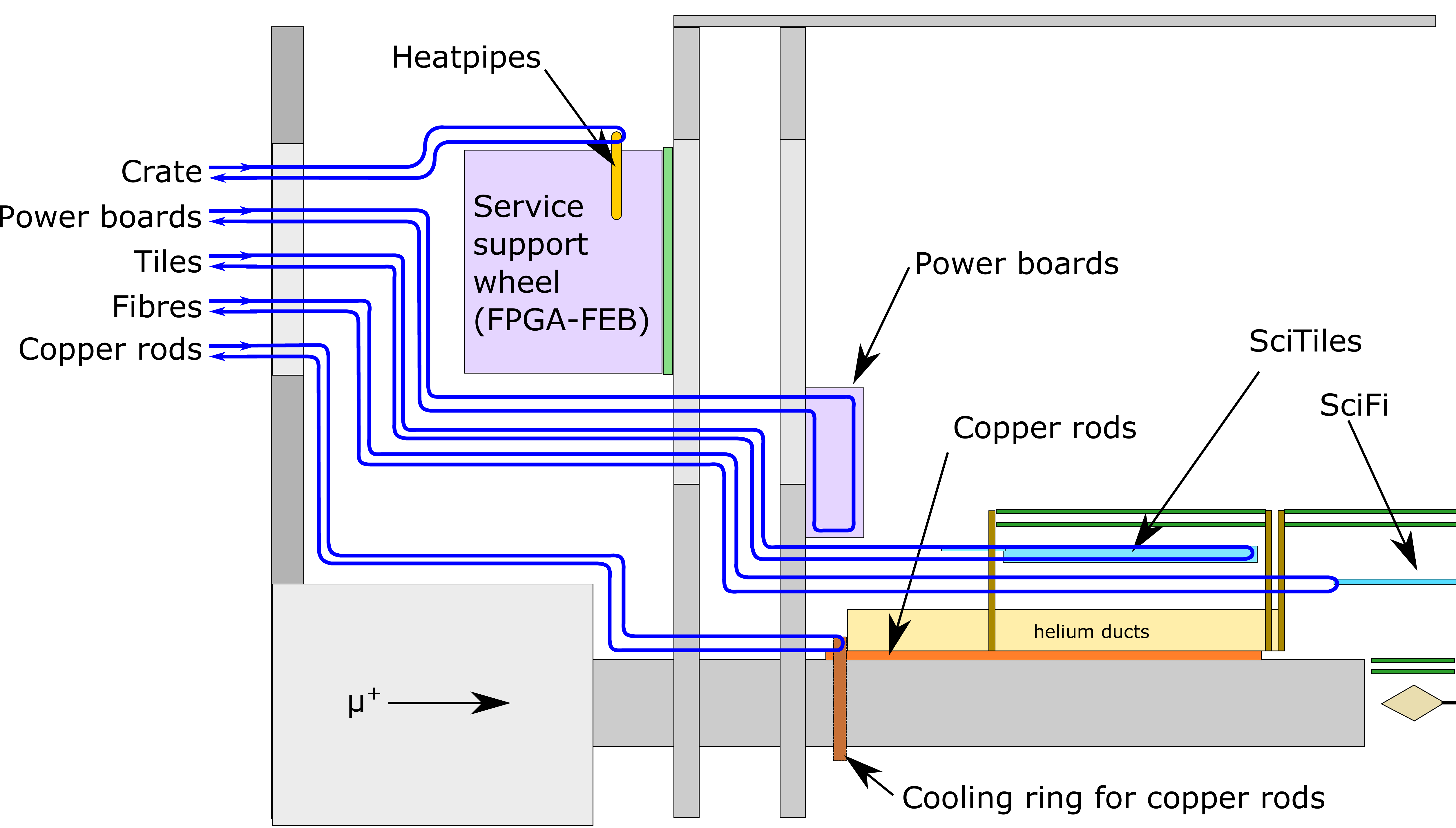}
  \caption{Schematic view of the water cooling topology for one quadrant of the experiment inside the magnet.}
  \label{fig:watercooling_scheme}
\end{figure*}

Water cooling is used to cool all the front-end electronics which
are located outside the active volume of the detector, i.e.~the front-end ASICs of the
timing systems, the front-end FPGA-boards, the DC-DC converters, voltage regulators, etc.
The anticipated heat load per source is listed in \autoref{tab:waterCooledSystems} and totals to
about \SI{5}{\kilo\watt}. To protect the detector from
ice buildup, the water inlet temperature is required to be above $\SI{2}{\degreeCelsius}$, although the helium atmosphere provides a dry environment with a dew point below \SI{-40}{\celsius}.
Pipe systems inside the experiment distribute the water to the heat sinks, see \autoref{fig:watercooling_scheme}. The FPGA boards are cooled via a manifold embedded
into the circularly shaped crates. The low-voltage DC-DC converters for the pixel powering are directly
connected to a cooling loop. Heat dissipation for the  power distribution between
the DC-DC converters and the front-end electronics (\mutrig{} and \mupix{} ASICs) is a potential issue for the copper rods around
the beam pipe. Due to this issue active cooling of them is provided through a dedicated cooling ring thermally
coupled to the rods. The timing detectors have their own cooling loops to dissipate the heat
from the front-end ASIC and to keep the SiPM at a controlled low temperature.
Further details on detector cooling of the timing systems can be found in \Autoref{sec:Fibre,sec:Tiles}, and on cooling of the FPGA boards inside the crate in \autoref{sec:DAQ}.

Chilled water will be used from the PSI main supply via heat exchangers. Additional chillers are in place for circuits requiring lower set temperatures. The timing detectors will receive their independent chilled water loops for enhanced control of their temperatures.

\begin{table}
\centering
\begin{tabular}{lc}
  \toprule
  System & Est. power \\
         & \si{\watt} \\
  \midrule
  Crate (front end FPGA boards) & 2700 \\ 
  DC-DC converters              & 1500 \\ 
  Copper rods                   &  200 \\ 
  Fibre detector (\mutrig, SiPM) &  120 \\ 
  Tile detector (\mutrig, SiPM)  &  420 \\ 
  \midrule
  Total                         & 4940  \\
  \bottomrule
\end{tabular}
\caption{List of systems requiring water cooling inside the experiment, with a conservative estimate of the heat dissipation. All circuits will be run independently.}
\label{tab:waterCooledSystems}
\end{table}

\begin{figure*}
  \centering
  \includegraphics[angle=90,width=0.9\textwidth]{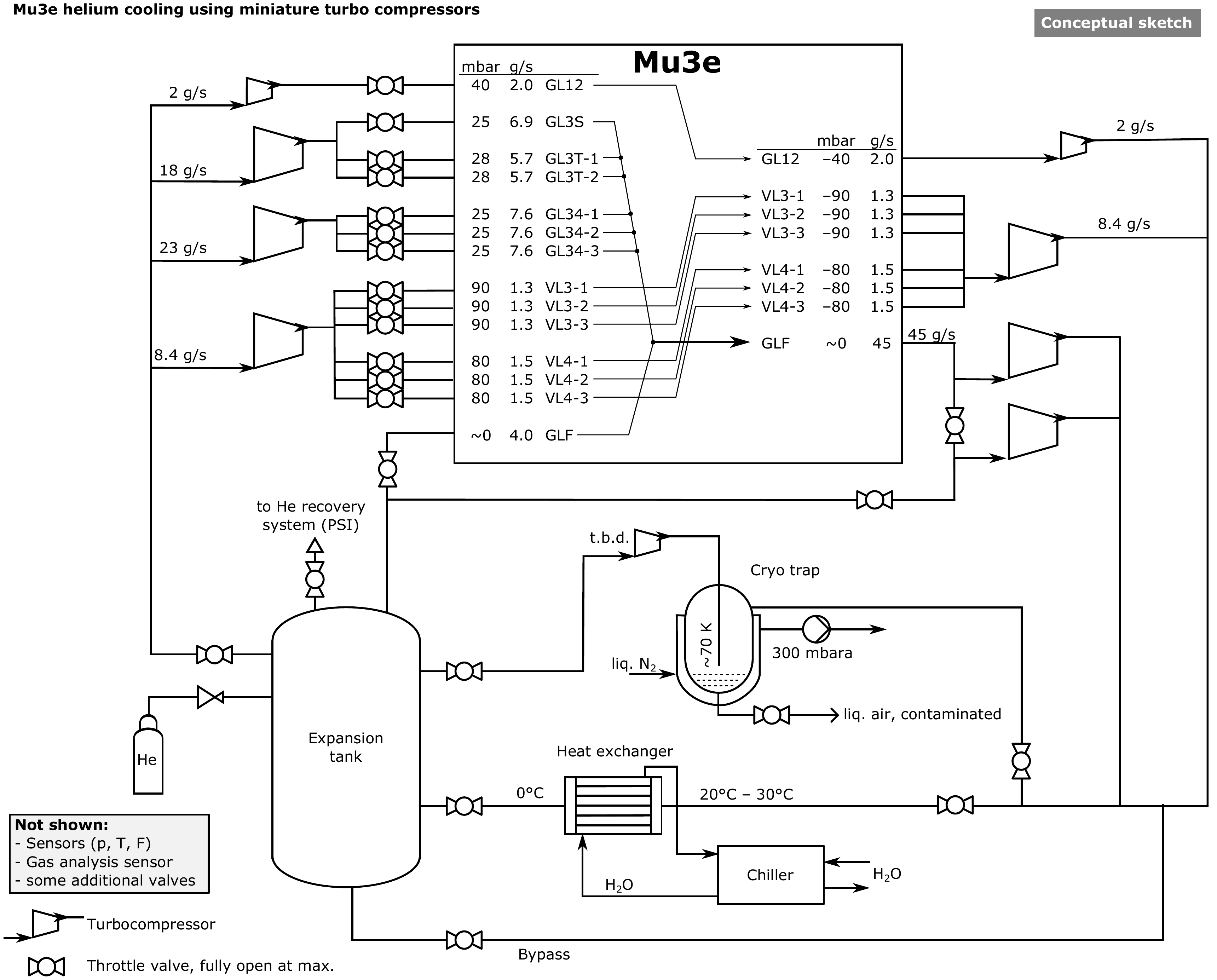}
  \caption{Conceptual process flow diagram of the Mu3e helium cooling infrastructure. Miniature turbo compressors in the circuit may be implemented using multiple units operated in parallel or in series, depending on needs.}
  \label{fig:Mu3eHe_PFDsketch}
\end{figure*}

\section{Helium cooling}
All \mupix chips of the pixel tracker are cooled by gaseous helium of $T_{\textrm{He},\textrm{in}}\gtrsim \SI{0}{\degreeCelsius}$ at approximately ambient pressure. Assuming a maximum power consumption of the pixel sensors of $\SI{400}{\milli\watt\per\cm\squared}$ the helium gas system is designed for a total heat transfer of $\SI{5.2}{kW}$, which increases the average gas temperature by about $\SI{18}{\degreeCelsius}$.\footnote{The pixel detector consists of 2844~chips (108 in the vertex detector, $3 \times 912$ in the outer layers), giving about \SI{1.14}{\metre\squared} of active instrumented surface ($\SI[parse-numbers = false]{20\times 20}{\milli\metre\squared}$ active area per chip, neglecting the chip periphery) or about \SI{1.3}{\metre\squared} including chip peripheries. The conservative (optimistic) scenario leads to about \SI{5.2}{\kilo\watt} (\SI{3.3}{\kilo\watt}) of dissipated heat. The specific heat capacity of gaseous helium is \SI{5.2}{\kilo\joule\per\kg\per\kelvin}.}
For this, the helium cooling system has to provide a flow of about \SI{20}{\metre\cubed\per\minute} (equal to \SI{56}{\gram\per\second} of helium) under controlled conditions split between several cooling circuits (see \autoref{sec:tracker_cooling}).

A process flow diagram for the helium plant is shown in \autoref{fig:Mu3eHe_PFDsketch}. Helium is pumped using miniature turbo compressors run at turbine speeds of up to \SI{240}{\kilo rpm}. These units provide compression ratios up to $\approx 1.2$ at mass flows in the range up to \SI{25}{\gram\per\second}, depending on supplier and model. The energy consumption of the compressors for the full system is estimated to be around \SI{6}{\kilo\watt} in total. The helium circuits are designed with minimised pressure drops for a most economic system layout. The combination of a compressor and a valve for every circuit allows the control of the mass flow and the pressure differential applied individually. Compact, custom made Venturi tubes will be used to monitor the mass flows of every circuit. Leaks lead to losses and will contaminate the helium with air. In addition, outgassing organic residues from electronic components and adhesives need to be removed. Hence a cold trap is included in a by-pass configuration to keep the helium pure enough. An expansion volume will be present to compensate for the compression and expansion of the gas volume during ramp-up and ramp-down of the gas flows. A low pressure drop shell-and-tube heat exchanger is used to remove the heat from the helium.


\chapter{Mechanical Integration}
\label{sec:MechanicalIntegration}
\nobalance
The detector is maintained at its nominal position inside the magnet by a removable frame called the \emph{detector cage}. 
The cage also carries infrastructure such as crates for the power converters and the front-end FPGA boards, and provides support for all cabling and piping.

\section{Detector Cage and Rail System}
\label{sec:DetectorCage}
\begin{figure}
    \centering
    \includegraphics[width=0.49\textwidth]{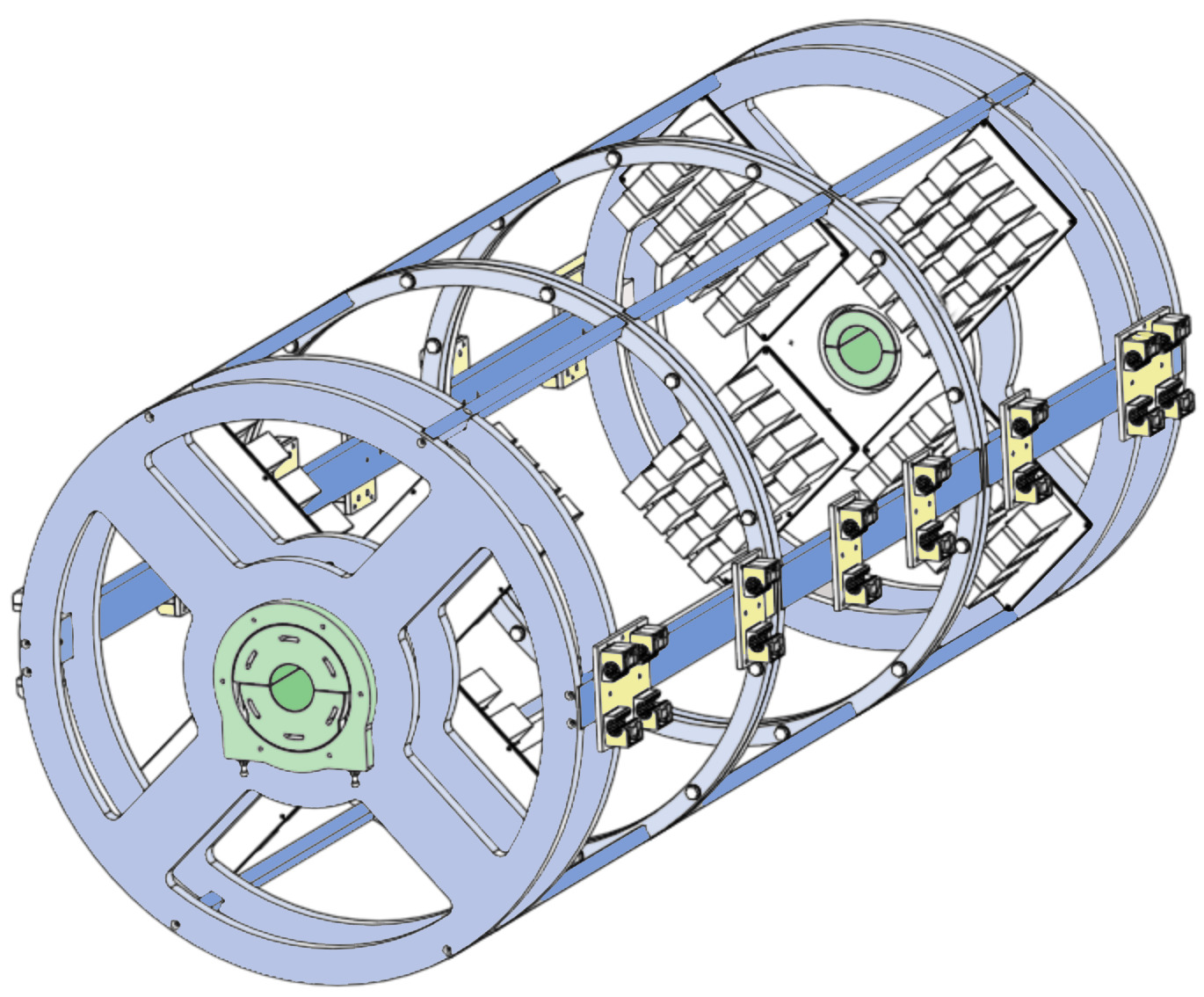}
    \caption{Detector cage structure consisting of two ring frames (light blue) connected by struts (dark blue). The clamps holding the beam pipes are inside the ring pairs at either end (shaded green). The gliders (yellow) allow the cage structure to be moved into the magnet on the rail system.}
    \label{fig:DetectorCageStructure}
\end{figure}

The detector cage has the shape of a hollow cylinder with its axis horizontal, as shown in \autoref{fig:DetectorCageStructure}.
At each end, a ring frame made of pairs of glass-fibre reinforced polymer wheels has a clamp at its centre for the beam pipe. 
Aluminium struts connect the two ring frames and form the cylinder. 
Gliders on the wider struts (at the 3- and 9~o'clock positions) guide the cage on the rail system inside the magnet.

To compensate for possible thermal expansion in the $x$ (horizontal, perpendicular to the beam pipe) direction, the gliders on the left rail are floating whilst on the other rail they are kept at a defined position. 
In the $y$ (vertical) direction the position is defined by the top surface of the rail. The $z$ position is kept fixed by screws.

The clamps in the centre of the rings at either end hold the two beam pipes in position and take all the weight of the detector. 
Mechanisms to fine-adjust the beam pipe pointing angles are built into the clamps. 
Finite element simulations were performed to test the sturdiness of the design. 
Load tests have been carried out on a full-scale mock-up, confirming the simulation results of a deflection of \SI{0.3}{\mm} under a typical detector load of \SI{10}{\kg} at the beam pipe tips. 
The connection of the beam pipes to the beam line is described in \autoref{sec:Area}.

\section{Mechanical support of detector stations}
\label{sec:SupportOfDetectorStations}
\begin{figure*}
	\centering
		\includegraphics[width=0.9\textwidth]{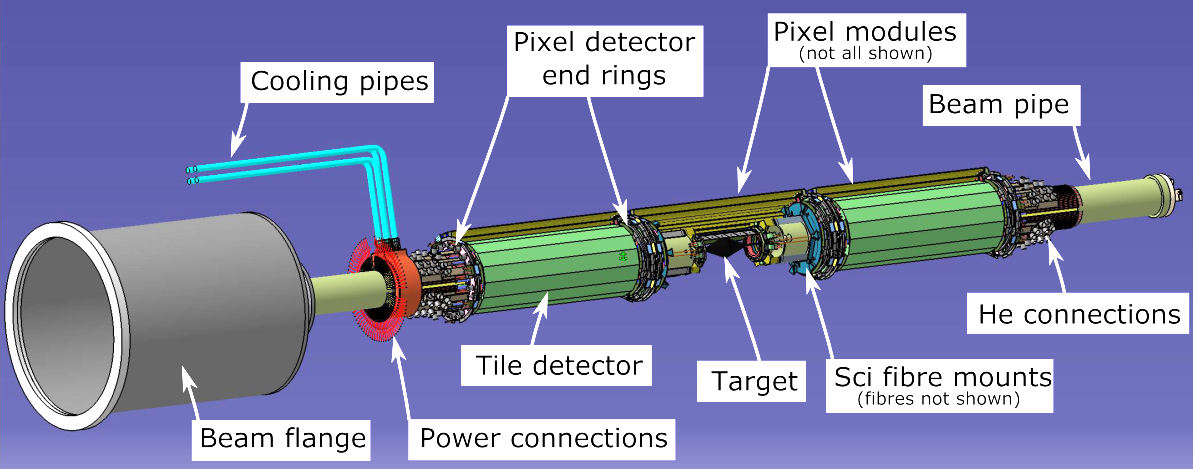}
	\caption{The Mu3e experiment mounted on the beam pipes. Not shown are the detector cage and supplies. Some parts have been partially removed for visibility.}
	\label{fig:Mu3eIntegrationView}
\end{figure*}

The detector components are mounted on the upstream and downstream beam pipe sections, see \autoref{fig:Mu3eIntegrationView}. 
As shown in the previous chapters, both pixel and timing detectors follow a barrel concept. 
They are mounted on pairs of end rings, supported on the beam pipes. 
Whilst the recurl stations have their support on one beam pipe, the central barrel has one mechanical support on the upstream beam-pipe and the other on the downstream beam-pipe. 
To compensate for any tilt of the end rings and movements due to thermal expansion, the detector mounts are spring-loaded at one end.

Detectors can be mounted and dismounted in sequence from inner to outer without the need to retract the beam pipes. 
For example to mount the central barrels, the vertex half-shells of layers 1 and 2 will be installed first, followed by the fibre ribbons. 
Finally, the pixel modules for layers 3 and 4 will be mounted. 
For this sequence, the cage can be placed on a special extraction cart on wheels. 
It has the same rail system as that inside the magnet. 
For better access, the cart has rollers allowing the rotation of the cage around its own $z$-axis in a safe manner.

The beam pipes also provide support for other services. 
The copper bars to supply power to the central detectors (\autoref{sec:Cabling}) are glued onto the beam pipe with a custom procedure to ensure proper electrical insulation, and manifolds for the helium cooling system are attached to the ends of the beam-pipes.

\section{Supply systems and cable routing} \label{sec:mechintSupplyAndRouting}
\begin{figure*}[t!]
    \centering
        \includegraphics[width=0.7\textwidth]{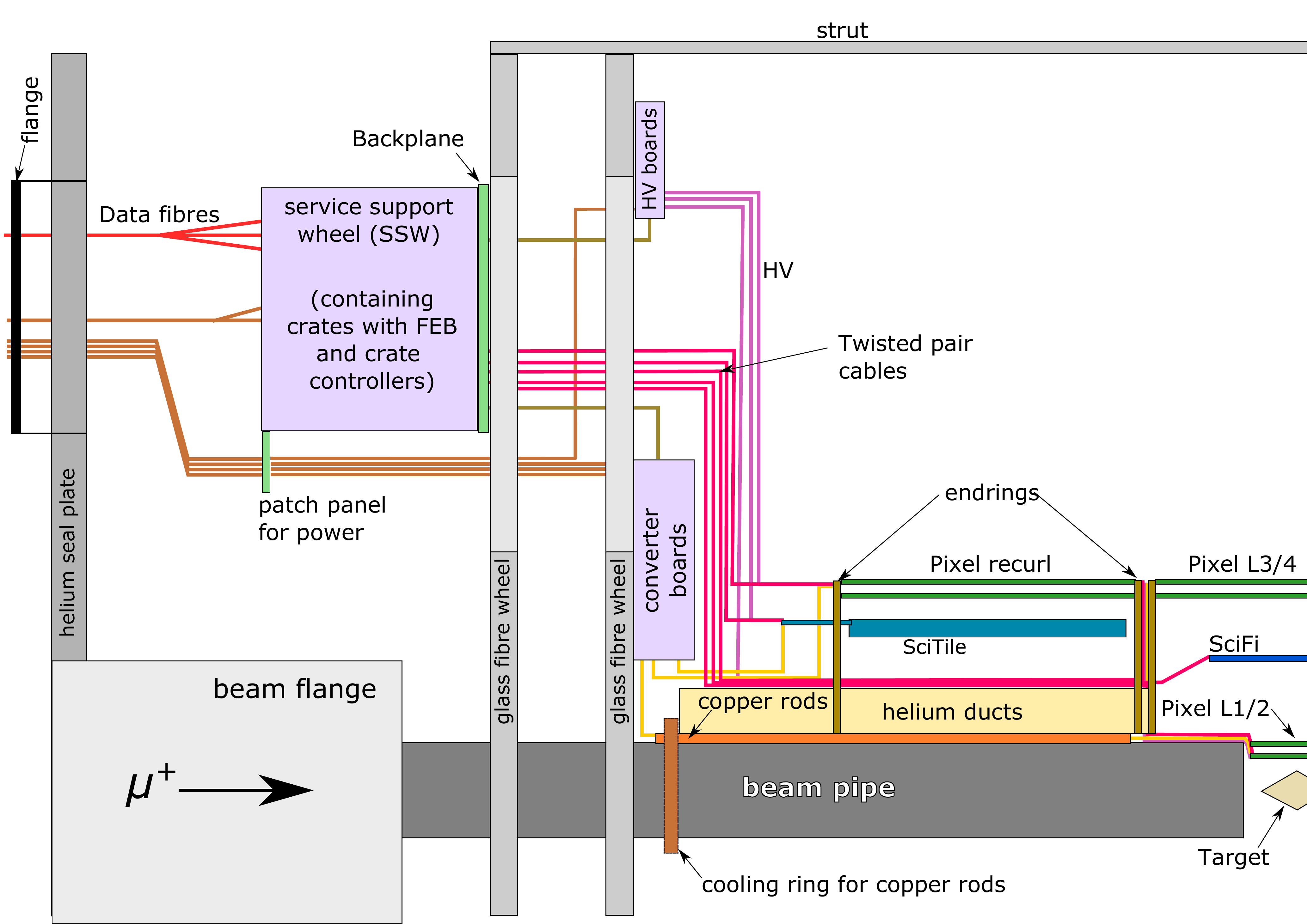}
    \caption{Conceptual view of supply system positioning and cable routing ($rz$-view, not to scale). All
    supplies can be disconnected for the extraction of the experiment. The feed-throughs on the helium seal plate are gas-tight.}
	\label{fig:Cablingscheme}
\end{figure*}

Service support wheels (SSW) are situated outside either end of the detector cage. 
They are loosely coupled to
the cage in the $z$ direction and have their own gliders to decouple mechanical forces 
from the cage. 
The SSWs hold crates for the front end boards, patch panels for the power connections
and routing for the cooling pipes (water and helium). 
The DC-DC converter boards (low voltage supply) and the bias voltage generators are mounted on the inner side of the glass-fibre wheels.
All services have connections at the outward facing planes of the SSWs. 
\autoref{fig:Cablingscheme} shows a conceptual view.

\begin{figure}
    \centering
    \includegraphics[width=0.45\textwidth]{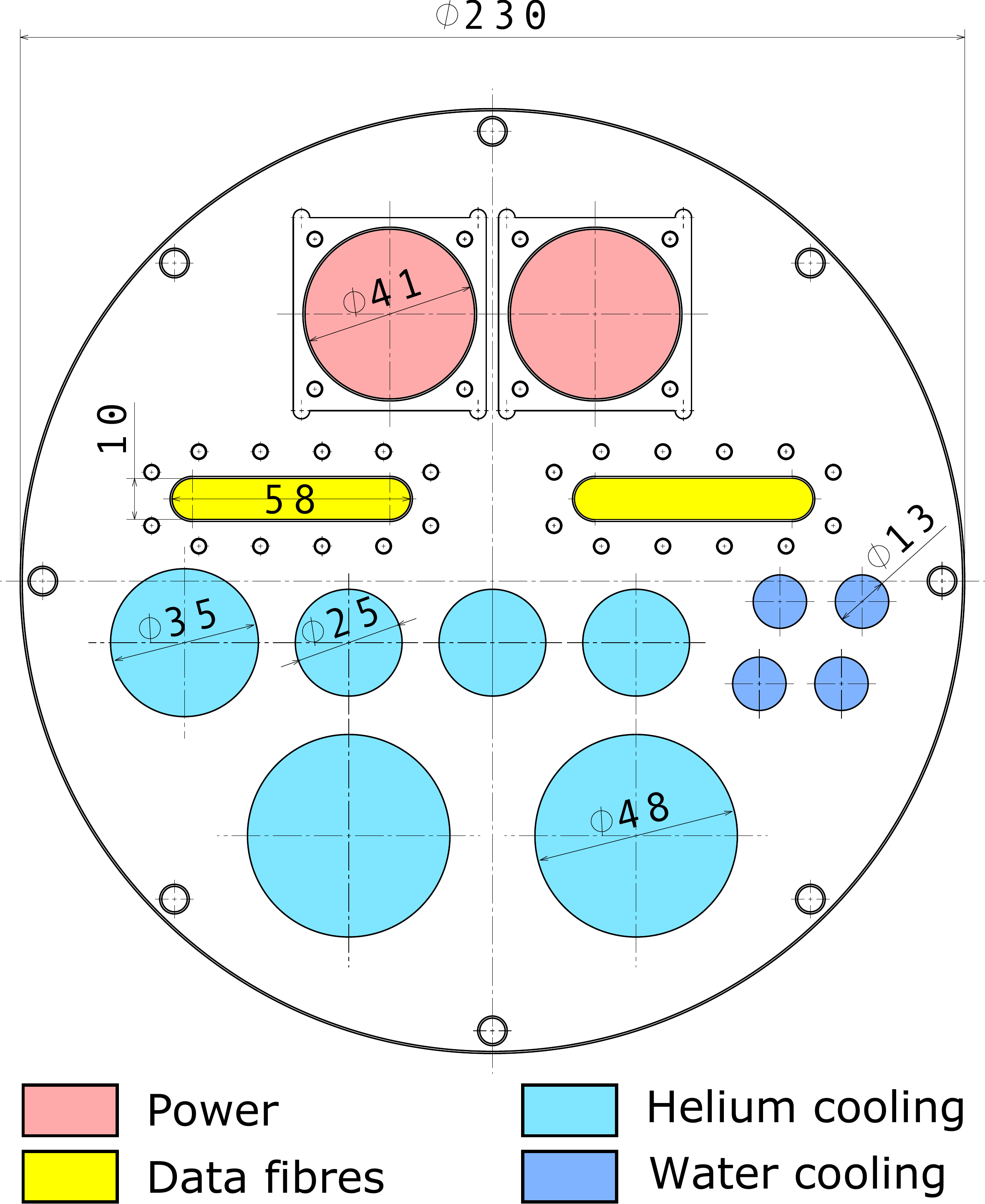}
    \caption{Drawing of the cabling flange. Additional ports will be added for auxiliary use. Dimensions in \si{\mm}.}
    \label{fig:CablingFlangeMu3e}
\end{figure}

Services have to be routed from the inside to the outside of the experiment through flanges sealing the internal dry helium atmosphere from the ambient environment. 
Four identical flange plates are mounted on four turrets at the end plates of the magnet, two at either end. 
Ports for all media are present and provide suitable connectors. 
For the power connections, sealed heavy-duty double-sided 56~pin connector assemblies are used\footnote{Supplier: Souriau-Sunbank}. 
Tubes for the helium and water coolants are welded into the flange and will use industry standard fluid connectors. 
The fibre bundles are sealed with epoxy into brackets that are sealed with an O-ring to the flange. 
A drawing is shown in \autoref{fig:CablingFlangeMu3e}.

\section{Access to the Mu3e detector}
Extracting the experiment from the warm bore of the magnet requires an orchestrated
procedure, which essentially looks as follows:
\begin{sloppypar}
\begin{enumerate}
    \item Detach the beam line, secure cables and hoses. Temporarily remove beam line parts as needed to make space.
    \item Move the magnet into the extraction position.
    \item Open the magnet doors. Remove access plates from the helium sealing plate.
    \item Disconnect all cables and hoses though access holes.
    \item Safely remove the sealing plates, secure cables and hoses while doing this.
    \item Place extraction cart in front of experiment. Engage rail coupling. Carefully remove experiment, guided by the rails.
\end{enumerate}
\end{sloppypar}
For detector insertion, the procedure is reversed. 
The extraction cart is the same as described in the previous section. 
Guide pins and clamps help to safely couple the cart to the rail system in front of the magnet.
 
For servicing the detector, a protective tent will be available that can be used either inside the area for quick work, or outside the area in a secure space. 
External crane attachment points are provided for transferring the experiment to outside of the beam area.


\chapter{Power Distribution and Cabling}
\label{sec:PowerDistribution}

\nobalance

\chapterresponsible{Frank and Stefan and Frederik}

\label{sec:Power}

\begin{figure*}[tb!]
        \centering
                \includegraphics[width=0.8\textwidth]{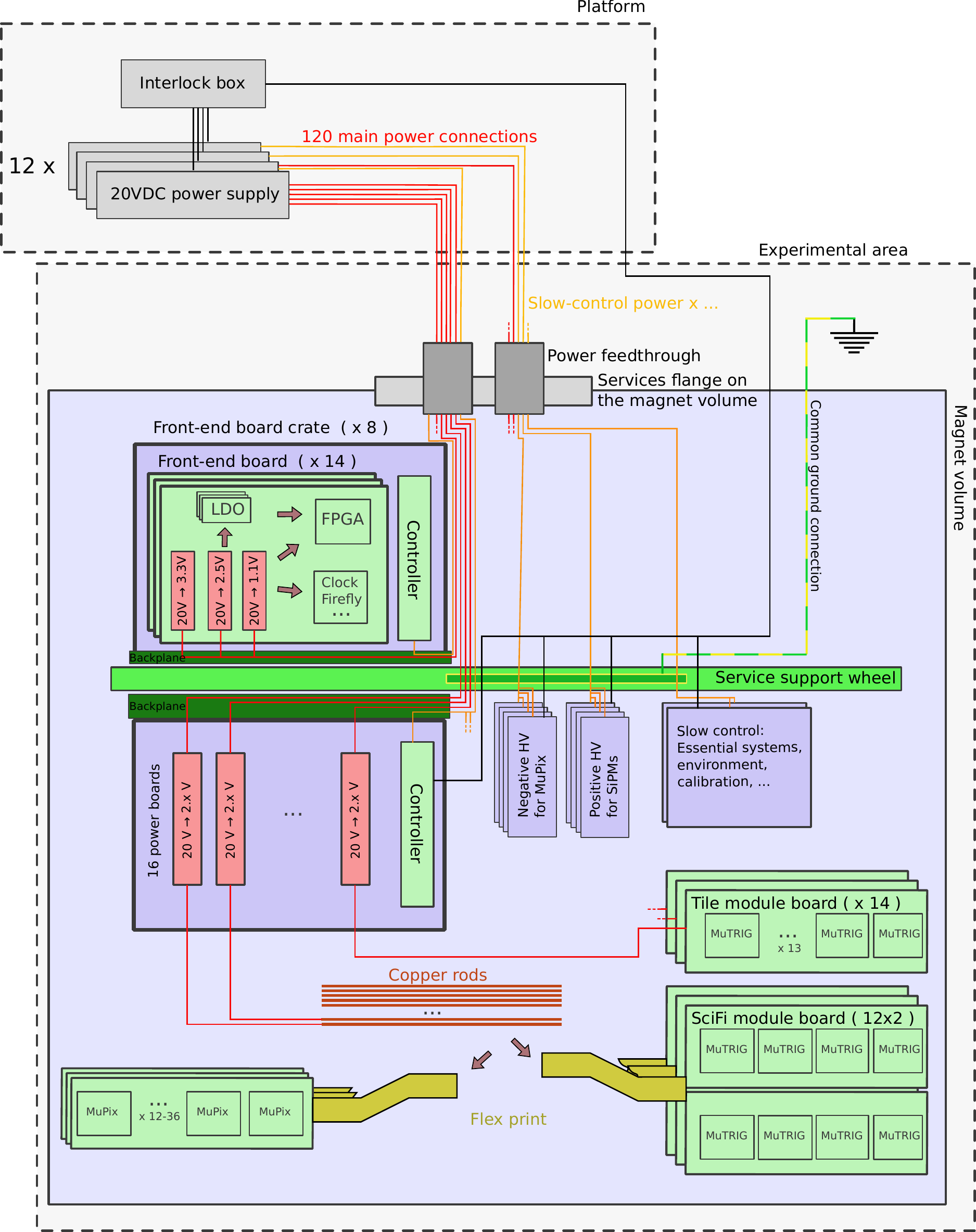}
        \caption{Schematic view of the power distribution inside the detector. Floating \SI{20}{\volt DC} supply lines provide up to \SI{12}{\ampere} of current each (red lines). Custom DC-DC converters on the front-end board and close to the active detector step this \SI{20}{\volt DC} down to the required voltages. Separate power is provided to the slow control systems (orange lines), which need to run independently from the main detector power. Note that these services are distributed over the upstream and downstream Service Support Wheel (SSW). This SSW also acts as a common ground plane, with a single ground connection to the outside.}
        \label{fig:PowerDistribution}
\end{figure*}

With a power consumption from the pixel tracker, the SiPM readout electronics, front-end board, and step-down converters (see \autoref{tab:waterCooledSystems}) of up to~\SI{10}{\kilo\watt}, the Mu3e detector needs a robust but also compact power-distribution system. The conceptual design for such a system is shown in \autoref{fig:PowerDistribution}. Power supplies located on the lower infrastructure platform deliver \SI{20}{\volt}~DC, a voltage high enough to allow for a compact and flexible set of power cables, which are brought into the experiment through a high-density power connector. From there, the power is distributed to either the front-end board crates with embedded buck converters, or to the power boards which step down the voltage for the \mupix chips, and the tile and fibre readout boards. In addition, separate power is provided to the slow control systems which need to run when the main detector power is switched off.

\section{Power Partitions and Grounding}

The Mu3e experiment is divided into 112 detector partitions, which also act as independently controlled power partitions (see~\autoref{tab:PowerPartitions}).  The DC power supplies for these partitions will be the TDK-Lambda GENESYS low-voltage power supply, which are known to be reliable, for example they are being used in the MEG experiment. Each supply can provide up to \SI{90}{\ampere} / \SI{2700}{\watt}, which is distributed to several power partitions via a power relay bank. A massive common return line per supply minimises the voltage drop. Each power supply output is floating, and the return line is referenced to the common ground inside the experimental cage. Slow control systems such as the alignment system, environment monitoring, the controller boards regulating the detector power, and all safety critical systems are powered separately. This enables the powering of all diagnostic tools of the experiment prior to the turn on of the high-power detector electronics.

This powering scheme means that care has to be taken to not introduce ground loops when connecting the various detector partitions to e.g.~a slow control bus or a high-voltage input. To avoid this all data connections to the outside go via optical fibres, the readout is therefore fully electrically decoupled.

\begin{table*}[tb!]
\begin{center}
\small
\begin{tabular}{l l c c  c c c}
\toprule
Partition type  & \#partitions  & \#ASICS/  & \multicolumn{2}{c}{Maximum power per partition	[\SI{}{\watt}]} & Total Power including		        \\
(ASIC)		      &		       			&	partition	&	Excluding	    & Including DC-DC	 		                            &	DC-DC losses [\SI{}{\watt}]   	\\
\midrule
Pixel(\mupix)		        &					    	&		       	&               &                                									&					                         \\
~layer 1 	 			&	4		          &	12		    &	19.2          &	25.6					                                  &	102				                       \\
~layer 2 				&	4		          &	15		    &	24.0		      &	32					                                    &	128                      				\\
~layer 3 	 			&	$3 \times 12$	&	32, 36		&	51.2, 57.6	  &	68.3, 76.8				                              &	2660				\\
~layer 4 				&	$3 \times 14$	&	36		    &	57.6		      &	76.8					                                  &	3230					\\
\midrule
Fibre(\mutrig)				    &	12		        &	8		      &	9.6		        &	12.8					                                  & 	153				\\
\midrule
Tile(\mutrig)					  &	14		        &	13		    &	15.6		      &	20.8					                                  &	291				\\
\midrule
Front-end board	&		8		        &	14 boards	&	266		        &	350					                                    &	2800				\\
\midrule
Total						&			          &			      &			          &					                                      	&	\bf{9370} 			\\
\bottomrule
\end{tabular}
\end{center}
\caption{Power partitions for the Mu3e detector ASICs and electronics inside the magnet bore. The high-power elements on the front-end board are the Arria~V FPGA, clock chip, and the transceivers. A respective maximum power consumption of \SI{1.2}{\watt} and \SI{1.6}{\watt} for the \mutrig and \mupix chips is assumed. The upper limits on the power figures are driven by the cooling system, and depend on power losses in the entire power distribution system. For the total power budget, a 75\% efficiency of the DC-DC converters is assumed.}
\label{tab:PowerPartitions}
\end{table*}

\section{DC-DC conversion}

Switching power converters will be used to step down the 20~V to the voltages needed by the detector and electronics (\autoref{tab:DCDCConverterSpecs}) Typical efficiencies are of the order of to 70\%~to~90\%, depending on the current and the voltage step. Compact high-power converters typically used for FPGA boards such as the LTM4601 (Analogue Devices) have a ferrite core inductor, with is incompatible with the high magnetic field environment of the experiment. Mu3e has selected the following solution: a commercial synchronous buck converter combined with a custom air coil, where the coil properties and the switching frequency are optimised for the required output voltage and current. As they are mounted outside the active area of the detector, these converters don't have to be radiation hard.

\subsection{Front-end board converters}

\begin{figure}[tb!]
        \centering
                \includegraphics[width=0.8\columnwidth]{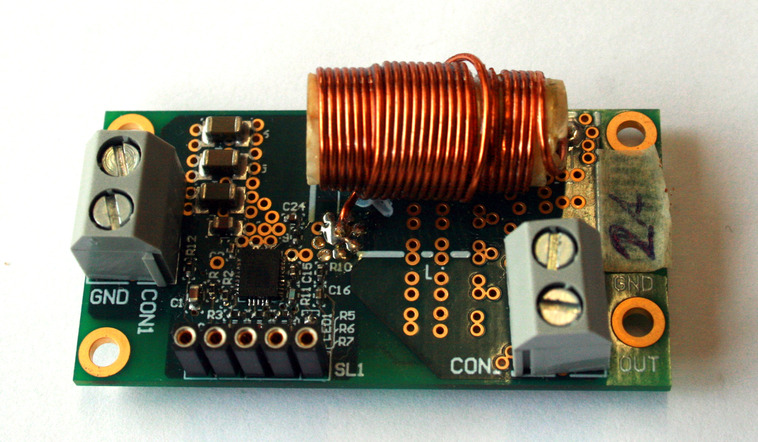}
        \caption{The second prototype for the buck converters for the frontend board. Good performance with efficiencies $>$75\% in a 0.7~T magnetic field was demonstrated.}
        \label{fig:fe_converter}
\end{figure}

The front-end boards with an Arria~V FPGA, and the LVDS and optical tranceivers (\autoref{sec:FrontEndFPGABoards}) require several DC voltages at typical currents of 1-3~\SI{}{A}. Three switching DC-DC converters will generate 1.1, 1.8, and 3.3~VDC with Peak-Peak ripple below \SI{10}{mV}  (see \autoref{tab:DCDCConverterSpecs}). Passive filters and active filtering with devices such as the LT3086 (Analogue Devices) further reduce the voltage ripple, and allow intermediate voltages to be generated.
The switching converters on the front-end board are based on a compact TPS548A20 synchronous buck converter with integrated switches (Texas Instruments), combined with single layer cylindrical air coils. \autoref{fig:fe_converter} shows a stand alone 2x4~cm \SI{1.8}{\volt} prototype, which has demonstrated good performance at operating conditions. The converter embedded on the front-end board will have a similar footprint, with an additional copper shielding box covering the coil to reduce EMI and improve mechanical stability~\cite{Hesping2019}.

\subsection{Power boards}
\label{sec:Powerboards}

\begin{figure}[hbt!]
        \centering
                \includegraphics[width=0.8\columnwidth]{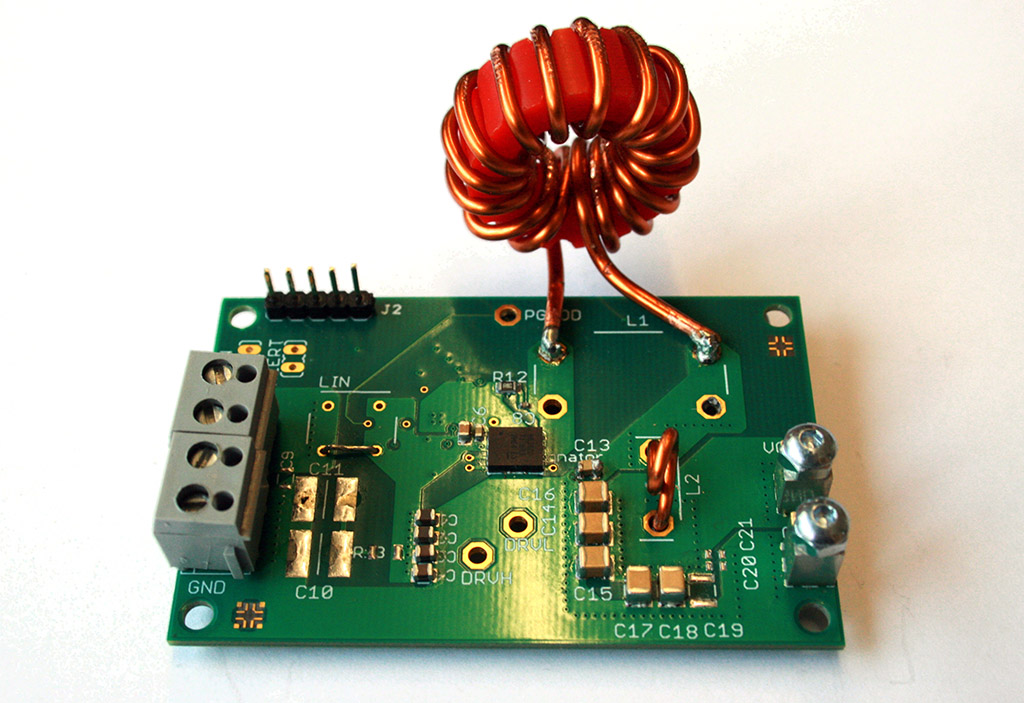}
        \caption{A 4.5~x~7~cm prototype for the power board, with a \SI{0.5}{\micro\henry} toroidal inductor and a secondary LC filter at the output. With an output ripple of 10-20~mV, this has been used to succesfully power \mupix 8 sensors.}
        \label{fig:power_board}
\end{figure}

With currents potentially up to \SI{30}{\ampere} and very few options for additional filtering further down the line, the requirements for the active detector DC-DC converters are more challenging. The TPS53219 buck controller and CSD86350Q5D power MOSFET switch from Texas Instruments were identified as meeting these requirements. A first prototype was developed (see \autoref{fig:power_board}). This board has space for various input and output filter configurations, and has dimensions close to the final form factor. The configuration shown, with a toroidal coil in combination with a secondary LC filter, has the best noise figure with a Peak-Peak ripple of approximately \SI{10}{\milli\volt}. The board was stress tested in a magnetic field, and successfully used to power a \mupix~8 pixel detector during a DESY testbeam campaign.

The final power board has a secondary output filter, and several additional features such as current monitoring, and interface connector for the back plane, and an embedded temperature interlock connected to a temperature diode on the \mupix sensor~\cite{Gagneur2020}.

In the experiment, 16 power boards are mounted in a crate on the SSW (see \Autoref{fig:PowerDistribution,fig:Cablingscheme}), with a MSCB slave (\autoref{sec:SlowControl}) as controller. This controller adjusts the output voltage, switching frequency, and monitors several parameters.  It also interfaces the DC-DC converters with an external interlock system. 

\begin{table*}[tb!]
\small
\begin{center}
    \begin{tabular}{lrrrc}
    \toprule
    Component     & Voltage [\SI{}{\volt}]  & Typical                   & Min. inductance                   & coil design    \\
                  &                         & current [\SI{}{\ampere}]  & air coil [\SI{}{\micro\henry}]    &                 \\
    \midrule
    Front-end board			&	1.1		&	2	&	2	&	cylindrical	\\
    Front-end board			&	1.8 		&	1.7	&	6	&	cylindrical	\\
    Front-end board			&	3.3 		&	2.2	&	4	&	cylindrical	\\
    \mupix partition (layer 1,2)	&	ca. 2.3  	&	10	&	0.5	&	toroid				\\
    \mupix partition (layer 3,4)	&	ca. 2.3  	&	21	&	0.4	&	toroid				\\
    Fibre partition			&	ca. 2.0  	&	7	&	0.7	&	toroid				\\
    Tile partition			&	ca. 2.0  	&	9	&	0.7	&	toroid				\\
    Tile partition			&	3.3	  	&	3	&	3.3	&	toroid				\\
    \bottomrule
    \end{tabular}

    \caption{Specifications for different buck converter channels stepping down the voltage from 20~V with an efficiency $>$70\%. The quoted \mupix voltage takes into account an anticipated voltage drop of 200 to 300~mV between the converter and the chip.}
    \label{tab:DCDCConverterSpecs}

    \end{center}
\end{table*}

\section{Bias voltage}
\label{sec:HighVoltage}
\label{sec:PixelHighVoltage}
Bias voltages between \SI{50}{\volt} and \SI{120}{\volt} are required for the SiPMs used in the fibre tracker and the tile detector as well as for the \mupix chip.
As only moderate currents of few \SI{}{\micro\ampere} per channel are needed, these voltages can be generated with a Cockroft-Walton chain. 
Converters supplying positive voltages have been developed and optimised in the context of the MEGII experiment. For Mu3e this design is carried over to a new board which will be mounted inside the magnet volume.

\begin{figure}[tb!]
	\centering
		\includegraphics[width=0.9\columnwidth]{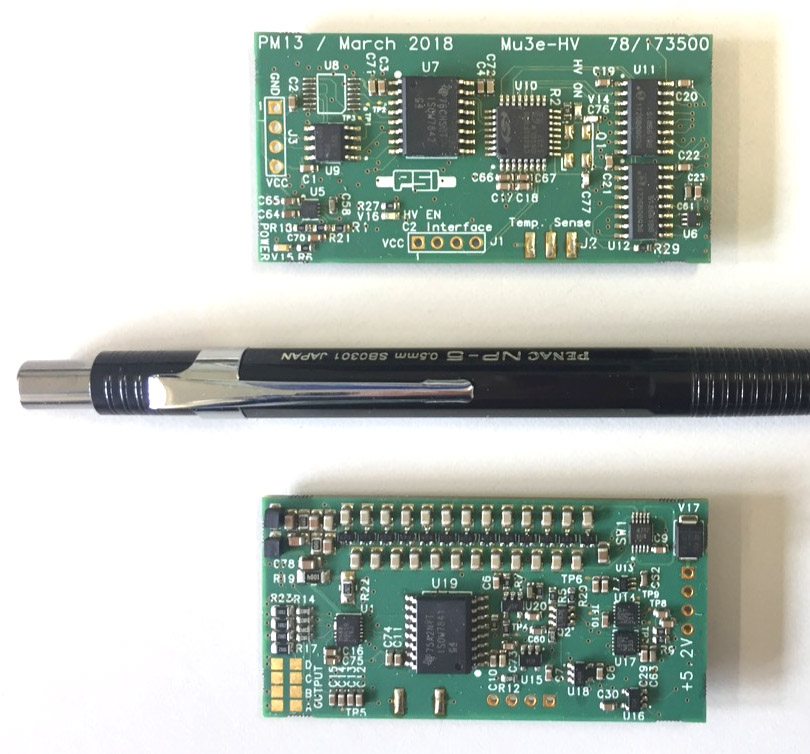}
	\caption{Prototype of the high-voltage generator board with top side (upper picture) and bottom side (lower picture).}
	\label{fig:PixelHighBoard}
\end{figure}

\begin{figure*}[tb!]
	\centering
		\includegraphics[width=1.00\textwidth]{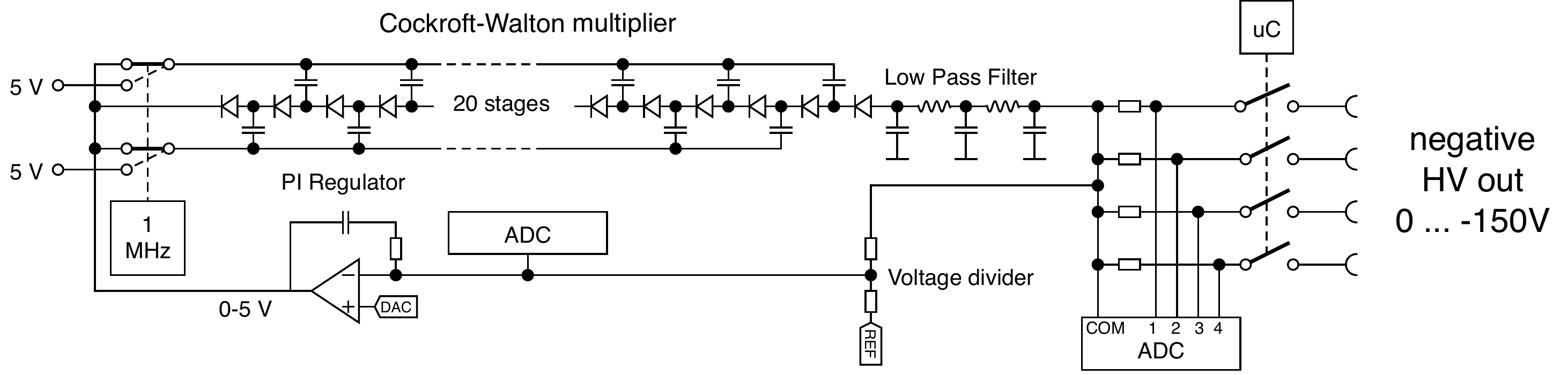}
	\caption{Block schematic of the pixel high voltage generation.}
	\label{fig:PixelHighVoltage}
\end{figure*}

The pixel tracker requires a negative bias voltage of up to -100 V for each chip.
For economic reasons, a set of four power groups is provided with a common voltage with dedicated current
measurements and the possibility to turn off each power group individually.
Since voltage generators which run at the high magnetic 
field are not available commercially, a custom board
based on the Cockroft-Walton voltage multiplier design has been created.  \autoref{fig:PixelHighVoltage}
shows the simplified block schematic of this device. A micro-controller
connected to the MSCB slow control system operates the DAC, ADC and switches
of the voltage generator. It is capable of generating a bias voltage from $\SI[parse-numbers=false]{0\ldots-150}{\volt}$
out of a single power supply of \SI{5}{\volt}. First tests with a prototype indicated that 
an absolute voltage accuracy of $\SI{\pm 1}{\milli\volt}$ at a current of \SI{2}{\milli\ampere} can be achieved with a residual
ripple below \SI{10}{\milli\volt}. Each channel contains a shunt resistor and an ADC, which can measure
the individual current. High voltage CMOS switches operated by the micro-controller
can switch off individual channels in case the corresponding pixel chips would
have a problem.

\autoref{fig:PixelHighBoard} shows the top and bottom sides of a prototype of the high voltage board. 
It has a size of $\SI[parse-numbers=false]{30 \times 60}{\mm\squared}$. The Cockroft-Walton chain can be identified on the bottom side of the board. No magnetic components have been used in the design, making it possible to operate the board in magnetic fields of up to \SI{2}{\tesla}.

%
%

\begin{figure*}[tb!]
	\centering
		\includegraphics[width=0.6\textwidth]{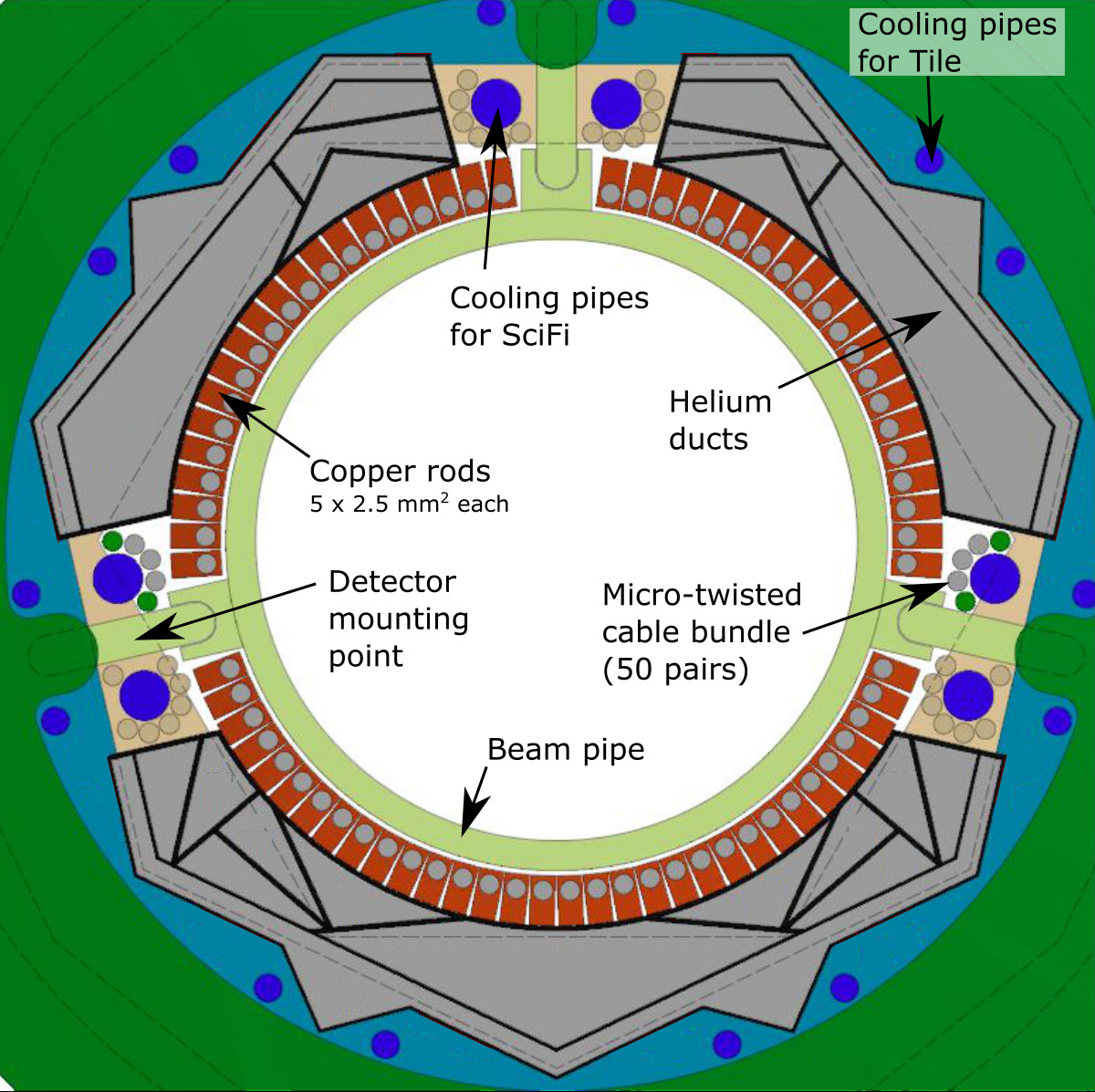}
		\caption{Cross section of a recurl station. The micro-twisted pair cables come in bundles and are shown as circles around the cooling pipes for the fibre detector (the colour code shows detector assignment: green for fibres, grey for vertex layers, brown for pixel outer layers). The helium ducts have separated channels for different destinations. Their cross sections are optimised for minimal pressure drop.}
		\label{fig:cablingXsecBeampipe}
\end{figure*}

\section{Cabling}
\label{sec:Cabling}
The basic concept of the cabling inside the detector is shown in \autoref{fig:Cablingscheme}. From the power boards, the connections to the detector components are carried out with minimal possible length using solid copper cable of \SI{2.5}{\mm\squared} gauge. Because all connections have to be done outside the detector acceptance, only the space around the beam pipes is left. Copper rods with a cross-section of $\SI[parse-numbers = false]{5 \times 2.5}{\mm\squared}$ are used to bridge the connection between the detector endring mount and the outer end of the beam pipe. These rods are individually insulated using a polyimide foil wrap, and held in place by epoxy. The rods are in a densely packed environment, hence the dissipated power will be actively removed using a copper cooling ring (see \autoref{sec:WaterCooling}). The cables are connected using screw-mounted copper clamps.

Data cables between detectors and the front-end boards are micro-twisted pair wires: AWG 36 wires with a Polyimide isolation and an impedance of \SI{90}{\ohm} from \emph{Heermann GmbH}. Each bundle of up to 50 pairs has a typical outer diameter of \SI{2}{\mm}. The bundles are arranged around the water cooling pipes:  see \autoref{fig:cablingXsecBeampipe} for a sketch of the arrangement. The data cables are attached to the detector elements using soldered connections on flexible printed circuit boards, which connect to the PCB or HDI via interposers. The attachment to the frontend boards takes place on the patch panel of the SSW using zero-force connectors.



\chapter{Clock Distribution}
\label{sec:ClockDistribution}

\nobalance

\chapterresponsible{Gavin}

\begin{figure*}[tb!]
	\centering
		\includegraphics[width=1.0\textwidth]{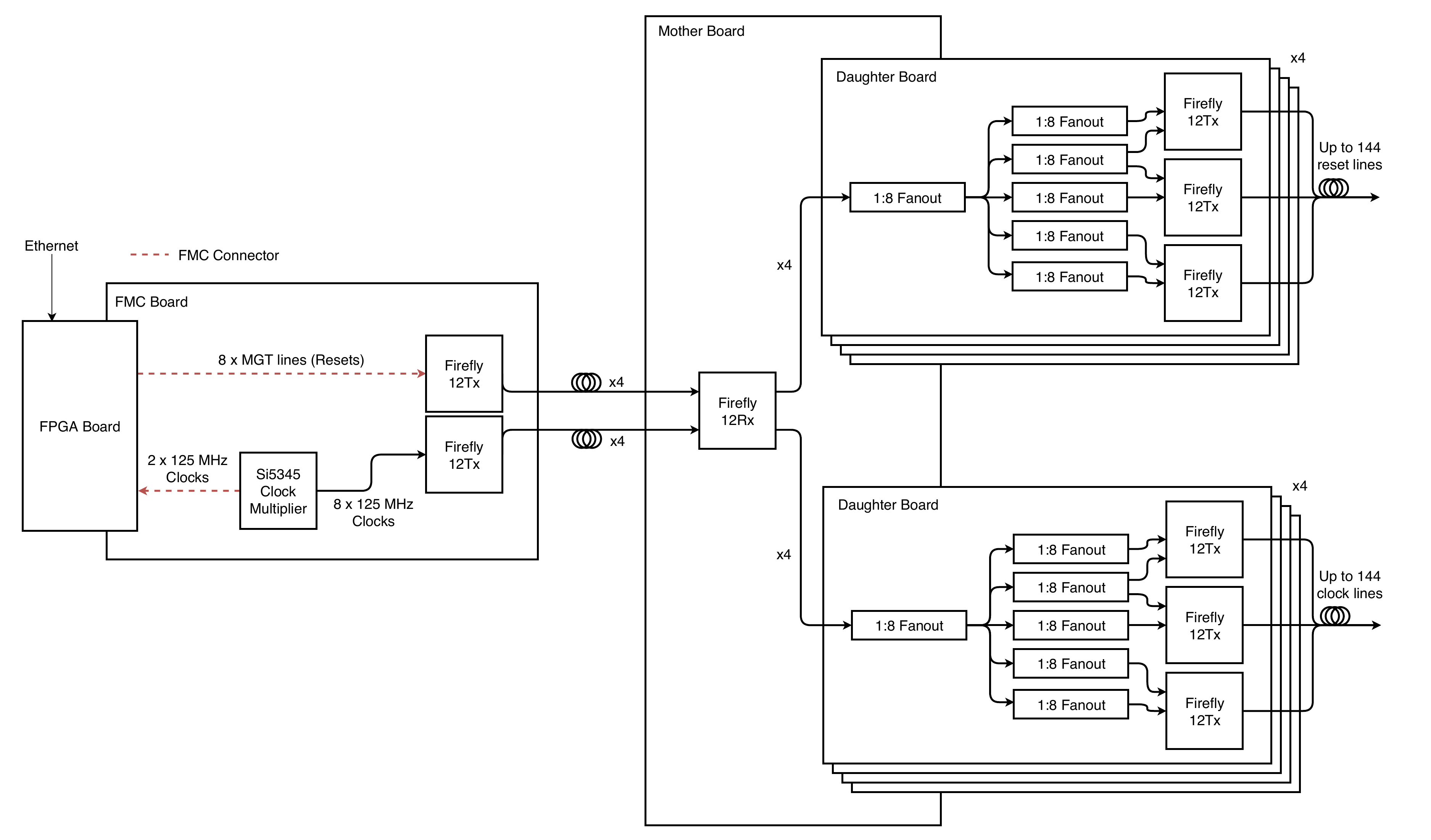}
	\caption{A block diagram illustrating the connections and relationships between the various boards of the clock and reset distribution system. The block diagram shows how the network-controlled FPGA can be paired with a bespoke FMC board to create 8 reset and 8 clock optical lines (only 4 of each are used in the Mu3e phase~I DAQ). These are then split with bespoke fan-out boards, effectively creating a 1:36 active optical splitter.}
	\label{fig:ClockDistribution}
\end{figure*}

The precise timing measurement as well as the operation of many \si{\giga\bit\per\second}
links in the experiment requires a very stable clock distribution.
In order to ensure synchronisation between the timestamps of all sub-detectors, 
a global synchronous reset is also required.

The frequency of the clock distribution system is chosen to be
$\SI{125}{\mega\hertz}$; other frequencies can be derived locally by
phase-locked loops. 
To meet the timing resolution requirements for all of the detector subsystems, the phase stability of the clock
distribution has to be better than $\SI{10}{\pico\second}$ over the
complete system.  
The jitter requirements of the global reset ($\sim$ half a clock period) are more relaxed.

The overall block diagram of the clock and reset system can be seen in
\autoref{fig:ClockDistribution}.  The $\SI{125}{\mega\hertz}$ clock
is generated by a commercial low-jitter clock oscillator on a
dedicated board. This board is controlled by an FPGA which also
controls the reset signal. The board provides both clock and reset
signals to optical transmitters, which are then passed to a number of
bespoke boards. These actively split the signal and supply the clock and reset lines
on optical fibres to local clock distribution boards inside the warm bore of the magnet,
as well as to the DAQ switching boards (see
\autoref{sec:SwitchingBoards}).
Inside the magnet, optical receivers forward the signals to a jitter
cleaner and fan-out chip on the front-end board, which then drives
LVDS signals to the FPGA and front-end ASICs. For the filter farm PCs, each
FPGA board inside a PC receives the global clock via a Clock
Transmission Board (CTB), a small custom board which converts the
optical clock signal to an electronic one. Reset and state changes are
communicated to the farm via ethernet.

The reset signal is implemented as a \SI{1.25}{Gbit/s} serial data stream 
synchronised to the clock and uses \SI{8}{bit} datagrams
in 8~bit/10~bit encoding \cite{Widmer1983, franaszek1984byte},  to induce not only
resets but also changes of operation mode. 
In idle mode, a comma word is sent, allowing for word alignment in the reset bit stream.
Resets of different subsystems and changes between idle and running modes are 
triggered by sending one of the 256 possible data words. 

The front-end boards then have the task of distributing the clock and reset 
(here the reset is an on/off signal) to all ASICS (\mupix and \mutrig).
A dedicated jitter cleaner and fan-out component is used, which will also be 
used to generate the $\SI{625}{\mega\hertz}$ clock needed by the \mutrig Phase-Locked Loop (PLL).

Slower clocks, required e.g.~for slow control and configuration signals, will be
generated by clock dividers and/or PLLs in the FPGAs.

\section{FMC Distribution Board}
\begin{figure}[tb!]
	\centering
		\includegraphics[width=0.33\textwidth]{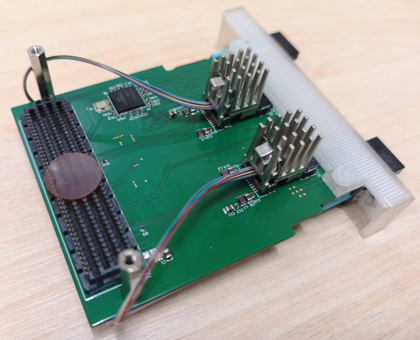}
	\caption{The Clock and Reset FMC distribution board has two Firefly optical 
	transceivers. The transceiver on the top connects directly to the Silicon 
	Labs Si5345 clock generator IC. The bottom optical transceiver connects to the 
	MGT lines coming from the FPGA via the FMC connector. Both Firefly transceivers 
	have 12 optical transmit lines but only 8 of each are used.}
	\label{fig:fmc_clock_board}
\end{figure}

\begin{sloppypar}
At the heart of the clock and reset system is the FPGA Mezzanine Card
(FMC) distribution board, shown in \autoref{fig:fmc_clock_board}. The
distribution board connects to the FPGA development board via an FMC
connector~\cite{FMC} which allows access to the 10 Multi-Gigabit
Transceiver (MGT) lines the Xilinx Kintex-7 FPGA offers. The low-jitter
IC which is used to generate the \SI{125}{MHz} clock is the Silicon
Labs Si5345 which can generate up to 10 any-frequency clock outputs
with an ultra-low jitter of \SI{90}{fs} RMS. The SI5345 also offers
in-circuit non-volatile programming which ensures a regular power up
with a known frequency.  The distribution board uses 8~MGTs for the
reset lines and all 10 clock outputs from the clock generator, two of
which are used to generate the MGT lines and the remaining 8 are used
for the clock lines. The clock and reset lines are routed to the
inputs of two optical Firefly transceivers. Such a configuration
allows the individual control of the 8 pairs of the clock and reset
lines.  In the experiment, these can be translated as 8 individually
controllable partitions, though only 4 are needed for the Mu3e phase~I 
DAQ. In addition, the distribution board provides an I$^2$C interface
which is used to communicate with the active splitting boards, which
among many other functions, can provide the ability to disable/enable
individual clock and reset lines.
\end{sloppypar}

\section{Active Splitting}
\begin{figure*}
  \centering
  \includegraphics[width=0.63\textwidth]{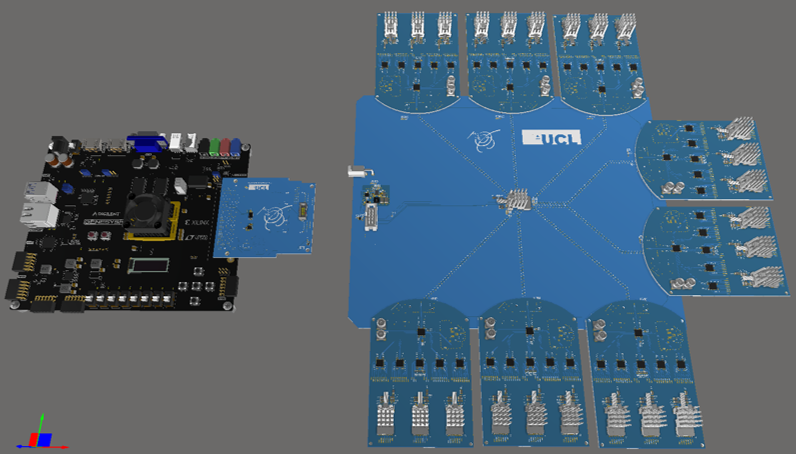}
  \includegraphics[width=0.36\textwidth]{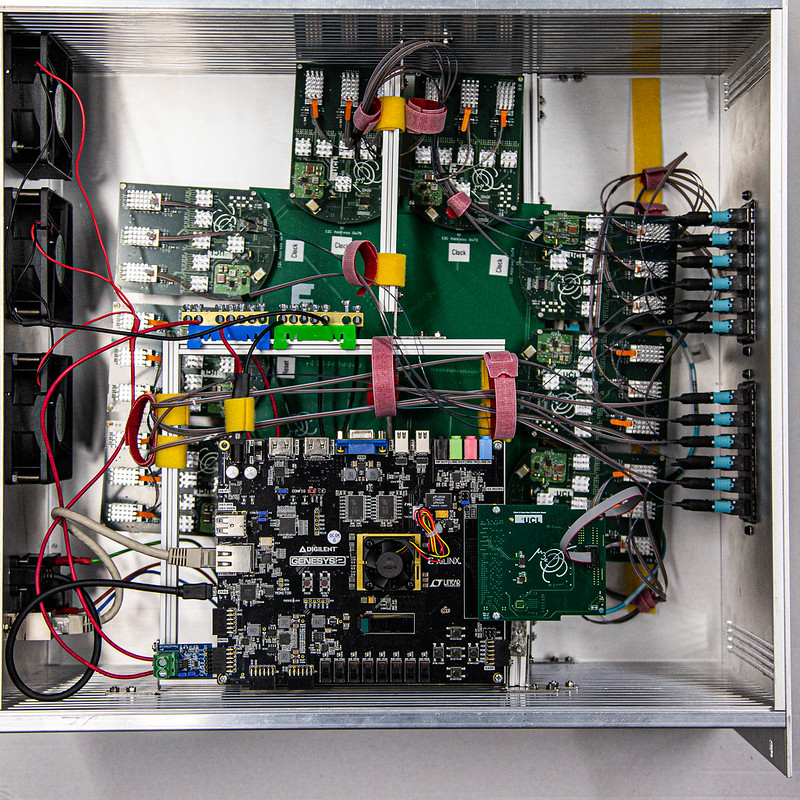}
  \caption{Left: A 3D representation of the full clock and reset distribution system. On the left side is the Genesys 2 board with the clock and reset FMC distribution board. On the right side is the active splitting motherboard with has an optical receiver (centre of the motherboard) that accepts the clock or reset lines from the FMC board via an optical fibre. The motherboard electrically routes the 8 signals to the fan-out daughter boards where each board generates 36 optical copies of the routed signal. Right: The full clock and reset system in a 19-inch rack-mountable box.}
  \label{fig:full_clock_3d}
\end{figure*}

The optical splitting of the clock and reset lines is modular in design. 
It allows the active splitting to be versatile and can be potentially used as a 
generic active optical splitting solution. The system unit comprises one 
motherboard and 8 daughter boards. The motherboard takes 8 optical inputs and 
electrically routes each of the 8 signals to one of the eight daughter boards. 
The daughter board, which connects to the motherboard via a high-speed mezzanine 
connector, creates 36 copies of the input signal for a total of 288 optical 
copies per system unit. One system unit is sufficient for all clock reset lines
required by the Mu3e Phase 1  DAQ. A 3D representation of the system can be seen in 
\autoref{fig:full_clock_3d}.

\subsection{Board designs}
\begin{figure}[t!]
	\centering
		\includegraphics[width=0.48\textwidth]{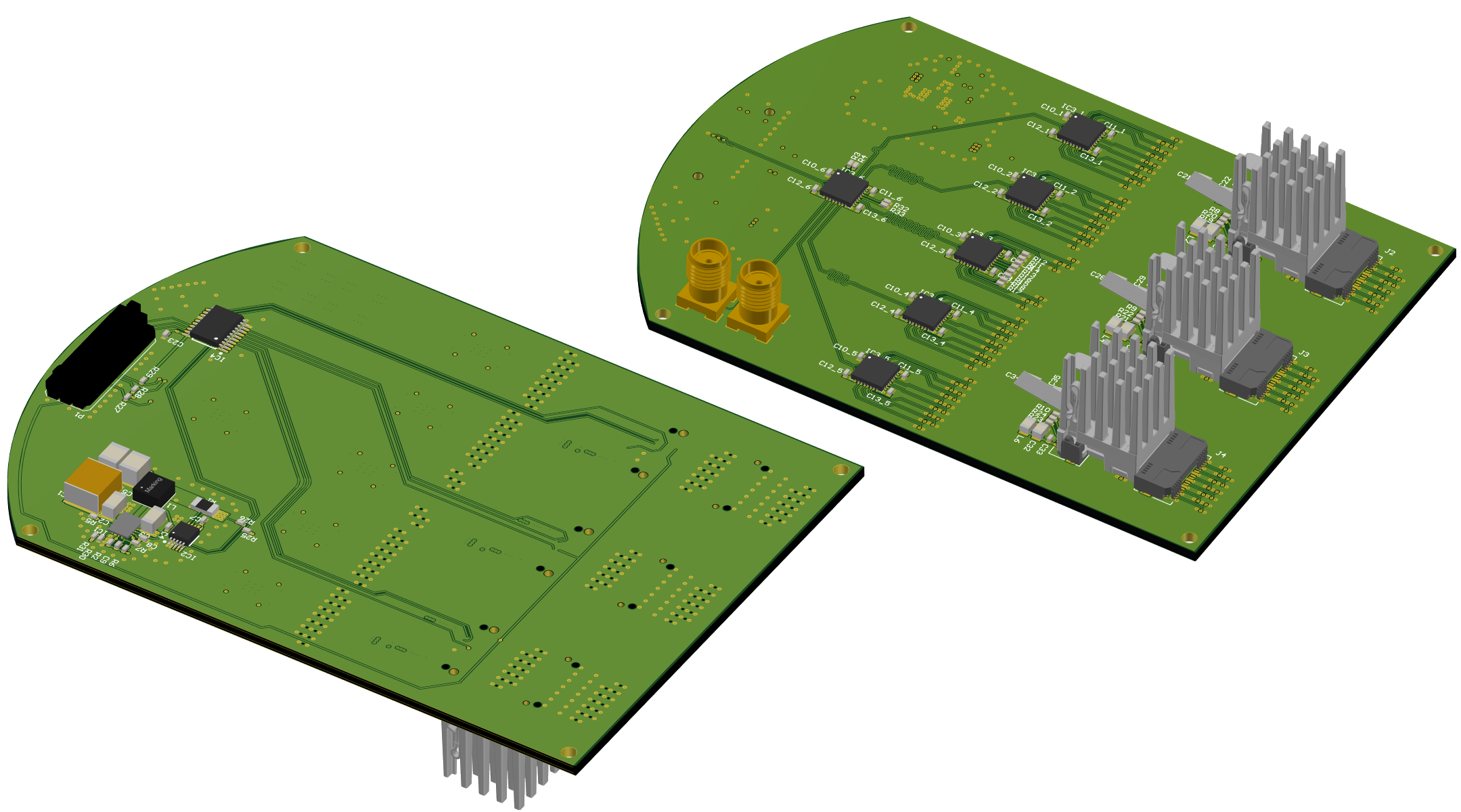}
	\caption{A 3D representation of the clock and reset daughter board. 
	Left: Bottom view of the board showing the mezzanine connector and DC-DC 
	circuitry. Right: Top view of the board showing the fan-out ICs and the three 
	Firefly optical transceivers.}
	\label{fig:daughter_clock_board}
\end{figure}

The daughter board, shown in \autoref{fig:daughter_clock_board}, utilises the 
OnSemi NB7L1008M fan-out chip; a 1:8 6.5 Gbit/s differential fan-out buffer with a 
random clock jitter < \SI{0.8}{ps} RMS. 
Overall, 5 fan-out chips are used per board, creating a total of 40 replicated signals. 
However, only 36 are used as the three on-board Firefly transceivers have a total of 36 optical transmitters. 
The 36 optical lines are carried by three 12-fibre OM3 MTP cables. The board 
also has a low-noise DC-DC converter with power monitoring and is connected to 
an I$^2$C bus which allows this power monitor to be read, in addition to 
controlling the Firefly transceivers (e.g.~disabling and enabling 
individual optical channels).

The daughter board receives its high-speed signals, communication bus
and power from the motherboard, as has been described and seen in
\autoref{fig:full_clock_3d}, where the full system integrated into
a 19 inch rack-mountable box is also shown.

\section{Clock and Reset Operation}

\begin{figure}
	\centering
		\includegraphics[width=0.4\textwidth]{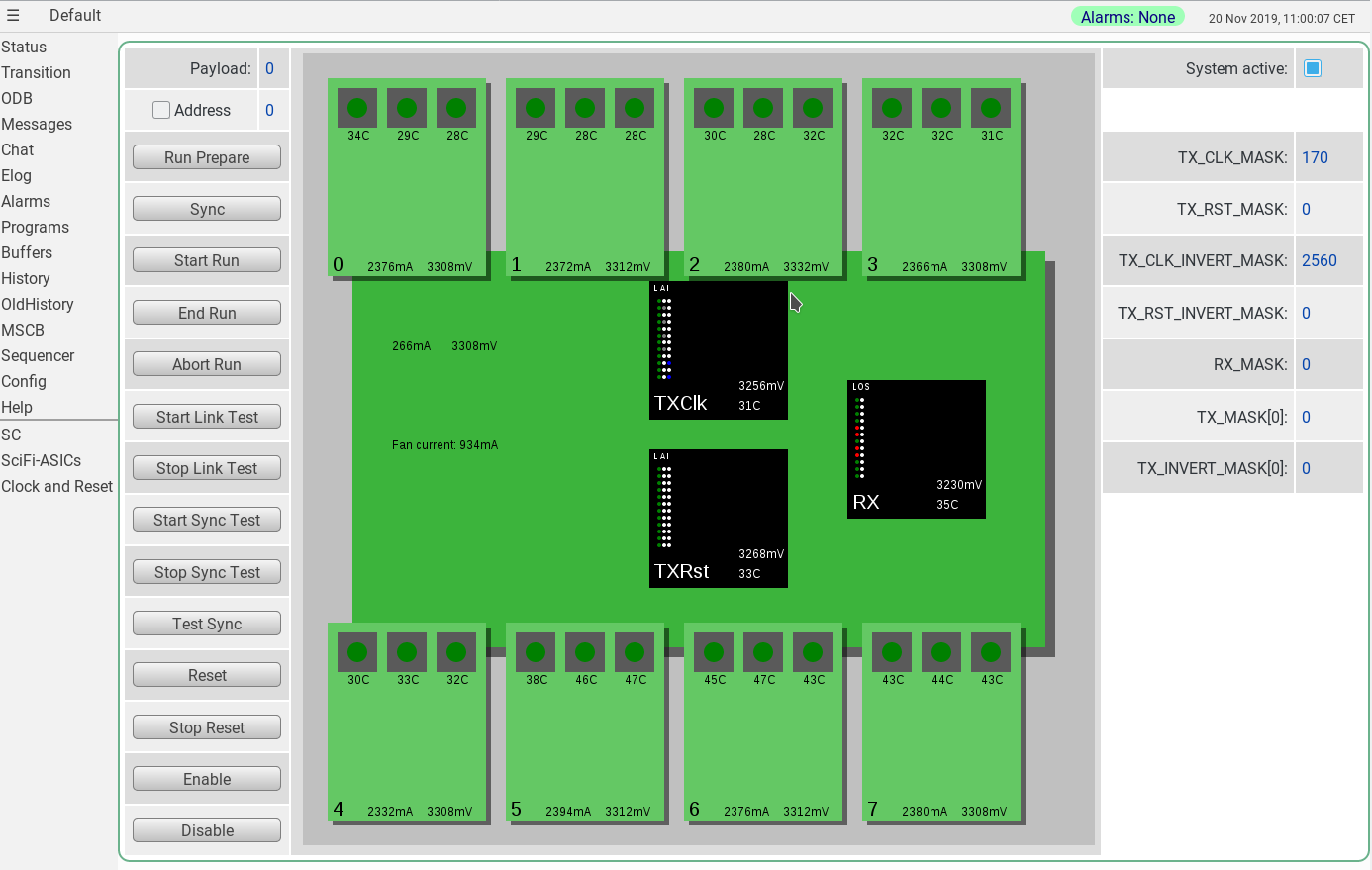}
	\caption{MIDAS page for control and monitoring of the clock and reset distribution system.}
	\label{fig:MIDASPage}
\end{figure}

The clock and reset firmware is developed on the Digilent Genesys~2
board~\cite{GENESYS}, a Xilinx Kintex-7 evaluation board. The custom
firmware implements the IPBUS protocol~\cite{IPBUS}, allowing the
end-user to modify registers through a network connection and thereby
control the clock and reset system. The firmware provides two
IPBUS-based interfaces: to the FPGA MGT 8b/10b transceivers; and
to the I$^2$C control of the Firefly transceivers, clock generator, and the
power and cooling systems. The IPBUS protocol is also implemented
in the MIDAS DAQ system, which provides a control page for the clock
and reset distribution system, see \autoref{fig:MIDASPage}.

\section{Performance}

\begin{figure}
	\centering
		\includegraphics[width=0.38\textwidth]{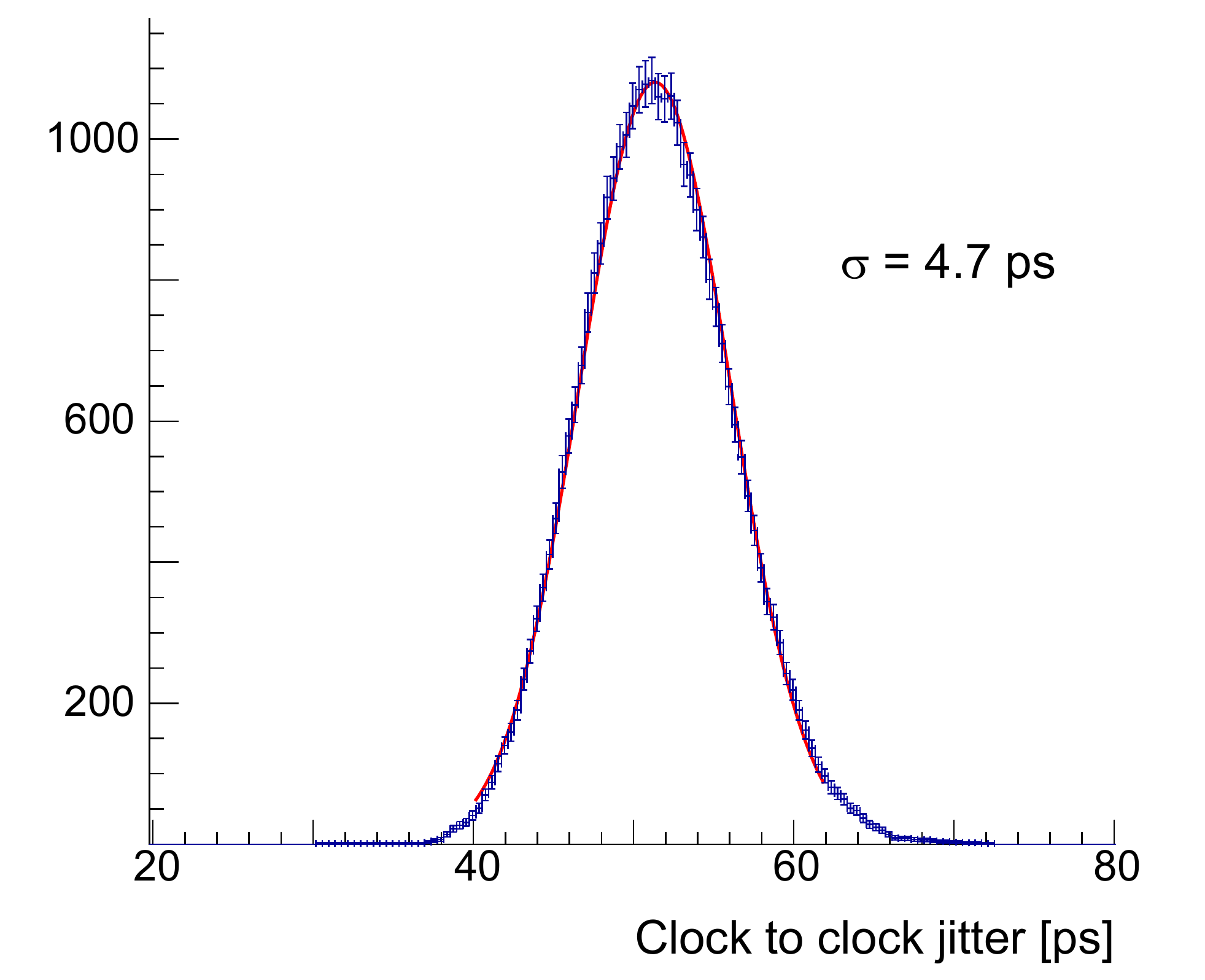}
	\caption{Rising edge time difference between two different clocks distributed via different daughter boards and different optical cable assemblies (leading to the \SI{50}{ps} offset) as determined via a fast oscilloscope. The fit is a simple
	Gaussian.}
	\label{fig:clockjitter}
\end{figure}

The full clock and reset system have undergone extensive testing. All
optical outputs are fully operational, as are the configuration and
monitoring of the system. The Firefly transceivers have good
thermal performance and stability with cooling provided by fans in the
rack-mountable box. The relative phase of clocks from different
daughter boards transmitted via separate optical fibre assemblies to
two different receivers has been measured to have a jitter of less than
\SI{5}{\pico\second}, see~\autoref{fig:clockjitter}.
Note that the jitter includes a sizeable contribution from the measurement set-up.



\chapter{Slow Control}
\label{sec:SlowControl}

\nobalance

\chapterresponsible{Stefan}

The slow control system deals with all ``slow'' data such as high voltages for the SiPMs and silicon sensors, ambient temperatures and pressures, the beam line magnet settings and parameters of the cooling system. 
The configuration of the \mupix and \mutrig ASICs is handled separately, as described in \autoref{sec:PixelSlowControl}.

For the slow control parameters it is important to have all data and control functionality in a homogeneous single system.
This makes the maintenance of the system much simpler, since only a limited number of different hardware standards have to be taken care of.
The integration of all data enables us to define control loops between otherwise completely different subsystems.
Examples are regulating or switching off the detector power in the case of overheating of the pixel sensors or irregularities appearing in the helium cooling system, or adjusting the high-voltage on the basis of detector data such as energy spectra or hit rates.

The integration of all systems will be done through the MIDAS DAQ system, and as much as possible combined with the associated MIDAS Slow Control Bus (MSCB) system \cite{MSCB:2001}, which is discussed in \autoref{sec:MSCB}.
In addition to the MSCB system, the MIDAS DAQ system receives and sends slow control data to the various layers of FPGAs and GPUs through the main fast data links (\autoref{sec:DAQ}).
The slow control system also contains interfaces to the PSI beamline elements via the EPICS system \cite{EPICS}.
This allows monitoring and control of the beamline from the main DAQ system, which has proven very versatile in other experiments using this scheme.

The full state of the system is kept in the MIDAS Online Data Base (ODB), and all slow control data is stored in the history system of the MIDAS system, so that the long term stability of the experiment can be effectively verified. The slow control data is also fed into the main event data stream, to be available for offline analysis.

All data fed into the MIDAS system is accessible by the MIDAS distributed alarm system. This system allows upper or lower limits to be set on all slow control data in a flexible way through the MIDAS web interface. 
In the event of an alarm, shift crews can be notified through spoken alarm messages and contacted via mobile phones. Scripts can be triggered which put the whole experiment in a safe state in order to avoid damage from excessive temperatures or other dangerous conditions.
In addition to this MIDAS-based alarm system, an interlock system that is fully independent from the DAQ handles the most critical parameters of the apparatus (see \autoref{sec:InterlockSystem}). 

\section{Midas Slow Control Bus}
\label{sec:MSCB}

The MSCB system uses a serial differential bus for communication, with two data lines (positive and negative polarity) and a common ground. Over long distances, such as between crates, the physical standard for this bus 
is RS-485, running at a relatively low speed of \SI{115.2}{kbit/s} in half-duplex mode.
The slow speed makes this bus highly immune against improper termination or electrical interference, while the short commands of the MSCB protocol still allow the readout of many hundreds of nodes per second.
This optimised protocol allows the monitoring of many thousands of channels with repetition periods in the \SI{100}{ms} range, which is more than sufficient.

\begin{figure}
	\centering
		\includegraphics[width=0.48\textwidth]{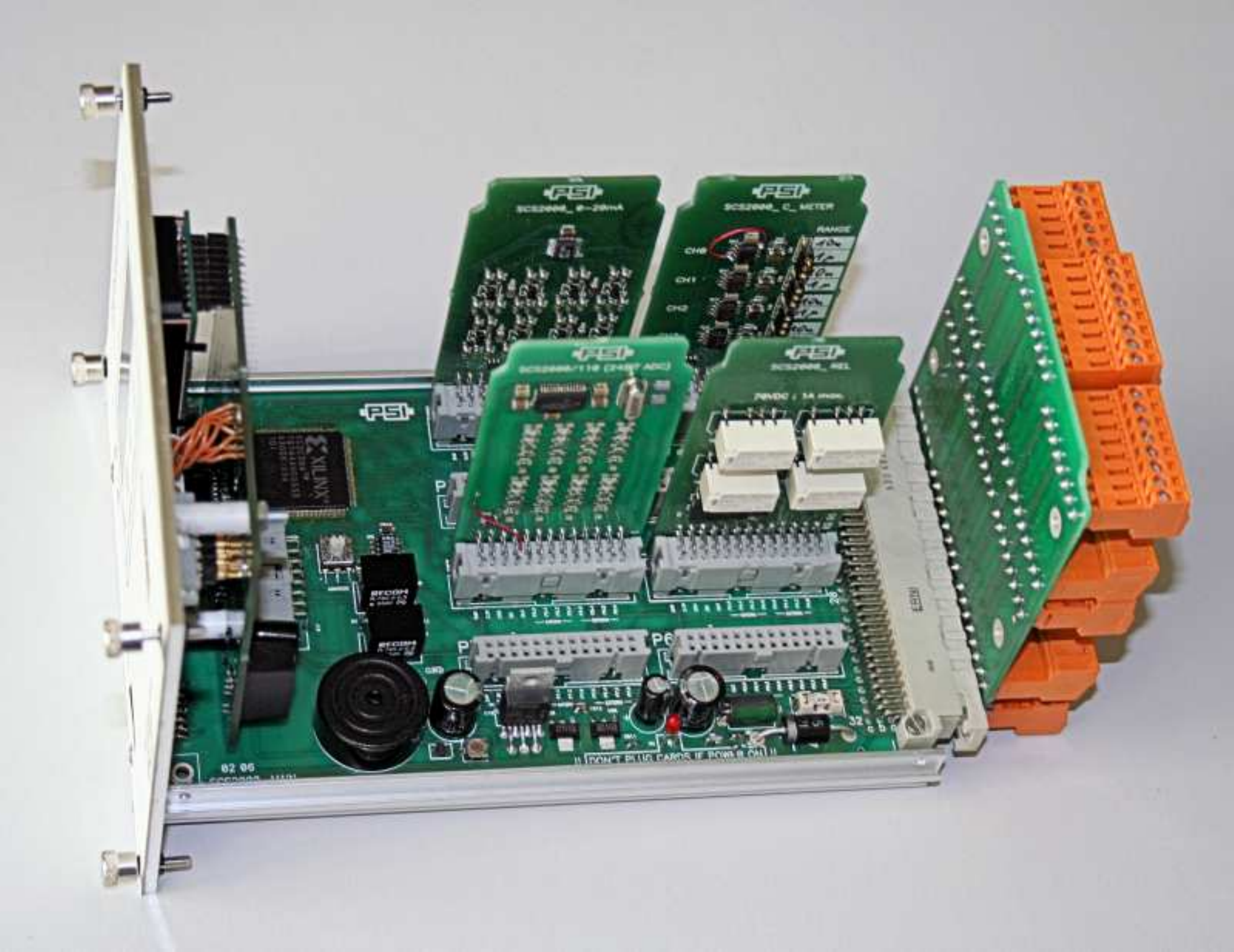}
	\caption{SCS-3000 unit as part of the MSCB slow control system. This unit
	has 64 input/output channels, which can be configured via plug-in boards as
	digital or analogue channels. Many plug-in boards exist already such as PT100
	temperature sensor readout cards, analogue high resolution inputs (24 bit
	resolution), valve control outputs and many more.}
	\label{fig:SCS3000}
\end{figure}

The MSCB bus uses a single-master, multiple-slave architecture, where all slave nodes on the bus only have to reply to requests sent by the master node, thus making the bus arbitration very simple. Many devices already exist for this system, such as the SCS-3000 units, as shown in \autoref{fig:SCS3000}. Since the system was developed at PSI, it can be quickly adapted to new hardware.

The MSCB nodes inside the experiment are either dedicated 8-bit microcontrollers or soft-core microcontrollers instantiated on the FPGAs, connected to the RS-485 bus via insulated transceivers to avoid ground loops and noise.
These microcontrollers perform local control loops, such as high-voltage stabilisation, power conversion or environmental control, and send measured values to the central DAQ system for monitoring.
Custom high-voltage boards mounted inside the magnet have an embedded microcontroller acting as an MSCB node, thus no high-voltage cables have to be fed from outside the magnet into the experimental volume (\autoref{sec:PixelHighVoltage}). 
The DC-DC converter controller boards (\autoref{sec:Powerboards}) also act as an MSCB node.

A dedicated slow control segment is connected to the environment sensors inside the magnet, and monitoring parameters such as the temperature and pressure of the helium flows, the humidity inside the cage, the magnetic field at various positions and the temperature of various detector components. The monitoring data processed by the detector ASICs and the FPGAs will primarily be read out through the main data stream, with the MSCB-based readout as a backup system.

\begin{sloppypar}
Microcontroller-based bus adapters are used to bridge between each RS-485 segment and optical fibres, which allows us to route all segments from inside the magnet to the outside world via optical fibres, where they are connected to the experiment Ethernet network. In this way, all MSCB nodes can be accessed from any computer connected to the experiment's Ethernet network. 
\end{sloppypar}

\subsection{Frontend board control}

All front-end boards are connected to the MSCB bus via 3.3~V RS-485 optically isolated transceivers.
Since the MSCB protocol is very simple, using only a few bytes for addressing, data and 
redundancy, its implementation requires less than 700 lines of C code.
This makes it possible to run the MSCB core inside a NIOS~II soft-processor on every FPGA 
used in the experiment.

Test implementations have shown that this needs only a few percent of the available FPGA 
resources, which can be easily accommodated.
Having a dedicated slow control link to all FPGAs in the experiment is a
powerful tool for debugging and configuration, since this allows the management
of the FPGAs even if the optical data links are down.

\section{ASIC Configuration}
\label{sec:PixelSlowControl}

\begin{figure}[tb!]
	\centering
		\includegraphics[width=0.46\textwidth]{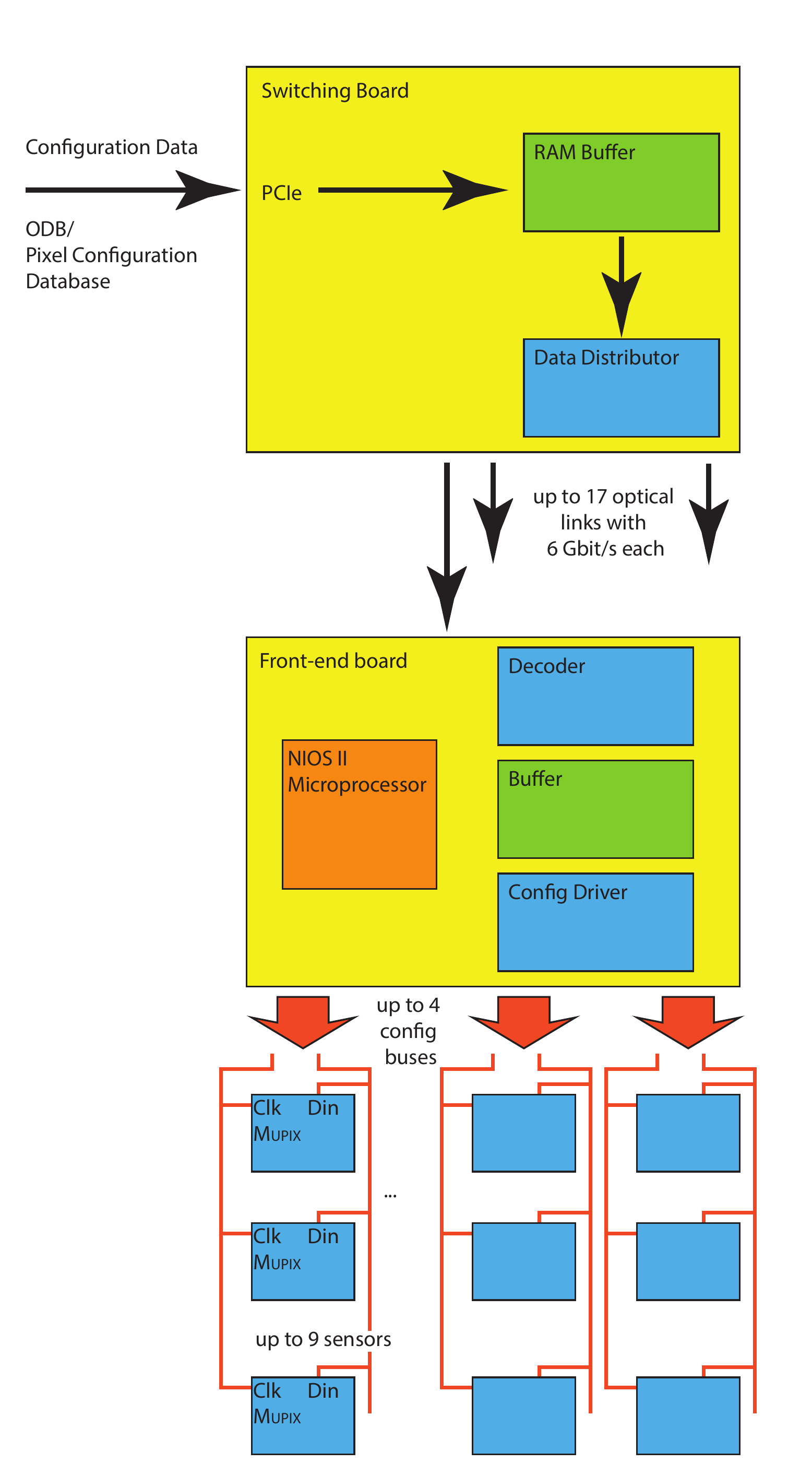}
	\caption{Data flow for the front-end ASIC configuration for the pixel detectors.}
	\label{fig:PixelConfigDataFlow}
\end{figure}

\begin{figure*}[t!]
	\centering
		\includegraphics[width=0.8\textwidth]{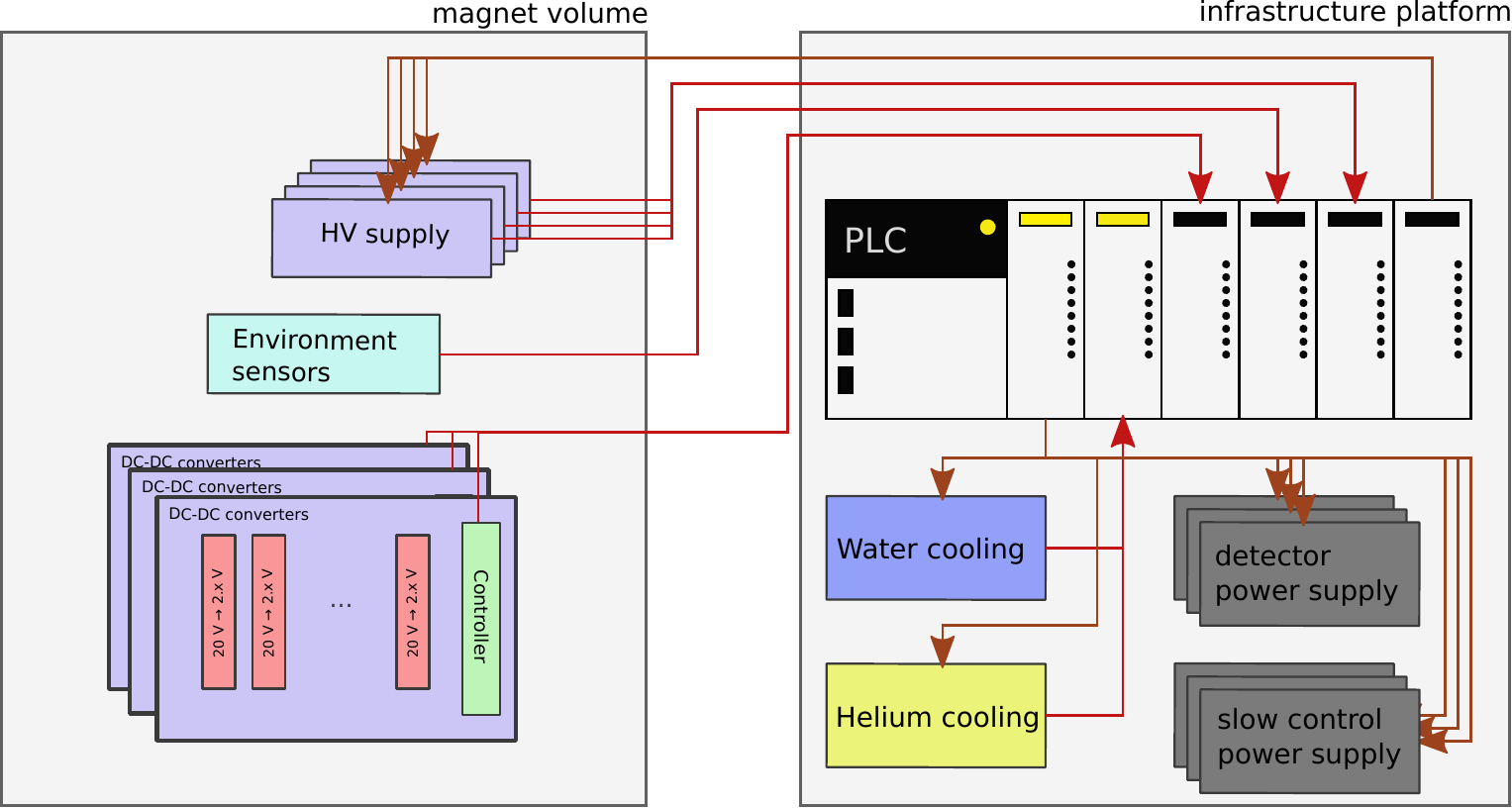}
	\caption{Layout of the Mu3e interlock system, directly interconnecting the safety critical sub-systems of the experiment. The PLC is a Siemens SIMATIC S7-series controller.}
	\label{fig:Interlock}
\end{figure*}

The configuration of the pixel detectors is a special case as it requires many millions of parameters, e.g.~the tune-DAC values for each pixel. 
Since this amount of data is considerably larger than the total for all other systems
($\sim$\SI{120}{MB} for the full phase~I detector),
an extension of the slow control system is implemented. 
A dedicated program manages, visualises and exchanges the pixel detector
configuration parameters between an optimised database and the pixel hardware. 
In this way the time required to configure the pixel detectors can be minimised,
while this program is still connected to the main DAQ system. 
It can be synchronised with run starts and stops, and can inject pixel
monitoring data periodically into the event data stream for offline analysis.
The regular slow control data stream contains a pointer to the relevant state
of the pixel configuration database.

The configuration of the individual pixel sensors is written via a dedicated differential
configuration bus with up to 9 sensors (one electrical group in the outer layers)
connected in parallel. 
This corresponds to approximately \SI{20}{Mbit} of configuration data, 
which in turn dictates the need
for configuration speeds above \SI{10}{MHz} in order to guarantee fast run starts.
Sensors on different ladders (different configuration buses) can and have to be 
programmed in parallel.
Slow control data output from the sensors is sent using the fast LVDS data 
link.
As the on-chip memory of the front-end FPGA is too small to hold the complete
configuration data for all connected sensors, it has to be delivered just in time
from the switching boards (see \autoref{sec:SwitchingBoards}). 
Bandwidth is not an issue (a
\SI{6}{Gbit\per s} optical downlink is available), but the data stream has
to be synchronised such that no buffering on the front-end is required, which
necessitates a careful interplay between the software driving the data into the
switching boards, the switching board firmware and the front-end firmware.
An overview of the data flow for the pixel configuration is shown in 
\autoref{fig:PixelConfigDataFlow}.

For the timing detector \mutrig ASICs, the same configuration path is used - the configuration
data is however much more compact than in the pixel case and can be stored in the MIDAS
ODB. 


\section{Interlock System}
\label{sec:InterlockSystem}

As \SI{10}{\kilo\watt} is dissipated in a small volume, continuously carried away by water and helium cooling systems, an additional interlock system fully independent from the MIDAS DAQ controls the critical parameters of the experiment.
This system returns the experiment to a safe state in case of an emergency or critical failure, and prevents unsafe transitions between operating modes requested by the user.
For example, the detector power can only be turned on when the helium and water cooling systems are fully functional.
The central controller of the interlock system is a commercial programmable Logic Controller (PLC) with fail-safe IO.
\autoref{fig:Interlock} shows a conceptual wiring diagram of this system.
It is located on one of the infrastructure platforms, and connected to the subsystems via closed-loop electrical circuits, which are decoupled from the detector and other slow control power circuits.
This system is designed as an additional safeguard; during normal operation all transitions are instigated by the MIDAS DAQ system.


\chapter{Data Acquisition}
\label{sec:DAQ}

\chapterresponsible{Nik}

\nobalance

\begin{figure*}
	\centering
		\includegraphics[width=1.00\textwidth]{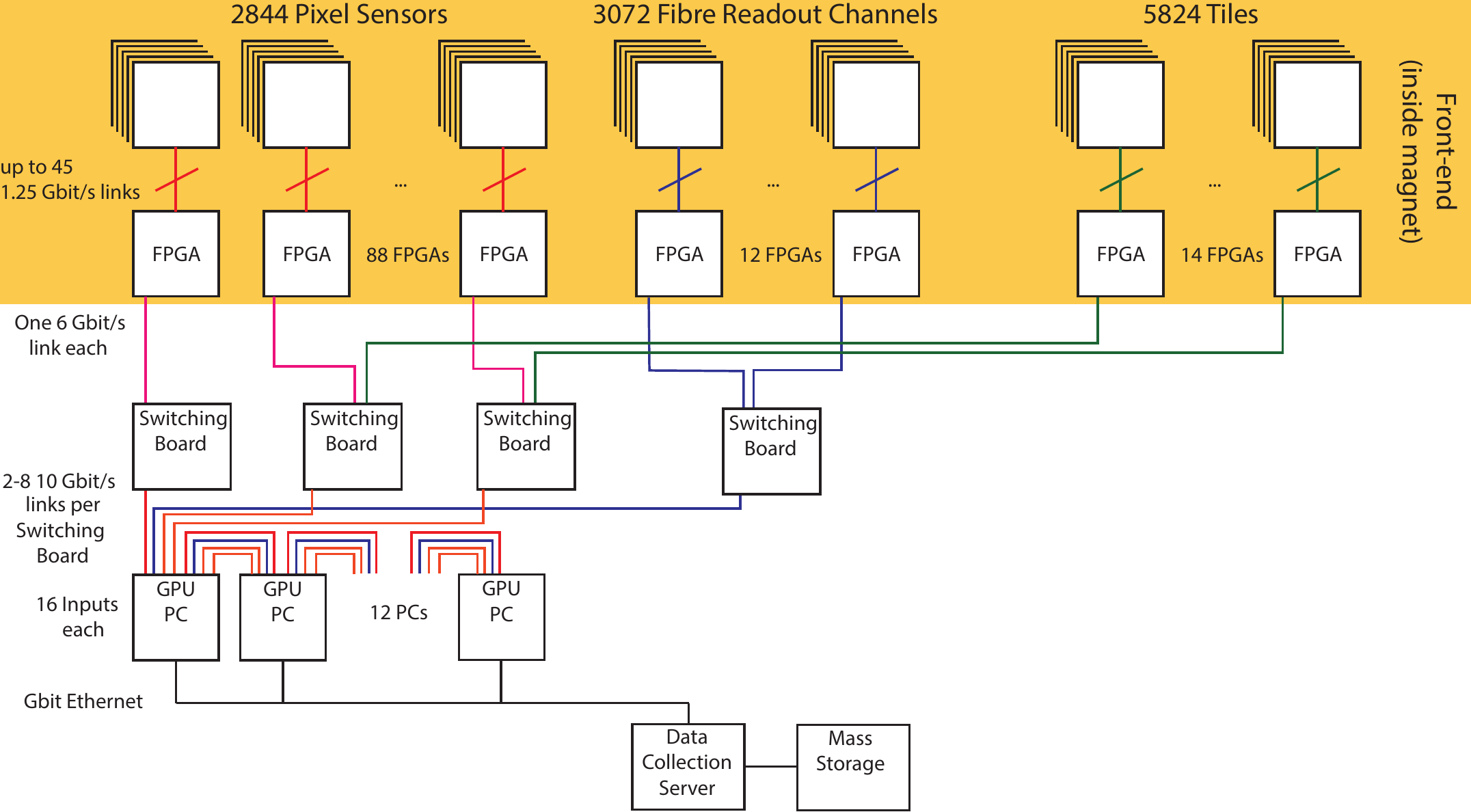}
	\caption{Overall Mu3e readout scheme.}
	\label{fig:RO_Scheme}
\end{figure*}

The Mu3e data acquisition (DAQ) system works without a hardware trigger on a push
basis, i.e.~the detector elements continuously send zero-suppressed hit 
information. The DAQ consists of three layers, namely front-end FPGAs, 
switching boards and the filter farm. 
The topology of interconnects is such that every farm PC receives the 
complete detector information for a certain time slice. 
See \autoref{fig:RO_Scheme} for an overview of the readout scheme.

\begin{figure*}
	\centering
		\includegraphics[width=1.00\textwidth]{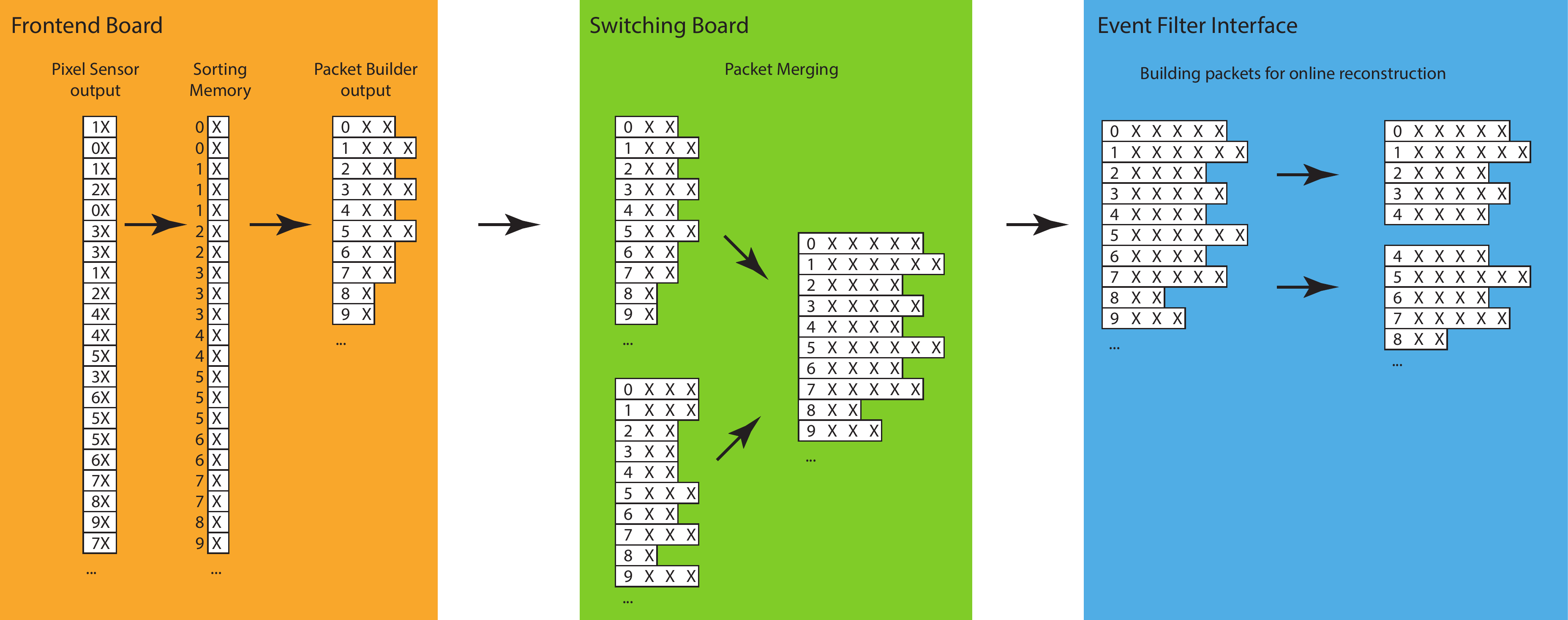}
	\caption{Schematic flow of pixel time information through the Mu3e readout system.
	Numbers stand for hit timestamps, X stands for the remaining hit information (address, charge).}
	\label{fig:Timesorting}
\end{figure*}

Hits in all subsystems are timestamped and the front-ends ensure that 
time-ordered information is forwarded to the rest of the readout system.
At the input to the farm PCs, data from several timestamps is merged to form
overlapping reconstruction frames, as shown in \autoref{fig:Timesorting}.
In this scheme, the latency of individual detector elements is not critical,
as long as the latency differences do not exceed the buffering capacity at
each step.

\section{Bandwidth Requirements}
\label{sec:Occupancy}

The bandwidth requirements of the data acquisition are largely determined by the
expected detector occupancy, as all the Mu3e subdetectors produce
zero-suppressed output.

\begin{sloppypar}
Occupancies have been estimated with the full simulation for a rate of muons
stopping on target of $\SI{1e8}{Hz}$, and pessimistically estimating the beam-related
background by assuming another $\SI{0.9e8}{Hz}$ of muons stopping along the
last metre of beam line.
\end{sloppypar}

\subsection{Front-end bandwidth requirements}
\label{sec:FrontEndBandwidthRequirements}

The pixel sensors contain electronics for hit detection, as well as time and
address encoding. 
The hits are then serialised and sent to the front-end FPGA board via a 
\SI{1250}{Mbit/s} low voltage differential signalling (LVDS) link. 

The sensors at the centre of the innermost layer have the highest occupancy, about
\SI{1.3}{MHz/cm^2}  or \SI{5.2}{MHz} per sensor.
The protocol implemented in the \mupix 8 prototype and all subsequent chips 
allows a maximum of 74\% of the sent words for hit information, the remaining
words are used for counters and status information. 
With 8~bit/10~bit encoding, this leads to a maximum hit bandwidth 
of \SI{740}{Mbit/s}, equivalent to \num{23d6} \SI{32}{bit} hits per link per second. 
This gives a safety factor of four even for the busiest sensors, which will use 
three parallel links.
The total bandwidth requirements for the phase~I pixel detector up to the
front-end boards are shown in \autoref{tab:PixelRONBumbersPhaseI}.

The average occupancy determines the bandwidth requirements, but 
fluctuations are also modelled in the simulation in order to optimise 
the system design.
In particular, online buffer sizes must be large enough to allow the 
latency required to absorb the highest expected peaks in hit rate.
We design the system to loose less than \SI{0.1}{\percent} of all hits
due to buffer overflows and detect and flag all possible overflows.

The \mutrig ASIC foreseen for both timing detectors will also output zero-suppressed
hit data with timestamps over a \SI{1250}{Mbit/s} LVDS link.
The average hit rate per channel of the fibre detector is estimated from the simulation
as \SI{620}{kHz}, with a hit size of \SI{28}{bits} \cite{Corrodi2018}.
With 32 channels per ASIC, this uses about \SI{700}{Mbit/s} of the link bandwidth,
limiting the acceptable dark count rate to roughly \SI{300}{kHz} per channel.

\begin{sloppypar}
The tile detector, operating at a relatively high threshold and an expected total hit
rate of roughly \SI{180}{MHz}, will contribute very little to 
the overall bandwidth requirements and is very far from saturating single channel
limits.
\end{sloppypar}

\begin{table*}[tb!]
\begin{center}

\small
\begin{tabular}{lrrlrrr}

\toprule
 & \#Sensor & Max hit & \multicolumn{1}{r}{Average hit} & \multicolumn{1}{l}{Chip$\rightarrow$FPGA} & \multicolumn{1}{l}{Chip$\rightarrow$FPGA} & \multicolumn{1}{l}{Front-end} \\
 & chips & rate & \multicolumn{1}{r}{rate} & link capacity & layer capacity & FPGAs \\
 & & \multicolumn{1}{l}{\num{d6}/Chip/s} & \multicolumn{1}{l}{\num{d6}/Layer/s}   & \multicolumn{1}{r}{Mbit/s} & \multicolumn{1}{r}{Gbit/s} & \multicolumn{1}{r}{} \\
 & &  &    & needed/available & needed/available & \\
\midrule
central station\\
~Layer 1 & 48 & 5.2 & \multicolumn{1}{r}{194} & 281/3750 & 10.5/180 & 4 \\
~Layer 2 & 60 & 5.2 & \multicolumn{1}{r}{195} & 281/3750 & 10.5/225 & 6 \\
~Layer 3 & 408 & 1.2 & \multicolumn{1}{r}{266} & 65/1250 & 14.4/510 & 12 \\
~Layer 4 & 504 & 1.2 & \multicolumn{1}{r}{248} & 65/1250 & 13.4/630 & 14 \\
recurl station\\
~Layer 3 U & 408 & 0.15 & \multicolumn{1}{r}{41} & 8.1/1250 & 2.2/510 & 12 \\
~Layer 4 U & 504 & 0.14 & \multicolumn{1}{r}{44} & 7.6/1250 & 2.4/630 & 14 \\
~Layer 3 D & 408 & 0.11 & \multicolumn{1}{r}{28} & 5.9/1250 & 1.5/510 & 12 \\
~Layer 4 D & 504 & 0.10 & \multicolumn{1}{r}{29} & 5.4/1250 & 1.6/630 & 14 \\
\midrule
Total & 2844 & \multicolumn{1}{l}{} & \multicolumn{1}{r}{1045} & \multicolumn{1}{l}{} & 56.4/3825 & 88\\
\bottomrule
\end{tabular}
\end{center}
\caption{Pixel front-end readout requirements ($10^8$ muon stops/s). The recurl station layers are labelled up- and downstream (U/D). The rates include protocol overhead and 8~bit/10~bit encoding, and assume \SI{32}{bit} hit size.}
\label{tab:PixelRONBumbersPhaseI}
\end{table*}

\subsection{Optical link bandwidth requirements}
\label{sec:OpticalLinkBandwidthRequirements}

\begin{table*}[tb!]
	\centering
		\begin{tabular}{lrrrr}
			\toprule
			Subdetector & Max. hit rate/FPGA & Hit size & Bandwidth needed  & FPGAs\\
									& MHz               & Bits     & Gbit/s 					 & \\
			\midrule
			Pixels 			& 58								& 48			 & 4.6 							 & 88\\
			Fibres			& 28								& 48			 & 2.3 							 & 12\\
			Tiles				& 15								& 48			 & 1.2               & 14\\
			\bottomrule
		\end{tabular}
	\caption{FPGA bandwidth requirements. For the fibre detector, clustering in the
	front-end FPGA is performed. For the bandwidth, \SI{75}{\percent} protocol efficiency and 
	8~bit/10~bit encoding are assumed. The pixel hit size assumes, conservatively, that
	the full hit and address information including time is transmitted for each hit.
	This can be reduced by time-grouping hits and encoding parts of the address in
	the link.}
	\label{tab:FPGABandwidthRequirements}
\end{table*}

The hits are collected on a front-end FPGA and transmitted off the detector using optical
links. The corresponding bandwidth requirements are listed in 
\autoref{tab:FPGABandwidthRequirements}. For the fibre detector, clustering is assumed to take place
on the front-end FPGA, although unclustered data could be sent out using twice the number of front-end boards with
two optical links each.

\begin{table}
	\centering
		\begin{tabular}{lrr}
			\toprule
													& Rate     & Bandwidth\\
													& MHz      & Gbit/s\\
			\midrule
			Central Pixels 			& 905	 		 & 58 \\
			Upstream Recurl 		&	191 & 12 \\
			Downstream Recurl 	& 131  & 8.4\\
			Fibres							&	337      & 21.5 \\
			\midrule
			Total								& 1564 		 & 100 \\ 
			\bottomrule
		\end{tabular}
	\caption{Switching board bandwidth requirements. \SI{48}{bit} hit size and
	\SI{75}{\percent} protocol efficiency are assumed.}
	\label{tab:SwitchingBoardBandwidthRequirements}
\end{table}

Four switching boards collect the data from the front-ends, one for the
central pixel detector, one each for the up- and downstream recurl stations
(pixels and tiles) and one for the fibre detector. The corresponding bandwidths
passing through these boards are listed in \autoref{tab:SwitchingBoardBandwidthRequirements}.

\section{Front-end FPGA Boards}
\label{sec:FrontEndFPGABoards}

\begin{figure}[tb!]
	\centering
		\includegraphics[width=0.49\textwidth]{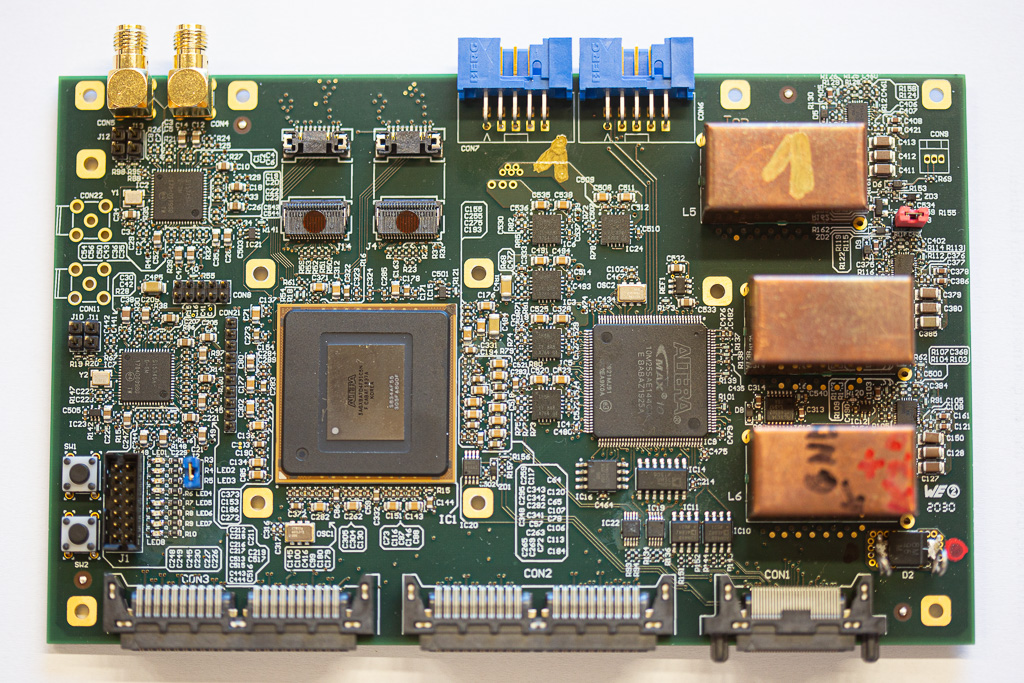}
	\caption{Prototype front-end board based on an Intel Arria~V FPGA.
	The FPGA in the centre left is surrounded by connectors to the crate backplane
	(leading to the detector ASICs) at the bottom, a Intel MAX10 CPLD for configuration
	and monitoring in the centre right, clocking circuitry at the left, two
	connectors for Firefly optical transceivers on the top left, blue JTAG connectors
	for programming on top and the DC/DC converter circuitry on the right. The copper
	boxes contain and shield the air coils.}
	\label{fig:frontendboard}
\end{figure}

The front-end boards have to collect the data sent from either the \mupix or the 
\mutrig chip, sort and package it, and then forward it to the switching boards
on a fast optical link.
In the case of the fibre \mutrig data, preliminary clustering will be applied
in order to reduce the data rate taken up by dark counts.
In addition, the boards have to provide the sensors with control signals and monitor the
environment.
The space constraints inside the magnet necessitate small, highly
integrated boards incorporating FPGAs and optical modules with a small footprint
and limited power consumption. A working prototype can be seen in \autoref{fig:frontendboard}.

The boards feature an Intel Arria V FPGA\footnote{Model 5AGXBA7D4F31C5} for data processing
as well as a flash-based Intel MAX10 FPGA\footnote{Model 10M25SAE144C8G} for configuration
and monitoring. For the optical data transmission we use Firefly transceivers by Samtec
(ECUO-B04-14), each of which provides four transmitting and four receiving links
at up to \SI{14}{Gbit/s} in a very small footprint 
(\SI{20.3}{mm}$\times$\SI{11.25}{mm}) at a power consumption of roughly
\SI{1}{W}. A single link per board is sufficient for the bandwidth requirements
of phase~I; we nevertheless foresee the option to install two Fireflys and thus
obtain 8 outgoing links. The incoming links are used for the clock and reset
distribution (see \autoref{sec:ClockDistribution}) as well as the slow control and pixel configuration.
Clocks received by the Fireflys are conditioned by two Si5345 jitter attenuator/
clock multiplier chips and forwarded to the FPGAs as well as the detector ASICs.
The front-end firmware receives detector data, performs time-sorting and multiplexes
the data from all connected ASICs to an optical link. Synchronisation and run transitions
are controlled by the reset link, as described in~\autoref{sec:RunStartSynchronisation}.

The boards are connected to a backplane, which forwards the detector signals and
provides control and monitoring signals via a separately powered crate controller.
The boards are cooled by a custom-made aluminium cooling plate connected to the
water-cooled frame of the crate via a heat pipe.

\subsection{Slow-control and configuration integration}
\label{sec:SlowControlIntegration}

For slow control and pixel configuration data, two paths are foreseen.
Firstly, surplus bandwidth
on the optical links to and from the switching board can be used, which is
especially useful for large volume data such as pixel tune values.
Secondly, a separate differential line for use of the MSCB protocol
(see \autoref{sec:SlowControl}) is foreseen for
monitoring the status of optical links, switching power, etc..
The interface to MSCB and the slow control-related tasks on the FPGA will be
implemented in a NIOS~II soft processor core \cite{Altera2015} on the FPGA~\cite{Mueller2019}.

The FPGA firmware can be updated by writing a Serial Peripheral Interface (SPI)
flash memory from either the optical slow control link or the MSCB connection
via the MAX10 FPGA. On power-up or a reconfiguration command, the MAX10 then
reprograms the Arria~V FPGA.

\section{Read-out Links}
\label{sec:ReadOutLinks}

\begin{sloppypar}
Electrical links are used between the detector ASICs and the front-end FPGAs,
all other data links are optical. The data links are complemented by a (smaller) number 
of slow control links in the opposite direction~\cite{Koeppel2019}.
\end{sloppypar}

The data from the \mupix and \mutrig chips will be transmitted to the front-end 
FPGAs via LVDS links at \SI{1250}{Mbit\per\s}.
The link is physically implemented as a matched differential pair of 
aluminium traces on the sensor HDI (only pixel detector), followed by a micro twisted-pair cable connected
to the detector side of the backplane, see \autoref{sec:signalPath}.

In the system, there are two types of optical high speed data links.
The first one goes from the front-end FPGAs to the switching boards, the
second from the switching boards to the FPGA PCIe boards in the event filter farm
PCs.
The optical links from the front-end FPGAs to the switching boards have a
bandwidth of \SI{6}{Gbit\per\s}, which fits well with the FPGA specifications.
Each FPGA has nine fast transceiver blocks, which connect to the
Firefly optical assemblies.
The laser has a wavelength of \SI{850}{\nm} and the optical fibre is
of 50/125 multi-mode OM3 type, since this is a standard both in industry and in
particle physics detector readout.

The links from the switching boards to the filter farm are implemented as \SI{10}{Gbit\per\s} high speed links.
The PCIe FPGA board is fitted with four quad small form-factor pluggable (QSFP+)
optical modules.

All the links have been tested using the development hardware and were
found to have bit error rates low enough for stable and consistent running of the
experiment \cite{Corrodi2014, Grzesik2014}; typically, no errors were found in
a few days of running, leading to upper limits on the bit error rate of \num{1e-14}
down to \num{1e-16}.

\section{Switching Boards}
\label{sec:SwitchingBoards}

\begin{figure}[tb!]
	\centering
		\includegraphics[width=0.49\textwidth]{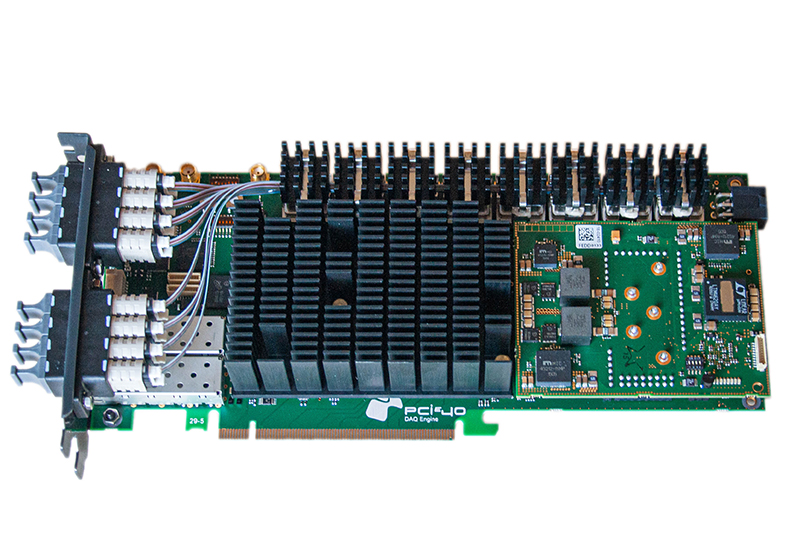}
	\caption{\emph{PCIe40} board as developed for the LHCb
and ALICE upgrades \cite{Durante:2014nva, Durante:2015gza} 
and employed as switching board in Mu3e. The large
Arria~10 FPGA with two PCIe Gen3 8 lane interfaces is complemented by
48 fast optical receivers and 48 fast optical transmitters.}
	\label{fig:switchingboard}
\end{figure}

The main task of the switching boards is to act as switches between the front-end
FPGAs on the detector and the online reconstruction farm, thus allowing each
farm PC to see data from the complete detector.
The board design and choice of FPGAs is dominated by the number of fast links
required.

We use four \emph{PCIe40} boards (see \autoref{fig:switchingboard}) developed for the LHCb
and ALICE upgrades at the LHC \cite{Durante:2014nva, Durante:2015gza, Cachemiche2015}.
These boards provide up to 48 full duplex optical links at up to \SI{10}{Gbit/s},
plus two eight lane PCIe 3.0 interfaces bundled to a sixteen lane interface by a
switch. The FPGA is an Altera Arria~10.
The PCs hosting the boards are used to store and transmit the extensive pixel
configuration and tuning data as well as the timing detector ASIC configuration
via PCI express, and link the boards to the experiment control and
monitoring system via standard Ethernet. 

The firmware for the switching board has to
receive several data streams in parallel, merge them synchronously and then
forward them to the event filter. Additional firmware is needed in the case of the fibre tracker, where hits
from both ends of the fibre have to be matched in order to suppress dark counts.

\section{Event Filter Interface}
\label{sec:EventFilterInterface}

\begin{figure}[tb!]
	\centering
		\includegraphics[width=0.49\textwidth]{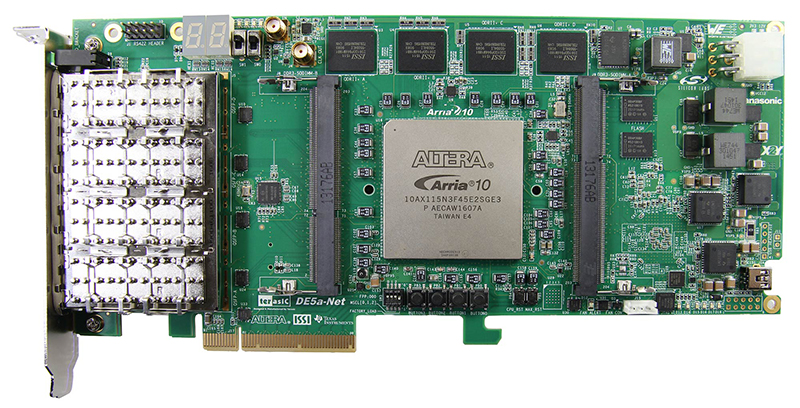}
	\caption{The DE5a NET board built by Terasic Inc.~and used as a 
	receiving board in the filter farm PCs. The Arria~10 FPGA with a
	PCIe Gen3 8 lane interface is complemented by a 16 duplex fast optical links
	and several GB of DDR4 memory.}
	\label{fig:receivingboard}
\end{figure}

The filter farm PCs are equipped with FPGA boards in PCIe slots and optical
receivers. The boards are commercial DE5a NET boards 
(see \autoref{fig:receivingboard}) built by Terasic Inc.
They are equipped with four QSFP+ quad optical transmitters/receivers, 
two laptop-compatible DDR3/DDR4 memory interfaces and a large Altera Arria~10 FPGA
with an 8 lane PCIe 3.0 interface.
This FPGA performs the event building and buffering, and also allows simple
clustering, sorting and selection algorithms to be run.
The event data is then transferred via Direct Memory Access (DMA) over the
PCIe bus\footnote{Note that PCIe is actually not a bus protocol, but offers switched point-to-point connections.
The \emph{bus} designation is due to
the software-side backwards compatibility to the original PCI bus interface.}
to the main memory of the filter farm PC and subsequently copied to
the memory of a GPU, where the fitting and vertex selection algorithms
are run.
The GPU then posts IDs of selected events to the main memory of the PC, which
triggers a transfer of the respective data from the FPGA buffer memory via 
the PC main memory and Ethernet to the central DAQ computer running the MIDAS software.
At that computer, the data streams from the farm PCs are combined into a 
single data stream, combined with
various slow control data, compressed and stored.
The maximum data rate over an eight-lane PCIe 3.0 bus is
$\SI{7.88}{Gbyte/s}$ of which we are able to use $\SI{4.8}{Gbyte/s}$ for user
data, amply sufficient for phase~I.

The GPU boards will be obtained commercially as late as possible in order to
profit from the ongoing rapid development and sinking prices.
Current high-end GPUs already have enough floating point capability
for high rate running.
Newer boards are, however, expected to offer higher memory bandwidth and better caching.
For example, between the GTX~680 and the GTX~980 GeForce GPUs, both the compute 
power and the copy speed increased by 30~\% \cite{NVIDIAGTX680,AMDGCN}.
The GTX~1080Ti cards we obtained in 2017 were sufficient to run the full Phase~I
selection load on 12 nodes~\cite{DissVomBruch2017}.

The farm PCs are hosted in individual rack-mounted tower casings,
ensuring enough space for the FPGA board, the high-end GPU 
and a custom clock receiver board~\cite{Wagner2018}
whilst allowing 
for air cooling. Each tower consumes around $\SI{0.7}{kW}$, so active
cooling of the racks and the counting house is necessary.

\section[Run Control, Data Storage]{Run Control, Data Collection and Storage}
\label{sec:DataCollectionAndStorage}

\subsection{The MIDAS System}
\label{sec:TheMIDASSystem}

The filter farm outputs selected events at a data rate of the order of 
\num{50}-\SI{100}{MBytes/s} in total. 
This data rate is low enough to be collected by a single PC
connected to the filter farm by common Gbit Ethernet and written to local disks.
Then the data is transferred to the central PSI computing centre, where it
is stored and analyzed. 
For the central DAQ, the well-established
MIDAS (Maximum Integrated Data Acquisition System) \cite{MIDAS:2001}
software package is used. This software is currently used in several
major experiments such as the T2K ND280 detector in Japan \cite{Abe:2011ks}, 
ALPHA at CERN and the MEG experiment at PSI \cite{MEG}. 
It can easily handle the required data rate, and
contains all necessary tools such as event building, a slow control system
including a history database and an alarm system. 
A web interface allows the control and monitoring of the experiment through the 
Internet. 
The farm PCs use MIDAS library calls to ship the data to the central DAQ PC. 
The framework also offers facilities to send configuration parameters from a central
database (the ``Online DataBase'' or ODB) to all connected farm PCs and to
coordinate common starts and stops of acquisition (run control).

For the purpose of monitoring and data quality control of the experiment, the
MIDAS system offers the capability to tap into the data stream to connect analysis and
graphical display programs. 

\subsection[Run start/stop]{Run start/stop synchronisation}
\label{sec:RunStartSynchronisation}

In traditional DAQ systems, starting and stopping
is controlled by enabling and disabling trigger signals.
In a streaming system such as in Mu3e this is not an option, and great care has
to be taken to synchronise data across the complete detector at run
start and ensure that the frame numbers in all subsystems are in agreement.

To this end, a global reset signal is distributed together with the global clock.
At the front-end, the reset signal is forwarded to the pixel sensors and there
sets the timestamp counters to zero as long as it is on (note that the pixel
sensors cannot be inactivated, so even during a reset they will still collect,
process and send hits, however all with timestamp zero).
At the start of a run, the reset signal is released synchronously for all sensors, which
then start counting timestamps.
The front-end firmware will ignore all hits with timestamp zero at the beginning
of the run and start sending packets into the switching network as soon as
non-zero timestamps arrive.
All subsequent stages in the network then synchronise on the first packet and
from then on stay in sync using consistent packet numbering.
A similar synchronization mechanism is implemented for the \mutrig ASICs.

At the end of the run, the global reset goes high and the front-end continues
forwarding packets until timestamp zero is detected.


\chapter{Simulation}
\label{sec:Simulation}

\chapterresponsible{Nik}

\nobalance

This chapter describes the Geant4 \cite{Allison:2006ve, Agostinelli2003250} based
simulation used to study and optimise the detector design, to develop the
reconstruction code and to estimate signal efficiency and background rates.

The Mu3e software stack consists of the simulation described here, which includes
generators for many different muon decays, the track reconstruction described in the
following chapter, a vertex fit program and a range of analysis codes. Besides
Geant4, ROOT \cite{Brun1997}, which is used for storage, histogramming and 
related analysis tasks, is the other major external library used. Core code is 
written in C++, python is used for some of the analysis and plotting code.

\section{Detector Geometry}

\begin{figure*}[t!]
	\centering
		\includegraphics[width=1.00\textwidth]{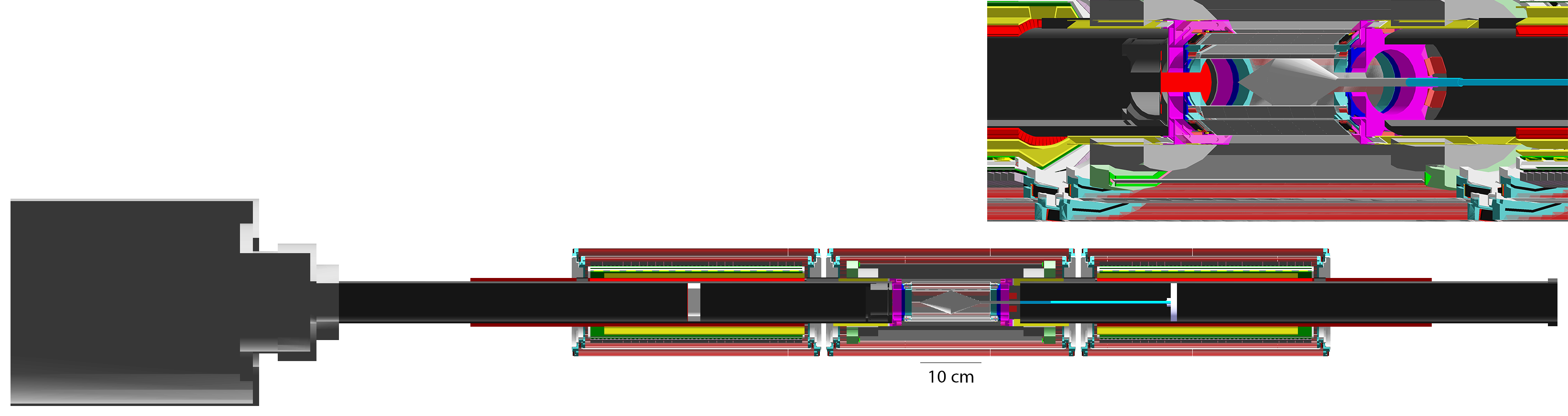}
	\caption{Side view of the simulated phase~I detector cut along the beam axis,
			displaying the volume with detailed geometry simulation. The insert
			in the top right shows a perspective view of the central part of the 
			simulated phase~I detector cut open at $x =$\SI{-19.0}{mm} to give
			an impression of the level of details implemented. Colours are used to
			differentiate simulation volumes.}
	\label{fig:phaseIb_side_cut}
\end{figure*}

The simulated detector geometry closely follows the planned detector geometry
described in earlier chapters.
The simulated volume extends for three metres in all directions from the target
centre. The magnet metal and the surrounding volume is only used in the
cosmic ray simulation.
\autoref{fig:phaseIb_side_cut} shows the
simulated detector geometry.

\subsection{Beam Delivery}
\label{sec:SimBeamDelivery}

In the detector simulation, the beam starts $\SI{1}{m}$ in front of the target
inside the beam pipe. Beam particles are generated with a profile and
momentum spectrum taken from the beam simulation at the same point.
The beam passes a \SI{600}{\micro\meter} Mylar moderator followed by a thick lead 
collimator removing particles undergoing large angle scattering in the moderator. It then
exits the beam vacuum through a \SI{35}{\micro\meter} vacuum window, the holding
structure of which serves as the final collimator.

\subsection{Target}

The target is simulated as a hollow Mylar double cone supported from the 
downstream side by a thin carbon fibre tube, see also 
\autoref{sec:Target}.

\subsection{Pixel Detector}

The simulated geometry of the pixel detector includes the sensor, the HDIs
(with an average trace density assumed and represented as thinner metal layers)
and the polyimide support structure.
The plastic endpieces and support wheels are also simulated in detail, including
flex prints, interposers and screws.

\subsection{Scintillating Fibres}
\label{sec:SimulationDetectorFibres}
The fibre ribbon simulation implements fibre shape, cladding thickness, 
staggering as well as optional fibre coatings and glue.
The fibres are matched to SiPM arrays at both ends.
Parameters of the geometry can be easily changed to study different options.
 
The baseline setting consists of 12 ribbons in 3 layers. Each layer consists 
of 128 round \SI{250}{\micro\meter} thick fibres read out by SiPM arrays with 
\SI{250}{\micro\meter} column width.

\subsection{Tile Detector}

In the simulation, the tile detector consists of the scintillating tiles, 
SiPMs, two layers of PCBs hosting the SiPMs and the readout chips, respectively, 
as well as the support structure.

\section{Magnetic Field}

The magnetic field in the simulation is taken from a cylindrically symmetric field
map with \SI{10}{mm} step size calculated by the magnet manufacturer.
Linear interpolation is used between the field map grid points.
It will be replaced by a measured map as soon as this becomes available.

\section{Physics Processes}

\subsection{Multiple Coulomb Scattering}

Multiple Coulomb scattering is the main limiting factor for the resolution of
the experiment; an accurate simulation is thus crucial.
The Urban model \cite{Urban:2006kd}, as implemented and recently 
improved \cite{Ivanchenko:2019ovi} in Geant4, is used. Alternative models are also
implemented, including one derived from the results of a dedicated study
of multiple Coulomb scattering in thin silicon at the DESY electron
test beam \cite{Berger:2014fsa}. This however still needs to be validated
at the low momenta expected in Mu3e.

\subsection{Muon Decays}
\label{sec:Simulation:MuonDecays}

Geant4 implements the Michel decay including polarization of both the muon and
the positron based on Scheck and Fischer \cite{Scheck:1977yg, Fischer:1974zh}. 
The radiative decay of the muon in Geant4 was implemented by the TWIST
collaboration \cite{Depommier2001} based on Fronsdal et al.~\cite{Fronsdal:1959zzb}.
This code has been adapted for the simulation of the Mu3e experiment
using the differential branching fraction provided by Kuno and Okada~\cite{Kuno:1999jp}.

An unified description of radiative corrections
for photons below a soft cut-off and of photons that are tracked within Geant4
above a cut-off  has been implemented based on calculations by Fael, Mercolli and Passera 
\cite{Fael:2015gua}.

The radiative decay with internal conversion is simulated using the
matrix element of Signer et al.~\cite{Pruna:2016spf}, with the option of
using the NLO version \cite{Pruna:2016spf, Fael:2016yle} when the accuracy 
is required.

\subsubsection{Signal}

\begin{sloppypar}
The signal kinematics are highly model-dependent, see \autoref{sec:DecayMu3e}. 
If not stated otherwise, we have used three particle phase
space distributions in the simulation, following the practice of SINDRUM and
earlier experiments.
We have also implemented the general matrix element by Kuno et al.~\cite{Kuno:1999jp}
in order to study the kinematics of different decay dynamics.
\end{sloppypar}

\subsubsection{Special Decays}
In order to study accidental background whilst factoring out timing and vertex suppression, the simulation code allows for more than one muon to decay at a single vertex. This is beneficial for studying the overlap of an internal conversion and a Michel decay.


\subsubsection{Cosmic Muons}

As the detector alignment will rely in part on the high momentum tracks of
cosmic ray muons, we have implemented a cosmic muon generator
based on the spectrum and angle parametrisation of Biallass and Hebekker 
\cite{Biallass:2009ev}.

\section[Time Structure and MC Truth]{Time Structure and Truth Information}
\label{sec:Simulation:TimeStructure}

The Mu3e experiment operates with a quasi continuous beam, which has
necessitated adaptations of the Geant4 package in order to take into account
particles crossing boundaries of reconstruction frames.
For every interaction with active detector material, both the particle of
origin and the sequence of interactions is saved, we thus have the full 
simulation truth available at every level of reconstruction and analysis.

\section{Detector Response}
\label{sec:Simulation:DetectorResponse}

\subsection{Pixel Detector Response}
\label{sec:PixelDetectorResponse}

The response of the pixel detector is simulated by either setting a
threshold on the charge deposited or by defining a single hit efficiency, which
is then applied by randomly discarding hits.
Noise is simulated by randomly creating extra hits at an adjustable rate.

The simulation does include effects of charge sharing between
pixels. $\delta$-electrons are simulated if they have a range above \SI{50}{\micro \meter},
the associated (large) clusters should thus be correctly simulated.
The sensor response has been modeled according to the measured properties in pixel sensors characterisation studies.

\subsubsection{Pixel Readout Simulation}
\label{sec:ReadoutSimulation}

The readout of the pixel detector is not strictly in order of timestamp (see \autoref{sec:mupix_design}) and
very large clusters of hits can lead to overflows in the sorting algorithm on the front-end FPGA (see \autoref{sec:FrontEndFPGABoards}).
These effects are simulated by treating each column as a queue, into
which hits are pushed at creation.
A fixed number of hits is then removed from these column queues for every time slot.
Hits are time-sorted in a separate programme and those that are too far out-of-time
are dropped; alternatively we can run a bit-accurate simulation of the front-end board firmware. 
We do currently not simulate the dead-time caused by hits stored in the pixel
cell.

 \subsection{Fibre Detector Response}
 \label{sec:FibreDetectorResponse}

In a first step the response of the scintillating fibres to an incident particle is simulated.
Since simulating single photon propagation inside fibres is not feasible in the main simulation, the response of the scintillating fibres is parametrised.
The number of arriving photons at both fibre ends can either be parametrised in deposited energy ($E_{dep}$) and hit position or simply generated according to measured efficiencies.

In a second step the SiPM response to the arriving photons is simulated and
the distribution of photons into the different SiPM cells is modelled.
The main parameter for this process is the photon distribution at the fibre ends and propagation in the epoxy layers before the SiPM active layer as well as optical cross-talk. 
The SiPM response depends on an adjustable photon detection efficiency (PDE) and is mixed with a constantly present dark rate and its own pixel to pixel cross-talk.
The time distribution of the detected events is based on the measured time resolution 
of fibre ribbons and photon time of flight in the fibres.
In a last step pile-up events are merged.

\subsection{Tile Detector Response}
\label{sec:TileDetectorResponse}

The tile detector will record the timestamps of the scintillation signals, as well as the energy deposition in the tiles, which is proportional to the number of scintillation photons.
The scintillation process and photon propagation are not simulated, in order to maintain a reasonable computation time.
Instead, the response characteristics of the tile, including the readout electronics, is parametrised, using the true timestamp and energy deposition 
of a hit as input. 
The response is described by the following parameters: time resolution, energy resolution, jitter of the readout electronics, channel dead-time and energy threshold.

A minimum energy deposition is required for a signal in the tiles to be detected by 
the readout electronics.
This corresponds to the energy threshold of the \mutrig chip, which is assumed to be roughly $E_{thresh} = \SI{0.1}{\MeV}$.
Due to the linearised ToT method implemented in the \mutrig chip, the digitised energy information is approximately proportional to the energy deposition in the tile.
The energy deposition of consecutive hits (pile-up events) which occur within the dead-time of the channel is assigned to the original hit.
This reflects the behaviour of the \mutrig chip.
The channel dead-time is determined by two parameters: the intrinsic dead-time of the \mutrig TDC and the dead-time related to the ToT of the analogue input signal.
The time resolution is parametrised by the intrinsic jitter of the \mutrig chip and the energy dependent resolution of the tile.


\chapter{Reconstruction}
\label{sec:Reconstruction}

\chapterresponsible{Alex}

\nobalance

\renewcommand {\phi}{\varphi}
\newcommand   {\lam}{\lambda}

\def\phims {\phi_{\mathsmaller{MS}}}
\def\lamms {\lam_{\mathsmaller{MS}}}
\def\sigms {\sigma_{\mathsmaller{MS}}}

The reconstruction algorithm has to efficiently identify the tracks
of particles from muon decays, while dealing effectively with the combinatorial background 
to keep the rate of incorrectly reconstructed tracks to an acceptable level.
The main challenges are the high event rate and resulting occupancy, and the 
curvature of trajectories of  low momentum particles in the $\SI{1}{\tesla}$ magnetic field.
Particle trajectories can make several turns in the detector,
and hit combinations can span distances of more than 0.5 meter with
hits on opposite sides of the detector.
This is of particular importance for the determination of the direction of motion
and therefore the charge of the particle.
Furthermore, fully reconstructing tracks is critical in correctly applying the information from the timing systems.

As the detector readout is triggerless, all muon decays have to be fully
reconstructed in the filter farm, setting high demands on the speed
of the online track reconstruction algorithm.

Multiple Coulomb scattering (MS) in the detector layers
is the dominant uncertainty.
The track finding and initial fitting is thus built around
a fast three-dimensional MS fit,
which is based on fitting the multiple scattering angles
at the middle hit position in a hit triplet combination
(see \cite{Berger:2016vak} for a detailed description).
In the most basic implementation of the fit,
spatial uncertainties of the hit positions are ignored.
This is a good approximation in the case of Mu3e, as the pixel
resolution uncertainty (${80 / \sqrt{12}}\approx \SI{25}{\um}$)
is much smaller than that from multiple scattering (typically several hundred \si{\um}).

To achieve the best possible resolution,
a general broken line
(GBL) fit~\cite{Blobel:2011az,Kleinwort:2012np} can be used.
This technique determines hit positions and scattering angles simultaneously
and also incorporates energy loss in the detector material, 
but requires knowledge of the assignments of hits to tracks
from a preceding linking step as well as an approximate track trajectory.
Therefore, it can only be used as a final step.
Currently a GBL fit is used for detector alignment
(see \autoref{sec:DetectorAlignment}), and it will be used in offline analysis.

The track finding and fitting studies presented in the following are all
based on a fast MS fit that also implements energy-loss corrections
and takes into account hit position uncertainties.
Events are generated with the full Geant4 simulation (see \autoref{sec:Simulation}).
The beam intensity is set such that \num{1E8} muons decay in the target
region per second, corresponding to an optimistic estimate for the rate
achievable at $\pi$E5.
The continuous data stream is divided into consecutive time frames
that are reconstructed independently.
In these studies a frame size of \SI{50}{\ns} is used.
Studies of the tracking acceptance and efficiency are performed on a sample
without simulated signal decays.

\section{Track finding}

\begin{figure}
\centering
    \includegraphics[width=1.0\linewidth]{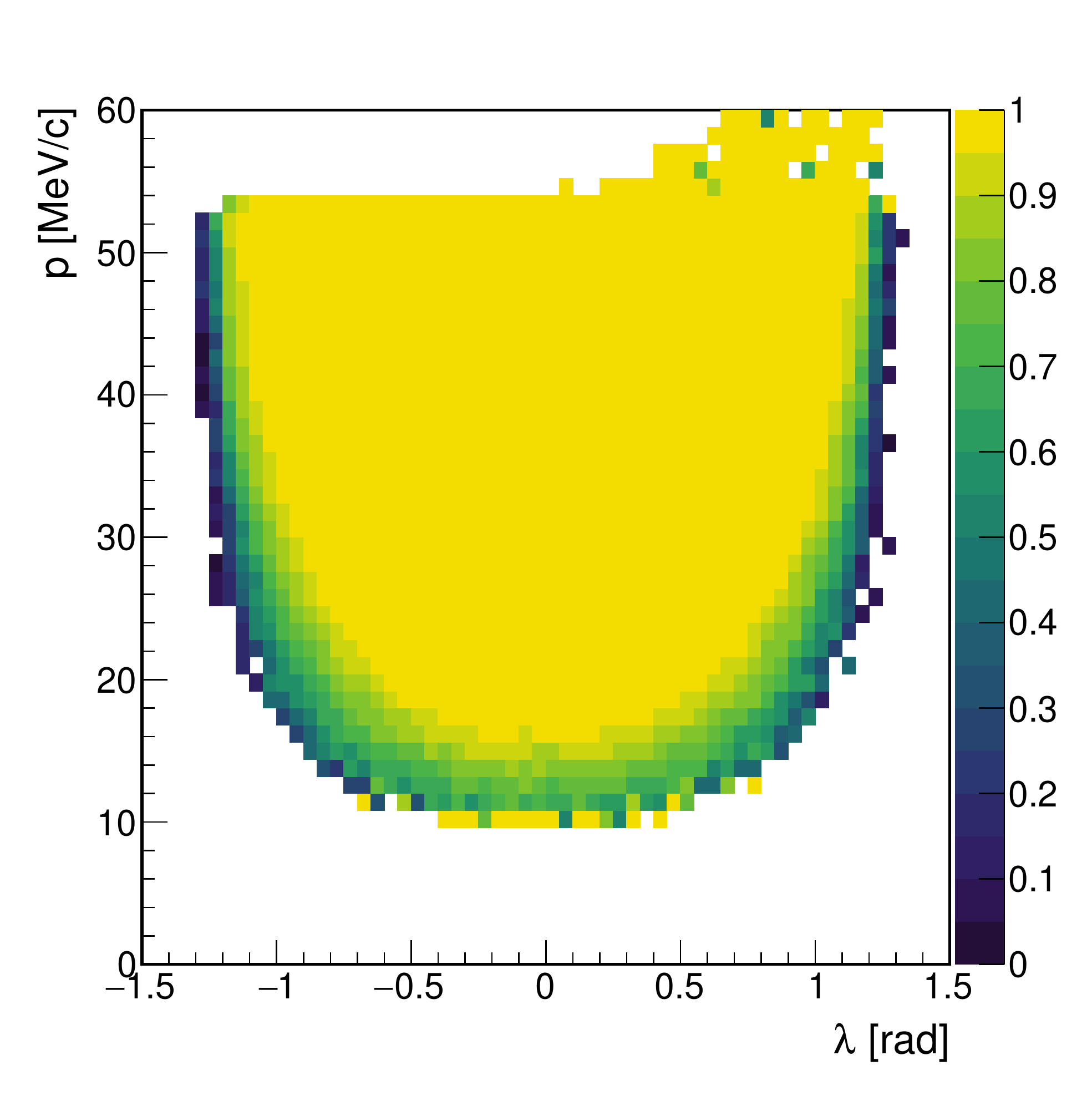}
\caption{
    Ratio of reconstructed short tracks
    to generated particles producing a hit in each of the four detector layers
    as a function of momentum $p$ and inclination angle $\lam$ (see 
    \autoref{sec:CoordinateSystem} for a definition of the coordinate system).
    The entries at high momentum in the forward direction ($\lam > 0$)
    are from decays in flight.
}
    \label{fig:rec:eff_S4}
\end{figure}

\begin{figure}[bt!]
\centering
    \includegraphics[width=1.0\linewidth]{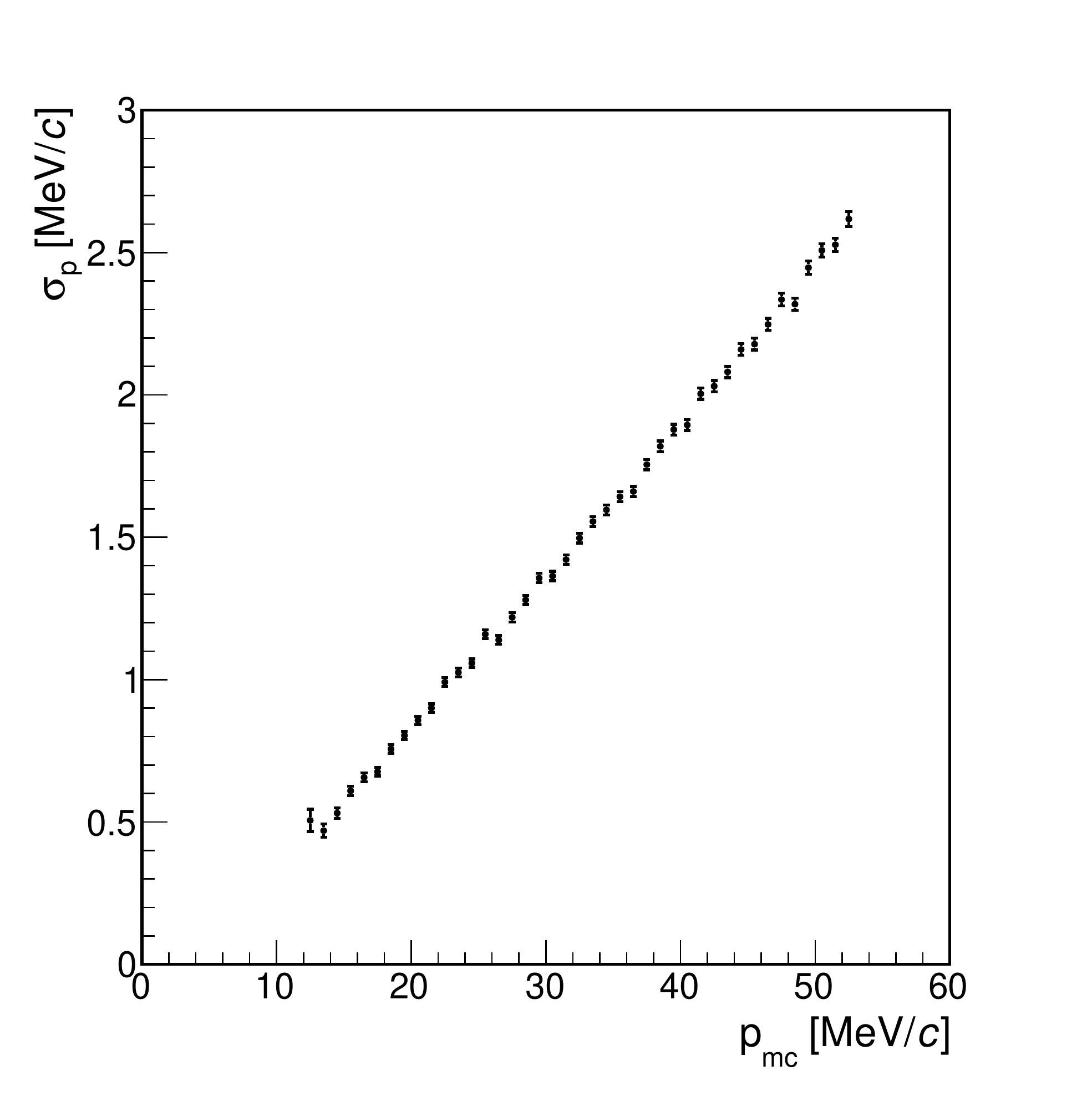}
\caption{
    Momentum resolution $\sigma_p$
    as a function of the generated momentum $p_{mc}$ of short tracks.
}
    \label{fig:rec:p2dp_S4}
\end{figure}

\begin{figure}[bt!]
\centering
    \includegraphics[width=1.0\linewidth]{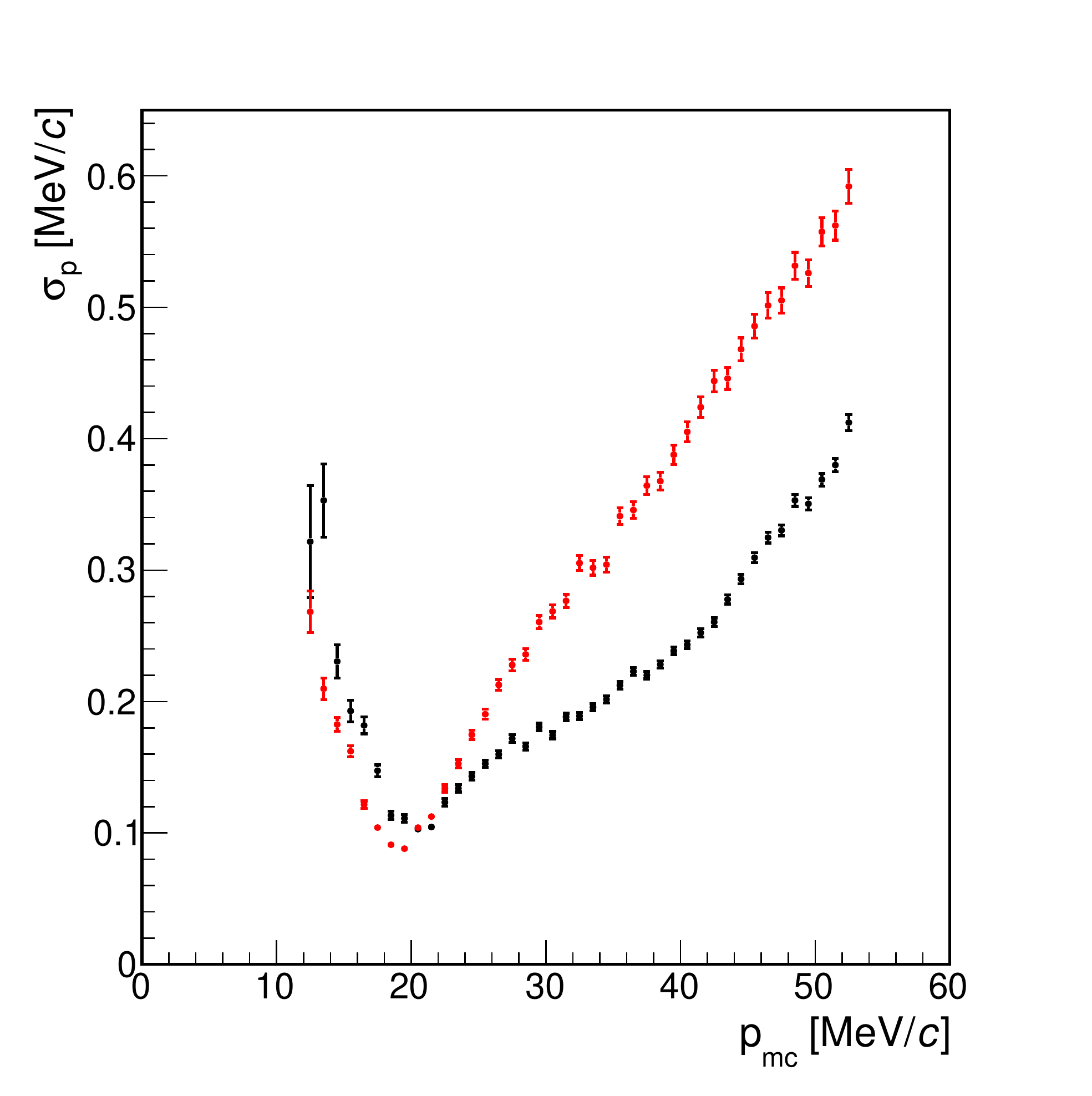}
\caption{
    Momentum resolution $\sigma_p$
    as a function of the generated total momentum $p_{mc}$
    for 6-hit (black) and 8-hit (red) long tracks.
    The momentum resolution has a minimum for tracks
    that traverse half a circle outside the outermost pixel layer.
}
    \label{fig:rec:p2dp_L}
\end{figure}

\begin{figure}
\centering
    \includegraphics[width=1.0\linewidth]{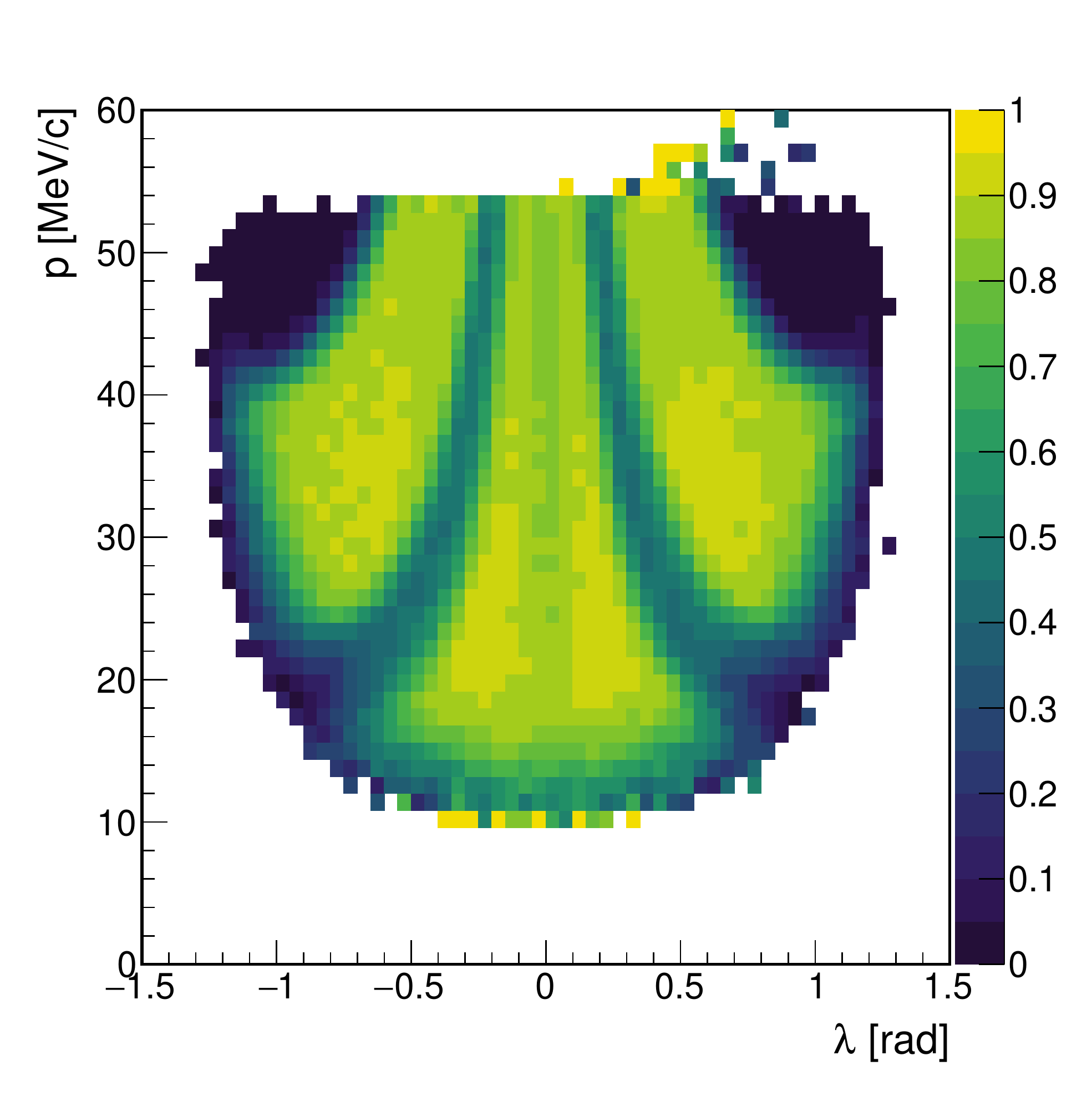}
\caption{
    Number of reconstructed long tracks relative to the number of short tracks
    as a function of momentum $p$ and inclination angle $\lam$.
    The entries at high momentum in the forward direction ($\lam > 0$)
    are from decays in flight.
}
    \label{fig:rec:eff_L}
\end{figure}

In the first step, triplets of hits in the first three layers consistent with
tracks originating from the target are identified.
These triplets are fit with the fast MS fit and if the fit $\chi^2$ is 
sufficiently good, they are extrapolated to the fourth layer, where the presence 
of an additional hit compatible with the triplet is required.
Again a fast MS fit is performed and a $\chi^2$ cut applied.

The resulting short tracks, with four hits each, are the input for the vertex fit
in the online reconstruction; see \autoref{fig:rec:eff_S4} for the
reconstruction efficiency and \autoref{fig:rec:p2dp_S4} for the momentum
resolution of 4-hit tracks.
The entries at high momentum ($p > m_\mu/2$) and in forward direction
are from decays in flight; their fraction relative to decays at rest is less then $10^{-4}$.

For the full offline reconstruction, the short tracks are extended to long tracks
with 6 hits (forward and backward going tracks)
or 8 hits (tracks close to perpendicular to the beam,
passing the vertex layers repeatedly)
incorporating the recurling parts of the track.
These long tracks have a much larger lever arm for momentum measurement
and thus provide a much enhanced momentum precision, as shown in \autoref{fig:rec:p2dp_L}.
With the phase~I detector setup, there is however a limited acceptance
to see the recurling part of the track for large inclination angles
and also in the gaps between the central part and the recurl stations,
as shown in \autoref{fig:rec:eff_L}.

\begin{sloppypar}
Additional algorithms are designed to remove incorrectly reconstructed tracks.
One algorithm is performed after the reconstruction of short tracks.
A graph is constructed where nodes represent tracks
and edges correspond to intersections (common hits) between those tracks.
A subset of nodes is selected that
maximises the number of unconnected nodes
(i.e.~the maximum number of tracks that do not share hits).
A second algorithm is run after the reconstruction of long tracks.
Chains of long tracks are constructed
where each pair of long tracks shares a short segment.
These chains are required to have no intersections,
with a maximum length and minimum total $\chi^2$.
In regions with many recurling tracks, machine learning
techniques can be used to correctly identify the sequence of
track segments \cite{Liechti2018}.
\end{sloppypar}

\section{Energy loss}

\begin{figure}
\centering
    \includegraphics[width=1.0\linewidth]{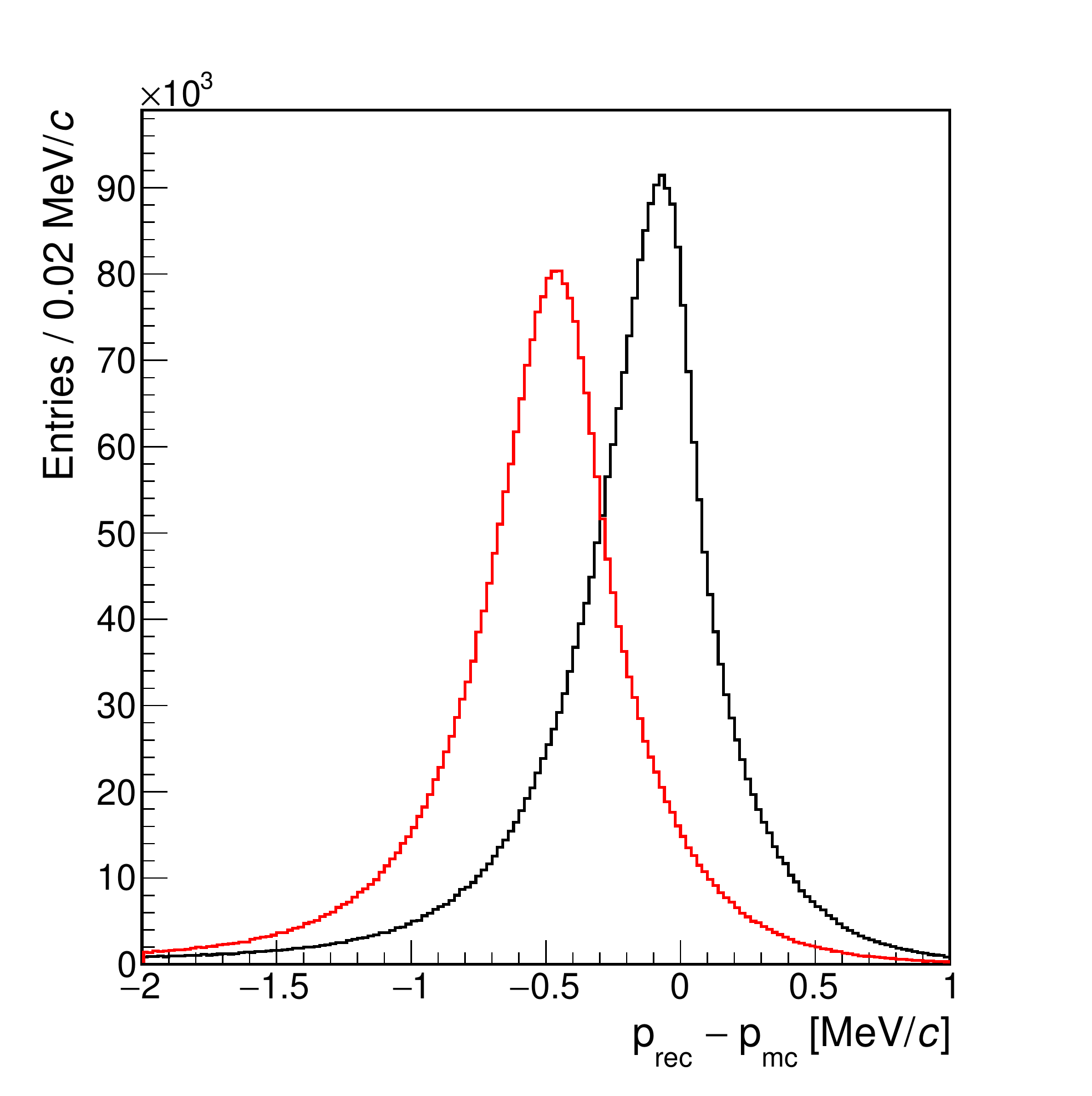}
\caption{
    Difference between reconstructed and generated momentum after the first detector layer,
    for long tracks with (black) and without (red) energy loss correction.
    The average momentum resolution is about \SI{0.2}{MeV/c}, the correction it improves by 10\%.
}
    \label{fig:rec:dp_L}
\end{figure}

For long tracks, the momentum resolution becomes comparable to the total
energy loss suffered by particles traversing the detector, resulting in an observable shift in the momentum.
The energy loss correction is implemented by adjusting the
curvature of each helix according to the sum of the most probable energy losses
in each layer passed by the particle up to that point in the helix.
\autoref{fig:rec:dp_L} shows the momentum resolution for long tracks
before and after applying the energy loss correction.

\section{Timing Detectors}

\begin{figure}
\centering
    \includegraphics[width=1.0\linewidth]{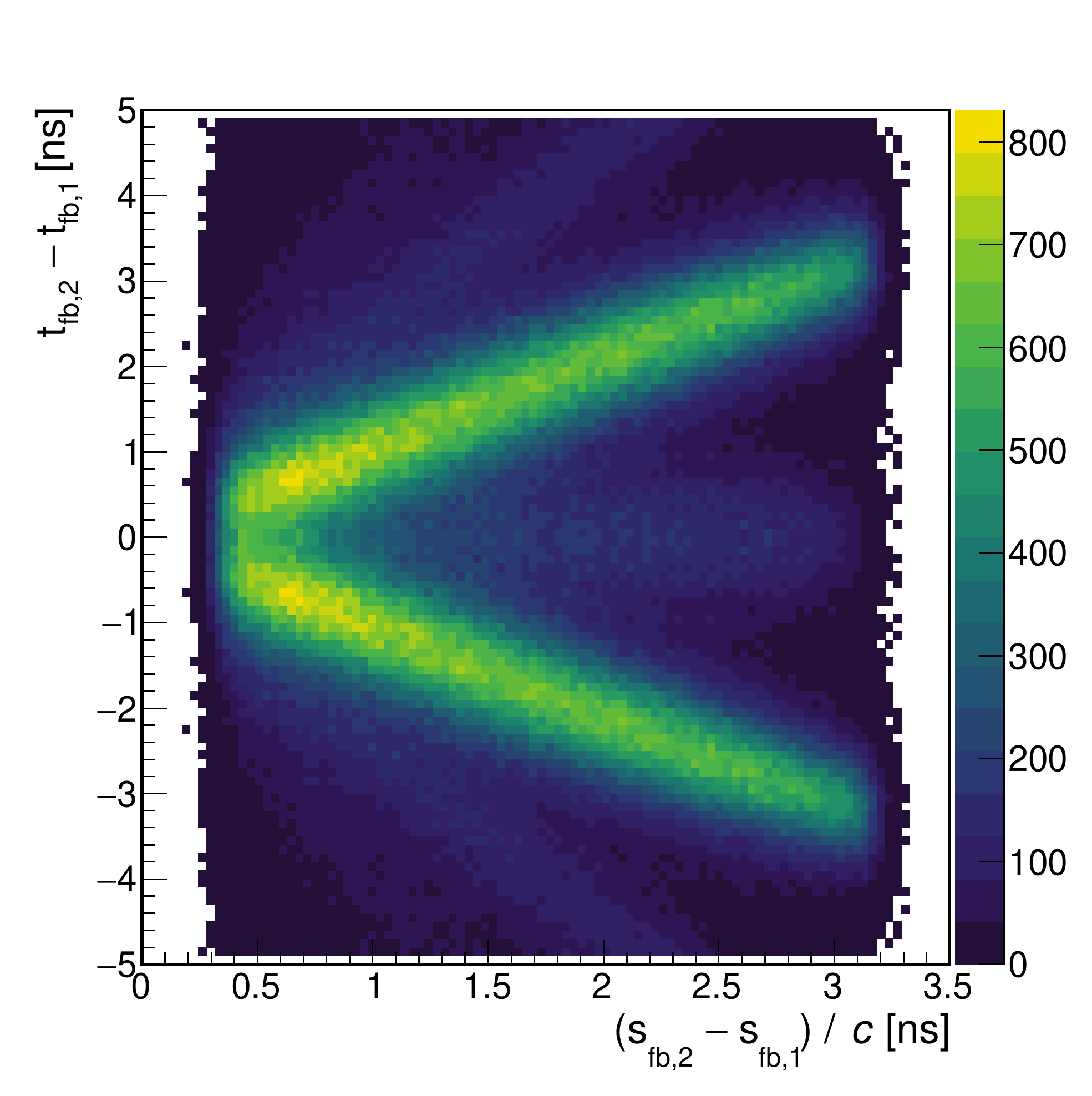}
\caption{
    Time difference between fibre clusters assigned to 8-hit long tracks
    as a function of distance along the trajectory.
    The upper branch corresponds to the correct charge assignment
    and direction of rotation,
    and the lower branch to the wrong charge assignment.
}
    \label{fig:rec:fb_ds2dt}
\end{figure}

Reconstructed tracks are extrapolated to the fibre and tile detectors
and the closest hits (within a maximum distance) are assigned to the tracks.
The timing from fibre hits allows the determination of
the direction of rotation (and thus the charge) of recurling particles.
\autoref{fig:rec:fb_ds2dt} shows the time versus distance
determined from two linked clusters of fibre hits for all 8-hit tracks,
and demonstrates the potential for charge identification by timing.

The tile hits have the best timing resolution,
providing an important constraint on the accidental combinatorial background
by allowing timing information to be compared for different tracks assigned to a single vertex.
The efficiency of the assignment of tile hits to tracks that pass through the tile detector is 98\%.


\chapter{Online Event Selection}
\label{sec:EventFilter}

\nobalance

\chapterresponsible{Nik}

The full data rate of the detector needs to be greatly
reduced before permanent storage -- only physically relevant event candidates 
can be kept.
Given the particle rates in Mu3e, requiring three tracks coincident in time is not sufficient 
to reduce the amount of data to be read out significantly.
Consequently no hardware trigger is employed and instead the online filter farm reconstructs all tracks and applies a selection algorithm in software.
The selection requires three tracks coincident in time, and consistent with originating at a common vertex and with the expected kinematic properties of signal events.
The computing power required for this is provided by \emph{Graphics Processing Units} (GPUs), where we profit from the very high rate of technological advances driven by the gaming and deep learning markets.

\begin{figure*}[t!]
	\centering
		\includegraphics[width=0.7\textwidth]{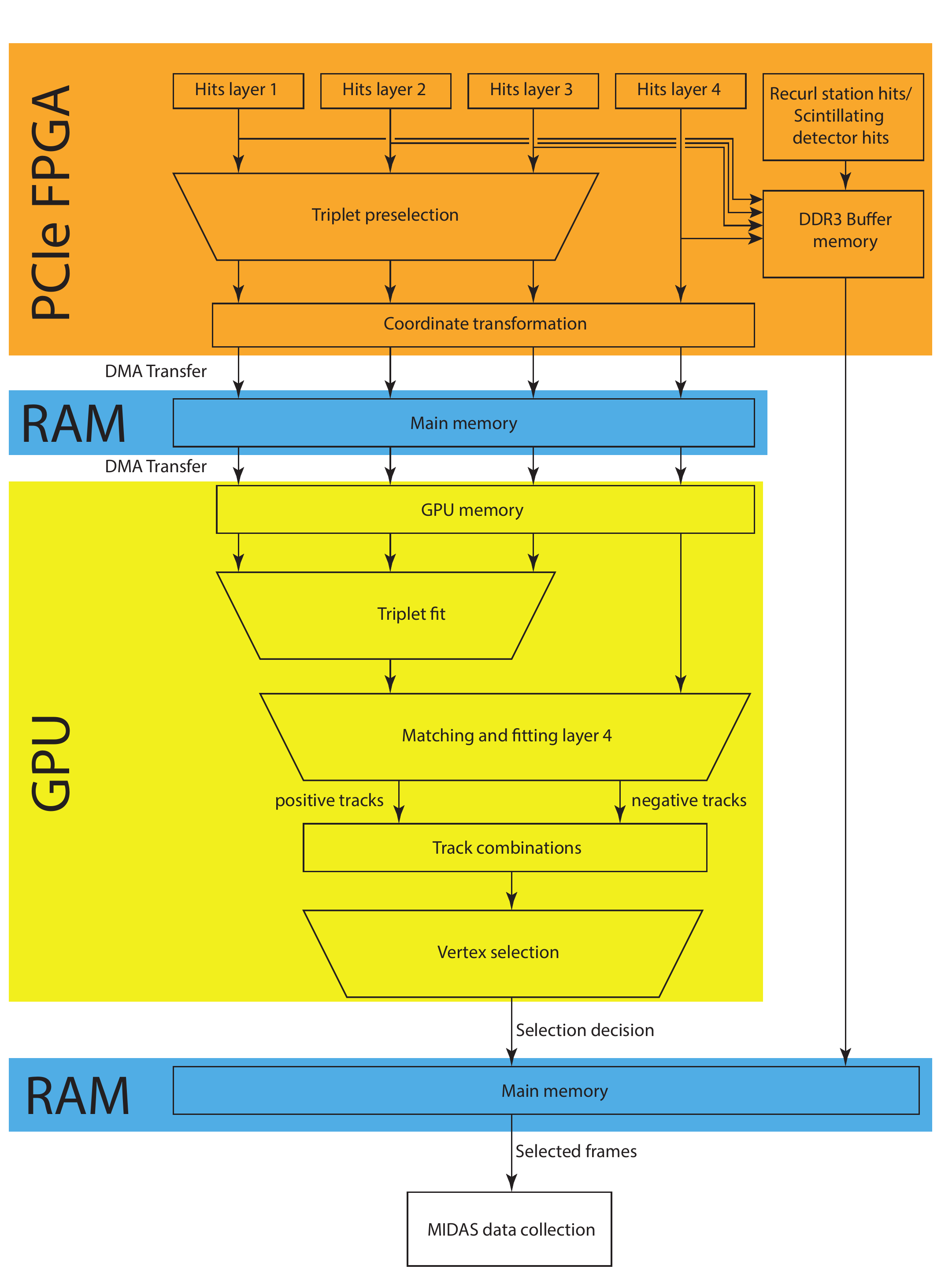}
	\caption{Flow diagram of the online reconstruction software and firmware.}
	\label{fig:OnlineRecFlow}
\end{figure*}

A simple version of the fast linear fit based on multiple
scattering (see \autoref{sec:Reconstruction}) is implemented on the GPUs for
quick track fitting.
In addition, events with at least two positive and one
negative electron tracks are checked for a common vertex and signal-like kinematics.
This selection is applied on a frame by frame basis on individual
farm PCs and the selected frames are merged into the global data stream, see \autoref{fig:OnlineRecFlow} for an overview.
The technical implementation of the event filter is described in
\autoref{sec:EventFilterInterface}. 

For the online reconstruction, only hits from the central station of the pixel
detector are considered, since matching recurling tracks and time information
from the tiles and fibres is computationally too expensive and also not necessary
for a first selection.

Combinations of hits from the first three detector layers are matched to form triplets.
Before the actual fitting procedure, a number of simple geometrical selection
cuts are applied at the FPGA stage in order to reduce the number of combinations by a
factor of about 50.

The fitting of triplets is non-recursive and linear, and hence can be done in parallel
for all hit combinations.
It is therefore an ideal candidate for parallelisation on GPUs.
With their many computing kernels but small memories, they perform well at
tasks where many similar computations are performed on the same memory content.
For a muon rate of $\SI{e8}{Hz}$ and \SI{50}{ns} time frames, we expect $\mathcal{O}(10)$ hits per layer  leading to $\mathcal{O}(10^3)$ combinations.
With code optimised for these conditions, we have achieved \SI{1e9}{fits/s} on a NVIDIA GTX 980 GPU (released 2014), which is sufficient to handle this level of combinations.

For each triplet passing the $\chi^2$ and radius cuts, the track is
extrapolated to the fourth detector layer. If at least one hit exists within a certain transverse radius and distance in
$z$, the hit closest to the extrapolated position is used to form a second
triplet from hits in layers two, three and four, which is then fitted. 
An improved value for the curvature of the track is then obtained from 
averaging the results of the two triplet fits.
Finally the charge of the particle is derived from the track curvature and all
combinations of two positive tracks and one negative track are examined with
respect to a common vertex.

The vertex position is calculated from the mean of two-circle
intersections of the tracks in the transverse plane (perpendicular to the magnetic field), weighted by the
uncertainty from multiple scattering in the first layer and hit resolution.
A $\chi^2$-like variable is defined using the distances of
closest approach of each track to the mean intersection position and their
uncertainties, both in the transverse and the $r-z$ plane.
Vertices are selected based on their proximity to the target and the $\chi^2$
value. In addition, cuts on the total kinetic energy and combined momentum of
the three tracks at the points of closest approach are applied.
After all cuts, the frame rate is reduced by a factor $\approx$\num{200}, which
is further reduced by the full online reconstruction as described in \autoref{sec:Reconstruction}.

In addition to signal candidate events, cosmic ray muon candidates and random frames will be saved for calibration, alignment and studies of the selection efficiency.
The parameters of all reconstructed tracks are histogrammed for monitoring as well as for searches e.g.~for two-body muon decays \cite{Perrevoort:2018okj}.

The triplet fit, the propagation to the fourth layer and the vertex fit as well as the monitoring have been implemented, optimised for performance and tested on GTX~1080Ti cards (of 2016/17 vintage).
It was shown that 12 of these cards are sufficient for the phase~I load~\cite{DissVomBruch2017}.

\chapter{Detector Alignment}
\label{sec:DetectorAlignment}

\chapterresponsible{Nik, Uli}

\nobalance

For the reconstruction algorithms to work optimally, the position and
orientation of all active detector elements and the stopping target have
to be known to good precision.
The position of the pixels inside a sensor is given by the tolerances of the
manufacturing process, which are much better than the minimal feature size of
\SI{180}{nm}; compared to all other sources of misalignment, this is completely
negligible.
The task of detector alignment is thus to determine the position, orientation
and deformation
of all active detector parts (HV-MAPS chips, fibres, tiles).

The first step in ensuring a well-aligned detector is the careful assembly of
modules and layers using precision tools, followed by a detailed survey.
After detector installation, movements of larger detector parts (e.g.~the
recurl stations with regards to the central detector) can be followed by a 
system of alignment markers observed by digital cameras inside the magnet.
The ultimate alignment precision is however only reached with track-based
alignment methods, starting with cosmic ray tracks, refining using
beam data.

\section{Effects of Misalignment}
\label{sec:EffectsOfMisalignment}

We have studied the effects of a misaligned pixel detector on the reconstruction 
efficiency and tracking resolution using the full detector simulation \cite{Hartenstein19}.
For technical reasons the sensors are all in their nominal position for the
Geant4 simulation, and the reconstruction is then performed with
different sensor positions.

The hierarchical mechanical structure of the pixel detector with stations, 
modules, ladders and sensors is expected to be reflected in the misplacements of 
all detector parts after assembly.
To reproduce this, the size of various misalignment modes (i.e.~rotations 
and shifts of all structures, and deformations of individual sensors) are 
estimated and then applied in a randomised way.
The result is an average absolute offset of about $\SI{450}{\micro \metre}$ of 
single sensor corners\footnote{Studying the sensor corners has the advantage of 
covering shifts as well as rotations of sensors with respect to their nominal 
position.} with respect to their nominal position for the estimated misalignment 
after detector construction.
This leads to a worsening of the reconstruction efficiency; there is however an 
efficiency plateau if the overall error on the corners of the sensors is less 
than $\SI{100}{\micro \metre}$.
Far more relaxed criteria apply to global movements of detector stations 
(e.g.~recurl stations, vertex layers) with regards to each other.

The constraints on the alignment accuracy for achieving optimal momentum 
resolution are much tighter than for the efficiency -- positions and rotations 
should be known well enough to achieve an error smaller than 
$\SI{50}{\micro\metre}$ for the sensor edge positions.

\section{Position Monitoring System}
\label{sec:PositionMonitoringSystem}

\begin{figure}[tb!]
	\centering
		\includegraphics[width=0.49\textwidth]{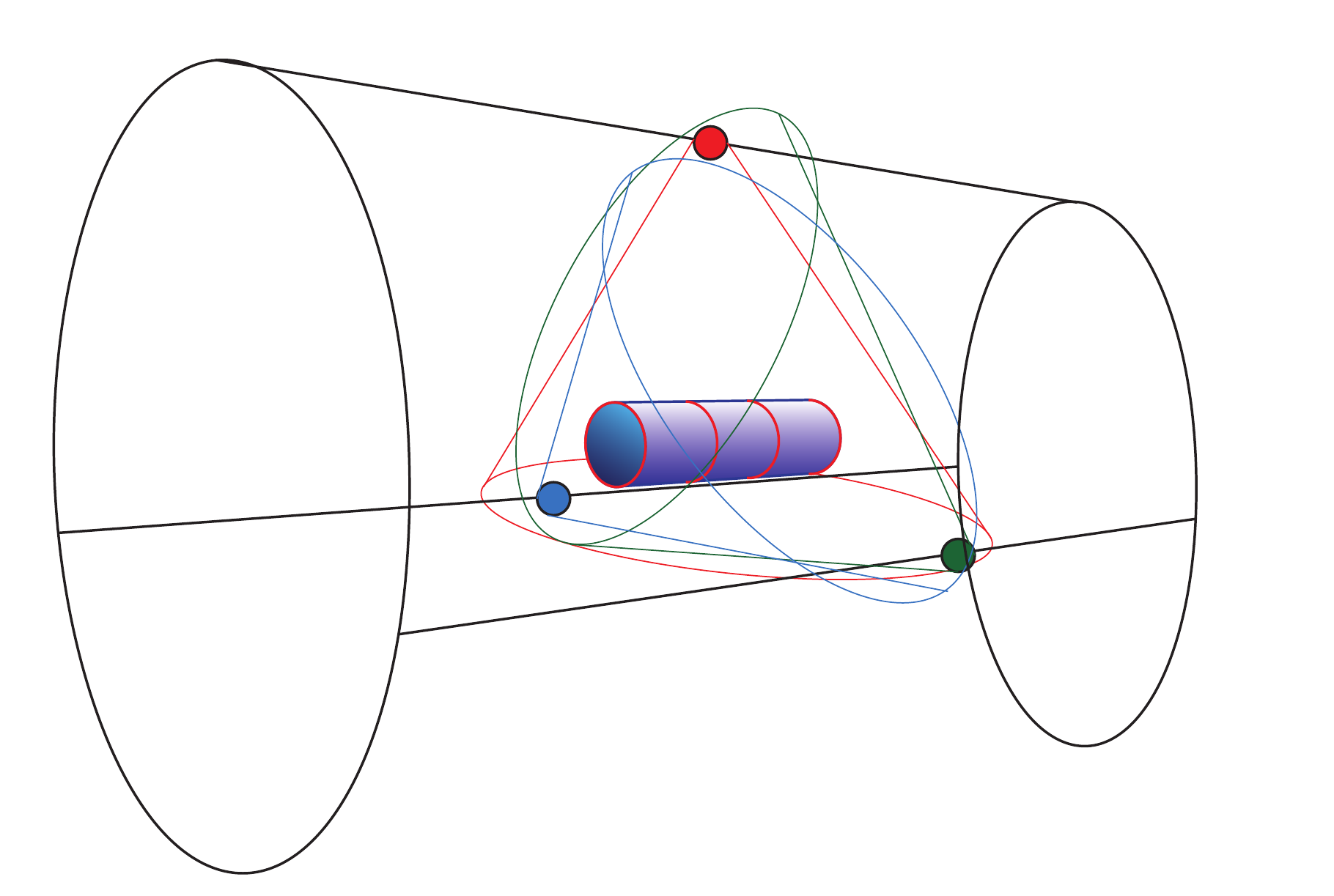}
	\caption{Schematic view of a possible alignment system using three cameras.
	The detector support cage is shown in black, the blue tube represents the detector
	stations with the endrings shown in red. The three cameras and their fields of
	view are shown in red, green and blue.}
	\label{fig:CameraAlignment}
\end{figure}

The positions of the detector stations relative to each other are monitored
by a series of cameras mounted to the detector cage.
They are complemented by light sources (the detector is usually operated in 
the dark) and alignment marks on the endrings of the detector stations.
Cameras with a \SI{85}{\degree} field of view are sufficient to view all end 
rings in the phase~I detectors when mounted to the detector cage. 
 A system of three cameras (e.g.~top and 
$\pm$ \SI{60}{\degree} from the
bottom) also allows tracking the relative movements of the cameras, as they can see
each other. 
Sub-millimetre resolution requires fairly high resolution cameras (2K or 4K)
or the use of separate cameras with zoom lenses focused on the station-station
transitions.
A possible three-camera system is shown in \autoref{fig:CameraAlignment}.

\section{Track-Based Alignment}
\label{sec:TrackBasedAlignment}

\begin{table*}[t!]
\begin{center}
\small
\begin{tabular}{lrrr}

\toprule
    Parameter & Nominal & Misaligned & Aligned \\
    \midrule
    Efficiency (short) [$\SI{}{\percent}$] & $100.00$ & $59.09 \pm 0.08$ & $100.01 \pm 0.03$ \\
    Efficiency (long) [$\SI{}{\percent}$] & $100.00$ & $46.72 \pm 0.12$ & $100.05 \pm 0.14$ \\
    Momentum resolution (short) & $ 2.628 \pm 0.003$ & $ 4.271 \pm 0.006$ & $ 2.635 \pm 0.003$ \\
    Momentum resolution (long) & $ 1.341 \pm 0.002$ & $ 1.645 \pm 0.003$ & $ 1.337 \pm 0.002$ \\
\bottomrule
\end{tabular}
\end{center}
\caption{
    Tracking performance (using Michel positrons) for nominal,
    misaligned and aligned configurations of the pixel detector.
    The efficiencies are relative to the nominal configuration.
    The misaligned version corresponds to an estimate
    of the expected sensor misplacements after assembly.
    Momentum resolutions show the RMS of the distributions.
}
\label{tab:track_reco_efficiency}
\end{table*}

\begin{table*}[t!]
\begin{center}
\small
\begin{tabular}{llrrr}

\toprule
    Parameter & & Nominal & Misaligned & Aligned \\
    \midrule
    \multicolumn{2}{l}{Efficiency (short) [$\SI{}{\percent}$]} & $100.0$ & $5.9$ & $99.7$ \\
    \multicolumn{2}{l}{Efficiency (long) [$\SI{}{\percent}$]} & $100.0$ & $2.2$ & $100.1$ \\
    \multirow{2}{*}{$x_{rec}-x_{true}$ [$\SI{}{\milli \metre}$]}
        & Mean & $-0.002 \pm 0.011$ & $0.029 \pm 0.068$ & $-0.021 \pm 0.011$ \\
        & RMS & $0.553 \pm 0.008$ & $0.724  \pm 0.048$ & $0.550 \pm 0.008$ \\
    \multirow{2}{*}{$y_{rec}-y_{true}$ [$\SI{}{\milli \metre}$]}
        & Mean & $-0.010 \pm 0.012$ & $-0.188 \pm 0.050$ & $0.048 \pm 0.012$ \\
        & RMS & $0.555 \pm 0.008$ & $0.687 \pm 0.035$ & $0.552 \pm 0.008$ \\
    \multirow{2}{*}{$z_{rec}-z_{true}$ [$\SI{}{\milli \metre}$]}
        & Mean & $0.003 \pm 0.009$ & $0.105 \pm 0.067$ & $-0.005 \pm 0.009$ \\
        & RMS & $0.356 \pm 0.006$ & $0.813 \pm 0.048$ & $0.355 \pm 0.006$ \\
\bottomrule
\end{tabular}
\end{center}
\caption{
    Signal reconstruction efficiency and vertex resolution for nominal,
    misaligned and aligned configurations of the pixel detector.
    The efficiencies are relative to the nominal configuration.
    The misaligned version corresponds to an estimate
    of the expected sensor misplacements after assembly.
}
\label{tab:signal_reco_efficiency}
\end{table*}

\begin{figure}
	\centering
		\includegraphics[width=0.49\textwidth]{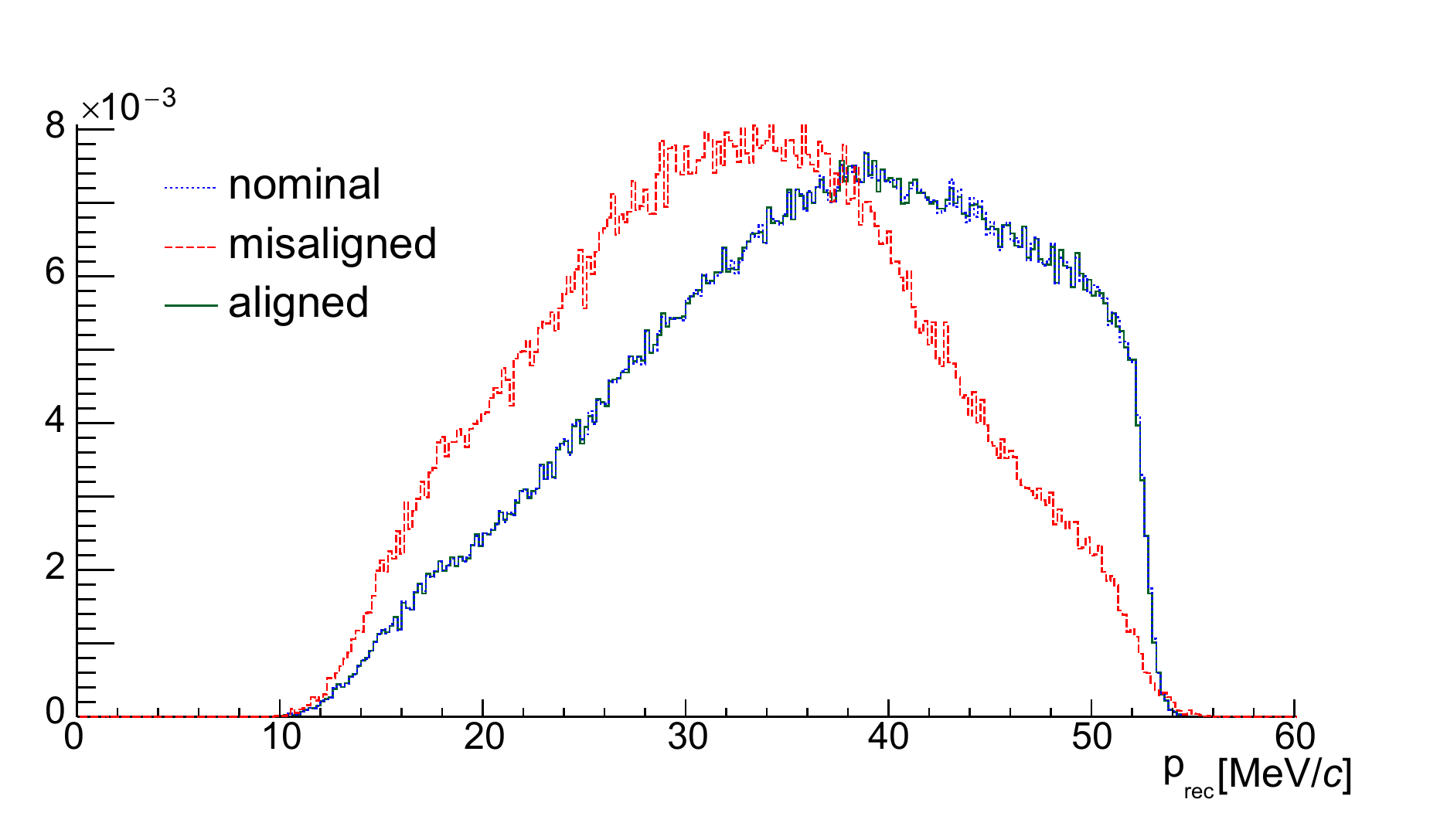}
    \caption{Reconstructed momentum of Michel positrons (using only long 
        tracks) for the nominal detector versus the (estimated) detector 
        after assembly and for the aligned detector. Note that 
        nominal and aligned histograms are very similar and thus hard to 
        discern visually. }
	\label{fig:michel_mis}
\end{figure}

The fine alignment of the silicon sensors (as well as the fibres and tiles) will
be performed using track-based methods initially developed in the H1 
experiment \cite{Kleinwort:2006zz} and subsequently successfully applied to a variety 
of large and very large tracking systems, e.g.~CMS at the LHC 
\cite{Blobel:2011az,Lampen:2014eya, Chatrchyan:2014wfa, BARTOSIK:2014hqa, 
Flucke:2012px,Behr:2012gf,Draeger:2011yda, Weber:2011zza}.

The alignment of the complete detector is a large minimisation problem, where,
for a very large sample of tracks, the residuals from the measured hits to the
fitted tracks have to be minimised under variation of both all track and all
alignment parameters (suitably parametrised detector positions). 
If a rough detector alignment is known, corrections will
be small and the minimisation problem can be linearised.

To this end, tracks reconstructed with the standard reconstruction algorithms
described in \autoref{sec:Reconstruction} are re-fitted using the general
broken lines (GBL) algorithm \cite{Blobel:2011az, Kleinwort:2012np}.
The GBL software can calculate and output the complete covariance matrix between
track and alignment parameters.
As the track parameters are not correlated between tracks and only relate to
the alignment of the small subset of sensors which are hit by the track, the
resulting matrix is sparse.
There are efficient algorithms for the inversion of such gigantic sparse 
matrices, one of which is implemented in the \emph{Millepede II} programme 
\cite{Blobel:2006yh}, which we are using.

\begin{sloppypar}
Whilst the sensor alignment is locally well constrained via the overlap of the
sensors in the azimuthal direction and the closeness of the double layers, overall
deformations such as shifts of the top part with regards to the bottom part, an
overall torsion or the position of the recurl stations are only weakly 
constrained by using tracks from muon decays.
These so-called \emph{weak modes} need additional input from tracks which
correlate distant parts of the detector.
These tracks are provided by cosmic ray muons.
As the cosmic rate is tiny compared to the beam muon rate, it is imperative to 
have a special trigger stream to collect enough cosmics for alignment.
\end{sloppypar}

\begin{figure}
	\centering
		\includegraphics[width=0.49\textwidth]{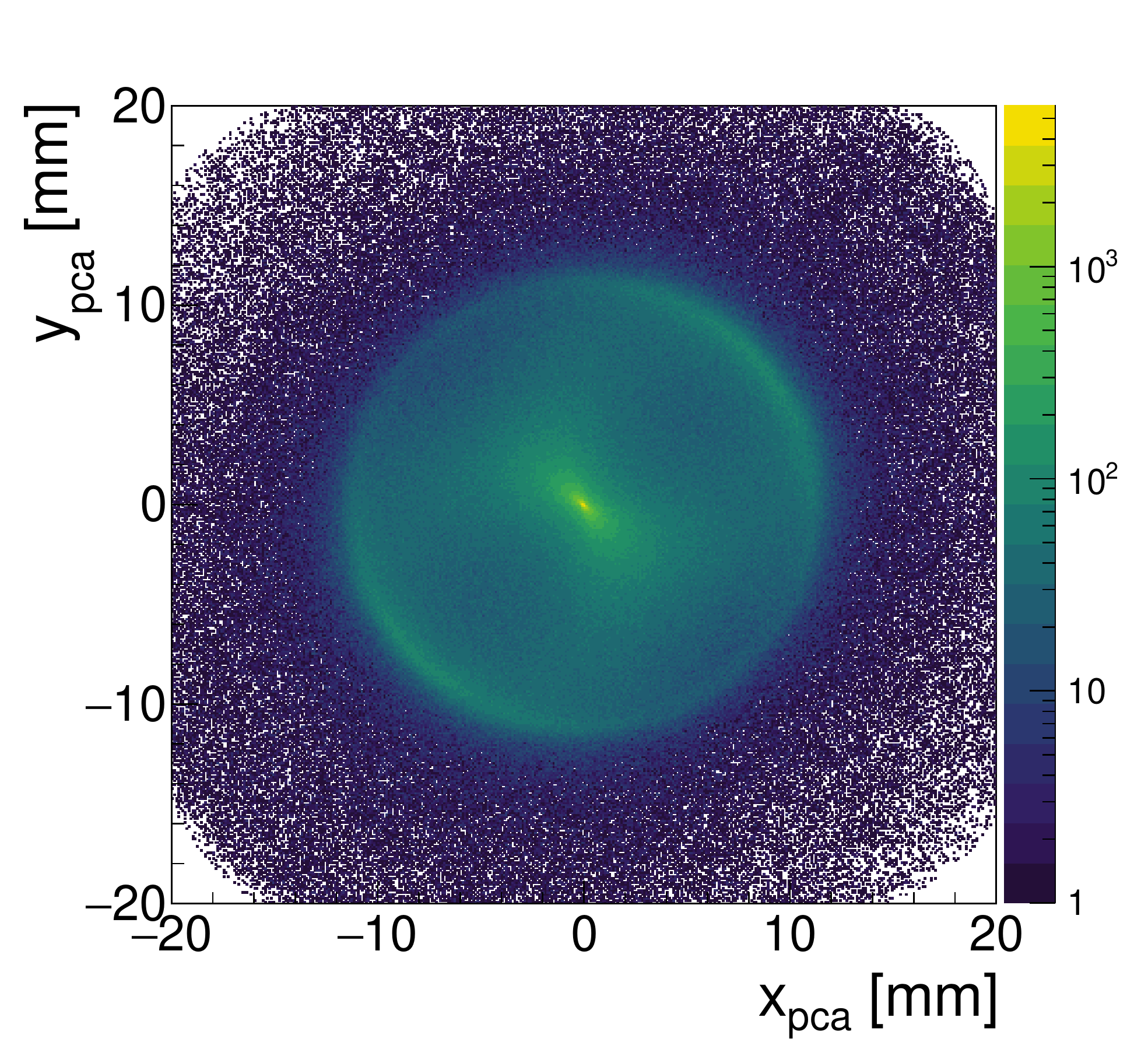}
		\caption{Position in $x$ and $y$ of the point of closest approach to the beam 
	line for a \SI{1}{mm} slice in $z$ at \SI{-20}{mm} for \num{3.84E8} stopped muons. 
	The target is clearly visible at its nominal position. The accumulation of entries towards the origin is a feature of the reconstruction method.}
	\label{fig:target_xy6}
\end{figure}

\begin{figure}
	\centering
		\includegraphics[width=0.49\textwidth]{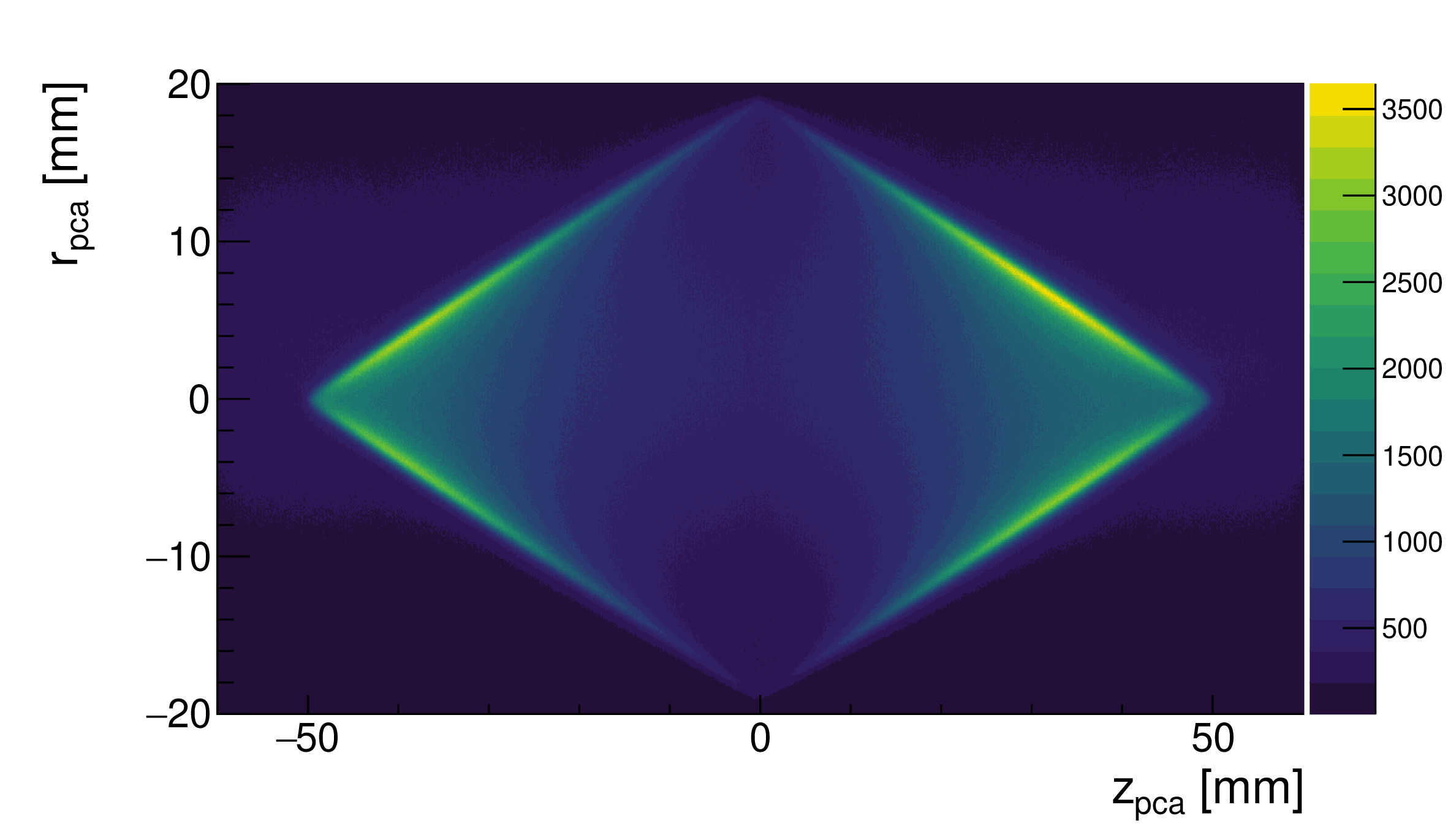}
	\caption{Position in $r$ and $z$ of the point of closest approach to the beam 
	line for \num{3.84E8} stopped muons; the target is clearly visible.
	Negative radius is defined to be when the beam line is inside the track circle, 
	positive is outside.}
	\label{fig:target_rz}
\end{figure}

Our strategy is to perform a preliminary alignment of the detector using cosmic 
muons, which will have to be good enought to allow for an efficient and clean reconstruction of
Michel tracks at large muon stopping rates. 
These Michel tracks can then be used for refinements until the required resolution is reached.
The requirements seem well within reach: in simulation, 
an average error on the sensor edge position of about $\SI{110}{\micro \metre}$
 has already been achieved\footnote{These results are based on an estimate of the misalignment right after detector assembly.}.
In addition the effects of the residual misalignments do not significantly deteriorate 
the performance of the tracking detector (see~\autoref{tab:track_reco_efficiency} and \autoref{tab:signal_reco_efficiency}).
The general reconstruction efficiencies and momentum resolutions for short and long tracks, as well as the signal reconstruction efficiencies for short and long tracks, in the re-aligned detector are almost identical to the values for the nominal detector.

In \autoref{fig:michel_mis} the distributions of the measured momenta of positrons originating from a Michel decay for the misaligned and aligned pixel detector are compared to the result for the nominal detector.
The misalignment applied in \autoref{fig:michel_mis} corresponds to expectations about detector misplacements right after assembly.
Where a misaligned detector geometry causes a clear distortion of the momentum distribution, we are able to recreate the nominal spectrum almost perfectly by applying the track-based alignment.

We have also implemented sensor deformations and temperature dependent sensor expansion in the alignment software.
The fibre and tile detectors will also be aligned using track-based methods, using the pixel detector as a reference.
The Millipede~II algorithm can then also be used for a precise time alignment of all detector parts.

\section{Target Alignment}
\label{sec:TargetAlignment}

The position of the target needs to be known with very high accuracy to allow placing requirements on the distance between the vertex and target.
As the target is passive, the residual-based method described in the previous
sections is not applicable.
The overwhelming majority of decay positrons originate
on the target surface, however, and will thus have a point of closest approach (POCA) to 
the beam axis inside the target.

This can be used to determine the target position by plotting the distribution 
of the POCAs in the transverse plane in slices of $z$ for many tracks, which
will give an accurate determination of the position of the outer target edge,
as shown in \autoref{fig:target_xy6} and \autoref{fig:target_rz}.
The target thickness has to be determined during manufacture or using photon
conversions.


\nobalance

\chapter{Performance Simulation}
\label{sec:SensitivityStudy}

\chapterresponsible{Nik}

The performance of the detector described in the preceding parts is studied by
running the Geant4 based Mu3e simulation and the reconstruction programme.
Even under optimistic assumptions, only a handful of signal decays are
expected in the data.
Nonetheless, we use relatively large signal samples to study the detector
performance in terms of resolution and efficiency and deduce preliminary 
event selection criteria in \autoref{sec:SignalPerformance}.

For the various expected backgrounds, the simulation of several
times the expected number of decays in the experiment is required, in principle.
This is impractical both in terms of processing time and available storage
space.
Instead we identify the important sources of background either from general
considerations (radiative decay with internal conversion) or from simulating 
a few seconds of run time (accidentals).
From these starting points we generate special simulation samples,
namely internal conversion decays with large visible energy and combinations
of Michel decays and Michel decays followed by Bhabha scattering, which
are then subjected to the event selection in order to show that the backgrounds
can be suppressed to a level that background-free running is possible with the
decay rates provided by the $\pi$E5 beamline. The background simulation is 
described in \autoref{sec:SimBackgrounds} and the experiment sensitivity is 
derived in \autoref{sec:Sensitivity}.

Note that in this section, a simple cut based analysis is used to show
that background-free running with the phase~I Mu3e experiment is possible.
An analysis with optimised cuts or based on likelihoods can likely deliver
a higher signal efficiency and thus final sensitivity per running time.

\section{Signal Performance}
\label{sec:SignalPerformance}

The nominal performance of the detector setup is studied using
about 8.5~million signal decays.
The decay electrons are generated with a phase space distribution.
Efficiencies are determined relative to all muons decaying inside a cylinder
with the outer dimensions of the stopping target.

In the first step, all three tracks from the signal decay have to be reconstructed
to at least short (4-hit) tracks; for the efficiency and resolution of the track
reconstruction, see \autoref{sec:Reconstruction}.

\subsection{Vertex Fit}
\label{sec:VertexFit}

\begin{figure*}[tb!]
	\centering
		\includegraphics[width=0.4\textwidth]{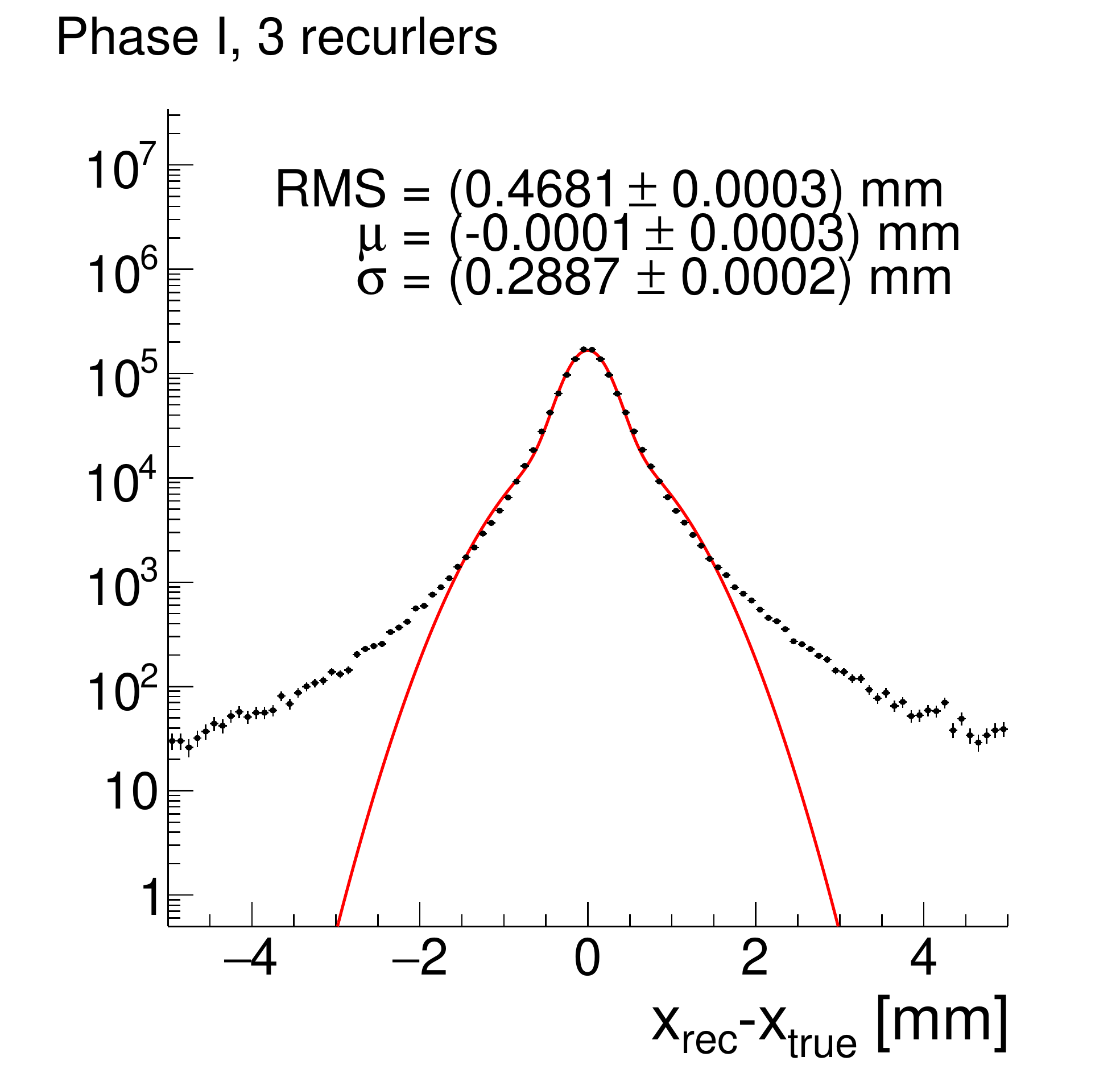}
		\includegraphics[width=0.4\textwidth]{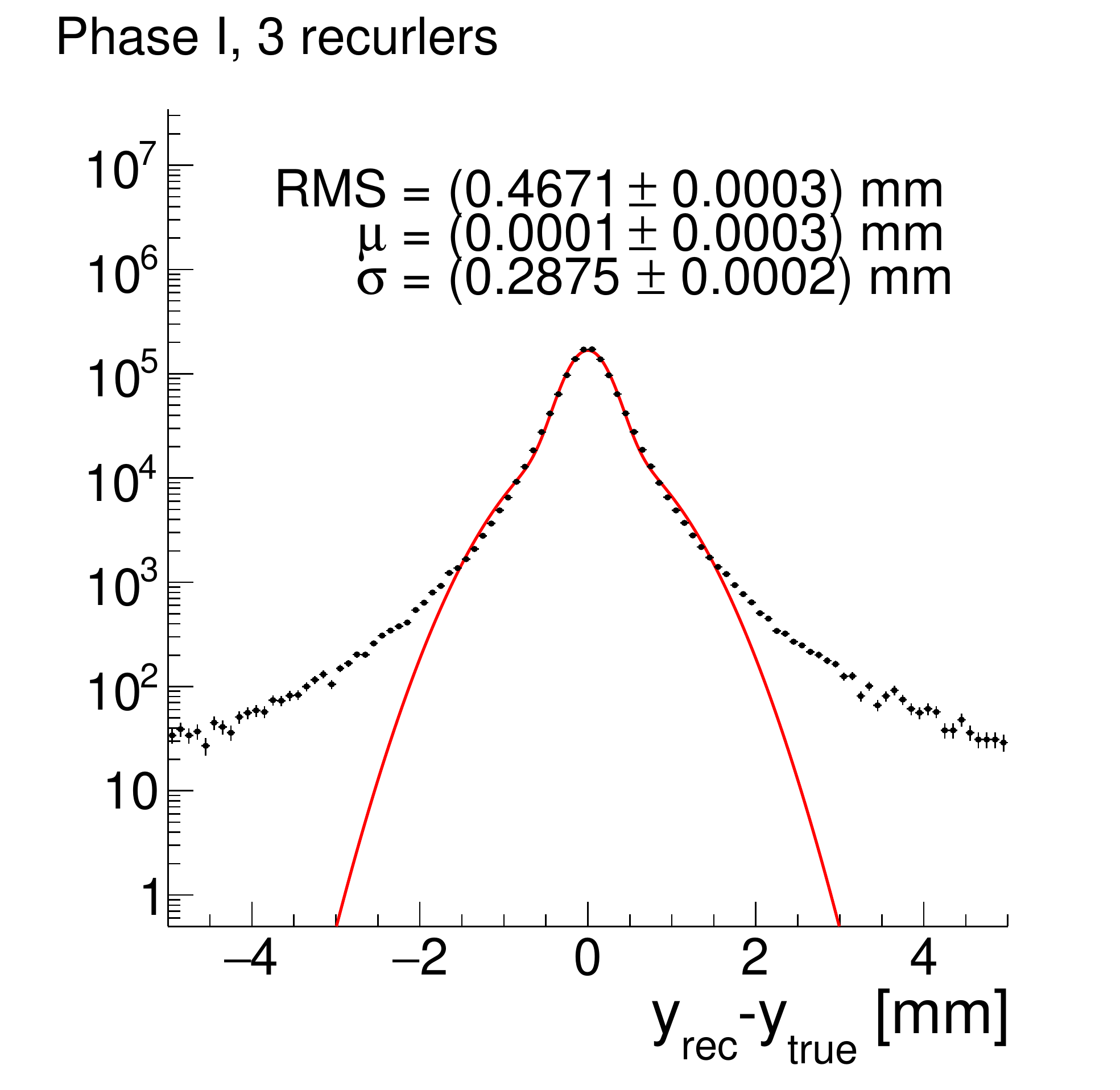}
		\includegraphics[width=0.4\textwidth]{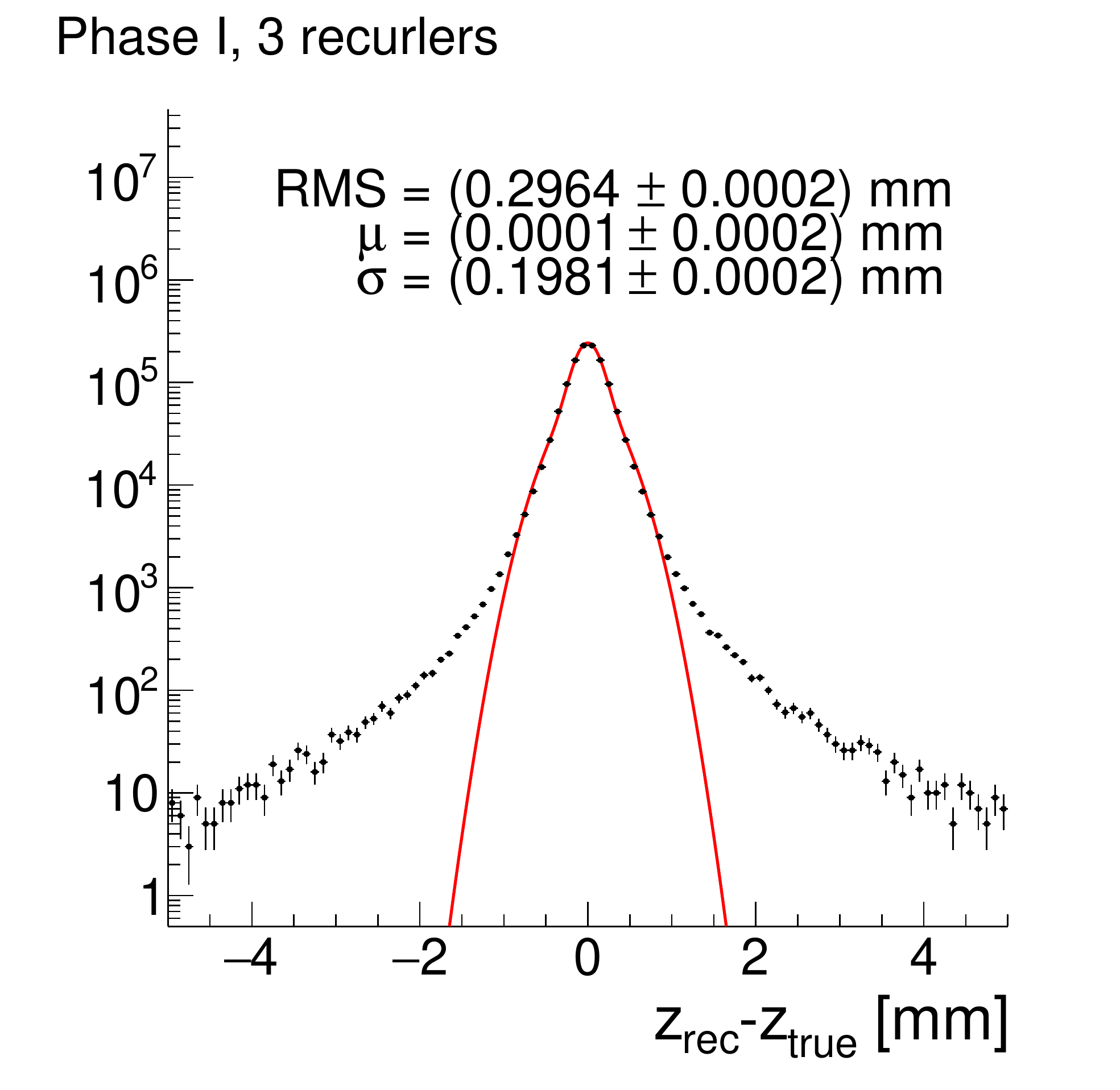}
		\includegraphics[width=0.4\textwidth]{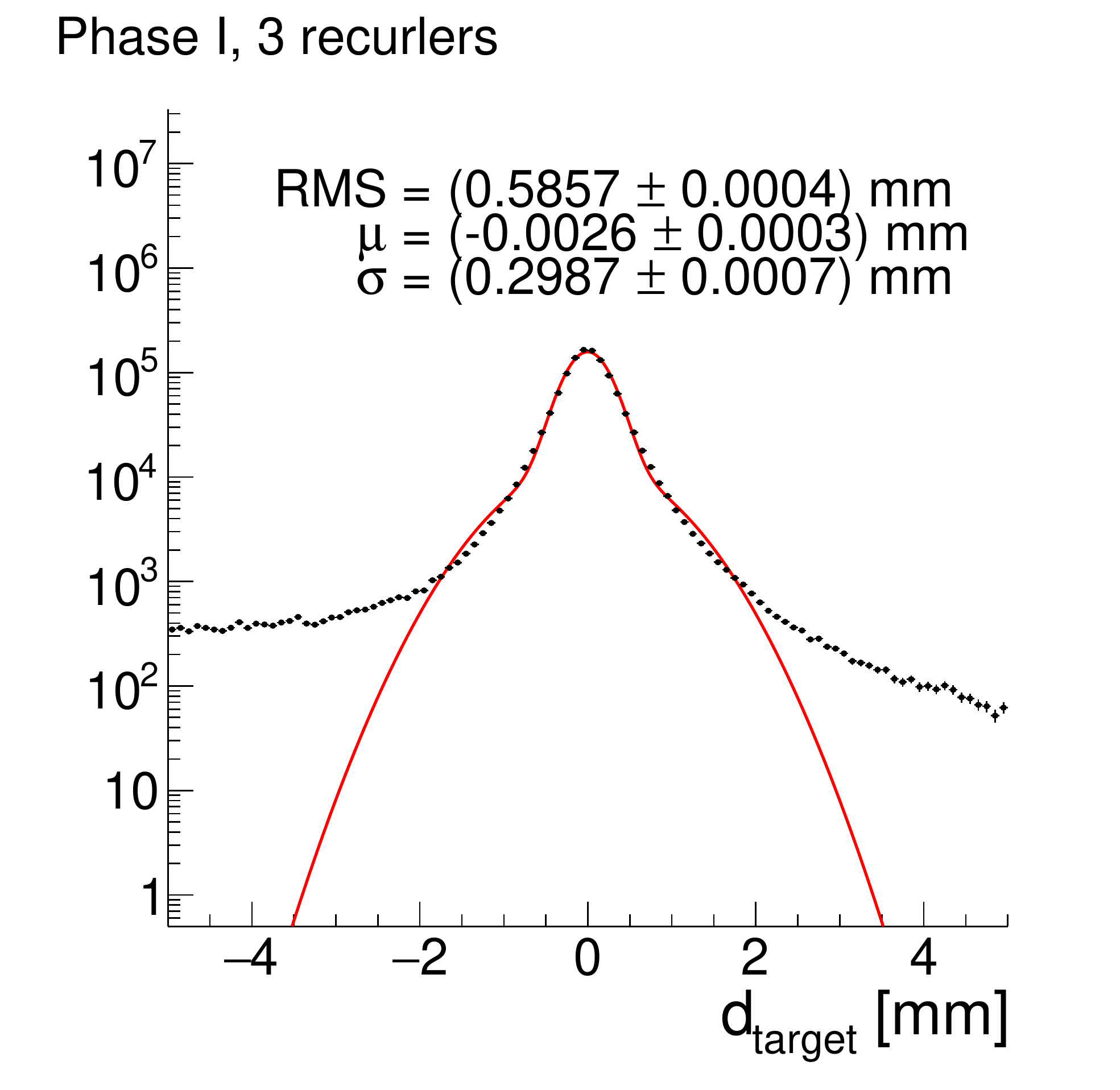}
	\caption{Vertex resolution for simulated signal decays (points). Three tracks with 
	recurlers are selected. The fits (lines) are the sum of two
	Gaussian distributions and the quoted $\sigma$ is the area-weighted mean.
	Top left in $x$, top right
	in $y$, bottom left in $z$ and bottom right in the distance to the target;
	negative target distances denote a reconstructed vertex position inside the target.}
	\label{fig:VertexX_Ib}
\end{figure*}

\begin{figure}[tb!]
	\centering
		\includegraphics[width=0.4\textwidth]{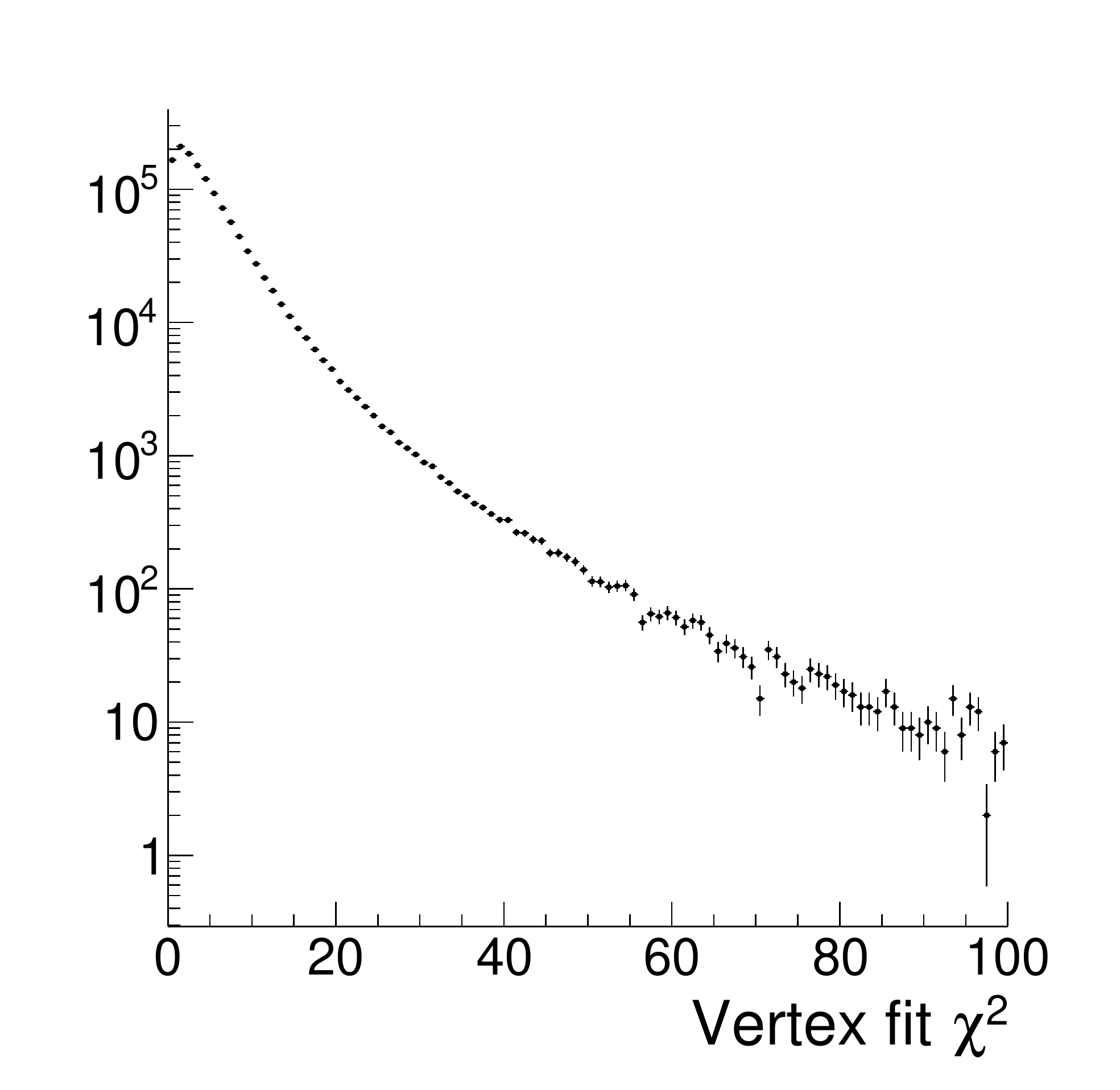}
	\caption{$\chi^2$ of the the vertex fit for signal events. The number of degrees
	of freedom is three~\cite{BachelorSchenk}, note however that the finite resolution
	of the pixel detector is ignored in the current implementation of the vertex fit and
	the distribution is thus not expected to follow a $\chi^2$ distribution with three
	degrees of freedom.}
	\label{fig:VertexChi2}
\end{figure}

The three tracks from signal decays should intercept at a common point on the
surface of the target.
We look at all combinations of a track with negative charge and two positively 
charged tracks.
In order not to fit recurling tracks with themselves, the track tangent vector
at the point of closest approach is determined. If the cosine of the opening
angle between two tracks is more than \num{0.99} for same sign combinations or 
less than \num{-0.99} for opposite sign combinations and the momentum difference is
less than \SI{1}{MeV/c}, the combination is not further considered.

Starting from the track positions and directions in the first detector plane,
we perform a vertex fit by forcing three tracks to intersect in a common point 
in space, taking multiple scattering in the first detector layer as the only
degree of freedom \cite{BachelorSchenk}.
The $\chi^2$ of the fit and the distance of the vertex to the target surface
are two handles for suppressing accidental background; 
the performance of the vertex reconstruction is
illustrated in \autoref{fig:VertexX_Ib}, and the $\chi^2$ distribution is shown in 
\autoref{fig:VertexChi2}. The target constraint is currently not used in the
analysis.

\subsection[Mass Reconstruction]{Mass and Momentum Reconstruction}
\label{sec:MassAndMomentumReconstruction}

\begin{figure*}[tb!]
	\centering
		\includegraphics[width=0.32\textwidth]{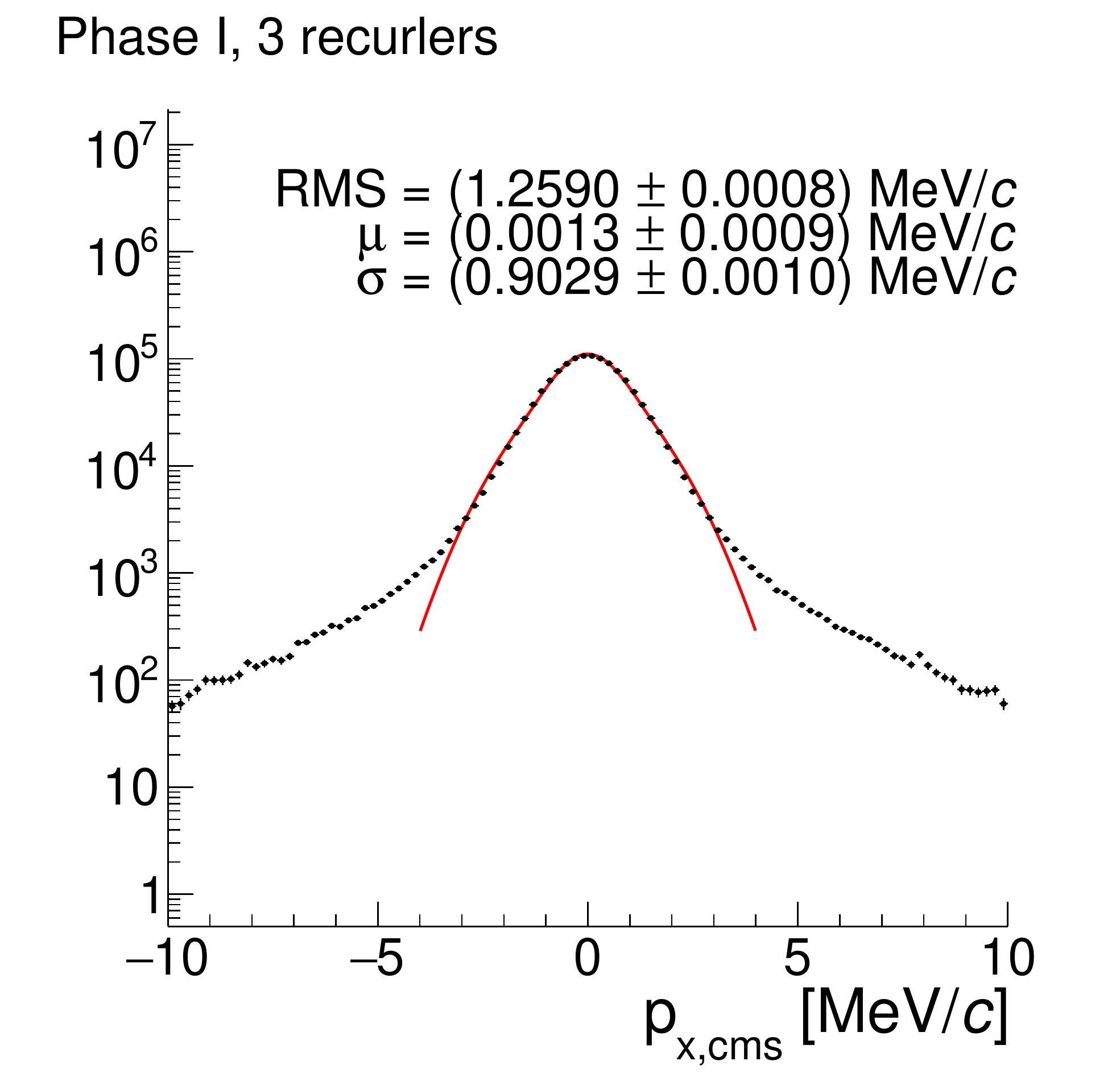}
		\includegraphics[width=0.32\textwidth]{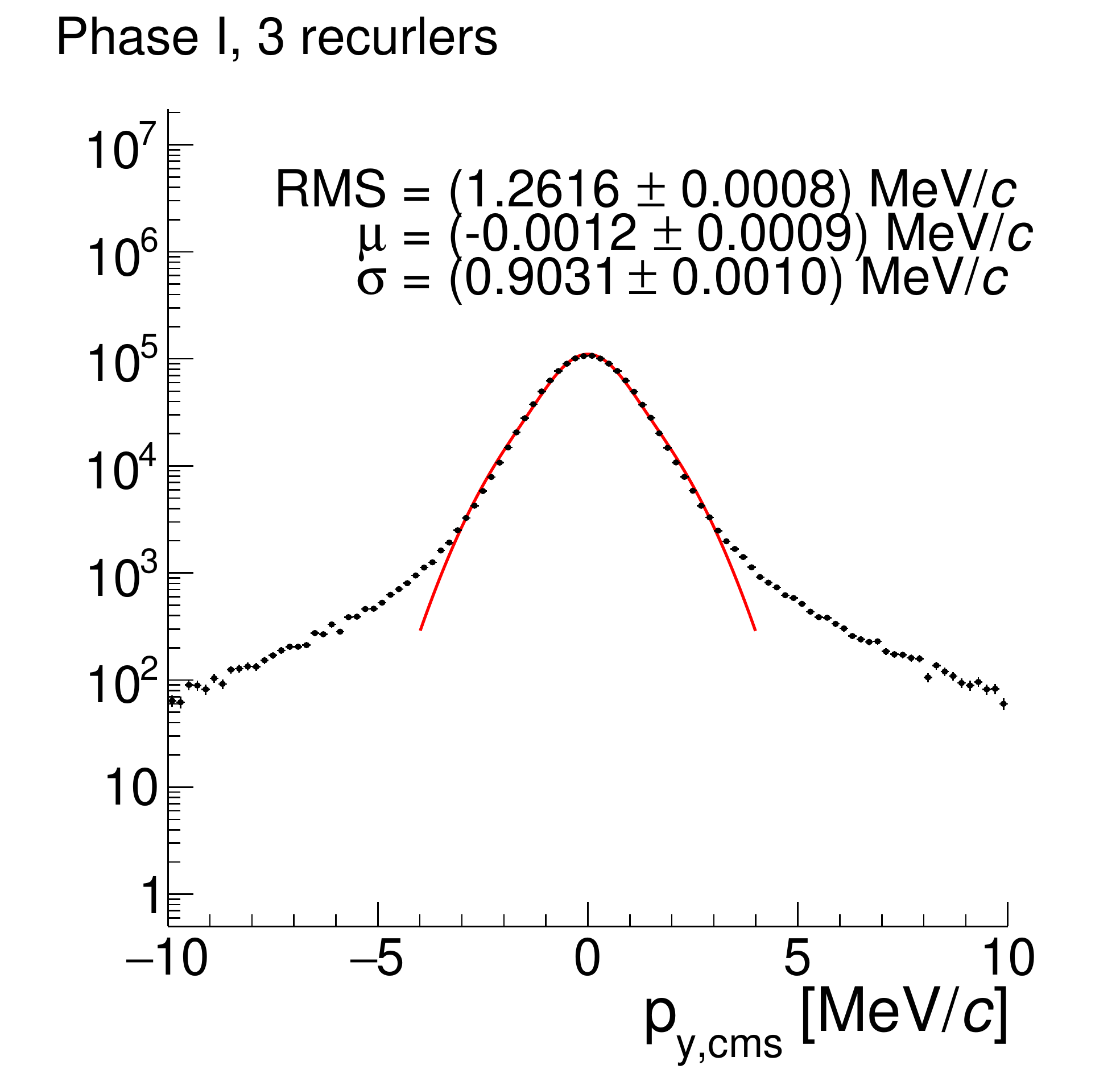}
		\includegraphics[width=0.32\textwidth]{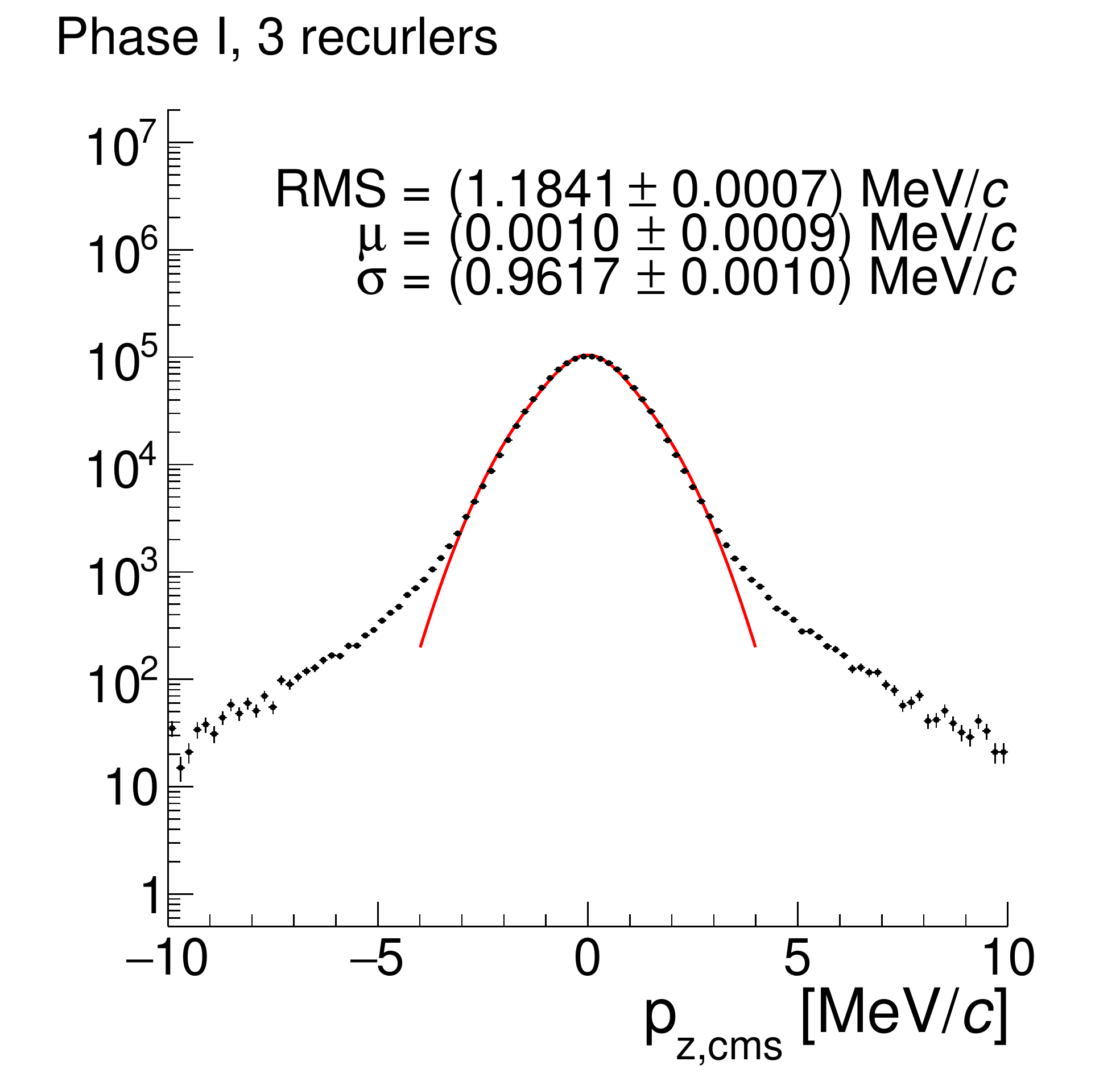}
	\caption{Reconstructed decay muon momentum (points) in $x$, $y$ and $z$ direction
	(which corresponds to the resolution for $p_x$, $p_y$ and $p_z$ for muons
	decaying at rest). Only long tracks enter
	the analysis. The lines are fits of Gaussian distributions used to estimate the core 
	resolution  and to identify biases.}
	\label{fig:momentumreso_3recurl_Ib}
\end{figure*}

\begin{figure}[tb!]
	\centering
		\includegraphics[width=0.4\textwidth]{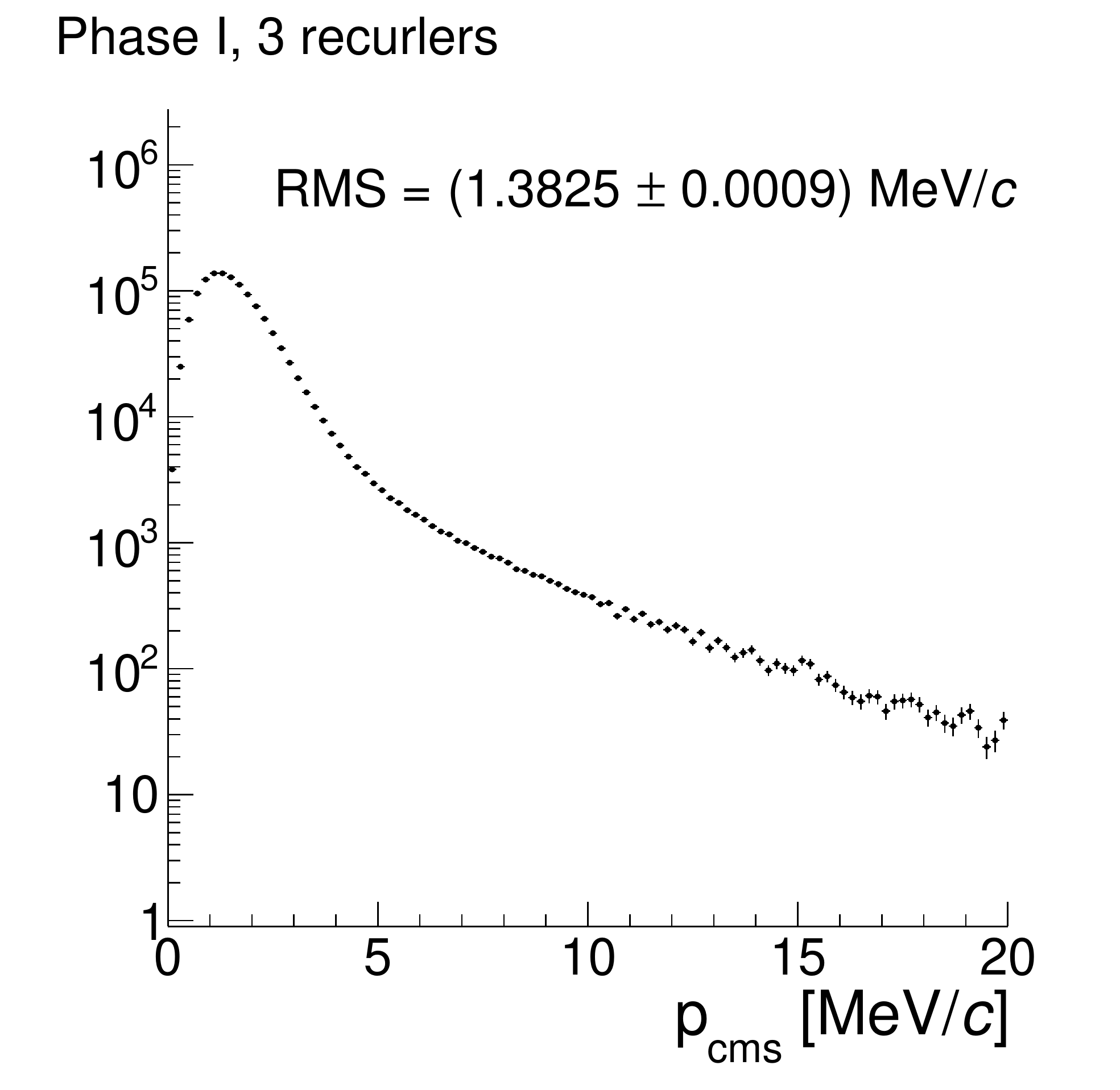}
	\caption{Magnitude of the center of mass system momentum reconstructed
	for signal events with three recurlers required.}
	\label{fig:ptot}
\end{figure}


\begin{figure*}
	\centering
		\includegraphics[width=0.4\textwidth]{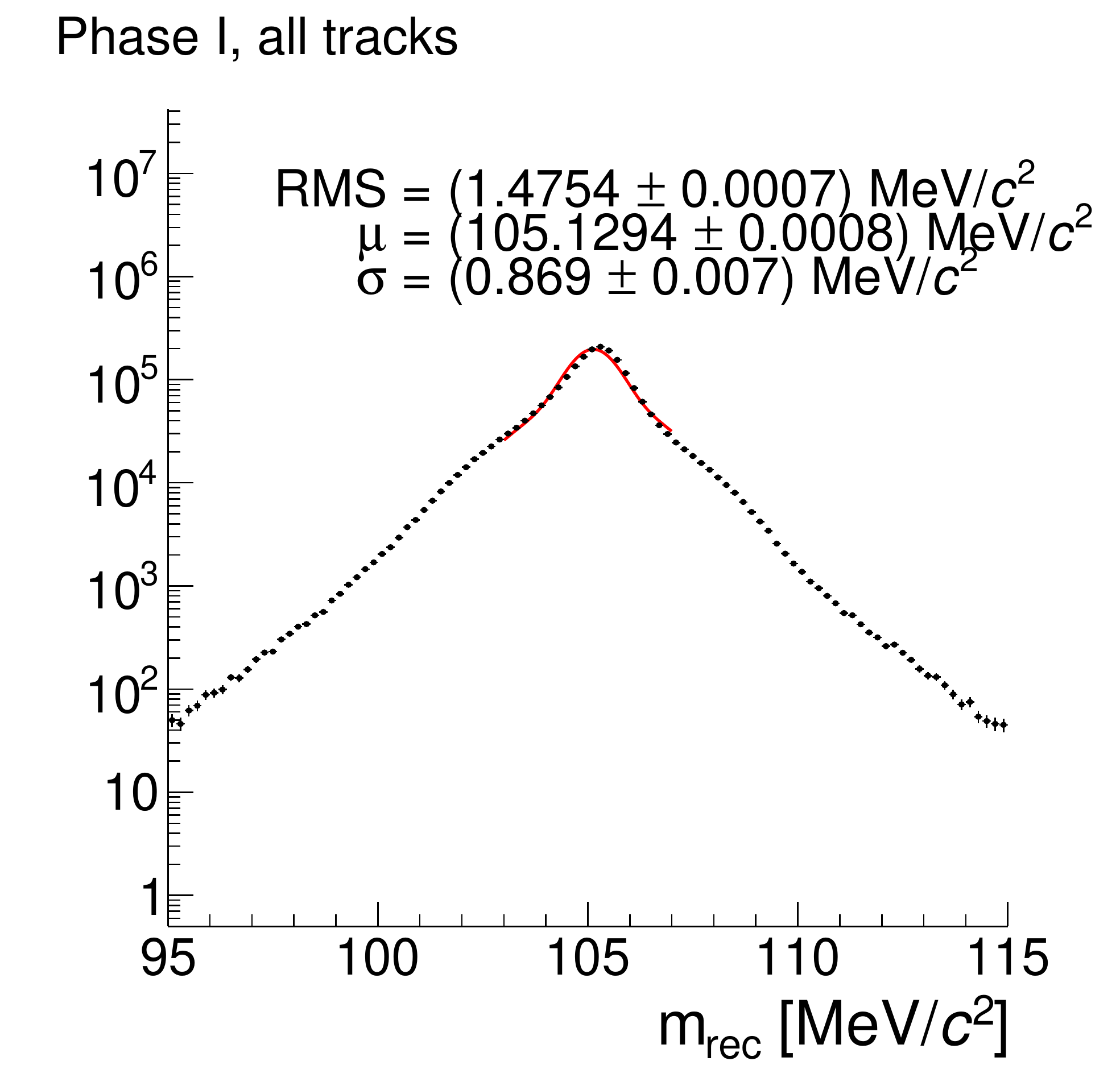}
		\includegraphics[width=0.4\textwidth]{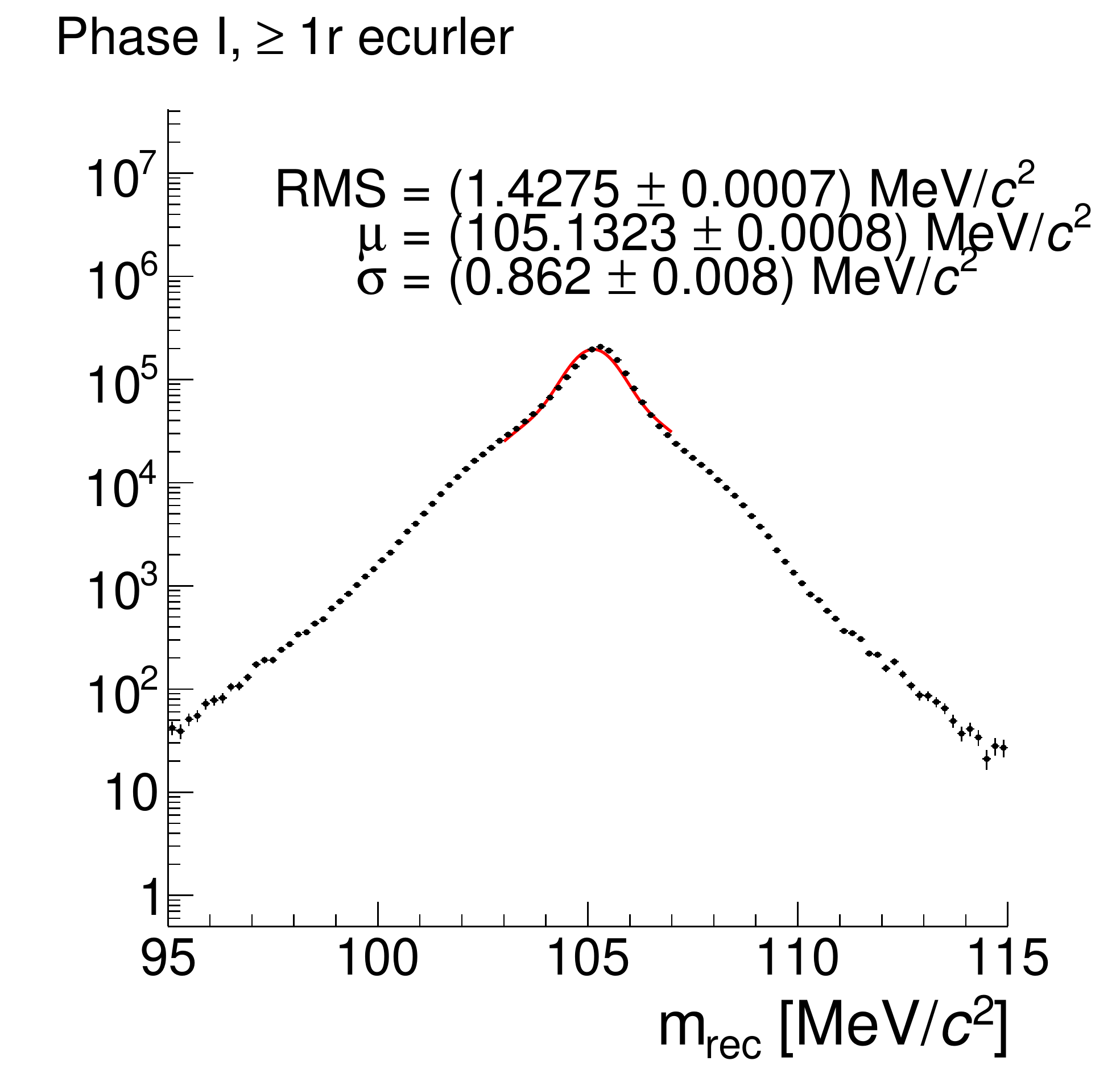}
		\includegraphics[width=0.4\textwidth]{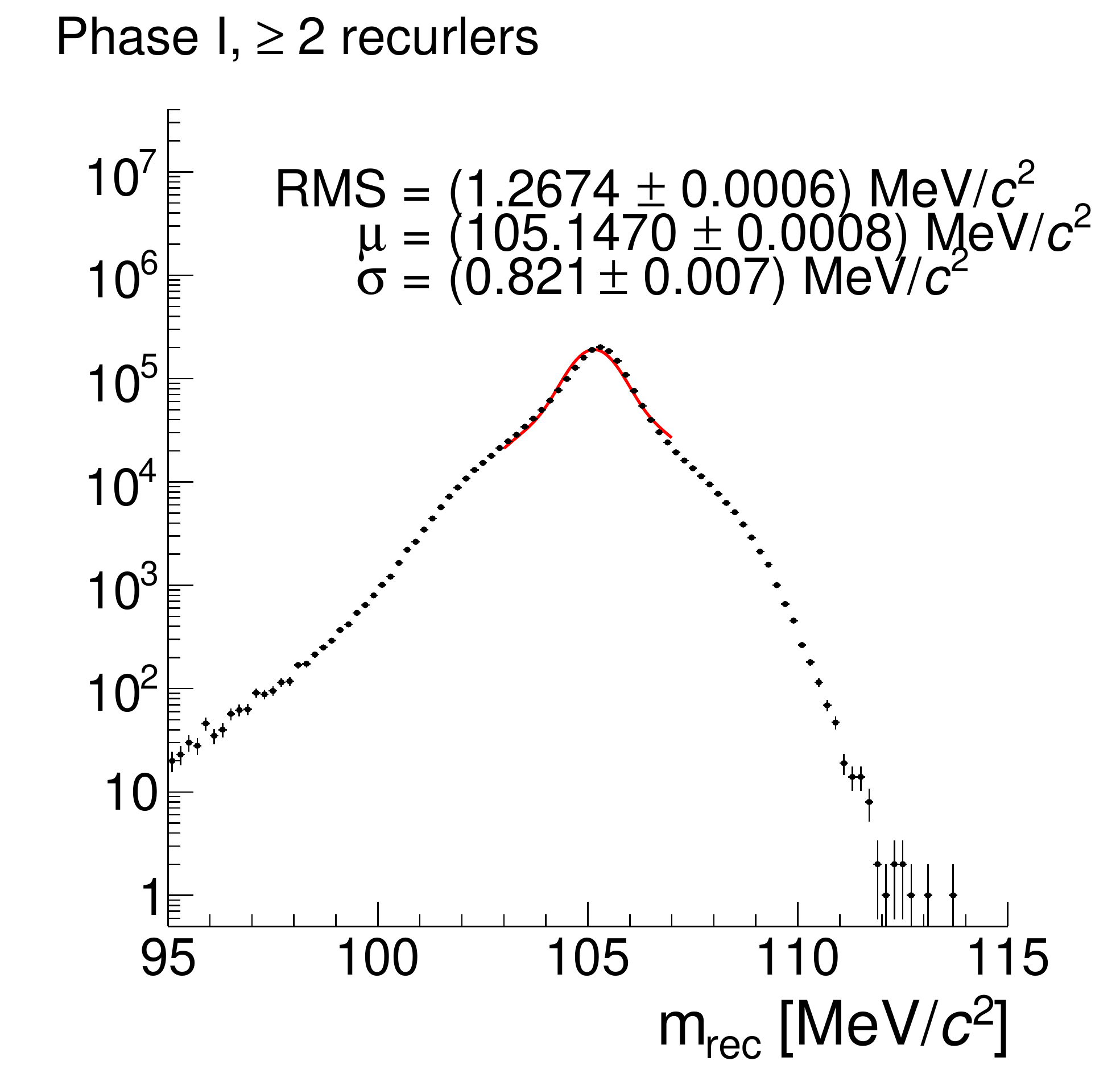}
		\includegraphics[width=0.4\textwidth]{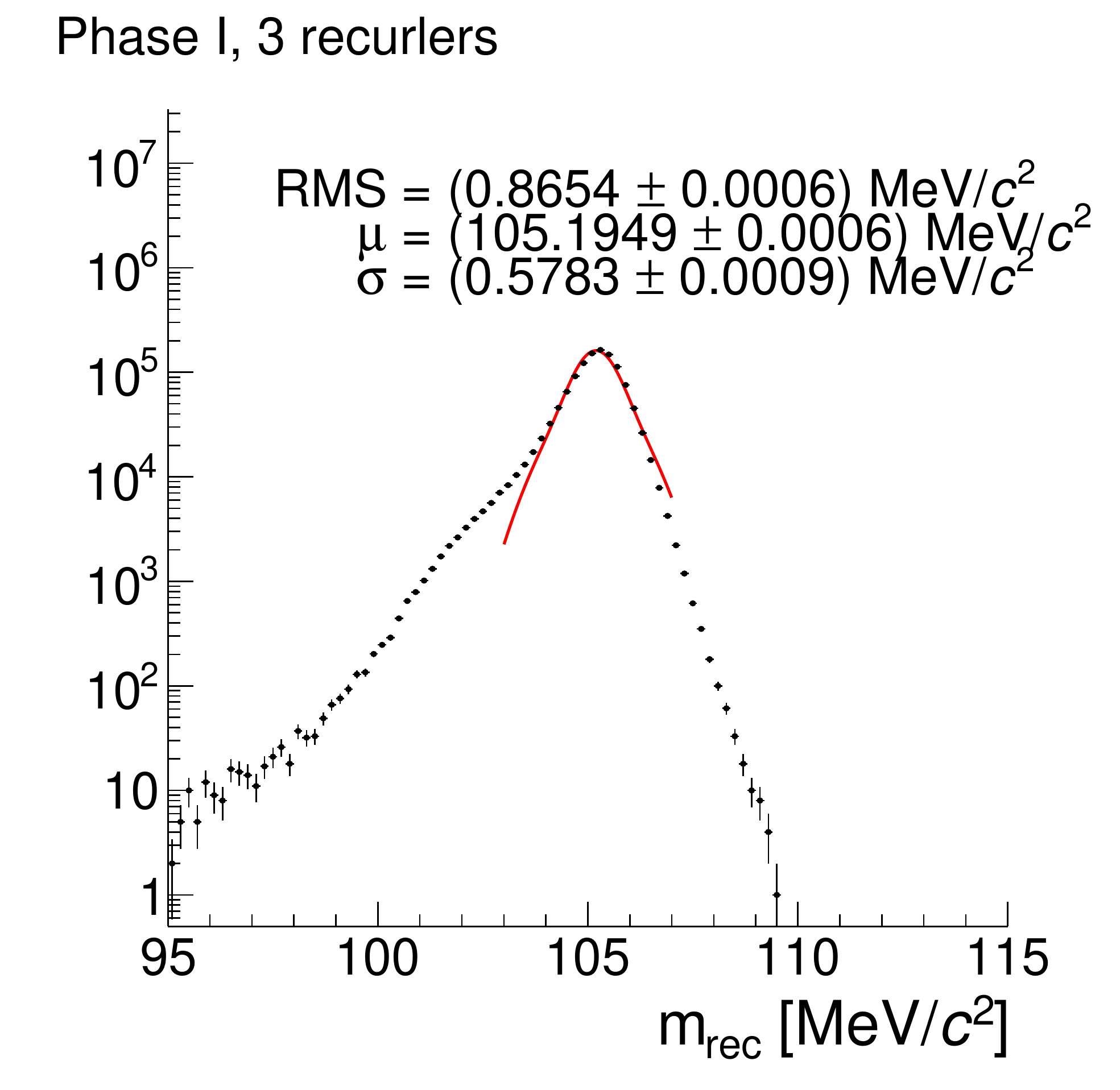}
	\caption{Reconstructed muon mass $m_{rec}$ for signal decays for all tracks (top left),
	at least one recurler (top right), at least two recurlers (bottom left) and three
	recurlers (bottom right).  The fits are the
	sum of two Gaussian distributions and the quoted $\sigma$ is the area-weighted mean;
	the main purpose of the fit is to guide the eye and highlight the non-symmetric
	resolution distribution.}
	\label{fig:mmass_all_Ib}
\end{figure*}

\begin{sloppypar}
For all candidates with a vertex fit $\chi^2 < 15$ 
the tracks are
extrapolated to the vertex and four-vectors are constructed with an electron
mass assumption.
From the three four-vectors, the mass of the decaying particle (should
correspond to the muon mass) and the momentum of the center-of-mass system (CMS)
in the detector frame of reference (should be zero for decays at rest) are determined.
\end{sloppypar}

The resolution for the muon momentum is depicted in \autoref{fig:momentumreso_3recurl_Ib}. 
The magnitude of the reconstructed momentum $p_{cms}$ is shown in \autoref{fig:ptot}.
No detailed optimization of the momentum selection has been performed for Mu3e to date, 
so for the distributions shown in this report, we used the requirement of
$p_{CMS} < \SI{4}{MeV/c}$, which loses less than 5\% of the signal.

In order to suppress background from combinations of Michel decays with Michel electrons 
undergoing Bhabha scattering, the lower of the two electron-positron invariant masses
$m_{ee,low}$ 
is required to lie outside a window ranging from \SI{5}{MeV/c^2} to \SI{10}{MeV/c^2},
see also section~\ref{sec:BhabhaPairSuppression}. 

The resolution for the reconstructed mass $m_{rec}$ is shown in \autoref{fig:mmass_all_Ib}.
As the distributions show, the central peak of the mass resolution fulfils the criteria
set out in \autoref{sec:DecayMu3e}, especially if requiring recurling 
tracks. Sizeable Landau-like tails only appear on the low mass side.

\subsection{Signal Efficiency}
\label{sec:SignalEfficiency}

\begin{figure}
	\centering
		\includegraphics[width=0.4\textwidth]{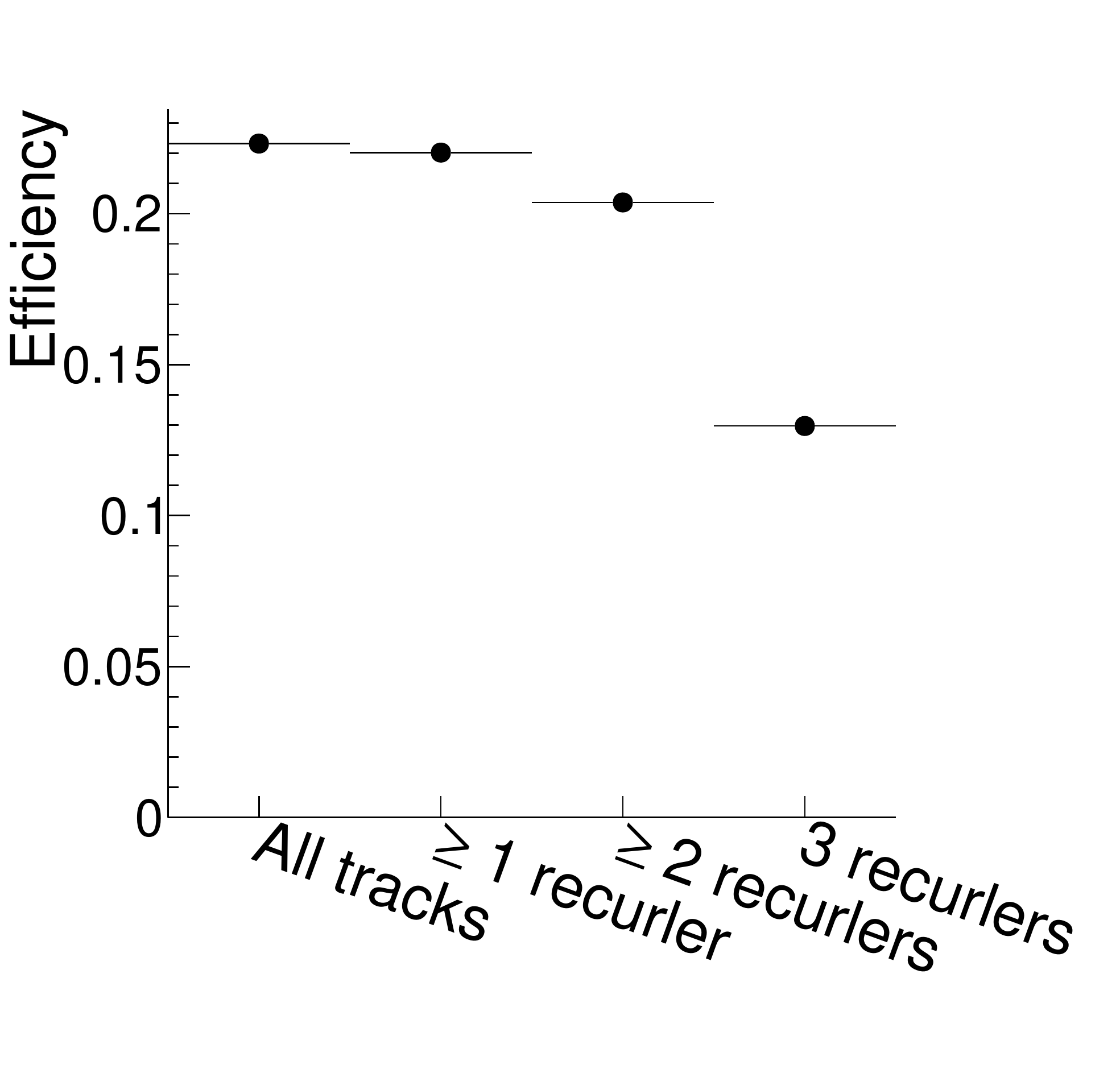}
	\caption{Total efficiency for reconstructing phase-space signal events
	as a function of the required number of recurling tracks.
	This includes the geometrical detector acceptance, track and vertex
	reconstruction and selection inefficiencies.}
	\label{fig:effhisto}
\end{figure}

\begin{figure}
	\centering
		\includegraphics[width=0.4\textwidth]{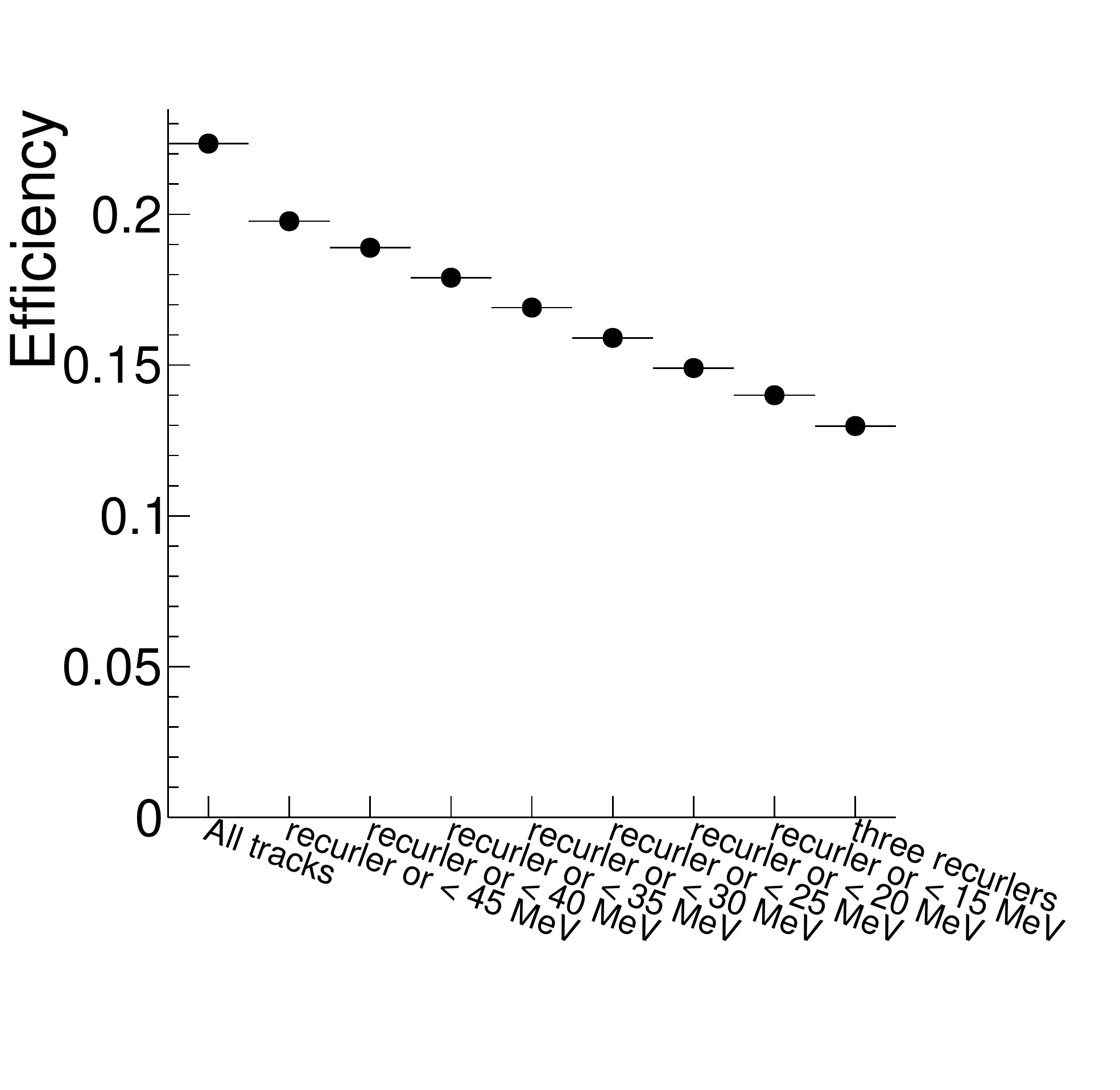}
		\includegraphics[width=0.4\textwidth]{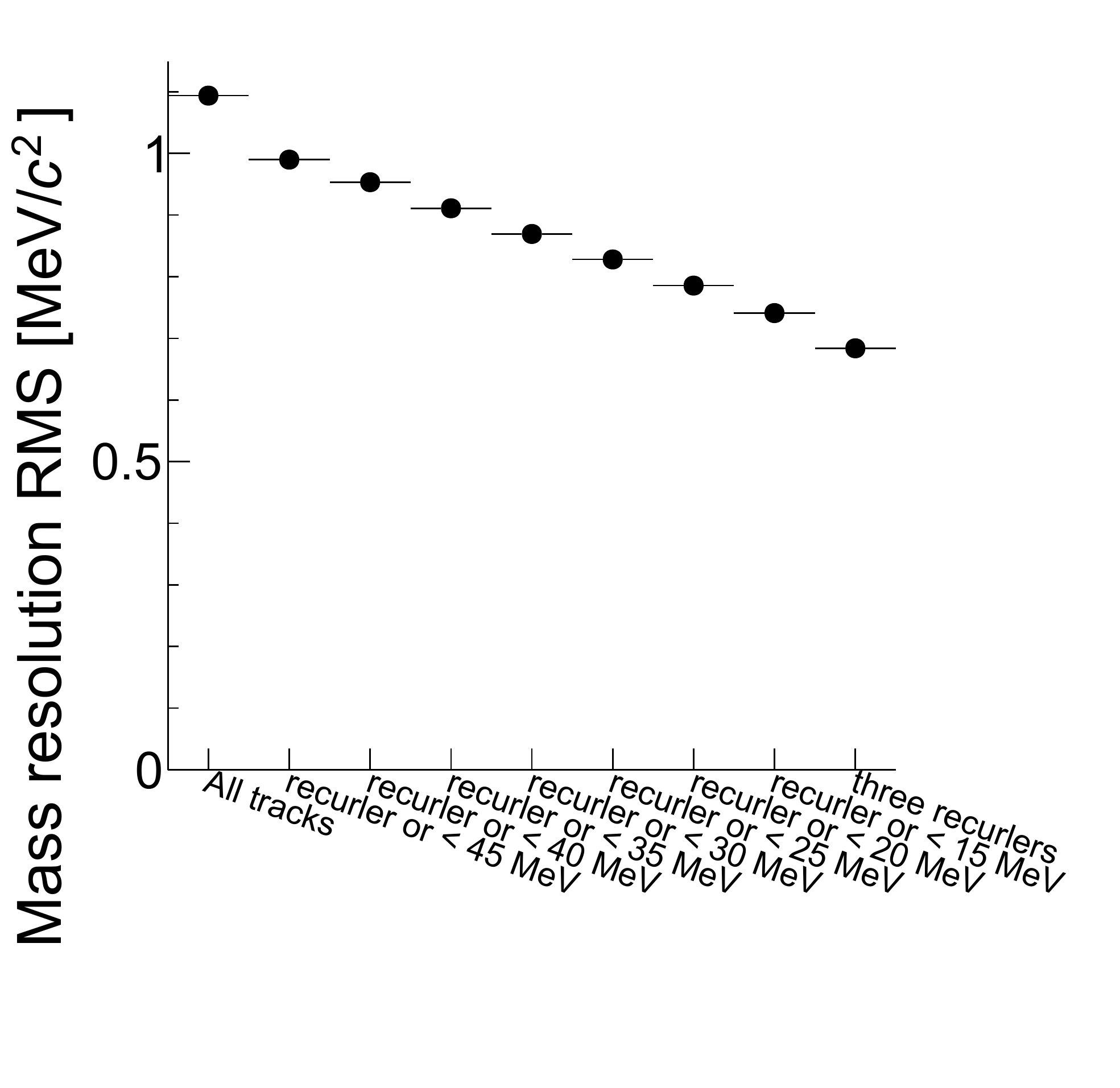}
	\caption{Efficiency before timing selection for reconstructing phase-space signal events (top) 
	and the $RMS$ of the corresponding three-particle invariant mass distribution (bottom). 
	Both use the same selection criteria. }
	\label{fig:effhisto_pcuts}
\end{figure}

\begin{table*}[tb!]
	\centering
		\begin{tabular}{lrr}
		\toprule
		Step 												&	Step efficiency  & Total efficiency \\
		\midrule
		Muon stops											&	100\%			 & 100\% \\
		Geometrical acceptance, short tracks				&	38.1\%			 & 38.1\% \\
		Geometrical acceptance, long tracks					&   68.0\%			 & 25.9\% \\
		Short track reconstruction							&   89.5\%			 & 34.1\% \\
		Long track reconstruction$^1$						&   67.2\%			 & 17.4\% \\
		Recurler rejection/Vertex fit convergence			&   99.4\%			 & 17.3\% \\
		Vertex fit $\chi^2 < 15$							& 	91.3\%			 & 15.8\% \\
		CMS momentum $<$ \SI{4}{MeV/c} 						& 	95.6\%		 	 & 15.1\% \\
		$m_{ee,low}$ < \SI{5}{MeV/c^2} or > \SI{10}{MeV/c^2}&   98.0\%			 & 14.9\% \\
		\SI{103}{MeV/c^2} < $m_{rec}$ < \SI{110}{MeV/c^2}	&	97.0\%		     & 14.4\% \\
		Timing 												&   90.0\% 			 & 13.0\%\\
		\bottomrule
		\end{tabular}
	\caption{Efficiency of the various reconstruction and analysis steps.\\
	$^1$ Note that the efficiency of this step is quoted relative to the acceptance for long tracks.}
	\label{tab:AnaEfficiency}
\end{table*}

For every reconstruction step, there is a possibility of signal loss; the
largest loss is due to the geometrical acceptance of the detector.
For signal decays in the target and evenly distributed in phase space, 
approximately \num{38.1}\% have all 
three electrons traverse the four layers of the central detector in the active region.
If recurling tracks are required, the acceptance is further reduced.
There are also inefficiencies in the reconstruction and vertex fits,
especially due to the $\chi^2$ cuts, which mostly get rid of tracks with large
angle scattering, that cannot be reliably and precisely reconstructed.
The signal losses in the CMS momentum and $m_{rec}$ cuts are mainly due
to events where one of the decay particles undergoes a large energy loss or 
radiates a Bremsstrahlung photon.
The overall efficiency after applying all mentioned cuts as well as
a veto on events where the tracks have inconsistent timing is shown in
\autoref{fig:effhisto} as a function of the required number of recurling 
tracks.

With the selection criteria used, the overall efficiency is \num{13.0}\% when
three recurling tracks are required.
The efficiency losses are listed in \autoref{tab:AnaEfficiency}.
Further gains are expected from a thorough optimisation of the cuts; on the other
hand, imperfections of the real detector will likely lead to some additional losses.
The selection efficiency can be increased (at the cost of a deterioration of the 
mass resolution) e.g.~by requiring recurling tracks only for high momentum tracks, 
see \autoref{fig:effhisto_pcuts}.

\section{Backgrounds}
\label{sec:SimBackgrounds}

\subsection[Internal Conversion]{Background from Radiative Decays with Internal Conversion}
\label{sec:InternalConversionBackgroundStudy}

\begin{figure}[tb!]
	\centering
		\includegraphics[width=0.49\textwidth]{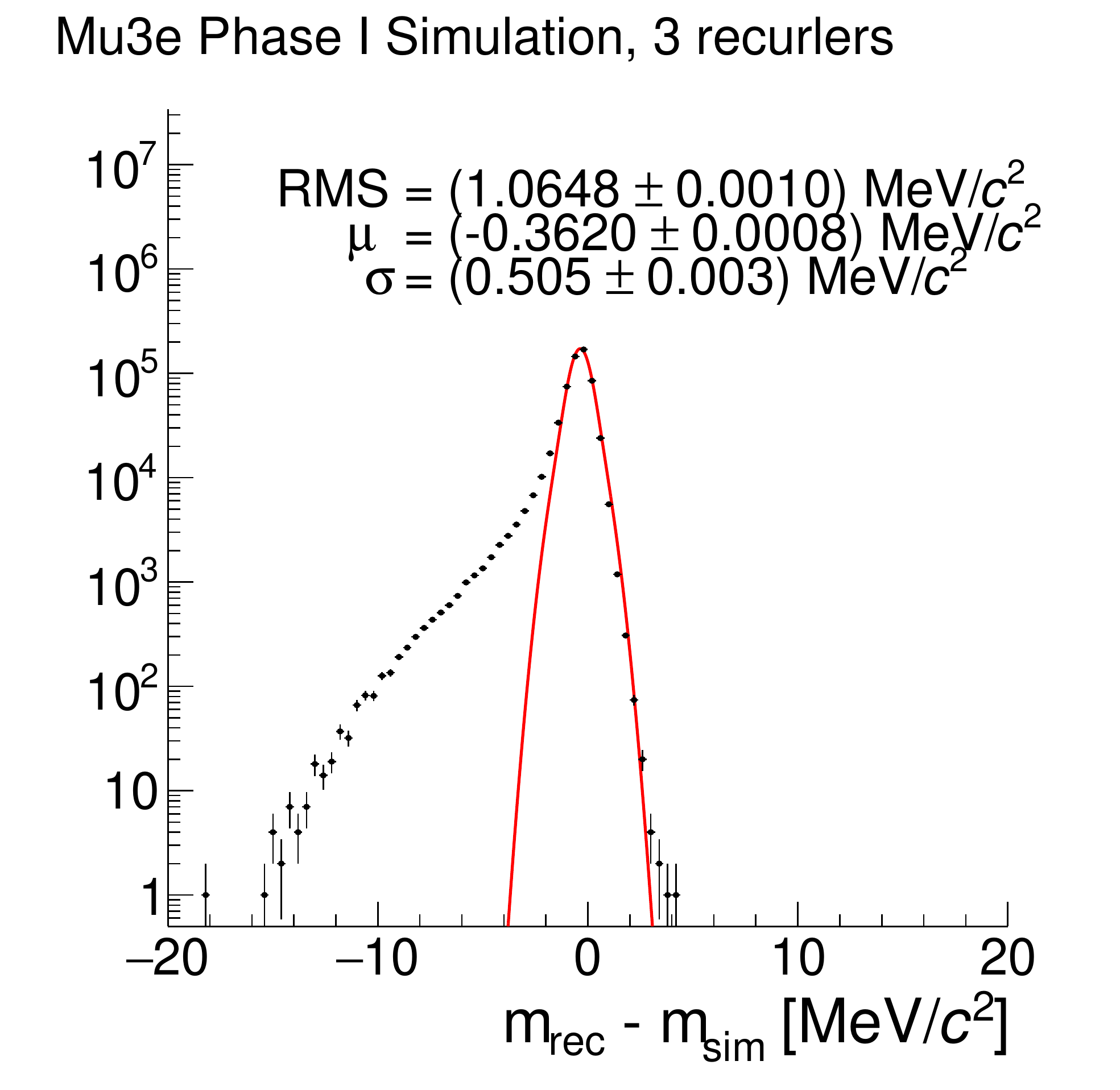}
	\caption{Resolution of the mass reconstruction for internal conversion events
	with a visible mass above \SI{90}{MeV/c^2} for three recurling tracks and a CMS momentum
	of the three particle system of less than \SI{8}{MeV}.}
	\label{fig:mmassreso}
\end{figure}

The background from radiative decays with internal conversion of the photon,
(\mtenunuSigned, referred to in short as \emph{internal conversion}) 
is simulated as described in \autoref{sec:Simulation:MuonDecays}
using the matrix element provided by Signer et al.~\cite{Pruna:2016spf}.
The total branching fraction for this decay is \num{3.4e-5} \cite{Bertl19851},
so a complete simulation is challenging in terms of computing time and storage space.
We are however mostly interested in the region of phase space where the neutrinos
carry little momentum, characterised by a large invariant mass (the \emph{visible mass})
of the $e^+e^-e^+$ system. The branching fraction for the high visible mass region
(a lower cutoff of \SI{90}{MeV/c^2} was used for the studies presented here),
is strongly suppressed and many more decays than expected in the entire run time
can be easily simulated. We generate the events evenly distributed in this
restricted region of phase space and weight them with the squared matrix element. 
Migrations from lower masses into the signal region are
very strongly suppressed if three recurling tracks are required, 
see \autoref{fig:mmassreso}.

\subsection{Accidental Background}
\label{sec:AccidentalBackgroundStudy}

Accidental background arises from the combination of two Michel positrons with
an electron.
It is thus important to understand and limit electron production in the target
region. This is of particular importance for processes such as Bhabha-scattering, where the electron
and positron tracks intersect in space and time and only the separation from
the second positron remains as a suppression criterion.

\subsubsection{Electron Production in the Target}
\label{sec:TargetSimulation}

\begin{figure*}
	\centering
		\includegraphics[width=0.55\textwidth]{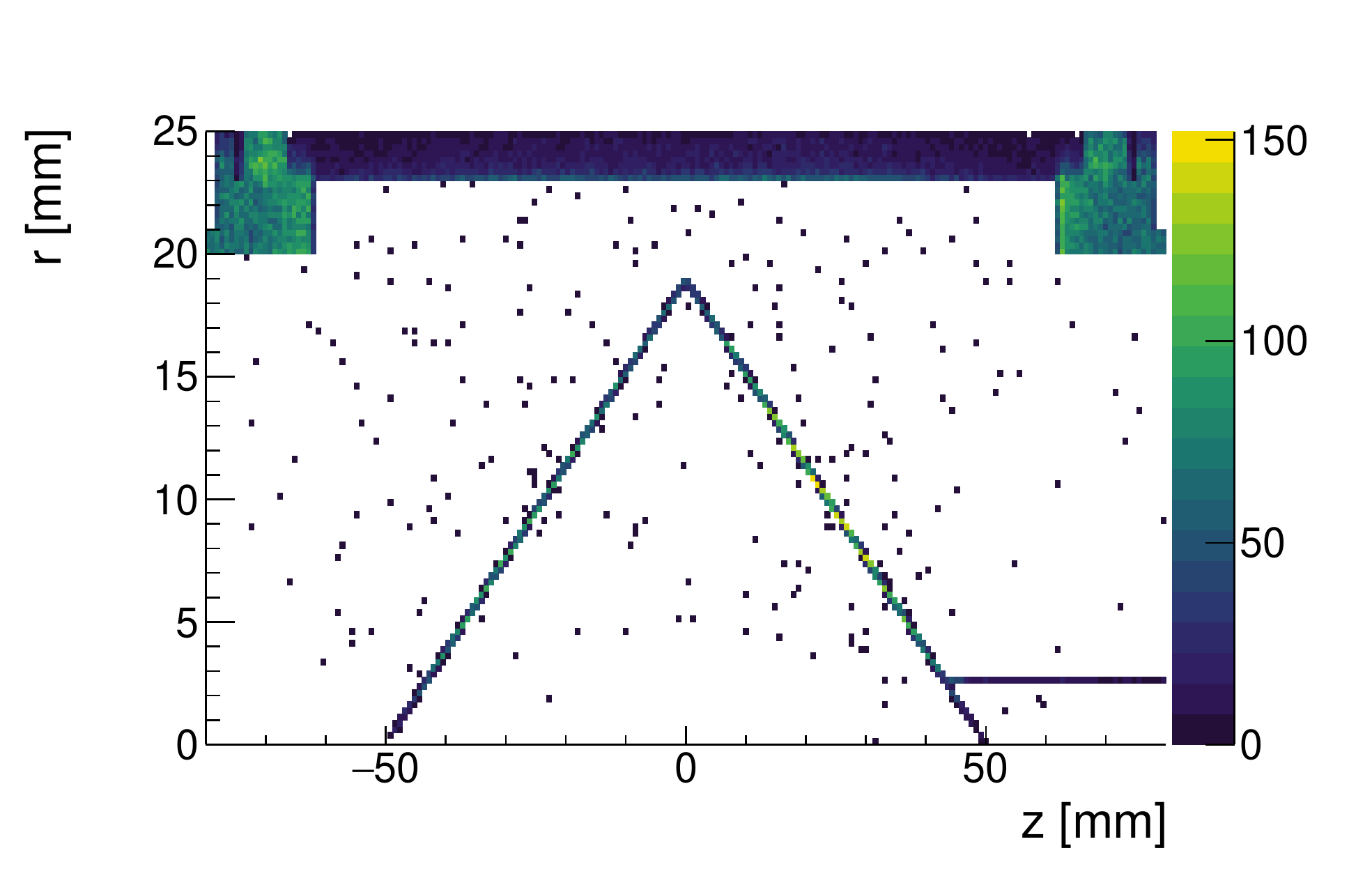}
		\includegraphics[width=0.35\textwidth]{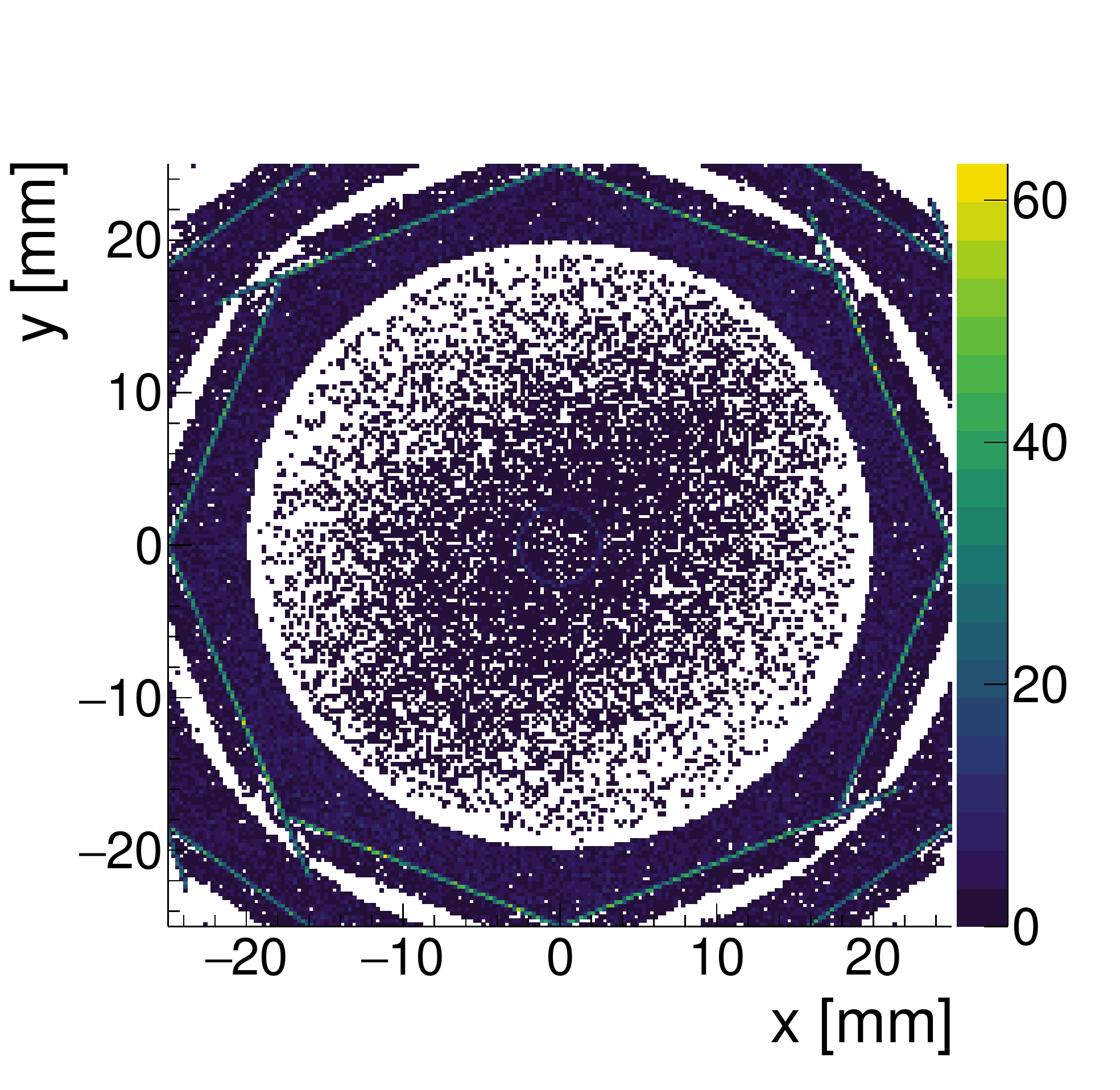}
	\caption{Longitudinal view (left) and transverse view (right) of the loci of Bhabha scattering producing an
	electron and a positron both in the detector acceptance in the target
	region for \SI{1.9}{s} of running at \num{1e8} muon stops per second. The distribution mirrors
	the material distribution in the centre of the detector.}
	\label{fig:bhabhasel_vertices_rz_zoom}
\end{figure*}

\begin{figure*}
	\centering
		\includegraphics[width=0.55\textwidth]{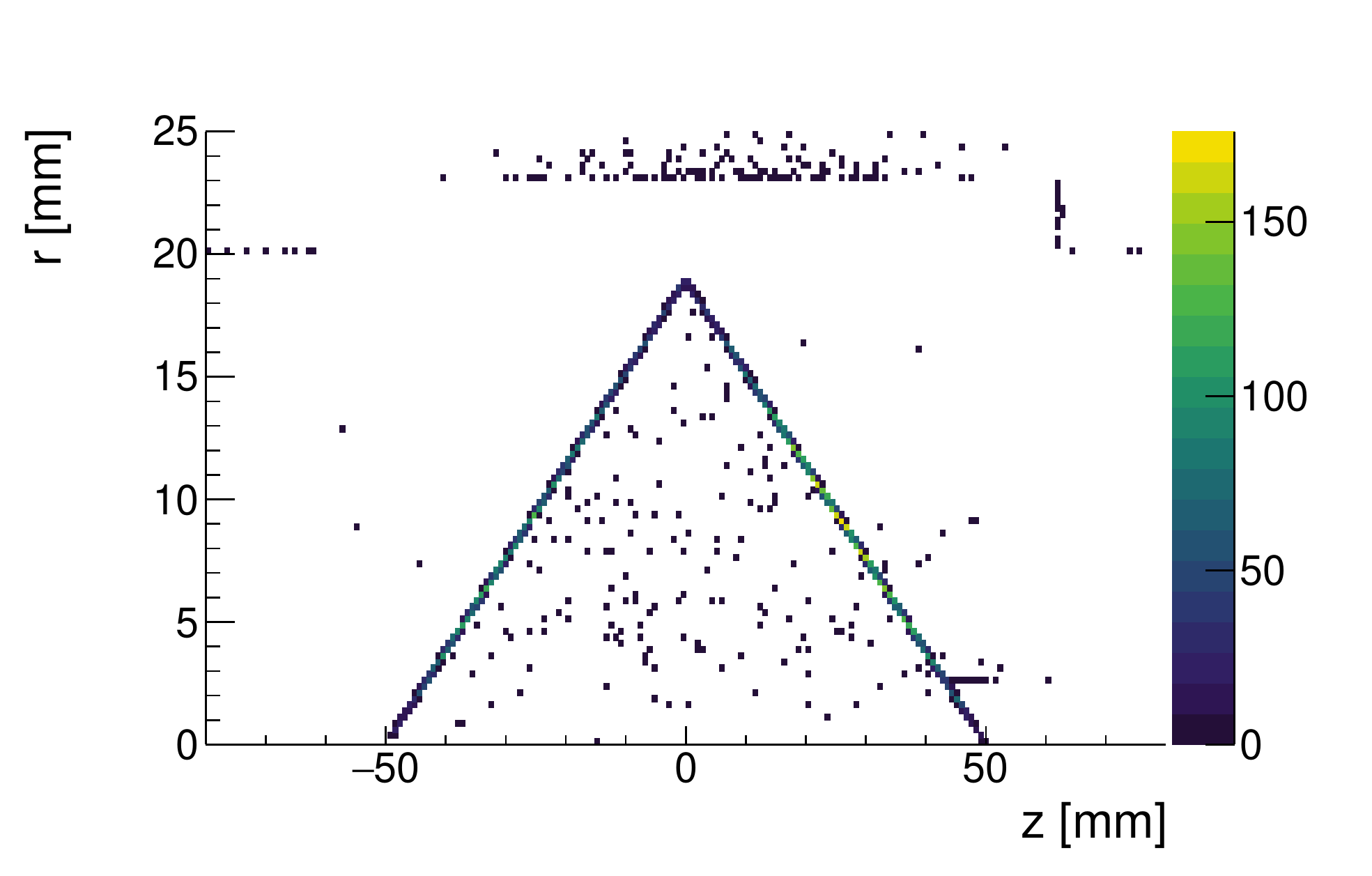}
		\includegraphics[width=0.35\textwidth]{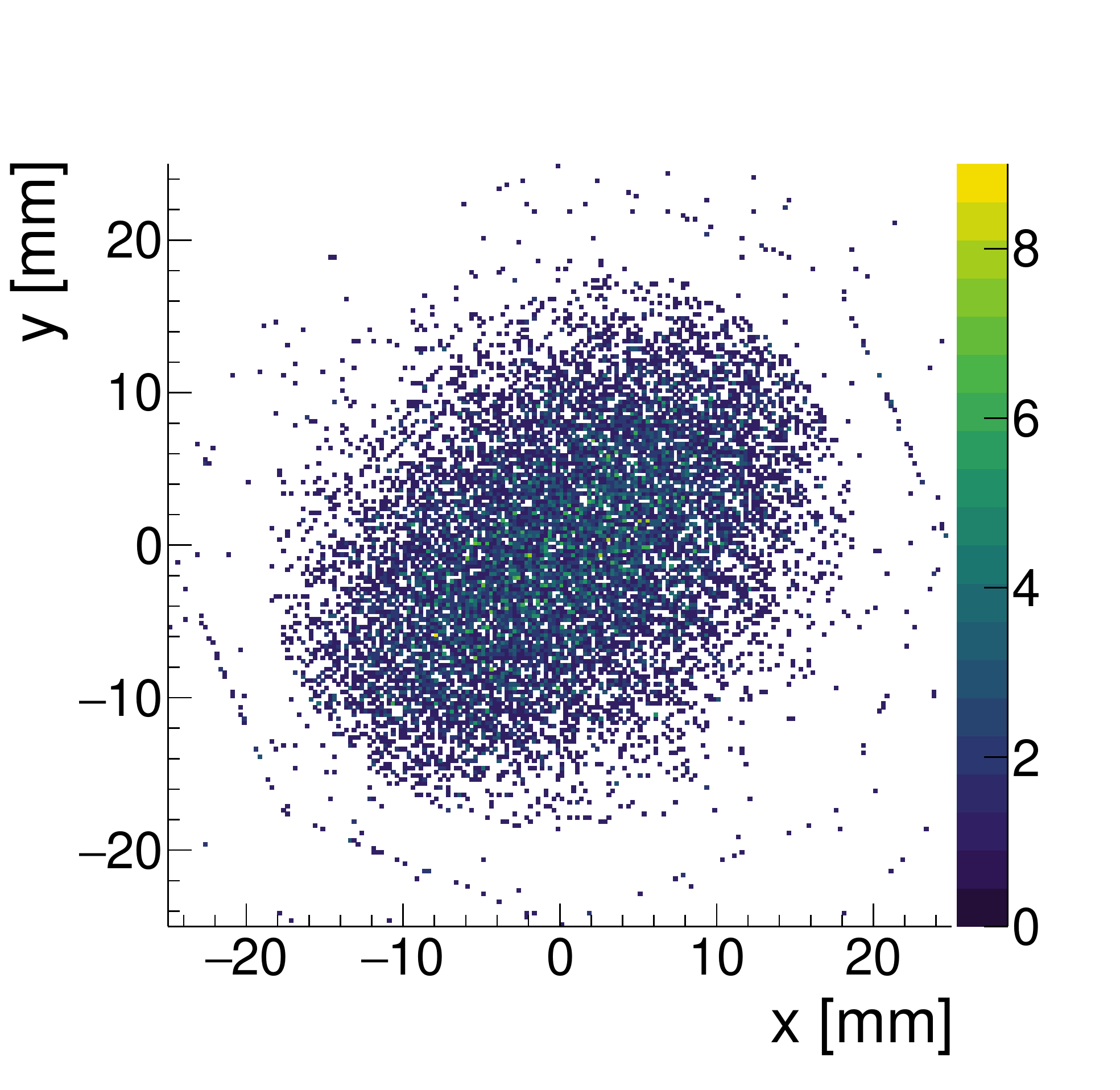}
	\caption{Longitudinal view (left) and transverse view (right) of muon decay vertices leading to a positron that
	then interacts via Bhabha scattering in the target resulting in an
	electron and a positron both in the detector acceptance in the target
	region for \SI{1.9}{s} of running at \num{1e8} muon stops per second. The distribution corresponds
	to the muon stopping distribution in the target region, showing no unexpected source of positrons.}
	\label{fig:targetbhabhasel_positron_vertices_rz_zoom}
\end{figure*}

\begin{table*}
	\centering
\footnotesize 
\begin{tabular}{lrrrrrr}
\toprule
                & Produced in    & Produced in   & Reconstructed  & Reconstructed & Reconstructed  & Reconstructed\\ 
Electron source & inner 				 & target         & inner detector & target region & inner detector & target region\\ 
                &  detector      & region         & short tracks   & short tracks  & long tracks    & long tracks\\ 
\midrule 
Bhabha scattering & \num{5.5E-04} & \num{1.1E-04} & \num{2.7E-04} & \num{5.7E-05} & \num{2.3E-04} & \num{4.4E-05} \\ 
~~~$e^+e^-$ visible   & \num{4.3E-04} & \num{7.7E-05} & \num{1.5E-04} & \num{2.6E-05} & \num{1.1E-04} & \num{1.7E-05} \\ 
Photon conversion & \num{2.3E-05} & \num{2.1E-06} & \num{1.1E-05} & \num{1.0E-06} & \num{9.2E-06} & \num{8.0E-07} \\ 
~~~$e^+e^-$ visible   & \num{5.7E-06} & \num{4.6E-07} & \num{1.5E-06} & \num{1.3E-07} & \num{1.2E-06} & \num{9.3E-08} \\ 
Compton scattering & \num{3.6E-05} & \num{4.3E-06} & \num{1.7E-05} & \num{2.2E-06} & \num{1.4E-05} & \num{1.7E-06} \\ 
Internal conversion & \num{3.1E-05} & \num{2.9E-05} & \num{1.7E-05} & \num{1.6E-05} & \num{1.3E-05} & \num{1.3E-05} \\ 
~~~$e^+e^-$ visible    & \num{1.1E-06} & \num{1.0E-06} & \num{3.6E-07} & \num{3.3E-07} & \num{2.3E-07} & \num{2.2E-07} \\ 
\midrule 
Total & \num{6.4E-04} & \num{1.5E-04} & \num{3.2E-04} & \num{7.6E-05} & \num{2.6E-04} & \num{5.9E-05} \\ 
\bottomrule 
\end{tabular}


	\caption{Electrons with transverse momentum larger than \SI{10}{MeV} created
	in the target region, relative to the number of muon stops.
	The inner detector region is a cylinder including the vacuum window and the
	first pixel layer, the target region is a cylinder just containing the target.}
	\label{tab:electrons_Ib}
\end{table*}

\begin{figure*}
	\centering
		\includegraphics[width=0.49\textwidth]{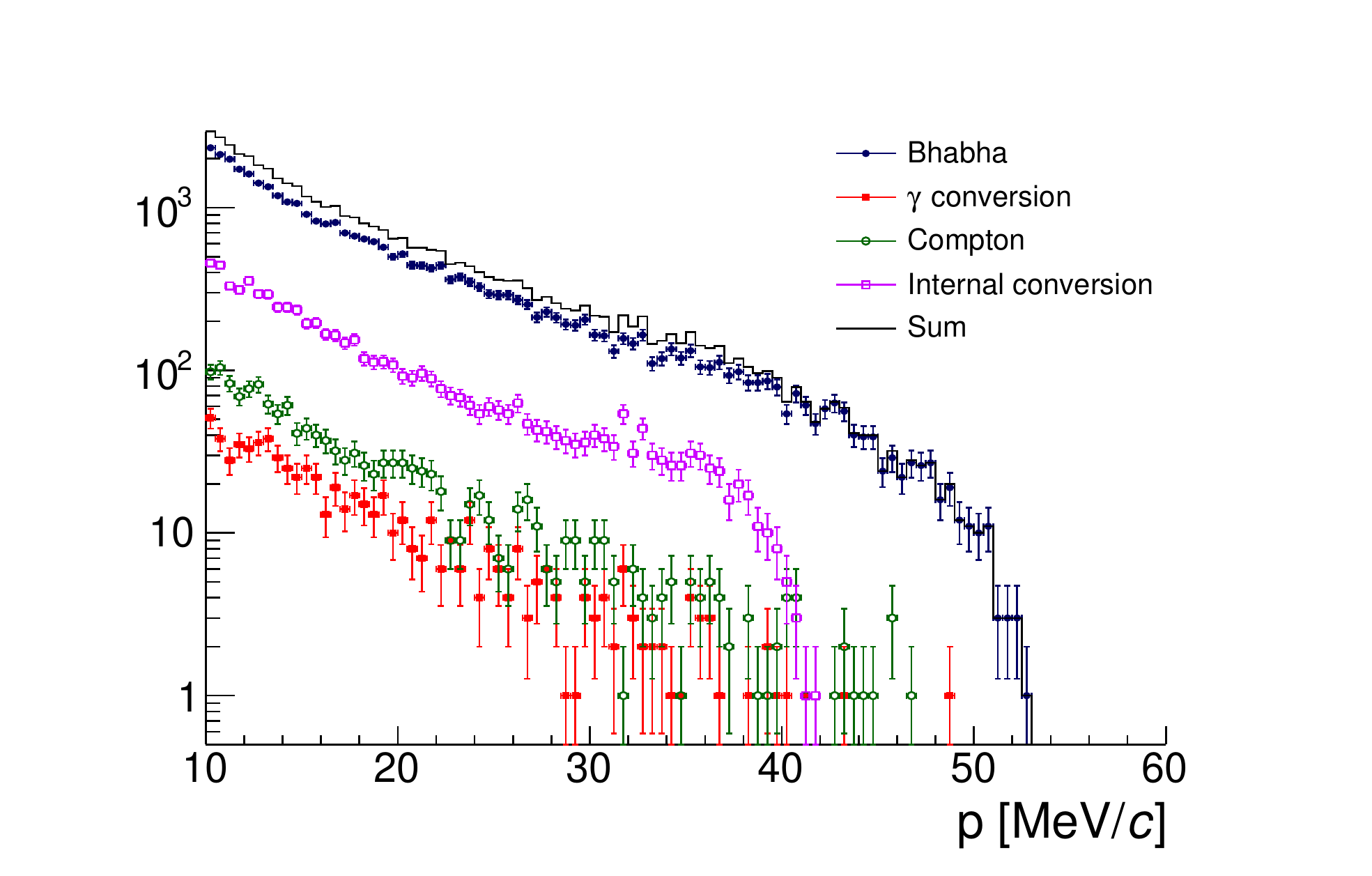}
		\includegraphics[width=0.49\textwidth]{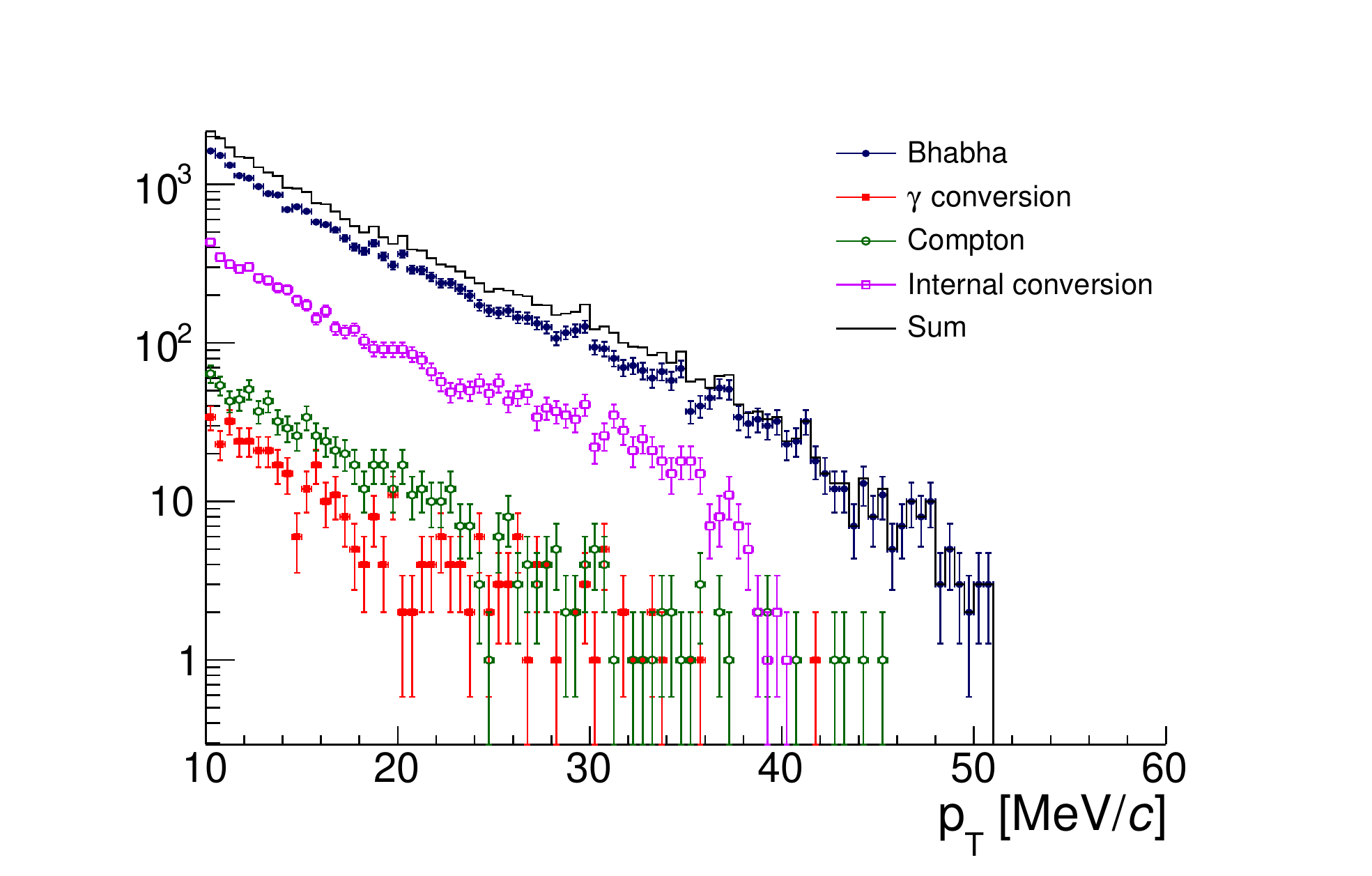}
	\caption{Momentum spectrum (left) and transverse momentum spectrum (right) of
	electrons produced in the target region.}
	\label{fig:ElectronMomentumSpectrum}
\end{figure*}

The default target is part of the Geant4 detector simulation as described in
\autoref{sec:Simulation}.
The material of the target is a place where electrons from Bhabha and Compton
scattering as well as from photon conversion can be produced and contribute to
accidental background.
Bhabha scattering needs special attention, as very often both the electron and the positron
partaking in the scattering process end up in the detector acceptance; the
corresponding vertices are shown in \autoref{fig:bhabhasel_vertices_rz_zoom}.
As shown in \autoref{fig:targetbhabhasel_positron_vertices_rz_zoom}, almost all the corresponding
primary positrons come from muon decays in the target and can thus not be further
reduced or shielded.

The total number of electrons produced per Michel decay is shown in
\autoref{tab:electrons_Ib}.
As can be seen, Bhabha scattering is the most important background process. 
The reason that there are significantly fewer electrons reconstructed than produced is because of the 
the steeply falling momentum spectrum, see \autoref{fig:ElectronMomentumSpectrum}. 
This means that many of the electrons end up at or below the detector and reconstruction acceptance.


\subsubsection{Timing Suppression}
\label{sec:TimingSuppression}

Time information from hits in the fibre and tile detectors provides an important
handle for the suppression of accidental backgrounds. For the purpose of this study,
the accidental background is estimated using the track combinations in \SI{50}{ns}
reconstruction frames (achievable with pixel detector timing only). The additional
suppression by the dedicated timing detectors is then expressed relative to these
combinations.

The precise timing of a track is determined by the number of assignable hits in
the fibre detector and the existence of a matched tile hit.
If a track reaches the recurl stations, the timing is
dominated by the tile detector, which is more accurate. Detailed studies of the signal 
efficiency and background suppression of the timing detectors are described in
\cite{Corrodi2018} and summarised in \autoref{sec:Fibre}.
Using this we have a working point of \SI{90}{\percent} efficiency for coincident tracks (signal), 
a timing suppression of approximately 70 for the dominant accidental background with two 
tracks correlated and one uncorrelated in time, and a suppression of more than 
three orders of magnitude for three uncorrelated tracks.

\subsubsection{Kinematic Suppression}
\label{sec:KinematicSuppression}

\begin{figure}
	\centering
		\includegraphics[width=0.49\textwidth]{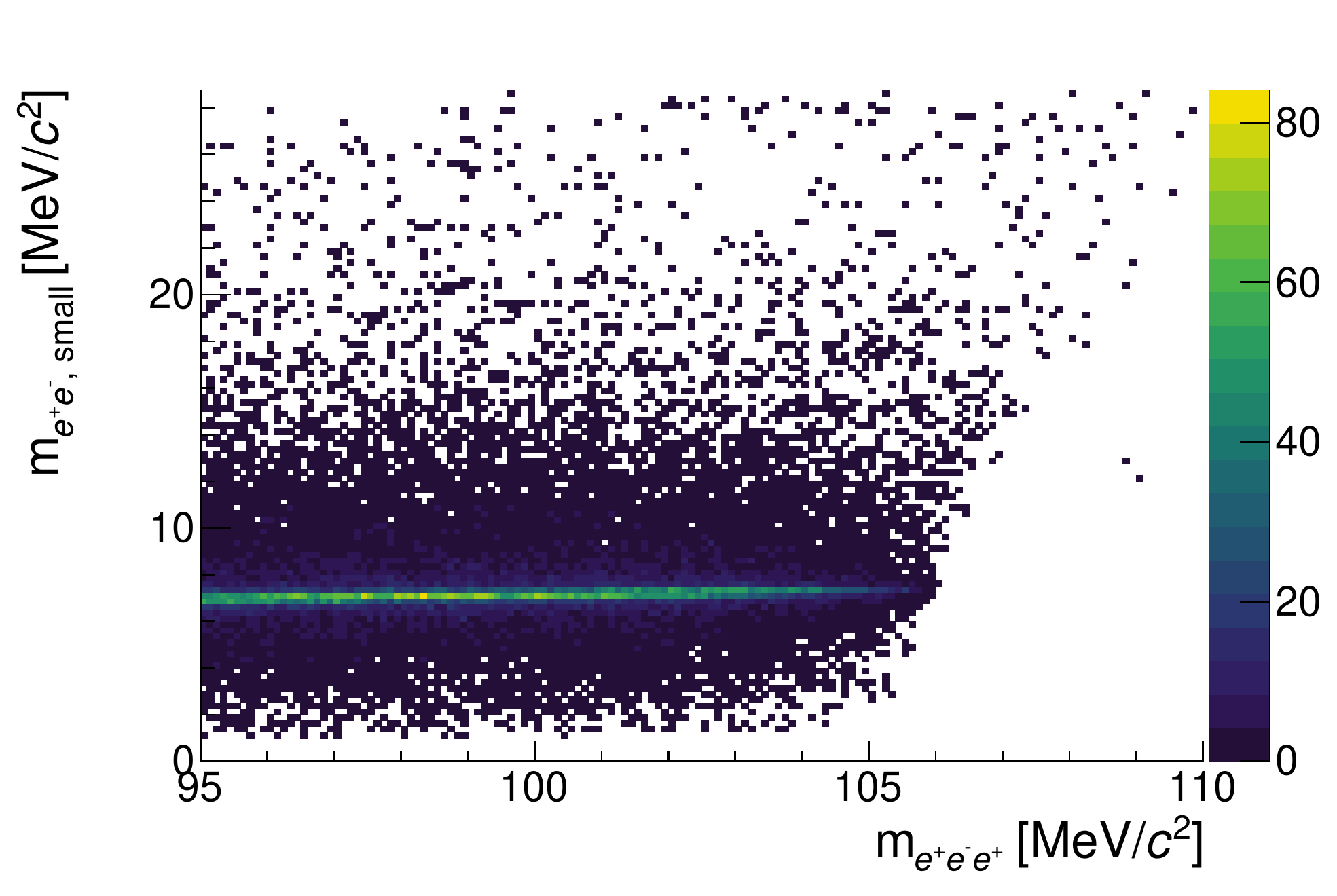}
	\caption{Small invariant mass of $e^+e^-$ pairs versus  $e^+e^-e^+$ invariant 
	mass for accidental combinations of a Bhabha $e^+e^-$ pair with a Michel
	positron. Simulated kinematics weighted with the track reconstruction efficiency.}
	\label{fig:bhabhakinematics}
\end{figure}

The largest suppression factors for accidental background come from kinematics,
i.e.~the requirement that the three momenta sum up to zero (enforced by the
total momentum selection) and a mass window around the muon mass.
Typical suppression factors are of the order of one million.
The kinematics of the event however also strongly affect the suppression power
of the vertex fit; the corresponding requirements unfortunately do not factorise
and large simulated samples are required.

\subsubsection{Bhabha Pair Suppression}
\label{sec:BhabhaPairSuppression}

\begin{sloppypar}
The suppression of accidental background by requiring a common vertex of three 
tracks is highly dependent on the kinematics, or rather the event topology.
In the interesting cases of Bhabha scattering or photon conversion, there is
an electron-positron pair with a small opening angle balanced with a positron
close to the Michel edge going in the opposite direction.
This case is favourable for vertex-based background suppression.
As vertex and kinematic suppression do not factorise, we have simulated the most
common accidental background, Bhabha scattering plus a Michel electron with almost 
full statistics.

\end{sloppypar}

The Bhabha events with signal-like kinematics almost all have an $e^+e^-$ 
invariant mass of around \SI{7}{MeV/c^2}, as they originate from a positron
close to the Michel edge scattering on an electron at rest:
\[ m_{e^+e^-} \approx \sqrt{2 m_e m_{\mu}/2},
	\]
see \autoref{fig:bhabhakinematics}. 
A requirement on this mass (here we exclude the region from \SI{5}{MeV/c^2}
to \SI{10}{MeV/c^2}) is used to reduce the Bhabha background, does however
also remove a specific part of the signal phase space. Without this requirement,
accidential Bhabha background would limit the reach of the Mu3e detector in 
the phase~I configuration.

\num{2.8E9} frames of \SI{50}{ns} duration and with the nominal muon stop rate
of \SI{1e8}{Hz} were simulated. In each frame, one muon decay where the positron
undergoes Bhabha scattering immediately after the decay was included
(which assumes that the geometrical distribution of muon stops corresponds to 
the distribution of Bhabha scatters). The Bhabha scattering events are generated
evenly in the part of the phase space where both decay particles have an energy
of at least \SI{10}{MeV} and are then weighted by the Bhabha matrix element.

It is expected that a fraction of \num{7.7e-5} of all muon stops produce Bhabha
scattering in the target with both the electron and the positron inside the detector
acceptance. Assuming a timing suppression factor of 70, the \num{2.8E9} simulated frames 
then correspond to Bhabha scattering from \num{2.5e15} muon stops.
After reconstruction and applying all cuts, two simulated Bhabha events with 
reconstructed masses above \SI{103}{MeV/c^2} are left and zero above \SI{104}{MeV/c^2}.

\begin{sloppypar}
A similar simulation study for accidental background from combinations of internal
conversion decays and Michel decays indicates that this background contributes an
expectation of \num{8e-4} events in the signal region for \num{e16} muon decays
\cite{GrovesPhD}.
\end{sloppypar}

\section{Sensitivity}
\label{sec:Sensitivity}

\begin{figure}
	\centering
		\includegraphics[width=\columnwidth]{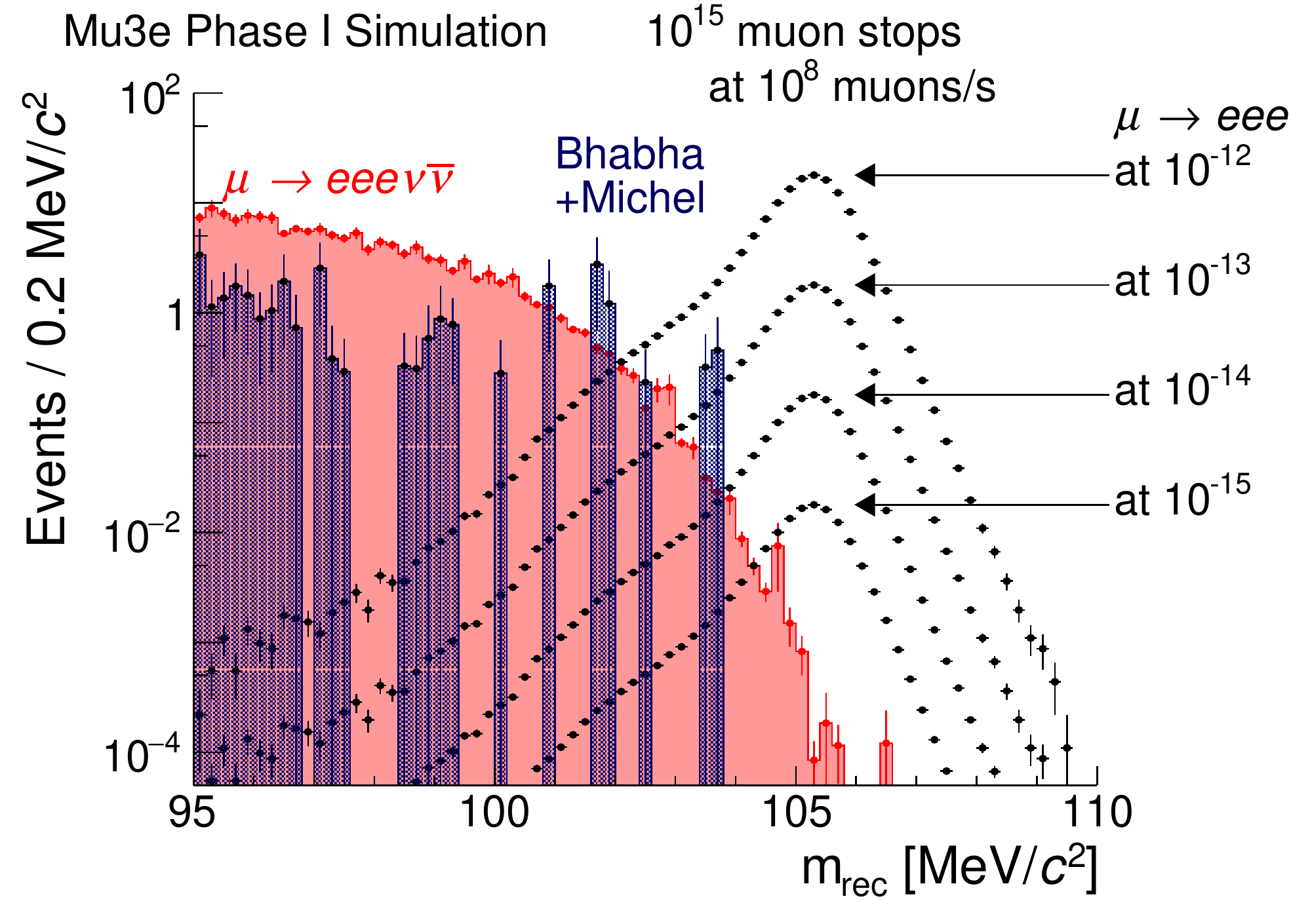}
	\caption{Reconstructed invariant mass for signal events at various branching fractions and
	events from radiative decays with internal conversion. Accidental background from combinations
	of Bhabha pairs and Michel electrons is also shown. The center-of-mass momentum is required to be less than
	\SI{4}{MeV/c}. Note that both the internal conversion and Michel and Bhabha simulation use weighted events.
	}
	\label{fig:SigICBhabha}
\end{figure}

\begin{figure}
	\centering
		\includegraphics[width=\columnwidth]{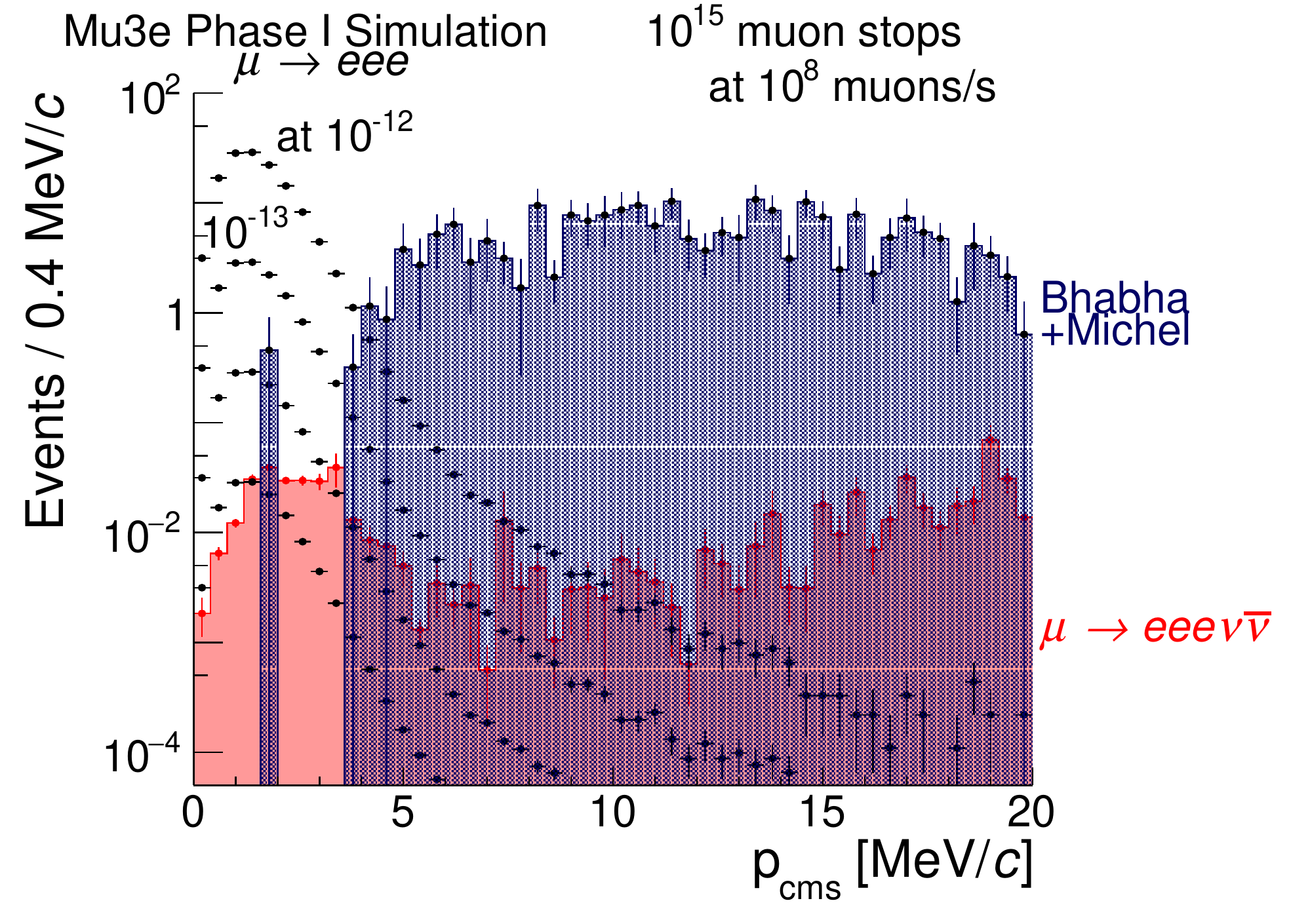}
	\caption{Reconstructed center-of-mass system momentum for signal events at various branching fractions 
	(\num{E14} and \num{E15} not labelled) and
	events from radiative decays with internal conversion. Accidental background from combinations
	of Bhabha pairs and Michel electrons is also shown. The reconstructed three-particle invariant mass 
	is required to be above \SI{103}{MeV/c^2} and below \SI{110}{MeV/c^2}. Note that both the internal 
	conversion and Michel and Bhabha simulation use weighted events.
	}
	\label{fig:SigICBhabhaPtot}
\end{figure}

\begin{figure}
	\centering
		\includegraphics[width=\columnwidth]{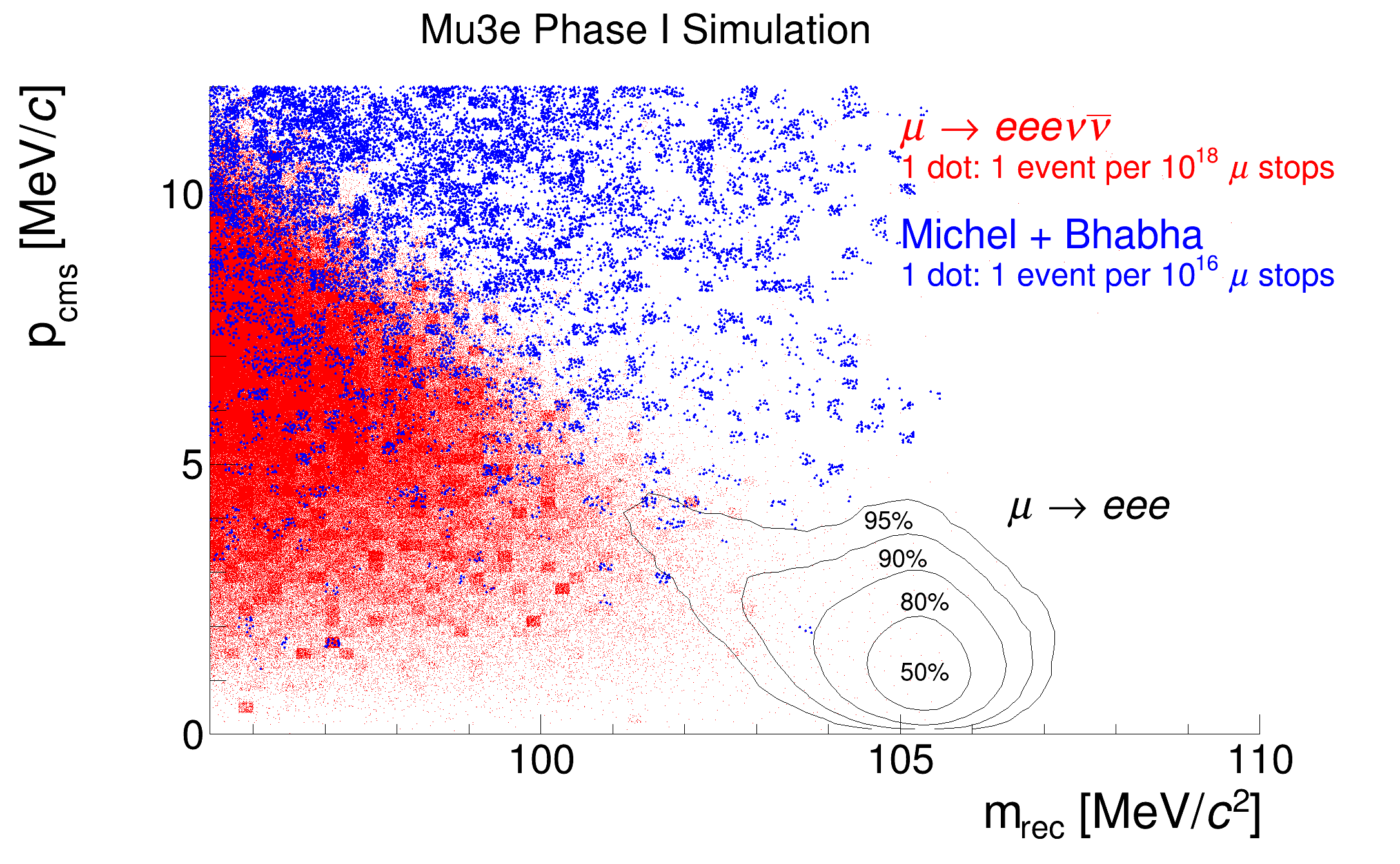}
	\caption{Reconstructed invariant mass versus the CMS momentum for signal events,
	events from radiative decay with internal conversion and accidental background from combinations
	of Bhabha pairs and Michel electrons. Note that both the internal conversion and 
	Michel and Bhabha simulation use weighted events.
	The shape of the signal contour at \SI{90}{\percent} and \SI{95}{\percent} comes from 
	events where one of the track has an upward fluctuation of the energy loss in the target 
	or the first tracker layers -- this leads to a lower reconstructed invariant mass and a larger
	reconstructed center-of-mass momentum due to the imbalance. }
	\label{fig:SigICBhabha2D}
\end{figure}

\begin{figure}
	\centering
		\includegraphics[width=\columnwidth]{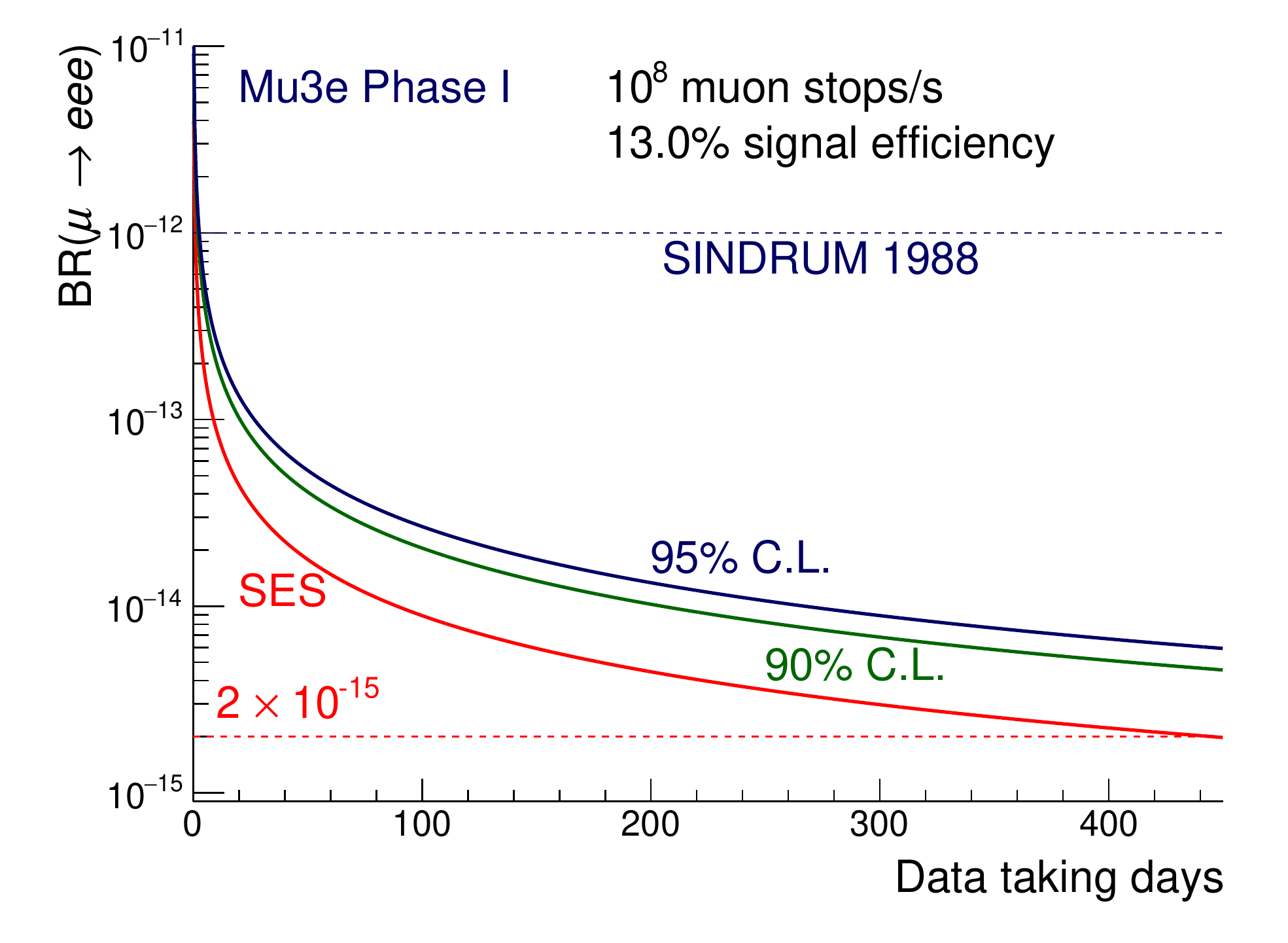}
	\caption{Single event sensitivity (SES) and the corresponding 90\% and 95\% C.L.
	upper limits versus data taking days for the phase~I Mu3e detector.}
	\label{fig:sensistivity}
\end{figure}

The simulated invariant mass distribution is shown in \autoref{fig:SigICBhabha},
the CMS momentum distribution is shown in \autoref{fig:SigICBhabhaPtot},
and the correlation of invariant mass and CMS momentum is shown in \autoref{fig:SigICBhabha2D}.

A study for defining an optimal signal region would require larger background samples and has not been done yet.
In the following, results are presented based on simple cut based signal region definitions.
If a wide signal box in the mass range from \SI{103}{MeV/c^2} to
\SI{110}{MeV/c^2} in reconstructed mass and with $p_{cms} < \SI{4}{MeV/c}$ is chosen,
$0.57 \pm 0.04$ internal conversion events and $1.9 \pm 1.4$ Michel plus Bhabha events are
expected in the signal region for \num{2.5e15} muon stops. For $m_{rec} > \SI{104}{MeV/c^2}$, 
no simulated Bhabha plus Michel events remain and the expectation for internal conversion
is reduced to $0.068 \pm 0.013$. With the phase~I Mu3e detector we thus 
have the capability of suppressing both accidental backgrounds and internal conversion 
events to a level that allows for a background free measurement for at least 
\num{2.5e15} muon stops.  This corresponds to about \num{300}~days of continuous 
running at \num{1e8} stops per second.
The sensitivity versus running time is shown in \autoref{fig:sensistivity}.



\titleformat{\chapter}[display]{\sc}{}{20.0pt}{\includegraphics[width=0.2\textwidth]{common/figures/logo_drawing} \Huge}

\appendix

\chapter{Acknowledgements}
\label{sec:Acknowledgements}

We gratefully acknowledge the beamtimes provided by the following facilities: 
test beam facility at DESY Hamburg (Germany), a member of the Helmholtz 
Association (HGF),
$\pi$M1 at Paul Scherrer Institut, Villigen (Switzerland), 
PS and SPS at CERN, Geneva, (Switzerland),
and MAMI A2 and X1 
at Institut f\"ur Kernphysik at the JGU Mainz (Germany).

We gratefully acknowledge important contributions to the experiment, made by staff in the mechanical and electrical workshops and cleanroom facilities in all collaborating institutes.

The Heidelberg groups acknowledge the support by 
   the German Research Foundation (DFG) funded Research Training Groups HighRR (GK
   2058) and ``Particle Physics beyond the Standard Model'' (GK 1994), 
   by the EU International Training Network PicoSec (grant no. PITN-GA-2011-289355-PicoSEC-MCNet),
   by the International Max Planck Research School for Precision Tests of Fundamental Symmetries (IMPRS-PTFS)
   and the Heinz-G\"otze-Stiftung.
   
\begin{sloppypar}
The work of the Mainz group has also been supported by the Cluster of Excellence ``Precision Physics,
Fundamental Interactions, and Structure of Matter'' (PRISMA EXC 1098 and
PRISMA+ EXC 2118/1) funded by the German Research Foundation (DFG) within the
German Excellence Strategy (Project ID 39083149). We acknowledge the contributions
of the PRISMA Detector laboratory to the development of the DC-DC converters.
\end{sloppypar}





The Swiss institutes acknowledge the funding support from
the Swiss National Science Foundation grants no.
200021\_137738, 200021\_165568, 200021\_172519, 200021\_182031 and 20020\_172706.

The Particle Physics Department (DPNC) of the University of Geneva gratefully acknowledges support from
from the Ernest Boninchi Foundation in Geneva.

\begin{sloppypar}
The UK institutes thank the Science and Technology Facilities Council 
for funding their work through the Large Projects scheme, under grant numbers:
ST/P00282X/1, ST/P002765/1, ST/P002730/1, ST/P002870/1.
\end{sloppypar}

N.~Berger, S.~Shrestha, A.~Kozlinskiy, A.-K.~Perrevoort, D.~vom~Bruch, 
Q.~H.~Huang, U.~Hartenstein and F.~Wauters 
thank the DFG
for funding their work on the Mu3e experiment through the Emmy Noether programme.

F.~Meier Aeschbacher, I.~Peri\'c, A.~Sch\"oning and D.~Wiedner thank the DFG
for funding their work under grant no. SCHO 1443/2-1.

G.~Hesketh gratefully acknowledges the support of the Royal Society through
grant numbers UF140598 and RGF\textbackslash EA\textbackslash 180081.






\addcontentsline{toc}{chapter}{Bibliography}
\small

\textbf{Note:} Theses and publicly available documentation is available on the project website at \url{https://www.psi.ch/de/mu3e/documents}.
\newpage
\end{document}